\documentclass{aa}  

\usepackage{graphicx}

\usepackage[]{hyperref}
\usepackage{amsmath}	
\usepackage{amssymb}	

\usepackage[below]{placeins}

\usepackage[varg]{txfonts}

\usepackage[usenames]{color} 
\usepackage{color}

\newcommand{\im}{i_\mathrm{m}}

\begin{document} 

\title{Then and now: A new look at the eclipse timing variations of hierarchical triple star candidates in the primordial \textit{Kepler}-field, revisited by TESS}
\titlerunning{ETV study of \textit{Kepler} triples reobserved with TESS}
\author{T. Borkovits\inst{1,2,3}\fnmsep\thanks{E-mail: borko@bajaobs.hu}
\and
 S. A. Rappaport\inst{2,4}
 \and
  T. Mitnyan\inst{1,2}
 \and
  I. B. B\'\i r\'o\inst{1,2}
 \and
  I. Cs\'anyi\inst{1} 
 \and
  E. Forg\'acs-Dajka\inst{1,2,5}
 \and
  A. Forr\'o\inst{3,6}
 \and 
 T. Hajdu\inst{3,6,7} 
 \and
 B. Seli\inst{3}
 \and
 J. Sztakovics\inst{7}
  \and
 A. G\"obly\"os\inst{8}
 \and
  A. P\'al\inst{3}
  }

\institute{Baja Astronomical Observatory of University of Szeged, H-6500 Baja, Szegedi \'ut, Kt. 766, Hungary
   \and
   HUN-REN-SZTE Stellar Astrophysics Research Group, H-6500 Baja, Szegedi \'ut, Kt.
766, Hungary
\and
Konkoly Observatory,HUN-REN Research Centre for Astronomy and Earth Sciences, H-1121 Budapest, Konkoly Thege Mikl\'os \'ut 15-17, Hungary
\and
Department of Physics, Kavli Institute for Astrophysics and Space Research, M.I.T., Cambridge, MA 02139, USA
\and
 Department of Astronomy, Institute of Physics and Astronomy, ELTE E\"otv\"os Lor\'and University, H-1117 Budapest, P\'azm\'any P\'eter s\'et\'any 1/A, Hungary
\and
 MTA CSFK Lend\"ulet Near-Field Cosmology Research Group 
\and
Eszterh\'azy K\'aroly Catholic University, Department of Physics, H-3300 Eszterh\'azy t\'er 1, Eger, Hungary
\and
 Institute of Physics, University of Szeged, H-6720 Szeged, D\'om t\'er 9., Hungary
}

   \date{Received ...; accepted ...}

 
  \abstract
   {A former analysis of eclipse timing variation (ETV) curves of eclipsing binaries (EB) observed by the \textit{Kepler} spacecraft during its $\sim4$-yr-long prime mission has led to the discovery and characterisation of 221 hierarchical triple star system with different confidence levels. Although the prime \textit{Kepler}-mission ended in 2013 (a little more than a decade ago), the TESS space telescope has revisited the original \textit{Kepler}-field on several occasions in between 2019 and 2024, thereby extending the time-base of high-precision eclipse timing observations for a substantially longer interval.}
       { In this paper we reanalyze the extended ETV curves of the formerly identified triple star candidates and many other \textit{Kepler} EBs. Besides the confirmations of the former findings and/or the improvements of the triple systems' orbital properties, the extended time-base allows us to identify several new, longer outer period triple systems, and it also makes possible a more detailed study of the dynamical perturbations in the tightest triple stars.}
       {We extend the ETV curves of the \textit{Kepler} triples with those mid-eclipse times which can be deduced from the TESS observations and, moreover, from targeted ground-based follow up observations for a number of the objects. In general, we use the same methods that were applied for the older studies, which are described in the literature. Due to the lower quality of the TESS observations, however, for the fainter systems we average light curves of the EBs for 5-20 consecutive cycles, and thereby calculate `normal' minima from these averaged light curves.}
       { In conclusion, we identified 243 hierarchical triple star candidates in the \textit{Kepler} sample. This sample strongly overlaps our former, nine-year-old sample, confirming the older results, or providing new solutions for 193 systems of the 2016 sample. For the remaining 28 hierarchical triple candidates of that former study, we were unable to find new solutions either because of the disappearance of the eclipses due to orbital plane precession, or due to instrumental reasons. On the other hand, due to the extended time series, we were able to identify 50 new, longer period triple star candidates, as well. We briefly discuss the main properties of each individual system and present statistical studies of the results, as well.}
   {}


   \maketitle
   
\nolinenumbers
   
\section{Introduction}
\label{sec:intro}

Hierarchical triple (and multiple) stellar systems are quite frequent objects both in our local and more distant neighbourhoods. For example, even the closest star (apart from, of course, our Sun) is likely a member of a wide triple stellar system. There are, however, much more compact and as well as tighter\footnote{Compactness' and `tightness' refer to two different properties of stellar triples and multiples. Compactness usually refers to cases where the characteristic size of the entire triple system does not exceed, let's say, a few AUs.  A triple system is said to be `tight' if the ratio of the instantaneous separation of the tertiary star from the inner binary members to the separation of the two inner binary stars is low enough to produce non-negligible gravitational perturbations in the motions of the three stars over and above the two (inner and outer) purely Keplerian two-body motions.} stellar triples than $\alpha$~Centauri. The most compact known stellar triples have dimensions that are comparable to the size of Mercury's orbit around our Sun. Currently, the shortest known outer period triple star system, TIC~290061484 has an outer period of only $P_2=24\fd5$ \citep{kostovetal24}, and is, incidentally, a subsystem of an at least a quadruple system.
 
The compact, or short outer period, triple systems carry exceptional interest and importance in regard to several different aspects of stellar astrophysics. These include different formation scenarios than for single and close binary stars (which scenarios can be substantially affected and modified by the presence of a third stellar mass object), or the final states of their stellar evolution, since triplicity will lead to even more varied and exotic stellar evolutionary paths than for binarity. \citep[For short summaries of these aspects of triple and multiple star studies, see, e.g., the reviews of][and further references therein]{tokovinin21,toonenetal20,toonenetal22,borkovits22}.

The two most effective traditional ways to discover compact or, what is nearly equivalent, short outer-period triple stars, are the radial velocity (RV) analysis of spectroscopic binaries, and the eclipse timing variation (ETV) study of eclipsing binary (EB) stars \citep[cf.][]{tokovinin14a}. Before the advent of the small, planet-hunter space telescopes, such as CoRoT \citep{auvergneetal09}, \textit{Kepler} \citep{boruckietal10}, TESS \citep{rickeretal15}, the vast majority of the shortest outer period triples were discovered spectroscopically. This is true for the very first known hierarchical triple stellar system, Algol \citep{belopolski906,belopolski911,curtiss908}, as well as for the long-known triple system $\lambda$~Tau \citep{ebbighausenstruve956} which, from 1956 until very recently, had the shortest known outer period.

Interestingly, both Algol and $\lambda$~Tau are long-known and (photometrically) frequently observed EB stars. 
Therefore, the presence of a third, distant companion, could have been discovered through periodic, quasi-sinusoidal shifts in their eclipse timings caused by the so-called light-travel-time effect (LTTE), i.e., the simple R{\o}mer-delay, which is the direct consequence of the varying distance of the EB from the Earth due to its revolution around the centre of mass of the entire triple system\footnote{Note, in the case of Algol, \citet{chandler892} already claimed in the $19^{th}$ century that some `inequalities' in the eclipsing periods of Algol might occur due to the effect of a third star. He proposed, however, this explanation for some other, longer timescale variations in the eclipse timings of Algol and not for the much later spectroscopically discovered $P_2=680$\,d-period third star.}.

The reason, however, why this ETV method was ineffective for the shortest outer period systems before the space-telescope era can readily be understood.  It arises primarily from the fact that the amplitude of a LTTE-dominated ETV curve scales as $P_2^{2/3}$ (see Sect.~\ref{sec:LTTE}) and generally remains below the detection limit of ground-based timing studies for outer periods shorter than $1-2$ years \citep[see Fig. 1. of][]{borkovits22}.  Moreover, the smaller amplitude LTTE variations can be hidden or suppressed by complex, sometimes erratic, and other poorly understood variations in the eclipse times (to be discussed in Sect.~\ref{sec:polynomial}). Therefore, it is not at all surprising that before the observations of \textit{Kepler} space telescope, IU~Aur, a semi-detached (SD) binary, formed by two bright, massive, B-type stars was the only EB where a third stellar companion with an outer period shorter than one year was discovered through LTTE, manifested in its ETVs \citep{mayer983,mayer990}. All the other (not too numerous) hierarchical triple star systems with outer periods shorter than one year, at that time, were discovered spectroscopically, through RV analyses.  Though, after the conclusive spectroscopic detection of third bodies in a few EBs, subsequent careful ETV analyses have also led to the identification of the LTTE, for example in the cases of DM~Per \citep{vanhammewilson07} and VW~LMi \citep{pribullaetal08}. In this context, it was noted by \citet{tokovinin14b} that the rarity of the most compact hierarchical triple star systems is most obvious in the volume-limited sample of FG-type stars.

This situation has changed dramatically in the last one and half decades, due to the above mentioned small, photometric space telescopes. They have led to unprecedentedly accurate, homogeneous, and quasi-continuous eclipse monitoring of millions of EBs. ETV curves formed from such observations, especially in the case of the nearly 4-year-long prime \textit{Kepler}-mission, were quite ideal for identifying hundreds of such non-linear ETVs, which exhibit low amplitude and short period LTTE signals of close, third stellar (or even, in a few cases, substellar) mass companions. And, besides these short-period, so-called LTTE triples, the signals of dozens of dynamically active, tight, and relatively compact triple stellar systems also became detectable, for the first time, through complex ETV analyses. 

Interestingly, and perhaps somewhat unexpectedly, the \textit{Kepler} spacecraft has introduced a new method for finding compact triple (and multiple) stars with the discovery of the first dozen triply eclipsing triple stars. In these systems the outer orbit is seen almost exactly edge-on and, hence, the third star periodically eclipses and/or is eclipsed by the inner binary members, producing variously shaped, characteristic extra fadings during their outer orbital revolution. (Note, the inner binary is not necessary itself eclipsing, however, most of the known, so-called triply eclipsing triples are actually triply eclipsers. i.e., the inner binary also exhibits regular, two-body eclipses.) As the eclipse probability decreases dramatically with the orbital separation, it is hardly surprising that these triply eclipsing systems are generally very compact. The above mentioned, shortest outer period triple, \object{TIC 290061484}, indeed, was also found through its extra, third-body eclipses. We also point out that the vast majority of the currently detected $\sim$100 triply eclipsing systems were discovered by TESS.  To our knowledge, only the discovery and analysis of 34 of these triply eclipsing systems have thus far been published \citep{borkovitsetal20b,borkovitsetal22a,mitnyanetal20,mitnyanetal24,rappaportetal22,rappaportetal23,rappaportetal24,powelletal22,czavalingaetal23a,kostovetal24}.

Turning back now to the stars in the prime \textit{Kepler} field, these have been intensively studied over the past ten years. Besides a copious comprehensive systematic set of survey papers, in-depth analyses of several particular interesting stars and star systems have been published. Here we mention only a few of these studies, in particular,  those which we found to be relevant and important from the perspective of the present work. 

First, spectroscopic surveys have monitored the vast majority of the \textit{Kepler} stars, including many of the currently studied triple-star candidate EBs. These studies have resulted in masses, temperatures, and luminosity classifications for a number of the current EBs. Here we refer, e.g., to the LAMOST \citep{qianetal18}, APOGEE \citep{jonssonetal20} and SDSS-HET \citep{mahadevanetal19} surveys. These surveys, however, in many cases, do not give binary solutions and, therefore, we decided not to use the given stellar masses as an estimation of the total mass of the inner binary.  But, as we will mention later in the text, we used other kinds of mass estimations or, in the few cases of SB2 systems, the results of dedicated spectroscopic studies were applied to obtain realistic binary masses. In this latter context, we refer especially to a series of papers by \citet{helminiaketal16,helminiaketal17a,helminiaketal17b,helminiaketal19}, in which the authors systematically observed and analyzed bright, detached, spectroscopically  single- double- and triple-lined EBs with the HIDES spectrograph. And, we also note the work of \citet{matsonetal17} who obtained radial velocity (RV) data for 41 \textit{Kepler} EBs, including several systems from our sample.

Besides the spectroscopic surveys, our present analyses also intensively utilize the results of \citet{kjurkchievaetal17} and \citet{windemuthetal19}, who  systematically analyzed most of the \textit{Kepler}-EB light curves, providing us with reliable input parameters, e.g., for the inner eccentricities and arguments of periastrons for eccentric EBs.

There have also been a few former ETV surveys, whose purposes were similar to that of the current work. One of the two most detailed studies was published by \citet{conroyetal14}, who investigated the ETVs of all \textit{Kepler} short-period ($P_1<3$\,d) EBs, and detected 236 hierarchical triple star candidates at different confidence levels. The other extensive work was the paper of \citet{borkovitsetal16} who extended the ETV studies to also include the longer period EBs. In that study, in contrast to the previous work of \citet{conroyetal14}, and similar to the current paper, the dynamical third-body effects (DE) were also considered. These authors identified 221 hierarchical triple star candidates, which largely cover the findings of \citet{conroyetal14}.

Besides these two larger surveys, smaller numbers of the ETVs of \textit{Kepler} potential triple star candidate EBs were investigated in several other papers, such as \citet{steffenetal11,giesetal12,giesetal15,rappaportetal13,borkovitsetal15,zascheetal15}, etc. Finally, we note the recent paper of \citet{inacioetal24}, who carried out a new ETV study of those short-period \textit{Kepler} EBs that were previously investigated in the \citet{conroyetal14} paper and, moreover, had available at least one sector of 2-min cadence TESS observations (up to Sector 55). Due to this second rather restrictive criterion (at least in our opinion), they sampled only 253 EBs, of which they finally found third-body LTTE solutions for only 75 systems.

As was mentioned above, our former comprehensive analysis of ETV curves of EBs observed by the \textit{Kepler} spacecraft during its $\sim4$-yr-long prime mission \citep{borkovitsetal16} has led to the discovery and characterization of 221 hierarchical triple star systems with different confidence levels for the outer orbital determinations. Although the prime \textit{Kepler}-mission ended in 2013 (more than ten years ago), the TESS space telescope has revisited the original \textit{Kepler}-field on several occasions between 2019 and 2024, allowing us to extend the time-base of high precision eclipse time observations for a substantially longer interval. Therefore, we believe the time is right to reanalyze the extended ETV curves of the formerly identified triple-star candidates and many other \textit{Kepler} EBs. Besides the confirmations of the former findings and/or the improvements of the triple systems' orbital properties, the extended time-base allows us to identify several new, longer outer period triple systems on the one hand, and also makes possible a more detailed study of the dynamical perturbations in the tightest triple stars, on the other hand.

In the forthcoming Section~\ref{sec:analysis} we briefly describe the theoretical background of our analyses and concentrate mainly on those issues which arise due to the substantially extended length of our dataset relative to the former \textit{Kepler} data. Then we discuss the questions of the selection of the systems and some specific technical issues concerning the handling of ETV data from the less accurate and sparse TESS data in Sect.~\ref{sec:Selection}. Our results are tabulated and thoroughly discussed both individually and statistically in Sect.~\ref{sec:Results}. Finally, in Appendix~\ref{app:Notesonindividualsystems} we discuss all of our investigated systems individually, in a few sentences each. (In these brief notes, we mention all the remaining papers which we used during our analyses.) Finally, in Appendix~\ref{app:ETVcurves} we present all the 243 ETV curves together with the accepted analytic solutions.

\section{Mathematical basis of our study}
\label{sec:analysis}

In this work, as was mentioned in the Introduction, we utilized, almost exclusively, the analyses of ETV curves formed largely from \textit{Kepler} and TESS spacecraft eclipse timing data to identify triple star candidates amongst EBs. These systems were observed sequentially by the two above mentioned space telescopes, separated by an interval of more than six years. Importantly, we used these ETV curves to determine the parameters of the third-body orbits for these candidates wherever possible. Analysis of ETV curves or, as they are also called `observed minus calculated' diagrams or `$O-C$' curves, is a powerful tool to detect, quantify, and study, all kinds of period variations in stars. This includes EBs and all manner of variable stars, such as  pulsating variables which exhibit more or less strictly periodic light variations \citep[see, e.g.,][and further articles in the same conference volume]{sterken05}. This is so, because the slope of the curve\footnote{Note, ETVs, at least in their traditional forms, are discretized samplings of continuous functions and, hence, strictly speaking, we cannot really talk about the slope of the curve at a given instant.  But, in most cases, and this is the situation in our current analysis as well, the slope of that continuous function with which the ETV can be described, or at least approximated, is what is meant by this somewhat imprecise term `slope of the ETV'.} at any given point gives the actual period of the measured phenomena (being either the eclipses in an EB or the pulsating maxima in a classic Cepheid or RR~Lyrae star\footnote{This latter case explains why the word `eclipse' or, in the case of transiting exoplanets, `transit', in the acronym E/TTV seems to be slightly restrictive in contrast to the former term $O-C$, though, naturally, the usage of this latter one also has several disadvantages, not necessary to mention here.}). As a direct consequence, since the slope varies only when the ETV curve has a non-zero curvature, any nonlinearity in the ETV curve reflects some period variation.

In what follows, restricting ourselves almost exclusively to EBs (but also allowing for ellipsoidal variables, `ELV's), these period variations traditionally divide into two large groups of (i) real (i.e., physical or intrinsic) and (ii) apparent period variations (e.g., LTTE). Due to experiences with the unprecedentedly accurate and quasi-continuous photometry of the dedicated planet-hunter space telescopes, we would also add to this division a third category, namely (iii) the `spurious' period variations. Regarding this last term we mean those detected non-linearities in the ETVs which occur due to certain types of light curve distortions which influence the determination of the mid-minima times of the eclipses instead of being caused by any real shift in the mid-minima times. Examples of these types of spurious non-linear ETVs (and, hence, implied period variations) are the effects caused by star spots \citep{kalimerisetal02,tranetal13,balajietal15}, as well as the distorting effects of stellar oscillations on the light curves \citep{borkovitsetal14}, or even the effects of aliases which may occur from a beating between the eclipsing and sampling periods \citep[see, e.g.,][for further details]{borkovitsetal15,borkovitsetal16}.

Returning to the two traditional categories, under the heading of real or physical period variations are meant those cases where the eclipsing period changes are due to actual variation(s) of the anomalistic orbital period. Several non-periodic and/or quasi-periodic phenomena can result from different forms of mass exchange, mass loss, magnetic effects (including, e.g., the so-called Applegate-mechanism, \citealp{applegate992} and its different, newer variants, like, e.~g., \citealp{volschowetal18}). Apart from this last, theoretically periodic effect, which is considered to be present in most of the binaries with late-type stars (such as the majority of the W UMa-type overcontact binaries) \citep[see, e.g.][]{lanzarodono999,szalaietal07}, these variations, especially the mass-transfer related ones, work on much longer timescales than what is available from an observational point of view. Therefore, it is a good approximation to consider such long-term period variations, whether stochastic or quasi-periodic in reality, to be linear (i.e. constant) during year to century timescales. And, it is well-known from the theory of ETV (or, $O-C$) curves, that a constant period variation manifests itself as a quadratic variation in the ETV curve\footnote{For a more realistic analytical description of the ETVs affected by some of these phenomena, see \citet{nanourisetal11,nanourisetal15}}.

From the point of view of the current study, these effects are interesting only in the sense that, as they are clearly present in a large portion of our sample, we had to model them, at least mathematically (i.e., to describe the non-third-body-caused variations of the ETV with some mathematical function irrespective of the physical origin[s] of the variations), as we will soon briefly discuss in Sect.~\ref{sec:polynomial}. There is, however, another class of the real or physical period variations which traditionally is only rarely considered, but it is definitely relevant in our study.  This is the class of the true (anomalistic) period variations caused by gravitational perturbations. As mentioned already in the Introduction, when the third stellar companion is close enough to the inner (eclipsing) binary (i.e., by definition, the triple can be considered to be tight), the gravitational perturbations of the third component can significantly influence the motions of the two components of the EB, and these perturbations manifest themselves in periodic or, quasi-periodic variations of the mid-eclipse times on different timescales. We discuss the modeling of these effects soon in Sects.~\ref{sec:dyn}--\ref{sec:AM}.

The other large category of the classic division is the class of the `apparent' period variations. These are those phenomena where the anomalistic period of the binary remains constant but, despite this, the eclipsing period is subject to changes. Traditionally the LTTE or, R\o mer-delay and the apsidal motion (AM) are classified here. Both of them are strictly periodic phenomena. The first case occurs due to the finite speed of light, when the EB revolves around a centre of mass of a triple (or, multiple) system and, hence, its distance from the observer varies periodically (assuming that the outer orbit is not seen from its pole). This is a very well known phenomenon which we discuss further very briefly in the forthcoming Sect.~\ref{sec:LTTE}. The second effect in this category is the AM caused apparent period variation which occurs in every sufficiently close, eccentric binary system, irrespective of whether it has any third companion, or not. We elaborate further on AM (concentrating mainly on its manifestation and mathematical description in ETVs) in Sect.~\ref{sec:AM}.

In conclusion, taking into account all of the above discussions in this work, we modeled the ETVs in a very similar manner as described in \citet{borkovitsetal15,borkovitsetal16}. The full analytic model takes the form:
\begin{equation}
\Delta=\sum_{\mathrm{i}=0}^{3}c_{\mathrm{i}}E^{\mathrm{i}}+[\Delta_{\mathrm{LTTE}}+\Delta_{\mathrm{dyn}}+\Delta_{\mathrm{apse}}]_{0}^{E},
\end{equation}
where the polynomial term contains corrections for the calculated eclipse times (in case of $i=0,1$), constant-rate and, linearly varying period variations (in case of $i=2,3$). The constant and linear ($i=0,1$) terms were used all over in our analyses. In contrast to this, the $i=2,3$ polynomial terms were fitted only in a few cases, where the presence of a quadratic or even more complicated additional variations in the ETV curves were evident. (We again return to this question in Sect.~\ref{sec:polynomial}).

Moreover, the ($\Delta_\mathrm{LTTE,dyn,apse}$) terms represent the contributions of LTTE, $P_2$-timescale dynamical perturbations and AM, respectively. Detailed descriptions of these terms can be found in \citet{borkovitsetal15,borkovitsetal16}. Here, for consistency, and similar to the recent paper of \citet{mitnyanetal24} which applies practically the same methodology to identify and quantify hierarchical triples amongst the EBs in the Northern Continuous Viewing Zone of TESS, we provide only very brief descriptions and additional remarks regarding the application of these effects.

\subsection{Light-travel time effect (LTTE)}
\label{sec:LTTE}

We took into account the LTTE contribution in the form
\begin{equation}
\Delta_\mathrm{LTTE}=-\frac{a_\mathrm{AB}\sin i_2}{c}\frac{\left(1-e_2\right)^2\sin(v_2+\omega_2)}{(1+e_2\cos v_2)},
\label{Eq:LTTEdef}
\end{equation}
where the negative sign on the right hand side arises from the fact that the argument of pericenter ($\omega_2$) refers to the tertiary component, which differs by $180\degr$ from the argument of pericenter of the orbit of the center of mass of the inner EB. 

With the use of eccentric anomalies instead of true anomalies, one can easily show that, as far as the orbital elements of the outer orbit remain constant, LTTE can be described as a pure sine (with eccentric anomaly in its argument; see, e.~g., eqns.~(4)-(8) of \citealt{borkovitsetal16}), of which the amplitude is
\begin{eqnarray}
\mathcal{A}_\mathrm{LTTE}&=&\frac{m_\mathrm{C}}{m_\mathrm{ABC}}\frac{a_2\sin i_2}{c}\sqrt{1-e_2^2\cos^2\omega_2}  \\
&=&\left(\frac{G}{4\pi^2}\right)^{1/3}f^{1/3}(m_\mathrm{C})\frac{P_2^{2/3}}{c}\sqrt{1-e_2^2\cos^2\omega_2}, \nonumber
\label{Eq:A_LTTE}
\end{eqnarray}
where, by the analogy of the mass function used for single-lined spectroscopic binaries (SB1) it is usual to introduce the same mass function as
\begin{equation}
f(m_\mathrm{C})=\frac{m_\mathrm{C}^3\sin^3i_2}{m_\mathrm{ABC}^2}.
\end{equation}

We fit the LTTE term of the ETVs of all of our candidates. By analogy to an SB1 system, a successful LTTE solution carries exactly the same information about the LTTE orbit (which is in general the wide or outer orbit in a hierarchical triple star system), as the radial velocity (RV) solution in an SB1 binary. It follows that we can give a robust minimum mass for the tertiary only in the case when the total mass of the inner binary is known. Despite this, we make efforts to estimate the possible minimum tertiary masses in all of our pure LTTE solution systems. When the total mass of the EB is not known from some dedicated RV studies (which are available only for double lined, i.~e., SB2 spectroscopic binary systems), we either (reasonably) estimate the total mass of the inner pair to be $m_\mathrm{AB}=2.0\,\mathrm{M}_\sun$ (for detached and semi-detached binaries) or, apply the empirical mass-period relations of \citet{gazeasstepien08} to obtain such a reliable total mass estimation in the case of W~UMa-type binaries.

\subsection{$P_2$-timescale dynamical perturbations}
\label{sec:dyn}

In contrast to the LTTE terms, which were fitted by default during our analyses, the $P_2$-timescale gravitational, or dynamical, effects (DE) were taken into account only when our previous estimation (see below) suggested that DE terms may give a substantial contribution to the ETV curve.  According to our knowledge, the currently available, most elaborate theory of the $P_2-$ (medium-) timescale perturbations and their manifestations in the ETV of a tight EB was presented and discussed in \citet{borkovitsetal15}. Excerpts of these formulae were published e.g., in \citet{hajduetal17} and, very recently, in \citet{mitnyanetal24} and, therefore, here we do not feel it necessary to repeat any parts of these quite long equations. Here we restrict ourselves only to providing two formulae, from which at least the likely magnitude of the amplitude of these non-strictly sinusoidal terms can be estimated. The magnitude of this DE contribution to the ETV curve can be, in general, estimated as
\begin{equation}
\mathcal{A}_\mathrm{dyn}=\frac{1}{2\pi}\frac{m_\mathrm{C}}{m_\mathrm{ABC}}\frac{P_1^2}{P_2}\left(1-e_2^2\right)^{-3/2},
\label{Eq:A_dyn}
\end{equation}
while, according to \citet{rappaportetal13}, in case of nearly coplanar inner and outer orbits, the expression below gives a better estimation:
\begin{equation}
\mathcal{A}^{\mathrm{coplanar}}_\mathrm{dyn}=\frac{3}{2\pi}\frac{m_\mathrm{C}}{m_\mathrm{ABC}}\frac{P_1^2}{P_2}\left(1-e_2^2\right)^{-3/2}e_2.
\label{Eq:A_dyn^cop}
\end{equation}
(Note, this latter form indicates, which is confirmed by the more detailed analytic formulae of \citet{borkovitsetal15}, and is also proven by numeric integrations, that in case of circular inner and outer orbits ($e_{1,2}=0$), and coplanar configuration ($\sin\im=0$), the $P_2$-timescale dynamical perturbations (and, hence, their contribution to the ETV curve), practically disappear.)

Here we mention that, in contrast to the pure LTTE-dominated ETVs, in the case of a significant DE contribution, the individual masses (or, more strictly, the total mass of the inner EB ($m_\mathrm{AB}$) and the individual mass of the tertiary ($m_\mathrm{C}$) can also be calculated, at least, in theory. This is so because (as was discussed in greater detail in \citealt{borkovitsetal15}), the DE analysis can yield both the outer mass ratio of $m_{\rm C}/m_{\rm ABC}$ and the outer inclination angle, $i_2$, while the amplitude of the LTTE term contains a product of this outer mass ratio squared and $m_{\rm C} \sin^3 i_2$ (see Eqs. [4]--[6]). Therefore, the values of both $m_{\rm C} $ and $m_{\rm AB}$ can be separately derived from the results. Note, as was also discussed in \citet{borkovitsetal15}, due to the substantial degeneracies of the LTTE and DE terms, especially in their amplitudes, the masses obtained in this way should be considered only with some caution.

As mentioned above, the DE terms were not included automatically in our ETV analyses. We switched on these formulae mainly in those systems where they had already been included in the analyses in \citet{borkovitsetal15,borkovitsetal16}. (These are mostly the tightest, eccentric and/or inclined triples, where the shapes of the ETV curves already reveal at a first glance that DE are important.) For other systems, however, the necessity of the DE terms was often far from evident. For this reason, in the case of all the pure LTTE-fitted ETVs the software automatically calculates the nominal amplitude ratio of $\mathcal{A}_\mathrm{dyn}/\mathcal{A}_\mathrm{LTTE}$ (Eq.~\ref{Eq:A_dyn}) and of its coplanar counterpart (Eq.~\ref{Eq:A_dyn^cop}) (adopting minimum masses according to the assumptions introduced above, in Sect.~\ref{sec:LTTE}). Then, for systems having a preliminary outer period of $P_2\lesssim1000$\,d, we checked the ratio calculated for $\mathcal{A}^{\mathrm{coplanar}}_\mathrm{dyn}$, while the non-coplanar ratio was checked for the wider triples.  When it was found that the given ratio is higher than $\approx0.3$, we repeated the fitting process, switching on DE.

\subsection{Apsidal motion (AM)}
\label{sec:AM}

Both LTTE (Sect.~\ref{sec:LTTE}) and DE (Sect.~\ref{sec:dyn}) have the characteristic periods of $P_2$, i.e., the orbital period of the wide orbit or, in other words, of the tertiary's orbital motion in a hierarchical triple stellar system.  As we are primarily interested in compact, i.e., short outer period triples this period or, `timescale', can be as short as a month to a few years. There are, however, other much longer timescale variations (either periodic or non-periodic) in our targets, which also manifest themselves as additional non-linearities in the ETVs of our investigated systems. For some of the systems under consideration, these longer-term non-linearities were already important  on the $\sim4$-year-long timescale of the prime \textit{Kepler}-observations.  By contrast, others did not manifest themselves at all (or only by a very small amount) on the \textit{Kepler} ETV curves and, hence, they could have been safely neglected in the case of the \textit{Kepler}-only ETV studies, as was done in \citet{rappaportetal13,conroyetal14} and, in its greatest part, in \citet{borkovitsetal16}.

The most typical, well-known, and in most cases strictly periodic, longer timescale variation is due to AM. This effect, which occurs in eccentric EBs, may have different origins. There are three such origins: (i) general relativistic, (ii) classical tidal (arising due to the non-spherical mass distribution of the tidally deformed bodies) and, (iii) dynamical (third-body perturbed) AM \citep[see, e.g.,][for a short review]{borkovitsetal19a}. Irrespective of the origin of this phenomenon, its effect on the ETV curves  can be described mathematically as follows:
\begin{eqnarray}
\Delta_\mathrm{apse}&=&\frac{P_1}{2\pi}\left[2\arctan\left(\frac{\pm e_1\cos\omega_1}{1+\sqrt{1-e_1^2}\mp e_1\sin\omega_1}\right)\right. \nonumber \\
&&\left.\pm\sqrt{1-e_1^2}\frac{e_1\cos\omega_1}{1\mp e_1\sin\omega_1}\right]. 
\label{Eq:apse-def}
\end{eqnarray}
(The more widely known form of this equation, given in trigonometric series of $\omega_1$ and, taking into account the slight inclination dependence, too, can be found e.g. in \citealt{gimenezgarcia83}.)

In general, in the case of such tight eccentric triples, where the $P_2$-timescale dynamical effects are significant (see above, in Sect.~\ref{sec:dyn}), the dynamically forced apsidal motion substantially dominates over the other tidal contributions. As was already shown in \citet{borkovitsetal15,borkovitsetal16}, for the majority of the currently investigated triples with eccentric inner binaries, this is the true scenario for the AM. Therefore, instead of fitting a (linear) apsidal advance rate ($\Delta\omega_1$), hidden in Eq.~\ref{Eq:apse-def}, as an additional free parameter, we calculated the inherent current values of $\omega_1$ (and also of $e_1$) for each moment during the fitting process, with the use of the perturbation equations, and in the manner described in \citet{borkovitsetal15}.

One should keep in mind, however, that for calculating the period of the dynamically forced apsidal motion, as well as modeling some other non-linearities in the apsidal motion rate, our ETV modeling code applies only the lowest, quadrupole-order analytic approximation \citep[for a more detailed description, see][]{borkovitsetal15}. This is true also for the secular (i.e. $P_\mathrm{apse}$-timescale) non-linear variations in some other orbital elements, which are also inherently built into the code. That approximation of the long timescale variations for the short, four-year-long interval of the \textit{Kepler} measurements was perfectly satisfactory, however, in the case of the currently analyzed extended ETV curves, which already cover one and a half decades, some substantial departures may and do occur in the perturbed motions of a few very tight triples. Therefore, in the near future we plan to build into our model higher-order secular perturbative terms as well. At this time, however, this is not the case. Instead, we found, that the departures from the quadrupole-order secular (primarily dynamical apsidal motion) terms, can be nicely modeled, at least mathematically, with the addition of a fictitious quadratic or cubic polynomial or, with a longer period, extra LTTE-like term, as we briefly discuss below.

\subsection{Other ETV terms}
\label{sec:polynomial}

A significant portion of the \textit{Kepler} EB sample and, similarly, the currently investigated systems, too, formed by late-type, magnetically active stars (i.e., masses $\lesssim 1.2$ M$_\odot$ and $T_{\rm eff} \lesssim 6000$ K). For example, $\sim$$30\%$ of the systems in the Villanova \textit{Kepler} EB catalog have an orbital period shorter than 1\,day. These EBs are either overcontact, W~UMa-type (EW) binary systems or semi-detached or detached binaries which are formed by two low-mass stars.  It is a well known fact that the ETV curves of such EBs are far from smooth, and not an easy task to interpret.  In the case of EW systems, it is widely accepted to model the longer term period variations in such EBs with a parabolic term, which is often interpreted as mass transfer amongst the components.  Other interpretations, however, such as angular momentum loss due to a magnetically constrained stellar wind -- AML, and/or, the consequences of so-called thermal relaxation oscillations -- TRO, have also been introduced \citep[see, e.~g.][]{qian01}. Moreover, quasi-periodic variations are widely interpreted as an LTTE orbit, or as a consequence of the formerly mentioned Applegate-effect, or even the combination of these two \citep[see, e.g., amongst all][]{borkovitsetal05,yangetal07}.

Regarding not only overcontact systems, but also other late-type stars such as the former donor stars in Algol systems, the picture becomes more complicated than in the case of the overcontact systems. There are dozens (if not hundreds) of long-known EBs for which the mid-eclipse times were/are observed by several professional, but mostly amateur, astronomers over many decades or, even more than a century. These observations are generally highly inhomogeneous, especially before the CCD era, quite low in quality, and sparse in time.  In spite of this, the ETV curves that can be formed from these data make it quite clear that most of the ETV curves cannot be described simply by a combination of a few simple functions (e.g., a parabola and/or a sine-like, periodic function).  Instead, most of the longer timescale ETV curves suggest erratic or stochastic variations, seemingly sudden jumps in period \citep[see, e.~g.][for a typical and well-documented case]{maxtedetal994}, arcs with different curvatures, and so forth. Several examples of such ETV curves can be found, e.g., in the compilation of \citet{kreineretal01}, and efforts to interpret such variations are also very numerous \citep[see, e.~g.,][]{biermannhall973}.

From our perspective, the presence of such additional period variations is important in the sense that we need to model them in some way in order to mine out more accurate parameters for the ETV variations caused by the third-body. This is one of the reasons why we introduced extra quadratic or cubic polynomial terms, or even a second extra LTTE-term into our fits.  For the purposes of this current study, however, there is no particular importance to the question of what these other ETV variations are, or even if there is any physical reality behind these extra terms that we had to introduce for an accurate modeling of the third-body effects.

There might be, however, another origin for such extra polynomial or, even cyclic (e.g., LTTE) terms. In the case of tight, dynamically active triples, the gravitational third-body perturbations may be non-negligible, and the secular, $P_\mathrm{apse}$-timescale perturbations might also be effective within our $\sim$$1.5$-decade-long observing window.  It may then happen that the higher than quadruple order, and, hence, unmodeled perturbations manifest themselves in such a manner that, in the absence of a satisfactory analytic perturbation model, we have to describe them mathematically by introducing such extra, non-linear terms.

Finally note that, naturally, it may also happen that what we currently model with a quadratic or cubic polynomial, is nothing other than an arc of an LTTE caused by a longer period fourth component. This is another reason why future, targeted follow-up observations would be very important.

\section{System selection and data collection}
\label{sec:Selection}

We primarily concentrated on those 221 hierarchical triple star candidates amongst the \textit{Kepler} EBs, the multiplicity of which we confirmed with either lower or higher confidence in \citet{borkovitsetal16}. Besides these systems, we followed several other similar \textit{Kepler} EBs, which exhibited suspicious non-linear ETVs during the era of the \textit{Kepler}-observations, but where the relatively short, four-year-long dataset did not make it possible to obtain reasonable third-body solutions, at least not at the level of the lowest-confidence group studied in \citet{borkovitsetal16}.  Note that many of these latter systems were included in the collection of \citet{conroyetal14} as EBs with either very long period, or purely parabolic, ETVs.  We have also taken some further EBs with varying eclipse depths, from the list of \citet{kirketal16}, Table~8.\footnote{In this regard, we should note that in the case of a few systems from this latter list, the third-body influenced ETVs are quite clear, and the properties of the third-body's orbit can easily be determined robustly with the exclusive use of the \textit{Kepler}-data. So, these systems should have been listed in the 2016 compilation of \citep{borkovitsetal16}, but at that time we simply overlooked them.} Moreover, we have checked some additional EBs where the presence of a third companion has been reported, or at least hypothetized, in the literature since the end of the \textit{Kepler}-era.  Finally, in parallel with the current project, we have also followed all those eccentric EBs in the prime \textit{Kepler}-field which were thought to exhibit classical tidally forced, and/or relativistically induced, apsidal motion  (Sztakovics et al., in preparation). Then, in a few cases, the new eclipse times determined from the TESS data, revealed unexpected highly non-linear variations, which make these EBs likely triple star candidates instead of classical apsidal motion binaries.  Thus, these were also included in the present investigations.

Our work is mainly based on the \textit{Kepler} and TESS observations. For most of the previously investigated 221 systems, we used the very same \textit{Kepler} ETV curves as in \citet{borkovitsetal16}. For the other systems, the majority of the \textit{Kepler} light curve files were also downloaded from the Villanova website\footnote{\url{http://keplerebs.villanova.edu/}} of the \textit{Kepler} EB catalog, and they were prepared in an identical way to those of the 2016 dataset.  This can be read about in detail in \citet{borkovitsetal16}, Sect.~3.  There are, however, some EBs, where \citet{abdulmasihetal16} found that the true sources of the eclipsing light curves are not the originally designated \textit{Kepler} targets, but rather background objects. They calculated new, non- or less-contaminated light curves for these systems with, naturally, deeper eclipses. These new light curves, together with the list of such false positives are now available at the above mentioned Villanova catalog site. Hence, in a few cases, we downloaded these new light curves, and recalculated the eclipse times from them.  Finally, we also used two more, newer sets of \textit{Kepler} light curves.  First, for a few sources we did our own pixel-by-pixel photometry in the very same manner as was described in \citet{bienasetal21}, while for some others we used light curves which were generated and made publicly available from the \textit{Kepler}-bonus project \citep{keplerbonus}. Independent of the origin of the \textit{Kepler} light curves, for all the studied systems, we calculated or recalculated the mid-eclipse times and, hence, the ETV curves in the very same manner as was described in \citet{borkovitsetal16}.

Regarding the TESS-observations, for most of the investigated systems only full frame images (FFIs) are available.  During the TESS Cycle 2 re-observations of the \textit{Kepler}-field (Sectors 14, 15 and, in a small part, 26) these were sampled in 30-min cadence mode. Then in Cycle 4 (Sectors 53--55) 10-min cadence FFIs are available.  Finally, a smaller section of the original \textit{Kepler} field was also observed in Sector 56 with 200-sec FFI cadence rate. The same 200-sec FFI cadence rate is also available for the recent, Cycle 6 observations of our targets. (These new observations typically were carried out in the 2024 winter sectors 74, 75, and the summer sectors 81, 82; however, a smaller fraction of our target stars were also observed either in Sector 73 or 76 or, later, in 80 or 83.)  Up to Sector 76 we processed the original FFIs using a convolution-based differential photometric pipeline implemented in the {\sc Fitsh} package of \citet{pal12}.  The newest, 2024 summer sectors (80--83) were processed, however, with the use of the publicly available pipeline {\sc Lightkurve} \citep{lightkurve18}. Many of these raw light curves exhibited strong scattered light, which made it necessary to utilize some further detrending.   Because our main purpose was simply the determination of accurate individual mid-eclipse times or ELV minima, we found that simple local polynomial smoothing was fully satisfactory for our purposes. Note also that in a few cases, due to the large pixel size of TESS cameras, the signal of the investigated EB was blended with another EB in the same aperture. In these cases, before determining the mid-eclipse times, we disentangled the signals of the two EBs, e.g., with the application of principal component analysis (PCA).

For a smaller portion of our targets, 2-min cadence TESS photometry is also available for a few sectors.\footnote{Note, in the Cycle 6 sectors 20-sec cadence light curves are also available for some of our targets, however, we did not use those data.} In these cases we downloaded the simple aperture photometry (SAP) light curves directly from the MAST database\footnote{\url{https://mast.stsci.edu/portal/Mashup/Clients/Mast/Portal.html}}. We calculated times of minima and, hence, ETV curves from the TESS light curves (both from 2-min cadence and FFI data) in a very similar manner to what was done formerly for the \textit{Kepler} light curves, and explained in detail in Sect.~3 of \citet{borkovitsetal16}. In this context we refer to the recent paper of \citet{marcadonprsa24}. These authors pointed out with detailed investigations, that the use of 2-min cadence data is superior relative to the FFI-derived ETVs in the accuracy of the LTTE solutions. Naturally, we did not know of their results at the time of our studies, but we had similar experiences and therefore, in the case of any available 2-min cadence observations, we preferred their use in contrast to the FFI data.

In a significant fraction of the newer segments of the ETV curves, based on TESS data, we found that the scatter of the ETV points substantially exceeds  the amplitudes of the \textit{Kepler} ETV-detected cyclic variations.  This is due to either the faintness of the systems or to the shallowness of the eclipses, or  both. Therefore, we found it useful to fold and average 5-20 consecutive EB-cycles, forming in such a manner `normal' light curves, and calculating from these folded, phase-binned, averaged light curves, so called `normal' eclipse times for our analysis. In such a manner, we were able to substantially reduce the scatter in the new ETV sections of many targets.

We also used further, supplemental, ground-based eclipse timing data.  Some of these came from targeted direct follow-up eclipse observations carried out by us with the 50-cm and, later 80-cm RC telescopes of Baja Astronomical Observatory (BAO), Hungary, and the 1-m RCC telescope of Konkoly Observatory, located at its Piszk\'es-tet\H o Mountain Station in Hungary. The instruments and the details of these targeted photometric observations are described e.g., in \citet{borkovitsetal19b,borkovitsetal22a}. We tabulate these new not-yet-published eclipse times in Table~\ref{tab:KICToM}. Moreover, we further collected times of minima determined from targeted eclipse observations from the literature.

Finally note that, similar to our former study \citep{borkovitsetal16}, we determined further mid-minima times from SWASP \citep{pollaccoetal06} survey observations. In most cases, however, these latter data were less useful for our purposes, due to the large scatter of the ETV points determined from these light curves.

At the end of this paper, in Appendix~\ref{app:Notesonindividualsystems} we briefly discuss for each individual system the main parts of the actual ETV analyses.

								      \section{Results}
\label{sec:Results}

In what follows, first we give a brief overview about our results and then we present a statistical summary of our sample.

We obtained new ETV solutions with higher or lower confidence levels for 243 \textit{Kepler} EBs. Similar to the former work of \citet{borkovitsetal16} we divided our solutions into pure LTTE and LTTE+DE systems, and within these two main categories, we grouped our solutions into three confidence levels.\footnote{Categorizing our solutions as \emph{pure} LTTE or LTTE+DE solutions, we do not consider the possible presence of any additional non-linear polynomial or fourth-body terms. Therefore the word `pure' simply indicates the absence of any DE terms in our solutions.} Regarding the pure LTTE systems, we found 63 certain, 45 moderately certain, and 73 uncertain solutions, while from the LTTE+DE systems, we classified 37 certain, 7 moderately certain, and 18 uncertain solutions. We classified those solutions as `certain' where (i) the third-body period does not exceed one third of the entire, usually $\sim5500-5600$\,d-long, dataset, and (ii) the TESS-originated ETV points, which generally have much larger scatter, either confirm, or at least do not contradict, the former pure \textit{Kepler} or \textit{Kepler}+ground-based ETV solutions.  In the case of the moderately certain solutions, the inferred outer periods are generally longer than one-third of the entire data train, but shorter than the duration of the dataset, and/or, other kinds of non-linearities in the ETV curves make the LTTE or LTTE+DE solutions less certain. Finally, under the subgroups of the uncertain solutions, we tabulate all those systems where we were able to find only such third-body solutions from the ETV curves where the period exceeds the length of the entire dataset, or where we cannot even be certain of the simple existence of a third companion in the system. (Or, in other words, it is uncertain whether the periodicity is real and, if it is, whether its origin is a third body or not).

The numerical results for the pure LTTE systems are given in Tables~\ref{Tab:OrbelemLTTE1}--\ref{Tab:OrbelemLTTE3}, while the LTTE+DE systems are reported in Tables~\ref{Tab:Orbelemdyn1}--\ref{Tab:Orbelemdyn3}. (The KIC~IDs of those ``new'' hierarchical triple star candidates, which were not listed in the corresponding tables of \citealt{borkovitsetal16} are typeset with \emph{italic} fonts.) The structure of these two sets of tables differs slightly from each other, in accord with the fact that in the case of LTTE+DE systems more parameters can be deduced from the analytic solution than in the case of pure LTTE systems \citep[see][for a brief overview]{borkovitsetal15}. Finally, in Table~\ref{Tab:AMEparam} we give additional parameters which are mostly related to the apsidal motion and the nodal regression of the studied systems. This latter table gives further parameters for all the LTTE+DE systems, which come naturally from the analytic LTTE+DE solutions, as was discussed in detail \citet{borkovitsetal15}. From the 181 pure LTTE systems, however, we also list in Table \ref{Tab:AMEparam} only the five systems for which the inner binaries were found to be eccentric, and we were able to (and had to) fit the apsidal motion period as an extra parameter. Moreover, we briefly discuss the analyses for each individual system in Appendix~\ref{app:Notesonindividualsystems}, while all 243 ETV curves, together with the best analytic fits, are plotted in Appendix~\ref{app:ETVcurves}. 
 
\subsection{Formerly analyzed triples without new ETV solutions.}

As was mentioned above, in this work we find third-body ETV solutions for 243 systems. This number formally exceeds the number of 221 third-body ETV solutions presented in \citet{borkovitsetal16}. Despite this, for 28 triple star candidates analyzed in \citet{borkovitsetal16} we did not calculate new improved hierarchical triple system solutions. This has three principal reasons, as follows.
\begin{itemize}

\item[i)]\emph{The lack of new ETV points is due to the disappearance of the eclipses.} Tight, compact, non-coplanar triple stars may exhibit large amplitude orbital inclination angle changes and, hence, eclipse depth variations (EDV) --- even the disappearance of the eclipses on timescales which can be such short as a couple of years to decades. The most remarkable situation in the \textit{Kepler}-field is that of \object{KIC 10319590}, whose eclipses completely disappeared after $\sim400$\,days from the beginning of the \textit{Kepler} observations \citep[see, e.g.,][]{rappaportetal13}. Besides this particular triple system, the eclipses of 8 other EBs have also disappeared in between the \textit{Kepler} and the TESS eras. Their KIC numbers are as follows: 4055092, 4078157, 4769799, 4948863, 5003117, 7670617, 7955301, 8938628.

Most of the above listed nine EBs exhibited strong EDV already during the four years of the \textit{Kepler} observations and, hence, the disappearance of the eclipses are far from surprising. Note that recently, \citet{gaulmeetal22} have published a complex, joint spectro-photodynamical \textit{Kepler}-light curve, ETV curve, ground-based follow up radial velocity (RV) curve and, SED analysis for \object{KIC 7955301}.  This study reveals that the binary eclipses most likely disappeared around the beginning of 2016, and they will not return until 2028.

Here we also mean to include \object{KIC 10223616}, one of the triple star candidates that we simply missed in our 2016 paper. Similar to the other former EBs listed above, the eclipses also completely disappeared by the time of the TESS observations. For this system we give our new analytic triple star solution which is based exclusively on the \textit{Kepler} ETV time series.
\item[ii]\emph{The lack of new ETV points is due to sensitivity limitations.} In this category we count those EBs for which we were either unable to determine new eclipses from the TESS data or, they were determined only with unusably large scatter. This typically occurred in the case of faint EBs with very shallow eclipses, where they simply disappeared in the much noisier TESS light curves. Examples for these effects are: 2856960, 3245776, 3766353, 4174507, 6964043, 7552344, 7593110, 7811211, 9084778, 9140402, 9353234, 9472174, 9574614, 9596187, 9706078, 9715925.

Note, that \citet{borkovitsetal22b} have published a photodynamical analysis for \object{KIC 6964043}. In that paper the authors discuss that while the regular eclipses are certainly present in the TESS data, due to their very shallows depths, useful eclipse times cannot be determined.
\item[iii]\emph{The lack of any third-body solutions on the new, extended ETV dataset.} There are some triple star candidates, for most (but not all) of which, very uncertain solutions were given in \citet{borkovitsetal16}, but the new measurements clearly contradict the older solutions. Partly as a result of this situation, we were unable to find any realistic third-body solutions. Hence, we removed the following systems from the list: KICs 3839964, 3853259, 8690104.

\end{itemize}

\subsection{EBs with LTTE solution}
\label{Subsect:LTTEgeneral}

In summary, we obtained pure LTTE solutions for 181 EBs in the original \textit{Kepler}-field. For most of the Group I and many of the Group II (robust and moderately certain) systems of \citet{borkovitsetal16} the TESS and ground-based follow up measurements confirm the former LTTE solutions. However, in many cases it was also necessary to add further, non-linear polynomial terms to describe other longer time-scale variations in the extended ETV dataset.

While for the majority of the robustly confirmed triple star candidates of \citet{borkovitsetal16} (see their Table~3) our new solutions give quantitatively very similar results, the importance of the current work starts with the longer-period triple star candidates. The $\sim1470$-day-duration of the original \textit{Kepler}-mission did not make it possible to confidently detect hierarchical triple systems with periods longer than 2-3 years. Accordingly, there were several EBs in our 2016 sample where, despite the fact that we were convinced that the ETV clearly showed longer period LTTE signals, we had to rank these systems into our Group~III (the most uncertain cases) even though we were able to give a formal LTTE solution. Now, TESS observations allow us to extend the length of the data-train sufficiently to obtain new, robust LTTE solutions. Therefore, besides the short outer-period, robust systems of \citet{borkovitsetal16}, we are now in a position to give robust pure LTTE solutions for 24 such triple candidates which, in \citet{borkovitsetal16} were categorized only as Group~II (i.e., moderately certain) systems.  Additionally, now we have even found robust solutions for seven former Group~III systems.

We also note that we found two additional EBs in the \textit{Kepler} sample which were not considered in \citet{borkovitsetal16}, but now we are able to provide robust pure LTTE solutions for them. These were background EBs which were not assigned their own postage stamps in the \textit{Kepler}-mission, but they were observed serendipitously. Similar to all the other EBs studied in this work, we discuss their details in Appendix~\ref{app:Notesonindividualsystems}.

Turning to the group of the moderately certain pure LTTE solution systems, the vast majority of the targets categorized in this way had been previously ranked in Group~III (uncertain LTTE solutions) by \citet{borkovitsetal16}. This is a natural (upgraded) shift in the categories which is due to the typically more than three times longer datasets, and reveals that most of the uncertain pure LTTE solutions of \citet{borkovitsetal16} were actually at least qualitatively real or, in other words, they revealed actual triple star systems. On the other hand, a more detailed comparison reveals that the new solutions in this group strongly depart quantitatively from the Group~III LTTE solutions of \citet{borkovitsetal16}. This fact emphasizes that it was a good choice previously, and should also be generally followed in the future, that one not take too seriously any quantitative third-body results which were obtained with the use of a dataset that is shorter than the period of the third-body.  Note, this is one of the principal reasons why we decided not to consider the most uncertain systems in our statistical studies below.

Finally, in the relatively large category of uncertain pure LTTE solutions, we include those EBs for which we were able to find any astrophysically more or less realistic LTTE solution, but the third-body period we obtained either exceeds the length of the entire dataset, or some segments of the ETV data clearly contradict the accepted solution. This is a very heterogeneous group of targets.  In many cases, especially when the ETV curve clearly exhibits a section of a longer period, quasi-sinusoidal curve, it is very likely that the ETV actually represents a section of an LTTE caused by a longer period third component. Typical examples are the ETV curves of 466952, 4937217, 6103049, 9532219, 10686876.  In other systems, some segment(s) of the ETV points are clearly outliers to the accepted solution, or we had to force quadratic or cubic polynomial fits in addition to the LTTE term in order to obtain any acceptable LTTE solutions.  In these latter situations, we are not even fully convinced that what we see is a real third-body effect as opposed to some other periodic, or quasi-periodic, variation as was briefly discussed above in Sect.~\ref{sec:polynomial}. A few examples are: 5621294, 6050116, 6265720, 7440742, 8953296, 9159301, 10557008, 12554536, etc.

\subsection{EBs with combined LTTE + dynamical solution}
\label{Subsect:LTTEdyngeneral}

Besides the 181 pure LTTE systems discussed above, we found 62 triple star candidates where the gravitational perturbations of the close, third companion made it necessary to include the dynamical effects in the ETV analysis.  By chance, this number is equal to the number of the LTTE+DE systems in \citet{borkovitsetal16}.  We mentioned earlier, however, that we had to omit some of the LTTE+DE systems of \citet{borkovitsetal16} due to the disappearance of the eclipses, because of either orbital plane precession driven by an inclined third component, or the faintness of the system. Incidentally, we were able to find an equal number of new LTTE+DE solution triple star candidates to those that we had to neglect. Similar to the pure LTTE systems (and also akin to \citealt{borkovitsetal16}) we ranked these findings into three sub-categories. In accord with these three subgroups we list the inferred orbital parameters of these systems in Tables~\ref{Tab:Orbelemdyn1} -- \ref{Tab:Orbelemdyn3} and, moreover, some additional parameters are also listed in Table~\ref{Tab:AMEparam}.

Similar to the pure LTTE triples, for most of the Group I (robust), combined LTTE+DE solution systems of \citet{borkovitsetal16} the TESS and ground-based follow up measurements confirm the former solutions.  Moreover, the new data reveal that there are no further, longer-term trends in the ETVs. A few important exceptions, however, should be mentioned. First, two short period systems, KICs 6545018 ($P_2=91$\,d) and 9714358 ($P_2=104$\,d) clearly reveal the inadequacy of the applied quadratic order approximations of the apse-node, or secular, timescale perturbation terms. Moreover, it is also likely that in the case of the somewhat longer period LTTE+DE triples KICs 5255552 and 8143170 the additional quadratic polynomial terms that we had to introduce for an improved, satisfactory solution have also arisen from the additional, not-considered octupole or quadrupole-squared perturbations terms.  We discuss this question in greater detail in Appendix~\ref{app:Notesonindividualsystems} separately for each system.

We note also \object{KIC 7289157}, one of the first discovered triply eclipsing triple star systems, for which two new third-body eclipses were also observed by TESS. It was mentioned already in \citet{borkovitsetal15} that the \textit{Kepler}-ETV residuals of this systems show non-linear, perhaps quadratic characteristics, however, according to the authors it was not clear whether that residual non-linearity was caused by some extra effects (e.g., the presence of a fourth, more distant component in the system) or was simply due to insufficient mathematical modeling of the perturbations. Now, taking into account the TESS eclipse times, we found that a longer period, second sine-like term gives a better description of this residual non-linearity and, hence, we fitted a longer period ($P_3=6950$\,d) LTTE orbit in addition to the $P_2=244$\,d LTTE+DE third-body solution. Therefore, this system might be a (2+1)+1-type hierarchical quadruple system, as is also discussed in Appendix~\ref{app:Notesonindividualsystems}.

In another similarity with the robust pure LTTE triples, thanks to the substantially extended timescale of the available ETV points, we were able to add robust LTTE+DE solutions for 9 somewhat longer period triple candidates which were categorized by \citet{borkovitsetal16} as only moderately certain systems. One candidate, the above mentioned \object{KIC 8143170}, jumped ahead two categories (from the uncertain to the robust) at once.

Considering, however, the moderately certain, and especially the uncertain groups of LTTE+DE triple star candidates, the similarity to the pure LTTE systems disappears. These is so mainly because, in contrast to the pure LTTE systems, of which the majority consist of short inner period EBs with periods from hours to a few days, there is one technical problem with the longer period, i.e. less compact, but tight (i.e. dominated by third-body perturbations) systems. It is very easy to see that if one increases the outer period by any given factor, the inner period should also be increased by the same factor to retain the tightness of the triple. Hence, the longer outer period tight triples, naturally, also contain longer period inner binaries. Then, in the case of a 5-10\,yr outer period tight hierarchical triple, dominated by dynamical perturbations, the inner pair should have orbital periods in the range of months. Thus, even if the inner binary remains continuously eclipsing, in a $\sim$$27$\,day-long TESS sector one cannot expect more than 1 or 2 new eclipses, or even less. In these circumstances we cannot expect very robust new LTTE + DE solutions for such systems.  Nonetheless, in a few cases, the new data do make it more or less likely that the non-linear ETV signals are coming from the gravitational perturbations of a third star.  As a consequence, the number of the moderately certain LTTE+DE systems in the current paper is only 7.   Even from this very narrow sample, the two shortest outer period (and hence, also the shortest inner period) members of this group (KICs 5080652 and 9664215) were ranked in this sub-category only due to the faintness (of the EBs) and shallowness (of the eclipses), which caused a large scatter in the new ETV points.

Despite the facts discussed in the last paragraph, out of the 17 targets, ranked in the uncertain subgroup of LTTE+DE systems, there are seven completely new triple system candidates. There are also three additional systems, which were listed in Table~11 of \citet{borkovitsetal16}, i.e., amongst those systems for which we were unable to give any third-body solution then, but we supposed that what was seen in the \textit{Kepler} ETV data was nothing other than a segment of a long-period LTTE+DE orbit. There is another substantial difference, however, amongst these uncertain LTTE+DE solutions and the uncertain pure LTTE solutions. In particular, for the uncertain LTTE+DE systems, due to the decidedly characteristic shape of an LTTE+DE ETV pattern, we can be convinced with high confidence that the ETV really results from a third, gravitationally interacting companion. Therefore, the very  existence of the third body is not in question, but only the quantitative parameters of the perturber's orbit (including even the correct period) are uncertain.

\subsection{EBs with ``four-body'' solutions}
\label{sec:fourbody}

There are 12 triple star candidates in our sample for which the ETVs exhibit double periodicities or, at least, we were able to obtain an acceptable solution by fitting a second, longer period LTTE curve together with the prime LTTE or LTTE+DE solution. Mathematically, such solutions imply a (2+1)+1 type quadruple star system configuration. There are examples for either robustly known such close quadruples (e.g. \object{HIP 41431}, \citealt{borkovitsetal19b}, \object{TIC~114936199}, \citealt{powelletal22}) or very likely quadruple systems (as e.g., the previously mentioned, currently shortest known outer-period triple system \object{TIC 290061484}, \citealt{kostovetal24} and the recently analysed triply eclipsing system, \object{TIC 321978218}, \citealt{rappaportetal24}). Doubly periodic ETV curves were also interpreted by others \citep[see, e.g.][]{zascheetal17,hajduetal19} as being caused by two distant companions.  Despite these facts, we are convinced that, in most of our doubly periodic cases, one of the two periodicities is likely to have a different origin than a distant body, and therefore, the doubly periodic description should be considered as a mathematical approximation. We discuss all these individual cases separately in the Appendix~\ref{app:Notesonindividualsystems}. Here we simply note that in the forthcoming statistical studies, we consider only the shorter period LTTE or LTTE+DE terms of these doubly periodic solutions.

\subsection{Statistical results}

We have studied the ETV curves of some 243 compact hierarchical triple star systems in this work.  Some have only LTTE orbits, while others have combined LTTE + DE orbital solutions.  As discussed in detail above, we classify the quality of the system solutions in three categories: (i) Robust, (ii) Moderately Secure, and (iii) Uncertain.  The numbers of each type of solution are summarized in Table \ref{tbl:classifications}.  These represent a substantial fraction of all the well studied compact triple systems that are currently known.  In this section we review some of the global parameters of these systems.

\begin{table*}[h]
\centering
\caption{Numbers of Systems with Different Types of Solutions}
\begin{tabular}{cccc}
\hline
\hline
Classification & LTTE only  & LTTE +dyn solutions &  Combined \\
\hline
   Robust & 63 & 37 & 100  \\
   Moderately Secure & 45  & 7 & 52  \\
  Uncertain  & 73  & 18 & 91  \\
  Totals & 181 & 62 & 243  \\
 \hline
\label{tbl:classifications} 
\end{tabular}
\end{table*} 

\begin{figure}[h]
\centering
\includegraphics[width=0.47\textwidth]{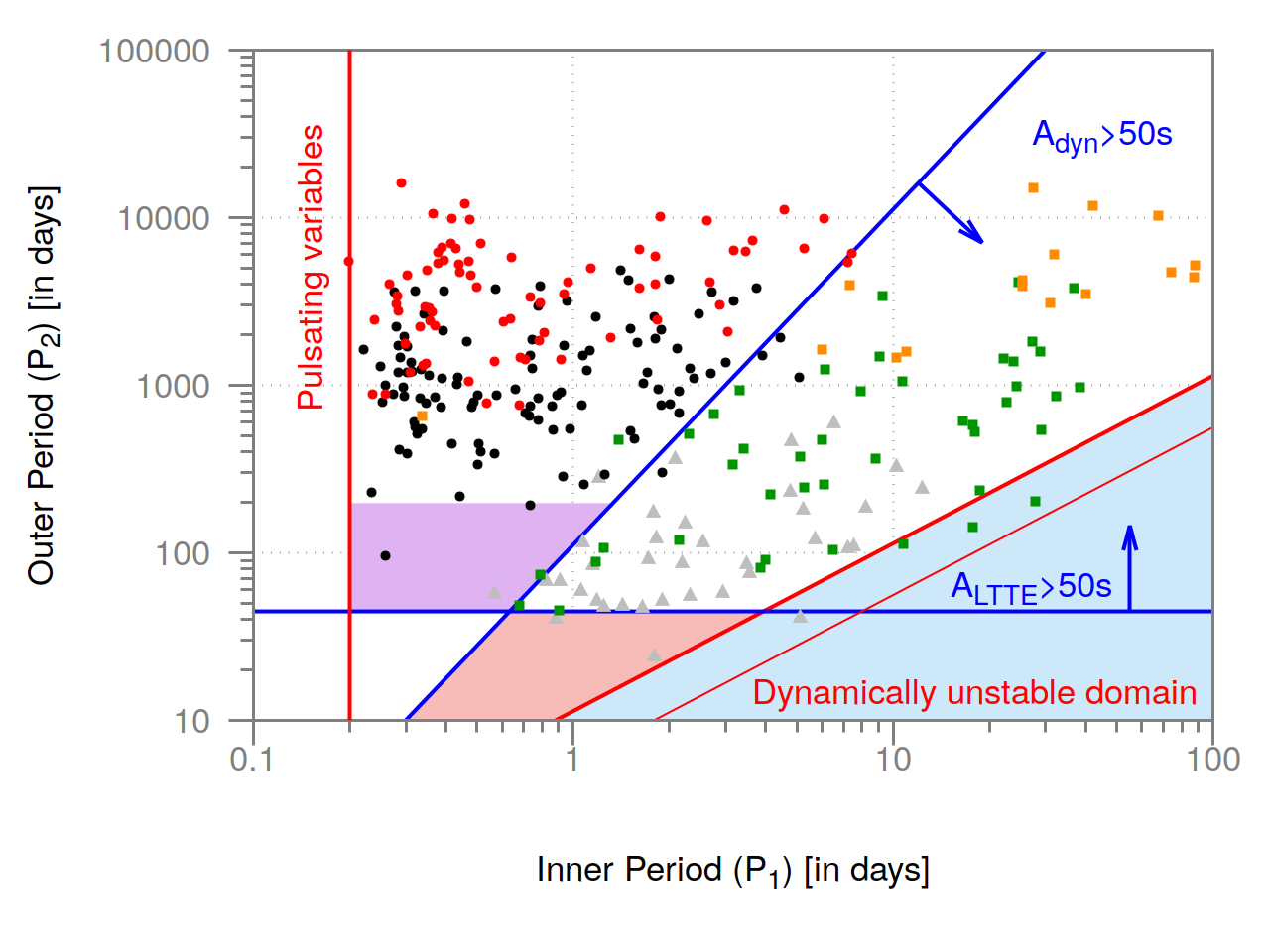}
\caption{Locations of the analysed hierarchical triple star candidates, together with other triply eclipsing triples, detected with space telescopes \textit{Kepler} and TESS in the $P_1-P_2$ plane. Black and red dots and green and orange squares represent the current TESS revisited \textit{Kepler} candidates. For the black and red systems simple LTTE solutions were satisfactory, while for the green and orange ones, the dynamical effects were also taken into account. (The red and orange symbols, however, denote the most uncertain solutions which were largely omitted from the statistical investigations.) Gray triangles represent recently discovered \textit{K2} and TESS systems with accurately known photodynamical solutions. (These are mostly triply eclipsing triple systems.) The vertical red line at the left shows the lower limit of the period of overcontact binaries. The horizontal and sloped blue lines are boundaries that roughly separate detectable ETVs from the undetectable ones. The detection limits again, were set to 50\,sec. These amplitudes were calculated following the same assumptions as in Fig.~8 of \citet{borkovitsetal16}, i.e., $m_\mathrm{A}=m_\mathrm{B}=m_\mathrm{C}=1\,\mathrm{M}_\odot$, $e_2=0.35$, $i_2=60\degr$, $\omega_2=90\degr$. The arrows indicate the directions of increasing LTTE and dynamical amplitudes. The shaded regions from left to right represent $(i)$ the W~UMa desert, i.e., the (almost) empty domain (purple) where a tight third companion of a short-period EB would certainly be detectable through its LTTE, even in the absence of measurable dynamical delays; $(ii)$ the purely dynamical region (pale orange), i.e., where the dynamical effect should be detectable, while the LTTE not and; $(iii)$ the dynamically unstable region (light blue) in the sense of the \citet{mardlingaarseth01} formula. Note, that while the border of this latter shaded area was also calculated with $e_2=0.35$, we give the limit for $e_2=0.1$, as the thinner red line within this (light blue) region, as well.}
 \label{fig:P1vsP2}
\end{figure}

In Figure \ref{fig:P1vsP2} we show the outer orbital period, $P_2$, vs.~the inner binary period $P_1$ of all the systems discussed in this work.  The black and red circles are the LTTE-only solutions, with the black being the robust and moderately secure solutions, with red the more uncertain solutions.  The green squares are more secure LTTE + DE solutions, while the orange squares are the more uncertain ones in this category.  The blue lines plus arrows point out regions where LTTE effects and/or where dynamical effects should be detectable.  The gray triangles are other well studied triples from \textit{K2} and TESS.  

The shaded purple region is the W~UMa desert with only one system in it, where one could expect to detect tertiaries orbiting in 40-150 day periods around tight binaries with periods between $\sim$0.2 and 0.6 days---if they existed.  There is also an interesting ``dynamical-only" region (pale orange) which likewise has no systems.  The blue shaded region is where triple systems are expected to be dynamically unstable (see e.g., \citealt{mardlingaarseth01}).  In a confirming way, we see only a few systems beginning to encroach on that region, but none firmly in it.

Systems with trustworthy solutions typically have outer periods between 40 and 3000 days, and binary periods between $\sim$0.2 d and 30 days.  Note that the longer period systems here are twice that of the entire \textit{Kepler} mission.  This is made possible by the extension of the ETV curve that TESS provides. Fortunately, a fair fraction of the systems have dynamical effects that are detected in addition to the LTTE orbits.  This provides crucial additional information about the system, e.g., mutual inclination angles and certain mass ratios.   

\begin{figure}
\centering
\includegraphics[width=0.47\textwidth]{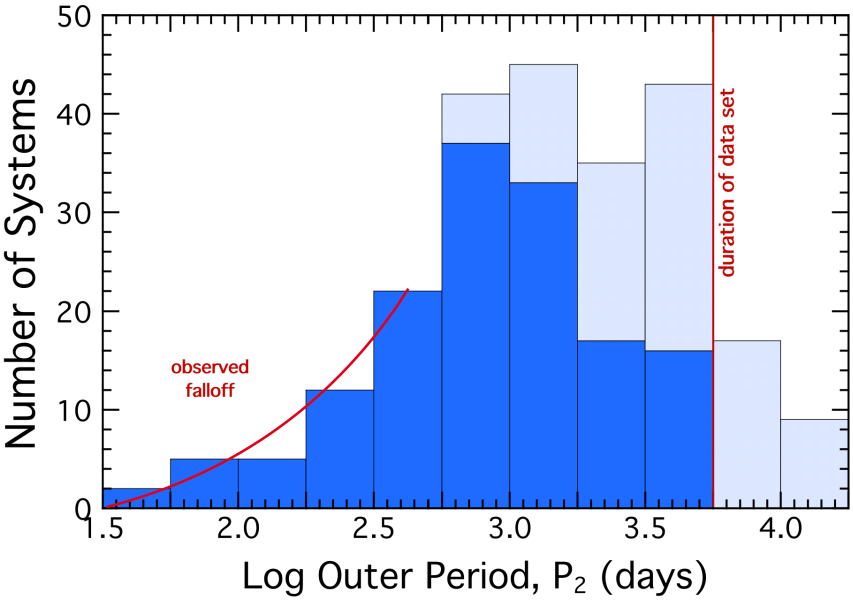}
\caption{Distribution of the logarithm of the outer orbital periods, $P_2$. Solid blue portions of the histogram are the secure and moderately secure LTTE and LTTE + DE solutions.  The faintly shaded regions are the uncertain systems from both the LTTE and LTTE + DE systems. The red vertical line marks the approximate duration of the combined span of the \textit{Kepler} and TESS data sets.}
\label{fig:logP2}
\end{figure}

We show in Figure \ref{fig:logP2} the distribution of outer orbital periods in logarithmic bins from 33 to $\sim$$15000$ days.  Quite naturally, the less certain systems tend to be the ones with the longer periods which are either a substantial fraction of the duration of the observations, or even longer.  If we count the uncertain systems, which fill in this longer period range, then the histogram can be considered roughly flat (per logarithmic interval) over the range $300 \lesssim P_2 \lesssim 5000$ days.  For outer periods $\lesssim 300$ days, however, the distribution is clearly falling off toward shorter $P_2$.  Taking a glance again at Fig.~\ref{fig:P1vsP2}, one can see, that this fact might not be a selection effect but, instead, it might be due in part to the W~UMa desert (the light purple region at the left side of the plot). This reflects primarily the rarity of short outer period ternary components around W~UMa-type overcontact binaries. On the other hand, however, one should notice that there are several gray triangles in the short outer period region of Fig.~\ref{fig:P1vsP2}. Those triangular symbols represent short outer period triples that were, in fact, discovered mostly in the TESS (and partly in the \textit{K2}) data through third-body eclipses of the outer components. (None of them consists of an overcontact inner binary.)  This fact, however suggests that in partial contrast to our above statement about the absence of observational selection effects, the small number of short outer period triples might also come, at least partly, from some kind of not well understood observational selection effects related to the ETV curves. For example, one may assume that in the case of the short outer period third-bodies which, in general, produce low amplitude ETV signals\footnote{In the case of pure LTTE orbits this statement is trivial, since the LTTE amplitude depends primarily on $P_2^{2/3}$ (see Eq.~[\ref{Eq:A_LTTE}]). For the DE terms this is not so evident, but one should consider that in the case of a small $P_2$, the inner period $P_1$ must also be sufficiently small to satisfy dynamical stability criteria, and the amplitude of the DE terms are scaled with $P_1$ (see Eqs.~[\ref{Eq:A_dyn}] or [\ref{Eq:A_dyn^cop}]), which also results in a smaller amplitude.}, variations can be easily hidden by other kinds of quasi-cyclic distortions on the ETVs, which might be caused by e.g., rotational or spot modulations \citep[see, e.g.][]{tranetal13}.
 
\begin{figure}
\centering
\includegraphics[width=0.47\textwidth]{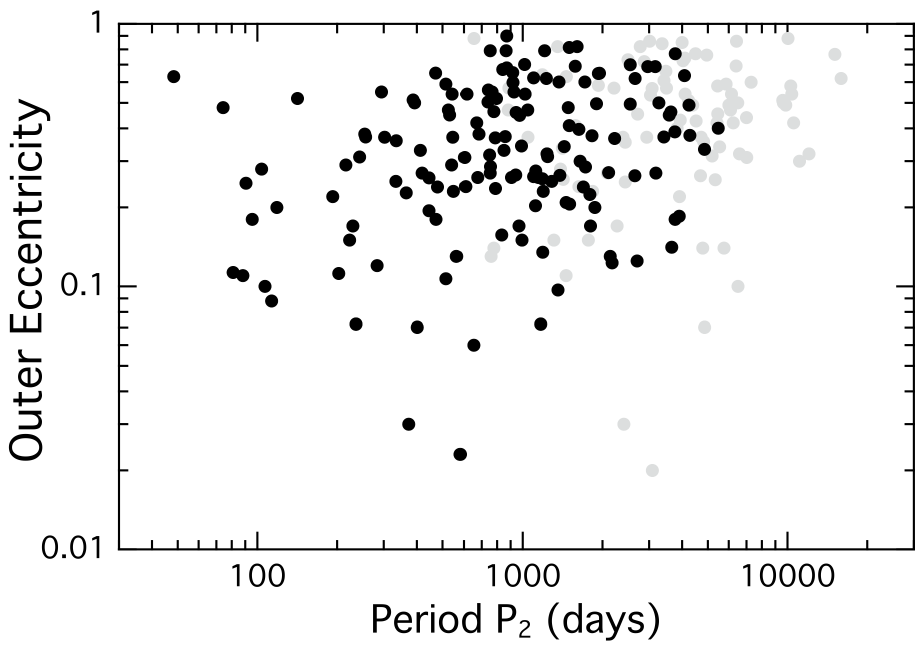}
\caption{The eccentricity of the outer orbit, $e_2$, plotted against the period of the outer orbit, $P_2$.  The black circles are all the secure and moderately secure LTTE and LTTE + DE solutions.  The faint grey dots are the uncertain systems from both the LTTE and LTTE + DE systems.}
\label{fig:e2vsP2}
\end{figure}

In Figure \ref{fig:e2vsP2} we plot the eccentricity of the outer orbit, $e_2$, against the period of the outer orbit, $P_2$, for all of our systems.  The range of $P_2$ is $\approx50-5000$ days, and the range of $e_2$ goes from nearly circular to $\sim$0.9. The median value of $e_2$ is 0.33. There is no significant correlation between these two parameters.  In other words, nearly any outer eccentricity can be found for any given $P_2$ between $\approx50-5000$ days.  It is highly likely that these eccentricities are primordial, and therefore they should ultimately tell us something about the formation scenario of these systems.

\begin{figure}
\centering
\includegraphics[width=0.47\textwidth]{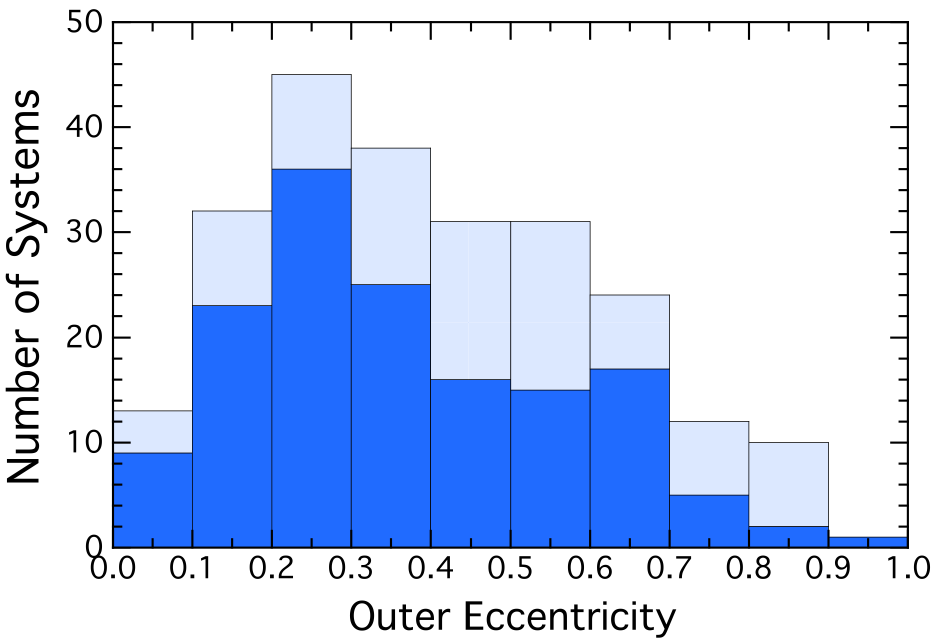}
\caption{Distribution of outer eccentricities, $e_2$. Solid blue portions of the histogram are the secure and moderately secure LTTE and LTTE + DE solutions.  The faintly shaded regions are the uncertain systems from both the LTTE and LTTE + DE systems.}
\label{fig:e2}
\end{figure}

Figure \ref{fig:e2} is a histogram of the outer eccentricities of all our systems.  The contributions of the ones with less secure solutions are shown with faint shading.  The mean, the median, and the mode of this distribution are 0.36, 0.33, and 0.25, respectively (including only the secure systems).

\begin{figure}
\centering
\includegraphics[width=0.47\textwidth]{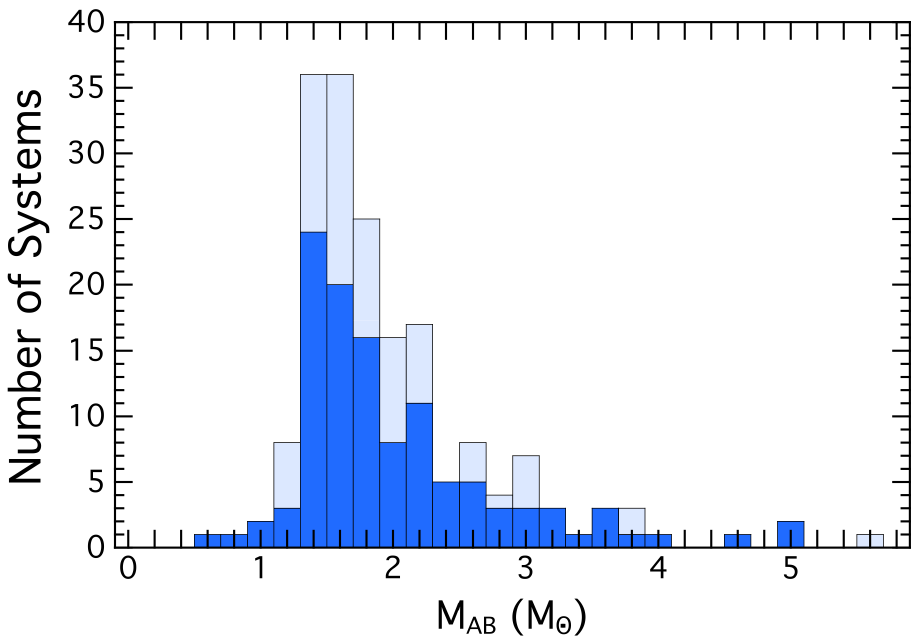}
\caption{The binary masses ($m_\mathrm{A}+m_\mathrm{B}$) for the secure and moderately secure LTTE and LTTE + DE solutions.  The faintly shaded regions are the uncertain systems from both the LTTE and LTTE + DE systems. We have eliminated all systems with a mass of exactly 2.0 that was used as a default in some systems.}
\label{fig:mbinary}
\end{figure}

The information that we were able to glean about the total mass of the inner binaries is summarized in Fig.~\ref{fig:mbinary}. This just emphasizes that the majority of binary stars we are dealing with are low to solar-mass stars.  Only three systems had binaries with combined masses above 3.8 M$_\odot$.

\begin{figure}
\centering
\includegraphics[width=0.47\textwidth]{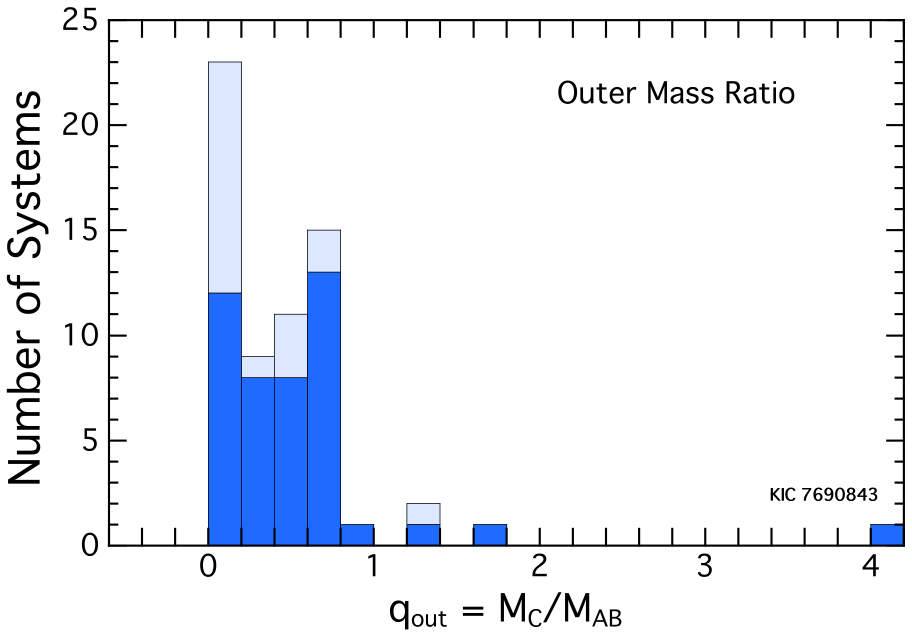}
\caption{The outer mass ratio, $q_2 \equiv m_\mathrm{C}/m_\mathrm{AB}$. Only LTTE + DE solutions were considered - the faintly shaded regions are the uncertain LTTE + DE systems. }
\label{fig:qout}
\end{figure}

An important parameter of triple star systems is the outer mass ratio, i.e., $q_2 \equiv m_\mathrm{C}/(m_\mathrm{A}+m_\mathrm{B})$.  The distribution of this mass ratio for our LTTE+DE solutions is shown in Fig.~\ref{fig:qout}. A substantial majority of the systems have $q_2 \lesssim 1$.  This is in keeping with our earlier findings \citep{kostovetal24} that in the formation scenario involving fragmentation followed by accretion, the tertiary star is usually less than or about equal to the mass of the inner binary.

\begin{figure}
\centering
\includegraphics[width=0.47\textwidth]{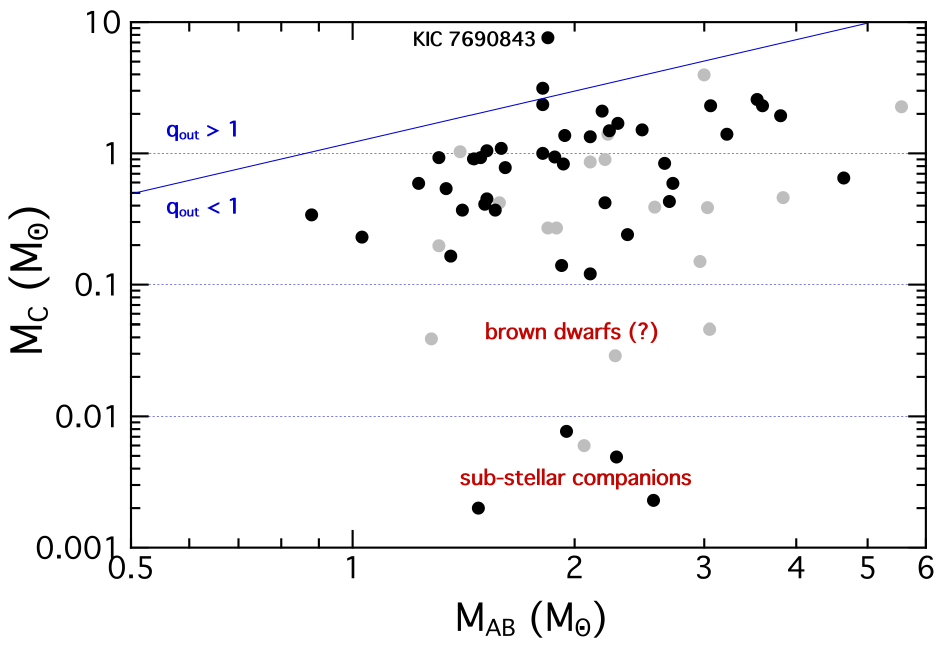}
\caption{Mass of the tertiary star, $m_\mathrm{C}$, plotted against the binary mass, $m_\mathrm{AB}$.  Only LTTE + DE solutions were considered - the faint gray circles are the uncertain LTTE + DE systems.  The blue line marks the $q_2 = 1$ boundary. }
\label{fig:Mc_Mab}
\end{figure}

Regarding the $q_2$ parameter, we show further information about this by plotting $m_\mathrm{C}$ vs.~$m_\mathrm{AB}\equiv m_\mathrm{A}+m_\mathrm{B}$ (see Fig.~\ref{fig:Mc_Mab}). Systems with $m_\mathrm{C}\gtrsim0.1 \, M_\odot$ form the bulk of the triple systems that we study,  There are three systems where we infer brown dwarf-like masses, but these systems are all among the uncertain classification, and therefore we do not take them too seriously.  Below the 10 $M_J$ mass level, there are four systems three of which are in the more secure category, and therefore we have reasonable confidence that these may well be massive planets with $m \simeq 2-9 \, M_J$.

\begin{figure}
\centering
\includegraphics[width=0.47\textwidth]{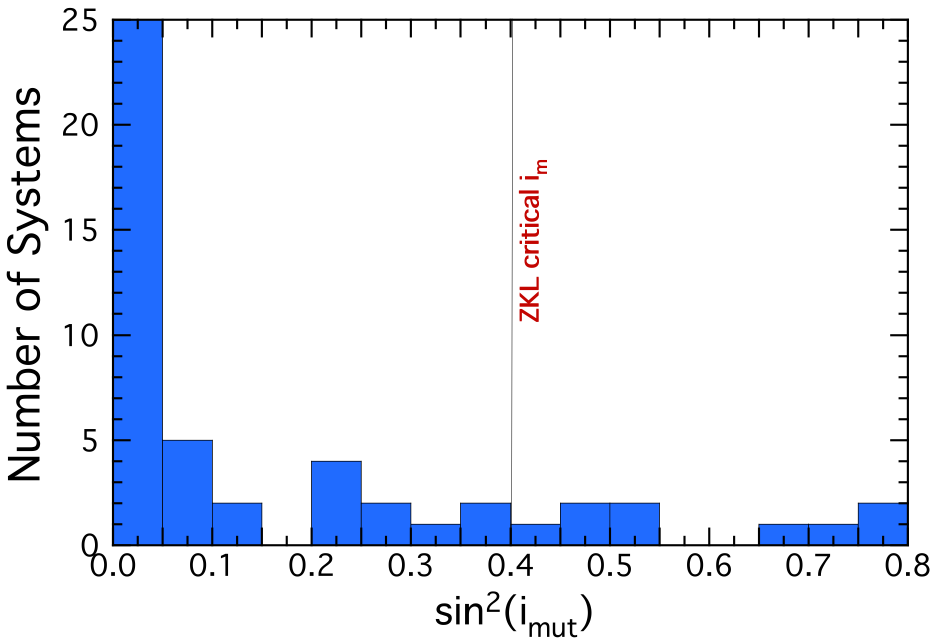}
\caption{Distribution of sine squared of the mutual inclination angle ($i_{\rm mut}$, i.e., the relative tilt of the planes of the inner binary with respect to that of the outer binary).  Only LTTE + DE solutions were considered.  (The first bin actually extends off scale to 37 systems.) Nine of the systems have $\sin^2i_{\rm mut} \gtrsim 0.4^\circ$, or $i_{\rm mut} \gtrsim 39.2^\circ$ where von Zeipel-Lidov-Kozai cycles could occur.} 
\label{fig:mut_incl}
\end{figure}

For the dynamically active systems we can also derive the mutual inclination angle, $i_{\rm mut}$.  The distribution of $\sin^2 i_{\rm mut}$ is shown in Fig.~\ref{fig:mut_incl}; it is this term that appears in the quadrupole order perturbations, and the elements of its distribution best indicate what can be inferred from the measurements.  Most of the systems (i.e., 44 of the 62) have $\sin \, i_{\rm mut} < 0.22$ corresponding to $i_{\rm mut} \lesssim 12\degr$, or nearly coplanar.  However, approximately nine of the systems have $ \sin^2 i_{\rm mut} \lesssim 0.4$ or $i_{\rm mut} \gtrsim 39\fdg2$, i.e., inclinations large enough, in principle, to drive a von Zeipel-Lidov-Kozai cycle.

\begin{figure}
\centering
\includegraphics[width=0.47\textwidth]{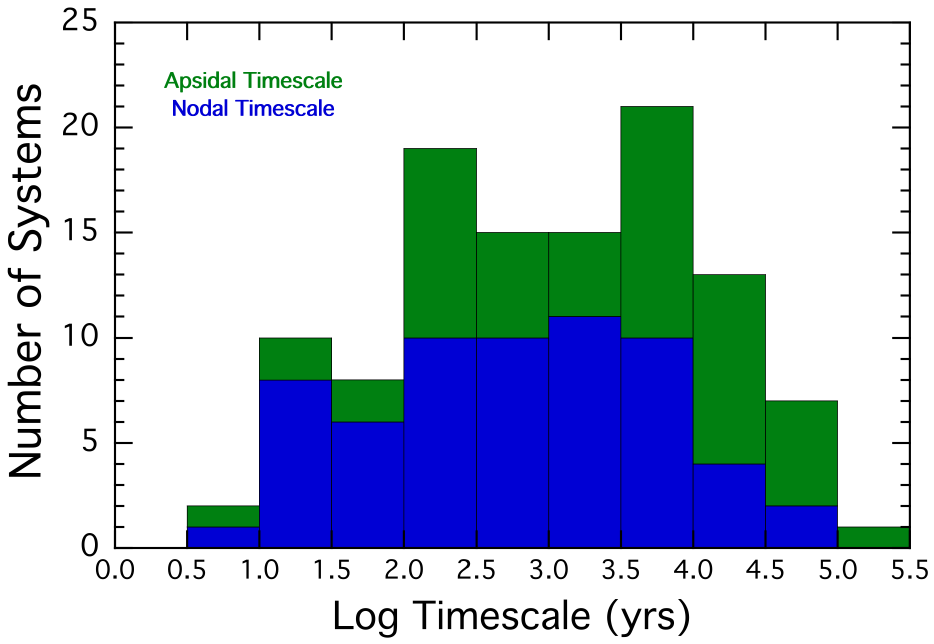}
\caption{Distributions of apsidal motion and nodal timescales for the secure LTTE + DE solutions.  Note that the times are plotted on a log scale.  The plus and minus signs (associated with prograde and retrograde motions, respectively), have been suppressed.  These times are inferred from the other system parameters rather than directly measured from the ETV curves.}
\label{fig:Papse}
\end{figure}

In Figure \ref{fig:Papse} we give the distributions of apsidal motion and nodal precession timescales for the secure LTTE + DE solutions.    The times are plotted on a log scale.  The plus and minus signs, associated with prograde and retrograde motions, respectively, have been removed in order to take logarithms.  These apse-nodal times are inferred from the other system parameters rather than measured directly from the ETV curves.  See the appendices in \citet{borkovitsetal15} for more details of the mathematical formulation.  These timescales are more or less uniformly distributed in log space between 10 and $10^5$ years.  We show these to illustrate the wide range over which apsidal motion and nodal precession can take place in compact triple systems.  

\begin{figure}
\centering
\includegraphics[width=0.47\textwidth]{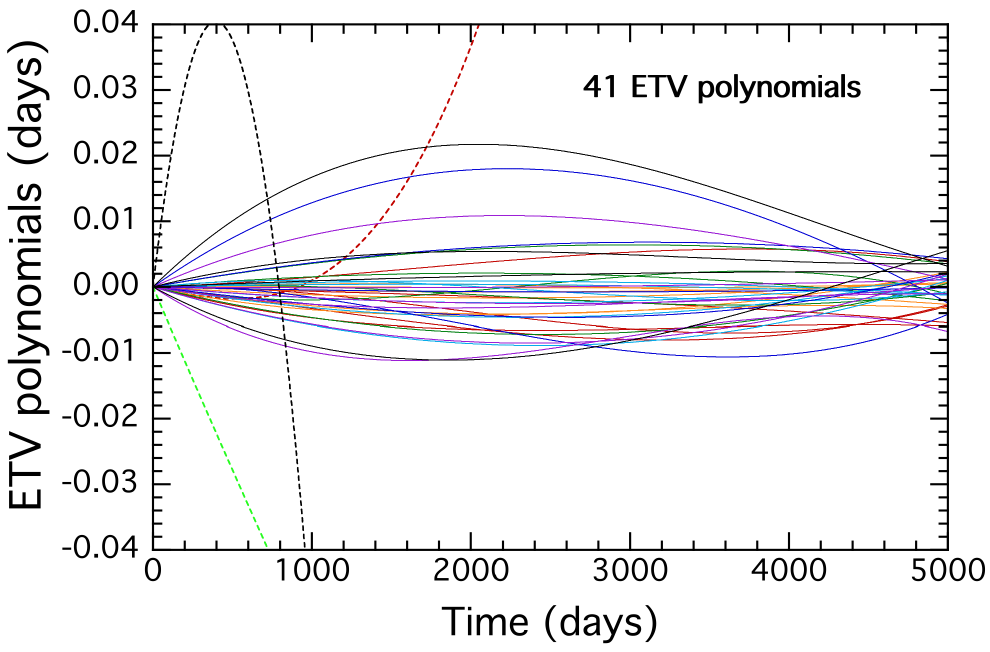}
\caption{ETV residuals fit to quadratic and cubic polynomials.  These are taken from 41 secure LTTE and LTTE+DE fits.  They are continuous reconstructions over a 5000 day interval (spanning the \textit{Kepler} and current TESS missions) using the polynomial coefficients that were part of the fitting process.  The colors are merely to help the eye separate the different curves.  The three curves shown as dashed lines that go off the scale of the plot likely represent higher-order terms that are not captured by the analytic formulae used to fit the ETV curves.}
\label{fig:polyetvs}
\end{figure}

Finally, of the one hundred secure LTTE and LTTE+DE fits to the \textit{Kepler}-TESS ETV curves, some 41 of them required an extra polynomial expression (i.e., quadratic, cubic, or both) to allow for an acceptable fit.  In order to get a feel for how large these extra delays are, we have plotted them all (superposed) in Fig.\ref{fig:polyetvs}.  They have been reconstructed over a 5000 day interval (13.7 years) from the polynomial coefficients found from the various fits.  As one can see, for all but three of the curves, the total extra deviations from the orbital fits amount to about 0.02 days or $\sim$30 minutes.  The formal rms deviations from zero for the 38 `well-behaved' polynomials are 0.0053 days or $\sim$8 minutes.  The three curves that go off the plot are cases where we believe there are significant higher-order terms in the analytic ETV models that have not been included, and therefore these are still in the category of ``understood'' astrophysical delays, while the meanderings in the remainder of the curves represent some not completely understood phenomenon that has been noticed by observers for decades while carrying out long-term timing of binaries.

One cautionary tale from these plots is that all of the solid curves could be unwittingly fit for the better part of an entire orbital cycle in the case of cubic polynomials, or about half an orbital cycle for the quadratic-dominated curves.  Thus, unless there is reason to believe there is indeed another body present in the system, it is wise to require at least two outer orbital cycles before accepting the presence of another body as at least somewhat conclusive.  

\section{Summary and Conclusions}
\label{Sect:Summary}

We have carried out ETV analyses for the complete EB sample of the original \textit{Kepler} mission after the revisit observations of its successor, the TESS space telescope. In such a manner we were able to extend the original four-year long, very precise ETV datasets of the \textit{Kepler} observed (and, mostly discovered) EBs up to one and a half decades. Moreover, for a smaller subset of the original \textit{Kepler} sample, for which the ``official'' targeted \textit{Kepler} photometry and, hence, eclipse timing data, were formerly available only for subsection(s) of the prime \textit{Kepler} mission, we were able to extend even the duration of the \textit{Kepler} data to the full extent (or, at least, a longer part) of the \textit{Kepler} mission. This was enabled either by our own efforts (e.g., Sect. \ref{sec:Selection} and Appendix \ref{app:Notesonindividualsystems}), or to new extended photometry, which is now available in the literature (e.g., Sect.~\ref{sec:Selection} and Appendix \ref{app:Notesonindividualsystems} and \citealt{abdulmasihetal16,keplerbonus}).  Furthermore, in the case of some targets, we carried out additional, targeted eclipse observations the results of which were also included in our analyses. For determining precise times of eclipses we have used mainly those techniques (e.g., averaging of primary and secondary ETVs, and fitting and subtracting smoothing polynomials around each individual eclipse in the light curves), which were described in \citet{borkovitsetal16} in some detail. Due to the larger scatter of the eclipse timings obtained from the analysis of the TESS observations of some fainter and/or lower amplitude EBs, we had to form phase folded, averaged light curves from 5-20 consecutive orbital cycles of these targets, and thereby calculated so called ``normal'' eclipse timings (or minima) for these EBs to reduce the scatter in the new sections of the ETVs.

In such a manner, altogether we found 243 systems ($\sim$$9\%$ of the entire \textit{Kepler} sample) as potential triple star candidates.  Our hierarchical triple star candidate sample strongly overlaps the former sample of \citet{borkovitsetal16}, as we confirm the former solutions, or give new solutions for 193 systems ($\sim$$87\%$) of the former sample. From the omitted 28 formerly listed hierarchical triple star candidate targets, in nine cases we were unable to add new solutions because, by the time of the new, TESS observations, the eclipses had disappeared because of the decreasing inclination angles, due to the precession of the orbital planes forced by an inclined third companion. (Therefore, in these cases, the disappearance of the eclipses confirms the presence of a third star with high confidence but, naturally, in the absence of eclipses and, hence, of new eclipse timing data, we were not in a position to calculate any new or refined ETV solutions.) In the case of 16 of the other `missing' targets, we were unable to calculate new ETV points only due to instrumental reasons (faintness, shallowness).  For the remaining 3 of these systems we were able to calculate new eclipse timings, but the new ETV sections clearly refuted the former third-body solutions of \citet{borkovitsetal16}, and we were unable to calculate any new solutions.
  
According to the results of our investigations we have classified our 243 triple star candidates into two main groups, as follows.
\begin{itemize}
\item[]{\emph{Group I}: These are the triple star candidates for which the ETVs of the inner, eclipsing binary are dominated by the LTTE delays, and the dynamical contribution to the ETVs can be neglected. We classified 181 targets in this group. The third companions of these systems have an outer period range of $95\lesssim P_2\lesssim12057$\,days or, neglecting the most uncertain cases, the longest outer period system amongst the -- at least -- moderately secure candidates is $P_2=4859\pm11$\,d. In 39 secure or moderately secure (and 25 uncertain) cases, an additional quadratic term was fitted simultaneously to the ETV curve, while a cubic polynomial was required for 43 (14) of the EBs. Furthermore, for 10 (4) of these 181 ETV curves we added four-body solutions. However, as we discussed in Sect.~\ref{sec:fourbody} we assume that these four body solutions, most likely are only mathematical approximations of doubly periodic ETVs and, at least one of the two periodic terms does not arise from an additional body, but has some other, different origin.}
\item[]{\emph{Group II}: Similar to \citet{borkovitsetal16}, 62 EBs exhibit a remarkable dynamical contribution to their ETV curve. Note, however, that the equal number of such systems in the two papers is only by chance, as the current work gives dynamically affected third-body solutions for 11 systems for the first time. For these 62 targets, we fit for a combined LTTE plus dynamical ETV solution (which includes apsidal motion terms for eccentric EBs, as well) and, in such a way we have obtained several additional parameters to those which can be obtained from a pure LTTE solution. As was mentioned already in the former papers of \citet{borkovitsetal15,borkovitsetal16}, the most relevant extra parameters are the total mass of the EB ($m_\mathrm{AB}$) and the individual mass of the ternary component ($m_\mathrm{C}$), and the relative, or mutual, inclination angle ($\im$) of the inner and outer orbits.  However, in most cases, the masses themselves can be obtained only with limited accuracy, not appropriate for deeper astrophysical considerations. Moreover, the mutual inclinations should also be considered with some additional caution, because the dynamical ETV terms depend only upon $\sin^2\im$. Therefore, it is difficult to separate mathematically an exactly flat configuration (where $\im=0\degr$, i.e. $\sin^2\im=0$) from one where, for example, $\im=20\degr$ (i.e., $\sin^2\im\approx0.12$) \citep[see further discussions in Sect.~5 of][]{borkovitsetal15}. In addition, quadratic and cubic polynomials were also fitted for one and three systems, respectively. Moreover, we added a second LTTE solution of a fourth hypothetical body in the case of two candidate systems. The reliability of such four-body solutions in these latter two cases is discussed in Appendix~\ref{app:Notesonindividualsystems}. The outer period range for the Group II systems is $45\lesssim P_2\lesssim15087$\,days (or, omitting the most uncertain cases, the longest deduced outer period remains $P_2=4084$\,days).}
\end{itemize}
These two large categories were also divided into subgroups according to the robustness, or reliability, of their hierarchical triple star candidate models. The majority of those candidates for which the observational data (including, in a few cases where additional ground-based eclipse timing observations extend the durations of the data trains) cover more than three orbital periods were selected to be in the first subgroups of both large categories. One can consider these systems to be the most secure triple candidates. Many of them were put into the most secure subgroups of \citet{borkovitsetal16}; however, the longer outer period members among these new secure subgroups were originally considered to be in the less certain subgroups of \citet{borkovitsetal16} due to the shortness of the dataset. Therefore, these longer period systems demonstrate clearly the benefit of using considerably extended datasets. Naturally, these robust-solution triple star candidates are the ones for which we can expect the most accurate parameters. Most of the members of the medium subgroups (moderately certain solutions) have outer periods shorter than the length of the available data, but longer than one third of the duration of the data train. Moreover, we categorized a few shorter outer period systems into these subgroups, where either the newly obtained TESS ETV points had too large a scatter, and therefore, we were unable to verify securely the former solution of \citet{borkovitsetal16} or, some sections of the ETV points were found to be as outliers. In sum total, the robust and moderately certain subgroups of hierarchical triple star candidates contain 152 (108 LTTE and 44 LTTE+DE) systems. For the remaining 91 EBs (73 and 18) we found only very uncertain solutions or, the period of our third-body solution was longer than the length of the ETV data. Naturally, these are the least certain cases. 

In the process of our analysis of the 243 relatively compact triple systems studied with \textit{Kepler} and TESS, we have collected some interesting general statistics on this type of system. The results are summarized in Figs.~\ref{fig:P1vsP2} to \ref{fig:Papse}.

The securely determined outer periods, $P_2$, ranged from 45 to $\approx 5000$ days (Fig.~\ref{fig:P1vsP2}).  Approximately 3/4 of the systems exhibit dominant LTTE orbits, while the other 1/4 also have nicely detectable dynamical effects.  The latter help lead to more extractable information about the systems, including more and better mass determinations as well as the dynamically important mutual inclination. The outer period distribution is approximately flat per logarithmic interval from about 300 to 5000 days, but distinctly falls off toward shorter $P_2$ (Fig.~\ref{fig:logP2}). This lack of very short outer period triples is likely due in part to a manifestation of the W~UMa desert, i.e. the absence of short outer period ternary components around W~UMa-type overcontact binaries. On the other hand, however, we cannot exclude the possibility, that some spurious ETV distortions might also hide the low amplitude signals of the shortest outer period ternary companions.  Regardless of the outer period over two orders of magnitude, almost any outer orbital eccentricity can be found (Figs.~\ref{fig:e2vsP2} and \ref{fig:e2}),  The vast majority of systems, though, have $e_2 \gtrsim 0.1$, and the mean value is close to $e_2 \simeq 0.3$. 

The masses of the inner binaries ($m_\mathrm{AB}$) are mostly in the range of $1.3-3 \,$M$_\odot$ (see Fig.~\ref{fig:mbinary}).  To an extent, this is something of a selection effect in that there are simply more low-mass binaries because the majority of the \textit{Kepler} targets were chosen to be lower-mass systems, that might be more hospitable to life.  Also, in the accretion scenario formation of triples (see, e.g., \citealt{kostovetal24}, and references therein), the hierarchical components tend toward equal masses, i.e., $q_1 \approx q_2 \approx 1$, and thus the masses of the individual binary stars may be only about 1/4 of the system mass.

The relation between the mass of the third body ($m_\mathrm{C}$) and the binary mass is explored in Figs.~\ref{fig:qout} and \ref{fig:Mc_Mab}, and continues to follow the `rule' that $q_2$ generally lies between 0 and 1, with only a few systems where the tertiary is more massive than the inner binary.  Accretion onto multistellar systems during their formation, naturally tends toward more mass accumulation onto the lower mass object at each hierarchical level, thereby leading to nearly equal masses \citep[see, e.~g.,][and further references therein]{tokovinin21}.

The orbits of the LTTE+DE systems tend to be at least somewhat coplanar, with $i_{\rm m} \lesssim 20\degr$ (see Fig.~\ref{fig:mut_incl}).  This may also be a signature of a formation scenario involving accretion onto the triple.  However, some 14 systems have $i_{\rm m} \gtrsim 30\degr$, and a few may even have $i_{\rm m}$ large enough to drive von Zeipel-Lidov-Kozai cycles.

We also found that the inferred apse-nodal timescales in our our systems (Fig.~\ref{fig:Papse}) are distributed roughly evenly in log space between 10 and $10^5$ years. Since these scale roughly as $P_2^2/P_1$ we do indeed anticipate a rather wide range of these long-term secular timescales.

Finally, we find that some 9\% of all the \textit{Kepler} binaries are quite reasonable candidates for triple star systems.  These relatively compact triples span some 1.7 decades in semimajor axes.  If all triples span the range of $\sim$$1/3$ AU \citep{kostovetal24} to about 0.1 pc (20,000 AU), then this represents about 4.8 decades in log space.  Thus, we might conclude that there are approximately three times (i.e., 4.8/1.7) as many binaries that actually have third body companions, or perhaps up to 25\% are triples! Here we have specifically adopted the assumption of an approximately logarithmically flat distribution for the outer periods of triples, as is often taken to be the case for binaries.

\begin{acknowledgements}

This project has received funding from the HUN-REN Hungarian Research Network.

T.\,B., T.\,M., I.\,B.\,B., E.\,F.-D, B.\,S. and A.\,P. acknowledge the financial support of the Hungarian National Research, Development and Innovation Office -- NKFIH Grants K-147131, K-143986 and K-138962.

This research has made use of data collected by the \textit{Kepler} mission, which is funded by the NASA Science Mission directorate.  

This paper makes extensive use of data collected by the TESS mission. Funding for the TESS mission is provided by the NASA Science Mission directorate. Some of the data presented in this paper were obtained from the Mikulski Archive for Space Telescopes (MAST). STScI is operated by the Association of Universities for Research in Astronomy, Inc., under NASA contract NAS5-26555. Support for MAST for non-HST data is provided by the NASA Office of Space Science via grant NNX09AF08G and by other grants and contracts.

We  used the  Simbad  service  operated by  the  Centre des  Donn\'ees Stellaires (Strasbourg,  France) and the ESO  Science Archive Facility services (data  obtained under request number 396301).  

This research has also made use of the VizieR catalogue access tool, CDS, Strasbourg, France (DOI : 10.26093/cds/vizier). The original description of the VizieR service was published in \citet{ochsenbein00}.

This research made use of Lightkurve, a Python package for Kepler and TESS data analysis \citep{lightkurve18}.

\end{acknowledgements}

\begin{appendix}

\onecolumn

\section{Tabulated results}
\FloatBarrier

\begin{table*}[h!]
\begin{center}
\caption{Ground-based follow up mid-eclipse-times for KIC targets}
 \label{tab:KICToM}
{\tiny
\begin{tabular}{@{}lrllrllrllrl}
\hline\hline
BJD & Type  & std. dev. & BJD & Type  & std. dev. & BJD & Type  & std. dev. & BJD & Type  & std. dev. \\ 
$-2\,400\,000$ &     &   \multicolumn{1}{c}{$(d)$} & $-2\,400\,000$ &     &   \multicolumn{1}{c}{$(d)$} & $-2\,400\,000$ &     &   \multicolumn{1}{c}{$(d)$} & $-2\,400\,000$ &     &   \multicolumn{1}{c}{$(d)$}\\ 
\hline
\multicolumn{3}{c}{\object{KIC 1873918}} &  \multicolumn{3}{c}{\object{KIC 4848423}} &  \multicolumn{3}{c}{\object{KIC 5962716}} & \multicolumn{3}{c}{\object{KIC 6543674}}  \\
58275.4451  & s & 0.0001 	&  58250.49338  & p &      0.00003 &  58574.5617   & p &      0.0002  & 57913.44125  & p &	0.00007  \\
58337.4361  & p & 0.0001 	&  58621.43765  & s &      0.00006 &  58695.4683   & p &      0.0002  & 58347.42088  & s &	0.00003  \\   
58645.4393  & s & 0.0001 	&  58642.46303  & s &      0.00003 &  58780.2845   & p &      0.0001  & 58366.54929  & s &	0.00009  \\
58688.4876  & p & 0.0001 	&  58663.48887  & s &      0.0000  &  \multicolumn{3}{c}{KIC 6265720} & 58396.43497  & p &	0.00003  \\		       
58710.4287  & p & 0.0001 	&  \multicolumn{3}{c}{\object{KIC 4859432}} &  58246.4782   & p &      0.0004  & \multicolumn{3}{c}{\object{KIC 8043961}}  \\		     
\multicolumn{3}{c}{\object{KIC 2302092}} &  58774.34707  & p &      0.00009 &  58339.4264   & s &      0.0003  & 56497.39813  & s &      0.00006  \\ 
58340.4714  & p & 0.0001	&  58779.35794  & p &      0.00007 &  58339.5839   & p &      0.0008  & 56511.43111  & s &      0.00007  \\
58393.3626  & s & 0.0003        &  \multicolumn{3}{c}{\object{KIC 4945857}} &  58405.3461   & s &      0.0002  & 56539.49708  & s &      0.00004  \\  
58403.3789  & s & 0.0003	&  58315.47527  & s &      0.00009 &  58758.3837   & s &      0.0003  & 56543.39568  & p &      0.00003  \\
58669.4686  & s & 0.0002	&  58370.4822	& s &      0.0002  &  \multicolumn{3}{c}\object{{KIC 6525196}} & 56593.28992  & p &      0.00005  \\		  
\multicolumn{3}{c}{\object{KIC 2305372}} &  58590.5146	& s &      0.0006  &  56464.5508   & s &      0.0001  & 57176.43565  & p &      0.00008  \\
58289.50040 & p & 0.00004	&  58661.4584	& p &      0.0002  &  56476.52231  & p &      0.00002 & 57915.5006   & p &      0.0002   \\     
58360.4413  & s & 0.0006        &  \multicolumn{3}{c}{\object{KIC 5039441}} &  56837.39400  & s &      0.00003 & 57989.56568  & s &      0.00007  \\   
58778.34299 & p & 0.00003	&  58352.38589  & p &      0.00007 &  57208.42765  & p &      0.00004 & 58258.52952  & p &      0.00006  \\
\multicolumn{3}{c}{\object{KIC 3228863}} &  58666.48991  & p &      0.00009 &  57232.47194  & p &      0.00004 & 58343.5031   & s &      0.0001   \\
58217.56430 & s & 0.00009	&  \multicolumn{3}{c}{\object{KIC 5269407}} &  57340.22376  & s &      0.00005 & 58569.5916   & s &      0.0001   \\
58228.52691 & s & 0.00004	&  58374.37855  & p &      0.00009 &  57574.52843  & p &      0.00004 & 58644.4293   & s &      0.0002   \\
58338.53168 & p & 0.00002       &  \multicolumn{3}{c}{\object{KIC 5478466}} &  58643.47098  & s &      0.00004 & 58665.4834   & p &      0.0002   \\
58359.36546 & s & 0.00003	&  58662.4929   & s &      0.0004  &  58655.44256  & p &      0.00006 & \multicolumn{3}{c}{\object{KIC 10275197}} \\
58401.39363 & p & 0.00002       &  \multicolumn{3}{c}{\object{KIC 5513861}} &  58708.45968  & s &      0.00004 & 60436.4157   & p &	0.0002   \\
58592.53866 & s & 0.00006	&  58280.48273  & p &      0.00005 &  58732.40347  & s &      0.00004 & 60441.49727  & p &	0.00003  \\		    
58731.41832 & s & 0.00006	&  58379.40199  & s &      0.00008 &  58756.34875  & s &      0.00005 & 60443.45062  & p &	0.00002  \\ 
58768.33109 & p & 0.00002       &  58647.46959  & p &      0.00007 &  58708.45968  & s &      0.00004 & 60444.42812  & s &	0.00006  \\
\multicolumn{3}{c}{\object{KIC 4244929}} &  58718.45047  & p &      0.00006 &  58732.40347  & s &      0.00004 & 60451.46374  & s &	0.00003  \\
58769.4034  & p & 0.0004        &  \multicolumn{3}{c}{\object{KIC 5621294}} &  58756.34875  & s &      0.00005 & 60498.55788  & p &	0.00004  \\
\multicolumn{3}{c}{\object{KIC 4647652}} &  58320.47631  & p &      0.00007 &               &   &              & 60510.47887  & s &	0.00004  \\			 
58675.5072  & p & 0.0002	&  \multicolumn{3}{c}{\object{KIC 5731312}} &               &   &              & 60517.51454  & s &	0.00002  \\
            &   &               &  58782.3491	& p &	   0.0002  &               &   &              & \multicolumn{3}{c}{\object{KIC 10296163}} \\
	    &   &               &  59346.5425	& p &	   0.0001  &               &   &              & 59068.56425  & p &	0.00006  \\
\hline								     
\end{tabular}}
\end{center}
\tablefoot{Observations before JD~2\,458\,915 were taken with the former 50-cm telescope of the Baja Astronomical Observatory, Hungary (BAO), while the newer ones were obtained with the new 80-cm RC telescope of BAO. The only exception is the sole primary eclipse of \object{KIC~10296163} which was observed with the 1-m RCC telescope of the Piszk\'es-tet\H o Station of the Konkoly Observatory, Hungary. Letters `p' and `s' refer to primary and secondary eclipses, respectively.}
\end{table*}

\setlength{\tabcolsep}{3.3pt}
\begin{table*}[h!]
\begin{center}
\caption{New, significantly improved, or confirmed, certain LTTE solutions.} 
\label{Tab:OrbelemLTTE1}  
\scalebox{0.91}{\begin{tabular}{lccccccccccc} 
\hline\hline
KIC No. & $P_1$ & $\Delta P_1$ & $P_2$ & $a_\mathrm{AB}\sin i_2$ & $e_2$ & $\omega_2$ & $\tau_2$ & $f(m_\mathrm{C})$ & $(m_\mathrm{C})_\mathrm{min}$ & $\frac{{\cal{A}}_\mathrm{dyn}}{{\cal{A}}_\mathrm{LTTE}}$ & $m_\mathrm{AB}$\\
        & (day) &$\times10^{-10}$ (d/c)&(day)&(R$_\odot$)  &       &   (deg)    &   (MBJD) & (M$_\odot$)       & (M$_\odot$)            & &  (M$_\odot$)    \\
\hline
1873918\tablefootmark{a}&0.33243308(5)&$-0.9$(1)&840.4(7)&98(2)& 0.67(2)& 84(2)   & 55064(5)   & 0.0177(8)   & 0.41 & 0.008& 1.56* \\
2302092\tablefootmark{b}&0.29467345(1)&$-3.23$(4)&973.1(4)&171.9(4)&0.448(5)&114.5(6)&55127(2) & 0.0719(6)   & 0.69 & 0.001& 1.44* \\
2835289  & 0.85776033(6)&$-$ & 754(2)   & 153(10)& 0.79(6) & 298(4)  & 54932(11)  & 0.08(2)     & 0.89 & 0.13 & 2.0: \\
3228863\tablefootmark{c}&0.73094352(2)&0.150(5)&653.3(4)&82.0(5)& 0.06(1)& 317(13)& 54862(23)  & 0.0173(3)   & 0.68 & 0.002& 3.59 \\
\emph{3430883}&0.370604859(5)&$-$&849.9(8)&97.4(5)& 0.33(1) & 261(2)  & 54550(5)   & 0.0171(3)   & 0.42 & 0.002& 1.67* \\
4069063  & 0.50429446(4)&4.84(7)&870.4(4)& 243(4)& 0.68(2) & 129(1)  & 55012(3)   & 0.25(1)     & 1.37 & 0.02 & 1.82 \\
4138301\tablefootmark{d}&0.25337853(2)&2.69(5)&788.5(6)&117.7(7)&0.368(9)& 222(1) & 55083(3)   & 0.0352(6)   & 0.48 & 0.002& 1.31* \\ 
4451148\tablefootmark{e}&0.73598113(4)&8.5(3)&749.8(2)&140.9(6)&0.317(7)& 52(1)   & 55019(3)   & 0.0667(9)   & 0.81 & 0.01 & 2.0: \\
4547308  &0.57692819(2) &$-1.61$(4)&871.1(3)&103(3)&0.90(2)& 56(2)   & 55079(7)   & 0.019(2)    & 0.53 & 0.06 & 2.23* \\
4647652  &1.064824942(6)&$-$ & 755.5(3) &105.9(2)&0.286(2) & 29.7(5) & 54757(1)   & 0.0279(1)   & 0.57 & 0.02 & 2.0: \\
4670267\tablefootmark{f}&2.00609767(8)&$-12$(2)&767(1)&27.8(6)& 0.55(3) & 74(4)   & 55066(9)   & 0.00049(3)  & 0.13 & 0.19 & 2.0: \\
4848423  & 3.00360202(3)&$-$ & 1359.1(6)&267.7(8)&0.097(5) & 56(5)   & 55598(18)  & 0.139(1)    & 1.19 & 0.05 & 2.30 \\
4945857\tablefootmark{g}&0.33541574(1)&4.48(3)&1242.1(2)&303.0(4)&0.312(2)&329.0(4)&54431(2)   & 0.242(1)    & 1.24 & 0.001& 1.57* \\
5039441  &2.151382814(6)&$-$ & 676.8(7) & 87.3(6)& 0.26(1) & 158(3)  & 55209(6)   & 0.0195(4)   & 0.41 & 0.09 & 1.46 \\
5128972  & 0.50532315(1)&0.81(2)&444.64(6)&117.3(4)&0.259(7)&280(1)  & 54929(2)   & 0.109(1)    & 1.01 & 0.008& 2.05* \\
5216727  &1.513022966(8)&$-$ & 532.2(3) & 29.1(5)& 0.45(3) & 128(3)  & 55158(5)   & 0.00116(6)  & 0.18 & 0.14 & 2.0: \\
5264818  & 1.90506184(2)&$-$ & 302.1(3) & 73(2)  & 0.37(4) & 188(8)  & 54928(7)   & 0.056(6)    & 1.31 & 0.30 & 5.0: \\
         &              &    & 3627(234)& 407(49)& 0.84(5) & 177(7)  & 53940(258) & 0.07(2)     & 2.08 &      & 6.0: \\
5310387  &0.441668563(5)&3.243(8)&215.68(8)&12.6(2)&0.29(3)& 204(6)  & 55043(4)   & 0.00058(3)  & 0.13 & 0.03 & 1.87* \\
5376552  &0.503818523(4)&1.806(7)&334.64(4)&42.0(2)&0.359(6)&353.6(9)& 54874.1(9) & 0.0089(1)   & 0.34 & 0.02 & 1.73 \\
5459373  &0.286608880(6)&$-0.805$(7)&412.6(1)&100.2(4)&0.330(8)&266(1)&55046(2)   & 0.079(1)    & 0.71 & 0.005& 1.42* \\
5478466  & 0.48250107(1)&$-3.34$(2)&739.0(4)&90.7(7)&0.504(9)& 15(1) & 54958(2)   & 0.0183(4)   & 0.48 & 0.01 & 1.98* \\
5513861  & 1.51020968(1)&$-$ & 2176.1(7)&303.2(9)& 0.123(4)& 190(2)  & 54069(14)  & 0.0789(7)   & 1.06 & 0.006& 2.82 \\
5962716  & 1.80458299(5)&$-$ & 1900(6)  &205.1(5)& 0.497(3)& 254.6(4)& 55807(4)   & 0.0321(3)   & 0.60 & 0.02 & 2.0: \\
6144827  &0.234650014(2)&$-$ & 229.4(4) & 33.3(2)& 0.17(6) & 34(18)  & 54907(12)  & 0.009(1)    & 0.28 & 0.004& 1.25* \\
         &              &    & 3719(42) & 98(9)  & 0.35(7) & 281(10) & 55684(106) & 0.0009(3)   & 0.21 &      & 3.0: \\   
6370665  & 0.93231405(3)&20.7(1)&283.3(3)&27.3(7)& 0.12(5) & 113(22) & 54846(17)  & 0.0034(3)   & 0.34 & 0.03 & 3.08* \\
6543674  &2.391030595(7)&$-$ &1100.2(3) &115.5(3)& 0.623(4)& 266.4(4)& 55037(1)   & 0.0171(1)   & 0.51 & 0.19 & 2.30 \\
6669809\tablefootmark{h}&0.733738633(2)&$-$& 192.6(1)& 26(1)  & 0.22(5) & 116(6)  & 54984(4)   & 0.0062(9)   & 0.34 & 0.06 & 2.16 \\
         &              &    & 4467(22) & 53(9)  & 0.98(2) & 105(15) & 56058(243) & 0.00010(5)  & 0.10 &      & 3.0: \\ 
6965293  & 5.0777444(1) &$-$ & 1118(1)  &197.6(6)& 0.203(6)& 313(2)  & 54719(6)   & 0.0826(8)   & 0.88 & 0.23 & 2.0: \\
7362751\tablefootmark{i}&0.33824935(2)&$-2.7$(1)&549.9(2)&111.3(8)&0.23(1)&125(3) & 54963(5)   & 0.061(1)    & 0.68 & 0.003& 1.57* \\
7385478  & 1.65547248(2)&$-$ & 1020(2)  & 68(1)  & 0.54(3) & 120(4)  & 55058(11)  & 0.0040(3)   & 0.28 & 0.05 & 2.08 \\
7630658  & 2.15115570(2)&$-$ & 917.3(6) & 178(1) & 0.652(6)& 325.9(5)& 55355(2)   & 0.090(2)    & 0.91 & 0.14 & 2.0: \\
7685689\tablefootmark{j}&0.325158458(9)&6.09(4)&514.6(2)&81.8(3)&0.107(6)& 171(3) & 54779(5)   & 0.0277(3)   & 0.48 & 0.001& 1.53* \\
8043961  &1.559212810(7)&$-$ & 478.74(9)& 82.4(4)& 0.239(9)& 11(2)   & 54815(3)   & 0.0327(5)   & 0.84 & 0.06 & 3.41 \\
\emph{8047291}&0.77544590(3)&$-2.71$(8)&834.1(5)&124.6(5)&0.157(8)&322(3)& 54601(7)& 0.0373(5)   & 0.64 & 0.004& 2.0: \\
8094140  &0.706428451(3)&$-$ & 684.2(5) & 56.9(5)& 0.38(1) & 168(2)  & 54755(4)   & 0.0053(1)   & 0.21 & 0.02 & 1.08 \\
8145477  & 0.56578338(3)&5.57(7)&392.2(2)& 62(1) & 0.50(2) & 185(2)  & 54880(2)   & 0.021(1)    & 0.54 & 0.04 & 2.20* \\
8190491  & 0.77787564(7)&18.1(2)&616(1) & 68(2)  & 0.54(5) & 60(5)   & 54790(10)  & 0.011(1)    & 0.40 & 0.04 & 2.0: \\
8330092\tablefootmark{k}&0.321723711(8)&$-0.31$(1)&563.6(5)&52.8(5)&0.13(2)& 24(7)& 55184(12)  & 0.0062(2)   & 0.27 & 0.003& 1.52* \\
8386865\tablefootmark{l}&1.2580440(2)&$-51$(3)&293.9(1)&96(3) & 0.55(4) & 322(4)  & 55033(3)   & 0.14(1)     & 1.10 & 0.22 & 2.0: \\
8394040  &0.302126048(7)&0.882(8)& 387.50(7)&123.6(5)&0.513(6)&293.4(7)&54810.1(8)& 0.169(2)    & 1.01 & 0.01 & 1.46* \\
8904448  &0.865982603(3)&3.2(1)&543.6(4)& 77(1)  & 0.54(2) & 305(2)  & 54786(4)   & 0.0204(9)   & 0.50 & 0.06 & 2.0: \\
8957887\tablefootmark{m}&0.34735527(2)&$-6.91$(7)&778.3(6)&185.2(8)&0.463(6)&144.7(7)&55262(2) & 0.141(2)    & 0.98 & 0.004& 1.60* \\ 
\hline
\end{tabular}}
\end{center}
\tablefoot{ Note, in this and also the forthcoming five tables \emph{italic KIC numbers} refer to those systems which were not included in the study of \citet{borkovitsetal16}. In the last column of the current and the forthcoming two tables, the $m_\mathrm{AB}$ values denoted with $*$ and $:$ refer to estimated values with the use of either the formulae of \citet{gazeasstepien08} or our own reasonable estimations, respectively. Values without such additional characters are taken from the literature, and referenced in Appendix~\ref{app:Notesonindividualsystems}.\\
\tablefoottext{a}{Cubic coefficient: $c_3=6.7(2)\times10^{-15}$\,d/c$^2$;}\tablefoottext{b}{Cubic coefficient: $c_3=8.4(1)\times10^{-15}$\,d/c$^2$;} \tablefoottext{c}{Cubic coefficient: $c_3=-7.8(5)\times10^{-15}$\,d/c$^2$;} \tablefoottext{d}{Cubic coefficient: $c_3=-3.3(1)\times10^{-15}$\,d/c$^2$;} \tablefoottext{e}{Cubic coefficient: $c_3=-6.4(2)\times10^{-14}$\,d/c$^2$;} \tablefoottext{f}{Cubic coefficient: $c_3=2.4(2)\times10^{-13}$\,d/c$^2$;} \tablefoottext{g}{Cubic coefficient: $c_3=-8.1(1)\times10^{-15}$\,d/c$^2$;} \tablefoottext{h}{The fourth-body orbit looks unphysical, it should be considered only as a mathematical description of the longer timescale non-linearities in the ETV;} \tablefoottext{i}{Cubic coefficient: $c_3=7.3(2)\times10^{-15}$\,d/c$^2$, see Appendix~\ref{app:Notesonindividualsystems} about the uncertain identification of the target;} \tablefoottext{j}{Cubic coefficient: $c_3=-1.40(1)\times10^{-14}$\,d/c$^2$;} \tablefoottext{k}{The third star is a $\delta$ Scuti/$\gamma$~Dor pulsator, which makes it possible to estimate the mutual inclination of the inner and outert orbits. See detailed discussion in Appendix~\ref{app:Notesonindividualsystems};} \tablefoottext{l}{Cubic coefficient: $c_3=4.4(2)\times10^{-13}$\,d/c$^2$;} \tablefoottext{m}{Cubic coefficient: $c_3=2.13(2)\times10^{-14}$\,d/c$^2$.}}
\end{table*}

\addtocounter{table}{-1}

\setlength{\tabcolsep}{3.3pt}
\begin{table*}
\begin{center}
\caption{continued} 
\scalebox{0.91}{\begin{tabular}{lccccccccccc} 
\hline\hline
KIC No. & $P_1$ & $\Delta P_1$ & $P_2$ & $a_\mathrm{AB}\sin i_2$ & $e_2$ & $\omega_2$ & $\tau_2$ & $f(m_\mathrm{C})$ & $(m_\mathrm{C})_\mathrm{min}$ & $\frac{{\cal{A}}_\mathrm{dyn}}{{\cal{A}}_\mathrm{LTTE}}$ & $m_\mathrm{AB}$\\
        & (day) &$\times10^{-10}$ (d/c)&(day)&(R$_\odot$)  &       &   (deg)    &   (MBJD) & (M$_\odot$)       & (M$_\odot$)            & &  (M$_\odot$)    \\
\hline
9075704  & 0.51315131(2)&0.56(4)&401.9(2)&68.8(6)& 0.07(2) & 217(15) & 55044(17)  & 0.0270(7)   & 0.56 & 0.003& 2.0: \\
9402652  &1.073106823(1)&$-$ & 1498.5(1)&162.7(2)& 0.813(2)&112.70(8)& 54844.7(6) & 0.0257(1)   & 0.74 & 0.03 & 3.21 \\
9665086  & 0.29653680(3)&$-1.24$(4)&854(1)&230(2)& 0.41(2) &  79(3)  & 54692(6)   & 0.225(7)    & 1.16 & 0.002& 1.48 \\ 
9711751  &1.711528191(8)&$-$ & 1185.5(4)&218.1(1)&0.2573(7)& 350.8(2)& 55384.5(7) & 0.0989(1)   & 0.95 & 0.02 & 2.0: \\
9722737\tablefootmark{a}&0.418528187(9)&0.61(5)&444.20(7)&102.9(2)&0.194(4)&220(1)& 54908(2)   & 0.0741(5)   & 0.79 & 0.004& 1.81* \\
9777984\tablefootmark{b}&0.258502522(8)&$-5.98$(3)&95.72(2)&21.6(3)&0.18(2)&255(8)& 54977(2)   & 0.0148(6)   & 0.35 & 0.02 & 1.32* \\
9838047\tablefootmark{c}&0.43615810(3)&25.8(1)&1119.3(6)&202.5(8)&0.275(7)& 169(1)& 55039(5)   & 0.089(1)    & 0.87 & 0.002& 1.86* \\
9912977  &1.887873127(6)&$-$ & 754.7(3) & 47.6(3)& 0.27(1) & 74(3)   & 55164(6)   & 0.00254(5)  & 0.19 & 0.06 & 1.42 \\ 
10226388\tablefootmark{d}&0.66065717(2)&13.0(2)&941.8(2)&206.8(4)&0.266(4)&108.8(9)& 54731(2) & 0.1336(8)   & 1.21 & 0.004& 2.44* \\
10275197\tablefootmark{e}&0.39084692(2)&$-9.45(8)$&2111.0(4)&594.7(7)&0.271(2)&215.8(3)&54868(2)& 0.633(2)   & 2.10 &0.0005& 1.73* \\
10383620 &0.734568334(1)&$-$ & 1504.8(1)&270.0(2)& 0.206(1)&  0.9(3) & 54286(1)   & 0.1152(2)   & 1.02 & 0.003& 2.0: \\ 
10583181 & 2.69635380(1)&$-$ & 1170.2(3)&154.6(3)& 0.072(4)& 93.7(3) & 54485(10)  & 0.0362(2)   & 0.73 & 0.05 & 2.52 \\
10724533 &0.74509315(3) &0.93(7)&1874(2)& 53.6(4)& 0.20(1) &  7(4)   & 54029(21)  & 0.00059(1)  & 0.14 & 0.002& 2.0: \\
10727655 &0.353364979(3)&1.024(4)&1145.9(2)&143.7(2)&0.260(3)&36.1(6)& 55057(2)   & 0.0303(1)   & 0.52 & 0.001& 1.62* \\
10991989 & 0.97447768(1)&$-$ & 546.7(3) & 106(1) & 0.37(2) & 34(3)   & 54970(5)   & 0.053(2)    & 0.71 & 0.04 & 1.90 \\
11042923 &0.390161206(7)&4.60(1)&1098.6(3)&125.6(4)&0.255(6)& 176(1) & 54438(4)   & 0.0220(2)   & 0.47 & 0.002& 1.73* \\
11234677 & 1.58741758(5)&$-$ & 1793(6)  &139.3(5)& 0.224(5)& 156(1)  & 55585(7)   & 0.0113(2)   & 0.40 & 0.01 & 2.0: \\
11968490 &1.078890278(9)&$-$ & 254.37(5)&111.5(8)& 0.38(1) & 282(2)  & 54859(2)   & 0.287(6)    & 1.53 & 0.15 & 2.0: \\
12019674 &0.354497306(2)&$-$ & 2700(2)  &411.2(5)& 0.121(3)& 187(2)  & 52957(15)  & 0.1279(5)   & 0.94 &0.0003& 1.62* \\
12055255\tablefootmark{f}&0.22094204(2)&$-2.04$(4)&1628.2(5)&252(1)&0.397(3)&272.1(7)& 55173(3)& 0.081(1)    & 0.65 &0.0003& 1.20* \\
12071741\tablefootmark{g}&0.31426748(3)&$-15.2$(1)&1207(2)&234(4)&0.79(1)&144.3(8)& 54644(4)   & 0.118(6)    & 0.87 & 0.004& 1.50* \\
\hline
\end{tabular}}
\end{center}
\tablefoot{\tablefoottext{a}{Cubic coefficient: $c_3=7.2(1)\times10^{-15}$\,d/c$^2$;}\tablefoottext{b}{Cubic coefficient: $c_3=3.3(1)\times10^{-15}$\,d/c$^2$;}\tablefoottext{c}{Cubic coefficient: $c_3=-6.45(4)\times10^{-14}$\,d/c$^2$;}\tablefoottext{d}{Cubic coefficient: $c_3=-4.24(6)\times10^{-14}$\,d/c$^2$;}\tablefoottext{e}{Cubic coefficient: $c_3=1.17(2)\times10^{-14}$\,d/c$^2$;}\tablefoottext{f}{Cubic coefficient: $c_3=2.2(1)\times10^{-15}$\,d/c$^2$;}\tablefoottext{g}{Cubic coefficient: $c_3=1.90(3)\times10^{-14}$\,d/c$^2$.} }
\end{table*}

\setlength{\tabcolsep}{3.3pt}
\begin{table*}
\begin{center}
\caption{New or improved, moderately certain LTTE solutions.} 
\label{Tab:OrbelemLTTE2}  
\scalebox{0.91}{\begin{tabular}{lccccccccccc} 
\hline\hline
KIC No. & $P_1$ & $\Delta P_1$ & $P_2$ & $a_\mathrm{AB}\sin i_2$ & $e_2$ & $\omega_2$ & $\tau_2$ & $f(m_\mathrm{C})$ & $(m_\mathrm{C})_\mathrm{min}$ & $\frac{{\cal{A}}_\mathrm{dyn}}{{\cal{A}}_\mathrm{LTTE}}$ & $m_\mathrm{AB}$\\
        & (day) &$\times10^{-10}$ (d/c)&(day)&(R$_\odot$)  &       &   (deg)    &   (MBJD) & (M$_\odot$)       & (M$_\odot$)            & &  (M$_\odot$)    \\
\hline
2305372\tablefootmark{a}&1.40471606(6)&$-200$(3)&4859(11)&582(5)&0.333(4)& 10(1)  & 55190(16)  & 0.112(3)    & 0.94 & 0.002& 1.80 \\
2450566  & 1.8445880(3) &$-$ & 944(4)   & 191(14)& 0.46(11)& 135(14) & 55033(38)  & 0.11(2)     & 0.98 & 0.08 & 2.0: \\
2715007  &0.297110446(4)&$-$ & 1949(2)  & 317(2) & 0.649(4)& 212.0(4)& 54423(2)   & 0.113(2)    & 0.84 &0.0009& 1.45* \\
2708156  &1.891268848(1)&$-$ & 2142(9)  &  48(1) & 0.13(1) & 330.5(9)& 54233(10)  & 0.00033(2)  & 0.21 & 0.008& 4.97 \\
         &              &    &35365(139)& 661(13)& 0.666(8)&  3(4)   & 58218(421) & 0.0031(2)   & 0.64 &      & 6.0: \\ 
3114667  & 0.88858371(2)&$-1.97$(7)&865(3)& 28(2)& 0.79(3) & 8(1)    & 55011(5)   & 0.00039(7)  & 0.12 & 0.18 & 2.0: \\
4074708\tablefootmark{b}&0.302116333(6)&1.31(1)&1196(1)&30.7(1)&0.23(1) & 359(2)  & 54644(7)   & 0.000270(4) & 0.09 &0.0009& 1.46* \\
\emph{4241946}&0.284384038(6)&1.218(6)&1718(1)&54.3(5)&0.601(9)&6.9(9)& 54279(5)   & 0.00073(2)  & 0.12 & 0.001& 1.41* \\
4244929\tablefootmark{c}&0.3414074(2)&13.4(3)&2652(3)& 431(7) & 0.264(6)& 288(2)  & 55196(13)  & 0.152(8)    & 1.01 &0.0003& 1.58* \\
4732015FP\tablefootmark{d}&0.9388524(2)&30(2)&1726(2)& 278(1) & 0.285(7)& 73(2)   & 55511(10)  & 0.097(1)    & 0.94 & 0.003& 2.0: \\
4758368  & 3.7499865(2) &$-$ & 3769(8)  & 459(9) & 0.77(2) & 320(2)  & 55574(24)  & 0.091(5)    & 1.05 & 0.07 & 2.50 \\
\emph{4851217}&2.47028868(3)&$-$& 2658(8)& 126(4) & 0.62(3) & 207(1)  & 54141(14)  & 0.0038(4)   & 0.43 & 0.03 & 4.09 \\ 
4859432\tablefootmark{e}&0.38547921(1)&1.11(1)&741.7(7)&66.8(4)& 0.56(1)& 274(1)  & 54647(3)   & 0.0073(1)   & 0.31 & 0.008&1.71* \\
4945588  & 1.12907978(2)&$-$ & 1606(2)  & 197(6) & 0.82(2) & 310(2)  & 54983(10)  & 0.040(3)    & 0.65 & 0.04 & 2.0: \\
\emph{5113053}&3.1850947(1)&$-$&3184(10) & 297(4) & 0.27(2) & 251(4)  & 55606(40)  & 0.035(1)    & 0.67 & 0.02 & 2.28 \\
5269407  & 0.9588568(1) &$-$ & 3165(36) & 384(4) & 0.687(6)& 121.9(8)& 55317(11)  & 0.076(3)    & 0.85 & 0.004& 2.0: \\
5353374\tablefootmark{f}&0.39332102(2)&4.11(9)&3629(18)&76.0(6)&0.462(9)& 330(1)  & 55263(17)  & 0.00045(1)  & 0.12 &0.0003& 1.74* \\
5611561\tablefootmark{g}&0.258694383(7)&1.51(2)&993(1)&40.0(2)& 0.15(1) & 318(5)  & 55304(13)  & 0.00087(2)  & 0.12 &0.0004& 1.33* \\
5903301  & 2.3203027(3) &$-$ & 1260(11) & 152(3) & 0.40(3) &  21(3)  & 54970(14)  & 0.030(2)    & 0.58 & 0.05 & 2.0: \\
6187893  & 0.78918203(2)&$-$ & 3901(6)  & 561(4) & 0.185(7)& 135(3)  & 53342(34)  & 0.155(3)    & 1.16 &0.0004& 2.0: \\
6265720\tablefootmark{h}&0.31242645(2)&6.78(5)&1372.2(4)&225(1)&0.601(5)& 185.1(6)& 55314(2)   & 0.081(1)    & 0.74 & 0.002& 1.50* \\
6516874  & 0.91632504(2)&$-$ & 908(1)   & 103(2) & 0.26(4) & 74(8)   & 54668(20)  & 0.018(1)    & 0.39 & 0.01 & 1.45 \\ 
6606282  & 2.10713173(7)&$-$ & 1647(3)  & 328(2) & 0.30(1) & 156(2)  & 54213(11)  & 0.175(4)    & 1.22 & 0.02 & 2.0: \\
7119757  & 0.74292031(3)&2.53(7)&1250(2)& 160(1) & 0.561(8)& 182.5(7)& 54603(3)   & 0.0351(8)   & 0.62 & 0.008& 2.0: \\
\emph{7431703}&0.572525296(2)&$-$&3756(5)& 49.9(3)& 0.388(7)& 309.9(9)& 55835(10)  & 0.000118(2) & 0.08 &0.0005& 2.0: \\ 
7518816  & 0.46658036(1)&2.33(2)&1804(2)& 51.7(4)& 0.17(1) & 241(4)  & 55300(21)  & 0.00057(1)  & 0.14 &0.0009& 2.0: \\
7680593  &0.276391471(3)&$-$ & 3597(3)  & 356(2) & 0.361(4)& 266.0(5)& 55659(6)   & 0.0466(6)   & 0.56 &0.0001& 1.38* \\
8081389  & 1.48944667(2)&$-$ & 4249(19) &  84(1) & 0.49(3) & 93(3)   & 56351(49)  & 0.00045(2)  & 0.13 & 0.003& 2.0: \\
8192840  & 0.43354738(1)&10.15(2)&1017.4(4)&102(2)&0.70(1) & 354.4(7)& 55458(2)   & 0.0137(8)   & 0.41 & 0.008& 1.85*\\ 
8242493\tablefootmark{i}&0.283285313(4)&1.10(1)&1191(1)&30.5(2)&0.135(9)& 133(4)  & 55422(12)  & 0.000269(4) & 0.08 &0.0008& 1.40* \\
8265951  & 0.77995359(1)&$-$ & 2970(5)  & 395(1) & 0.688(3)& 199.2(2)& 55340(2)   & 0.093(1)    & 1.11 & 0.003& 2.73* \\
8429450  & 2.70514407(3)&$-$ & 3575(5)  &142.6(8)& 0.449(8)& 180(1)  & 56109(11)  & 0.00304(5)  & 0.33 & 0.01 & 3.14 \\
8444552  &1.178077718(8)&$-$ & 2548.3(7)&391.4(5)& 0.495(2)& 106.7(4)& 55310(2)   & 0.1237(5)   & 1.05 & 0.004& 2.0: \\
\emph{8509014}&0.303511673(3)&$-$&1696(1)& 227(1) & 0.239(8)& 342(2)  & 54380(8)   & 0.0546(8)   & 0.62 &0.0005& 1.47* \\
8739802  &0.274512860(3)&$-$ &  879(2)  & 39.6(8)& 0.41(3) & 168(3)  & 55318(8)   & 0.00108(7)  & 0.14 & 0.004& 1.38* \\ 
8758161  &1.996416816(8)&$-$ & 4277(2)  &249.6(5)& 0.376(3)& 135.2(5)& 55520(7)   & 0.01140(7)  & 0.40 & 0.005& 2.0: \\ 
8868650  & 4.4474169(3) &$-715$(4)&1931(1)&339(3)& 0.647(8)& 240.1(7)& 55397(4)   & 0.140(3)    & 1.11 & 0.15 & 2.0: \\
9110346  & 1.79057425(2)&$-$ & 2549(2)  & 254(2) & 0.699(8)& 300.5(9)& 55397(7)   & 0.0340(6)   & 0.61 & 0.02 & 2.0: \\
9181877  &0.321014419(4)&$-$ & 3652(2)  & 644(6) & 0.141(8)& 205(2)  & 53563(25)  & 0.269(8)    & 1.28 &0.0001& 1.52*\\
9272276\tablefootmark{j}&0.28061872(2)&$-7.44$(4)&2223.0(3)&423.8(6)&0.366(1)&342.9(2)&55324(1)& 0.2065(8)   & 1.08 &0.0003& 1.40* \\
9392702  & 3.90939116(2)&$-$ & 1499(3)  & 271(3) & 0.41(4) & 273(3)  & 54995(16)  & 0.119(4)    & 1.03 & 0.10 & 2.0: \\
9412114  & 0.25025182(2)&2.70(2)&1286.6(7)&219.1(7)&0.251(5)& 342(1) & 55482(5)   & 0.0852(9)   & 0.70 &0.0005& 1.30* \\
9592145\tablefootmark{k}&0.4888674363(3)&$-$& 796(2) & 8.0(9) & 0.52(8) & 213(40) & 54881(88)  & 0.000011(4) & 0.04 & 0.01 & 2.0: \\
         &              &    & 3585(11) & 72(1)  & 0.900(1)& 198.0(7)& 56711(14)  & 0.00038(2)  & 0.16 &      & 3.0: \\ 
\emph{9882280}&0.289075993(1)&$-$&1458.7(4)&149.8(4)&0.209(5)&105(1)  & 54876(5)   & 0.0212(2)   & 0.42 &0.0006& 1.42* \\           
9994475\tablefootmark{k}&0.318406251(1)&$-$& 605(1)  & 87(2)  & 0.31(5) & 227(5)  & 54854(10)  & 0.024(2)    & 0.45 & 0.003& 1.51* \\
         &              &    & 3267(109)& 108(7) & 0.50(6) & 211(8)  & 55293(86)  & 0.0016(3)   & 0.26 &      & 3.0: \\ 
10268903 & 1.1039767(4) &$-$ & 1228(23) & 209(3) & 0.62(1) & 133(3)  & 54966(19)  & 0.081(4)    & 0.87 & 0.02 & 2.0: \\ 
\hline
\end{tabular}}
\end{center}
\tablefoot{\tablefoottext{a}{Cubic coefficient: $c_3=2.23(3)\times10^{-12}$\,d/c$^2$;}\tablefoottext{b}{Cubic coefficient: $c_3=-4.0(1)\times10^{-15}$\,d/c$^2$;}\tablefoottext{c}{Cubic coefficient: $c_3=-3.94(5)\times10^{-14}$\,d/c$^2$;}\tablefoottext{d}{The true target is TIC~1716037782, a fainter background source in the aperture of KIC~4732015. The cubic coefficient: $c_3=-2.49(9)\times10^{-13}$\,d/c$^2$;}\tablefoottext{e}{The cubic coefficient: $c_3=-1.09(2)\times10^{-14}$\,d/c$^2$;}\tablefoottext{f}{The cubic coefficient: $c_3=-1.06(1)\times10^{-14}$\,d/c$^2$;}\tablefoottext{g}{The cubic coefficient: $c_3=-3.1(1)\times10^{-15}$\,d/c$^2$;}\tablefoottext{h}{The cubic coefficient: $c_3=-1.87(1)\times10^{-14}$\,d/c$^2$;}\tablefoottext{i}{The cubic coefficient: $c_3=-1.6(1)\times10^{-15}$\,d/c$^2$;}\tablefoottext{j}{The cubic coefficient: $c_3=1.01(1)\times10^{-14}$\,d/c$^2$;}\tablefoottext{k}{The fourth-body orbit looks unphysical, it should be considered only as a mathematical description of the longer timescale non-linearities in the ETV.}}
\end{table*}

\setlength{\tabcolsep}{3.3pt}
\begin{table*}
\begin{center}
\caption{Very uncertain LTTE solutions.} 
\label{Tab:OrbelemLTTE3}  
\scalebox{0.91}{\begin{tabular}{lccccccccccc} 
\hline\hline
KIC No. & $P_1$ & $\Delta P_1$ & $P_2$ & $a_\mathrm{AB}\sin i_2$ & $e_2$ & $\omega_2$ & $\tau_2$ & $f(m_\mathrm{C})$ & $(m_\mathrm{C})_\mathrm{min}$ & $\frac{{\cal{A}}_\mathrm{dyn}}{{\cal{A}}_\mathrm{LTTE}}$ & $m_\mathrm{AB}$\\
        & (day) &$\times10^{-10}$ (d/c)&(day)&(R$_\odot$)  &       &   (deg)    &   (MBJD) & (M$_\odot$)       & (M$_\odot$)            & &  (M$_\odot$)    \\
\hline
\emph{2444187}&0.39015981(2)&$-$&6580(215) &  63(3) & 0.32(1)  & 67(6)   & 55774(136)& 0.00008(1) & 0.06 &0.0001& 1.73*\\ 
2715417\tablefootmark{a}&0.2364406260(8)&$-$&  882(28) &  21(1) & 0.47(9)  & 158(13) & 54793(68) & 0.00016(3) & 0.07 & 0.002& 1.25*\\
        &               &     & 2667(89)  &  81(4) & 0.57(7)  & 221(17) & 54951(226)& 0.0010(1)  & 0.22 &      & 3.0: \\
2983113 & 0.39516072(5) & $-$ & 5546(66)  & 284(8) & 0.34(1)  & 25(1)   & 54967(37) & 0.0100(8)  & 0.35 &0.0001& 1.74*\\
\emph{3221207}&0.4738284(1)&$-$& 9652(909) & 254(26)& 0.50(2)  & 155(1)  & 56055(168)& 0.0024(9)  & 0.22 &0.0001& 1.96* \\
\emph{3245644}&0.70880665(2)&$-2.7$(1)&1421(7)&53.3(3)&0.256(5)& 111(2)  & 54559(8)  & 0.00100(2) & 0.17 & 0.003& 2.0: \\
3248019 & 2.6681974(1)  & $-$ & 4106(17)  & 264(46)& 0.54(5)  & 315(15) & 56317(186)& 0.015(8)   & 0.44 & 0.01 & 2.0: \\
3335816 & 7.422096(1)   & $-$ & 6088(189) & 325(13)& 0.47(3)  & 317(5)  & 54950(106)& 0.012(2)   & 0.42 & 0.04 & 2.0: \\
3338660 & 1.8733925(2)  & $-$ &10043(1021)&1037(235)& 0.88(8) & 135(6)  & 59248(740)& 0.15(11)   & 0.87 & 0.01 & 1.24 \\
3440230 & 2.88106459(7) & $-$ & 3016(9)   & 364(8) & 0.86(2)  & 69(3)   & 54962(33) & 0.071(4)   & 0.84 & 0.10 & 2.02 \\
\emph{3936357}\tablefootmark{b}&0.36915379(1)&$-5.00$(4)&2272(6)&37.6(4)&0.17(2)&110(4)&55168(24) & 0.000138(4)& 0.07 &0.0004& 1.67*\\ 
4037163 & 0.635447042(7)& $-$ & 2498(15)  & 92(17) & 0.73(15) & 107(9)  & 54095(63) & 0.0017(9)  & 0.20 & 0.003& 2.0: \\
\emph{4074532}&0.353156917(4)&$-$&2880(2)  &438.0(8)& 0.718(5) & 264.5(2)& 54768(4)  & 0.1357(8)  & 0.97 &0.0007& 1.62* \\
4574310\tablefootmark{c}& 1.30622099(4) &$-23.3$(7)&1930(8)&22.5(2)&0.583(7)& 151(1) & 55906(8)  & 0.000041(1)& 0.05 & 0.01 & 1.69 \\
\emph{4669592}&0.37843058(1)&$-$&6150(81)  & 355(4) & 0.54(1)  & 239.4(4)& 54316(51) & 0.0158(7)  & 0.41 &0.0001& 1.69* \\
4681152 & 1.8359308(7) &118(5)& 2439(23)  & 105(5) & 0.25(3)  & 332(4)  & 55574(34) & 0.0026(4)  & 0.24 & 0.009& 2.0: \\
4762887 & 0.73657411(2) & $-$ & 3365(37)  & 62(7)  & 0.84(7)  & 18(5)   & 55344(54) & 0.0003(1)  & 0.11 & 0.009& 2.0: \\ 
4937217 & 0.42933648(3) & $-$ & 6505(107) & 540(8) & 0.10(1)  & 31(13)  & 53279(236)& 0.050(3)   & 0.67 &0.0001& 1.77 \\
\emph{4940226}&0.37873492(2)&$-$& 5333(47) & 173(2) & 0.255(8) &  71(4)  & 55427(65) & 0.0024(1)  & 0.21 &0.0001& 1.69* \\
\emph{5023948}&3.649282(2)& $-$ & 7295(367)&829(151)& 0.6(1)   & 336(6)  & 53690(169)& 0.14(8)    & 1.17 & 0.01 & 2.17 \\ 
\emph{5080671}&0.47304781(1)&$-$& 5446(85) & 111(2) & 0.58(5)  & 307(4)  & 55176(76) & 0.00061(4) & 0.14 &0.0003& 1.96* \\
\emph{5288543}&3.457101651(5)&$-$&6242(130)& 282(8) & 0.42(2)  & 262(1)  & 54324(91) & 0.0077(7)  & 0.37 & 0.008& 2.20 \\
5307780 & 0.30885011(1) &3.91(1)& 1187(2) &  34(1) & 0.64(3)  & 336(1)  & 55049(10) & 0.00038(4) & 0.10 & 0.002& 1.48* \\ 
5621294\tablefootmark{d}&0.93890770(7)&$-51.2$(7)&3476(46)&36.4(9)&0.66(2) & 305(2)  & 56149(40) & 0.000054(4)& 0.07 & 0.003& 2.64 \\ 
5956776 & 0.569115865(5)& $-$ & 1388(6)   & 29.1(7)& 0.28(5)  & 22(9)   & 54489(34) & 0.00017(1) & 0.09 & 0.002& 2.0: \\
5975712 & 1.13607609(4) & $-$ & 4959(8)   & 562(5) & 0.76(1)  & 126(1)  & 55564(19) & 0.097(2)   & 0.94 & 0.004& 2.0: \\
\emph{6050116}&0.239906813(4)&1.312(6)&2440(3)&82.1(5)&0.354(6)& 271.3(8)& 53814(7)  & 0.00125(2) & 0.13 &0.0002& 1.26* \\
6103049 & 0.6431774(2)  & $-$ & 5764(119) & 545(17)& 0.14(2)  & 119(7)  & 56739(130)& 0.065(7)   & 0.72 &0.0002& 1.68  \\
\emph{6118779}&0.3642382(4)&$-$&10542(882) &1627(125)&0.42(5)  & 218(2)  & 57241(262)& 0.52(15)   & 1.86 &0.00003&1.65* \\
6281103 & 0.363284978(6)& $-$ & 2717(3)   & 288(1) & 0.453(8) & 245.1(8)& 55046(7)  & 0.0435(6)  & 0.60 &0.0004& 1.65* \\
6615041\tablefootmark{e}&0.340085336(7)&2.82(2)&1310(1)& 33.5(2)& 0.15(1)  & 94(4)   & 54627(15) & 0.000295(6)& 0.09 &0.0009& 1.58* \\
6671698\tablefootmark{f}&0.47152269(2)&$-16.7$(1)&1047.4(7)&91.7(5)&0.37(1)& 131(2)  & 54949(5)  & 0.0094(2)  & 0.37 & 0.003& 1.95* \\
6766325 & 0.439968289(9)& $-$ & 5263(8)   & 800(4) & 0.584(9) & 282.9(5)& 53740(8)  & 0.248(4)   & 1.38 &0.0002& 1.87* \\   
6794131 & 1.6133259(1)  & $-$ & 3766(20)  & 364(22)& 0.83(3)  & 145(3)  & 55886(46) & 0.046(8)   & 1.05 & 0.02 & 4.0: \\
\emph{7031714FP}\tablefootmark{g}&0.8141396(3)&$-59$(3)&2056(11)&147(3)&0.422(7)&354(2)& 55297(12) & 0.0100(5)  & 0.38 & 0.003& 2.0: \\
7272739 &0.2811642561(8)& $-$ & 3415(3)   & 61.6(2)& 0.608(6) & 133.9(6)& 55804(6)  & 0.000268(3)& 0.08 &0.0003& 1.40* \\
7339345\tablefootmark{a}&0.2596622951(5)&$-$ & 886(4)  & 16(3)  & 0.57(5)  & 290(3)  & 55103(12) & 0.00007(4) & 0.05 & 0.003& 1.33* \\
        &               &    &  5076(54)  & 332(16)& 0.37(2)  & 200(11) & 56416(164)& 0.019(3)   & 0.63 &      & 3.0: \\
7440742\tablefootmark{h}&0.28399008(3)&6.96(6)&2780(11)&  63(4) & 0.82(3)  & 313(4)  & 56266(38) & 0.00043(8) & 0.10 & 0.001& 1.41*\\	 
7877062\tablefootmark{i}&0.30365600(6)&$-7.5$(2)&4520(6)&277(3) & 0.427(4) & 32.5(6) & 55891(8)  & 0.0140(4)  & 0.36 &0.0001& 1.47*\\ 
\emph{7938468}&7.2269895(7)&$-$& 5413(61)  & 198(5) & 0.42(2)  & 357(3)  & 55241(58) & 0.0035(3)  & 0.26 & 0.05 & 2.0: \\
8016214 &  3.1749314(2) & $-$ & 6401(217) & 435(10)& 0.691(7) & 171(2)  & 55320(56) & 0.027(3)   & 0.56 & 0.02 & 2.0: \\
\emph{8257903}&0.51506353(2)&$-$&7013(109) & 220(4) & 0.31(1)  & 189(1)  & 56741(47) & 0.0029(2)  & 0.25 &0.0001& 2.07* \\
8553788 & 1.60619507(3) & $-$ & 6425(30)  & 834(19)& 0.86(1)  & 68.7(4) & 56539(23) & 0.19(1)    & 1.14 & 0.01 & 1.67 \\ 
\emph{8953296}\tablefootmark{j}&0.78429167(8)&40.0(5)&1837(6)&45.0(4)&0.23(1)& 268(6) & 55652(35) & 0.000363(9)& 0.12 & 0.003& 2.0: \\
8982514 & 0.41448635(1) & $-$ & 7002(65)  & 403(2) & 0.439(5) & 213(1)  & 55868(29) & 0.0179(5)  & 0.45 &0.0001& 1.80* \\
9083523\tablefootmark{k}&0.91842011(3)&25.8(2)&1412.0(7)&53.6(3)& 0.336(7) & 102(1)  & 54853(5)  & 0.00104(2) & 0.17 & 0.006& 2:0: \\
9091810 & 0.47972069(1) &1.04(2)&4494(32) & 45.2(5)& 0.77(1)  & 217(2)  & 55682(26) & 0.000061(2)& 0.06 & 0.001& 2.0: \\
9101279 & 1.8114480(1)  &$-74.3$(8)&4001(14)&367(15)& 0.85(1) &  45(2)  & 54984(34) & 0.042(5)   & 0.67 & 0.03 & 2.0: \\
9159301\tablefootmark{l}&3.0447680(2)&$-38$(6)&2089(11)& 39.4(8)& 0.27(4)  & 275(6)  & 54178(40) & 0.00019(1) & 0.09 & 0.03 & 2.01 \\
9283826\tablefootmark{m}&0.35652427(6)&$-16.6$(1)&2410(6)&178(3)& 0.03(1)  &  69(18) & 54860(118)& 0.0129(6)  & 0.37 &0.0003& 1.63* \\
9532219 & 0.198153269(4)& $-$ & 5455(17)  &268.1(7)& 0.454(3) & 226.3(9)& 55927(17) & 0.00868(9) & 0.26 &0.00004& 1.12* \\
\emph{9541127}\tablefootmark{a}&0.5366500567(9)&$-$&778(3)&21.0(7)&0.14(3)  &  27(5)  & 54845(12) & 0.00021(2) & 0.10 &0.002 & 2.0: \\
        &               &     & 2796(11)  & 110(2) & 0.40(4)  & 305(5)  & 54306(44) & 0.0023(1)  & 0.29 &      & 3.0: \\
9788457 & 0.96333950(3) &0.50(8)&4095(9)  & 52.1(5)& 0.741(7) &  8.9(7) & 55711(10) & 0.000113(3)& 0.08 & 0.005& 2.0: \\ 
9821923 & 0.349531575(3)& $-$ & 4825(7)   & 303(2) & 0.356(6) & 121(1)  & 55814(20) & 0.0160(4)  & 0.40 &0.0001& 1.61* \\
\emph{9832227}&0.4579406(3)&$-$&12057(1077)&1164(87)& 0.32(4)  & 204(3)  & 55401(146)& 0.15(4)    & 1.03 &0.00004&1.71 \\
\hline
\end{tabular}}
\end{center}
\tablefoot{\tablefoottext{a}{See Appendix~\ref{app:Notesonindividualsystems} for a more detailed discussion;}\tablefoottext{b}{Cubic coefficient: $c_3=1.18(1)\times10^{-14}$\,d/c$^2$;}\tablefoottext{c}{Cubic coefficient: $c_3=1.38(6)\times10^{-13}$\,d/c$^2$;}\tablefoottext{d}{Cubic coefficient: $c_3=2.79(6)\times10^{-13}$\,d/c$^2$;}\tablefoottext{e}{Cubic coefficient: $c_3=-1.20(1)\times10^{-14}$\,d/c$^2$;}\tablefoottext{f}{Cubic coefficient: $c_3=4.30(3)\times10^{-14}$\,d/c$^2$;}\tablefoottext{g}{The true target is the background object TIC~1882359676; cubic coefficient: $c_3=2.7(2)\times10^{-13}$\,d/c$^2$;}\tablefoottext{h}{Cubic coefficient: $c_3=-6.8(1)\times10^{-15}$\,d/c$^2$;}\tablefoottext{i}{Cubic coefficient: $c_3=3..17(3)\times10^{-14}$\,d/c$^2$;}\tablefoottext{j}{Cubic coefficient: $c_3=-1.89(2)\times10^{-13}$\,d/c$^2$;}\tablefoottext{k}{Cubic coefficient: $c_3=-9.3(1)\times10^{-14}$\,d/c$^2$;}\tablefoottext{l}{Cubic coefficient: $c_3=-3.8(1)\times10^{-12}$\,d/c$^2$;}\tablefoottext{m}{Cubic coefficient: $c_3=7.9(3)\times10^{-15}$\,d/c$^2$.}}
\end{table*}

\addtocounter{table}{-1}

\setlength{\tabcolsep}{3.3pt}
\begin{table*}
\begin{center}
\caption{continued.} 
\scalebox{0.91}{\begin{tabular}{lccccccccccc} 
\hline\hline
KIC No. & $P_1$ & $\Delta P_1$ & $P_2$ & $a_\mathrm{AB}\sin i_2$ & $e_2$ & $\omega_2$ & $\tau_2$ & $f(m_\mathrm{C})$ & $(m_\mathrm{C})_\mathrm{min}$ & $\frac{{\cal{A}}_\mathrm{dyn}}{{\cal{A}}_\mathrm{LTTE}}$ & $m_\mathrm{AB}$\\
        & (day) &$\times10^{-10}$ (d/c)&(day)&(R$_\odot$)  &       &   (deg)    &   (MBJD) & (M$_\odot$)       & (M$_\odot$)            & &  (M$_\odot$)    \\
\hline
10095469\tablefootmark{a}&0.677759677(5)&$-$&  760(4)  &  27(2) & 0.13(6)  &   2(14) & 54990(31) & 0.00045(9) & 0.13 & 0.003& 2.0: \\
        &               &     & 4550(32)  & 365(34)& 0.56(8)  & 253(6)  & 52923(87) & 0.031(9)   & 0.76 &      & 3.0: \\
\emph{10208759}&4.57503(4)&$-$&11097(26774)&120(1064)& 0.3(9)  & 45(74)  &53434(4868)& 0.0002(52) & 0.09 & 0.005& 2.0: \\
\emph{10216186}&0.60593901(3)&$-0.78$(6)&2392(4)&49.8(4)&0.37(1)&  2(1)  & 55102(11) & 0.000289(7)& 0.11 & 0.001& 2.0: \\
\emph{10322582}&0.2912730(4)&$-$&15947(1924)&1593(199)&0.62(2) & 301(2)  & 55800(233)& 0.21(9)    & 1.11 &0.00002&1.43* \\
\emph{10388897}&0.34372477(1)&$-$& 2941(6) & 448(3) & 0.37(1)  & 100.6(8)& 54701(9)  & 0.139(3)   & 0.97 &0.0003& 1.59* \\
\emph{10481912}&0.442368086(1)&$-$& 4721(5)& 92.0(4)& 0.370(5) & 259(1)  & 55595(17) & 0.000468(6)& 0.12 &0.0002& 1.87* \\
\emph{10486425}&5.274835(2)&$-$& 6494(1098)& 124(37)& 0.5(1)   & 336(14) & 54946(513)& 0.0006(6)  & 0.17 & 0.02 & 2.65 \\
10557008& 0.265419020(9)&$-1.825$(9)&4016(9)&127.8(4)&0.726(4)& 224.3(5)& 55356(8)  & 0.00173(2) & 0.16 &0.0003& 1.35* \\
\emph{10581918}&1.80186478(6)&$-15.6$(4)&5869(151)&39(1)&0.59(4)& 154(3) & 53310(72) & 0.000024(3)& 0.04 & 0.005& 1.47 \\
10686876& 2.6183818(2)  & $-$ & 9600(139) & 757(9) & 0.511(3) & 178.8(6)& 54971(22) & 0.063(3)   & 0.98 & 0.003& 2.88 \\
10848807& 0.346246314(7)&2.517(8)& 1355(5)& 27(1)  & 0.82(2)  &   4(1)  & 54526(7)  & 0.00014(2) & 0.07 & 0.02 & 1.60* \\
10916675& 0.4188635(7)  &6.1(5)&9848(2754)&314(116)& 0.49(6)  & 171(2)  & 55478(246)& 0.004(5)   & 0.26 &0.00007& 1.81* \\
10934755& 0.78648758(1) & $-$ & 3089(12)  &104.6(4)& 0.577(4) & 19.2(5) & 55739(6)  & 0.00161(2) & 0.20 & 0.002& 2.0: \\
11246163\tablefootmark{b}&0.279226191(2)&0.85(2)&3037(4)&88.8(7)& 0.535(8) & 208.2(7)& 55835(6)  & 0.00102(2) & 0.13 &0.0003& 1.39* \\ 
11604958\tablefootmark{c}&0.29892819(2)&3.76(5)&1772(4)& 54.1(9)& 0.15(2)  & 79(7)   & 55061(36) & 0.00068(3) & 0.12 &0.0004& 1.45* \\
12055014& 0.4999044070(8)&$-5.06$(1)&3830(7)& 69(1)& 0.40(1)  & 321(2)  & 56569(24) & 0.00030(1) & 0.11 &0.0004& 2.03* \\
\emph{12071006}&6.095986(1)&$-$&9852(1264) & 383(35)& 0.52(5)  & 240(7)  & 55968(350)& 0.008(3)   & 0.52 & 0.01 & 3.70 \\
\emph{12458133}&0.332522595(6)&$-1.893(6)$&2223(4)&57(2)&0.62(2)&182(1)  & 54228(10) & 0.00050(5) & 0.11 &0.0009& 1.56* \\
12554536& 0.684496426(2)& $-$ & 1469(1)   & 42.8(6)& 0.61(2)  & 234(1)  & 54865(6)  & 0.00049(2) & 0.13 & 0.006& 2.0: \\
\hline
\end{tabular}}
\end{center}
\tablefoot{\tablefoottext{a}{See Appendix~\ref{app:Notesonindividualsystems} for a more detailed discussion;}\tablefoottext{b}{Cubic coefficient: $c_3=-1.1(1)\times10^{-15}$\,d/c$^2$;}\tablefoottext{c}{Cubic coefficient: $c_3=-5.0(1)\times10^{-15}$\,d/c$^2$.}}
\end{table*}

\setlength{\tabcolsep}{3.3pt}
\begin{table*}
\begin{center}
\caption{Orbital elements from certain combined dynamical and LTTE solutions for systems.} 
\label{Tab:Orbelemdyn1}  
\scalebox{0.91}{\begin{tabular}{lccccccccccc} 
\hline\hline
KIC No. & $P_1$ & $P_2$ & $a_2$ & $e_2$ & $\omega_2$ & $\tau_2$ & $f(m_\mathrm{C})$ & $\frac{m_\mathrm{C}}{m_\mathrm{ABC}}$ & $m_\mathrm{AB}$ & $m_\mathrm{C}$ & $\frac{{\cal{A}}^\mathrm{meas}_\mathrm{dyn}}{{\cal{A}}_\mathrm{LTTE}}$\\
        & (day) & (day) &(R$_\odot$)  &       &   (deg)    &   (MBJD) & (M$_\odot$)   &    & (M$_\odot$)            &  (M$_\odot$)  &   \\
\hline
3544694    & 3.84572314(8)& 80.92(3)& 97(4) &0.113(4)&327(2) &55723.9(4)& 0.043(8) & 0.29(1) & 1.34(20)& 0.54(8)& 4.13 \\
4079530    & 17.727144(1) & 141.7(9)& 130(5)&0.52(2)&  22(7) & 54987(8) &0.000000004(1)&0.0013(2)&1.48(16)&0.0020(3)&475\\
4909707    & 2.30236795(4)& 515.0(3)&489(23)&0.59(2)& 175(3) & 54848(5) &0.349(124)& 0.39(06)& 3.60(77)&2.31(35)& 0.52\\ 
4940201    & 8.8165585(3) & 364.6(2)& 303(4)&0.227(5)&258(3) & 54868(3) & 0.10(1)  & 0.33(1) & 1.88(9) &0.94(6) & 3.06 \\
5095269    & 18.6118308(1)&235.89(2)&219.32(2)&0.072(3)&328.4(6)&55004.2(5)&0.000000016(1)&0.00192(7)&2.28&0.0049(2)&133 \\ 
5255552\tablefootmark{a}& 32.46529(7)  & 860.1(6)&574(7)&0.372(2)& 40.2(4)& 54868(1) & 0.20(1)  & 0.390(5)& 2.10(10)& 1.34(7)& 35 \\
5384802    & 6.08309218(9)&256.08(5)&297(9)&0.371(4)&   9(2) & 54998(2) & 0.42(6)  & 0.43(1) & 3.06(39)&2.31(32)& 4.4 \\
\emph{5650317}&2.14054699(2)&118.38(2)&182(6)&0.20(2)& 228(7) & 55238(3) & 0.22(5)  & 0.34(2) & 3.81(48)&1.94(31)& 0.83\\
5653126    & 38.49130(2)  & 969.4(4)&687(5)&0.170(3)&322.2(8)& 54514(3) & 0.127(9) & 0.303(7)& 3.22(9) & 1.40(6)& 30 \\
5731312    & 7.94638807(8)& 919.7(8)&458(2)&0.598(2)&  19(1) & 54830(1) &0.00192(5)&0.1082(8)& 1.36(2) &0.165(3)& 4.69 \\
5771589    & 10.738237(2) &113.20(2)&168(2)&0.088(2)& 105(2) &54974.8(4)& 1.25(7)  & 0.634(4)& 1.81(9) &3.14(17)& 16.8 \\
5952403    & 0.90567813(1)&45.468(7)& 86(2)&0.0007(5)&131(20)& 55506(3) & 0.75(12) & 0.57(3) & 1.81(9) &2.35(30)& 0.036\\  
6525196    & 3.42059731(1)&418.90(8)&330(5)& 0.27(1)&  92(4) & 55063(5) & 0.075(6) & 0.301(7)& 1.93(11)& 0.83(6)& 0.41\\
6531485\tablefootmark{b}&0.676990690(7)&48.324(7)& 60(5)& 0.63(1)&  47(10)& 54982(2) & 0.025(8) & 0.276(8)& 0.88(28)&0.34(11)& 4.23\\
6545018    &3.99145675(2) &90.586(3)&106.3(8)&0.247(5)&233(1)&54970.3(4)& 0.023(1) & 0.228(5)& 1.52(4) & 0.45(2)& 8.33\\
6546508    & 6.1071013(2) & 1235(27)&722(35)&0.32(3)& 318(8) & 55088(28)& 0.23(7)  & 0.41(4) & 1.94(34)&1.37(31)& 0.29\\
7177553    & 17.9964523(4)& 526.5(4)&343(4)& 0.47(2)& 213(2) & 54714(2) &0.00000012(2)&0.0039(3)&1.95(6)&0.0077(6)&39 \\ 
7289157    & 5.2665486(2) & 244.1(1)&241(6)& 0.31(1)& 163(2) & 54941(2) & 0.17(2)  & 0.38(2) & 1.96(18)&1.18(14)& 4.08 \\
           &              &6950(424)&426(42)\tablefootmark{c}&0.26(6)&137(18)&53338(409)&0.021(6) & $-$     & 0.66$^c$& 3:     & \\
7668648    & 27.82517(4)  & 202.9(1)&180(4)&0.112(7)& 188(6) & 54931(2) & 0.018(2) & 0.213(1)& 1.51(13)& 0.41(4)& 78 \\
7690843\tablefootmark{d}& 0.78625579(4)& 74.21(3)&157(13)&0.48(3)& 358(2) & 54990(1) & 4.7(16)  & 0.81(5) & 1.84(9) & 7.6(24)& 0.77 \\
7812175    & 17.793579(1) & 582.0(3)&358(14)&0.023(2)&220(6) & 54799(10)& 0.061(1) & 0.323(4)& 1.23(19)& 0.59(9)& 6.27 \\
7821010    &24.23821642(9)& 990(1)  &572(1)&0.343(7)& 128(2) & 55122(5) &0.000000002(0)&0.00090(2)&2.56(1)&0.00229(6)&35.3\\
8023317    & 16.5789557(3)& 609.9(1)&385(11)&0.24(2)& 165(2) & 55010(4) & 0.0006(1)& 0.067(6)& 1.92(17)& 0.14(2)& 8.31 \\
8143170\tablefootmark{e}& 28.78582(1)  &1580.5(6)&995(9)&0.690(6)&107.3(4)&54541.1(6)& 0.0081(6)& 0.123(4)& 4.64(13)& 0.65(3)& 22.2\\
8210721    & 22.673093(2) & 789.1(2)&480(5)&0.236(2)&218.1(5)& 54634(1) & 0.082(4) & 0.325(4)& 1.61(6) & 0.78(3)& 9.90 \\
8719897    & 3.15142051(1)&333.30(4)&269(8)&0.251(4)& 130(1) &54997.1(8)& 0.13(2)  & 0.384(8)& 1.46(19)&0.91(12)& 0.56\\
9007918    &1.387206538(2)& 471.3(3)&328(7)& 0.65(1)& 286(2) & 54818(1) &0.00034(3)& 0.054(2)& 2.10(13)&0.121(8)& 0.34 \\ 
9451096    &1.250390741(3)&106.89(2)&144(5)& 0.10(1)& 163(6) & 54994(2) & 0.046(7) & 0.24(1) & 2.65(32)& 0.84(11)& 0.19\\
9714358    & 6.4742284(2) &103.76(1)&101(1)& 0.28(1)& 119(1) &54977.6(2)& 0.008(1) & 0.185(9)& 1.03(5) & 0.23(2)& 26.2\\
9850387    & 2.74849803(3)& 670.6(8)&479(1)& 0.42(3)& 112(4) & 54690(9) & 0.019(1) & 0.180(5)& 2.72    & 0.59(2)& 0.24 \\
10095512   & 6.0172058(1) & 472.8(2)&334(23)&0.18(1)& 232(6) & 54861(9) & 0.16(5)  & 0.42(2) & 1.31(36)&0.93(27)& 0.79 \\
\emph{10223616}&29.12926(8)&541.9(6) &395(14)&0.29(3)& 239(3) &55299.0(7)& 0.13(6)  & 0.36(7) & 1.81(8) &1.00(31)& 33 \\
10483644   & 5.11077040(5)& 372.8(6)&299(14)&0.03(3)& 320(9) & 55314(7) & 0.0019(8)& 0.09(2) & 2.36(37)&0.24(6) & 0.14 \\
10549576   & 9.0895323(8) & 1486(4) &1002(25)&0.48(4)&158(7) & 54970(32)& 0.35(7)  & 0.42(3) & 3.54(25)&2.58(38)& 0.60\\
10613718   & 1.17587798(2)& 88.18(5)&129(4)& 0.11(3)& 35(17) & 54989(4) & 0.24(6)  & 0.40(3) & 2.23(23)&1.49(25)& 0.25 \\
10979716   & 10.6840969(4)&1046.0(6)&602(16)&0.470(8)& 58(2) & 54513(7) & 0.15(2)  & 0.41(1) & 1.59(17)&1.09(13)& 2.29 \\ 
11519226   & 22.161754(1) & 1434(1) &873(13)&0.339(3)&321.9(6)&55010(2) & 0.33(3)  & 0.43(1) & 2.50(13)&1.84(13)& 4.90\\
\hline
\end{tabular}}
\end{center}
\tablefoot{\tablefoottext{a}{Quadratic and cubic ephemeris: $\Delta P=-5.5(2)\times10^{-5}$\,d/c$^2$, $c_3=1.15(5)\times10^{-7}$\,d/c$^3$. These non-linear polynomial coefficients are for the mathematical representations of additional perturbations, not considered in the analytic ETV three-body model (see Appendix~\ref{app:Notesonindividualsystems} for details);}\tablefoottext{b}{Quadratic ephemeris: $\Delta P=-1.02(2)\times10^{-10}$\,c/d;}\tablefoottext{c}{The ``fourth-body orbit'' is a pure LTTE solution, hence, the given value is $a_{ABC}\sin i_3$ instead of $a_3$ and, similarly, instead of $m_\mathrm{D}$, what is given is its minimum mass (for $i_3=90\degr$);}\tablefoottext{d}{Quadratic and cubic ephemeris: $\Delta P=53.8(3)\times10^{-10}$\,d/c$^2$, $c_3=-1.49(2)\times10^{-13}$\,d/c$^3$;}\tablefoottext{e}{Quadratic and cubic ephemeris: $\Delta P=2.4(4)\times10^{-6}$\,d/c$^2$, $c_3=1.16(8)\times10^{-8}$\,d/c$^3$. These non-linear polynomial coefficients are for the mathematical representations of additional perturbations, not considered in the analytic ETV three-body model (see Appendix~\ref{app:Notesonindividualsystems} for details).}}
\end{table*}

\setlength{\tabcolsep}{3.3pt}
\begin{table*}
\begin{center}
\caption{Orbital elements from moderately certain combined dynamical and LTTE solutions for systems.} 
\label{Tab:Orbelemdyn2}  
\scalebox{0.91}{\begin{tabular}{lccccccccccc} 
\hline\hline
KIC No. & $P_1$ & $P_2$ & $a_2$ & $e_2$ & $\omega_2$ & $\tau_2$ & $f(m_\mathrm{C})$ & $\frac{m_\mathrm{C}}{m_\mathrm{ABC}}$ & $m_\mathrm{AB}$ & $m_\mathrm{C}$ & $\frac{{\cal{A}}^\mathrm{meas}_\mathrm{dyn}}{{\cal{A}}_\mathrm{LTTE}}$\\
        & (day) & (day) &(R$_\odot$)  &       &   (deg)    &   (MBJD) & (M$_\odot$)   &    & (M$_\odot$)            &  (M$_\odot$)  &   \\
\hline
5080652    & 4.14435547(8)& 222.9(5)&251(28)&0.15(5)& 350(7) & 54865(15)& 0.50(44) & 0.49(15)& 2.18(45)&2.10(133)&0.89 \\ 
6877673    & 36.7593121(8)& 3765(48)&1268(11)&0.18(1)& 158(2)& 53869(39)& 0.0134(4)& 0.193(2)& 1.56(1) & 0.37(1) & 1.9\\
7837302    & 23.836787(2) &1381.3(9)&720(13)&0.265(8)& 155(7)& 56346(7) & 0.010(2) & 0.16(2) & 2.20(13)& 0.42(5) & 7.1 \\
9664215    & 3.3195332(1) &  928(2) &537(18)&0.55(2)& 182(3) & 55758(6) & 0.14(2)  & 0.39(1) & 1.49(20)& 0.93(14)& 0.45 \\
           &              & 5470(65)&321(31)\tablefootmark{a}&0.40(7)&254(16)&56414(252)&0.015(4)& $-$    & 0.57$^a$& 3:      & \\
10268809   & 24.708843(5) & 4084(6) &1473(27)&0.636(6)&290(2)& 56144(4) & 0.17(2)  & 0.41(1) & 1.52(11)& 1.05(9) & 4.61\\
10296163   &9.29678473(3) & 3408(3) &1274(68)&0.385(6)&2.6(8)& 55953(5) & 0.0036(8)& 0.142(6)& 2.05(38)& 0.34(6) & 0.31\\
12356914   & 27.308390(3) & 1824(6) &942(15)&0.384(9)& 106(3)& 55863(8) & 0.0052(5)& 0.135(5)& 2.92(15)& 0.46(3) & 7.8 \\ 
\hline
\end{tabular}}
\end{center}
\tablefoot{\tablefoottext{a}{The ``fourth-body orbit'' is a pure LTTE solution, hence, the given value is $a_{ABC}\sin i_3$ instead of $a_3$ and, similarly, instead of $m_\mathrm{D}$, what is given is its minimum mass (for $i_3=90\degr$).}} 
\end{table*}

\setlength{\tabcolsep}{3.3pt}
\begin{table*}
\begin{center}
\caption{Orbital elements from combined dynamical and LTTE solutions for systems with very uncertain solutions} 
\label{Tab:Orbelemdyn3}  
\scalebox{0.91}{\begin{tabular}{lccccccccccc} 
\hline\hline
KIC No. & $P_1$ & $P_2$ & $a_2$ & $e_2$ & $\omega_2$ & $\tau_2$ & $f(m_\mathrm{C})$ & $\frac{m_\mathrm{C}}{m_\mathrm{ABC}}$ & $m_\mathrm{AB}$ & $m_\mathrm{C}$ & $\frac{{\cal{A}}^\mathrm{meas}_\mathrm{dyn}}{{\cal{A}}_\mathrm{LTTE}}$\\
        & (day) & (day) &(R$_\odot$)  &       &   (deg)    &   (MBJD) & (M$_\odot$)   &    & (M$_\odot$)            &  (M$_\odot$)  &   \\
\hline
 2576692 & 87.87876(1) & 5196(3)  & 1843(3)  & 0.314(8)& 334(3)  & 52976(62)  &0.00035(3)&0.048(2) & 2.96(1) & 0.15(1) & 10 \\
\emph{3247294}&67.421632(3)&10287(37)&3055(44)& 0.58(2) & 359(1)  & 53456(50)  & 0.21(3) & 0.39(2)  & 2.22(5) & 1.40(15)& 7.57 \\
3345675  &120.009134(9)& 4865(8)  & 1384(14) & 0.07(2) & 185.8(3)& 57038(10)  &0.0027(1) &0.131(2) & 1.31(5) & 0.198(8)& 19 \\   
\emph{3938073}&31.0242625(6)&3091(76)&1301(57)& 0.02(3) & 50(20)  & 54524(200) &0.000010(3)&0.015(2)& 3.05(38)& 0.046(9)& 0.08 \\
 4753988 & 7.3045003(2)& 3934(36) & 1506(18) & 0.43(5) &  20(7)  & 55606(63)  &0.0055(6)& 0.131(6) & 2.57(9) & 0.39(3) & 0.07 \\
\emph{5393558}&10.2171100(3)&1458(15)& 594(66)& 0.11(5) &  74(8)  & 54303(39)  &0.00003(1)& 0.030(3)& 1.28(44)&0.039(14)& 1.06 \\
\emph{5632781}&11.02520064(1)&1584(2)& 728(3) & 0.41(9) & 144(2)  & 56110(2)   &0.00000006(3)&0.0031(6)&2.06(2)&0.006(1)& 2.04 \\ 
\emph{6146838}&27.4674428(4)&15087(280)&3745(59)&0.766(4)&  59(2) & 56496(5)   & 0.068(5)& 0.290(8) & 2.20(8) & 0.90(5) & 1.08 \\
6233903  & 5.9908193(5)& 1630(16) & 837(10)  & 0.24(6) & 34(9)   & 54554(44)  & 0.073(9)& 0.29(1)  & 2.10(7) & 0.86(6) & 0.13 \\
\emph{8211618}\tablefootmark{a}&0.33744694(4)&655.2(3)&407(10)&0.88(2)& 1(2)   & 55103(4)   & 0.004(1)& 0.13(2)  & 1.84(14)& 0.27(5) & 0.15 \\
\emph{8553907}&42.0323426(4)&11769(8)&3540(12)& 0.000(1)& 34.1(6) & 58134(23)  &0.00288(6)&0.1072(7)& 3.84(4) & 0.461(6)& 0.45 \\
\emph{8560861}&31.9731357(7)&6009(6)& 2093(3) & 0.49(1) & 261(3)  & 56824(3)   &0.0049(2)& 0.113(2) & 3.03(1) & 0.386(8)& 1.07 \\
9028474  & 124.93731(2)& 4786(5)  & 1576(8)  & 0.14(1) & 63(2)   & 56190(2)   &0.0000047(4)&0.0127(5)&2.27(3)& 0.029(1)& 35 \\
 9963009 & 40.073522(2)& 3492(87) & 1850(49) & 0.57(2) & 167(5)  & 53475(155) & 0.71(9) & 0.57(2)  & 3.00(13)& 3.97(42)& 7.79 \\
\emph{10666242}&87.245726(6)&4386(5)&1457(80) & 0.49(6) & 24(4)   & 54747(140) & 0.004(7)& 0.12(9)  & 1.89(27)& 0.27(22)& 19 \\
11502172 & 25.432011(3)& 3907(39) & 1403(55) & 0.22(2) & 80(13)  & 57743(158) & 0.18(4) & 0.42(2)  & 1.40(21)& 1.03(18)& 0.61 \\
11558882 & 73.913218(9)& 4692(42) & 1487(16) & 0.264(9)& 318(2)  & 54980.5(6) &0.018(2) & 0.211(8) & 1.58(5) & 0.42(2) & 6.7 \\
\emph{12302391}&25.321556(1)&4234(66)&2186(32)& 0.51(2) & 5(1)    & 53100(41)  & 0.08(1) & 0.29(2)  & 5.56(8) & 2.27(23)& 1.73 \\
\hline
\end{tabular}}
\end{center}
\tablefoot{\tablefoottext{a}{Cubic ephemeris: $\Delta P=-6.8(1)\times10^{-10}$\,d/c$^2$, $c_3=8.6(2)\times10^{-15}$\,d/c$^3$.}}
\end{table*}

\begin{table*}
\begin{center}
\caption{Apsidal motion and/or orientation parameters from AME and dynamical fits} 
\label{Tab:AMEparam}
\scalebox{0.91}{\begin{tabular}{lccccccccccc} 
\hline\hline
KIC No. & $P_\mathrm{anom}$ & $a_1$      & $e_1$ & $\omega_1$ & $\tau_1$ & $P_\mathrm{apse}$ & $\im$ & $i_1$ & $i_2$ & $\Delta\Omega$ & $P_\mathrm{node}$\\
        & (days)            &(R$_\odot$) &       & (deg)      & (MJD)    &   (years)         & (deg) & (deg) & (deg) &   (deg)   &  (years) \\
\hline
4758368 & 3.75048(6)  & $-$     & 0.004(1) & 340(7)  & 54958.92(7)& 77(10)   &&&&&\\
4851217 & 2.4703961(1)& $-$     &0.03179(2)& 348.9(3)&54954.418(2)& 153(2)   &&&&&\\
5039441 & 2.1513865(2)& $-$     & 0.007(5) & 294(15) & 54955.49(9)& 3573(185) &&&&&\\ 
5288543 & 3.4574(1)   & $-$     &0.00184(3)& 334(5)  &54965.42(4) & 118(48)   &&&&&\\  
6965293 & 5.07784(9)  & $-$     & 0.0105(6)& 192(16) & 54956.4(2) & 774(741)  &&&&&\\
\hline
2576692 & 87.87923(1) & 119.4(2)& 0.193(3) &317.4(9) & 55035.1(2) & 21158       & 7.2(7)    & 89 & 91 & 7.1(7)  & $-7729$ \\
3247294 & 67.422032(3)& 90.9(6) & 0.444(1) & 157.0(5)& 54955.0(1) & 17672       & 21(2)     & 89 & 95 & 20(2)   & $-3926$ \\
3345675 &119.992116(9)& 112(1)  & 0.4434(8)& 349.5(6)& 55094.4(1) & $-3315$     & 61.2(6)   & 89:& 68 &$-59.1$(6)& $-4585$ \\
3544694 &3.84763242(9)& 11.4(6) &0.00143(2)& 356(3)  & 55741.57(3)& 21.3        &  0        & 83 & 83 &   0     & $-17.0$\\
3938073 & 31.024399(7)& 60(2)   & 0.4337(2)& 5.1(4)  & 54964.79(3)& 14497(1566) &  0        & 88 & 88 &   0     & $-12570$\\
4079530 & 17.727605(1)& 33(1)   & 0.266(3) & 323(1)  & 54996.35(3)& 2022        &  0        & 86:& 86 &   0     & $-28$ \\
4753988 & 7.3044960(2)& 21.7(2) & 0.027(5) & 80(9)   & 54971.5(2) & 36689       & 44(12)    & 89 & 110& 40(11)  & $-52200$ \\
4909707 &2.30239303(4)& 11.2(8) & 0.0067(5)& 203(10) & 54953.50(6)& 569         & 10(5)     & 88:& 93 &   8(4)  & $-525$\\
4940201 & 8.8179382(4)& 22.2(4) &0.00132(3)& 187(5)  & 54965.3(1) & 153         &  0        & 86 & 86 &   0     & $-131$ \\
\hline
\end{tabular}}
\end{center}
\end{table*}

\addtocounter{table}{-1}

\begin{table*}
\begin{center}
\caption{continued} 
\scalebox{0.91}{\begin{tabular}{lccccccccccc} 
\hline\hline
KIC No. & $P_\mathrm{anom}$ & $a_1$      & $e_1$ & $\omega_1$ & $\tau_1$ & $P_\mathrm{apse}$ & $\im$ & $i_1$ & $i_2$ & $\Delta\Omega$ & $P_\mathrm{node}$\\
        & (days)            &(R$_\odot$) &       & (deg)      & (MJD)    &   (years)         & (deg) & (deg) & (deg) &   (deg)   &  (years) \\
\hline
5080652 &4.14362482(8)&  14(1)  & 0        & $-$     &  $-$       &  $-$        &  0        & 87 & 87 &   0     & $-79$ \\
5095269 &18.6121418(1)& 40.32102& 0.5032   & 288.34  &54967.138959(6)& 5121     & 31(11)    & 85 & 73 &  30(11) & $-153$ \\  
5255552 & 32.45864(7) & 54.8(9) & 0.237(2) & 110.2(1)& 54957.18(2)& 177         &  3(1)     & 88 & 90 &   3(1)  & $-131$ \\
5384802 &6.08125035(9)& 20.4(9) & 0        & $-$     &  $-$       &  $-$        &  0        & 83 & 83 &   0     & $-65$ \\
5393558 &10.2171140(3)& 22(2)   & 0.24(3)  & 275.4(6)&54970.34(1) & 25938       &  0        & 88 & 88 &   0     & $-9852$ \\
5632781 &11.0252224(2)& 26.52(9)& 0.2798(3)& 205.2(1)&54974.562(3)& 16445(175)  &  2(2)     & 89 & 90 & $-2$(2) & $-8640$ \\
5650317 &2.14029507(2)& 10.9(5) & 0        & $-$     &  $-$       &  $-$        &  0        & 88 & 88 &   0     & $-58$ \\     
5653126 & 38.50620(2) & 70.9(6) & 0.319(6) & 300.4(8)& 54987.51(5)& 316         & 15.3(9)   & 87 & 85 &$-15.2$(9)&$-210$ \\
5731312 &7.94638200(8)& 18.6(1) & 0.435(4) & 162(3)  & 54966.83(6)& $-6368$     & 38.15(6)  & 87 & 88 &$-38.2$(1)&$-911$ \\
5771589 & 10.784681(2)& 25.0(4) & 0.0134(3)& 242(3)  & 54961.26(8)& 6.79        & 1.3(5)    & 86 & 85 &   0.7(4)& $-6.21$ \\
5952403 &0.90547513(1)& 4.79(8) & 0        & $-$     & $-$        &  $-$        & 0.8       & 87 & 88 &   0     & $-13.8$ \\
6146838 &27.4677544(4)& 49.8(6) & 0.282(7) & 301(2)  & 54978.0(1) & 39000       & 14.9(8)   & 89 & 76 &$-8.0$(7)& $-18990$ \\
6233903 & 5.9908471(5)& 17.8(2) & 0.008(2) & 73(45)  & 55004.4(8) & 5064        &  0        & 87 & 87 &   0     & $-4574$ \\ 
6525196 &3.42052772(1)& 11.9(2) & 0        & $-$     &  $-$       &  $-$        & 15(4)\tablefootmark{a}& 85 & 95 & $-11$(6)& $-505$\\
6531485 &0.676913205(7)& 3.1(3) & 0        & $-$     &  $-$       &  $-$        &  0        & 80:& 80 &   0     & $-18$ \\
6545018 & 3.99241(1)  & 12.2(1) &0.00278(2)& 181(1)  & 54964.85(1)& 46.0(6)\tablefootmark{b}&  0        & 87 & 87 &   0     & $-22$ \\
6546508 & 6.1071516(2)&  18(1)  & 0.005(2) & 80(7)   & 55192.7(1) & 1869        &  0        & 85 & 85 &   0     & $-1750$ \\
6877673 &36.7592372(8)&53.92(9) & 0.1491(2)& 48.65(1)&55001.992(4)& 6613        & 28(3)     & 90 & 99 & 26(3)   &$-6505$\\
7177553 &17.9964861(4)& 36.1(4) &0.39185(7)& 182.1(5)& 54952.32(2)& 46335(192)  & 32(2)     & 88 & 85 & 32(2)   & $-304$ \\
7289157 & 5.2672593(2)& 15.9(5) & 0.095(2) & 68.5(6) & 54972.23(1)& 103         &  8(3)     & 81 & 90 &  0(3)   & $-81$ \\
7668648 & 27.91210(4) & 44(1)   & 0.0388(6)& 81.0(5) & 54976.53(4)& 24.8        &  4(1)     & 89 & 86 & $-3$(1) & $-21$ \\  
7690843 &0.78615097(4)& 4.40(7) & 0        & $-$     &  $-$       &  $-$        &  0        & 80 & 80 &   0     & $-21$ \\
7812175 & 17.797913(1)& 31(2)   & 0.1568(4)& 325.6(2)&55004.688(8)& 217         &  0        & 88 & 88 &   0     & $-169$ \\
7821010 &24.2382708(1)& 48.18(8)& 0.667(2) & 238.1(2)&54969.283(5)& 57120(128)  &  14(3)    & 90 & 91 &$-14$(3) & $-1098$ \\      
7837302 &23.838001(2) & 45.3(9) & 0.27(1): & 272(8): & 54983.0(3):& 1799:       &  8(2)     & 86:& 78 &   1(2)  & $-907$ \\
8023317 &16.5790020(3)& 34(1)   & 0.258(2) & 165(2)  & 54976.2(1) & $-637$      & 57(2)     & 87 &104 &  55(2)  & $-921$ \\
8143170 & 28.78644(1) & 65.9(6) & 0.153(2) & 290.4(3)& 54971.31(2)& 811         & 37.9(2)   & 89:&111 &$-31.6$(3)&$-781$ \\
8210721 & 22.678148(2)& 39.5(5) & 0.1363(4)& 162.3(4)& 54965.34(4)& 285         &  0.9(9)   & 88 & 89 &$-0.06$(7)&$-240$ \\  
8211618 &0.33744683(4)& 2.50(7) & 0        & $-$     &  $-$       &  $-$        &  0        & 82:& 82 &   0     & $-2827$ \\
8553907 &42.0321296(4)& 79.6(3) & 0.543(1) & 327.6(3)& 54959.78(1)& $-66015$    & 61.8(2)   & 90 & 55 &$-55.1$(2)&$-80375$ \\
8560861 &31.9732355(7)& 61.30(7)& 0.037(1) & 187(6)  & 54959.8(5) & 24236       &  0        & 87 & 87 &   0     & $-17520$\\
8719897 &3.15130127(1)& 10.3(4) & 0        & $-$     &  $-$       &  $-$        &  0        & 83 & 83 &   0     & $-276$   \\
9007918 &1.387205051(2)& 6.6(1) & 0        & $-$     &  $-$       &  $-$        &  0        & 81 & 81 &   0     & $-2706$\\      
9028474 & 124.93719(2)& 138.1(7)& 0.8139(4)& 345.6(3)& 55012.79(2)& $-109417$   &  46(1)    & 88 & 91 & $-47$(1)& $-3101$ \\
9451096 &1.250421903(3)&6.8(3)  &0.00052(4)& 149(13) & 54954.31(4)& 137         &  0        & 80 & 80 &   0     & $-115$  \\
9664215 & 3.3195521(1)& 10.7(5) & 0.0036(2)& 136(2)  & 54963.70(2)& 1424        &  0        & 86:& 86 &   0     & $-1313$ \\
9714358 & 6.47470(2)  & 14.8(2) & 0.031(1) & 115(1)  & 54964.62(2)& 300(18)     &  0.4(3)   & 88 & 88 &  0.4(3) & $-19.9$ \\
9850387 &2.74848703(3)& 11.5    & 0        & $-$     &  $-$       & $-$         &  6(9)     & 82 & 85 &   5(7)  & $-2060$ \\
9963009 & 40.070257(2)& 71(1)   & 0.27(3)  & 260(2)  & 54985.4(1) & $-3428$     &  36(4)    & 89:& 55 &  12(11) & $-946$ \\
10095512& 6.0175275(1)& 15(1)   &0.00120(5)& 204(6)  & 54952.8(1) & 309         &  0        & 85 & 85 &   0     & $-282$\\
10223616& 29.14894(8) & 48.6(7) & 0.08(1)  & 66(5)   & 54987.5(5) & 95          & 11(3)     & 91 & 95 & $-10$(3)& $-77$ \\
10268809& 24.708999(5)& 41(1)   & 0.321(6) & 141(2)  & 54965.4(2) & 5229        & 27(3)     & 89 & 89 & $-27$(3)& $-2232$ \\
10296163&9.29678447(3)& 24(1)   & 0.40(1)  & 44.2(1) & 54961.79(4)& $-15595$    & 56.1(9)   & 88 &126 &$-44.2$(8)&$-22266$\\
10483644&5.11068306(6)& 16.6(9) & 0        & $-$     &  $-$       & $-$         &  0        & 85 & 85 &   0     & $-704$ \\
10549576& 9.0895504(8)& 27.9(7) & 0.0104(8)& 66(2)   & 54976.00(4)& 15547       & 28(7)     & 87 &114 &$-10$(10)& $-1509$ \\
10613718& 1.1757927(2)& 6.1(2)  & 0        & $-$     &  $-$       & $-$         &  0        & 86:& 86 &   0     & $-54$ \\
10666242& 87.246994(6)& 102(5)  & 0.22(4)  & 139.8(3)& 54960(1)   & 13761       & 28(6)     & 90 & 97 &  27(6)  & $-2818$ \\
10979716&10.6844567(4)& 23.8(8) & 0.083(5) & 104.4(9)& 54962.25(3)& 867         & 20(2)     & 88 & 69 &  7(2)   & $-599$ \\
11502172& 25.431989(3)& 41(2)   & 0.0920(6)& 351(2)  & 54973.6(1) & 4795        &  4(4)     & 89:& 85 &  2(2)   & $-4520$ \\
11519226& 22.163248(1)& 45.0(8) &0.18729(5)& 357.4(3)& 54977.05(2)& 877         & 14(2)     & 87 & 90 & -14(2)  & $-540$ \\
11558882& 73.913277(9)& 86.3(9) & 0.31(1)  &185.69(7)& 54977.5(3) & $-5170$     & 46.1(8)   & 89 & 84 &$-45.9$(8)&$-4210$ \\
12302391& 25.321357(1)& 64.3(3) & 0.20(1)  & 299(2)  & 54974.04(8)& $-6602$     & 44(1)     & 90:& 48 & $-18$(2)& $-6152$ \\
12356914& 27.308455(3)& 55(1)   & 0.43(2)  & 253.8(8)& 54976.06(3)& $-6365$     & 42.0(4)   & 90 &121 &$-29.5$(7)&$-1614$ \\ 
\hline
\end{tabular}}
\end{center}
\tablefoot{\tablefoottext{a}{This would predict an inclination variation of $\Delta i_1\approx+2\degr$ for today from the beginning of the \textit{Kepler} observations;}\tablefoottext{b}{$P_\mathrm{apse}$ was set to be an adjusted parameter.}} 
\end{table*}

\FloatBarrier

\twocolumn

\section{Notes on individual systems}
\label{app:Notesonindividualsystems}

\emph{1873918}: Besides \textit{Kepler} and TESS (FFI) observations, five additional eclipse times were measured at BAO, in 2018-19. \\
\emph{2302092}: In addition to \textit{Kepler} and TESS observations, four additional eclipses were observed at BAO, in 2018. \\
\emph{2305372}: Aside from \textit{Kepler} and TESS observations, we also used SWASP data and, furthermore, three eclipses were observed at BAO in 2018 and 2019. Moreover,  we also used some ASAS and HAT eclipse timings from the compilation of \citet{zascheetal15}. Finally, one additional eclipse time was taken from \citet{zascheetal15}. This is a double-lined spectroscopic binary. The total mass of the inner EB was calculated from the RV amplitudes given in \citet{matsonetal17}.\\
\emph{2444187}: In the absence of any third-body solution, this one was not listed in \citet{borkovitsetal16}. \citet{conroyetal14} tabulate it amongst the purely parabolic ETV systems. \\
\emph{2450566}: This is a very low amplitude ELV binary. For this reason, despite the relative brightness of the target ($T=11.3$), the ETV points formed from TESS light curves (mostly FFIs, but including the only two-min cadence S14 dataset) have larger scatter (even after forming normal ETV points) than the amplitude of the periodic signal found in the \textit{Kepler} ETV points \citep{conroyetal14,borkovitsetal16}. The new ETV points, however, imply that there was no longer-term period variation over the last 15 years.\\
\emph{2576692}: This is one of the longest period EBs in our sample ($P_1=87\fd9$). TESS observed only four new eclipses, two primary (S41, 54) and two secondary (S55, 74) ones. Detailed spectroscopic analyses are given in \citet{jonssonetal20}.\\
\emph{2708156 = UZ Lyr}: This is a long-time known, relatively well studied, classic, semi-detached (SD) Algol-system (EA), which is also a double-lined spectroscopic binary (SB2). The most recent spectroscopic and photometric studies were carried out by \citet{matsonetal17} and \citet{roobiatpazhouhesh22}. The first visual times of eclipse minima date back to 1920. The century-long ETV displays a very complicated structure, which is far from unusual for this kind of system. \citet{borkovitsetal16} presented an LTTE solution, approximating the longer-term variation with a cubic polynomial. \citet{roobiatpazhouhesh22} present a four-body solution, fitting two long-period LTTE orbits (with periods of $P_2=23.1$\,yr and $P_3=360$\,yr, respectively). We found that the combination of \textit{Kepler} and TESS minima themselves could be described nicely with a $P_2\approx2130$\,d LTTE orbit. This already contradicts, however, those relatively accurate, ground-based CCD eclipse times which were gathered over the previous two decades. Finally we used all the available eclipse times (though, most of them were from visual estimations, which we took into account by using much lower weights), and fitted two LTTE orbits. The shorter period orbit is mainly determined by the accurate satellite data, while the longer period one serves as some mathematical model of all the other timing variations. We took the total mass of the inner binary from \citet{matsonetal17}.  \\
\emph{2715007}: This likely contact ELV system is very strongly contaminated by the nearby, 8-9 mag brighter star KIC~2715115. Therefore, it was one of those few targets where we were unable to apply the {\sc FITSH} pipeline for the TESS FFI observations. In contrast to this, we were able to get acceptable normal light curves and then corresponding normal ETV points,  with a combination of the {\sc Lightkurve} pipeline with the PCA method. Therefore, finally, we were able to get a moderately certain $P_2=1949\pm2$\,d-period LTTE solution for this system. \\
\emph{2715417}: This short-period eclipsing binary ($P_1=0\fd24$) shows an EW-like light curve, however, \citet{negmeldinetal19} found a semi-detached light curve solution. TESS observed this target only in FFI mode. Due to the large scatter of the ETVs calculated from these latter newer observations, we formed normal light curves and, hence, determined normal eclipse times, and used them for our current ETV analysis. The interpretation of the evident, complex, non-linear timing variations was not an easy task. We finally arrived at a double LTTE, i.e., a (2+1)+1 type hierarchical quadruple system model, however, one should keep in mind, that the ratio of the two outer periods is very low, ($P_3/P_2\approx3.0$) which cannot result in stable orbits for more or less similarly massive objects. Hence, we conclude that our solution must only be an unphysical mathematical description of the ETV and, at least one of the two periodic effects should be of a different origin instead of the effect of an unseen stellar component. Therefore, we tabulate our solution for this object amongst the most uncertain LTTE cases.\\
\emph{2835289}: This is an ELV system, which exhibits third-body eclipses \citep[see, e.g.][]{conroyetal14}. One new third-body eclipse occurred and was observed in Sector 40. Due to the very low amplitude, only highly scattered ETVs can be calculated from the TESS light curves. Hence, we used normal TESS eclipse points. The new LTTE solution is in perfect accord with that of \citet{borkovitsetal16} and no longer-term trend can be detected.\\
\emph{2983113}: This is an EW system, which was observed with TESS only in FFI mode. The ETV curve suggests a longer period LTTE solution, however, our results are quite uncertain.\\
\emph{3114667}: The \textit{Kepler} light curve of this short period ($P_1=0\fd89$) EA system is available in the Villanova catalog only from Quarter 10. The system was also observed, however, in the early quarters. Hence, we downloaded and used the newly processed light curves of the \textit{Kepler} bonus project \citep{keplerbonus}. We also used TESS observations (2-min data for Year 4 and FFI only data for Years 2 and 6 visits). Due to the large scatter of these new ETV points, we formed normal light curves and, hence, eclipse times from these  TESS data. Here we present a moderately certain LTTE + quadratic polynomial solution; however, one should keep in mind that the very scattered S14 points are completely off the fitted curve.\\ 
\emph{3221207}: This totally eclipsing EW binary was observed in two-min cadence mode in all but one sector when TESS revisited this area of the  sky. (The sole exception was Sector 74, for which only FFI data are available.) According to \citet{alicavussoydugan17}, this is a low mass ratio contact binary. We obtained only a very uncertain, long-period LTTE solution.\\
\emph{3228863 = V404 Lyr}: This Hipparcos-discovered SD EA is an SB2 spectroscopic binary, which also exhibits $\gamma$~Dor-type pulsations. A recent analysis  of this system was published by \citet{leeetal20}. Besides \textit{Kepler} and TESS eclipse times we used several ground-based eclipses (including from the SWASP survey, as well as several targeted observations). The $P_2=653$\,d-period LTTE orbit is clearly detectable. We tried to model the longer timescale variations with both a cubic polynomial and additional LTTE variations caused by a fourth body. We found the best fit in the context of an LTTE+cubic polynomial model, and hence we present this solution. The total mass of the inner EB was taken from \citet{leeetal20}. Note, TESS observed this target in two-min cadence mode during most of the sectors when this target was revisited. The sole exception was Sector 75 for which only FFIs are available. \\
\emph{3245644:} This short period ($P_1=0\fd71$) binary was initially identified with KIC~03256638 in the Villanova catalog, but later \citet{abdulmasihetal16} reported that the true source was this background object. We used the larger amplitude, new light curve of \citet{abdulmasihetal16} to obtain better quality \textit{Kepler} ETVs which had been derived formerly from the blended light curve of KIC~03256638. The system was unobserved in TESS Year 2, but was observed in FFI mode in Years 4 and 6. These latter observations resulted in ETVs with larger scatter, hence, we formed normal light curves and eclipse timings, and we used them for our new analysis. We found and list here an uncertain LTTE+quadruple polynomial solution, where ETV points obtained from S74, 75 observations, are clearly off. \\
\emph{3247294}: This is a very long period ($P_1=67\fd42$) EA system, hence, during its eight-sector revisit by TESS (four in Cycle 4 and the other four in Cycle 6) only three new primary and four new secondary eclipses were observed. (And, moreover, in the case of the Sector 75 primary eclipse, the entire egress phase is missing and, therefore, we could not use this latter event.)  While, at a glance, the ETV curves strongly suggest some kind of dynamical timing variation caused by a third-body, the currently available dataset is insufficient for any certain solution of this type. Here we give a very uncertain LTTE+DE solution, allowing for precession of the orbital planes, but this result should be adopted only with considerable caution. (Note, we took the initial inner inclination angle from \citep{windemuthetal19}. \\
\emph{3248019:} The Villanova \textit{Kepler} EB catalog contains observations of this source only for Q3--17, hence, \citet{borkovitsetal16} gave an ETV solution only for this interval. Now, due to the \textit{Kepler}-bonus project \citep{keplerbonus} the light curve for Q1--2 is also available, therefore, first we recalculated the \textit{Kepler} ETV curves extending them to these early quarters. This new ETV, however, does not show any significant new characteristics compared with the former one \citep{borkovitsetal16}. Regarding TESS data, this object was not revisited during Year 2. For Year  4 observations, only FFI data are available, while during the recent Year 6 visits, the target was observed in 2-min cadence mode with the sole exception of Sector 82 for which, again, only FFI data are available. Due to the very shallow eclipse depths, and the faintness of the target, the calculated ETV points show very large scatter. Therefore, we formed normal light curves and eclipse times. Using them, we obtained a quite uncertain LTTE solution with a period much longer than the one that was published in \citet{borkovitsetal16}.\\
\emph{3335816}: This EA system was observed in two-min cadence mode in all TESS sectors when it was revisited. The earlier, $P_2=2250$\,d LTTE period, found by \citet{borkovitsetal16}, is likely incorrect. Newer observations imply a longer period LTTE orbit, however, our more recent period of $P_2=6088$\,d is also quite uncertain, and the newest, Sector 80 and 81 observations might suggest a reversal in the ETV curve.\\ 
\emph{3338660}: The Villanova EB catalog contains \textit{Kepler} light curves only from Quarter 8 onwards. Light curves and, hence, eclipse timings for the earlier quarters were taken from the \textit{Kepler}-bonus project \citep{keplerbonus}. This short period ($P_1=1\fd87$), detached EA system was observed only in FFI mode by TESS. We used only the timings of the much deeper primary eclipses. The new TESS ETVs contradict the former third-body solution of \citet{borkovitsetal16}. We were able to obtain only a very uncertain LTTE model; however, future observations are needed to establish a viable solution. (Note, the mass of the inner, eclipsing binary was taken from \citealt{cruzetal22}.) \\
\emph{3345675}: This is second the longest (inner) period ($P_1=120\fd00$) EB in our sample. It exhibits shallow ($\sim$4-5\% in depth) eclipses every 120 days in the \textit{Kepler} data. Because of this very long period, \textit{Kepler} observed only 11 eclipses during its entire original mission, while TESS detected only one additional primary event (available both in two-min cadence and FFI mode, in Sector 40 data).  Due to the absence of secondary eclipses, this system was first cataloged as a planetary candidate, but, as pointed out by \citet{armstrongetal17}, it is very likely a false positive. Moreover, while \textit{Kepler} did not observe any secondary eclipses, we were able to identify two shallow secondary eclipses in the TESS light curves at photometric phase $\phi\sim0.223$, during Sectors 41 and 81!  Hence, we can reliably assume that the light curve and, therefore, the ETV belong to an eccentric, not exactly edge-on binary. During our fitting process, we assumed that the inner inclination of such a very long-period and, hence, relatively wide binary, departs only very slightly from a perfectly edge-on view ($i_1\sim89\degr$), and the observed eclipses likely occur closer to periastron passage ($180\degr\leq\omega_1\leq360\degr$). On the other hand, however, despite the fact that we fitted only 12 primary and 2 secondary data points, we had to allow for a non-zero mutual inclination and, therefore, we were unable to minimize the effects of unavoidable overfitting. Using our pre-assumptions we found a good-looking, satisfactory solution which suggests a low mass third body revolving on a $P_2\approx4900$\,day, very highly inclined ($i_\mathrm{mut}=61\degr$) orbit. We checked that such an orbital configuration would cause the inner inclination to grow by $\Delta i_1\approx0\fdg9$ from the beginning of the \textit{Kepler} era to the Cycle 6 TESS observations.  This may be in nice accord with the appearance of the secondary eclipses during the recent TESS era.  On the other hand, however, we emphasize again that the recent appearance of the secondary eclipses provides only the combined information about $e_1\cos\omega_1$ instead of the eccentricity and orientation of the inner orbit ($e_1$, $\omega_1$) separately.  Moreover, the orbital solution we obtained contains 13 fitted parameters ($c_0$, $c_1$, $e_1$, $\omega_1$, $P_2$, $e_2$, $\omega_2$, $\lambda_2$, $m_\mathrm{AB}$, $m_\mathrm{C}/m_\mathrm{ABC}$, $i_\mathrm{mut}$, $n_2$) and is based on only the 14 available ETV points.  Therefore, the solution should be considered only with considerable caution. \\
\emph{3430883}: This likely EW binary was first found by \citet{bienasetal21} in the background of the \textit{Kepler}-target KIC~3430893. We downloaded and used the \textit{Kepler}-bonus light curve \citep{keplerbonus} to calculate the \textit{Kepler} portion of the ETV curve. The target was observed only in FFI mode in TESS sectors. Due to the close proximity of the much brighter original \textit{Kepler}-target (3430893), we had to apply a PCA method to obtain usable TESS light curves. The eclipse times, calculated from these latter light curves had very large scatter. Thus, we formed normal light curves from 20 consecutive cycles, and calculated normal eclipse times from this material. Finally, our $P_2=850$\,d-period LTTE solution looks quite certain.\\
\emph{3440230}: This EA system is also an SB2-type spectroscopic binary \citep{matsonetal17}. TESS made observations of this target only in FFI mode. We used only the much deeper primary eclipses for computing the ETVs. While the large amplitude ETV is evident, and it is also clear that the former, shorter period LTTE+quadratic polynomial models of \citet{zascheetal15,borkovitsetal16} are untenable, we were unable to find any certain solution. Therefore, here we present only a very uncertain LTTE model. We calculated the total mass of the eclipsing binary from the RV amplitudes and inclination, given in \citet{matsonetal17}. \\
\emph{3544694}: Targeted \textit{Kepler}-observations for this system are available only for Quarters 10--17. This eclipsing system, however, was also observed as a background object during the first 9 quarters of that mission. Therefore, thanks to the \textit{Kepler} bonus project \citep{keplerbonus}, these early sector data also became available. We downloaded this dataset from the MAST server, and recalculated the ETV curve for the entire \textit{Kepler}-mission. TESS also observed this target in FFI mode. We processed these data as well, but we were unable to identify any eclipse events, even using pixel-by-pixel photometry. Therefore, we do not use the TESS data but, in contrast to \citet{borkovitsetal16}, which studied only ETVs from Quarters 10--17, we analysed the ETV curves for the entire $\sim1470$\,d duration of the \textit{Kepler}-mission. Our results are in accord with the former solution of \citet{borkovitsetal16}. Note, we took the inclination angle of the inner binary from \citet{windemuthetal19}. \\ 
\emph{3936357 = V865 Lyr}: Two-min cadence observations are available and used for Year 2, 4, and 6 observations.  We extracted additional eclipse timing data from the publicly available SWASP observations, and added one more individual eclipse time \citep{ibvs6149} from the literature. These were all used in the analysis. We give a cubic+LTTE solution.\\
\emph{3938073}: The presence of a $P_2=270$\,d-period, non-transiting circumbinary planet around this long-period ($P_1=31\fd02$) eclipsing binary was announced by Welsh in a conference presentation\footnote{\url{http://www.astro.up.pt/investigacao/conferencias/toe2014/files/wwelsh.pdf}}; however, we are unable to either confirm or refute the existence of such a periodic signal in the \textit{Kepler} ETV curve. Instead, adding the TESS ETV points (for which 2-min cadence data were used for Years 2 and 4, and FFIs for Year 6), we now present a much longer outer period ($P_2=3091\pm76$\,d), though quite uncertain, low-amplitude combined LTTE+DE solution.  This suggests the presence of a third companion in the brown-dwarf mass regime. Note, also, that the apsidal motion of the inner eclipsing binary looks somewhat faster than what could be explained purely due to the dynamical perturbations of the third object and, hence, we fitted the apsidal motion period as an additional independent parameter.\\ 
\emph{4037163}: In the original \textit{Kepler} data, only Q10--17 observations are available. Based on this shortened (approx.~2-year-long) dataset, both \citet{conroyetal14} and \citet{borkovitsetal16} reported the detections of a relatively short outer period ($P_2\approx267$\,d) third companion via an LTTE solution of the ETV curve. Due to the \textit{Kepler} bonus project \citep{keplerbonus}, now observations for the entire original \textit{Kepler} mission are available. Hence, we recalculated the ETV for this longer $\sim4$-yr-long interval. While the $\sim267$\,d-period cyclic variations are continuously present, now it seems evident that there are substantial differences both in amplitudes and shapes amongst the consecutive cycles, which make the LTTE explanation very questionable. On the other hand, an additional, longer timescale non-linear variation is clearly visible in the extended-length \textit{Kepler} ETV dataset. Combining this latter data set with the new ETV points obtained from the TESS FFI observations, we now find weak evidence for the presence of a longer outer period ($P_2=2498\pm15$\,d) third component. We present this latter solution amongst the less uncertain LTTE cases. (Note, due to the large scatter of the TESS ETV points, we formed and used normal eclipse timings from the TESS data.) \\
\emph{4069063}: The TESS light curves for this source are strongly contaminated by the nearby RR~Lyr star V799~Cyg, hence, we had to apply PCA in order to obtain usable light curves and eclipse timings from the new FFI observations. Due to the very shallow eclipses, we also formed  normal light curves and eclipses to avoid too large a scatter in the new ETV points. Finally, these new ETV points confirm the former findings of \citet{conroyetal14,borkovitsetal16} about the $P_2\approx900$\,d LTTE period induced by the third star, and we also detected further non-linearities in the ETV which we describe with a quadratic polynomial. We took the total mass of the inner binary from the estimation of \citet{cruzetal22}. The minimum mass of the tertiary was found to be at least 75\% of the total mass of the eclipsing pair, which predicts a significant third light and, hence, might explain the very small eclipse depths. \\
\emph{4074532}: This EW-type overcontact binary was observed only in FFI mode during the TESS sectors. While in the \textit{Kepler} era, the ETV had exhibited a clearly parabolic shape \citep{conroyetal14}, after adding the TESS observations, an LTTE interpretation looks more likely.  However, currently, our LTTE model is yet very uncertain.\\
\emph{4074708}: Due to the large scatter, we used normal light curves and, hence, normal eclipse times formed from the TESS FFI data. The longer term variation was approximated with a cubic polynomial.\\
\emph{4079530}: The Villanova Kepler EB catalog contains a light curve for this binary only from Quarter 11. That dataset was used for the ETV analysis of \citet{borkovitsetal16}. Observations for the earlier \textit{Kepler} quarters became available due to the \textit{Kepler} bonus project \citep{keplerbonus}. Therefore, we recalculated the ETV points for this longer \textit{Kepler}-dataset. The primary eclipses display a clear $P_2=142$\,d-period, low amplitude cyclic variation. The very shallow secondary eclipses were used to produce a secondary ETV curve, for which the scatter is substantially larger than the amplitude. Despite this, we kept these secondary ETV points, as they clearly diverge from the average of the primary ETV points, implying apsidal motion in the system. Moreover, we added TESS primary eclipse times (determined from PCA corrected FFI light curves). We were unable to calculate secondary eclipse times from the lower quality TESS light curves. Finally, using the most extended data series, we obtained a quite certain LTTE+DE solution with apsidal motion. The inferred mass of the tertiary is located in the substellar mass regime. Therefore, the third-body most likely is an unknown, non-transiting circumbinary planet. \\
\emph{4138301}: This is likely an ELV binary. TESS observed this system only in FFI mode. To obtain decent ETV points from these latter data, we first applied PCA cleanings for all sectors, and then formed normal light curves for the eclipse timing calculations. In such a manner we are in a position to confirm the former findings of \citet{conroyetal14} and \citet{borkovitsetal16}, who reported an $\approx800$\,d-period LTTE orbit. We also confirm some additional ETV variations which we model with a cubic polynomial. \\
\emph{4241946}: No two-min cadence TESS data are available for this EW binary. The inferred low third-body mass is in accord with the absence of significant contaminated (or, third) light \citep{kobulnickyetal22}.\\
\emph{4244929}: This is a low mass ratio \citep[$q_1=0.12$,][]{kobulnickyetal22} EW system. The ETV solution is quite uncertain; however, the large amplitude (and, hence, the inferred relatively large minimal mass for the tertiary) is in accord with the substantial amount of third light \citep[$\ell_3=0.486$,][]{kobulnickyetal22}. No two-min cadence data. One BAO eclipse time is added.\\
\emph{4451148}: TESS observed this target in two-min cadence mode during its Year 2 and 4 visits. For Year 6 observations only FFI mode data are available. SWASP data are available, but were not used for the analysis. The former $P_2\sim750$\,d-period LTTE solution of \citet{conroyetal14,borkovitsetal16} is confirmed, but an additional longer timescale non-linear trend (modeled here by a cubic polynomial) is evident.\\
\emph{4547308}: \citet{kobulnickyetal22} gives a third light of $\ell_3\approx0.21$ which appears to be in accord with the inferred third-body mass of our current solution. The outer eccentricity was found to be the highest amongst the certain solution systems with $e_2=0.90\pm0.02$. Besides the LTTE parameters, a quadratic polynomial was needed for the fit as well.\\
\emph{4574310}: This is an SB2 EA system, with deep primary and shallow secondary eclipses. Hence, we used only primary eclipses. The solution is very uncertain, and Year 2 TESS-points are clearly off the best-fitting solution. The total mass of the EB was taken from \citet{matsonetal17}. \\
\emph{4647652}: Two-min cadence TESS observations are available and were used for eclipse timing, in all but Sectors 74 and 75 for which only FFIs are available and, hence, used. Two additional primary eclipses were observed at BAO in 2019, however, the second one was not used because TESS Sector 14 observations are also available for that event. We used only the ETV data from primary eclipses.  Our results nicely confirm the former LTTE solutions of \citet{rappaportetal13,conroyetal14,borkovitsetal16}. \\
\emph{4669592}: This target was observed by TESS only in FFI mode, however, two-min cadence data are also available indirectly, since its signal contaminates the two-min cadence light curves of KIC 4576968. We calculated ETVs from both the current target's own FFI light curves, and the two-min cadence blended light curves, but found that the two kinds of data are of similar quality. \cite{conroyetal14} reported a pure quadratic ETV curve. For the present, longer observing window, departures from the quadratic ETV can evidently be detected. The ETV now can be modeled equally well with either a simple cubic polynomial or an LTTE orbit. Here, naturally, we present the LTTE solution amongst the most uncertain cases. \\
\emph{4670267}: The TESS ETV points have larger scatter than the amplitude of the $P_2\approx760$\,d-period LTTE signal, which was first detected by \citet{rappaportetal13} in the \textit{Kepler} data. Hence, we formed normal eclipse points from the TESS (FFI) data. There is likely an indication of further, longer-term period variations, which we model with a cubic polynomial. \\
\emph{4681152}: This EA system was observed only in FFI mode during all the TESS visits. The former, shorter period LTTE solution of \citet{borkovitsetal16} is clearly untenable. Our new, longer period LTTE+quadratic solution is very uncertain as well. Only primary eclipses were used.\\
\emph{TIC 1716037782 = 4732015FP}: This target has no KIC ID. Formerly the EB was assigned to KIC 4732015, but \citet{abdulmasihetal16} found that the true source is this background object. Hence, using the new photometry of \citet{abdulmasihetal16} we recalculated the \textit{Kepler} ETV curve as well. After adding the TESS ETVs (calculated from FFI light curves) the LTTE signal looks quite likely. Further, longer timescale non-linearities can also be detected in the ETV, which we model with a cubic polynomial. \\
\emph{4753988}: Only FFI data are available. We found a quite uncertain LTTE+DE solution, which yields $i_\mathrm{mut} \approx 44\degr$, but this is not to be taken too seriously, since the dynamical contribution to the ETV is only about $\sim7\%$. \\
\emph{4758368}: TESS observed this target in 2-min cadence mode during Cycles 2 and 4, while only FFIs are available for the Cycle 6 revisits. The tertiary star is most likely a red giant \citep{gaulmeetal13,nessetal16}. $\delta$ Sct-type pulsations are also detected \citep{gaulmeguzik19,chenetal22}. The lines of the RG star can be detected spectroscopically. SB1 RV analysis was carried out by \citet{helminiaketal16,helminiaketal19}, however, their deduced outer periods contradict our findings. We estimated the mass of the EB taking into account the masses deduced from the two RV solutions of \citet{helminiaketal16,helminiaketal19}.\\
\emph{4762887}: This is a likely ELV system, which was observed only in FFI mode with TESS. We formed normal light curves and, hence, normal eclipse times from these new data but, even so, the scatter is larger than the \textit{Kepler}-observed ETV amplitude. Therefore, our LTTE solution is very uncertain. \\
\emph{4848423}: Two-min cadence TESS observations are available and were used for the eclipse timing for the Cycle 2 and 4 revisits, while the system was observed only in FFI mode during Cycle 6. Four eclipses were observed at BAO, in 2018 and 2019. This is a double-lined spectroscopic binary. The total mass of the inner EB was calculated from the RV amplitudes given in \citet{matsonetal17}. \\
\emph{4851217}: This is an SB2 system. Detailed spectroscopic analyses were carried out by \citet{matsonetal17} and \citet{helminiaketal19}. We have taken the binary mass from this latter work. The EB stars also have a $\delta$ Sct pulsational component \citep{liakos20,chenetal22}. A recent photodynamical analysis was carried out by \citet{jenningsetal24}. \\
\emph{4859432}: This EW binary was only observed in FFI mode by TESS. Two BAO primary eclipses from 2019 were also added to the analysis. In addition to the cyclic ETV observed already in the \textit{Kepler} era \citep{conroyetal14,borkovitsetal16}, now a longer timescale, non-linear variation is also detectable. We model that with cubic polynomial.\\
\emph{4909707}: This is a totally eclipsing, and most likely a semi-detached system. The quadratures have slightly different brightness levels. which might be due to Doppler boosting. The secondary eclipse is slightly offset, indicating a small inner eccentricity. This is unexpected for an SD system, and it might be `virtual only' because of the small asymmetry in the out-of-eclipse light curve. Despite this fact, we modeled this system allowing for a non-zero inner eccentricity and, apsidal motion, as well. Note, this target was observed mostly in 2-min cadence mode with TESS; the only exceptions are Sectors 74 and 75, where only FFI observations are available.\\
\emph{4937217}: According to \citet{liliu21}, this EW binary may be a member of the open cluster NGC~6819. The TESS (FFI) eclipses contradict the former solution of \citet{borkovitsetal16}. Here we present a completely new, though quite uncertain, long-period LTTE solution. We took the EB's mass from \citet{liliu21}. Note, TESS observed this system only in FFI mode. \\
\emph{4940201}: The TESS light curves of this detached EA system are strongly contaminated with the signal of the EW star KIC 4940226 (see the next item, below) and, moreover, an additional, shorter period EA binary also contaminates some sectors. Despite this, with the application of the careful PCA process, we were able to obtain new eclipse times for this shallow-eclipsing, faint system from the TESS data.  (Note, it was observed in two-min cadence mode in Sectors 81 and 82, while only FFIs are available for the former sectors.) With the use of these new ETV points, we confirm the former LTTE+DE solutions of \citet{borkovitsetal15,borkovitsetal16}. The inclination of the eclipsing binary was taken from \citet{windemuthetal19}. \\
\emph{4940226}: Non-linear variations of ETVs of this EW binary were not reported by either \citet{conroyetal14} or \citet{borkovitsetal16}. Instead, the Villanova \textit{Kepler} EB catalog claims that the EB signal comes from KIC~4669592 (see above), but this is clearly an incorrect conclusion. We obtained a relatively long-period LTTE solution ($P_2=5333\pm47$\,d), but, this period is very close to the total length of the ETV dataset ($\sim5594$\,d). Therefore, despite the good-looking fit, we tabulate this system amongst the most uncertain cases, and flag it as a system where future eclipse monitoring may be important for any long-term, non-linear period variation.\\
\emph{4945588}: The source of this ELV-type light curve is most likely a low-inclination, detached binary system \citep{kobulnickyetal22}. \citet{conroyetal14} tabulates it as a possible triple system with a period, longer than the duration of the \textit{Kepler}-mission. It was observed in two-min cadence mode in TESS Cycles 2, 4, and also in FFI during Cycle 6 sectors. Our finding confirms the assumption of \citet{conroyetal14}, however, the very large outer eccentricity ($e_2=0.82\pm0.02$) should be taken with some caution. \\
\emph{4945857}: In addition to the TESS FFIs, four BAO eclipse times (from 2018-19) were also included. The large ETV amplitude implies a massive tertiary, which is in accord with the $\ell_3=0.656$ third-light fraction obtained by \citet{kobulnickyetal22}, hence, we ranked this system as being in the group of the certain LTTE identifications. Note, for purposes of the mathematical modelling of some other non-linearities in the ETV points, we added a cubic polynomial to the LTTE fit.\\
\emph{5023948}: This is a triple-lined spectroscopic triple system where the inner pair is a detached EA binary which is formed by two almost twin stars. According to the dedicated study of \citet{breweretal16}, this system is a likely member of the open cluster NGC~6819. They give a $P_2=8333$\,day-period, eccentric ($e_2=0.37$) spectroscopic solution for the outer orbit, which they find to be in accord with the \textit{Kepler} ETV points (after adding a few additional ground-based eclipse timings from their own observations to that dataset). We find, however, that after taking into account the new TESS FFI observations and the ETV points derived from these data, some of their own ETV points look clearly off. With the combination of the \textit{Kepler} and TESS ETV data we found a quite uncertain LTTE solution with a somewhat shorter period ($P_2=7295$\,d) and higher eccentricity ($e_2=0.58$). However, we claim that the ETV induced by the third star looks fairly certain. Despite this, due to the fact that the period is longer than the duration of the full dataset, we put this system into the most uncertain class. (Note, the total mass of the inner binary was taken from the above mentioned study of \citealt{breweretal16}). \\
\emph{5039441}: This totally eclipsing eEA binary exhibits a clear third-body eclipse at the beginning of TESS Sector 15, and another two-dipped, very uncertain, secondary third-body eclipse event near the beginning of the Sector 82 observations. For TESS Year 2 and 4 only FFIs are available, but {in the Year 6 sectors (75, 81, 82)} it was observed in two-minute cadence mode. Two additional primary eclipses were observed at BAO in 2019. These eclipse times are included in our analysis. The photometric total mass estimation of the binary was taken from \citet{cruzetal22}. We plan to carry out a complex photodynamical analysis in the near future.\\
\emph{5080652}: This EA system was observed in two-min cadence mode in Sectors 80, 81, and only in FFI mode in all the other sectors with TESS. The scatter of the new ETV points is much larger than the amplitude of the $P_2=223$\,d-period cyclic variation, reported first by \citet{borkovitsetal16}. We forced a coplanar LTTE+DE solution, without any additional period variations. We took the inner inclination from \citet{windemuthetal19}. Note, a former light curve analysis was presented in \citet{kjurkchievaatanasova16}. \\
\emph{5080671}: This EW-type overcontact binary, which was not a primary \textit{Kepler}-target, is only $\sim45''$ away from KIC~05080652 (see above). Hence, the TESS light curves of the two targets strongly contaminate each other, and therefore needed disentanglement. Due to its proximity to KIC~05080652, we were able to obtain its \textit{Kepler} light curve and, therefore, the corresponding ETV curves, with pixel photometry in the very same manner described in \citet{bienasetal21}. Our quite uncertain solution suggests the LTTE of a long period ($P_2=5446\pm85$\,d) third stellar component. Note, SWASP observations, though available, were not used due to the very large scatter of the ETVs obtained from that data set. \\
\emph{5095269}: The likely presence of a planet-mass, non-transiting, inclined circumbinary object (with $\im\approx40\degr$) was first found by \citet{borkovitsetal16} through their ETV analysis. Then \citet{getleyetal17} confirmed this finding, but they claimed that the third object revolves on a retrograde orbit ($\im=120\degr$). Recently \citet{goldbergetal23} carried out a comprehensive analysis combining photometry, spectroscopy, SED analysis and dynamical modelling. They also confirm the existence of this planetary mass body, however, they state that such an inclined massive planet would cause readily observable eclipse depth variations during the four years of the \textit{Kepler} observations. Therefore, they infer that the circumbinary planet is likely almost coplanar with the inner EB. We note, however, that in our analytic calculations, the published \citet{borkovitsetal16} LTTE+DE model, even with $\im=40\degr$, would cause an inclination variation of only $\Delta i_1\approx-0\fdg1$ during the entire 4\,years of \textit{Kepler} observations instead of an inclination variation rate of $\Delta i_1=-0\fdg144\,\mathrm{yr}^{-1}$, which \citet{goldbergetal23} claim the \citet{borkovitsetal16} solution. Now we add TESS ETV points (determined from two-min cadence data). In the case of our current analysis, we set and fix the total mass of the inner binary, as well as the orbital elements of the inner pair ($e_1$, $\omega_1$ and $i_1$) according to the results of \citet{goldbergetal23}, but we again allow for an arbitrary mutual inclination. Our results prefer a mutual inclination of $\im\approx30\degr$, which would produce a total visible inclination variation of $\Delta i_1\approx0\fdg4$ from the beginning of the \textit{Kepler} observations up to now. \\  
\emph{5113053}: This EA system is an SB2 spectroscopic binary, and a member of the open cluster NGC~6819. The triplicity of this target was first found spectroscopically by \citet{jeffriesetal13}. Later, \citet{breweretal16} gave an orbital solution for the wide orbit by analysing the systemic radial velocity variations of the eclipsing pair. Besides the \textit{Kepler} and TESS (FFI) points we took some further ground-based eclipse timings from this latter work. Our LTTE solution looks close to the spectroscopically determined orbit of \citet{breweretal16}, hence, we qualify our findings as moderately certain. We took the total mass of the inner pair from this latter work.\\
\emph{5128972}: This is an EW or $\beta$~Lyrae (E$\beta$\footnote{Usually the $\beta$ Lyrae-type subclass of eclipsing binaries is denoted by `EB' but, in the present paper, the abbreviation `EB', in general, stands for eclipsing binaries.  Hence, we decided to use the abbreviation `E$\beta$' for the $\beta$~Lyrae-type subclass.})-type EB. Its $P_2\approx444$\,d-period third component was first detected by \citet{rappaportetal13} and confirmed later by \citet{conroyetal14,borkovitsetal16}. Adding the TESS FFI ETV points, we also rigorously confirm this periodicity in the extended ETV curve; however, now one can detect further slight non-linear variations, which we (mathematically) interpret with a quadratic polynomial fitted simultaneously with the $P_2=444\fd64\pm0\fd06$ period LTTE term. Note, due to the large scatter of the TESS observationally derived ETV points, we formed and used normal light curves and, hence, normal eclipse timings for these data.\\
\emph{5216727}: This is a chromospherically active EA binary system with shallow eclipses. Interestingly, in the TESS FFI light curves, the eclipses appear to be substantially deeper than in both the \textit{Kepler} and the TESS Year 6 two-min cadence data (only FFIs are available for the former TESS observations). Normal eclipses were calculated from both the TESS FFI and two-min data. Besides the $P_2\approx532$\,d cyclic ETV variations, the very scattered new ETV points do not imply any further, longer term variations.\\
\emph{5255552}: This long-period ($P_1=32\fd46$), detached EB exhibited three sets of multiply dipped extra eclipses during the \textit{Kepler} mission. TESS did not observe any new extra eclipsing events and, because the inner EB period is longer than the duration of one TESS sector, we found only 14 new binary eclipses (one of which became unusable due to the missing ingress phase). (Note, TESS observed this target in Years 2 and 4 only in FFI mode, but two-min cadence data are available for the three Year 6 sectors.) The dynamically dominated ETV of the binary was analysed by \citet{borkovitsetal15,borkovitsetal16} and the analytic three-body LTTE+DE solution was found to be in accord with the times of the extra eclipses. On the other hand, however, due to the fact that the four extra eclipses were spread over a 25-day interval (between BJDs 2\,455\,509 and 2\,455\,536), it was speculated that the eclipsing (or, eclipsed) companion might be a second binary in and of itself \citep[see, e.g.,][]{getleyetal20}. Then, recently, \citet{orosz23} found a consistent, 2+2 quadruple system photodynamical model solution for this target where only one of the two binaries is eclipsing. In this work, similar to that of \citet{borkovitsetal15,borkovitsetal16} we consider this extra companion to be one body (as we assume that the effects of its binarity are negligible for such a short-term, analytic ETV analysis). We set, however, the total mass of the eclipsing pair, as well as the initial orbital parameters of both the inner and outer orbits according to the results of \citet{orosz23}. Moreover, we forced the mutual inclination of the inner eclipsing binary orbit and the wide orbit to remain under $\im=3\degr$, which was a necessary criterion to preserve both the triply eclipsing nature of the third star (or the members of the other binary), and to avoid any observable, but non-observed EDVs of the 32\,d-period EB. Even when we applied such a constraint, we found that adding a cubic-order polynomial to the LTTE+DE model substantially improves the fit. We explain these additional non-linear terms as a mathematical description of the higher order secular perturbations which are beyond the scope of our quadrupole-order analytic secular perturbation model (see Sect.~\ref{sec:polynomial} for details). \\
\emph{5264818}: This ELV binary contains a chemically peculiar Ap star \citep{shietal23}. It was observed in two-minute cadence mode in TESS Cycles 2 and 4, while only FFIs are available for Cycle 6 sectors. There is an evident jump between the \textit{Kepler} and TESS ETV data, which we modeled with a fourth, more distant stellar component on a quite eccentric orbit. \\
\emph{5269407}: This is a chromospherically active Algol-system with shallow secondary eclipses. Hence, we used only the primary eclipses. Besides TESS FFI data, we also used one targeted primary eclipse observation, carried out at BAO in 2018. Now, the LTTE solution appears to be reasonably robust, though its period is much longer than was given in \citet{conroyetal14} and \citet{borkovitsetal16}. \\
\emph{5288543}: This is an SB2 eEA system. All TESS observations were carried out in FFI mode. It is unclear, whether the ETV curve exhibits pure parabolic variations or, a segment of a long-period LTTE orbit, though this latter assumption provides a better fit. Hence, we give the numerical results for this latter model. The total mass of the binary was taken from \citet{hambletonetal22}.\\
\emph{5307780}: No two-minute cadence TESS observations exist, hence, the FFIs were used. The long-term variation looks certain, but the low-amplitude cyclic ETV, interpreted as an LTTE, is less than secure.\\
\emph{5310387}: Only FFIs are available for TESS sectors. The light curves are blended with the EA system KIC 5310435. We successfully disentangled the signals applying PCA. The new ETV points have too a large scatter for the independent detection of the low amplitude LTTE signal, but the quadratic term found already in \citet{borkovitsetal16} is nicely confirmed. \citet{kobulnickyetal22} list $\approx35\%$ of the flux as third light. If that is correct, the inferred third-body orbit must have very low inclination or, the source of the contaminated light could be a fourth stellar body.\\
\emph{5353374}: For the TESS sectors only FFIs are available. The shorter, but very uncertain, third-body period given in \citet{conroyetal14} and \citet{borkovitsetal16} appears to be incorrect. The current, longer period may be more likely but, especially due to the additional likely non-linear variations, which we currently modeled with a cubic polynomial, further ETV points are really required.\\
\emph{5376552}: The former third-body solutions \citep{rappaportetal13,conroyetal14,borkovitsetal16} are nicely confirmed. The substantially longer observing window also provides evidence for an additional longer-term period variation. This latter is modeled with a quadratic polynomial. We took the total mass from the estimation of \citet{liliu21}. No two-min cadence data are available.\\
\emph{5384802}: This detached EA system shows very shallow eclipses, which made it necessary to form normal light curves (and, hence, normal eclipse times) from the TESS FFI (Cycles 2 and 4) and two-min cadence (Cycle 6) observations. Using these normal eclipses we can confirm the former relatively short-period ($P_2\approx256$\,d) third-body solution of \citet{rappaportetal13} and \citet{borkovitsetal16}, while the new data also reveal that there were no other period variations between the observations of the two planet-hunting space telescopes. Note, $i_1$ was taken from \citet{windemuthetal19}. \\
\emph{5393558}: TESS observed this target in FFI mode only. We give an LTTE+DE solution, however, it is quite uncertain. We took the $i_1$ from \citet{windemuthetal19}. \\
\emph{5459373}: The TESS (FFI) ETV points have larger scatter than the amplitude of the $P_2\approx413$\,d LTTE orbit, found previously by \citet{conroyetal14,borkovitsetal16}. Hence, normal eclipses were formed. These indicate some longer timescale non-linear variations, which we modeled with a parabola. Note, however, TESS Year 2 ETV points are clearly off the model. \\
\emph{5478466}: Only TESS FFI observations are available and used. One BAO secondary eclipse just before S14 is also used. We formed normal eclipses from the TESS data. The resultant LTTE solution is quite robust. The additional non-linearity is modeled with a quadratic polynomial. \\
\emph{5513861}: This is an SB2 system. Two-min cadence TESS observations are available for all sectors. The EB's mass is taken from \citet{matsonetal17}. Four eclipses were observed at BAO in 2018-2019. Additional eclipse times were taken from \citet{zascheetal15}.\\
\emph{5611561}: This EW system was observed only in FFI mode with TESS, and from those observations we formed normal light curves and, hence, normal eclipse timings. Note, before the newest data from Sectors 80--82, a pure LTTE solution would have been  preferable with a period longer by $\sim$$150-200$ days than the period of the finally accepted LTTE+cubic model, but these last sectors clearly contradict that model. Hence, we finally settled on an LTTE+cubic polynomial model solution with $P_2=993\pm 1$ days.\\
\emph{5621294}: This detached EA system is also an SB2 spectroscopic binary \citep{matsonetal17}, hence, the masses of the inner pair can be inferred dynamically, and we used the total mass of the binary as it was determined by \citet{matsonetal17}. \citet{leeetal15} claimed, that the (\textit{Kepler}) ETV of the system can be described with a combination of a quadratic polynomial and the low amplitude LTTE of a likely substellar mass companion. Such a solution was also given in \citet{borkovitsetal16}, however, they ranked this solution as quite uncertain. The ETV points formed from the recent TESS FFI observations, as well as one additional primary eclipse time measured at BAO in 2018, do not verify such an assumption. Instead, we obtained a longer period LTTE plus cubic polynomial solution with a low mass, but likely stellar, companion; however, even these new results are far from being conclusive.\\
\emph{5632781}: This totally eclipsing, eccentric Algol-system (eEA) was observed in 2-min cadence mode over all its TESS visits. \citet{kjurkchievavasileva18} in their dedicated study report clearly detectable periastron bumps on the 4-year-long \textit{Kepler} light curves, which indicate that this system is an eclipsing heart-beat binary. We took the initial parameters for our Markov chain Monte Carlo (MCMC) runs from their study. We derived a $P_2=1584\pm2$ day period LTTE+DE solution together with an independently fitted $P_\mathrm{apse}=16\,445\pm175$\,yr long apsidal motion period. This latter effect cannot have a dynamical origin, and this is the  reason why the apsidal motion period was fitted independently, instead of constrained from the third-body parameters.  This conclusion stems from the fact that the derived third-body mass is clearly in the substellar domain ($M_\mathrm{C}=0.006\pm0.001\,\mathrm{M}_\sun$), the perturbations from which would produce dynamical apsidal motion only with an order of magnitude longer period.\\
\emph{5650317}: This detached EA system was listed in the Kepler EB catalog under the ID of KIC~5650314. It was shown, however, by \citet{abdulmasihetal16} that this was a false positive, as the true EB is 5650317 in the field of 5650314. Hence, we also recalculated the \textit{Kepler} eclipse times using the new photometry of the \textit{Kepler} bonus project \citep{keplerbonus}. As usual, we also used TESS FFI data, from which we were able to calculate readily usable light curves applying only the PCA method. The TESS ETV points have larger scatter than the amplitude of the short-period sinusoidal term, which is evident from the \textit{Kepler} ETV points. Hence, we formed normal eclipses from the TESS data. We conclude that this system is a very compact, and relatively tight, newly discovered clear triple system with an outer period of $P_2=118\fd4$, and period ratio of $P_2/P_1=118.4/2.14\approx55.3$. For our LTTE+DE solution we consider the inner orbit to be circular ($e_1=0$) and the two orbits as coplanar ($\im=0\degr$). \\
\emph{5653126}: This detached EA system displays clear eclipse depth variations. During the early \textit{Kepler} quarters no secondary eclipses were observable. They appeared only $\sim 1.5$\,yrs after the beginning of the \textit{Kepler}-mission. Toward the end of the \textit{Kepler} observations both kinds of eclipses became progressively deeper (in depth) and longer (in duration). Then, during the 2019 TESS observations, the eclipses changed to be U-shaped instead of V-shaped. But, in the newer TESS sectors, the eclipse shapes returned to V-like, and their depths and durations started to decrease again, indicating that the inner pair went through an exactly edge-on phase ($i_1=90\degr$) during its ongoing precession cycle. The ETV of the target was analysed first by \citet{borkovitsetal15,borkovitsetal16}. Recently a thorough photodynamical analysis was also carried out by \citet{borkovitsetal22b}, using observations up to TESS Sector 41. In the current analysis we naturally use all the currently available ETV points from TESS (FFIs for Cycle 2, 4 data, and 2-min cadence observations for Cycle 6). The initial values of most of the parameters to be fitted in our analytic ETV study were taken from the photodynamical results of \citet{borkovitsetal22b}. Our analytic LTTE+DE solution is in fine accord with the more complex photodynamical results of \citet{borkovitsetal22b}. \\
\emph{5731312}: This is another detached system with characteristic ETV and EDV. The secondary eclipses disappeared in the gap between the \textit{Kepler} and TESS observations. After the first ETV studies of \citet{borkovitsetal15,borkovitsetal16}, \citet{borkovitsetal22b} carried out a complex photodynamical analysis of the system. We use their results as initial parameters for most of the adjustable variables. Besides the ETVs from \textit{Kepler} and TESS FFIs, we also use two ground-based eclipse times obtained at BAO between the Cycle 2 and Cycle 4 TESS observations. \\
\emph{5771589}: This is one of the tightest EBs in the \textit{Kepler}-sample with $P_2/P_1\approx10.5$. The four years of \textit{Kepler} observations cover more than a half of the (dynamically forced) apsidal motion period. Following the ETV analyses of \citet{borkovitsetal15,borkovitsetal16}, the complex photodynamical analysis was published in \citet{borkovitsmitnyan23}. We used these latter results for the initial parameters for our analytic LTTE+DE ETV fits. (Note, TESS observed this target in two-min cadence mode during all sectors for which the source was in the camera fields.) \\
\emph{5903301}: Despite the fact that we were unable to detect the very shallow eclipses of this faint EA system in the TESS observations, we include here a new ETV analysis of the \textit{Kepler} ETV data which now, thanks to the \textit{Kepler}-Bonus project \citep{keplerbonus}, cover the full range of the \textit{Kepler} observations. The former analysis of \citet{borkovitsetal16} considered ETV data only from \textit{Kepler} quarters 3--17. The eclipse timings of the now available earliest \textit{Kepler} quarters, however, show a clear reversal in the ETV data, making necessary a new analysis. Now the $P_2=1260$\,d-period LTTE solution looks quite convincing but, due to the lack of any newer ETV data, we rank our solution only as moderately certain.\\ 
\emph{5952403}: This is one of the very first triply eclipsing triple star systems \citep{derekasetal11} that was discovered.  It is composed of a detached eclipsing pair of two low-mass, K-type stars ($P_1=0\fd906$), and a chromospherically active red giant tertiary revolving on a $P_2=45\fd47$-period outer orbit. Detailed analyses of this remarkable triple were presented in \citet{borkovitsetal13,fulleretal13,czeslaetal14}. As both orbits are circular, and the two orbital planes are practically coplanar, one cannot expect detectable third-body perturbation effects \citep{borkovitsetal03}. Despite this, we carried out both LTTE and LTTE+DE-model runs. We obtained essentially similar results to those in these earlier papers.  In the case of both the LTTE and LTTE+DE models, the analytic fits underestimate the amplitude of the $P_2=45$\,d-period ETV cycles, which leads to somewhat smaller third-body mass ($m_\mathrm{C}=2.3\,\mathrm{M}_\sun$) than the one obtained from the more complex analysis of \citet{borkovitsetal13} ($m_\mathrm{C}=3.0\,\mathrm{M}_\sun$). Note, this target was observed in two-min cadence mode during most of its TESS revisits. The three exceptions are Sectors 74, 75 and 82, where only FFIs are available.\\
\emph{5956776}: This is a faint (likely) EA system with very shallow secondary eclipses. We used only the primary ETV curve for all of our analysis. The Villanova EB catalog contains the EB's light curve only from Quarter 8 onward. We added, however, the light curves and, hence, the ETV points of the early \textit{Kepler} quarters Q0 and 1 from the \textit{Kepler}-bonus project \citep{keplerbonus}. TESS observed this system only in FFI mode. From these latter observations, we formed normal light curves and, therefore, normal eclipse timings to reduce the scatter of the new ETV points. Finally we got a slightly shorter outer period LTTE solution than what was published in \citet{borkovitsetal16}.  But, primarily due to the large scatter of the TESS normal ETV points, we rank our solution as uncertain. \\
\emph{5962716}: TESS observed this target only in FFI mode. Three primary eclipses were observed at BAO in 2019 and these were added to the ETV data. Only primary eclipses were used in the fit.\\
\emph{5975712}: This ELV binary was observed in 2-min cadence mode during TESS Year 2, while only FFI data are available for Years 4 and 6 sectors. We formed normal light curves and, hence, normal eclipse times from the TESS data. Our -- quite uncertain -- solution suggests a long period ($P_2=4959\pm8$\,d), highly eccentric ($e_2=0.76\pm0.01$) LTTE signal.\\
\emph{6050116}: This EW system was listed in \citet{conroyetal14} as a candidate host of a $P_2\approx1078$\,d-period third component. On the other hand, this target was left out of the \citet{borkovitsetal16} paper due to the remarkable difference amongst the averaged ETV and QTV patterns. Adding ETV points determined from TESS FFI observations reveal the presence of a $P_2\approx2440$\,d-period cyclic term, which we interpret as an LTTE signal, and some longer timescale non-linear variations which can nicely be modelled with a quadratic polynomial. Note, the folded \textit{Kepler} light curve was analysed in \citet{negmeldinetal19}.\\
\emph{6103049}: This short eclipsing period ($P_1=0\fd64$) likely detached EA-type system was observed only in FFI mode throughout the TESS visits. Due to the large scatter of the new eclipse timing data we formed and used normal light curves and, hence, normal ETV points from the TESS data. The ETV curve exhibits a clear, long-period ($P_2=5764$\,d) sinusoidal pattern for which a likely explanation is a LTTE orbit. However, as this outer period is slightly longer than the length of the full dataset, we mark this triple candidate as an uncertain case. (Note, the total mass of the inner, eclipsing binary was taken from \citealt{cruzetal22}.) \\
\emph{6118779}: This totally eclipsing EW-type binary exhibits large light curve variations on a short timescale and, even the depth ratio of the primary and secondary eclipses may reverse. This is likely due to a substantial starspot activity \citep[see, e.~g.][]{balajietal15}. TESS observed this target only in FFI mode. The eclipse times calculated from these data reveal large non-linear variations in contrast to the \textit{Kepler} timing data. We calculated a long period ($P_2=10542\pm882$\,d) LTTE model solution, however, the derived minimum third-body mass exceeds the estimated total mass of the eclipsing binary. In contrast to this, \citet{kobulnickyetal22} obtained only $\approx21$\% third light contribution. We rank our solution as highly uncertain.\\               
\emph{6144827}: The TESS ETV points (determined exclusively from FFI data) have larger scatter than the amplitude of the short period cyclic variations reported in \citet{conroyetal14,borkovitsetal16}. Hence, we formed normal eclipses. These new ETV points suggest the presence of an additional sinusoidal term. Hence, we modeled the ETV with a combination of two LTTE orbits, forming a (2+1)+1 quadruple configuration. One should keep in mind, however, that if this is the true system configuration, the outermost subsystem would be sufficiently tight, that outer dynamical terms should be considered for a correct analysis.\\
\emph{6146838}: \textit{Kepler} most likely caught a periastron event of the third star, orbiting around this long period $P_1=27\fd47$, detached, EA system. Due to the $\sim$one-sector-long eclipsing period, TESS gathered only very few new eclipses. Neither of the two space telescopes observed secondary eclipses, indicating an eccentric inner orbit. This fact makes the ETV solution more uncertain. The only certain thing is that the slope of the ETV points during the TESS era (or, in other words, the average eclipsing period) differs substantially from that of the \textit{Kepler} era. Therefore, our LTTE+DE solution should be considered only with considerable caution. \\
\emph{6187893}: Two-min cadence TESS light curves are available (S14, 26, 40, 41, 53, 54, 80, 81) and they were all used for calculating the ETVs. (For S75 only, FFIs were used). We also used SWASP data and, two eclipse times from \citep{zascheetal20}. The third star is most probably a red giant \citep[see, e.g.,][]{gaulmeetal20}.\\
\emph{6233903}: The \textit{Kepler} Villanova EB catalog contains the light curves of this eEA binary only from Quarter 8 onward. Light curves from the earlier quarters are, however, now available from the \textit{Kepler}-bonus \citep{keplerbonus} project. Therefore, in contrast to \citet{borkovitsetal16} we were already able to extend the \textit{Kepler} ETV section substantially. Regarding the TESS (FFI) measurements, however, we had to make extraordinary efforts to find the signal of this very faint and strongly contaminated source in all TESS sectors. Finally, to extract ETV data from these very distorted light curves we formed normal light curves for all TESS sectors separately. In this manner we obtained one primary and one secondary ETV point for each of the 8 TESS sectors individually, but even these ETV points were found to be nearly unusable. Consequently, our new LTTE+DE solution basically depends on the extended \textit{Kepler} ETV points and, therefore, we rank it as an uncertain solution.\\
\emph{6265720}: Aside from the TESS (FFI) ETV points, five BAO eclipses, obtained in 2018-19 were also used. The $P_2 \simeq 1372$ day cyclic ETV is clear, and the points since Kepler really help firm up this solution. The evident longer-term variation was modelled with a cubic polynomial.\\
\emph{6281103}: There are no two-min cadence TESS observations for this target.  Due to the large scatter of the TESS-derived ETV points, we formed normal light curves and, hence, normal times of minima and used them for our analysis. The LTTE-period appears to be much longer than the one previously reported in \citet{conroyetal14,borkovitsetal16}. \\
\emph{6370665}: The scatter of the TESS ETVs (obtained from FFI light curves) are larger than the full amplitude of the low amplitude LTTE detected in the \textit{Kepler} ETVs. The longer term quadratic behavior, however, is nicely confirmed. To reduce the scatter of the TESS eclipse times, we used normal points. \\
\emph{6516874}: The scatter of the TESS (FFI) ETVs is much higher than the amplitude of the cyclic ETV term. Therefore, we formed normal  eclipses. They demonstrate that there is no additional, longer-term period variation. The total mass of the inner EB was taken from \citet{cruzetal22}.\\
\emph{6525196}: This is a triple-lined RV system. A recent, dedicated study of this system was published by \citet{moharanaetal23}. We took the initial values for the MCMC runs from their study. The system was not observed in TESS Cycle 2, but 2-min cadence light curves are available for Sectors 40, 41, 54, 55, 74, 75, 81, 82. In addition to these data we also used 12 eclipse times obtained at BAO between 2013 and 2019. \\
\emph{6531485}: The short-period ($P_2=48$\,d) ETV of this detached eclipsing binary was first reported by \citet{rappaportetal13} and, later, confirmed in \citet{conroyetal14,borkovitsetal16}. TESS observed this target only in FFI mode. Due to its faintness, and the shallow eclipses, we had to form normal light curves and, hence, normal eclipse times. These show that there might be  further non-linear ETVs because adding a quadratic or cubic polynomial to the LTTE+DE fit would lead to a significantly smaller $\chi^2$ value.  We took the initial mass of the eclipsing binary from the mass estimation of \citet{cruzetal22}. \\
\emph{6543674}: This is a triply eclipsing triple star \citep[see, e.g.,][]{masudaetal15}. TESS observed this system only in FFI mode during Cycles 2 and 4, while 2-min (as well as 20-sec) cadence photometry is available for Cycle 6 sectors.  New third-body eclipses were detected by TESS at JDs~2\,459\,425 Sector 41 and, 2\,460\,527 (S81). Besides TESS data, four BAO eclipse observations were also added and used in the analysis. The EB mass was taken from \citet{masudaetal15}.\\
\emph{6545018}: This is an EA system with flat (total) eclipses. Two-min (and 20-sec) TESS data are available for S74, 75, 81, 82 but only FFIs are available for S14, 15, 41, 54, 55. One additional eclipse was observed at BAO in July 2022, but was not used since TESS observed the same event. Note, $i_1$ was taken from \citet{windemuthetal19}.\\
\emph{6546508}: This detached EA system is too faint for TESS ($T=15.152$) to obtain high-precision eclipse times. Therefore, we were able to obtain new mid-eclipse times only with a scatter which itself is larger than the amplitude of the cyclic ETV term, which was robustly detected with \textit{Kepler} \citep{borkovitsetal16}. Despite this, we used these low-accuracy TESS ETV-points, as they show that, apart from the formerly discovered periodic component, there are no additional period variations, at least not on a time scale of the past 1.5 decades. Note, the inner EB was analysed by \citet{windemuthetal19}, and we took $i_1=84\fdg6$ from their study. \\
\emph{6606282}: This is likely a detached EA system with very shallow eclipses. TESS observed this target only in FFI mode. We formed normal light curves for eclipse timing calculations. This extended ETV curve nicely confirms the former findings of \citet{borkovitsetal16} revealing a moderately certain, $P_2=1647\pm3$\,d-period LTTE solution.\\
\emph{6615041}: Only FFI light curves are available for the TESS data. We provide an LTTE plus cubic polynomial solution; however, the only thing that is certain is the period variation itself.\\
\emph{6669809 = V854 Lyr}: This relatively short outer period ($P_2=193$\,d) system contains an oEA binary with a $\delta$ Sct-component \citep{liakos17}. It was observed in two-min cadence mode in all TESS sectors where it was in the cameras' fields. The total mass of the inner binary is taken from \citet{liakos17}. The ETV, in addition to the $P_2=193$\,d likely third-body signal, clearly exhibits further, longer timescale non-linear variations. In \citet{borkovitsetal16} these variations were (mathematically) modeled with a cubic polynomial. By the TESS-era, however, this signal can be fitted much better with a second, $P_3=4467\pm22$\,d-period LTTE orbit. The corresponding outer eccentricity $e_3=0.98\pm0.02$, however, is so high, that it makes this (2+1)+1 assumption clearly unphysical. (With such a larger value of $e_3$, the orbit of the  fourth body would cross the orbit of the third star.) So, similar to the former cubic assumption, this description of the longer-term variations should be considered simply as a mathematical model instead of a physical reality. Independent of this, the existence of the third star and its orbital parameters look highly certain.\\
\emph{6671698}: This is another EW system, where only FFI data are available from TESS observations. The third-body forced period variation looks secure, but a robust explanation of the additional longer-term variations will require further follow up studies.\\
\emph{6766325}: This EW system was observed only in FFI mode during the TESS revisits. We obtained a good-looking pure LTTE solution with an outer period of $P_2=5263$\,d, which is nearly equal to the length of the entire ETV dataset (5594\,days). Therefore, we consider this solution as  uncertain.\\
\emph{6794131}: This is likely a hot, A-type ELV binary, which also exhibits rotational modulations \citep{balona17}. Due to the very shallow ellipsoidal variations, we formed normal light curves and, hence, eclipse times from the TESS FFI data, but even these ETV points show large scatter. Hence, we rank our LTTE solution as uncertain.\\
\emph{6877673}: The deduced masses and orbital elements appear to be realistic. One should keep in mind, however, that the TESS observations have added only twelve new points (six primary and six secondary eclipses) to the ETV curves and, moreover, there are no measurements around the predicted outer periastron passages, where the ETVs would better help define the outer orbit. Note also that both the \textit{Kepler} and TESS observations show total eclipses (primary transits and secondary occultations). These findings, however, do not contradict the inferred mutual inclination ($\im\sim28\degr$) due to the very slow orbital plane precession. \\
\emph{6965293}: This is an eEA with apsidal motion. Two-min data are available only for the Cycle 6 sectors (74, 75, 81, 82).  For the former sectors FFIs were used. The former ETV solution of \citet{borkovitsetal16} is solidly confirmed.\\
\emph{TIC 1882359676 = 7031714FP}: Another target, which has no \textit{Kepler} ID. Originally it was thought that the low amplitude EA-type light variation belongs to the original \textit{Kepler} target. It was also shown that the system contains an oscillating red giant star, however, it was already noted  by \citet{gaulmeetal13} that the eclipses might come from a faint background object.  Later, \citet{abdulmasihetal16} confirmed that the true source is a background object. Hence, we downloaded the \textit{Kepler}-Bonus \citep{keplerbonus} light curve of TIC~1882359676, and we obtained a much better ETV curve from this object than from the original \textit{Kepler} light curve of KIC~7031714. In the case of the TESS FFI data, however, due to the large pixel size, we had no chance of separating the fluxes of the two sources and, therefore, we obtained only low amplitude, highly contaminated light curves, from which we had to form normal light curves and, therefore, normal eclipse times in order to obtain ETV points with acceptable scatter. Here we give an LTTE+cubic polynomial solution, however, we consider it as highly uncertain.\\
\emph{7119757}: This EA system exhibits large likely spot modulations. It was observed only in FFI mode during all the TESS revisits. In order to reduce the scatter of the newly calculated ETV data, we formed normal eclipses and, therefore, normal eclipse timings from the TESS data. We obtained a slightly shorter period third-body solution than \citet{borkovitsetal16}. Moreover, we modelled the further, slight non-linearities with a quadratic polynomial. The new LTTE+quadratic polynomial solution looks moderately certain.\\
\emph{7177553}: \citet{borkovitsetal16} earlier reported a $P_2\approx1.4$\,yr-period planet-mass companion to this EA system. The spectroscopic follow up observations of \citet{lehmannetal16} led to the unexpected conclusion that KIC~7177553 indeed is a wide 2+2 quadruple system, where the second binary resembles the eclipsing one, but does not show eclipses. It was also discussed by these latter authors that the $\sim530$-day-period ETV signal cannot arise from the second binary. Hence, the findings of \citet{lehmannetal16} do not refute the hypothesis of a non-transiting circumbinary planet companion, though, they were unable to confirm this model. TESS observed this target in two-min cadence mode. The new ETV points have somewhat larger scatter than that of the \textit{Kepler} points; however, they support the former hypothesis and, perhaps, even dynamically forced apsidal motion is also manifested. For our eccentric LTTE+DE model we took the initial system parameters from the work of \citet{lehmannetal16}. \\
\emph{7272739}: TESS observed this EW binary only in FFI mode. The combination of \textit{Kepler} and TESS ETVs clearly reveals a cyclic feature, however, a secure interpretation as an LTTE orbit will require further studies. Hence we ranked this solution as uncertain. \\
\emph{7289157:} This is one of the first known triply eclipsing triple stars \citep{slawsonetal11}. ETV studies were carried out by \citet{rappaportetal13,borkovitsetal15,borkovitsetal16}. We are aware that spectro-photodynamical analyses were also carried out for this very interesting target \citep[see, e.g.][]{orosz15,shortetal18} but, unfortunately, none of them is available in the literature, at least, up to now. The depths of the regular eclipses were continuously decreasing during the entire \textit{Kepler} mission. TESS observed this target during 9 sectors. Cycle 2 and 4 observations were carried out only in FFI mode, while for the Cycle 6 measurements two-min (and 20-sec) cadence light curves are also available.  TESS has detected two third-body events in Sectors 54 and 81. The regular eclipses, however, are almost undetectable during the entire TESS mission. They are continuously present, but their depths are only $\sim0.2-0.4\%$ Therefore, we formed normal light curves and times of minima, averaging light curves separately for each physical TESS orbit. (In other words, we formed two folded [i.e. normal] light curves for each sector.)  But, even in this way, we were able to obtain usable (normal) eclipse timings only for some sectors. Despite this, these new ETV points are very important for a new ETV study because they carry substantial information about the dynamically forced apsidal motion.   Moreover, they also support the former findings of \citet{borkovitsetal15} that the ETV curve cannot be described perfectly with a simple, lowest order LTTE+DE model. Here we add a new LTTE+DE+LTTEout model, indicating a theoretical (2+1)+1 type quadruple star solution. Note, however, that while the inner triple subsystem and its solution look quite robust, the fourth-body solution is most likely only a mathematical approximation which describes the imperfectly modeled, secular timescale third-body perturbations.  Hence, we plan a complex spectro-photodynamical analysis of this remarkable system in the near future. \\
\emph{7339345}: The ETV of this EW binary is quite mysterious. At first sight, the combination of the \textit{Kepler} and TESS (FFI) points suggest a $P_2\approx5100$\,d-period cyclic variation. On the other hand, \citet{borkovitsetal16} found a much shorter period ($P_2=892$\,d), very low amplitude LTTE solution, superposed on a quadratic ETV. Since this latter, shorter period, low amplitude variation is clearly visible during the \textit{Kepler}-section of the ETV curve, we made another trial, fitting two LTTE orbits simultaneously--one with $P_2\approx886$\,d, and the other with $P_3\approx5076$\,d. In such a manner, we obtained a much better ETV solution (i.e. lower $\chi^2$ by about $60\%$) than the single long-period LTTE orbit solution. The problem, however, is that such a small ratio of $P_3/P_2\approx5.7$ would likely be dynamically unstable. Therefore, we assume, that the longer period might belong to a real third star companion, while the shorter period cyclic variation might have arisen for some different reason. Despite this, we tabulate the parameters of the formal (2+1)+1 solution.\\
\emph{7362751}: The true origin of the target's low amplitude EW signal is ambiguous. Neither the original Villanova catalog nor \citet{abdulmasihetal16} mention that the true eclipsing system might be some background object instead of the primary \textit{Kepler} target KIC~7362751.  However, the light curves produced by the \textit{Kepler} Bonus project \citep{keplerbonus} show that the EW signal comes from a star with the designation of Gaia DR3~2125824166586455296 (aka TIC~1882408344), a 3 magnitude fainter background object, instead of from the light curve of KIC~7362751. Independent of the question of the true host star of the EW binary, what was evident for us is that the TESS FFI light curves for this target are strongly contaminated by several other stars. Therefore, mining out the EW signal required careful sector-by-sector PCA processes first, then the construction of normal light curves and, finally, the calculation of normal eclipse times. Even using such procedures, we could obtain new ETV points only with relatively large scatter. The extended ETV curve does not contradict the former $P_2\approx550$\,d LTTE solutions of \citet{rappaportetal13,conroyetal14,borkovitsetal16}, but reveals that some further, longer timescale period variation is also present in the system. Thus, finally we arrived at an LTTE plus cubic polynomial model.\\
\emph{7385478}: This is an SB2 EA system, which exhibits $\gamma$~Dor-type pulsations \citep{ozdarcandal17,guoli19}. Due to its pulsating nature and relative brightness, it was observed in two-min cadence mode in TESS Cycles 2 and 4 (while it was measured in FFI mode in Cycle 6). We obtained a robust pure LTTE solution which confirms the former findings of \citet{borkovitsetal16}. We took the total mass of the binary from the dedicated analysis of \citet{ozdarcandal17}.\\
\emph{7431703 = V882 Lyr}: This is most likely a classic, SD Algol-system. Due to the very shallow secondary eclipses, only primary eclipses were used in the analysis. No ETV solution was given in \citet{conroyetal14,borkovitsetal16}. Two-min cadence light curves are available for almost all observing sectors, with the exceptions of Sectors 74 and 75. For these latter sectors, we naturally used the FFI observations. \\
\emph{7440742}: Two-min cadence data are available for TESS Cycles 2 and 4, and Sector 80 of Cycle 6, while for the other Cycle 6 observations (Sectors 74, 75, 81, 82) only FFIs are available. The large scale timing variation is evident, but the LTTE+cubic polynomial solution is far from compelling. \\
\emph{7518816}: This (most likely E$\beta$-type) system was observed in FFI mode in all TESS sectors. The LTTE+quadratic polynomial model appears to be moderately certain. \\
\emph{7630658}: This EA system was observed only in FFI mode during the TESS visits. Due to the large scatter in these ETV points we formed normal eclipses from the TESS data. The new ETV solution -- not surprisingly -- securely confirms the former findings of \citet{zascheetal15} and \citet{borkovitsetal16}. \\
\emph{7668648}: This is currently the tightest known stellar triple system with $P_2/P_1=203/27.8\approx7.3$. This is also a triply eclipsing triple star system \citep{orosz15,borkovitsetal15}. Moreover, rapid EDV can also be detected in the \textit{Kepler} observations. Former ETV solutions were given in \citet{rappaportetal13,borkovitsetal15,borkovitsetal16}. Then, recently, \citet{orosz23} published a complex photodynamical solution, using not only \textit{Kepler} and (early) TESS photometry, but RV measurements as well. In our current ETV study we use the results of \citet{orosz23} as input parameters and, moreover, in accord with his findings, we do not allow the mutual inclination ($\im$) to lie above $4\degr$. \\
\emph{7680593}: This likely overcontact binary was observed only in FFI mode during the TESS sectors. We formed normal light curves and, hence, normal ETV points from TESS data. Our LTTE solution looks moderately certain.\\
\emph{7685689}: The Villanova catalog contains \textit{Kepler} light curves only from Quarter 4 for this EW system, but Q2-Q3 light curves are also available from the \textit{Kepler}-bonus project \citep{keplerbonus}. Hence, we used this longer \textit{Kepler} light curve and the derived ETV dataset for the current analysis. The TESS FFI light curves are strongly contaminated with the $\sim0\fd6$-period EB system KIC~7685748 (a.k.a. TIC~240183397); therefore, we had to apply the PCA method to separate the signals of the two eclipsing binaries. (Note, this latter, contaminator system was not observed by \textit{Kepler} in any quarters.) After the PCA separation, due to the large scatter of the TESS derived ETV points, we also formed normal light curves and recalculated normal ETV points from these data. In such a manner, we are able to confirm the $P_2\approx515$\,d-period LTTE in the ETV curve, reported formerly by \citet{conroyetal14,borkovitsetal16}, however, the presence of further, longer times-scale, non-linear variations in the ETV curve are also evident, and we modeled them with cubic polynomial.\\
\emph{7690843}: The rather short, $P_2=74$\,d-period third star was already found very early during the original \textit{Kepler} mission by \citet{rappaportetal13} and \citet{gaulmeetal13}. These latter authors also pointed out that the third star was an oscillating red giant. Later, \citet{borkovitsetal16}, confirming the $\sim74$\,d cyclic variations, also reported for the first time that the EB displays some additional timing variations. In this latter paper, the extra timing variations were modeled with a cubic polynomial. TESS observed this target in two-min cadence mode in Cycles 2 and 4, and in FFI mode in Cycle 6. While the $\sim74$\,d-period cyclic variation is continuously present in the ETV data, the extended dataset also makes it very clear that the longer timescale, non-linear variation is also real. We attempted to model it with both a cubic polynomial and a second, $P_3\sim5500$\,d-period, wider LTTE orbit. While this latter model gives a somewhat better fit (i.e., lower $\chi^2$), here we present the cubic polynomial representation, naturally, in addition to the $P_2=74$\,d-period LTTE+DE third-body term. Further follow up observations are requested to help decide whether the four-body solution provides a better solution only due to the larger number of adjustable parameters, or if this does indeed describe the true system configuration.\\
\emph{7812175}: The ETV of this long-period ($P_1=17\fd79$) detached Algol-system is a typical example where the primary and secondary eclipses are anticorrelated on the time scale of the outer orbit of the third star (i.e., $P_2$), instead of the usual, much longer, apsidal motion time scales. Such a short-timescale anticorrelation is a clear indicator of a dynamically dominated ETV curve behaviour. This interesting ETV was found and analyzed for the first time by \citet{borkovitsetal15,borkovitsetal16}. \textit{Kepler}, made targeted observations only from Quarter 10 onward. Fortunately, however, this binary was also observed incidentally in the former quarters, as a background source to KIC~7812179. Thus, we were able to carry out pixel-by-pixel photometry for this target in the manner, described in \citet{bienasetal21}. In this manner we were able to extend the ETV determination for the entire four-year-long \textit{Kepler} mission. The same target was also revisited with TESS (in FFI-cadence mode) during ten sectors. Despite the faintness of this target, we were able to obtain moderately usable light curves, and to determine new eclipse times from these TESS data. We included these ETV points in our analysis as well. We assumed a coplanar configuration as there is no indication for EDVs, and took the inclination from the analysis of \citet{windemuthetal19}.\\
\emph{7821010}: This is a very eccentric ($e_1=0.68$), relatively bright ($T=10.4$), detached, eclipsing, and doubly-lined spectroscopic binary. The likely presence of a non-transiting circumbinary planet was first reported in 2014 in a conference  proceeding\footnote{\url{http://www.astro.up.pt/investigacao/conferencias/toe2014/files/wwelsh.pdf}}, but the promised detailed study has not been published since then. An analytic ETV analysis, confirming the presence of a non-transiting planetary mass companion in the system, was published in \citet{borkovitsetal16}. A detailed study of this system, using HIDES spectroscopy and \textit{Kepler} photometry, was then reported by \citet{helminiaketal19}. TESS observed this target in two-min cadence mode during Year 2 and 4 revisits (except for Sector 14, for which only FFI data are available), while only FFI data are available for Year 6 observations. Extending our ETV dataset with the new timing data from TESS, we reiterated our 2016 analytic ETV study. Note, that the convergence of the primary and secondary ETV curves reveals evident apsidal motion which, however, cannot be explained by the dynamical effects of the planetary-mass companion object (the effect of which would yield an order of magnitude longer dynamically forced apsidal motion period). Hence, we assume, that the origin of the apsidal motion should come mainly from general relativistic and classical tidal effects and, therefore, we adjusted the apsidal motion period as an additional independent parameter during our MCMC runs, instead of constraining it from the analytic hierarchical three-body perturbation theories. For our analysis we took the initial total mass of the binary, and also the (fixed) inner inclination, from the work of \citet{helminiaketal19}.\\
\emph{7837302}: This is a mono-eclipsing detached binary, implying more or less grazing eclipses on an eccentric orbit. TESS observed this target only in FFI mode during Cycles 2 and 4, but two-min cadence observations are also available (and used) in the Cycle 6 sectors. The shape of the ETV curve makes it very likely that this is a triple system with dominant dynamical perturbations. However, in the absence of secondary eclipses, the inferred orbital parameters are only weakly constrained. Despite this, the outer period of $P_2\approx1381$\,d appears to be quite secure. We give an LTTE+DE solution, where we constrained the mutual inclination angle to be $\im\leq10\degr$ due to the lack of any EDVs. \\
\emph{7877062}: This EW system was observed only in FFI mode with TESS. These new observations do not confirm the former ETV solutions of \citet{conroyetal14,borkovitsetal16}. Here we present an uncertain LTTE+cubic polynomial solution. What is certain, however is, the remarkable long-term period variation, which requires further observations for a real understanding. \\
\emph{7938468 = V481 Lyr}: The ETV of this EA system is a good example for the need of ever longer observations. The timing data have finally reversed direction around the time of the Sector 74 observations, implying a more than decade-long third-body orbital period. Note, TESS observed this target only in FFI mode in Years 2 and 4, but 2-min cadence data are available for the Year 6 revisits. \\
\emph{8016214}: The light curve of this totally eclipsing EA system shows chromospheric activity. The primary eclipse is the occultation, while the secondary is transit, implying, that the hotter star is the smaller one. The secondary ETV points have much larger scatter than the primary points; therefore, we used only the primary eclipses for our analysis. TESS observed this target only in FFI mode. \\
\emph{8023317}: This is one of the most inclined of the known, tight hierarchical triple star systems. According to the complex photodynamical analysis of \citet{borkovitsetal22b}, (which was carried out after TESS Sector 41 data), the mutual inclination of this system is $\im=55\fdg7\pm0\fdg8$ which was even higher by $\sim6\degr$ than the former analytic ETV results of \citet{borkovitsetal15,borkovitsetal16}. Now, besides the \textit{Kepler} data, we also used all the available TESS observations up to the Cycle 6 sectors. For the current analytic ETV analysis we naturally used the results of \citet{borkovitsetal22b} as input parameters, and we obtained a new LTTE+DE ETV solution which is in accord with both the \textit{Kepler}-only ETV results of \citet{borkovitsetal15,borkovitsetal16} and with the newer, complex photodynamical results. \\
\emph{8043961}: Two-min cadence TESS light curves are available (S14, 15, 41, 54, 55, 74, 75, 81, 82) and they were used for calculating the ETVs. The system has been intensively followed both photomoterically and spectroscopically by our group. We use 11 unpublished eclipse times obtained at BAO. Moreover, we took the total mass of the inner EB from our yet unpublished RV data. \\ 
\emph{8047291}: This chromospherically active EA system is located in the field of KIC~8047283 which, originally, was erroneously claimed to be the EB itself \citep{abdulmasihetal16}. We downloaded the re-extracted \textit{Kepler} light curve of \citet{abdulmasihetal16}, and redetermined the \textit{Kepler} ETV points. For the TESS ETV points we used the exclusively available FFI observations, and analyzed only the primary eclipse times. Because of the large scatter of the latter ETV points, we used normal TESS eclipses. Besides the evident $P_2=834$\,d-period cyclic variation, interpreted as an LTTE orbit, due to the tighter constraints when using the TESS data we had to add a quadratic term as well.\\
\emph{8081389}: This is another EA-system, where the deep primary eclipses allow us to calculate much more accurate ETV points than for the shallow secondary eclipses. Hence, only the former were used. Our solution clearly prefers a longer period and substantially larger amplitude ETV solution, compared to  the former results of \citet{borkovitsetal16}.\\
\emph{8094140}: After adding new eclipse times from the TESS FFI data, the former, short-period LTTE solution is robustly confirmed. The total mass of the binary was taken from \citet{cruzetal22}. \\
\emph{8143170}: This long-period EA system shows shallow-depth, but wide, total eclipses. According to the Gaia NSS catalog, this is an SB1-type spectroscopic binary \citep{GaiaNSS}. Moreover, \citet{gaulmeguzik19} found, that this EA system contains a red giant star. Notwithstanding these properties, the most striking feature of this EB is its enigmatic ETV curve. This is especially true for the primary ETV curve, which displays a sudden jump in amplitude of $\sim$0.1 days around BJD~2\,456\,200, implying the periastron passage of a highly eccentric third body in the system. The first ETV studies were carried out by \citet{borkovitsetal15,borkovitsetal16}. These former solutions were found, however, in the absence of any indication of the true long-period outer orbit, which was quite uncertain. Now we have added new eclipse times from TESS (FFI and, in the case of Cycle 6, two-minute cadence) observations, and have carried out a new analysis. Interestingly, we were able to obtain good fits only when a cubic polynomial was also fitted together with the LTTE+DE parameters. We explain this fact by invoking the presence of additional, higher order gravitational perturbations, which are not modeled in our analytic three-body ETV model \citep{borkovitsetal15}. Such extra effects have been previously detected in the case of TIC~167692429 \citep{borkovitsetal20a}, and BU~CMi \citep{pribullaetal23}, and occur naturally in photodynamical analyses, where the motion of the triple stars are numerically integrated. Hence, we conclude that a complex, photodynamical analysis of this triple system will be needed before a substantially improved solution can be found. \\
\emph{8145477}: This totally eclipsing EW system was observed with \textit{Kepler} only during the first half of the primary mission (Quarters 1--6). TESS observed this system in 8 sectors in FFI mode. Due to the large scatter of the TESS (FFI) ETVs we used normal eclipses for these new data. Besides the $\sim1$\,yr-long cyclic variation interpreted as an LTTE orbit by \citet{conroyetal14,borkovitsetal16} a longer-term nonlinearity is also clearly present in the ETV curve. The \textit{Kepler} light curve was analyzed by \citet{zolaetal17}. They found a third light of $\ell_3=0.16$ which appears to be in accord with the third-body mass inferred from the LTTE solution.\\
\emph{8190491}: This ELV system was observed only in FFI mode during all TESS visits. The new ETV points were determined from normal light curves, but even so, their scatter exceeds the amplitude of the $P_2\approx620$\,d variation found formerly by \citet{conroyetal14} and \citet{borkovitsetal16}. These new points, however, clearly reveal the presence of longer timescale variations. Therefore, we finally arrived at an LTTE+quadratic polynomial model with a likely third-body period of $P_2=616$\,d. \\
\emph{8192840}: TESS observed this EW or ELV system only in FFI mode. We found an LTTE+quadratic polynomial solution with a bit shorter period than was given in \citet{borkovitsetal16}.\\
\emph{8210721}: This is a long-period ($P_1=22\fd67$) eEA-system, with ETVs that are clearly dominated by third-body dynamical perturbations \citep{borkovitsetal15}. Unfortunately, the very shallow secondary eclipses cannot be detected in the noisier TESS (FFI) data. Despite the lack of recent secondary eclipse times, our new analysis practically confirms the former results of \citet{borkovitsetal15,borkovitsetal16}. We took the inner inclination from the study of \citet{windemuthetal19}.\\
\emph{8211618}: \textit{Kepler} observed this EW system only in Quarters 2-3 ($\approx 180$\,days). The ETV derived from this short dataset led \citet{conroyetal14} to report the presence of a very short, $P_2\approx127$\,d-period third component, which would be one of the shortest known outer period hierarchical triple systems, where the inner binary would be an overcontact system. We obtained eclipse timings (after forming normal light curves) from 8 sectors of TESS FFI observations. This extended, $\approx5500$\,day-long dataset reveals the presence of much longer timescale variations. These are superposed on a shorter period, smaller amplitude cyclic variation, for which the period was claimed by \citet{conroyetal14} to be $\sim$$4$ months. Finally, regarding this shorter timescale cyclic term, we obtained a very eccentric ($e_2=0.88\pm0.02$) $P_2=655\fd2\pm0\fd3$-period third-body solution (instead of $P_2\approx127$\,d) which, due to the large eccentricity, we modeled with a combined LTTE+DE terms, while the longer timescale period variation is described with a cubic polynomial. Further observations are critical to understanding the true nature of this system.\\
\emph{8242493}: All TESS measurements were carried out in FFI mode when it was being observed.  From the TESS data we formed normal light curves and, hence, calculated normal eclipse times. We evaluated an LTTE+cubic polynomial model.\\
\emph{8257903}: This EW-type system was observed only in FFI mode during all TESS sectors. We found an LTTE solution with a period much longer than the length of the dataset. Therefore, our solution is quite uncertain. \\
\emph{8265951}: This is a totally eclipsing EW binary. The combination of \textit{Kepler} and TESS (FFI-derived) ETVs reveal a large amplitude, likely cyclic timing variation with a period of $\sim8$\,years. The most plausible explanation would be an LTTE orbit caused by a quite massive ($m_\mathrm{C}\geq1.1\,\mathrm{M}_\sun$) third star. Interestingly, however, neither \citet{zolaetal17} nor \citet{kobulnickyetal22} report a significant contribution from contaminated light. Further studies of this object are clearly needed.\\
\emph{8330092}: This is an overcontact binary, where the tertiary star, which was first detected through ETV analysis by \citet{conroyetal14}, displays pulsations (both $\delta$~Scuti and $\gamma$~Dor types), as was shown by \citet{lietal20}. These latter authors analysed the timing delays in the pulsations of the third star, and found that ``the combined mass of the binary system is $0.68 \pm 0.04$ times the mass of the third component''. This allows us to make a crude estimation of the inclination ($i_2$) of the tertiary's orbit. This comes from the fact that, according to this statement, the outer mass ratio is $q_2=1.47$. Moreover, taking into account the fact that our LTTE solution resulted in a mass function value of $f(m_\mathrm{C})=0.0062\pm0.0002\,\mathrm{M}_\sun$ and, writing the mass function into the form of $f(m_\mathrm{C})=m_\mathrm{AB}\frac{q_2^3}{\left(1+q_2\right)^2}\sin^3i_2$, one can readily get the value of  $m_\mathrm{AB}\sin^3i_2\approx0.0119\,\mathrm{M}_\sun$. From this, and accepting the estimated mass of the inner, eclipsing binary to be $m_\mathrm{AB}\approx1.52\,\mathrm{M}_\sun$, one finds $i_2\approx11\fdg4$ (or, its retrograde counterpart, $i_2\approx168\fdg6$), i.e., the tertiary orbit could be seen nearly pole-on. The latter, of course, also implies a very large mutual inclination, since $\cos(i_1+i_2)\leq\cos\im\leq\cos(i_1-i_2)$. Note, the outer orbit was also detected astrometrically with the \textit{Gaia} space telescope, and listed in the \textit{Gaia} DR3 non-single-stars (NSS) catalog \citep{GaiaNSS}. This latter astrometric outer orbit solution finds $i_2 = 156\degr \pm 5\degr$. Considering the fact that our estimation was quite crude, and even the total mass of the inner binary was estimated only from a statistical, empirical relation, we can state that our finding is consistent with \textit{Gaia}'s result. Finally, we mention that due to the large scatter of the TESS FFI-derived ETV points, we used normal TESS eclipses.\\ 
\emph{8386865}: The likely presence of a relatively short-period $P_2\approx294$\,d) third stellar component in this ELV binary first was announced by \citet{rappaportetal13} and confirmed later by \citet{conroyetal14} and \citet{borkovitsetal16}. We extended the available ETV time series with new eclipse timings derived from TESS observations (two-min cadence in Year 2 and FFIs during Years 4 and 6 visits). Due to the very low amplitude of the ellipsoidal light variations, the TESS derived ETV data have very large scatter. Therefore, we formed normal ETV points, which, continue to have quite large scatter relative to the LTTE amplitude. We can state, however, that the new points do not contradict the former, short-period LTTE hypothesis, and reveal that there might be some additional longer timescale period variation, which we model (mathematically) with an additional cubic polynomial, above the LTTE term. In conclusion, similar to \citet{borkovitsetal16}, we tabulate this triple candidate amongst the most certain LTTE cases.\\
\emph{8394040}: This likely EW system very certainly hosts a $P_2\approx387$\,d-period third stellar component, as was first reported by \citet{rappaportetal13} and then confirmed by \citet{conroyetal14,borkovitsetal16}. Adding new normal eclipse times formed from TESS FFI observations, we confirm the former findings and also obtain some evidences for an additional, longer timescale period variation which we now describe with a quadratic polynomial. The inferred minimum mass of the tertiary seems to be quite large ($(q_2)_\mathrm{min}\approx0.69$) which might well be in accord with the $\approx80\%$ of third light found by \citet{kobulnickyetal22}.\\
\emph{8429450}: This EA system exhibits $\gamma$~Dor and $\delta$~Sct-type pulsations \citep{gaulmeguzik19,leeetal20}. All the TESS observations were carried out only in FFI mode. We found a moderately robust LTTE solution with $P_2=3575$\,d, which, interestingly, is longer only by $\sim16\%$ than the former solution of \citet{borkovitsetal16}, the latter of which was inferred from only a small fraction of one LTTE orbit. We took the total mass of the system from \citet{fullerfelce24}.\\
\emph{8444552}: This is a short-period EA binary system. Two-min cadence light curves are available for the Cycle 6 observations (for the earlier sectors there are only FFIs).  LAMOST spectral classification is available \citep{zhangetal19}. \\
\emph{8509014}: This EW system did not belong to the primary targets of the \textit{Kepler} mission. It was incidentally imaged in the aperture of KIC~8509015. Its variable nature, and fully parabola-like ETV was first reported within the framework of the \textit{Kepler} pixel photometry project \citep{bienasetal21}. To calculate the eclipse times we used the public \textit{Kepler} bonus light curves \citep{keplerbonus} on one hand, and the TESS FFI light curves. These latter ETV points reveal that the parabolic ETV during the primary \textit{Kepler} mission was merely a section of a cyclic variation which we interpret as an LTTE orbit with a period of $\sim$1696 days.\\
\emph{8553788}: A dedicated analysis of this oscillating semi-detached Algol system (and also an R CMa star) was carried out by \citet{liakos18}. We took the EB mass from that paper. Besides \textit{Kepler} and TESS (FFI) data, we used ground-based eclipse times (most of them derived from SWASP observations) taken from \citet{zascheetal15}. \\
\emph{8553907}: This long-period ($P_1=42\fd03$), eccentric, detached EB shows quite long-period sinusoidal primary and secondary ETV curves, which however are not in phase with each other, indicating a remarkable dynamical contribution. TESS observed only eight new eclipses (three primary and five secondary ones) in eight sectors, which were naturally added to the \textit{Kepler} eclipse times. We took the inner inclination from the analysis of \citet{windemuthetal19}.\\
\emph{8560861}: This is a new triple-candidate. (It is not listed in \citealt{borkovitsetal16}.) The inner EB, according to \citet{borkovitsetal14}, exhibits anomalous ellipsoidal variations and oscillations which are tidally excited -- at least in part. A complex analysis of the system was carried out by \citet{borkovitsetal14}. The initial orbital and dynamical parameters for our ETV analysis were taken from that work. (This system was observed in 2-min cadence mode during all the TESS revisits.) \\
\emph{8719897}: The Gaia NSS catalog gives an SB1 spectroscopic solution for the outer orbit \citep{GaiaNSS} with $P_2 = 333 \pm 1.5$ days, in nearly perfect agreement with our ETV result. Only TESS FFIs are available for Cycle 2 and 4 sectors, but TESS observed this target also in two-min cadence mode during the Cycle 6 sectors. \\ 
\emph{8739802}: The inner pair is likely an ELV binary. The Villanova Catalog contains the \textit{Kepler} light curves only from Quarter 4 onward. We added the earlier \textit{Kepler} light curves from the \textit{Kepler}-bonus \citep{keplerbonus} project. The newer ETV section added from the TESS (FFI) data has a scatter larger than the amplitude of the cyclic \textit{Kepler} ETV variations, even with the use of normal eclipse times. Despite this, our new LTTE solution confirms the former findings of \citet{conroyetal14} and \citet{borkovitsetal16} and, reveals that there are no significant other non-linearities in the ETV, at least during the currently $\sim5600$-day-long timescale of the total data set.\\ 
\emph{8758161}: Two-min cadence TESS light curves are available for S14, 15, 40, 41, 54, 55, and FFIs for Cycle 6 observations. ETV curves for Sectors 54 and 55 have reversed direction, making the inferred long period outer orbit of $\sim$4277\,d more robust. LAMOST spectroscopy is available.\\
\emph{8868650}: This is a totally eclipsing EA system. Two-min cadence data are available for all but Sector 75 (the latter of which was observed only in FFI mode). The ETV curve is quite complex. We interpret it as a combination of an LTTE orbit and a linear period variation (i.e. quadratic polynomial). Further ETV points are evidently needed.\\
\emph{8904448}: This is likely an E$\beta$-type EB (because of the highly unequal eclipse depths). We used only the deeper primary eclipses for the ETV analysis. Moreover, due to the large scatter of the TESS (FFI-derived) times of minima, we formed normal eclipses from the TESS primary eclipses. \\
\emph{8953296}: This EA system was listed in neither \citet{conroyetal14} nor \citet{borkovitsetal16}. TESS observed it only in FFI mode. We used only the much deeper primary eclipses for the ETV analysis.  Due to the large scatter of the new TESS ETV points, we formed normal light curves and, hence, normal eclipse times for these data. The only fact which is evident about this ETV curve is its non-linear nature. We give here an LTTE+cubic polynomial approximation; however, the solution is obviously quite uncertain.\\ 
\emph{8957887}: This EW system was observed only FFI mode by TESS. From these latter data we formed normal eclipses. Our solution confirms the former results of \citet{conroyetal14} and \citet{borkovitsetal16} and shows evidence for longer-term period variations as well. \\
\emph{8982514}: A typical EW system with FFI only TESS data. The inferred ETV period is much longer than that found in \citet{borkovitsetal16}.\\
\emph{9007918}: EA system with one suspicious extra dimming event in the \textit{Kepler} data \citep[see][]{borkovitsetal16}. Two-min cadence TESS observations are available for all but one sector and used for the analysis. Note, the sole exception is Sector 81, where we used FFI observations for the ETV analysis.) Spectroscopic data are also available from the LAMOST survey \citep{zhangetal19}. APOGEE-2 DR16 \citep{jonssonetal20} tabulates 16 individual RV data points.  While in \citet{borkovitsetal16} this system was marked as having a pure LTTE orbit, now we have reclassified it as a combined LTTE+DE ETV triple. \\
\emph{9028474}: This is the longest (inner) period ($P_1=124\fd94$) and most eccentric ($e_1=0.81$) EA-type system in our sample. Therefore, despite the fact that TESS revisited this target in seven sectors (always in FFI mode), only one new primary and secondary eclipse were spotted (both of them in Sector 81). Adding these two new times of minima to the 21 former \textit{Kepler} ETV points, naturally, we were able to find a very uncertain inclined LTTE+DE solution. The input parameters of the inner pair for our MCMC runs were taken from the recent, dedicated work of \citet{ozdarcan23}.\\
\emph{9075704}: This E$\beta$-type system was directly observed by \textit{Kepler} from Q08 onward. Indirectly, however, as a background source in the aperture of KIC~9075708, its brightness variations were recorded from the beginning of the \textit{Kepler}-mission. Hence, downloading the publicly available \textit{Kepler} Bonus photometry \citep{keplerbonus} we were able to extend the ETV series backward to the beginning of the \textit{Kepler}-mission. Moreover, as usual, we determined new eclipse times from the TESS FFI observations as well. These latter ETV points have a very large scatter, hence we used normal eclipses. Our results for the cyclic variation (LTTE) confirm the former findings of \citet{conroyetal14,borkovitsetal16}, while the extended TESS series reveal that there are no significant, longer-timescale period variations (though, a small, but non-zero second order polynomial coefficient was also found).\\
\emph{9083523}: This is a totally eclipsing, E$\beta$-type EB. Two-min cadence TESS observations are available for (almost) all TESS Sectors, with the exceptions of Sectors 74, 75 and 81, when the target was observed only in FFI mode. Prior to the latter Cycle 6 observations, the ETV suggested a pure LTTE orbit with a period of $\sim7.3$\,yr. The newest ETV points, however, refute this former model. The currently best-fit LTTE+cubic polynomial model appears quite uncertain; hence we rank this system in the less confident group.\\
\emph{9091810}: This semi-detached system was observed only in FFI mode in the TESS sectors. The new ETV points suggest a significantly longer third-body period than the formerly reported one \citep{conroyetal14,borkovitsetal16}. Here we tabulate a relatively uncertain LTTE+quadratic polynomial solution.\\
\emph{9101279 = V1580 Cyg}: Old ground-based and SWASP eclipses are available for this EA system and, hence, we made use of them for the ETV curve. Due to their considerable scatter, however, we formed normal eclipses from the SWASP data. TESS observed this system only in FFI mode. We used only the primary eclipses for the ETV analysis. Our result is far from convincing. Besides the LTTE term, we modelled the longer timescale variations with a quadratic polynomial.\\
\emph{9110346}: This is a faint, detached EB.  We recalculated the \textit{Kepler} ETV data with the use of the light curve of the \textit{Kepler} bonus project \citep{keplerbonus} which, in contrast to the formerly used Villanova catalog data, also contains the very early \textit{Kepler} quarters, Q1 and 2. We also added the new TESS (FFI) observed eclipse timings. Despite the relatively short outer period of $P_2=2549$\,d, we were able to rerank this ETV solution from the most uncertain category to a moderately certain one.  This occurred only at the last moment, since the ETV curve has undergone a clear reversal just in the very last sectors of S81, 82. \\
\emph{9159301}: This oEA system displays $\delta$~Scuti pulsations \citep{gaulmeguzik19}. This is also an SB2 system, hence the total mass of the EB is known \citep{matsonetal17}. It was observed in two-min cadence mode in TESS Cycles 2 and 4, and in FFI mode during Cycle 6. We also added a few SWASP eclipse time obtained from just before the \textit{Kepler} measurements. We used only the primary eclipses, and our solution is quite uncertain.\\
\emph{9181877}: This is an EW binary with a red giant tertiary \citep[see, e.g.,][]{nessetal16} on a most likely highly non-aligned outer orbit.  TESS observed this target in two-min cadence mode in Cycles 2 and 4, while FFIs are available for the Cycle 6 revisits. The $P_2=3652$\,d-period LTTE solution appears to be quite certain. \\
\emph{9272276}: Only FFI data are available for the TESS Sectors. The ETV is a clear superposition of sinusoidal and further variations for which the latter are modeled with a cubic polynomial. \citet{kobulnickyetal22} found an extra light of $\sim40\%$ which seems to be in good qualitative agreement with the inferred large third-body minimum mass. \\
\emph{9283826 = V2366 Cygni}: For this EW system, only FFI TESS observations are available. Two earthly observed eclipses can be found in \citet{ibvs6010}, but these were also recorded during the much more accurate \textit{Kepler} observations; therefore, we omitted them. The ETV curve has a large amplitude, which we model with a combination of an LTTE orbit and a cubic polynomial. The result is, however, rather uncertain.\\
\emph{9392702}: \textit{Kepler} observations are available only during the first half of the original mission (until Q10). TESS observed this target only in FFI mode. Our solution suggests a much longer-period cyclic ETV than was reported in \citet{borkovitsetal16}.\\
\emph{9402652 = V2281 Cygni}: Two-min cadence TESS light curves are available for S14, 15, 40, 41, 54, 55, 82 and FFIs for S74, 75, 81. Further ground-based survey and targeted follow up eclipse observations are available. Due to the large scatter of the ETV points calculated from SWASP and another survey observations, we omitted using them, and included only the newer, individual eclipse times listed in \citet{diethelm14,zascheetal15}. The total mass of the inner EB, formed by two twin stars, was taken from the dedicated study of \citet{kooetal17}. \\
\emph{9412114}: This faint likely EW system was observed only in FFI mode by TESS. We formed normal light curves and, hence, normal eclipse timing to reduce the scatter of the new ETV points. We found and present a moderately certain LTTE+quadratic polynomial solution.\\
\emph{9451096}: This was observed in two-min cadence mode in all TESS sectors (S14, 26, 40, 41, 53, 54, 55, 74, 75, 80, 81, 82). SWASP eclipse times are unusable due to  too large a scatter. APOGEE-2 DR16 \citep{jonssonetal20} spectra and RVs are available. We set $i_1=80.3\degr$ from \citet{windemuthetal19} and, our results prefer a coplanar LTTE+DE solution. \\
\emph{9532219}: This ultrashort-period ($P_1=0\fd198$) EW type contact binary was observed only in FFI mode during its TESS visits. Due to the large scatter of the TESS eclipse timings, we formed normal light curves and determined ETV points from these normal eclipses. \citet{leeetal16} claimed that the \textit{Kepler}-ETV dataset could be described by a combination of a low amplitude $P_2=1195\fd6$-period LTTE orbit, and a quadratic polynomial.  According to these authors, the latter term might be due to the LTTE orbit of a fourth body. In contrast to this claim, the extended current ETV curve can be described simply with a $P_2\approx5455$\,d-period, pure LTTE orbit. The fit appears quite convincing, but since the period is near the length of the full data-set ($\sim5465$\,days), we rank our solution as an uncertain one.\\
\emph{9541127}: The ETV of this EB-type system displays a double periodicity. Hence, we describe it with a (2+1)+1 four-body model. This solution, however, is clearly unphysical due to the unrealistic period ratio of $P_3/P_2=3.59$. Such a quadruple stellar system cannot be dynamically stable. We tabulate, however, both ``LTTE'' orbits, since we cannot judge which signal may come from an additional body (if any), and which has a different origin. Note, this target was observed only in FFI mode during the TESS mission. \\
\emph{9592145}: The inner EB exhibits an E$\beta$-type light curve. It was observed in FFI mode during all of the TESS sectors. Due to the large scatter of the new ETV points, we used only the much deeper primary eclipses and, moreover, we formed normal light curves and, hence, normal eclipse timings from the TESS data. Formerly \citet{borkovitsetal16} have reported a $P_2\sim730$\,d-period, low amplitude, sinusoidal ETV which they explained as an LTTE orbit caused by a likely substellar mass third component. The TESS ETV points reveal that such a short period, low amplitude variation is continuously present in the ETV but, furthermore, there is some additional much higher amplitude variation (including an abrupt period jump) in between the Cycle 4 and 6 observations. Finally we found a mathematically acceptable double LTTE, i.e., (2+1)+1 type quadruple system solution but, similar to some other double LTTE solutions in our sample, we assume that at least one of the two LTTE orbits has a different origin. \\
\emph{9664215}: This Algol-type system was observed in FFI mode during the entire TESS mission. Due to the shallowness of the primary and, especially the secondary eclipses, as well as the substantial continuous out-of-eclipse light variations which exceed in amplitude the depths of the secondary eclipses, we had to form normal light curves from the TESS material to mine out relatively usable (normal) eclipse timings. And, we were able to obtain such timing values only for the primary eclipses. Adding these new ETV points to the \textit{Kepler} ETV data, the $P_2\approx910$\,d third-body period of \citet{borkovitsetal16} appears to be confirmed, however, some other, longer timescale variation in the eclipse timings is also revealed. We made efforts to model it with quadratic and cubic polynomials; however, we found a satisfactory analytic ETV model with only the addition of the LTTE model for a fourth star. Hence, here we present a (2+1)+1 hierarchical quadruple star solution, where the inner orbit is described with a coplanar $P_2=928$\,d-period LTTE+DE model, while the additional variations are simultaneously described with a $P_3=5470$\,d-period, outer LTTE orbit. We note, however, that due to the very small outer period ratio of $P_3/P_2\approx5.9$, the interpretation of the longer period term as an LTTE orbit is quite unlikely.\\
\emph{9665086}: This short-period, likely, EA system ($P_1=0\fd30$), which was observed only in FFI mode during the TESS revisits, exhibits very shallow eclipses. This feature seems to be in accord (at least qualitatively), with our findings that the third star might be the most massive component of the triple. Due to the shallowness of the eclipses, even normal TESS eclipse timings show large scatter, but, despite this, our new ($P_2=854$\,d) LTTE+quadratic polynomial solution nicely confirms the former results of \citet{conroyetal14,borkovitsetal16,czavalingaetal23b}. It is also in accord with the $P=856$\,d astrometric solution of the Gaia NSS catalog \citep{GaiaNSS}. Note, the total mass of the inner binary was taken from \citet{cruzetal22}.\\ 
\emph{9711751}: This Algol-system showed rapid out of eclipse brightness variations during the \textit{Kepler} observations, but this is much less pronounced in the \text{TESS} light curves. (Two-min data are available only for Ss74, 75, 81.) \\
\emph{9714358}: This is one of the tightest triple systems with $P_2/P_1=16.01$. A recent, complex, spectro-photodynamical analysis was carried out by \citet{borkovitsmitnyan23}. They showed, that octupole-order secular perturbations are quite important in this triple system. This fact explains that our ETV model (which models the long-term or secular perturbations correctly only at the quadrupole level) does not result in correct results on longer timescales. When we did not constrain the apsidal motion period, however, we obtained a relatively satisfactory fit which resulted in a much longer apsidal motion period than what is found from the quadrupole order secular perturbation theory. For the LTTE+DE solution that we present currently, we forced $\im<1\degr$, i.e., we were looking for a practically flat (or, coplanar) solution, which is in accord with the results of \citet{borkovitsmitnyan23}. Note, the unconstrained mutual inclination runs resulted in much better fits around $\im=20\degr$, describing even the readily visible inflection points in the ETV curves around the beginning of the TESS observations.  But this would result in huge inclination variations, which is not the case. Finally, note that this system was measured in FFI mode during TESS Cycles 2 and 4, but 2-min cadence data are available for Cycle 6.\\
\emph{9722737}: The potential third-body induced $\sim445$\,d period LTTE in the ETV of this likely overcontact binary (either EW or ELV) was first reported by \citet{rappaportetal13} and was also confirmed by \citet{conroyetal14,borkovitsetal16}. Despite these earlier results, this outer orbit was also listed in \citet{murphyetal18} as a new single-pulsator binary, found by a phase modulation study of pulsating stars. Regarding the fact that the average of the odd and even minima are different in depth, the odd and even maxima are not the same height, and the ETVs of the odd and the even minima are also different, we claim, that the \textit{Kepler} and TESS light curves are coming from an EW or ELV system, instead of a single pulsating star. In spite of the challenges and methods for distinguishing between ELV, EW, and pulsating variables \citep{borkovitsetal16}, we are convinced that the $P_2=444$\,d-period ETV is actually due to a \emph{third} body. From the TESS FFI data we formed normal eclipses, which nicely confirm this periodicity and, moreover, display some further, longer timescale non-linearity, which we model with a cubic polynomial. \\
\emph{9777984}: This is the shortest outer-period pure LTTE system in our sample. In \citet{borkovitsetal16} this is tabulated under the false KIC ID of 9777987, but \citet{abdulmasihetal16} found that the true source was KIC 9777984. Here we used the \textit{Kepler} light curve from the \textit{Kepler} Bonus project \citep{keplerbonus} which contains time series for more \textit{Kepler}-quarters than the \textit{Kepler} EB catalog version. Hence, we also recalculated the \textit{Kepler} eclipse times. Due to the faintness of the target ($T=16.99$), we had to make a considerable effort to extract more or less usable light curves from the TESS FFIs, using PCAs for each sector. And then we formed normal light curves for 20-20 consecutive cycles to obtain normal eclipse points for the ETV. In this manner, however, we clearly detect the long-term, non-linear period variation, which already in the \textit{Kepler}-era was modeled with a quadratic polynomial \citep{borkovitsetal16}. Now, a cubic polynomial provides a much better fit, hence, we used that model. This may actually represent, however, a more distant fourth body. Thus, future monitoring would be important. \\
\emph{9788457}: The ETV of this EA or E$\beta$ system resembles that of KIC~8953296. Similar to that former system, we can give only a quite uncertain LTTE+quadratic polynomial solution. TESS observed this system only in FFI mode.\\
\emph{9821923}: Only FFIs are available for the TESS sectors. We find that the former solution of \citet{borkovitsetal16} is certainly incorrect. \\
\emph{9832227}: This EW system became popular and well-studied after the paper of \citet{molnaretal17} in which the authors claimed that KIC~9832227 is a red nova precursor, where the binary components will be subject to a merger around 2022. This statement was refuted shortly thereafter with ETV analyses by \citet{sociaetal18} and then \citet{kovacsetal19}. Besides \textit{Kepler} and TESS data (the latter observed this target in two-minute cadence mode in Cycles 2, 4 and, in FFI mode during Cycle 6) we also used publicly available WASP data and HAT measurements \citep[published in ][]{kovacsetal19}. We found a long period ($P2\sim33$\,yr) LTTE solution. The total mass of the overcontact binary was taken from \citet{molnaretal17}. \\
\emph{9838047}: The longer timescale ETV curves, extended by virtue of the TESS (FFI-derived) eclipse times, nicely confirm the results of \citet{borkovitsetal16}. They also imply some additional non-linearities, which we describe with a cubic polynomial, however, there are additional smaller discrepancies even with respect to this LTTE+cubic polynomial model.\\
\emph{9850387}: This is a relatively well-studied oEA system, and an SB2 spectroscopic binary. The primary exhibits hybrid $p-$ and $g$-mode pulsations \citep{sekaranetal20}. According to the asteroseismic studies of \citet{zhangetal20}, this is a pre-main-sequence system with extremely slow internal rotation \citep[see, also][]{fullerfelce24}. The ETV study of \citet{borkovitsetal16} revealed the presence of a $P_2=671$\,d-period tertiary with a non-negligible dynamical ETV component. Adding the eclipse times calculated from the two-min cadence TESS Cycles 2,4,6 observations, we confirm the former findings. \\
\emph{9882280}: This EW system was observed in FFI mode over the TESS mission. The ETV curve suggests a quasi-cyclic behavior, however, the departures from a strictly periodic variation cannot be described uniquely with polynomials of different degrees from 1 to 3. Hence, we rank our solution as a moderately certain one.\\
\emph{9912977}: This is a short-period detached system. It was observed in two-min cadence mode in TESS Cycle 6, while for the earlier sectors only FFI data are available. The total mass of the binary was taken from \citet{cruzetal22}.\\
\emph{9963009}: This is a long-period ($P_1=40\fd07$), detached binary. During the \textit{Kepler} observations, the depths of its secondary eclipses decreased continuously, while the primary eclipses showed a constant depth. This fact suggests that the (current) primary eclipses occur closer to the periastron passage of the binary (i.e., $180\degr<\omega_1<360\degr$). For the TESS-observations (FFIs in Year 2 and 4 data, and 2-min cadence in Year 6) the secondary eclipses became undetectable. Hence, we add only new primary eclipse times to the \textit{Kepler} ETV material. We found a quite uncertain, however physically realistic, inclined ($\im=36\degr$), $P_2=3492$\,d-period third-body solution. Further observations will be necessary to confirm or refute this result.\\
\emph{9994475}: This is an EW system with a very interesting ETV. The best mathematical model is a doubly periodic, double LTTE representation, which, formally, would imply a (2+1)+1-type hierarchical quadruple configuration. On the other hand, however, due to the very low period ratio of $P_3/P_2=3267/605\approx5.4$, such a configuration would likely be unphysical. Hence, we claim that the $P_2=605$\,d cyclic term actually describes an existing third stellar component, while the longer period ($P_3=3267$\,d) term must have some different origin instead of the effect of another body in the system. Note, TESS observed this system only in FFI mode.\\
\emph{10095469}:  In the Villanova catalog, \textit{Kepler} observations are available only starting from Q8. For earlier quarters we use the \textit{Kepler} bonus data \citep{keplerbonus}. Only FFIs are available for the TESS data. We used just the primary eclipses for our analysis. For the TESS data we formed and used normal eclipses.  Similar to the previous system, the eclipse timings also indicate a double periodicity. Therefore, to obtain a satisfactory solution, we had to fit two LTTE orbits simultaneously and, therefore, we found a (2+1)+1-type hierarchical quadruple solution with periods $P_2=760$\,d, $P_3=4550$\,d. This solution, however, should be accepted only with considerable caution. \\
\emph{10095512}: There are two-min cadence data for TESS Cycle 6 sectors, while FFIs are available for the earlier sectors. The coplanar ($i_\mathrm{m}:=0\degr$) LTTE+DE solution of \citet{borkovitsetal16} is confirmed on a longer timescale. Note, $i_1=i_2=85\degr$ was taken from \citet{windemuthetal19}.\\
\emph{10208759}: This is a totally eclipsing, detached EA system, revolving on a circular orbit \citep{windemuthetal19}, with almost equal primary and secondary eclipse depths. Interestingly, the hotter component appears to be somewhat smaller (primary eclipses are occultations, while secondary ones are transits), indicating a slightly evolved primary. We found a very low-amplitude, long-period, highly uncertain LTTE solution. No two-min cadence TESS observations are available. \\
\emph{10216186}: This is likely a totally eclipsing E$\beta$-type EB. We used only the deeper primary eclipses in the ETV analysis. TESS observed this target only in FFI mode. We formed normal eclipses from the TESS data.\\
\emph{10223616}: The eclipses disappeared during the \textit{Kepler}-era. No eclipses can be detected in TESS data. This source was not included in \citet{borkovitsetal16}. \\
\emph{10226388}: There are two-min cadence TESS light curves for Sectors 14, 15, 41, 54, 56 but FFIs only for Cycle 6 sectors. Besides the clear $P_2=942$\,d-period LTTE orbit, the ETVs suggest some non-linear variations as well.  These manifest themselves in a clearly non-quadratic way, but can be well approximated with a cubic polynomial..\\
\emph{10268809}: The \textit{Kepler} ETV curve for this EA system resembles that of KIC~813170. However, in contrast to this other system, the current eclipsing binary showed dramatic EDVs during the \textit{Kepler} mission. For the TESS (FFI) observations only the (originally) secondary eclipses have remained. We were unable to identify eclipses around photometric phase zero. Moreover, for Cycle 6 observations, these shallow secondary eclipses show a phase jump, indicating a periastron passage between Cycle 4 and 6 observations. Hence, due to the likely detection of two sequential outer periastron passages (one during the \textit{Kepler} observations, and the new one from TESS), our new ETV solution looks at least moderately certain, and implies a substantially shorter outer period than the one found in \citet{borkovitsetal16}. \\
\emph{10268903}: The Villanova EB catalog contains the light curve of this target only from Quarter 9 onward, hence only a $\sim700$\,d-long ETV curve was used for the analysis of \citet{borkovitsetal16}. Due to the \textit{Kepler}-bonus project \citep{keplerbonus}, additional \textit{Kepler} light curves for some of the earlier quarters are now also available, extending the length of the \textit{Kepler} ETV data train to $\sim$$1237$ days. Unfortunately, although TESS observed this very faint target during 11 sectors, and even 2-min cadence light curves are also available for 4 sectors, we were unable to obtain any usable eclipse times from the TESS observations (though, the very shallow eclipses are continuously visible at least in the sector by sector folded light curves). Hence, here we give a new LTTE solution only for the now extended, $\sim$$1237$\,d-long \textit{Kepler} ETV dataset.\\
\emph{10275197}: The combination of \textit{Kepler} and TESS ETVs (the latter from FFIs), earlier SWASP normal eclipses and eight recent eclipses from BAO reveal that a longer timescale period variation is superposed on the $P_2=2111$\,d periodic signal.  We interpret the periodic signal as an LTTE orbit, while the longer timescale effect is described with a cubic polynomial. The large amplitude of the LTTE term leads to a high minimum mass of ($M_\mathrm{C,min}=2.10\,\mathrm{M}_\sun$), which is larger than the total mass of the EB. In this regard, note that \citet{kobulnickyetal22} gives a third light of $\ell_3=0.468$, but even this value appears to be too low. (An explanation might be that the third star is itself a second binary.) \\
\emph{10296163}: This system has an uncertain, very inclined ($i_\mathrm{m}=56\degr\pm1\degr$) LTTE+DE solution. Our analytic results suggest a currently retrograde apsidal motion, but this is quite uncertain. Besides \textit{Kepler} and TESS FFI data, one additional primary eclipse obtained with the 1-m RCC telescope of Konkoly Observatory in 2019 was also used. The inner inclination ($i_1=88\degr$) was taken from \citet{windemuthetal19}. Numerical studies of this system were carried out by \citet{getleyetal20}. \\
\emph{10322582}: We formed normal eclipses both from the TESS FFI data and the publicly available SWASP data. The ETV curve displays large amplitude variations; however, the time-scale seems to be longer than the duration (6316\,days) of the whole dataset. Hence, our solution would be in the category of very uncertain.\\ 
\emph{10383620}: Four LAMOST RV points are available for the primary, making this likely SD EA-type EB into an SB1 spectroscopic binary \citep{zhangetal19}. This system is also an oscillating Algol \citep{gaulmeguzik19}.  TESS observed this system only in FFI mode. Despite this, our $P_2=1504\fd8\pm0\fd1$ LTTE solution is very robust. \\
\emph{10388897}: TESS observed this system only in FFI mode. We calculated and used normal eclipses for the TESS observations. No LTTE solution was published in \citet{borkovitsetal16} and, the current solution is also not very convincing. Further observations would be essential.\\
\emph{10481912}: This system was observed in two-min cadence mode during TESS Cycles 2, 4, and 5, while FFI observations are available for the Cycle 6 sectors. The ETV curve seems to be sinusoidal, but due to the long inferred outer period we rank it as an uncertain LTTE solution. \\
\emph{10483644}: The presence of the relatively short-period ($P_2=371$\,d) third stellar companion was first reported in \citet{borkovitsetal16}. The TESS observations (FFIs for Cycles 2, 4, 5), and two-min cadence data for Cycle 6 produce very scattered ETV points. We formed normal eclipses (one for each TESS sector), but even these data have larger scatter than the amplitude of the cyclic timing variations. These new points, however, reveal that there are no additional period variations within the duration of the extended dataset. Hence, our new LTTE+DE solution is very close to that of \citet{borkovitsetal16}. For this solution we assumed $e_1=0$, $\im=0\degr$. We took $i_1$ from \citet{windemuthetal19}. \\
\emph{10486425}: This is a detached EA and an SB2 system, as well, of which the primary star is likely a $\gamma$~Dor-type pulsator \citep{zhangetal18}. \citet{matsonetal17} also reported the presence of a third spectral component. The target was observed only in FFI mode with TESS. Due to the shallowness of the secondary eclipses we used only the primary eclipse timings for our analysis. We interpret the slight curvature of the $\sim1.5$ decade-long ETV as the LTTE signal of a $P_2\approx6500$\,day-period third component, but this interpretation is currently very uncertain. We took the total mass of the inner binary from the study of \citet{zhangetal18}.\\
\emph{10549576}: This slightly eEA system was observed only in FFI mode by TESS. The LTTE+DE solution is quite secure.  The inner inclination ($i_1=87\degr$) was taken from \citet{windemuthetal19}. \\
\emph{10557008}: This target was observed only in FFI mode during the TESS mission.  We formed normal eclipses from the TESS data. The non-linear ETV is evident; however, our LTTE+quadratic polynomial solution is very uncertain.\\
\emph{10581918 = WX Dra}: This is an oscillating Algol-system, which is also an SB2 binary \citep{matsonetal17}. An earlier ETV study was carried out by \citet{zascheetal15}, who, in addition to the \textit{Kepler} data, collected earlier eclipse times from the literature. We also used their collection and, moreover, added the new TESS eclipse times from the FFI data. We give an LTTE+quadratic polynomial solution, however, this is far from conclusive. Note, the total mass of the inner binary was taken from \citet{matsonetal17}.\\
\emph{10583181}: This detached EB is an SB2 spectroscopic binary as well. It was analyzed in detail by \citet{helminiaketal19}. They detected the effect of the third star in variations of the systemic radial velocity ($\gamma$) of the binary. This binary was observed in two-min cadence mode in all available TESS sectors. Our LTTE analysis essentially confirms the earlier results of \citet{borkovitsetal16}. We took the mass of the eclipsing binary from the paper of \citet{helminiaketal19}. \\
\emph{10613718}: The short period ($P_2=88$\,d) outer orbit was first noticed in the early ETV study of \citet{rappaportetal13} and, was confirmed by \citet{borkovitsetal16}. Later \citet{lurieetal17} found from likely spot modulation that this is a non-synchronously and slowly rotating EA system \citep[see, also][]{felcefuller23}. TESS observed this target mostly in FFI mode, but two-min cadence data are available for sector 41 only. Due to the very small depths of the eclipses, the new eclipse times of the ETV curve have a huge scatter, which exceeds the amplitude of the cyclic ETV; hence, we formed and used normal eclipses. The previously identified $\sim88$\,d-period cyclic ETV, which we model with an LTTE+DE solution, is continuously visible in the TESS portion of ETV curve, and no additional variation can be seen. \\
\emph{10666242}: This is a single eclipser (i.e. only primary eclipses can be observed), which makes the whole solution extremely uncertain. Because of the long EB period, only two eclipses were observed by TESS (in Sectors 41 and 55).\\
\emph{10686876}: This totally eclipsing, detached EA system was found to be an SB1 binary by \citet{matsonetal17}. The primary star also exhibits $\delta$~Scuti-type pulsations, which were analyzed by \citet{liakos20}. Previous ETV studies (besides \textit{Kepler} observations, included historical, ground-based eclipses as well) were carried out by \citet{zascheetal15,borkovitsetal16}. We use all of these data, and also add the new eclipse times based on TESS observations. (Note, that TESS observed this system in two-min cadence mode in Cycles 2, 4, 5, and Sectors 76 and 83 of Cycle 6, as well as in FFI mode in the Sectors 74, 81, 82 of Cycle 6.) The new eclipse times from TESS suggest a longer-period and more eccentric third-body solution than the earlier solutions. (For the minimum third-body mass calculation we use the realistic binary mass estimation of \citealt{liakos20}).\\
\emph{10724533}: This is an E$\beta$-type EB, and an SB1 spectroscopic binary as well \citep{GaiaNSS}. Two-min cadence data are available for TESS Cycles 2 and 4, while in Cycle 6, it was observed in FFI mode. The LTTE solution looks rather convincing, however, we had to add a quadratic polynomial, as well, for a satisfactory fit to all the TESS ETV sections. \\
\emph{10727655 = V2280 Cyg}: Besides the \textit{Kepler} and TESS FFI eclipse times, we gathered 22 additional ground-based points from the literature. Now we confirm the cyclic part of the LTTE orbit that was found in \citet{borkovitsetal16}. The extended duration of the time series clearly reveals some additional non-linearity, which we model with a quadratic polynomial.\\
\emph{10848807}: This EW system was observed only in FFI mode with TESS. We cannot confirm the earlier shorter period, low amplitude LTTE solution of \citet{conroyetal14,borkovitsetal16}. Here we give an uncertain, longer $P_2$-period LTTE+quadratic polynomial solution. \\  
\emph{10916675}: This EW, again, was observed only in FFI mode during the TESS mission. We formed normal eclipses from the TESS data. Our LTTE+quadratic polynomial solution is quite uncertain. \\
\emph{10934755}: This is a totally eclipsing, likely E$\beta$-type EB. TESS observed this target only in FFI mode. We formed normal eclipses from the TESS data and used only the much deeper primary eclipses to form the ETV curve. We give an LTTE solution which differs substantially from that of \citet{borkovitsetal16}. That former model is evidently refuted by the new eclipse timing results. \\
\emph{10979716}: The ETV of this detached system was analysed for the first time by \citet{borkovitsetal15}. TESS observed this target only in FFI mode. These observations have led to only quite scattered eclipse times (at least, relative to the amplitude of the ETV). Despite this, one can judge that these new data confirm the earlier LTTE+DE solution and reveal that there are no further period variations. For our new analysis we took the inner inclination from the work of \citet{windemuthetal19}. \\
\emph{10991989}: The inner eclipsing binary is likely a detached system, which provides only $\sim1\%$ of the total light of the system. The tertiary is a pulsating red giant \citep{gaulmeetal13}. An ETV solution was given for the first time by \citet{rappaportetal13}, while \citet{helminiaketal16} carried out an RV analysis of the tertiary, as a $P_2=548$\,d-period single-lined spectroscopic binary. Hence, with the combination of the RV solution of the tertiary, and the LTTE solution of the ETV of the inner binary, one can infer the outer mass ratio ($q_2=1.32\pm$) as well. Furthermore, \citet{gaulmeetal13} gives an astroseismic mass for the tertiary red giant ($m_\mathrm{C}=2.5\pm0.4\,\mathrm{M}_\sun$) and, hence, the outer inclination ($i_2\sim24\degr$) can also be inferred.  (Note, the system was observed in 2-min cadence mode during the entire TESS mission.) \\
\emph{11042923}: This EW-type system was observed only in FFI mode with TESS. While the \textit{Kepler} ETVs made it very likely that there is an LTTE orbit with $P_2\sim1042$\,days \citep{rappaportetal13,borkovitsetal16}, the new data makes evident the presence of some additional, longer timescale ETV, which we model with a quadratic polynomial. Our solution appears to be quite convincing.\\
\emph{11234677}: This is an EA system with very shallow secondary eclipses. Hence, similar to \citet{borkovitsetal16}, we used only the primary eclipses in the ETV analysis. This target was observed only in FFI mode with TESS. Our LTTE solution seems to be quite robust, and nicely confirms the results of \citet{borkovitsetal16}. \\ 
\emph{11246163}: This EW system, again, was observed only in FFI mode during the TESS survey. The non-linear period variation is evident, however, the pure LTTE and LTTE+quadratic polynomial solutions show systematic residuals. Hence, here we tabulate an LTTE+cubic polynomial solution. We rank our solution as rather uncertain.\\
\emph{11502172}: This long-period ($P_1=25\fd43$) EA system was observed in FFI mode during TESS Cycles 2 and 4, and in two-min cadence mode in Cycle 6. We give a non-coplanar, eccentric LTTE+DE solution. The initial value of $e_1$ for the MCMC runs were taken from \citet{kjurkchievaetal17}. \\ 
\emph{11519226}: TESS observed this system only in FFI mode.  The LTTE+DE solution is in fairly good agreement with the former model of \citet{borkovitsetal16}. The stability of the system was studied by \citet{getleyetal20}. \\
\emph{11558882}: Only three primary and one secondary eclipses can be detected in the TESS (FFI) data. {There are solo primary eclipses at the very end of S14, and another in S54, while both kinds of eclipses occurred in S81.} This system was also studied by \citet{getleyetal20}. \\ 
\emph{11604958}: This system was observed only in FFI mode with TESS. We formed and used normal eclipses not only for TESS, but even for the \textit{Kepler} data. The non-linearity in the ETV curve is evident, but we found only some very uncertain solutions.\\
\emph{11968490}: This is a chromospherically active EA system. Two-min cadence light curves are available for the TESS Cycle 6 sectors (in the earlier sectors there are FFIs only). The new solution confirms the findings of \citet{rappaportetal13} and \citet{borkovitsetal16}. \\
\emph{120196741 = V2294 Cygni}: TESS observed this EW system only in FFI mode. Similar to \citet{borkovitsetal16}, ground-based observations of eclipse times, made before the \textit{Kepler} era, were also used. \\
\emph{12055014}: This is a longer period, totally eclipsing EW-type overcontact binary, which was observed only in FFI mode with TESS. We found only a quite unconvincing LTTE+quadratic polynomial ETV solution. \\
\emph{12055255}: This likely ELV binary is close to the lower cut-off period of overcontact binaries. It was observed only in FFI mode with TESS. From these observations we calculated normal eclipses. The LTTE solution appears to be quite compelling, however, we had to also add a cubic polynomial (with small non-linear coefficients) to compensate for some further non-linearities in the ETV data. \\
\emph{12071006 = V379 Cyg}: This is a totally eclipsing EA system, and an SB2 spectroscopic binary as well \citep{matsonetal17}. It was observed only in FFI mode during all TESS revisits. We used only the primary eclipses (which are $\approx90\%$-deep), and found a quite uncertain, long-period $P_2=9852\pm1264$\,d-period ETV solution. (The total mass of the inner, eclipsing pair was taken from the RV analysis of \citealt{matsonetal17}.)\\
\emph{12071741}: This system is a low amplitude, likely overcontact binary with grazing eclipses. It was observed only in FFI mode during all TESS visits. Due to the large scatter of the new ETV points, we had to form normal light curves and, hence, normal eclipse times. The ETV is modeled with the combination of a $P_2\approx1207$\,d-period LTTE orbit and a cubic polynomial.  Note that this period is longer (by $\approx30\%$) than the one found formerly by \citet{conroyetal14} and \citet{borkovitsetal16} and, therefore, we cannot conclude that our solution confirms these former findings with substantial confidence.\\
\emph{12302391}: This is a long-period ($P_1=25\fd32$), detached EA system. In addition to the different curvatures of the \textit{Kepler} primary and secondary ETV curves, the TESS ETV points suggest some jumps in the timings which might be the consequence of a periastron passage of a third body. TESS observed this target in FFI mode during Cycles 2, 4, 5 and Sector 74 of Cycle 6, while two-min cadence data are available for Sectors 75, 80, 81, 82. Our inclined LTTE+DE solution should be considered only as a very preliminary and very uncertain first trial for the explanation of this enigmatic ETV signal. \\
\emph{12356914}: Long-period ($P_1=27\fd31$) EA system with very shallow secondary eclipses.  These secondary eclipses were flat during both the \textit{Kepler} and the beginning of the TESS eras, and they are much longer in duration than the primary eclipses. By the TESS Year 6 observations, these secondary eclipses became V-shaped, and somewhat shorter (but, continuously longer than the primary eclipses). These facts suggest some orbital plane precession and also imply that the primary eclipses should be closer to the periastron passage of the inner orbit (i.e., $180\degr<\omega_1<360\degr$). We constrained our solution according to this last condition. The inner inclination ($i_1=89.7\degr$) was taken from \citet{windemuthetal19}. It was also studied in \citet{getleyetal20}. No two-min cadence TESS observations. We plan to carry out a complex photodynamical analysis in the near future.\\
\emph{12458133}: The cyclic ETV of this EW binary was interpreted as the LTTE signal of a $P_2\approx1116$\,d-period third component in \citet{conroyetal14}. This object, however, was left out from the third-body candidate systems of \citet{borkovitsetal16} because of the different amplitudes of the averaged ETV and QTV curves, which fact made the LTTE interpretation questionable. Adding the new TESS FFI data determined ETV points, additional, larger amplitude, longer timescale variations are visible. We interpret the ETV curve as the combination of a $P_2=2223\pm4$\,d-period LTTE orbit and a quadratic polynomial.\\
\emph{12554536}: This is most likely an Algol-type SD system. The shallow secondary eclipses were not used. The system was observed only in FFI mode in all available TESS sectors. Our LTTE solution is mostly dominated by the \textit{Kepler} ETV points, but the new TESS points likely contradict our accepted, but very uncertain solution. \\

\onecolumn

\section{ETV curves}
\label{app:ETVcurves}


\FloatBarrier
\begin{figure*}[h!]
\includegraphics[width=60mm]{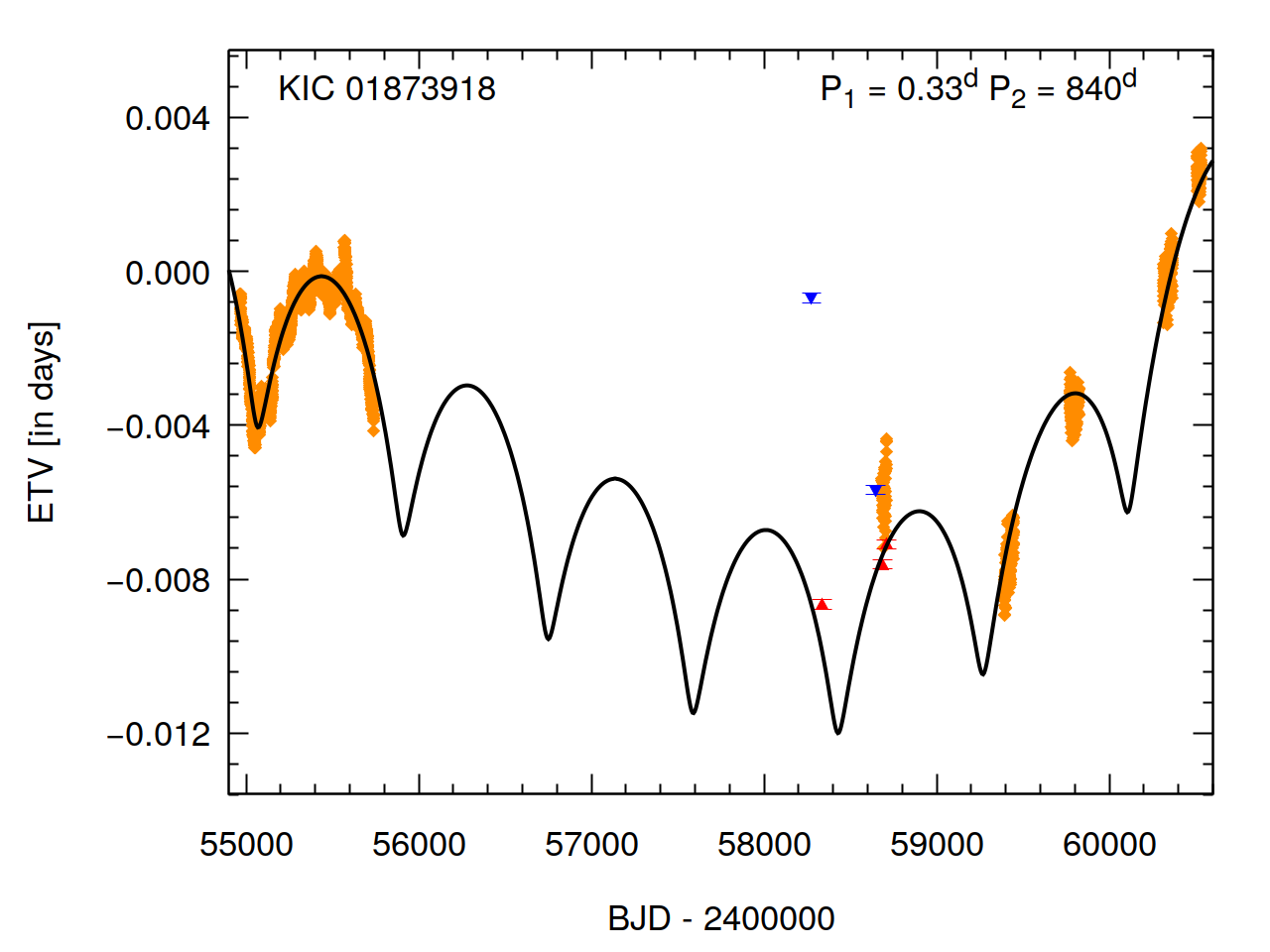}\includegraphics[width=60mm]{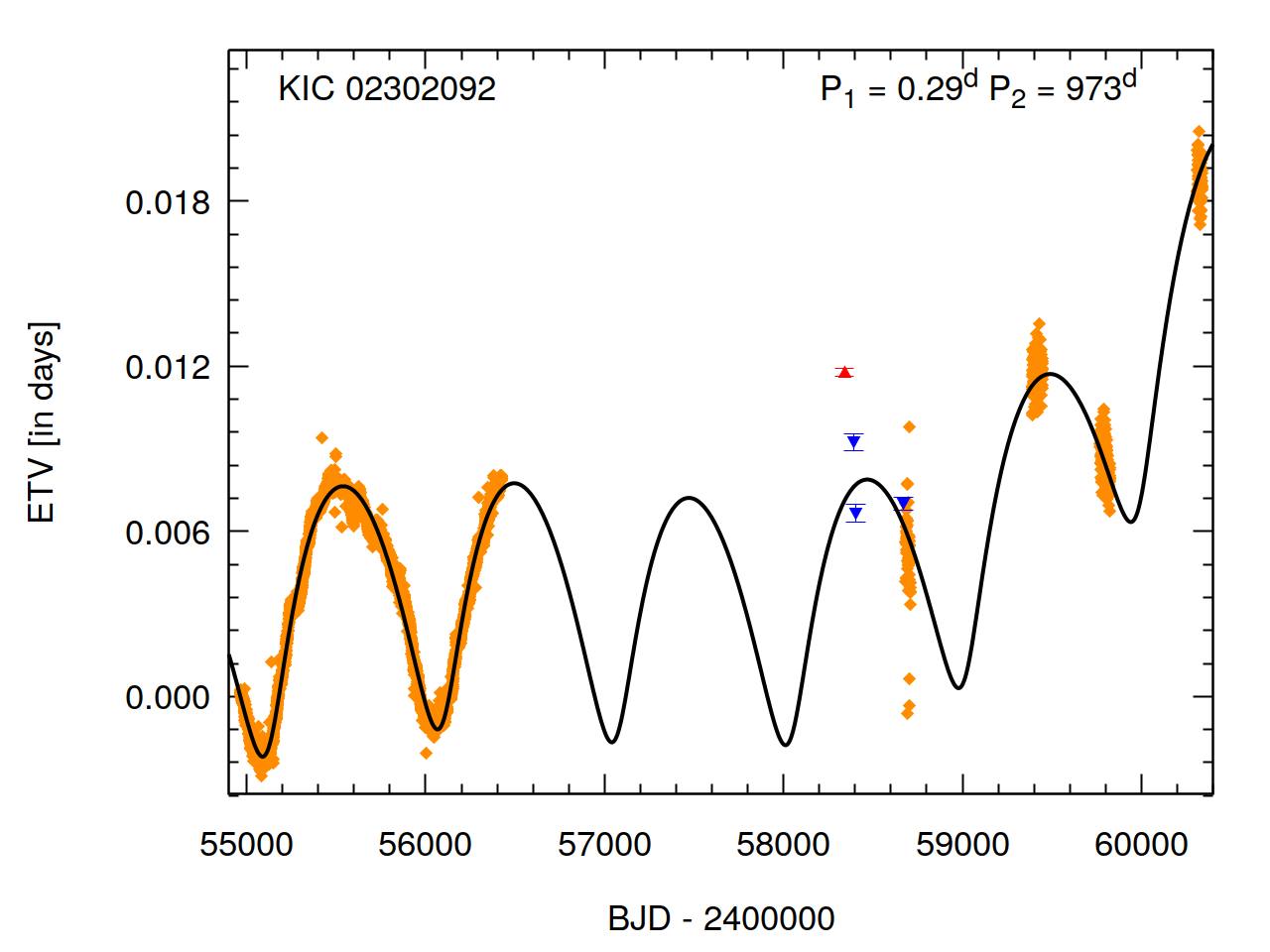}\includegraphics[width=60mm]{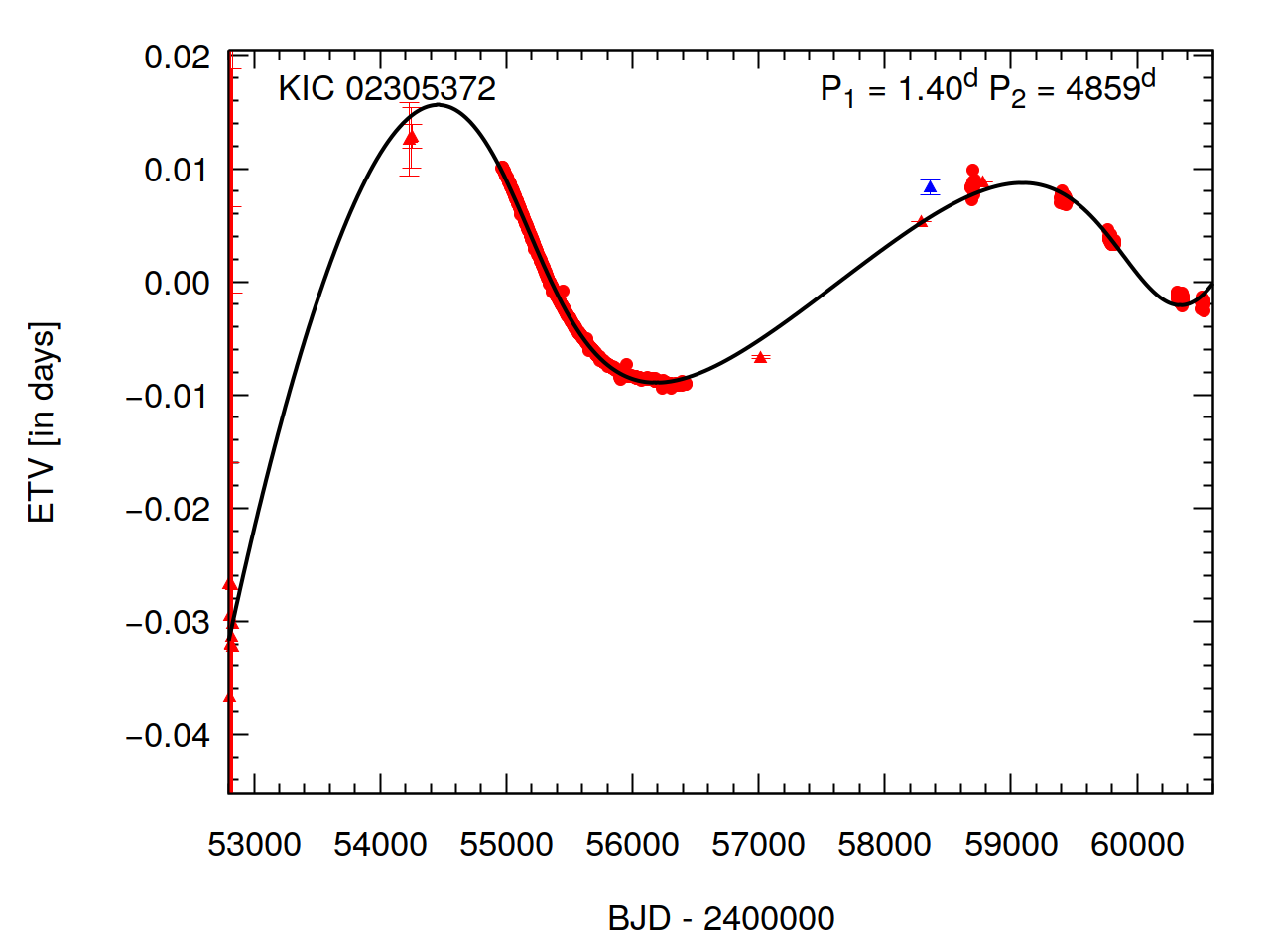}
\includegraphics[width=60mm]{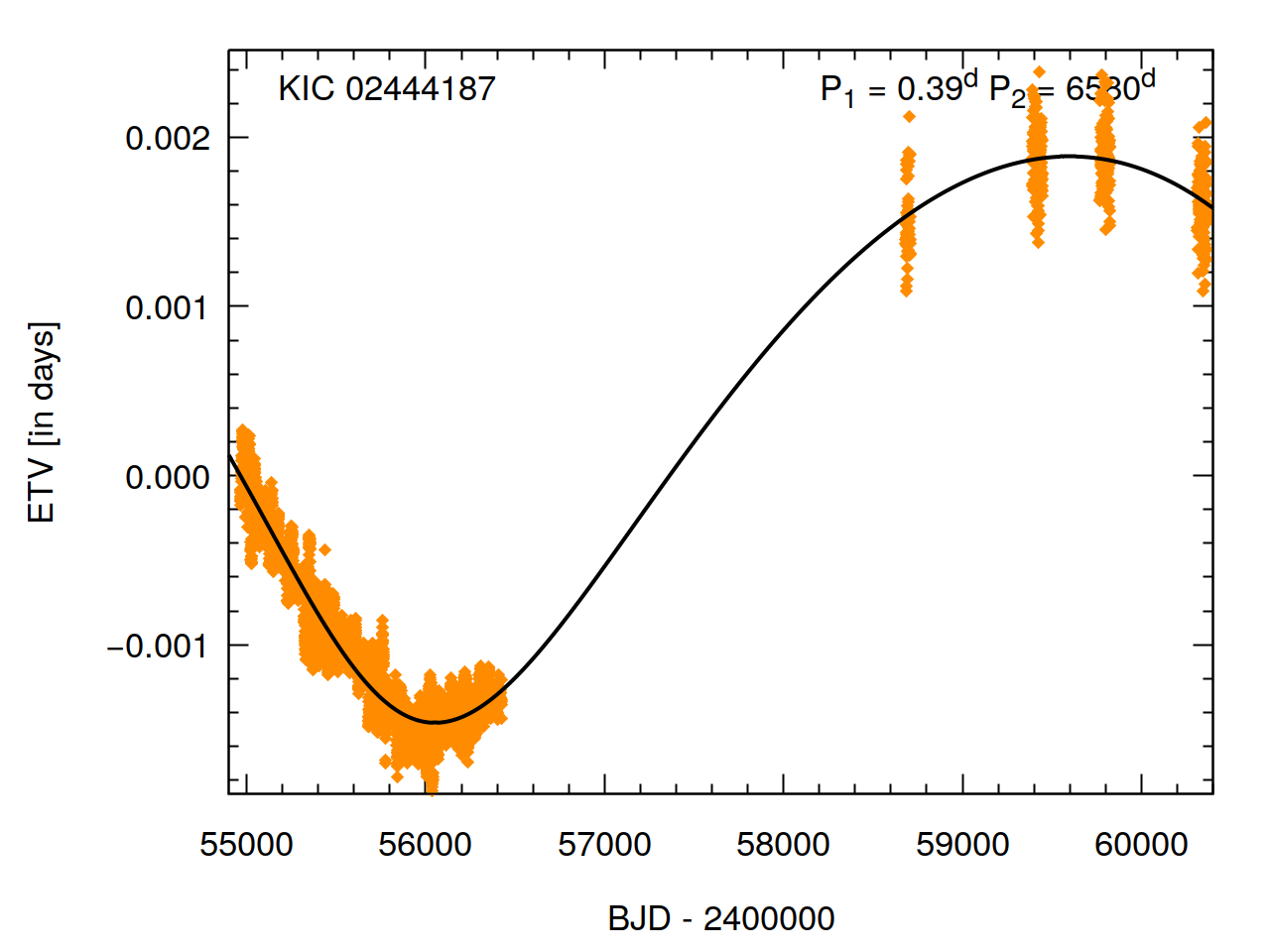}\includegraphics[width=60mm]{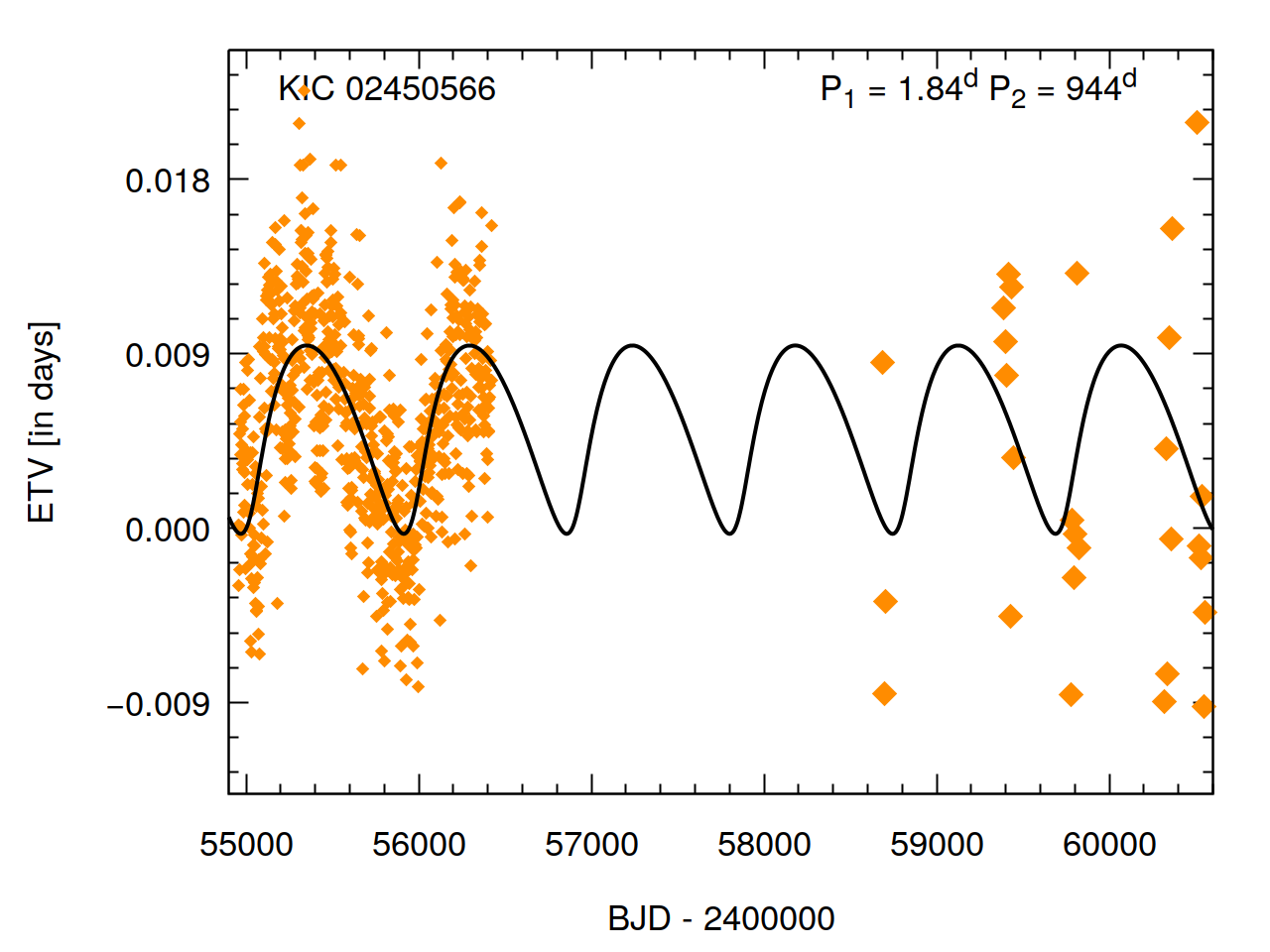}\includegraphics[width=60mm]{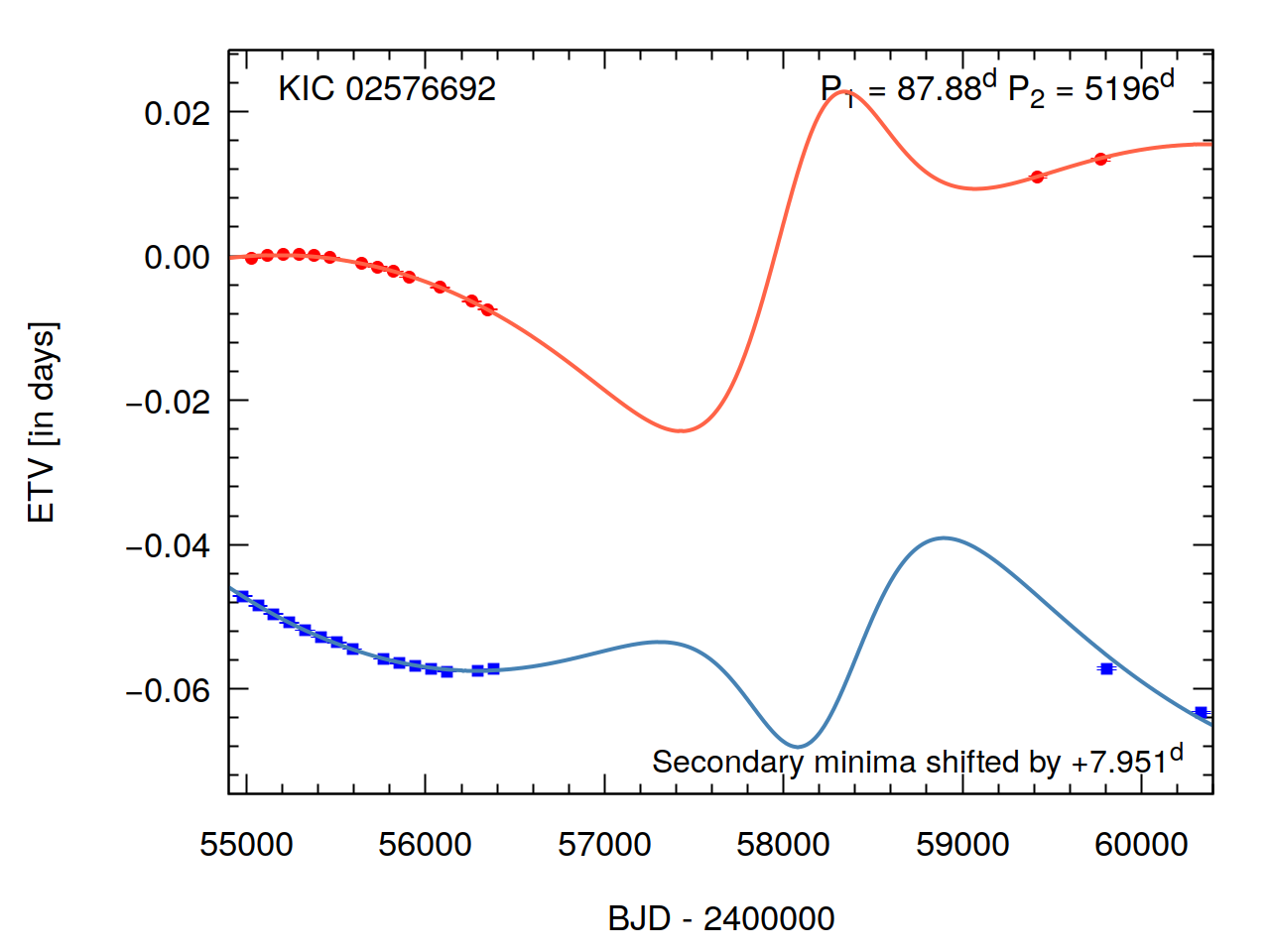}
\includegraphics[width=60mm]{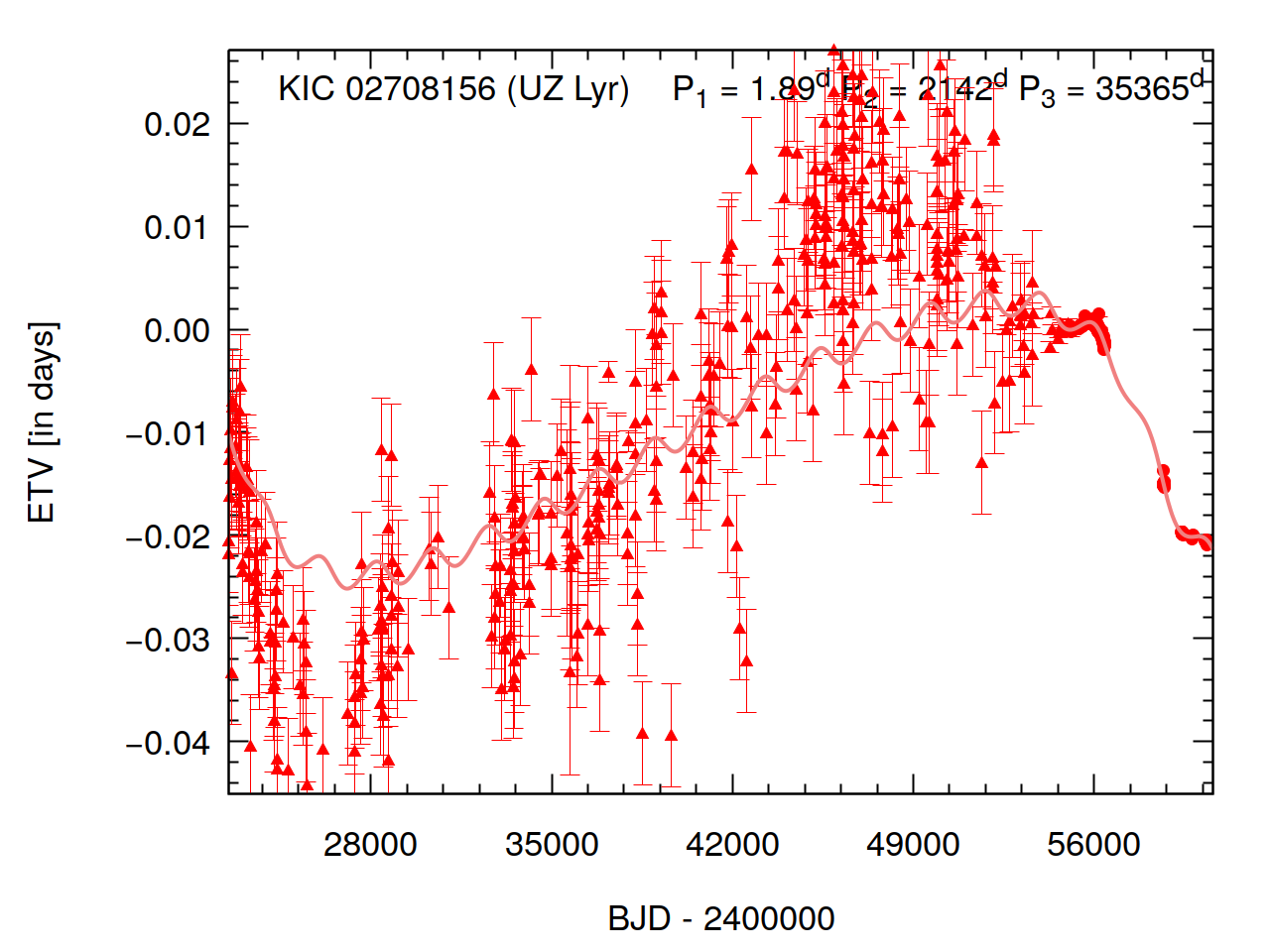}\includegraphics[width=60mm]{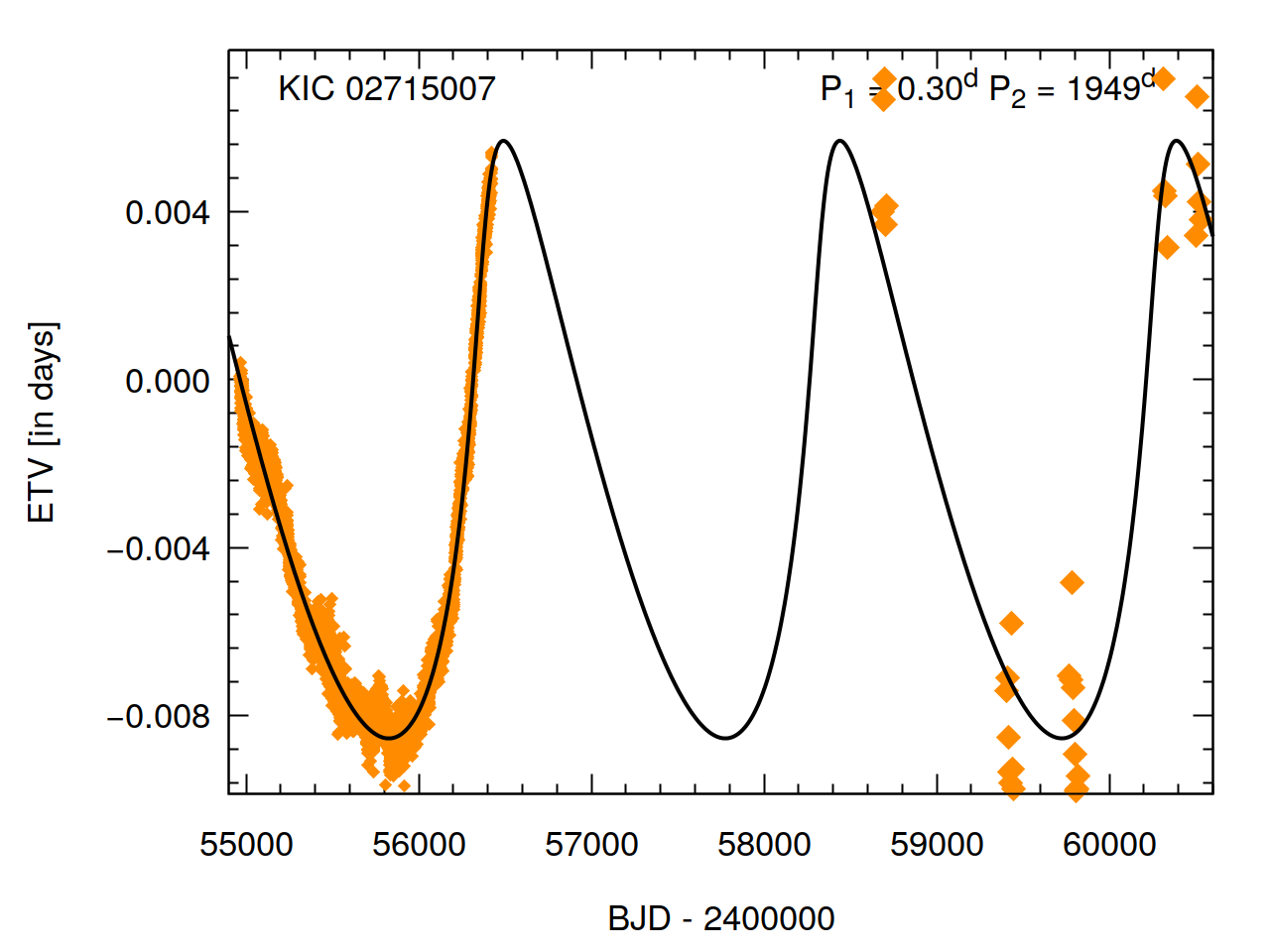}\includegraphics[width=60mm]{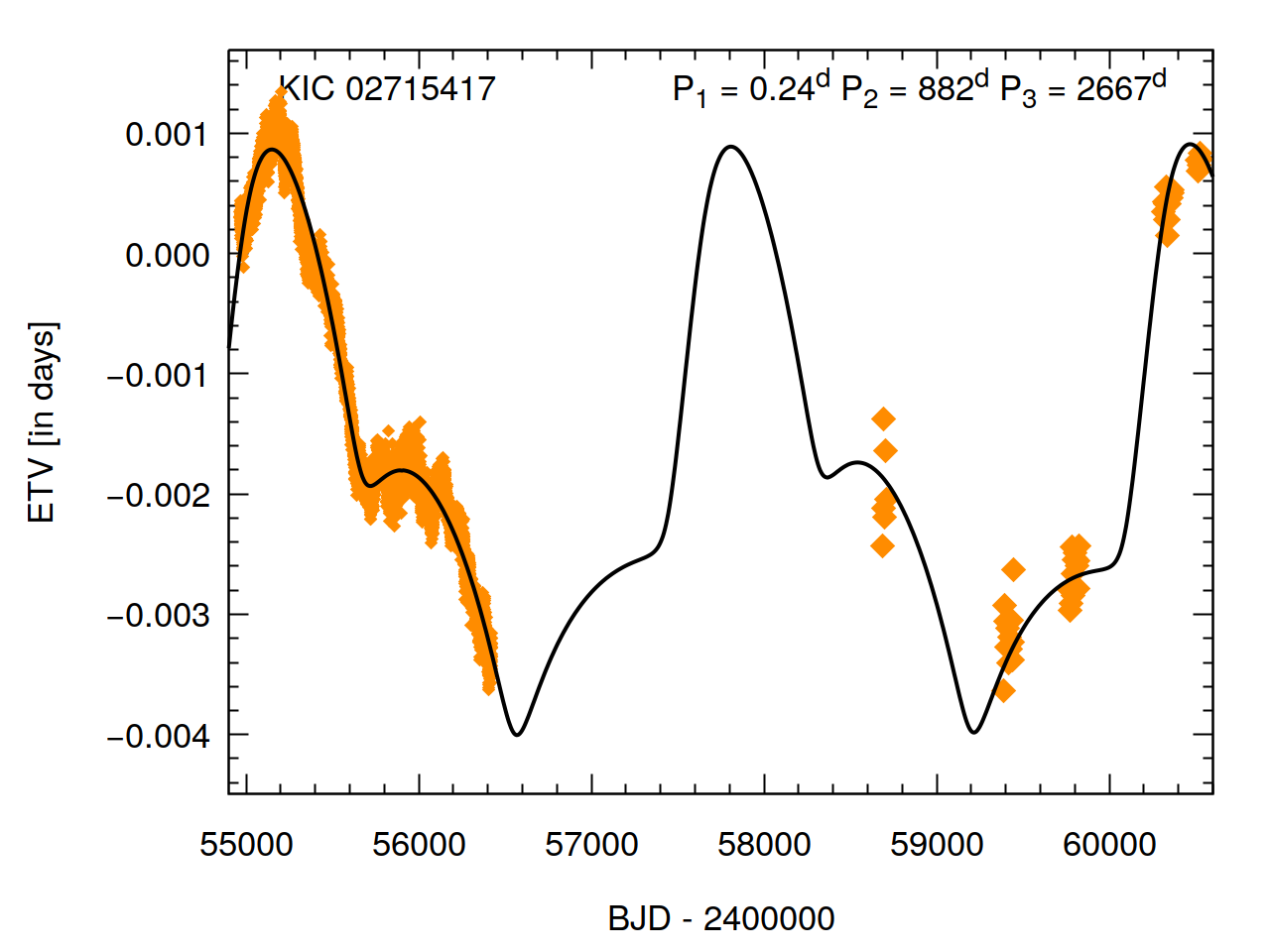}
\includegraphics[width=60mm]{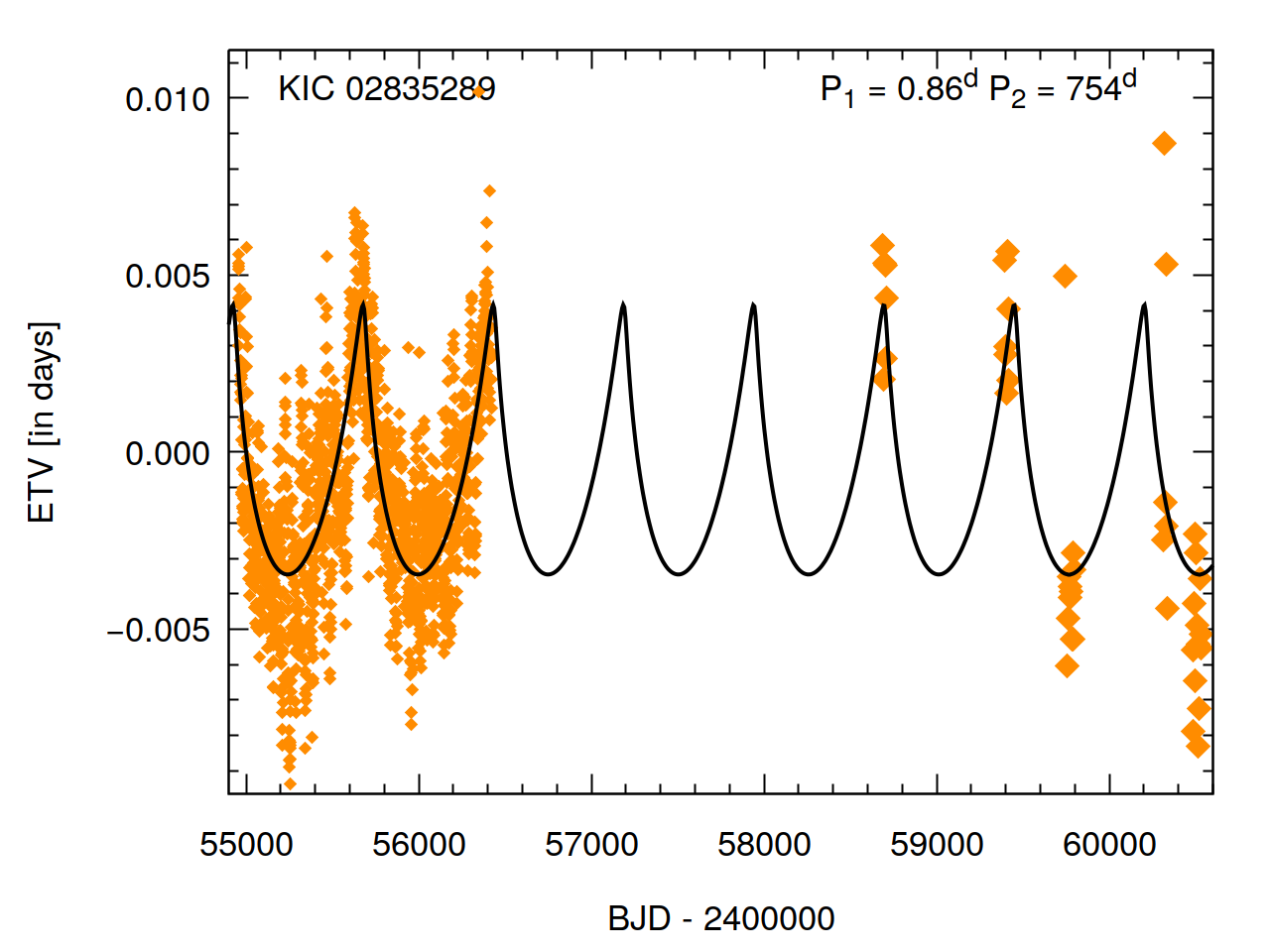}\includegraphics[width=60mm]{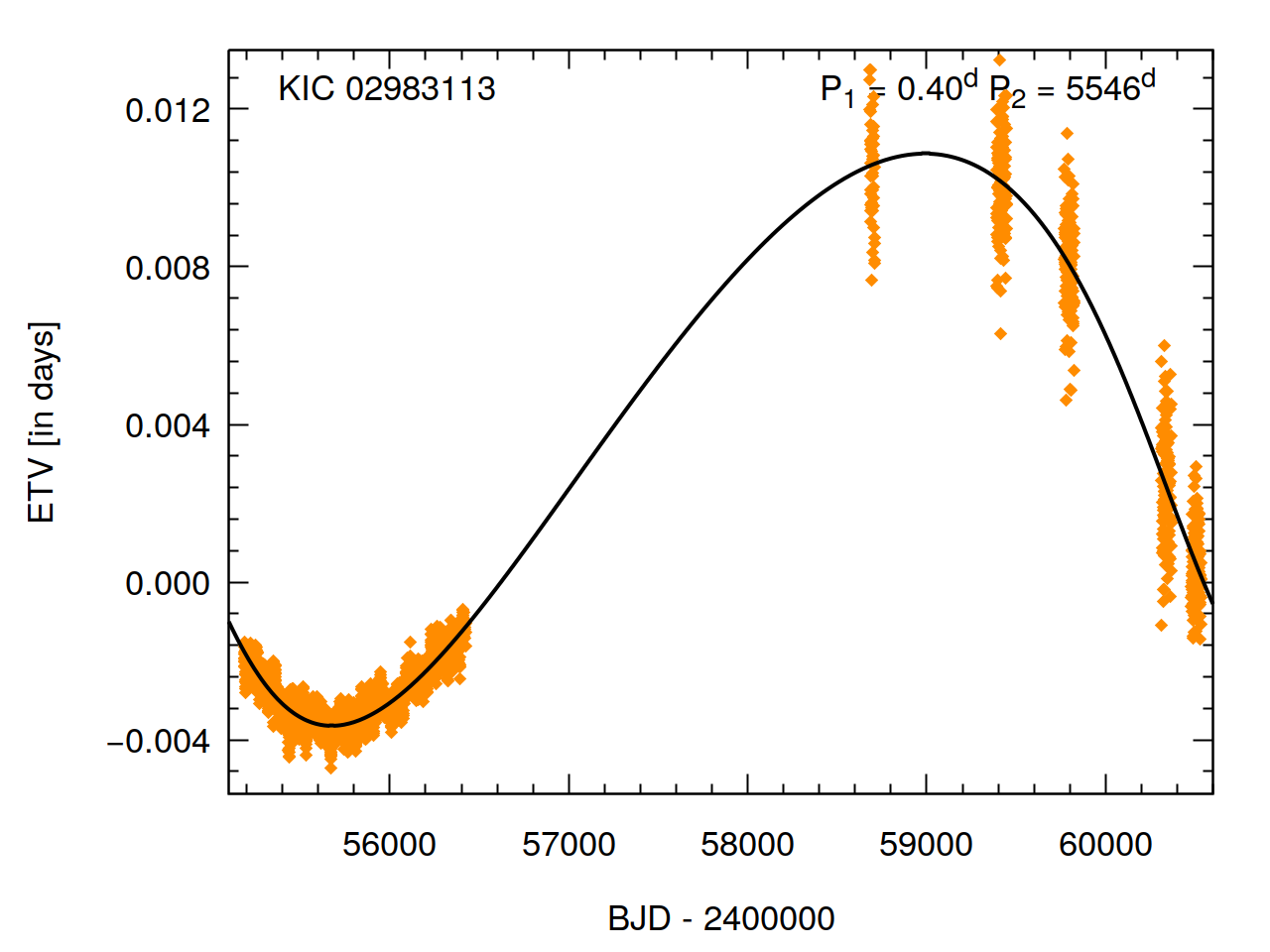}\includegraphics[width=60mm]{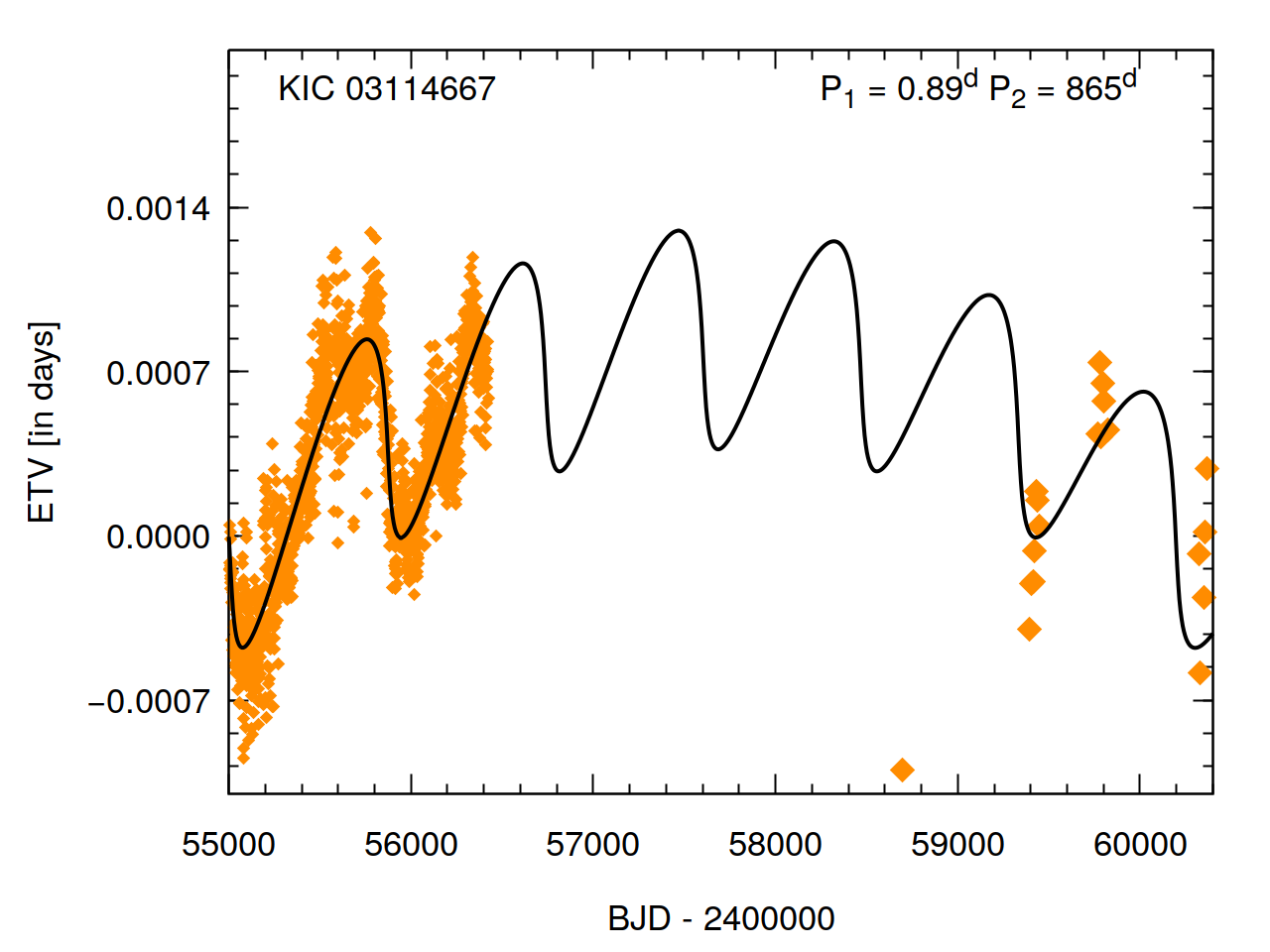}
\caption{ETVs with third body solutions. ETV curves calculated from satellite observations of primary and secondary minima, and the average of the two, are denoted by red circles, blue boxes, and orange diamonds, respectively. (Where normal minima were used, these are denoted with larger sized symbols.) We display and fit the ETV curves for both the primary and secondary eclipses only when the data quality warrant a joint analysis and the binary is eccentric. If the primary and secondary ETV curves are of comparable quality and the binary eccentricity is nearly zero, we display and fit only the average of the two ETV curves. If the quality of the primary ETV curve is significantly better than that of the secondary curve or, if only primary eclipses are present, we present only the plot and the fit for the primary eclipses. Ground-based minima (taken from either the literature, or our own follow-up observations, and available only for a few systems) are denoted by upward red triangles (primary) and downward blue triangles (secondary); their estimated uncertainties are also indicated.  Pure LTTE solutions are plotted with black lines, while combined dynamical and LTTE solutions are drawn with grey lines. (Note, the use of quadratic or cubic terms is not indicated; for these and other details, see Table~\ref{Tab:OrbelemLTTE1}--\ref{Tab:Orbelemdyn3}}
\label{Fig:ETVs}
\end{figure*}

\addtocounter{figure}{-1}

\begin{figure*}
\includegraphics[width=60mm]{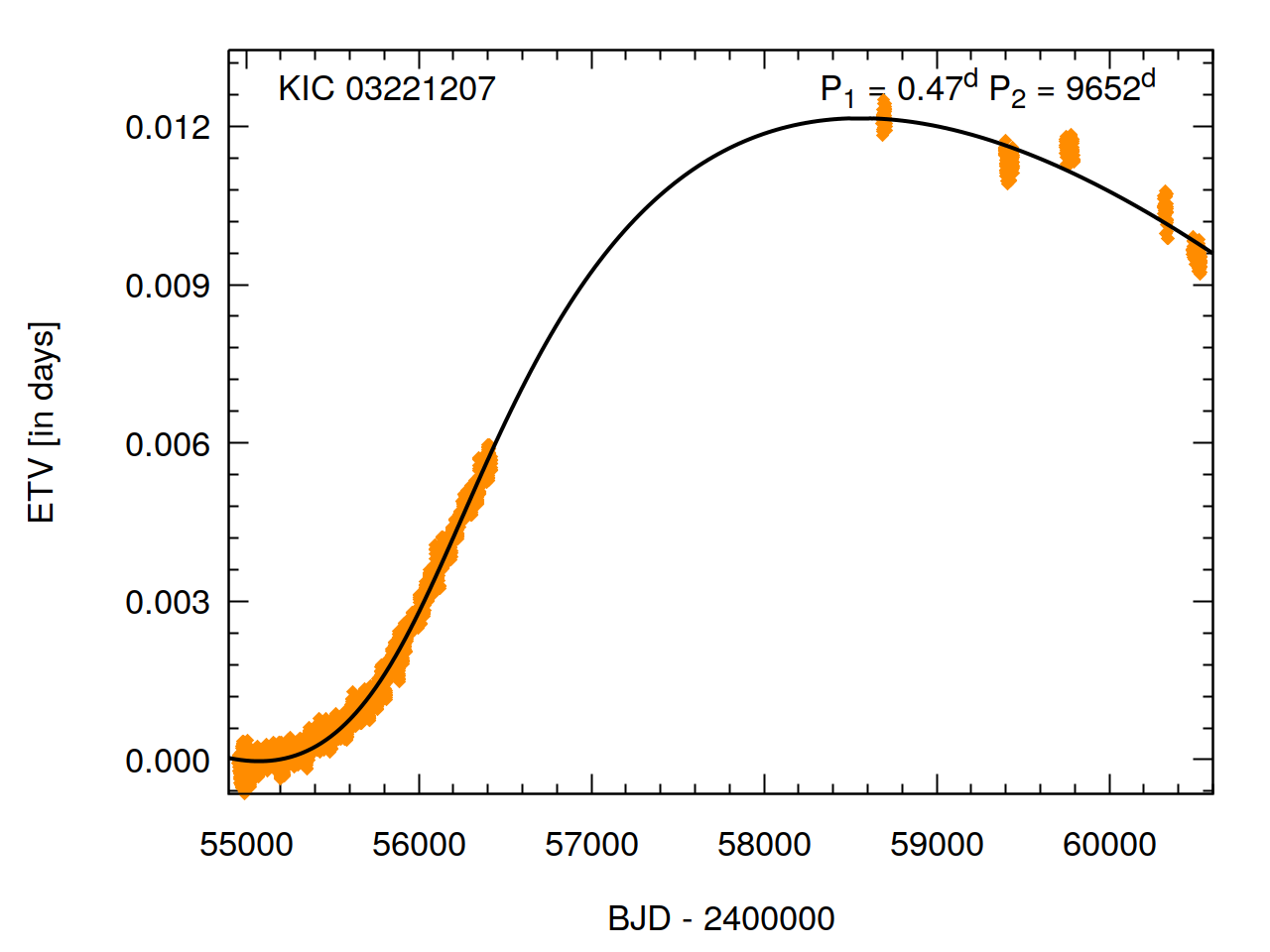}\includegraphics[width=60mm]{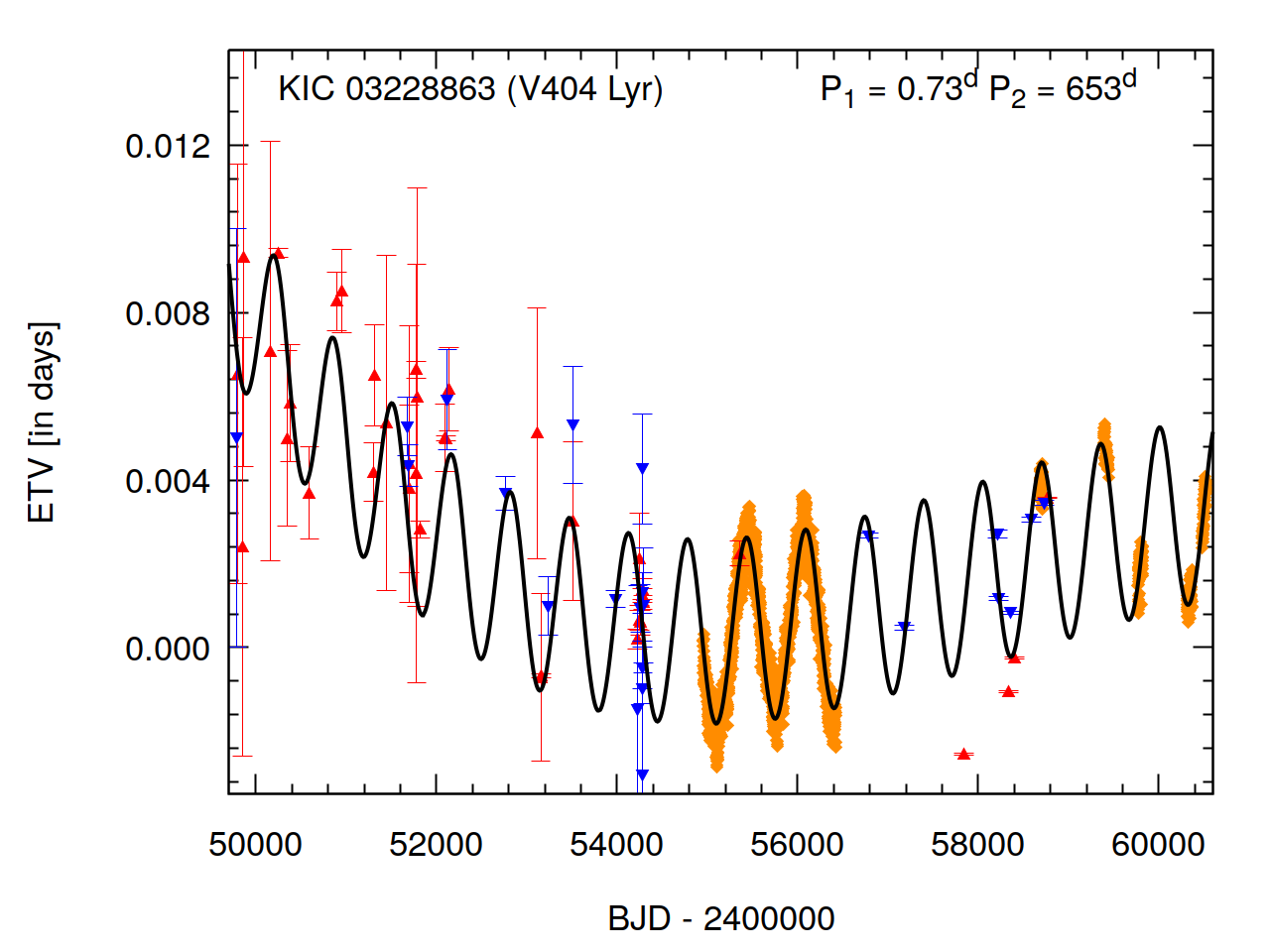}\includegraphics[width=60mm]{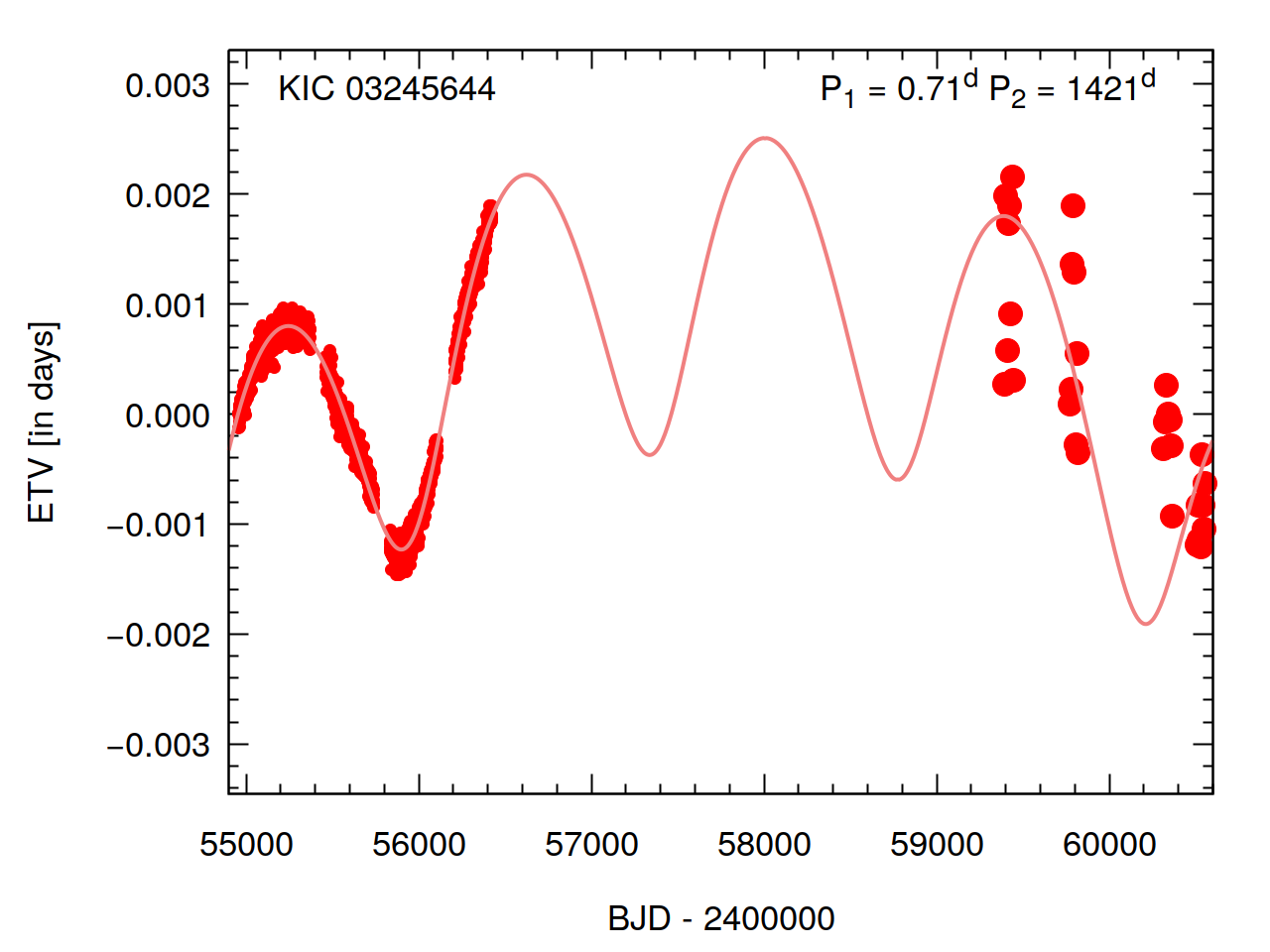}
\includegraphics[width=60mm]{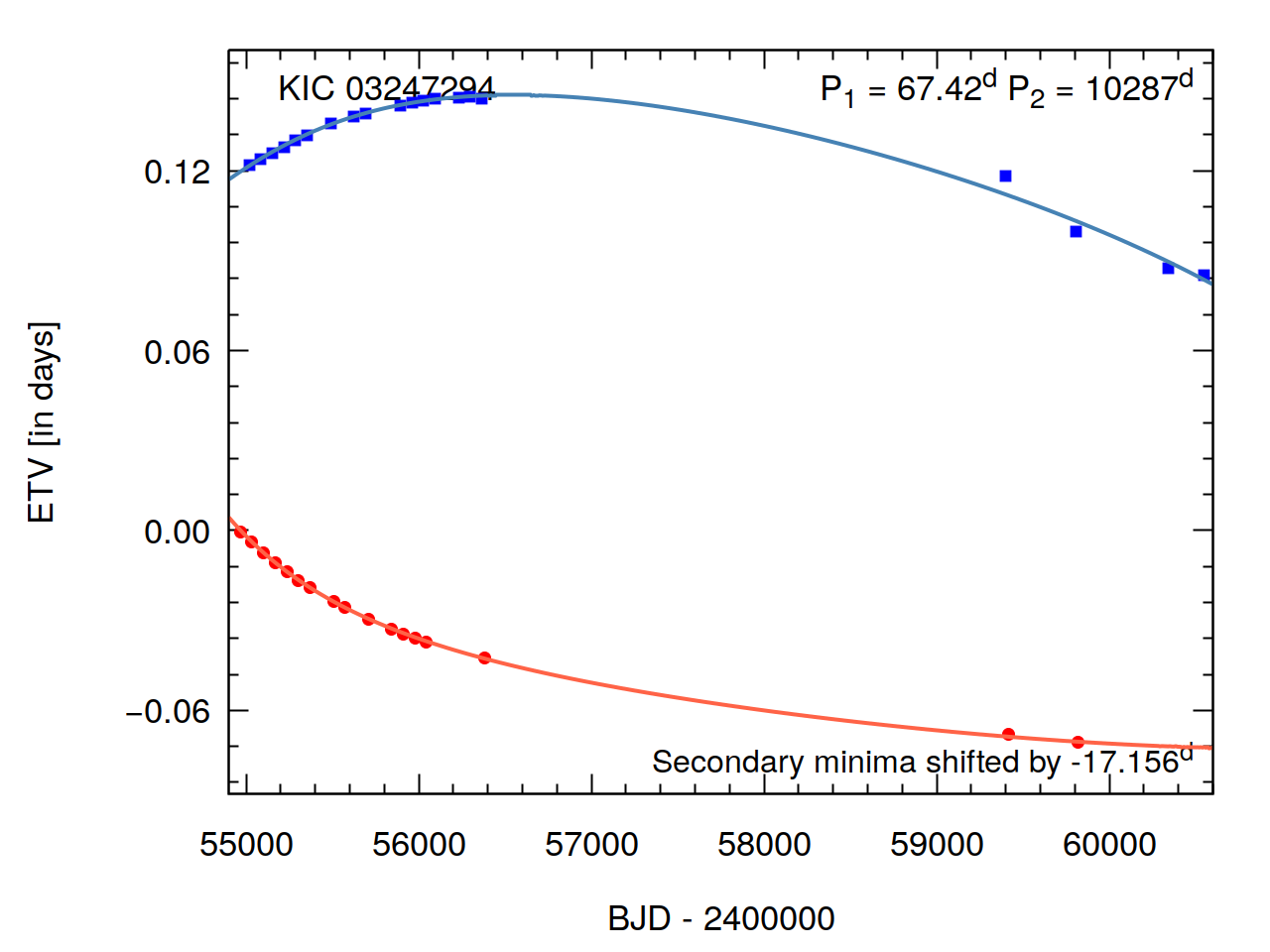}\includegraphics[width=60mm]{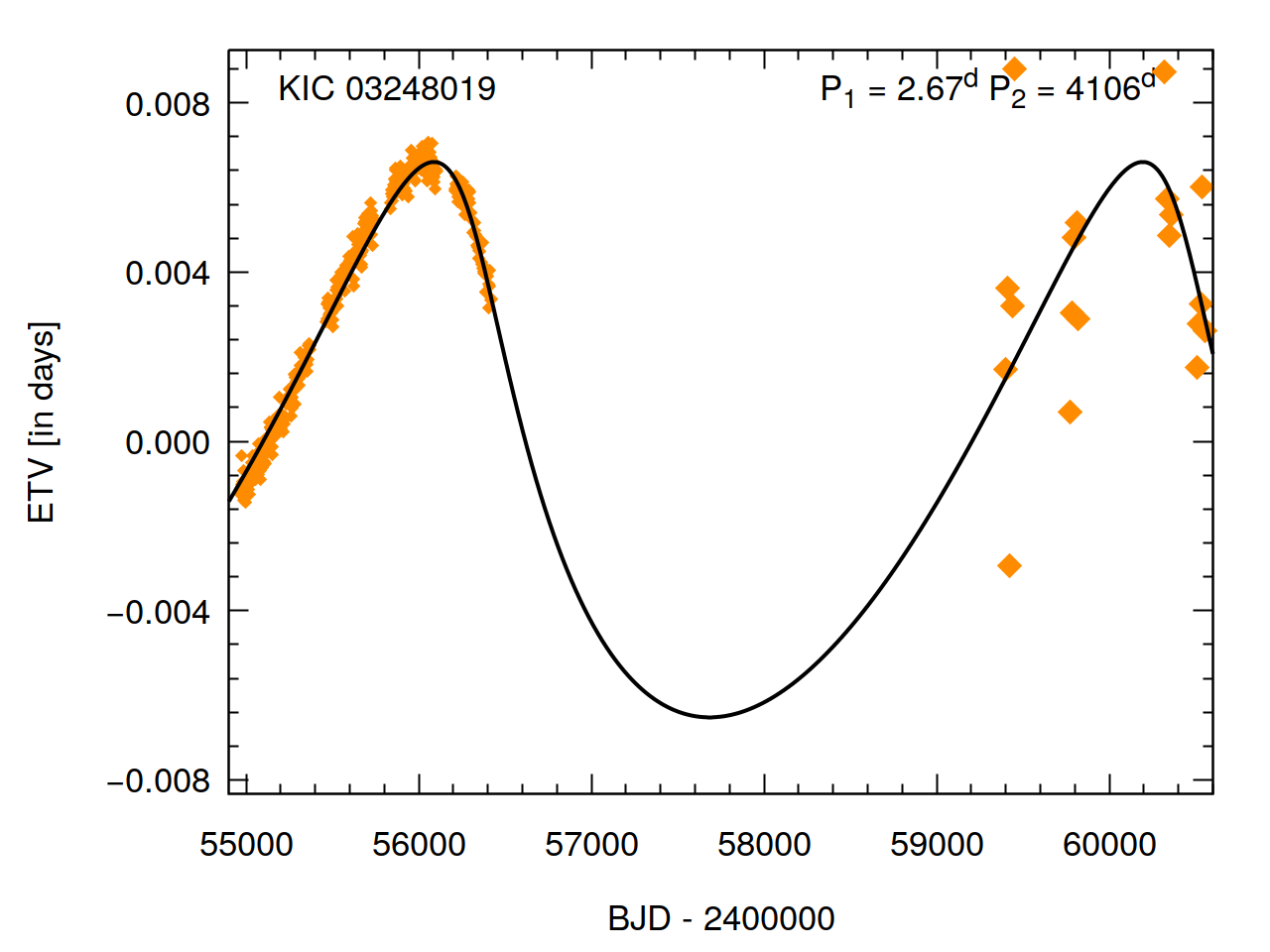}\includegraphics[width=60mm]{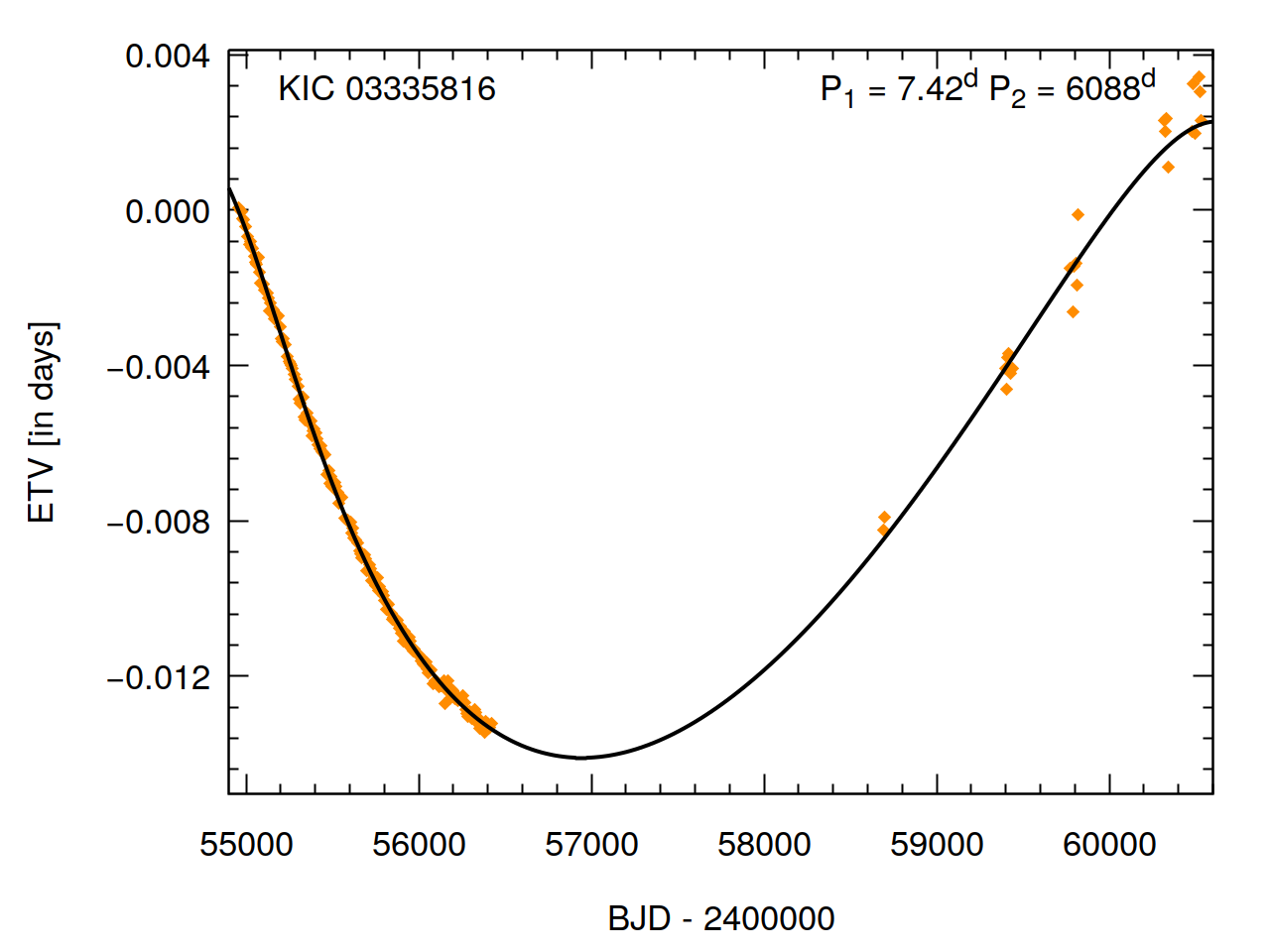}
\includegraphics[width=60mm]{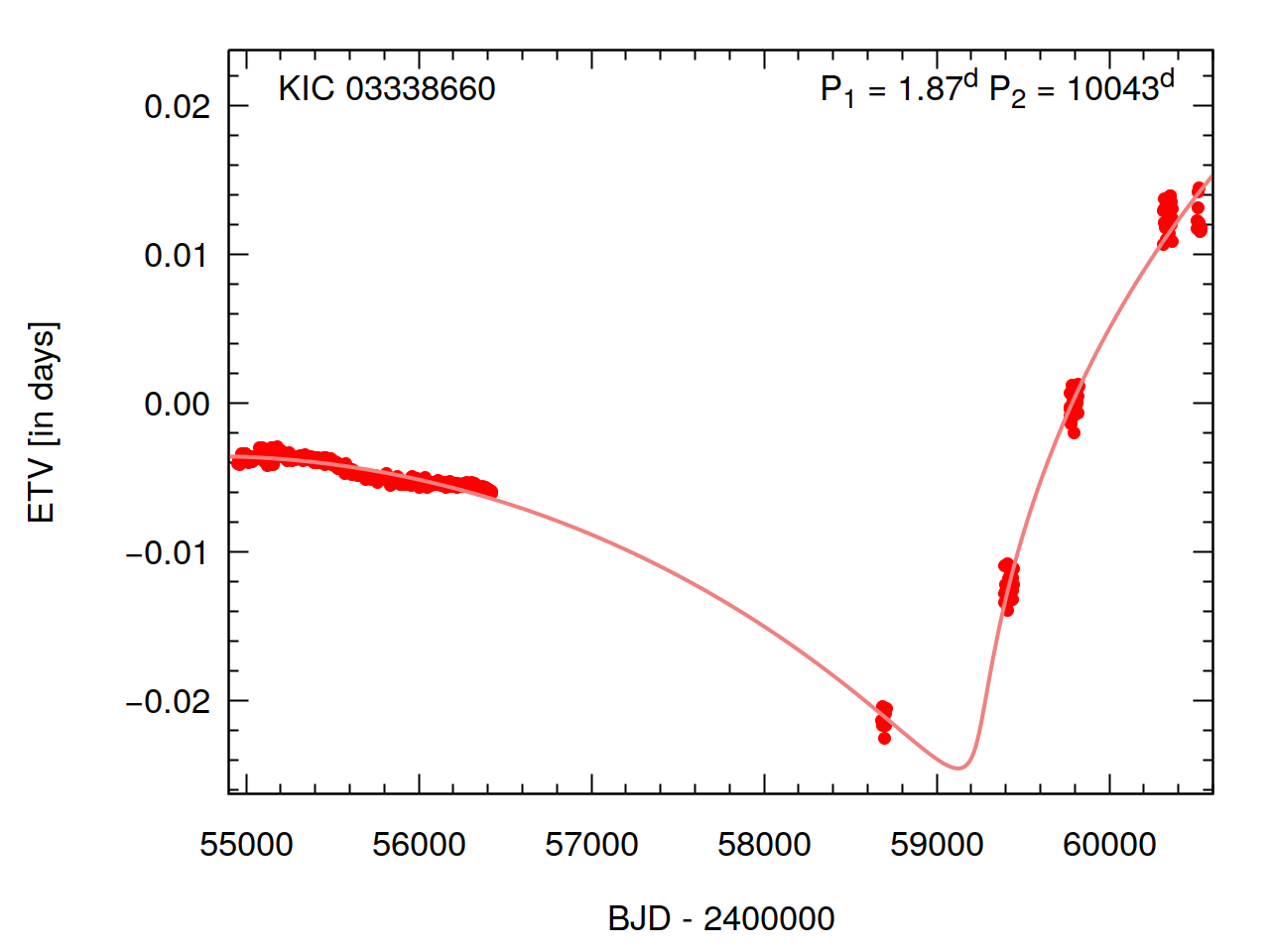}\includegraphics[width=60mm]{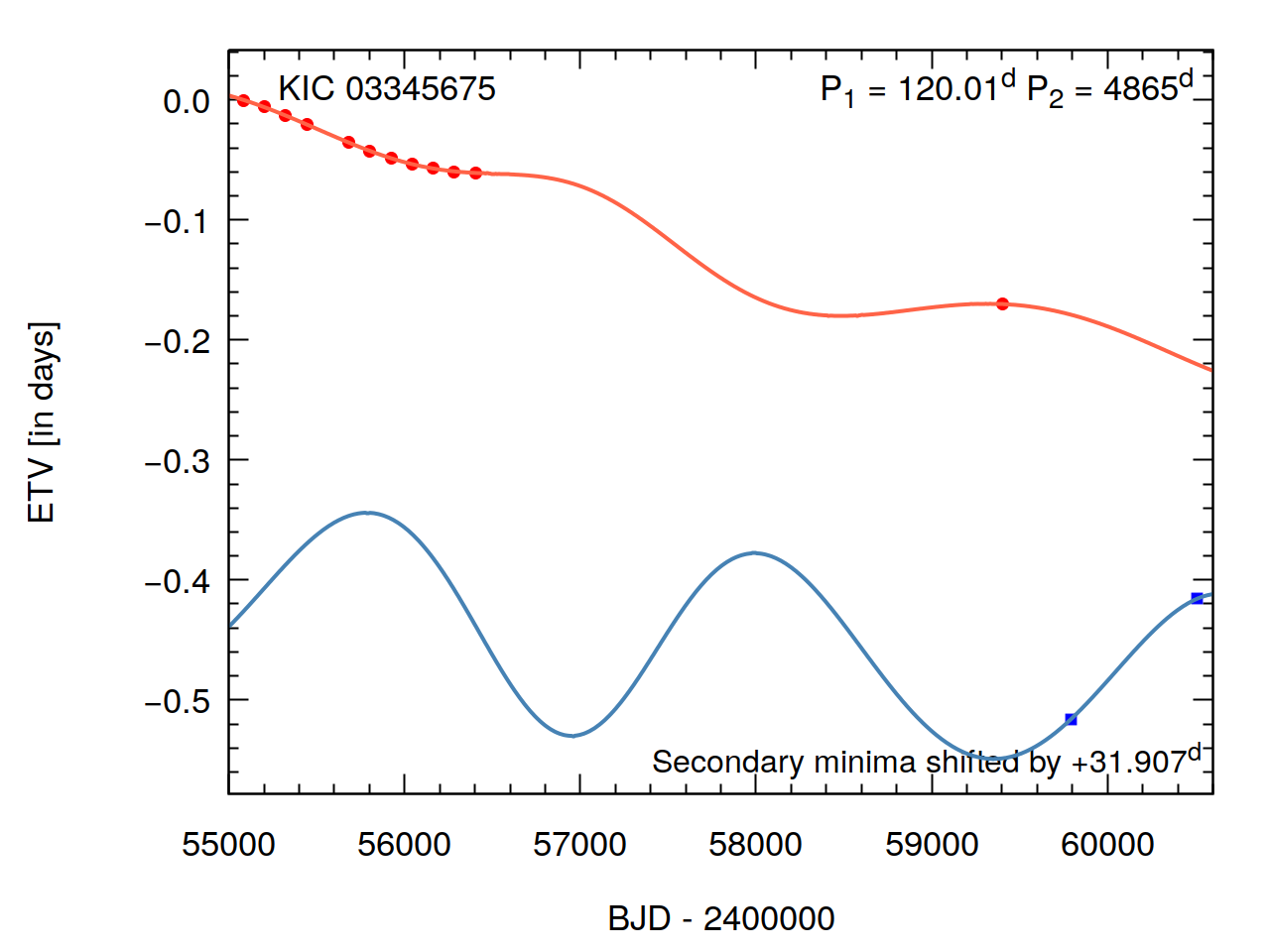}\includegraphics[width=60mm]{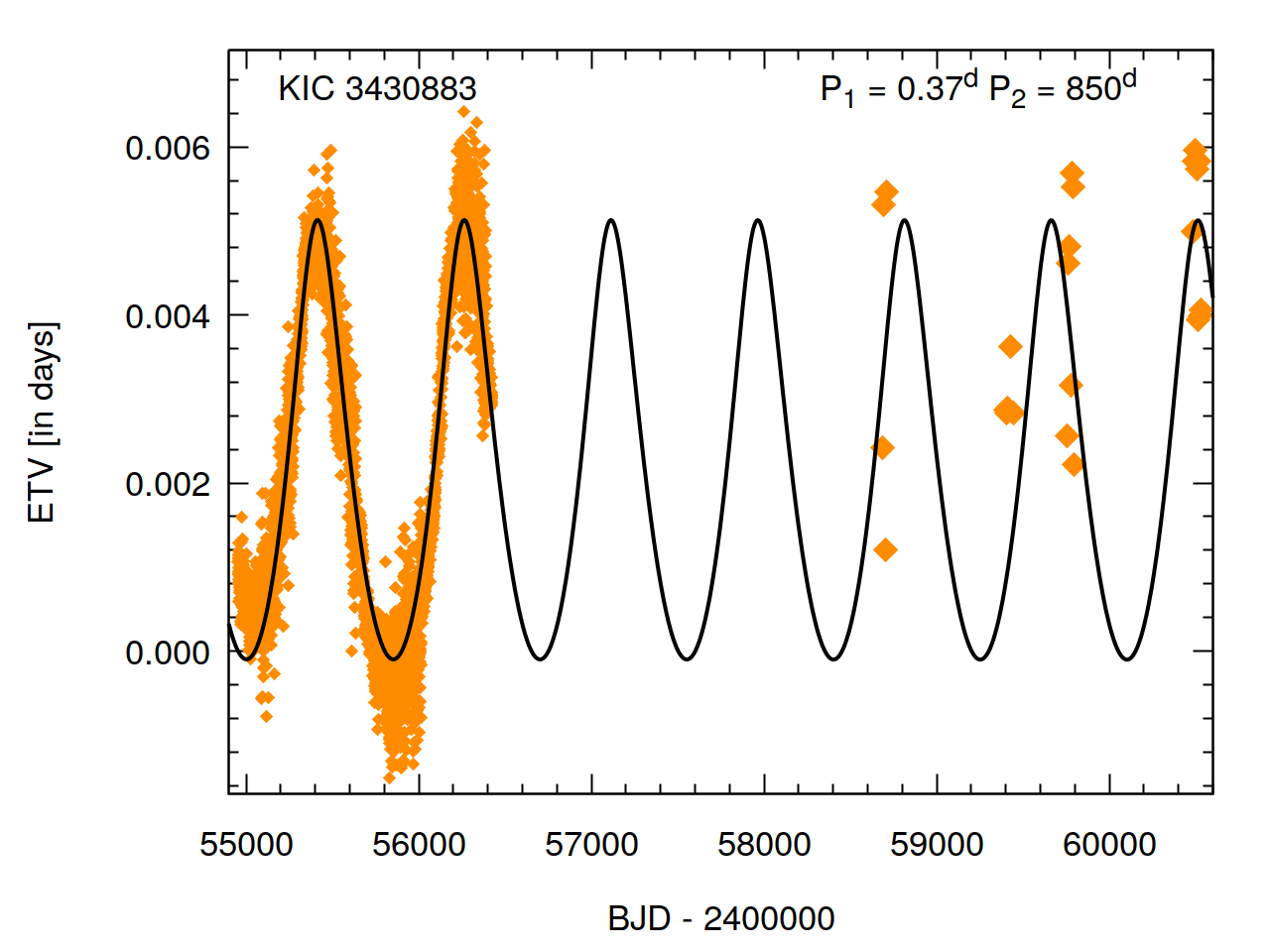}
\includegraphics[width=60mm]{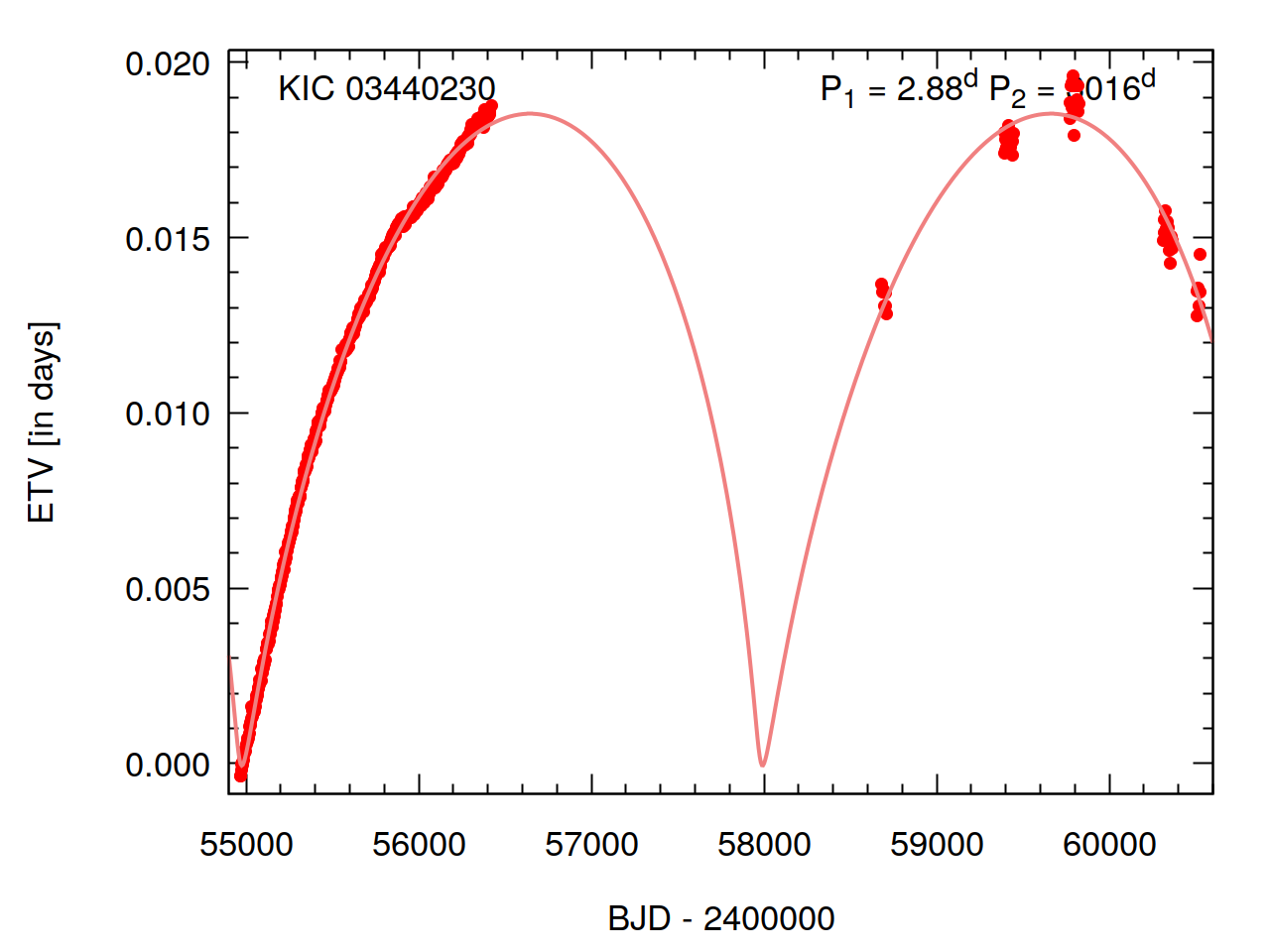}\includegraphics[width=60mm]{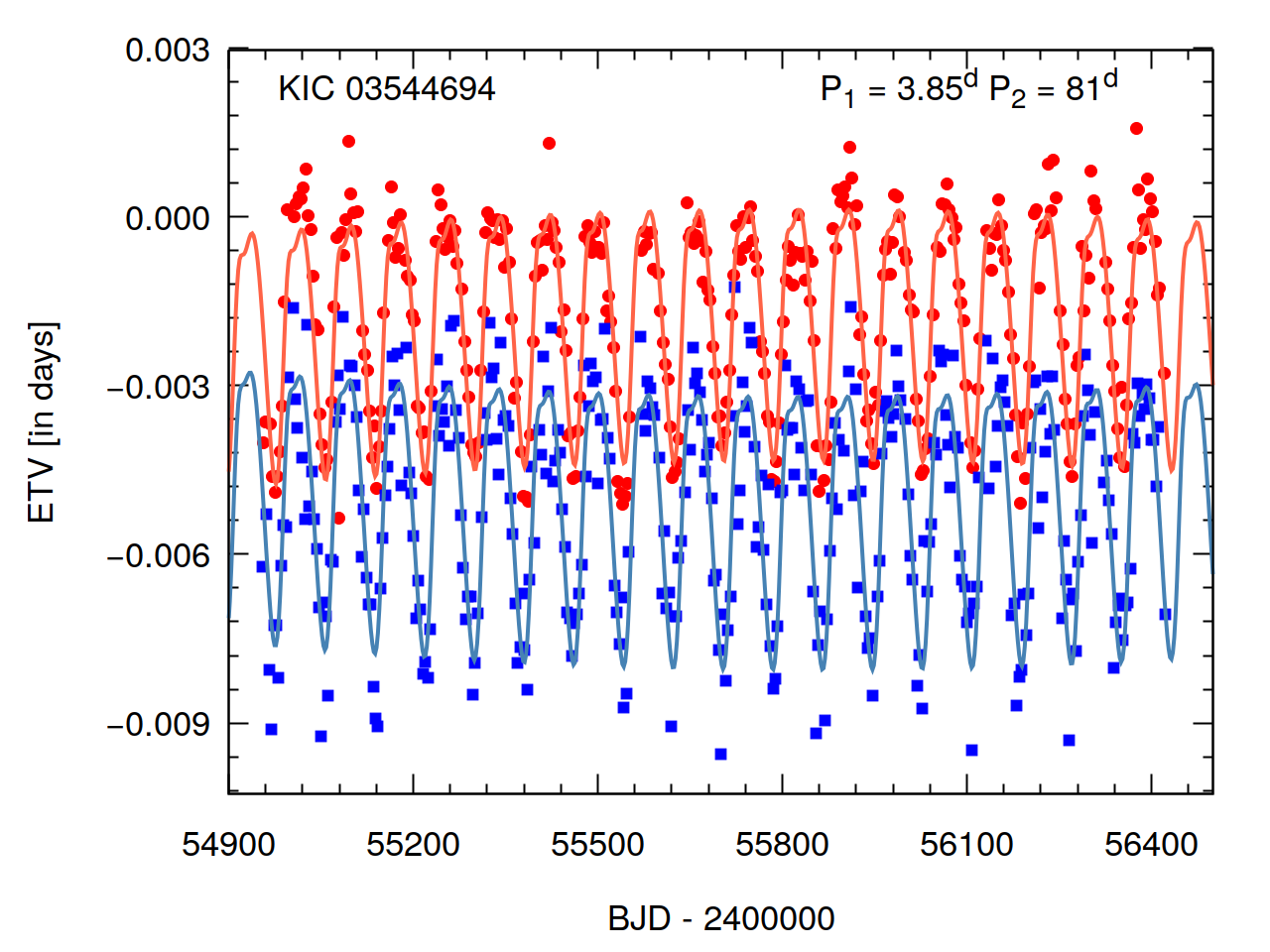}\includegraphics[width=60mm]{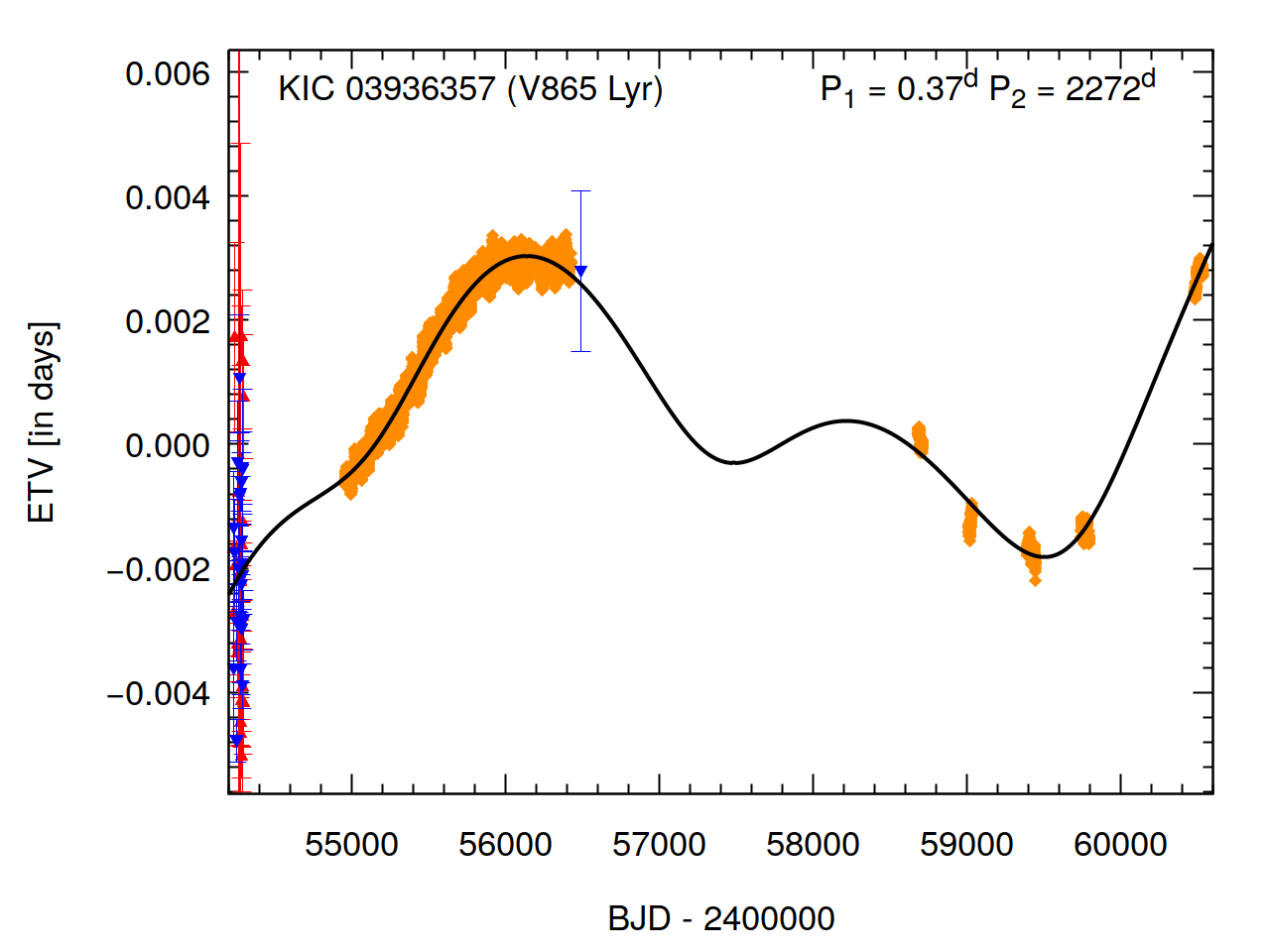}
\includegraphics[width=60mm]{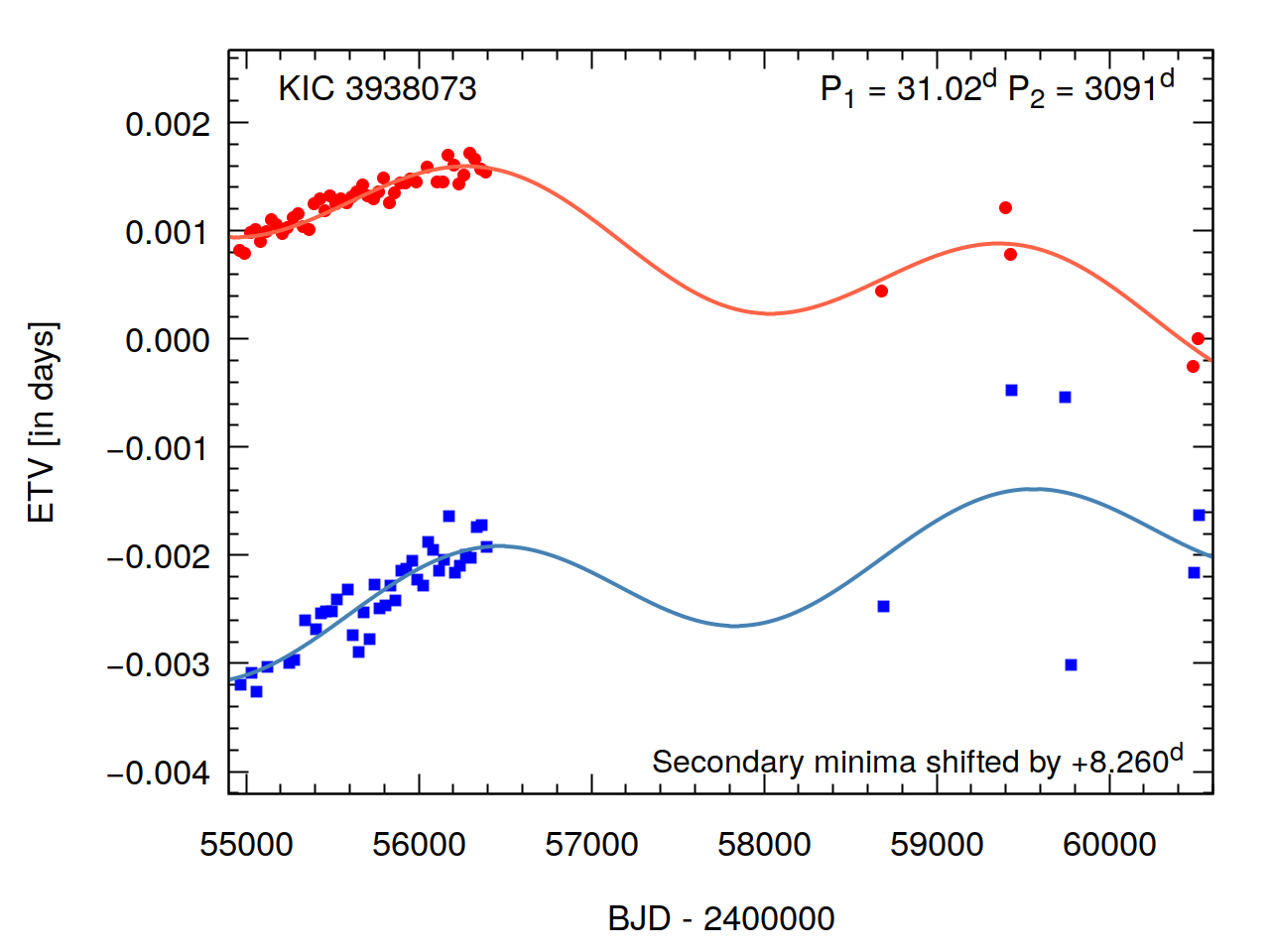}\includegraphics[width=60mm]{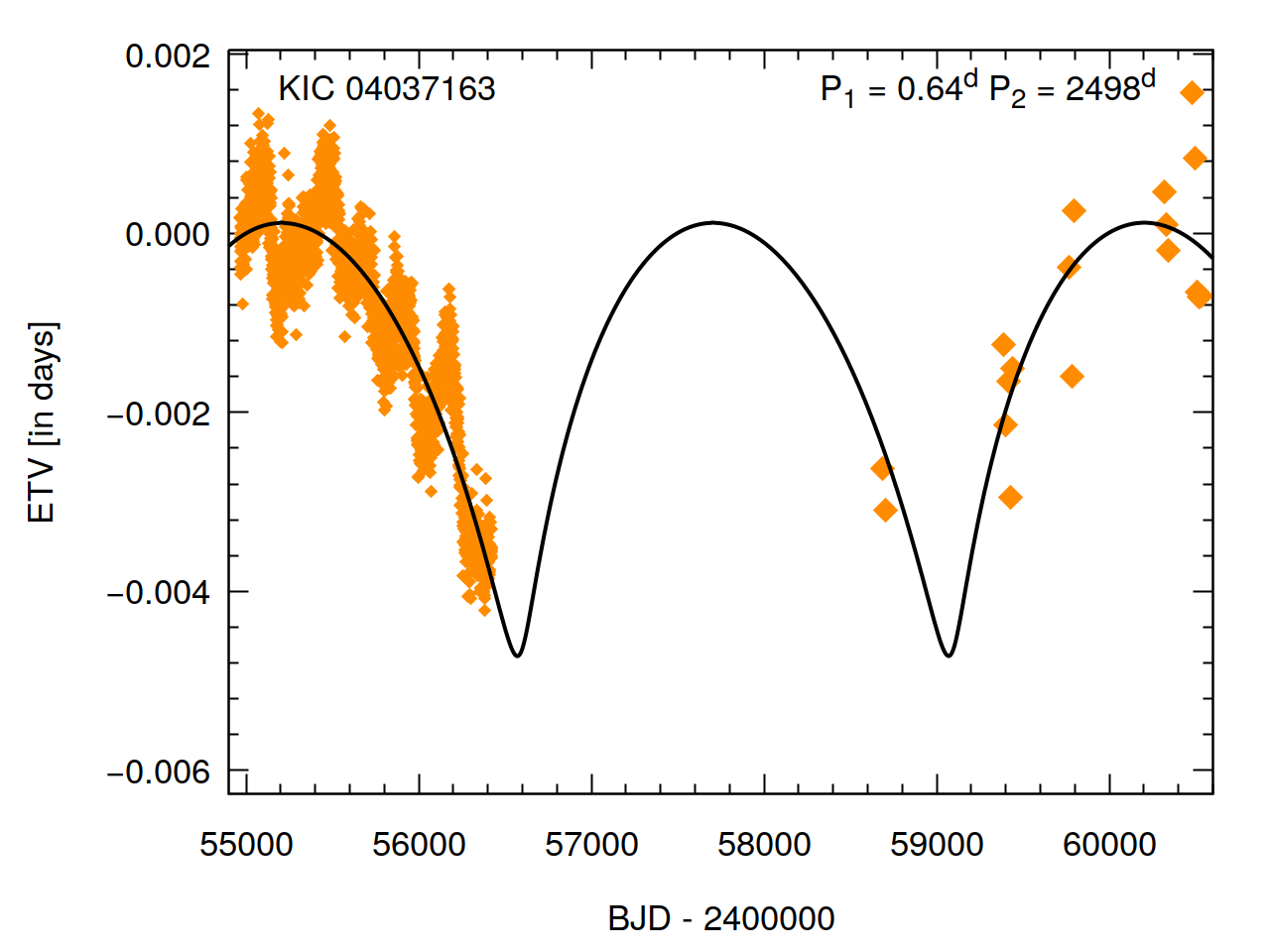}\includegraphics[width=60mm]{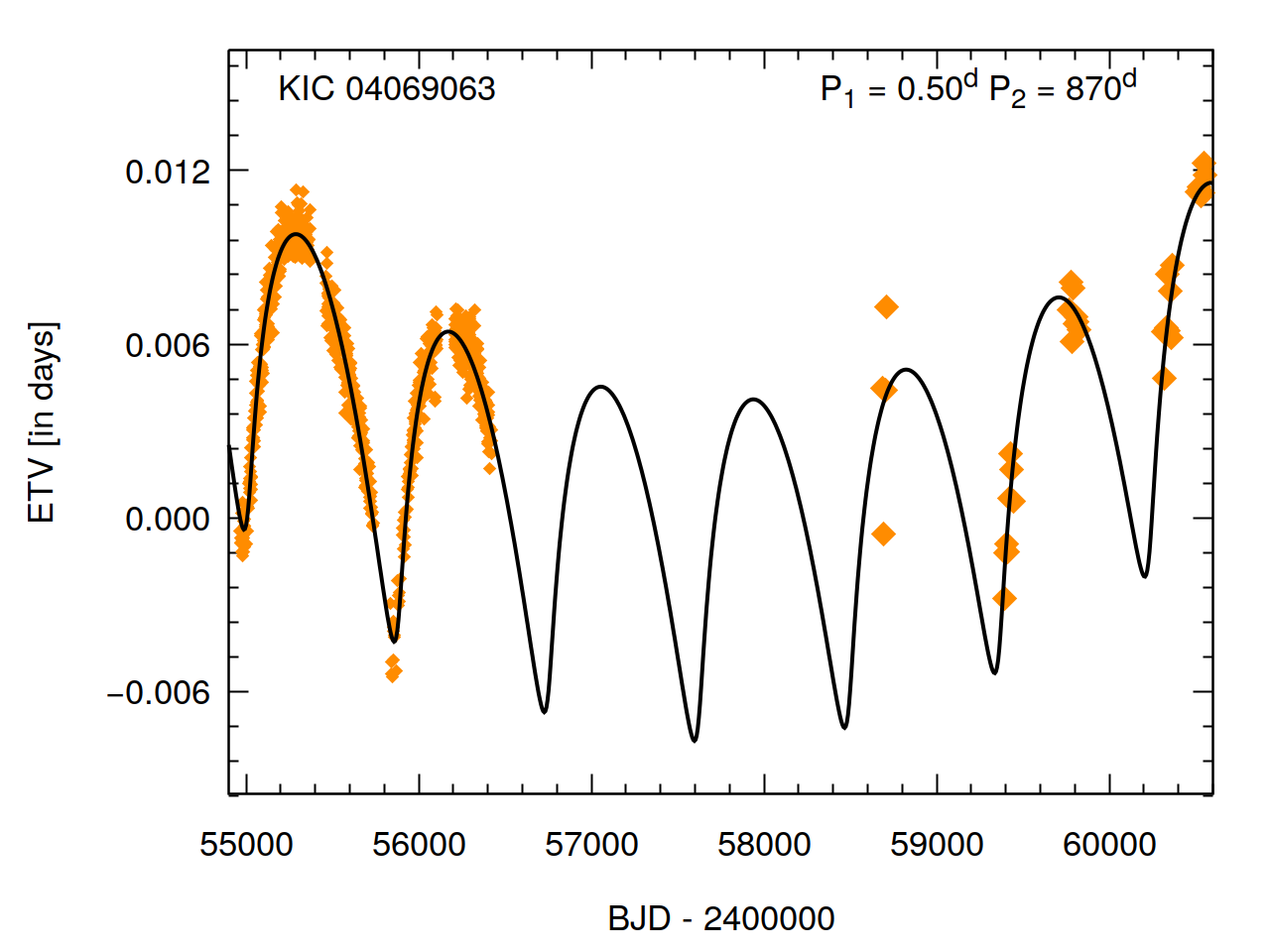}
\caption{continued}
\end{figure*}

\addtocounter{figure}{-1}

\begin{figure*}
\includegraphics[width=60mm]{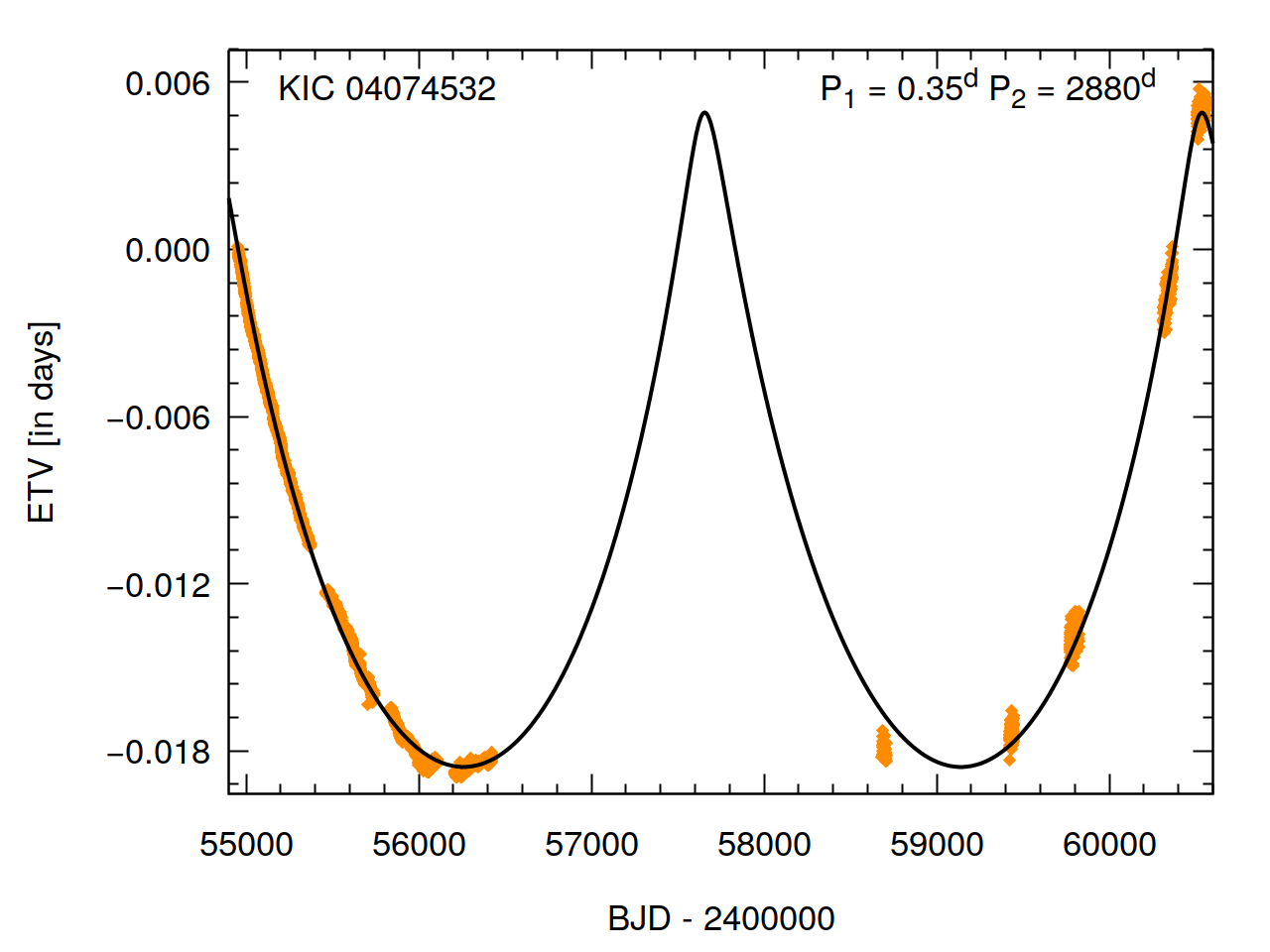}\includegraphics[width=60mm]{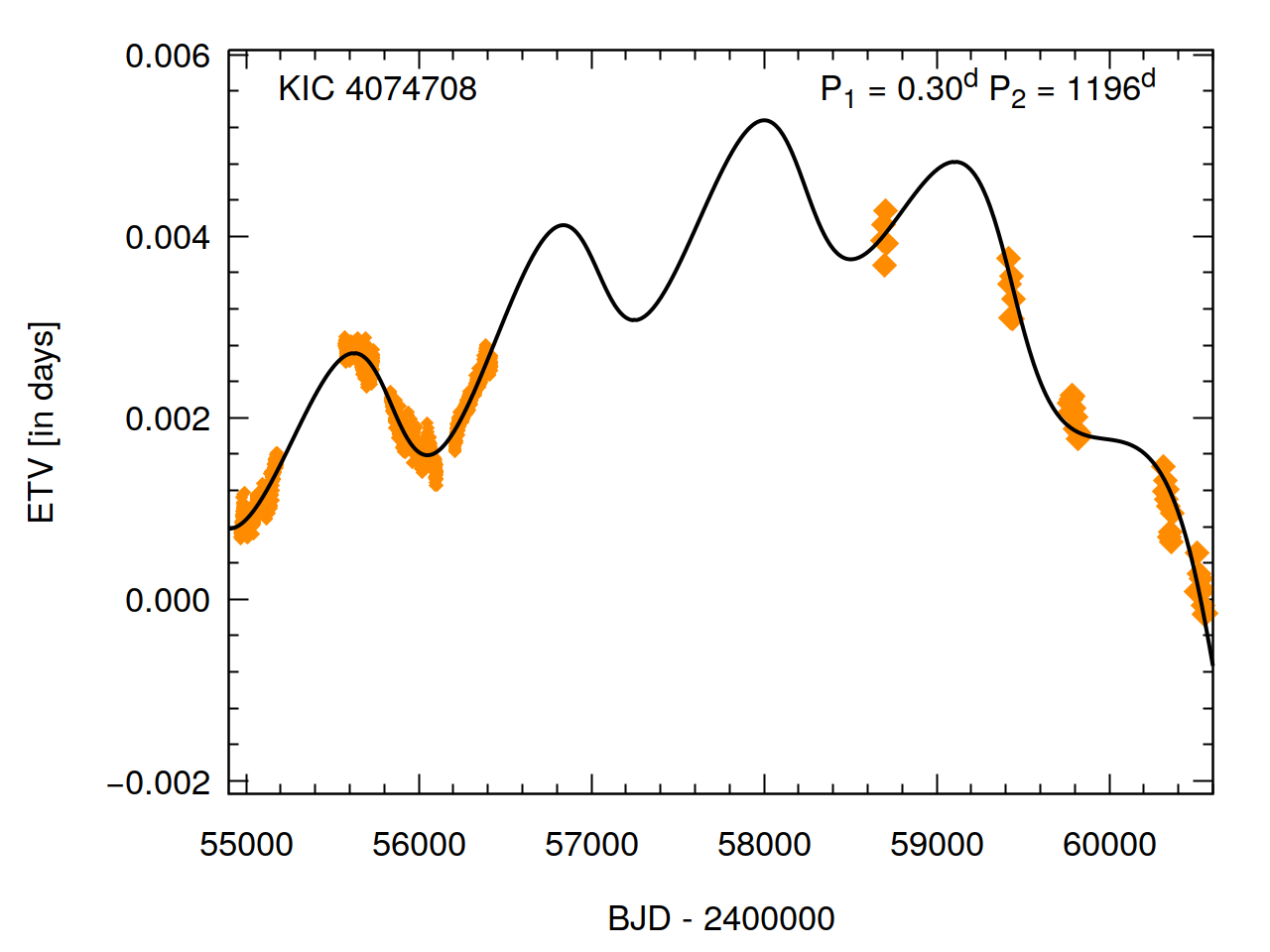}\includegraphics[width=60mm]{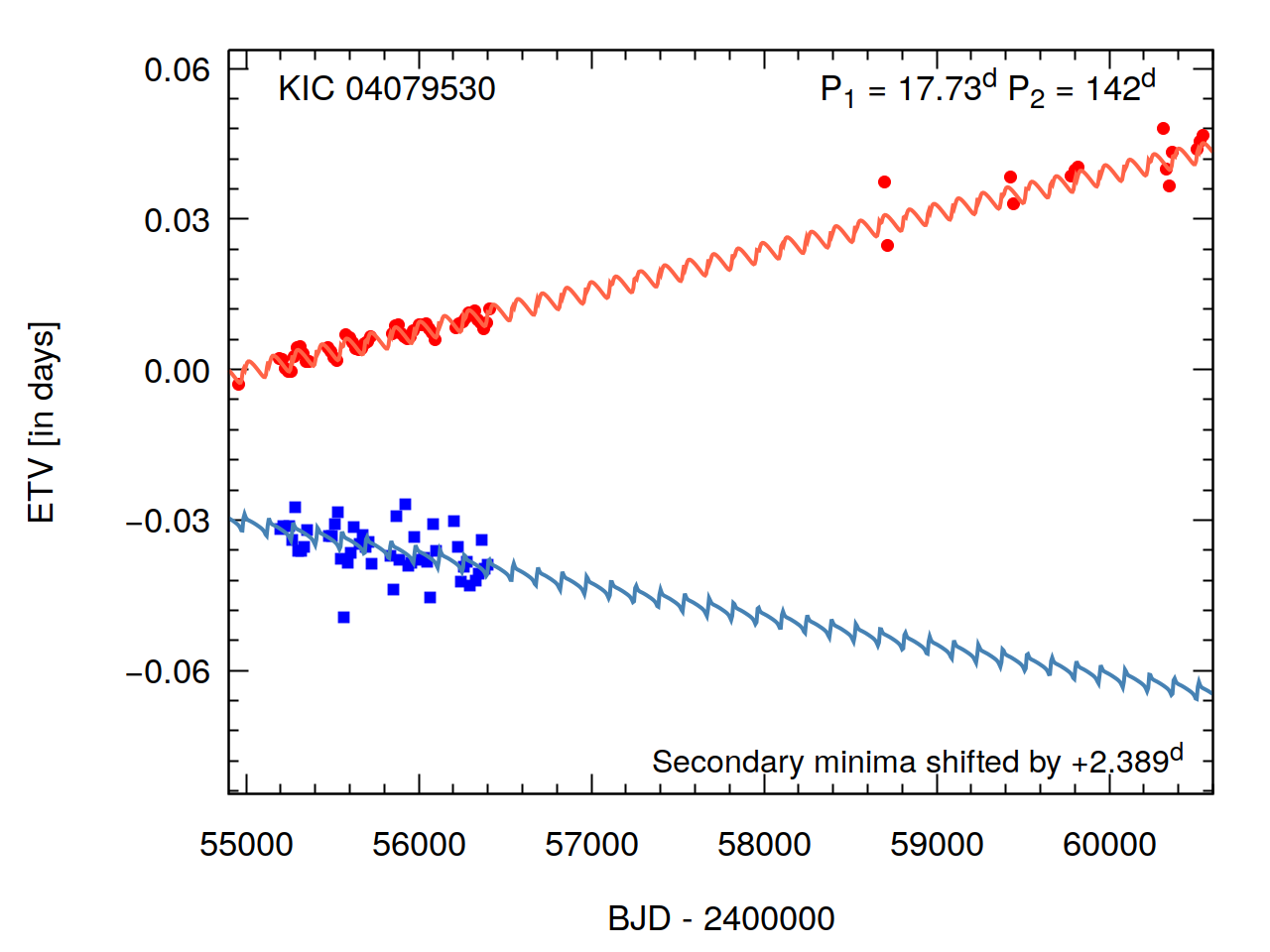}
\includegraphics[width=60mm]{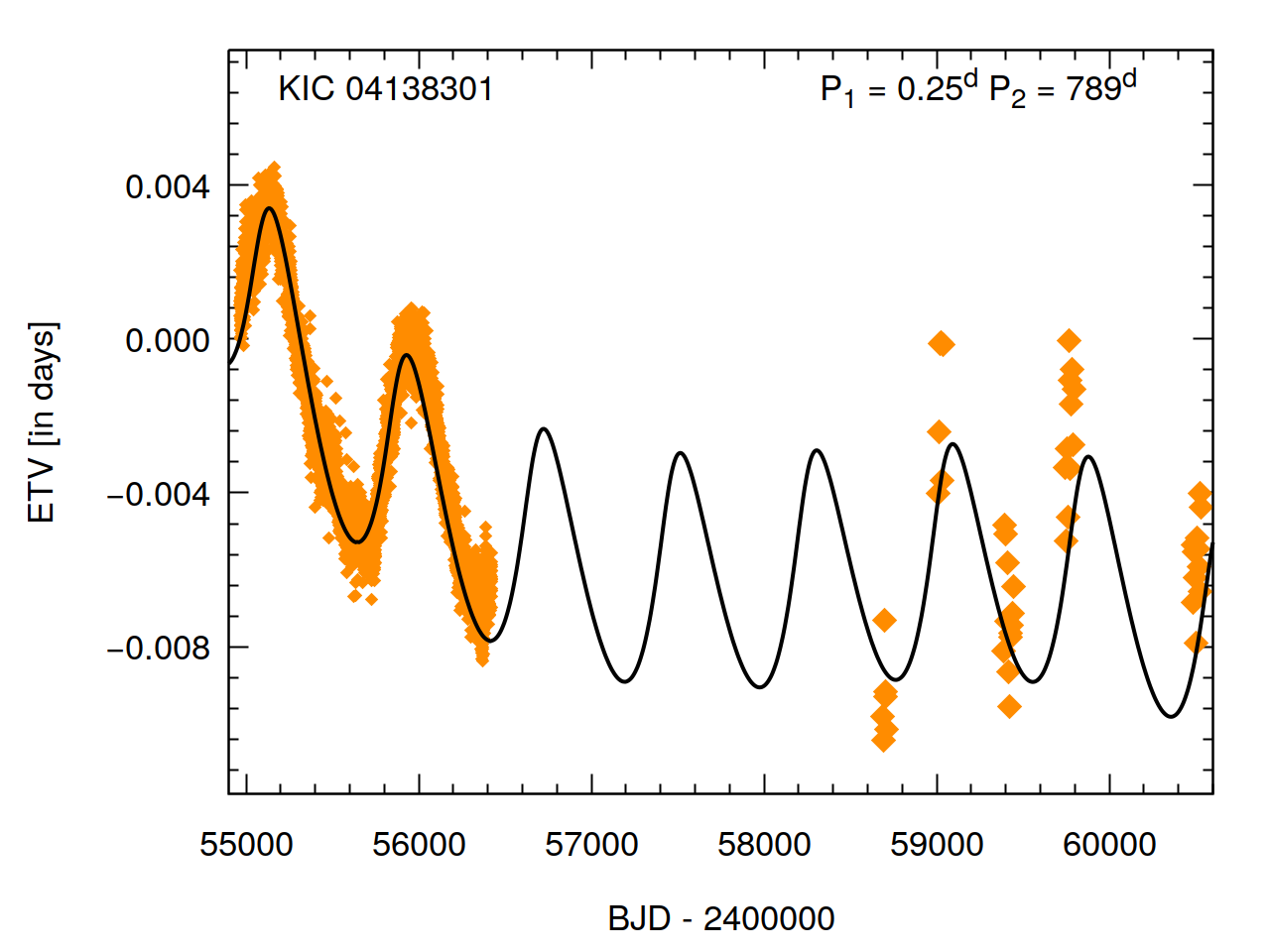}\includegraphics[width=60mm]{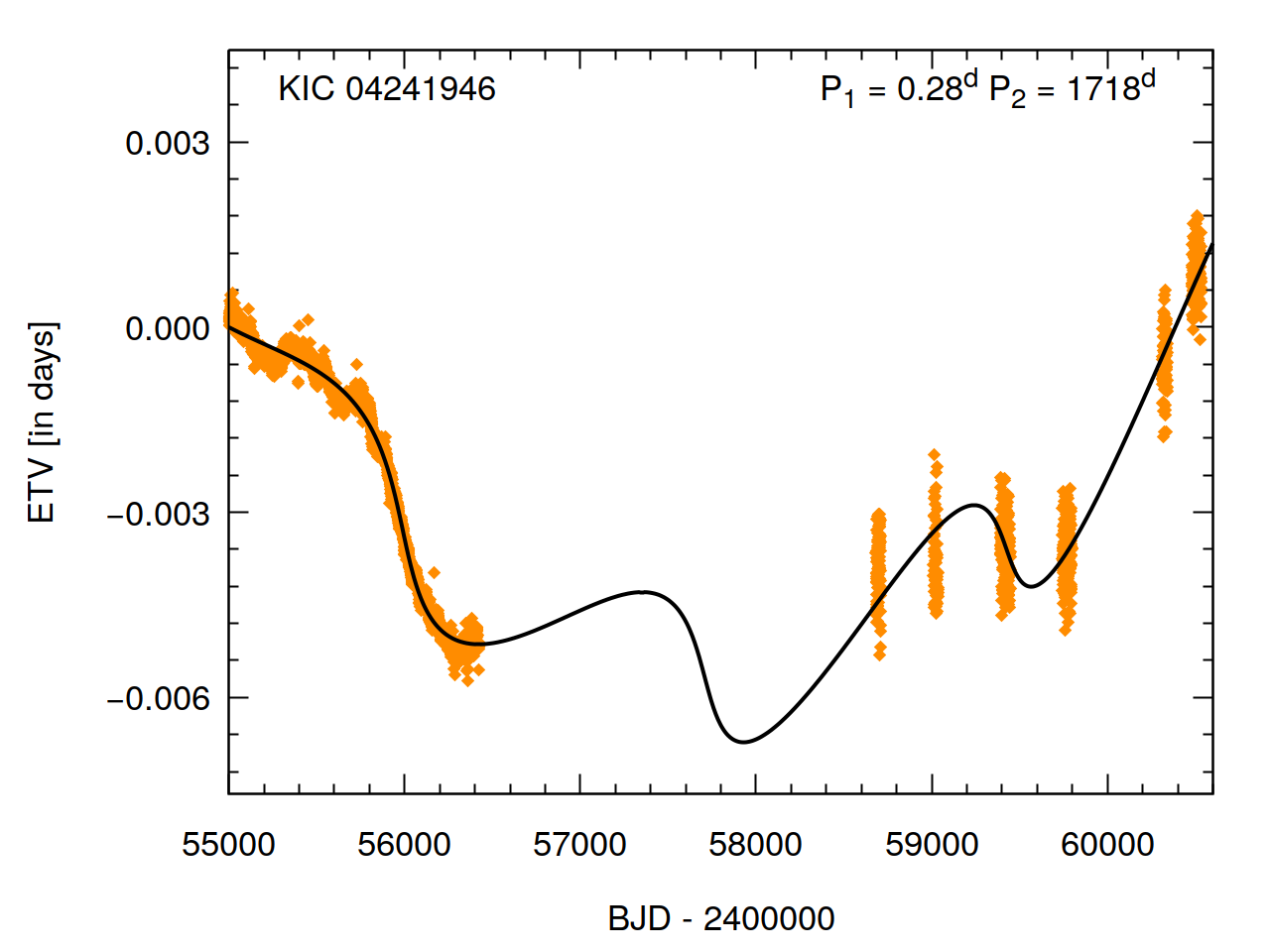}\includegraphics[width=60mm]{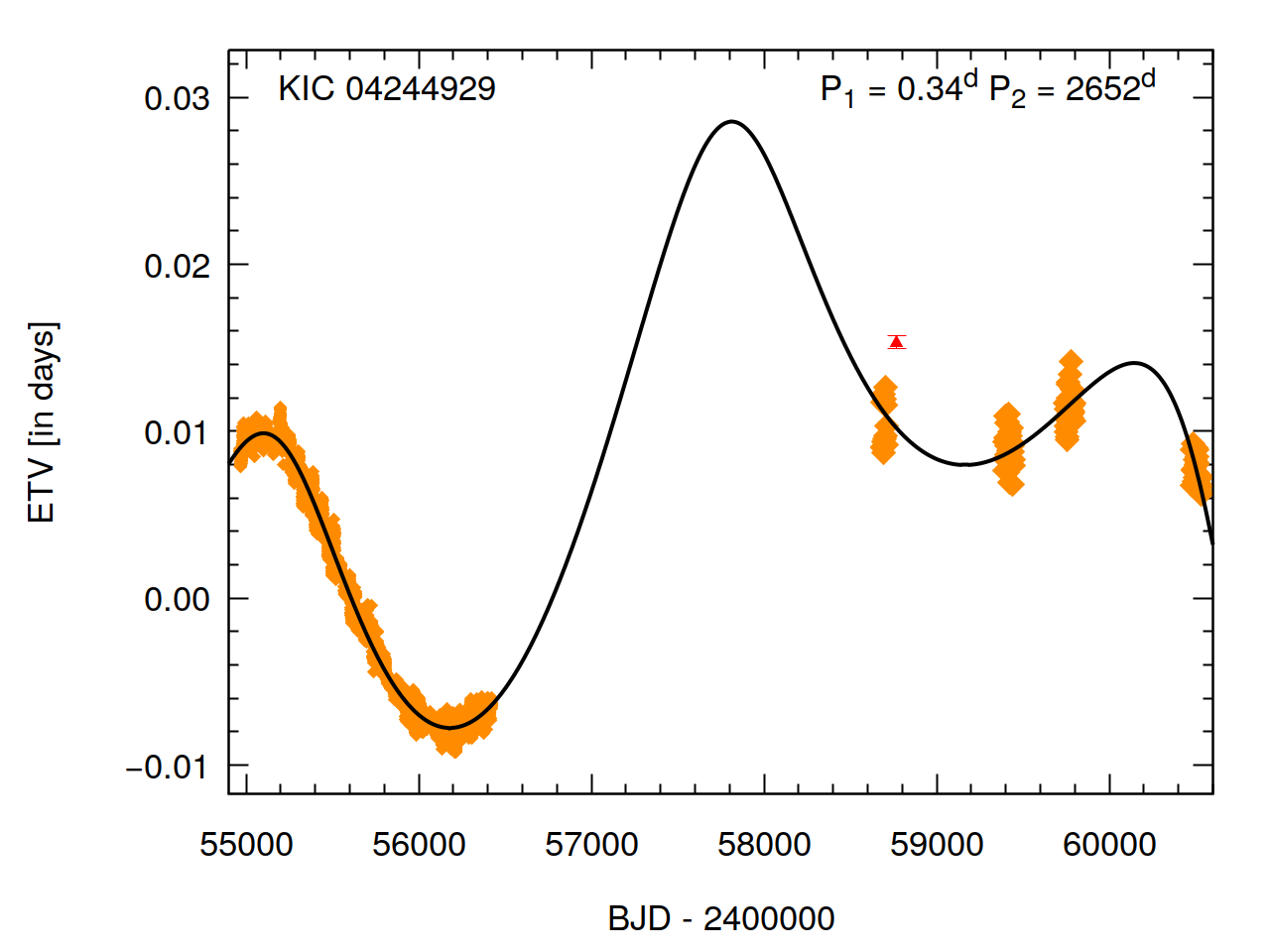}
\includegraphics[width=60mm]{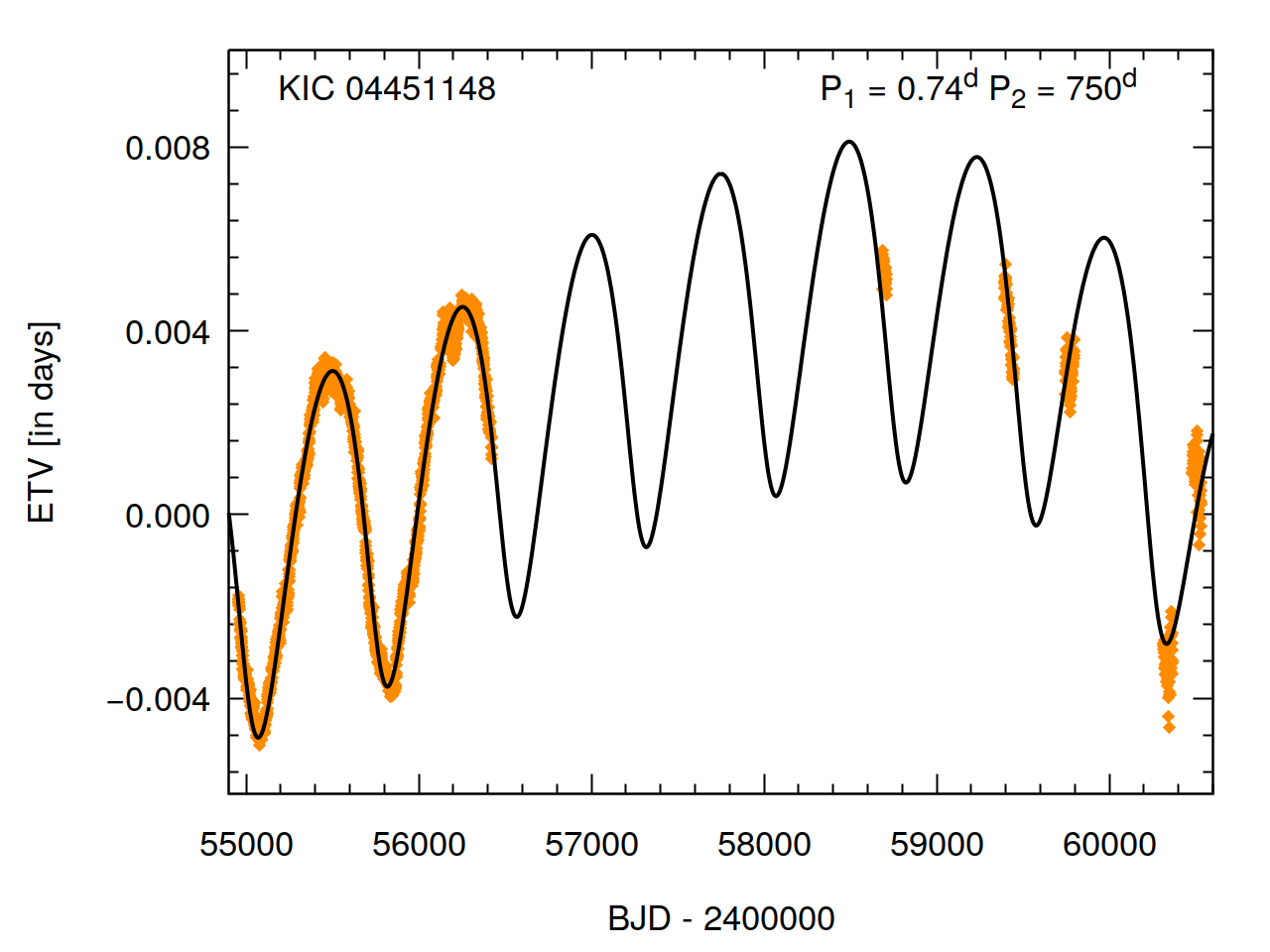}\includegraphics[width=60mm]{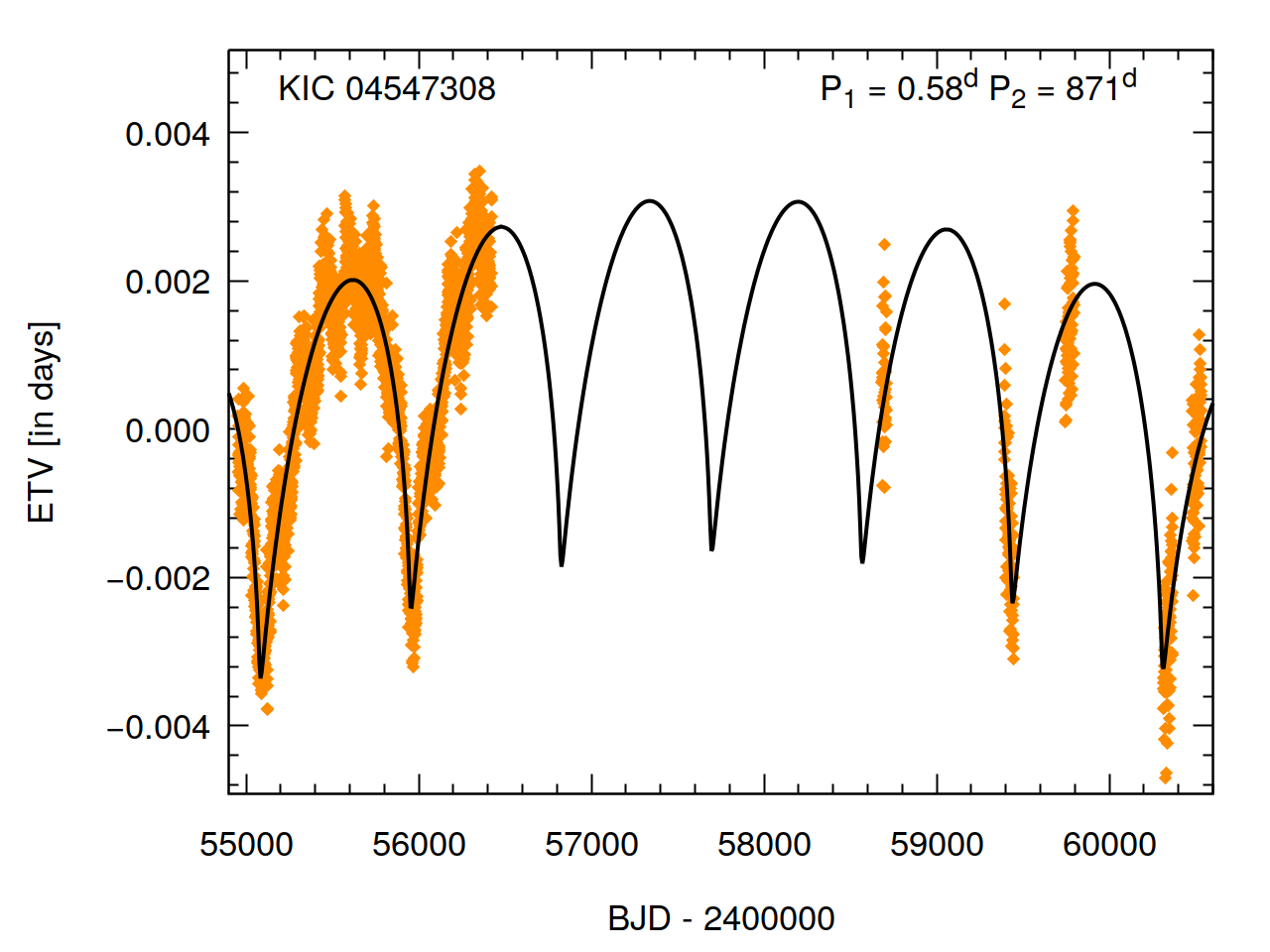}\includegraphics[width=60mm]{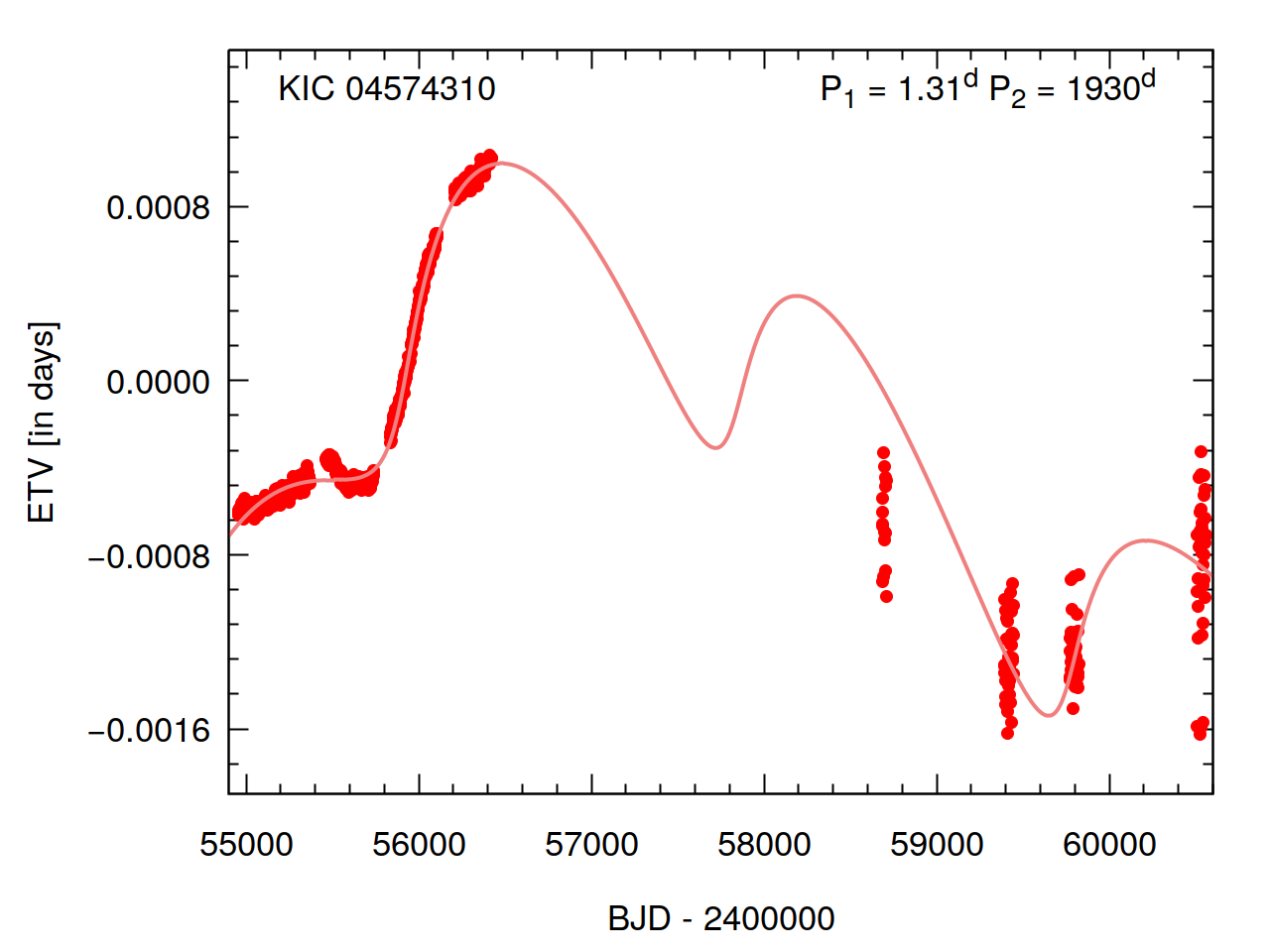}
\includegraphics[width=60mm]{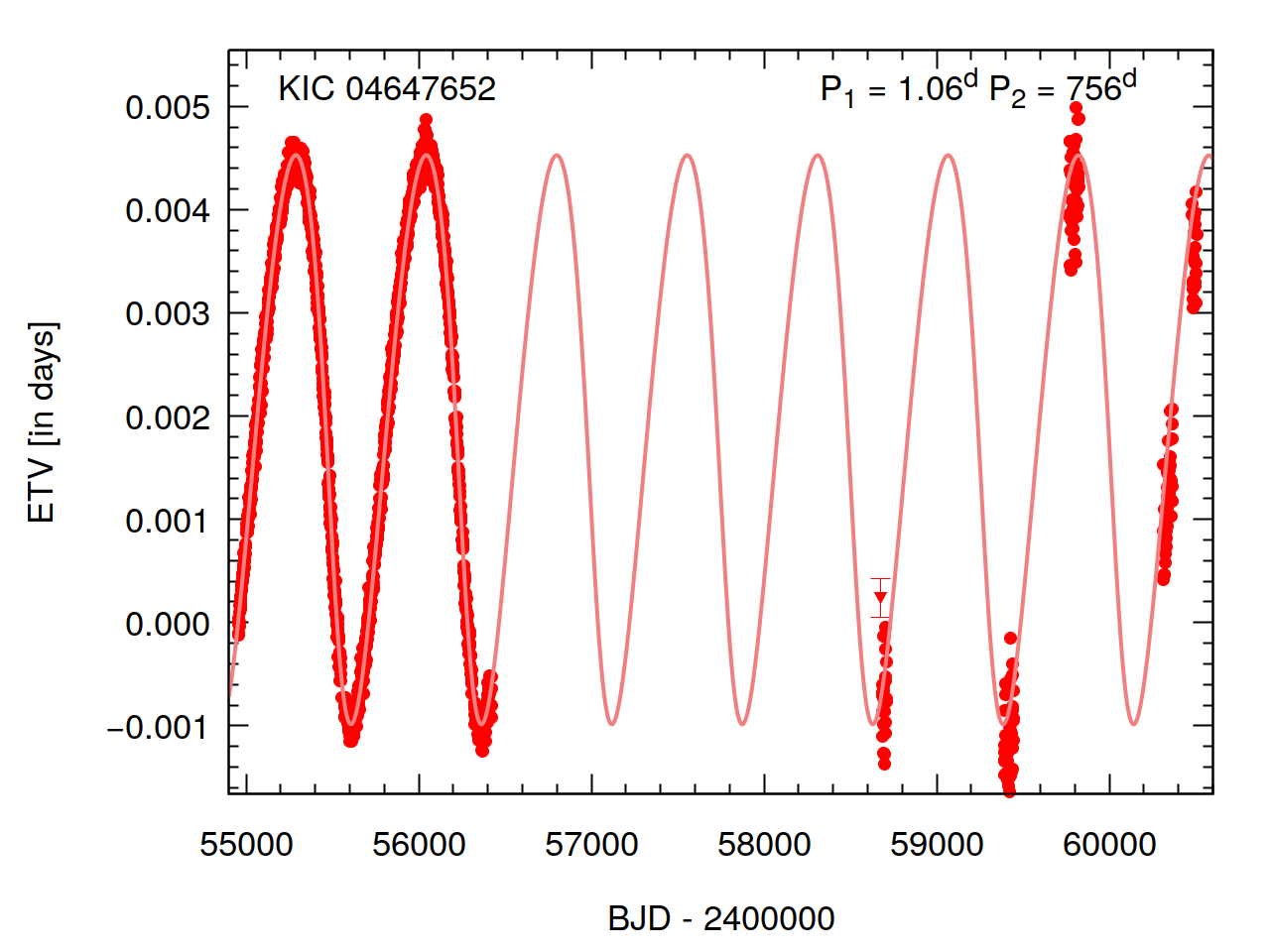}\includegraphics[width=60mm]{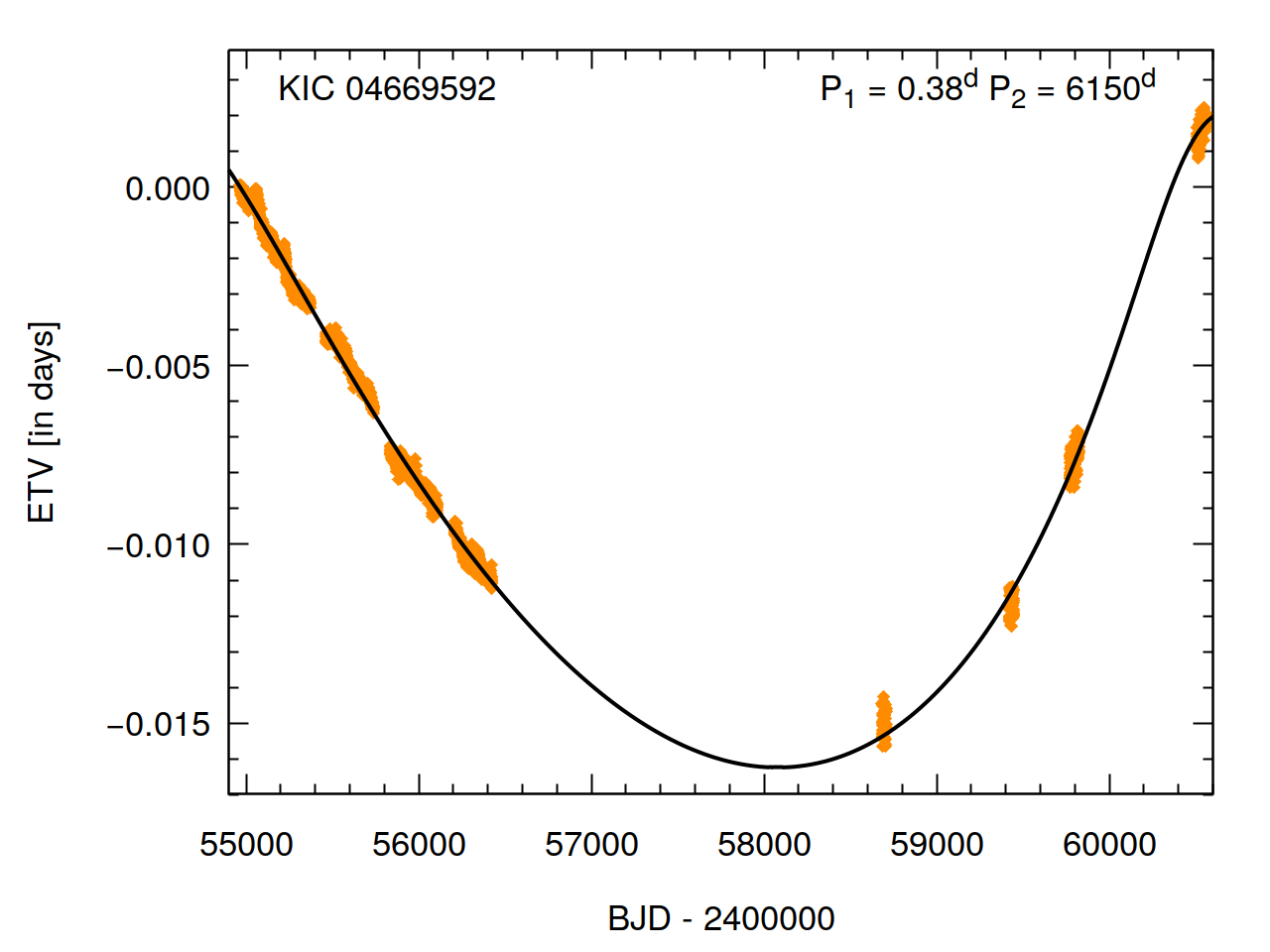}\includegraphics[width=60mm]{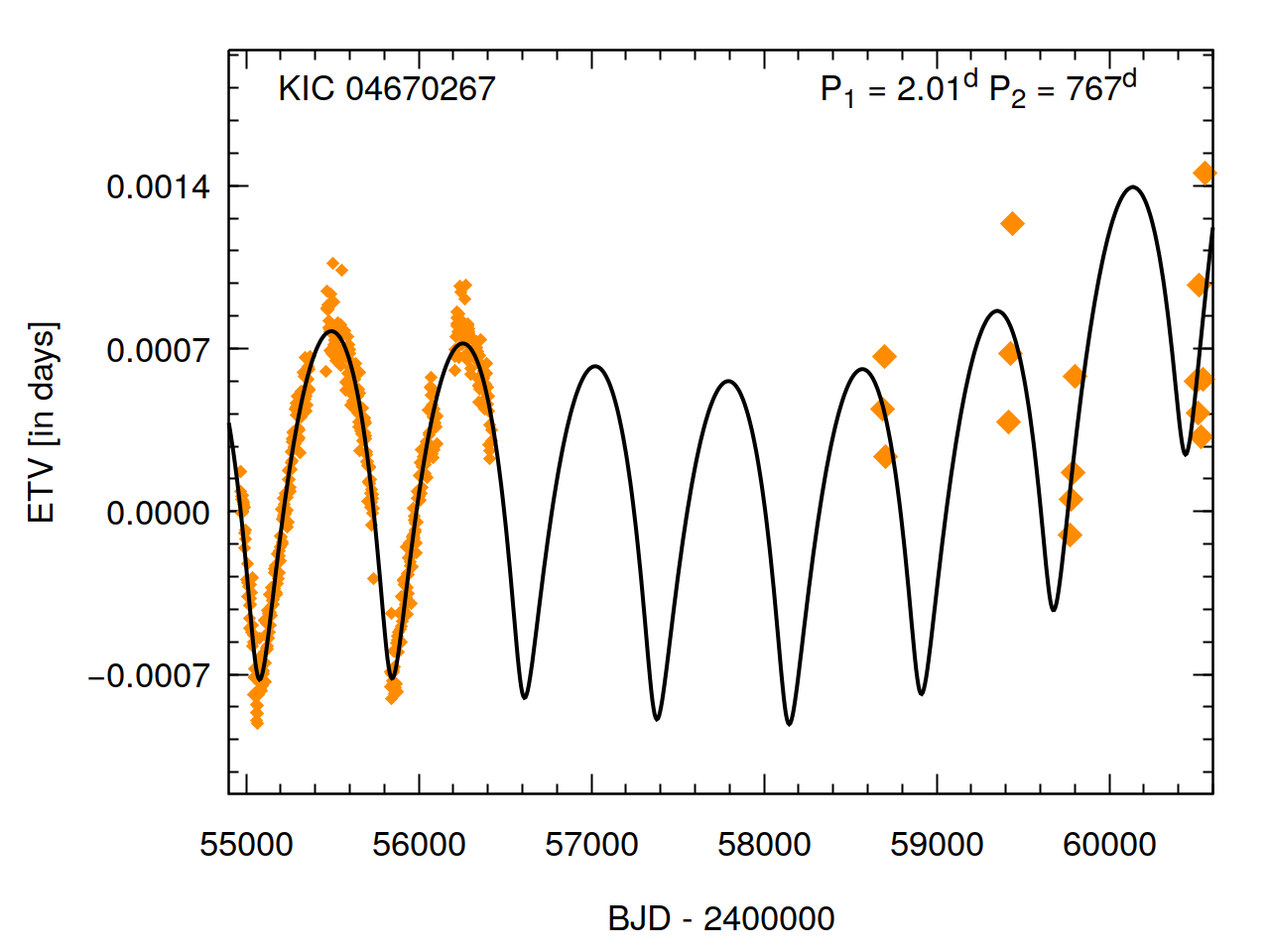}
\includegraphics[width=60mm]{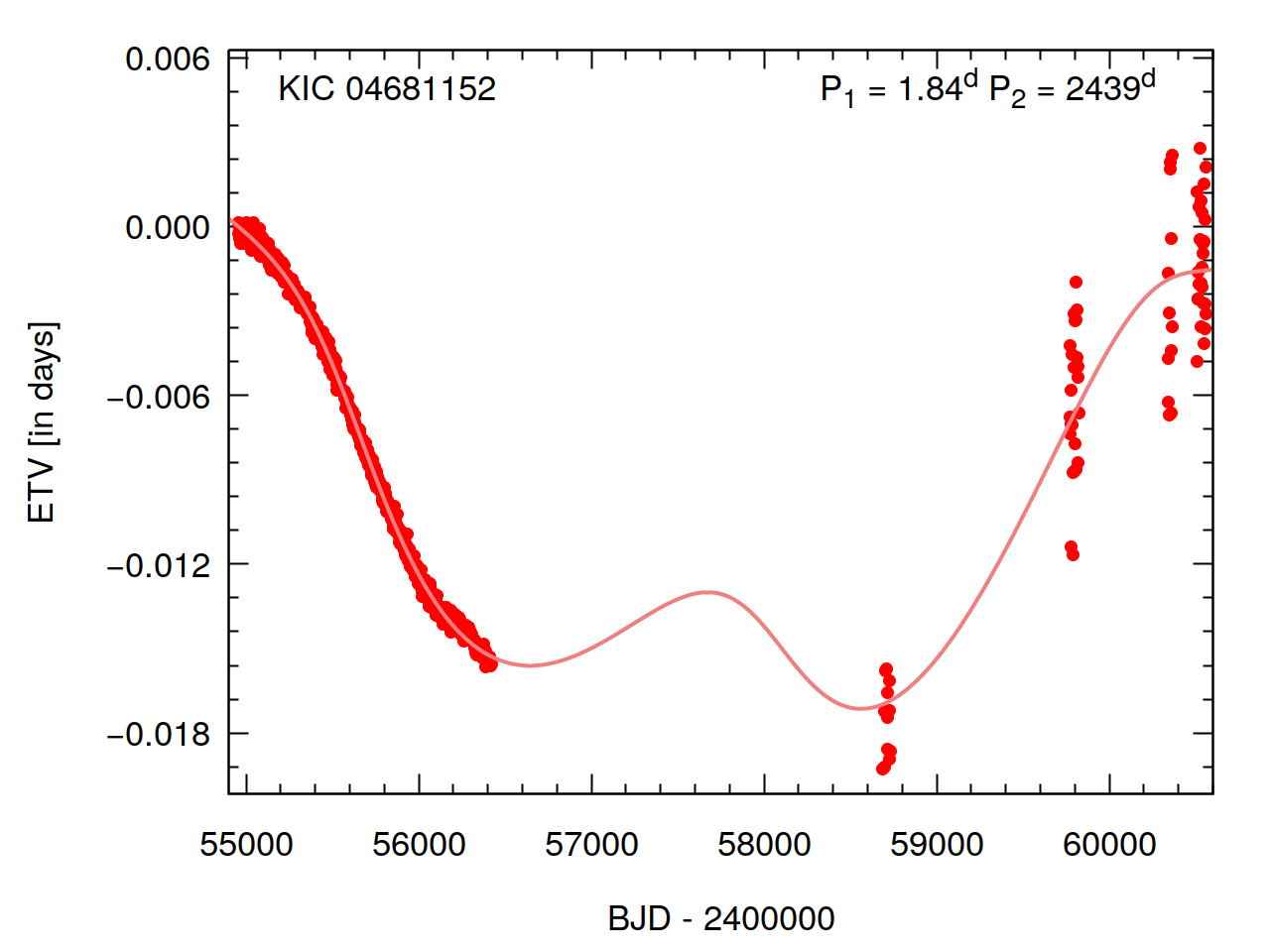}\includegraphics[width=60mm]{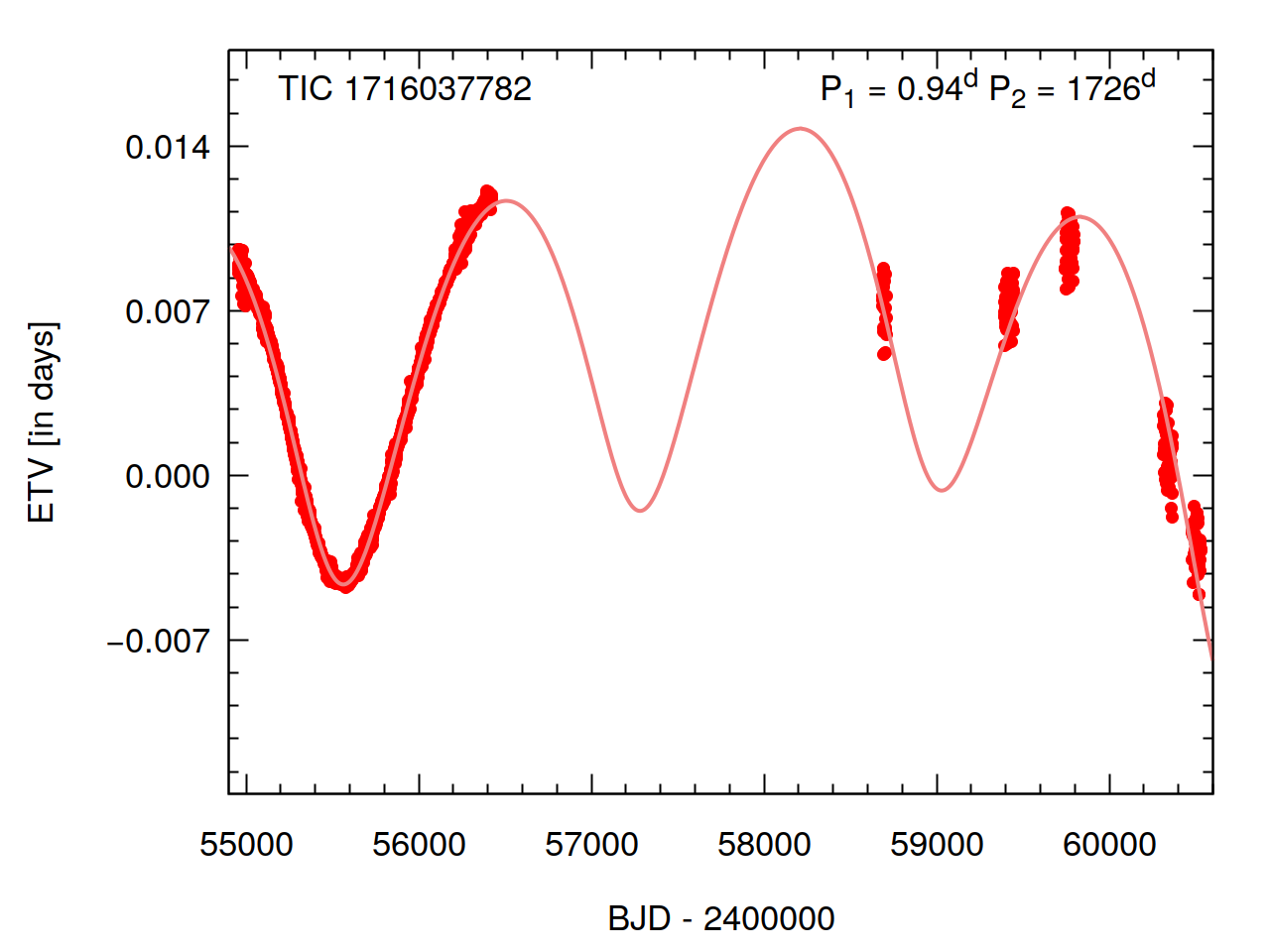}\includegraphics[width=60mm]{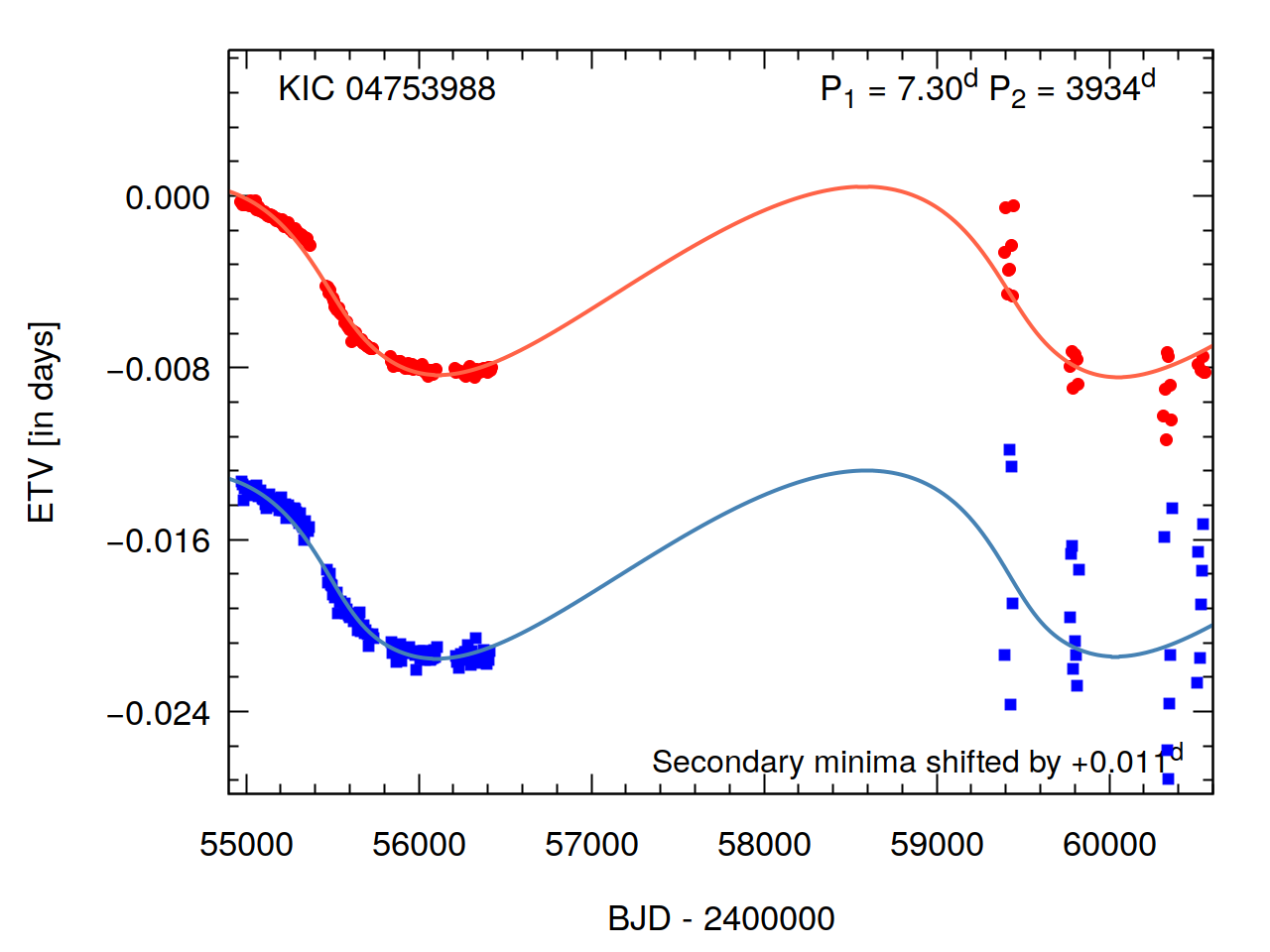}
\caption{continued.}
\end{figure*}

\addtocounter{figure}{-1}

\begin{figure*}
\includegraphics[width=60mm]{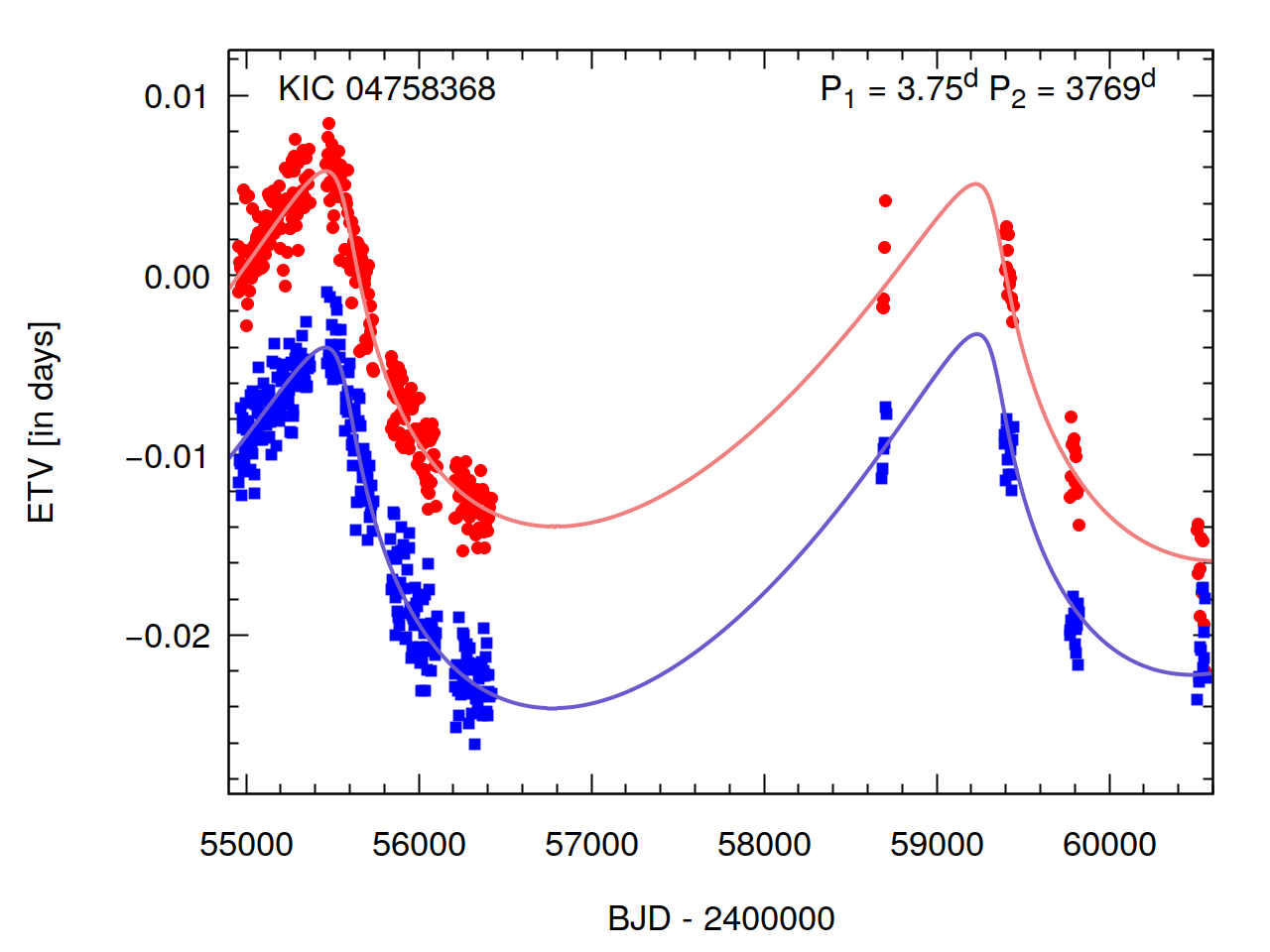}\includegraphics[width=60mm]{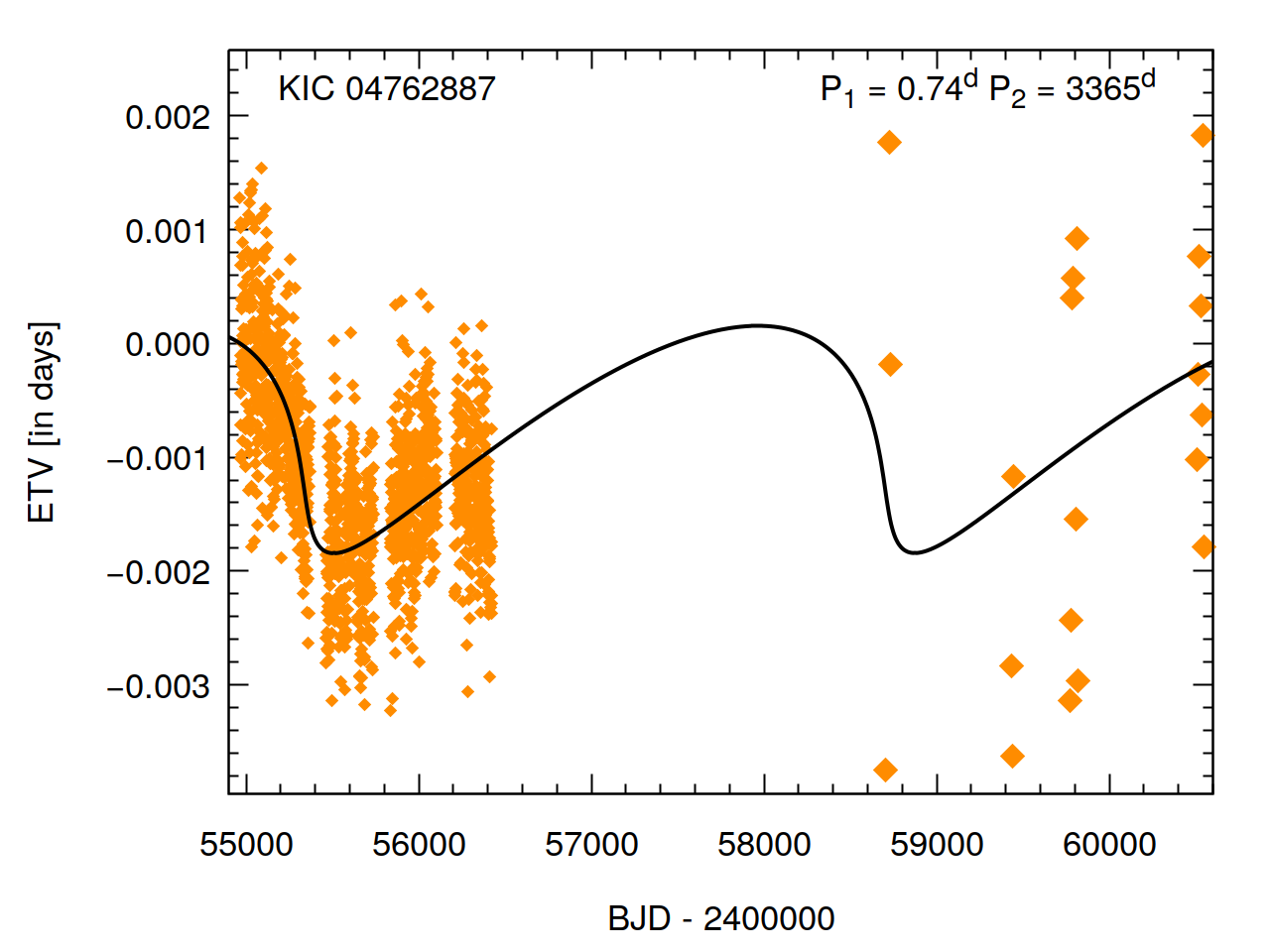}\includegraphics[width=60mm]{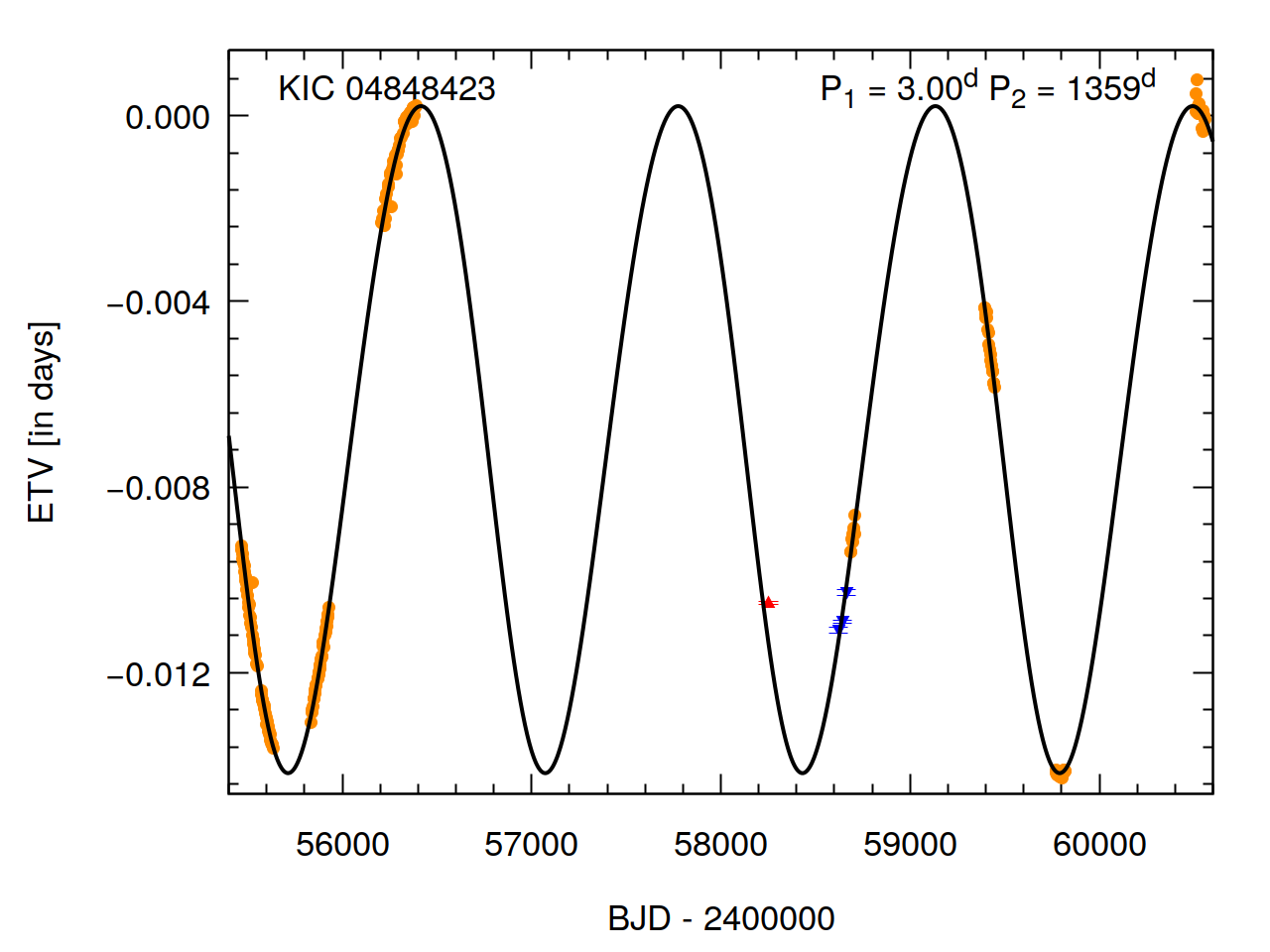}
\includegraphics[width=60mm]{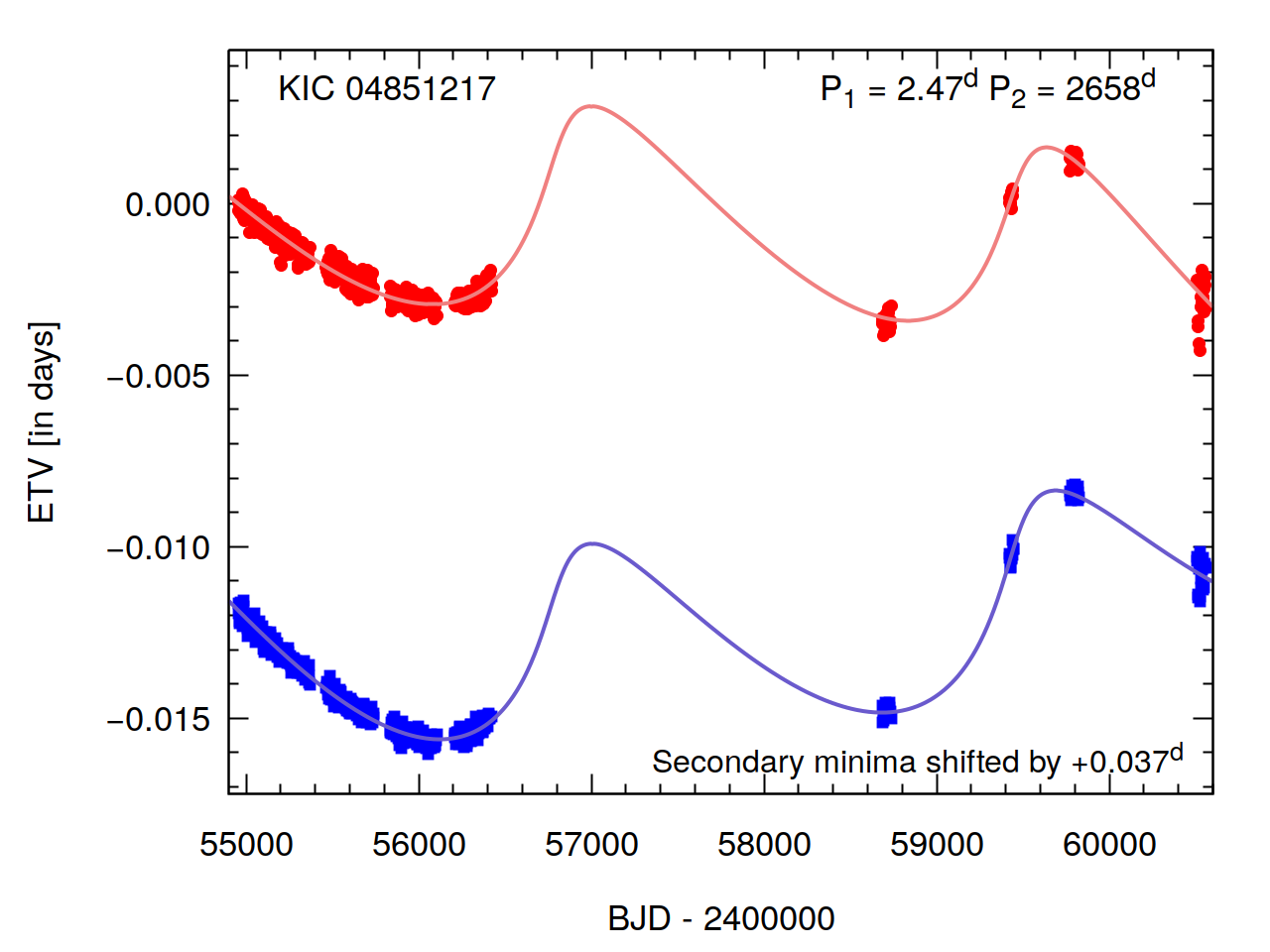}\includegraphics[width=60mm]{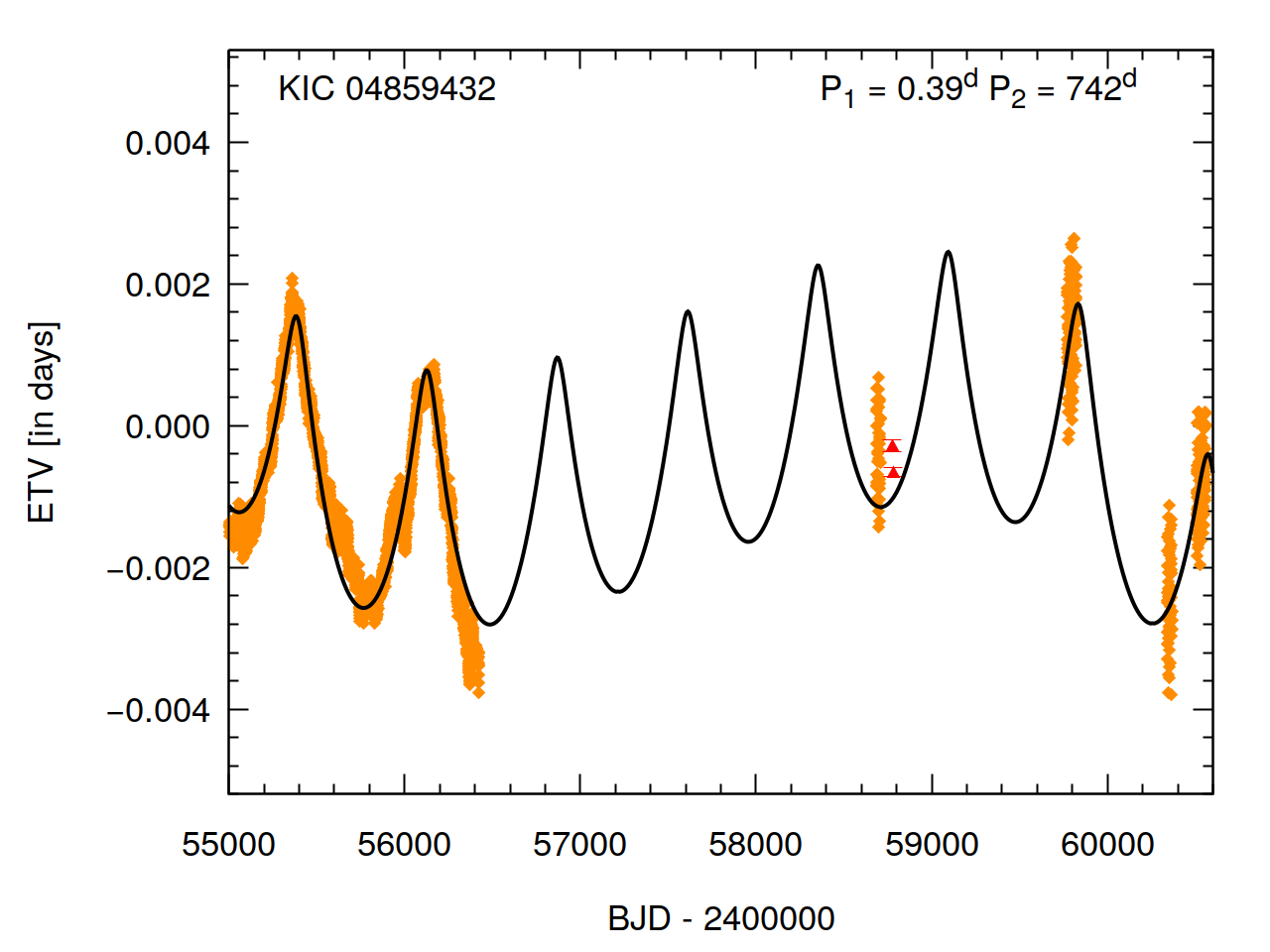}\includegraphics[width=60mm]{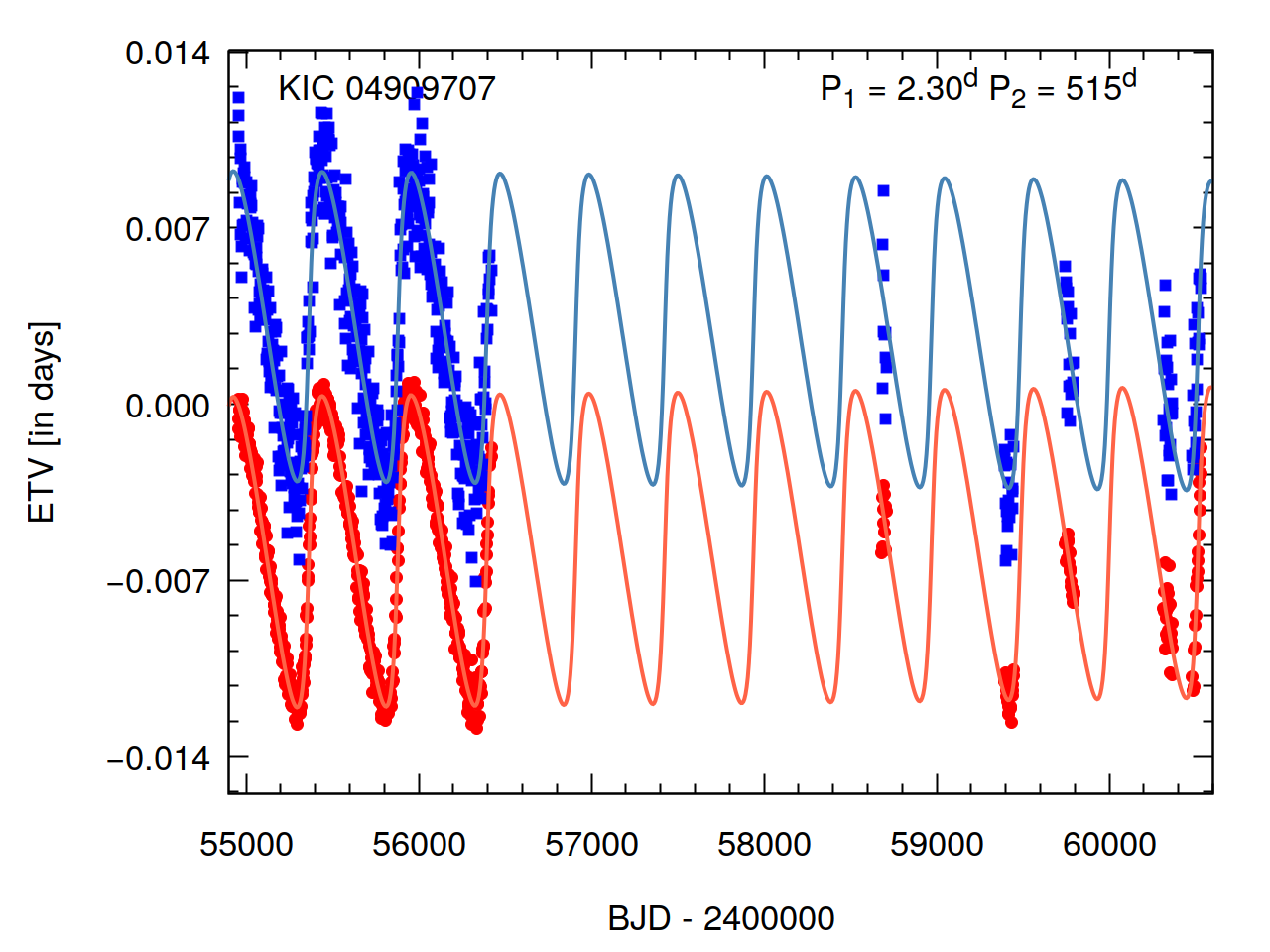}
\includegraphics[width=60mm]{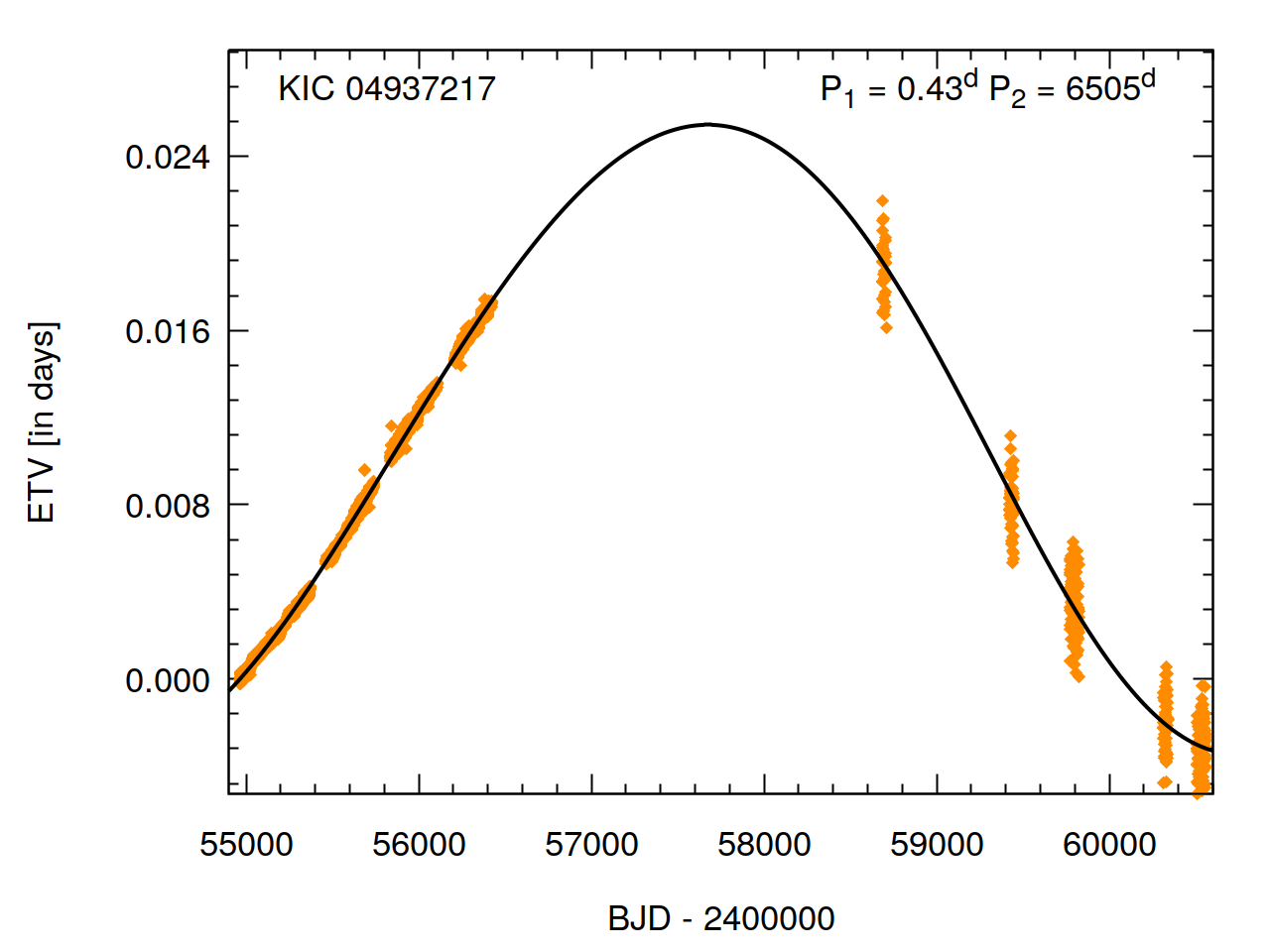}\includegraphics[width=60mm]{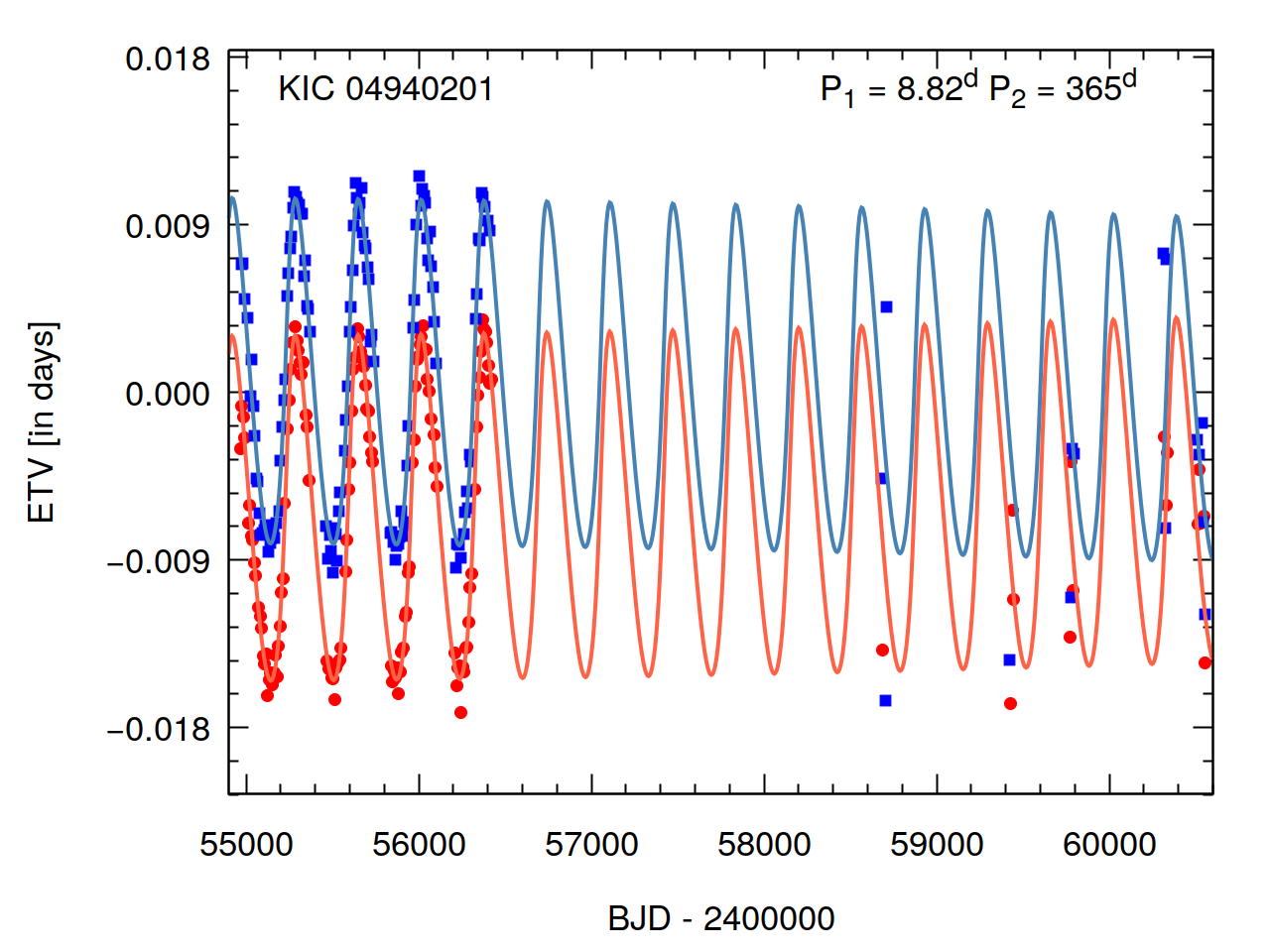}\includegraphics[width=60mm]{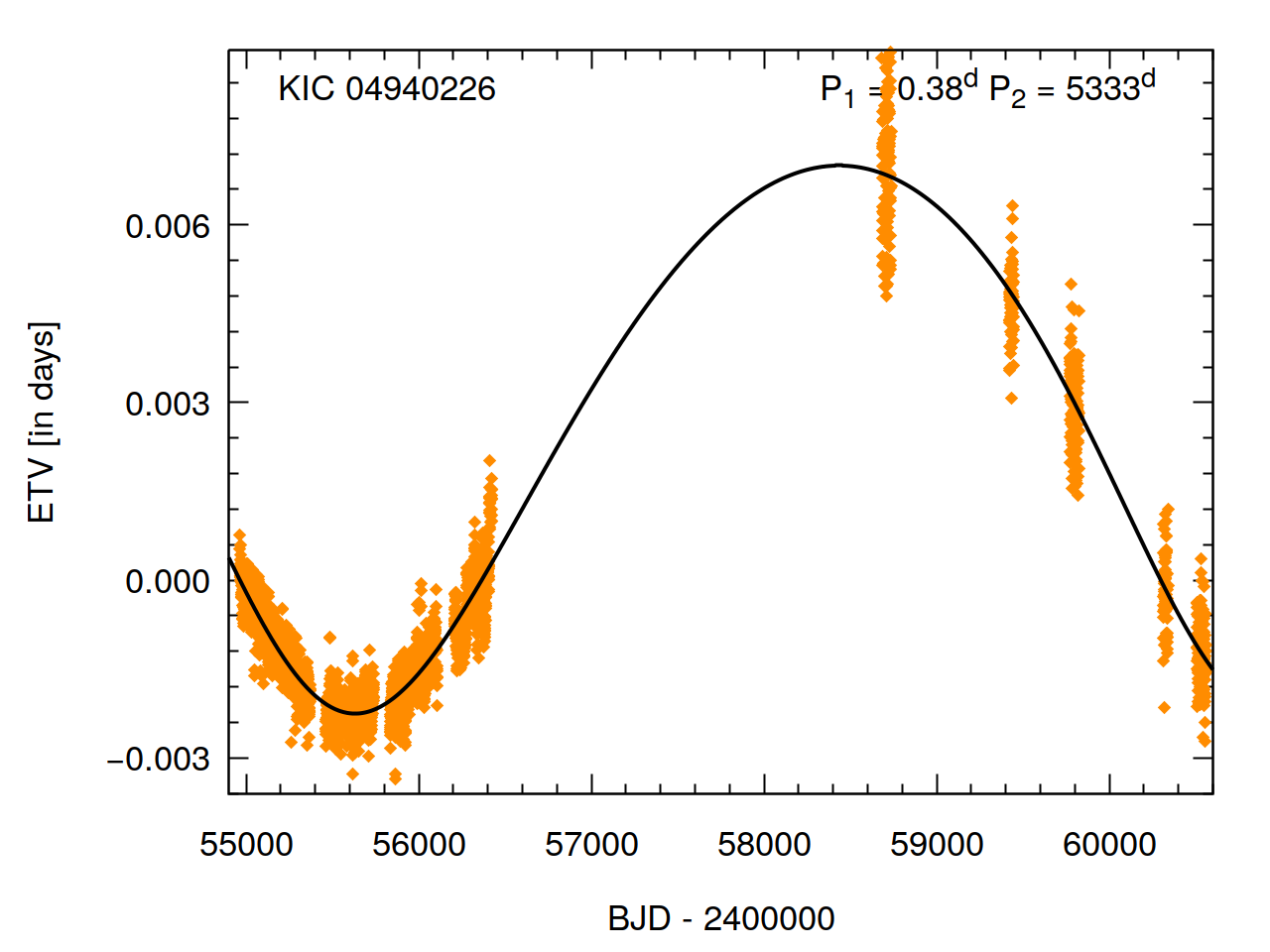}
\includegraphics[width=60mm]{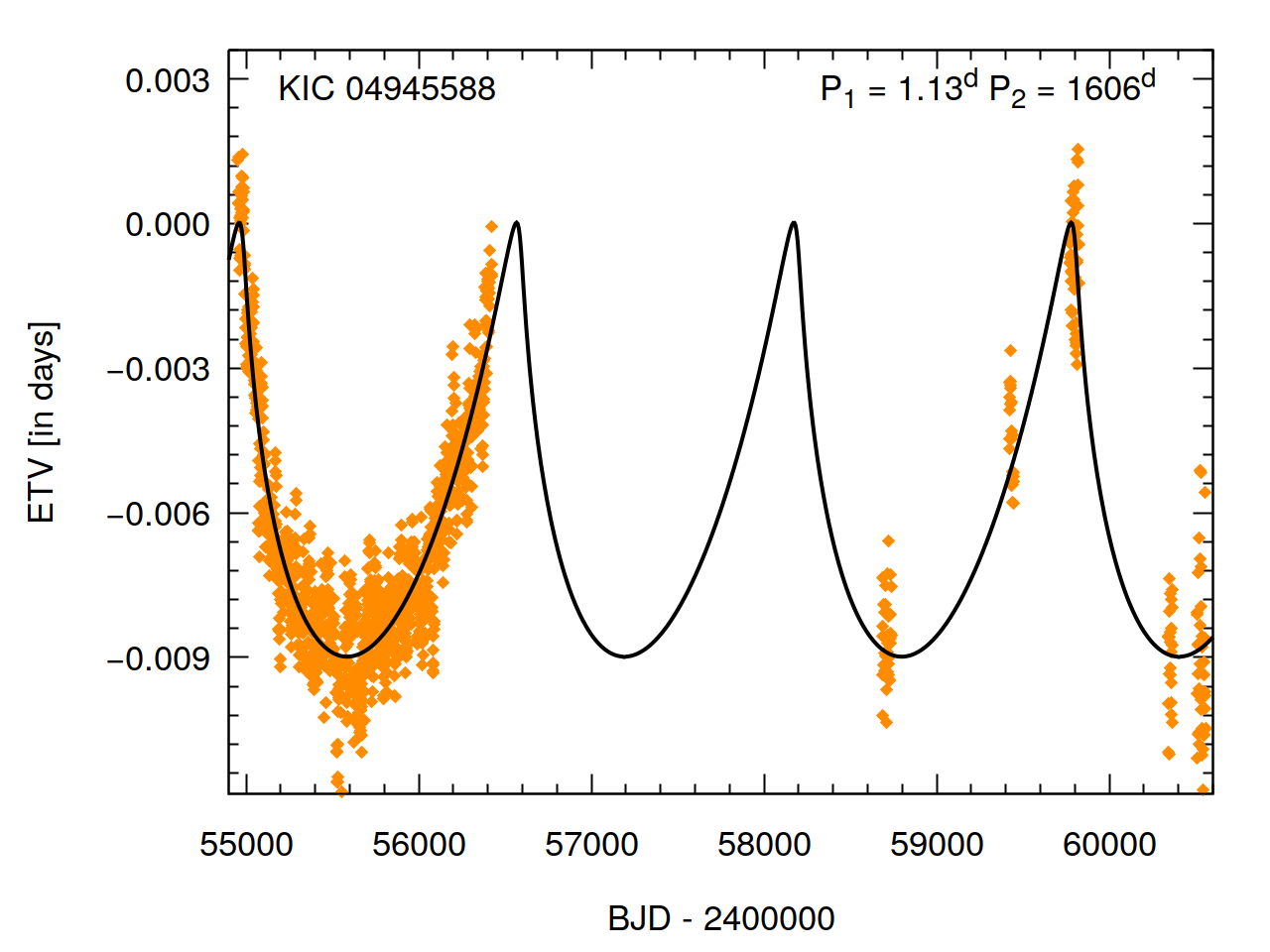}\includegraphics[width=60mm]{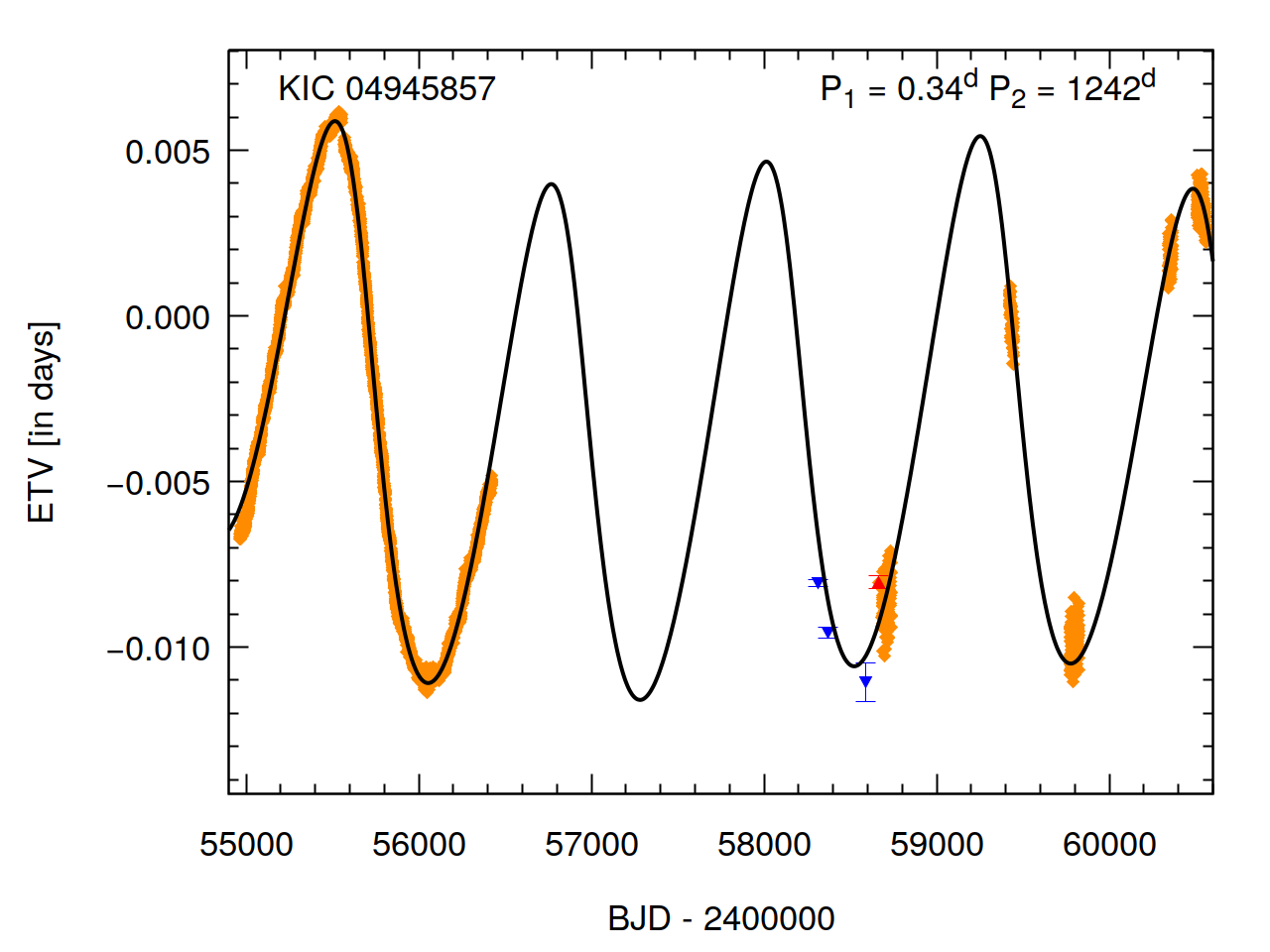}\includegraphics[width=60mm]{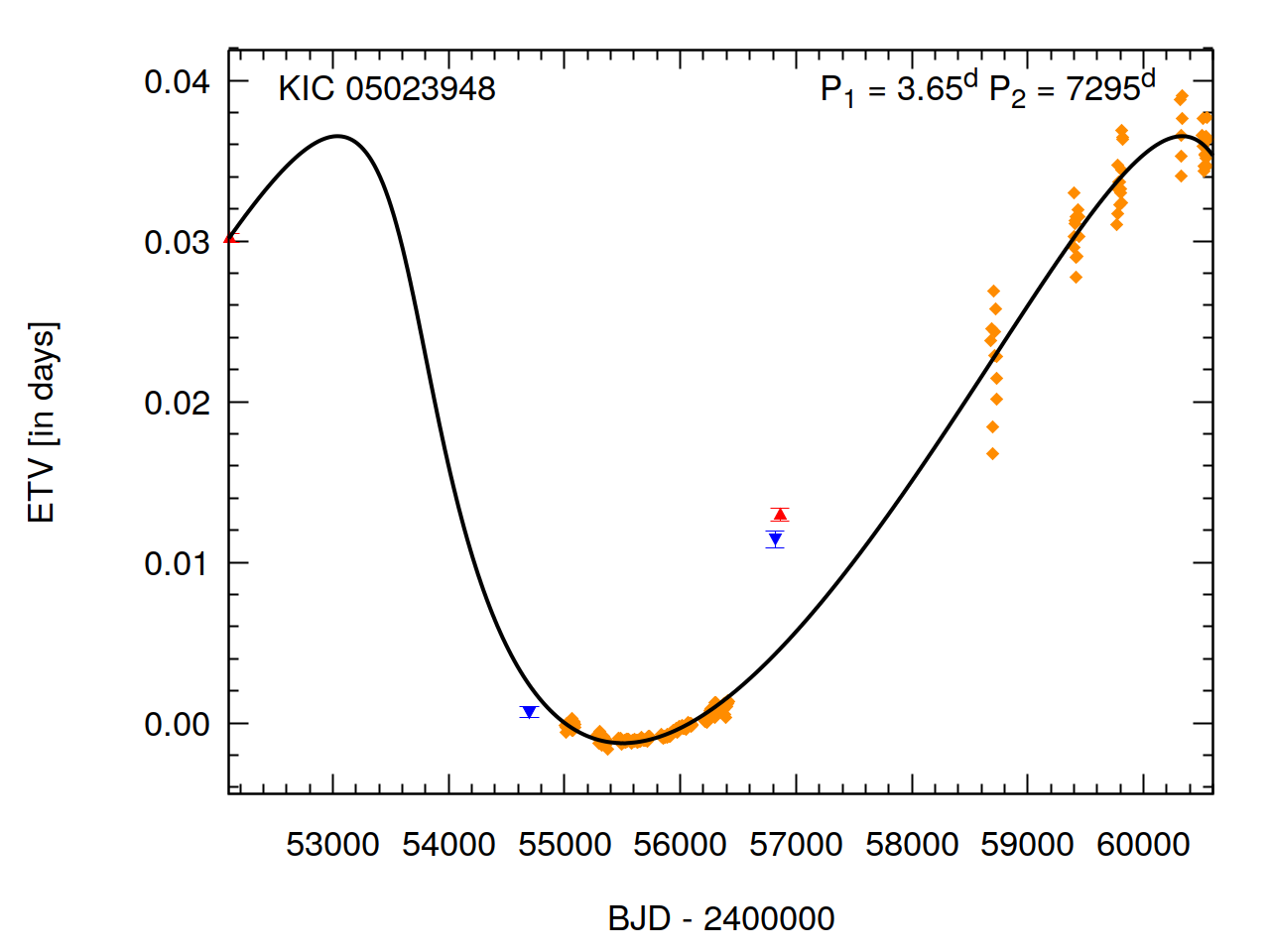}
\includegraphics[width=60mm]{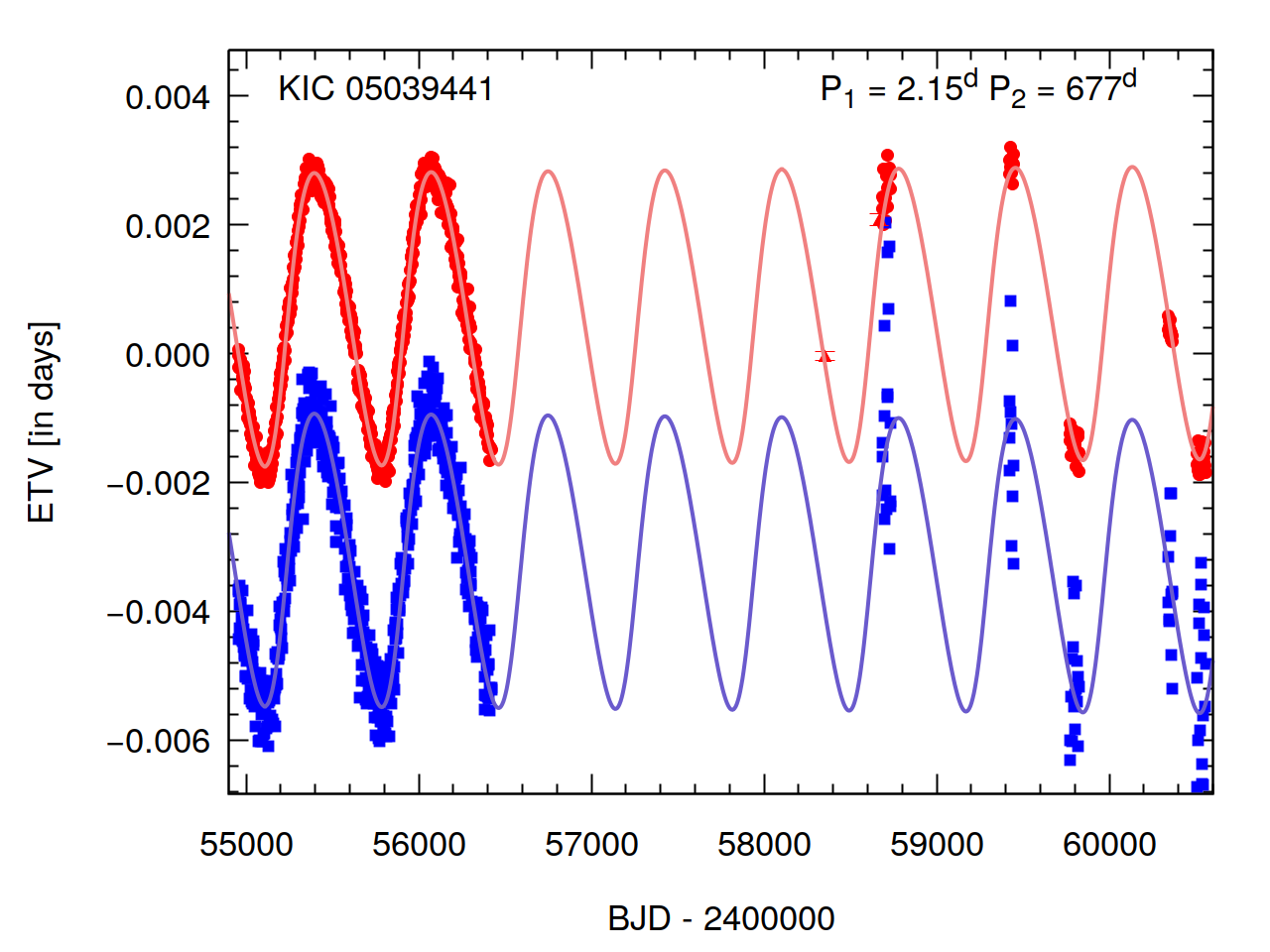}\includegraphics[width=60mm]{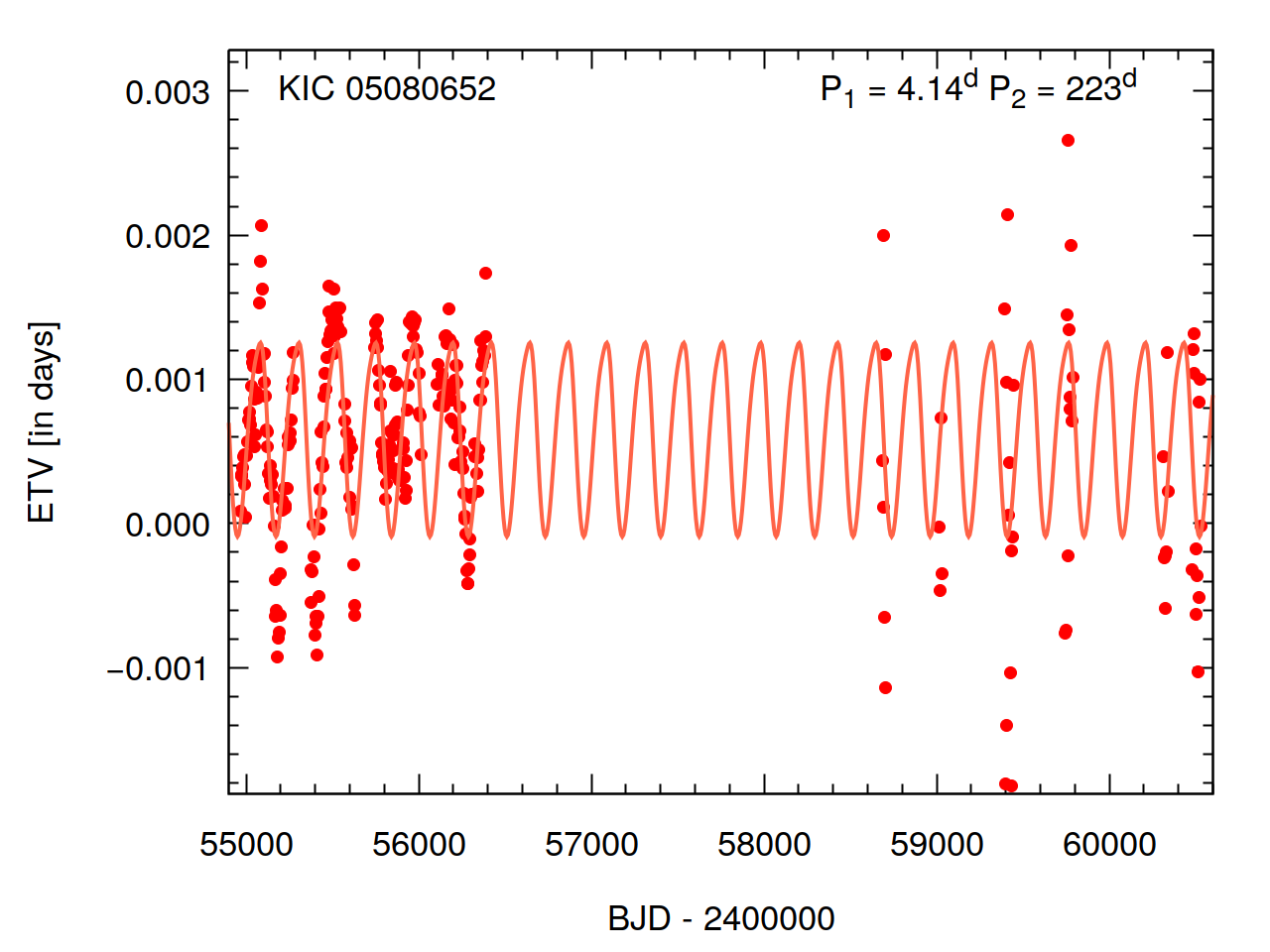}\includegraphics[width=60mm]{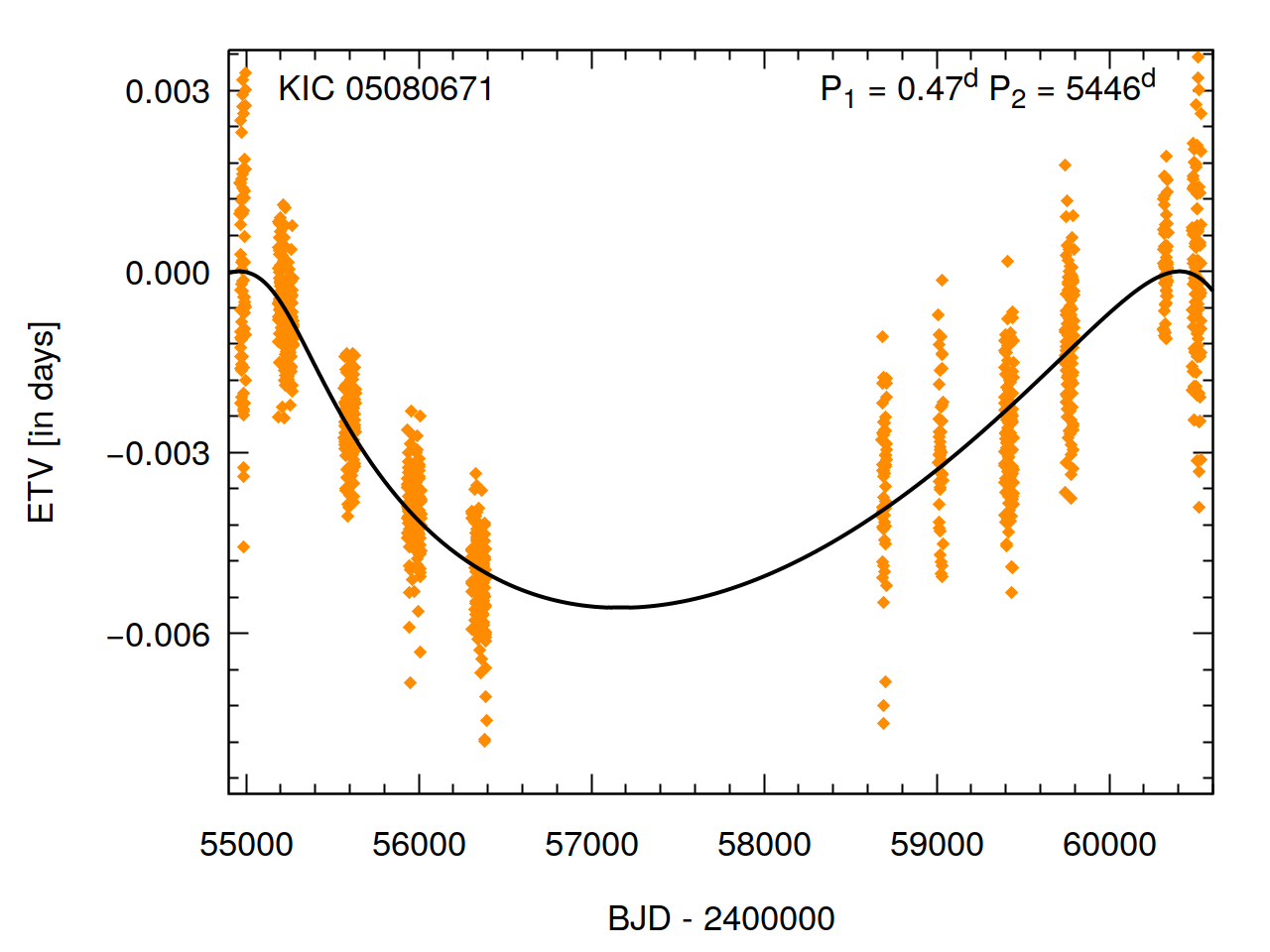}
\caption{continued.}
\end{figure*}

\addtocounter{figure}{-1}

\begin{figure*}
\includegraphics[width=60mm]{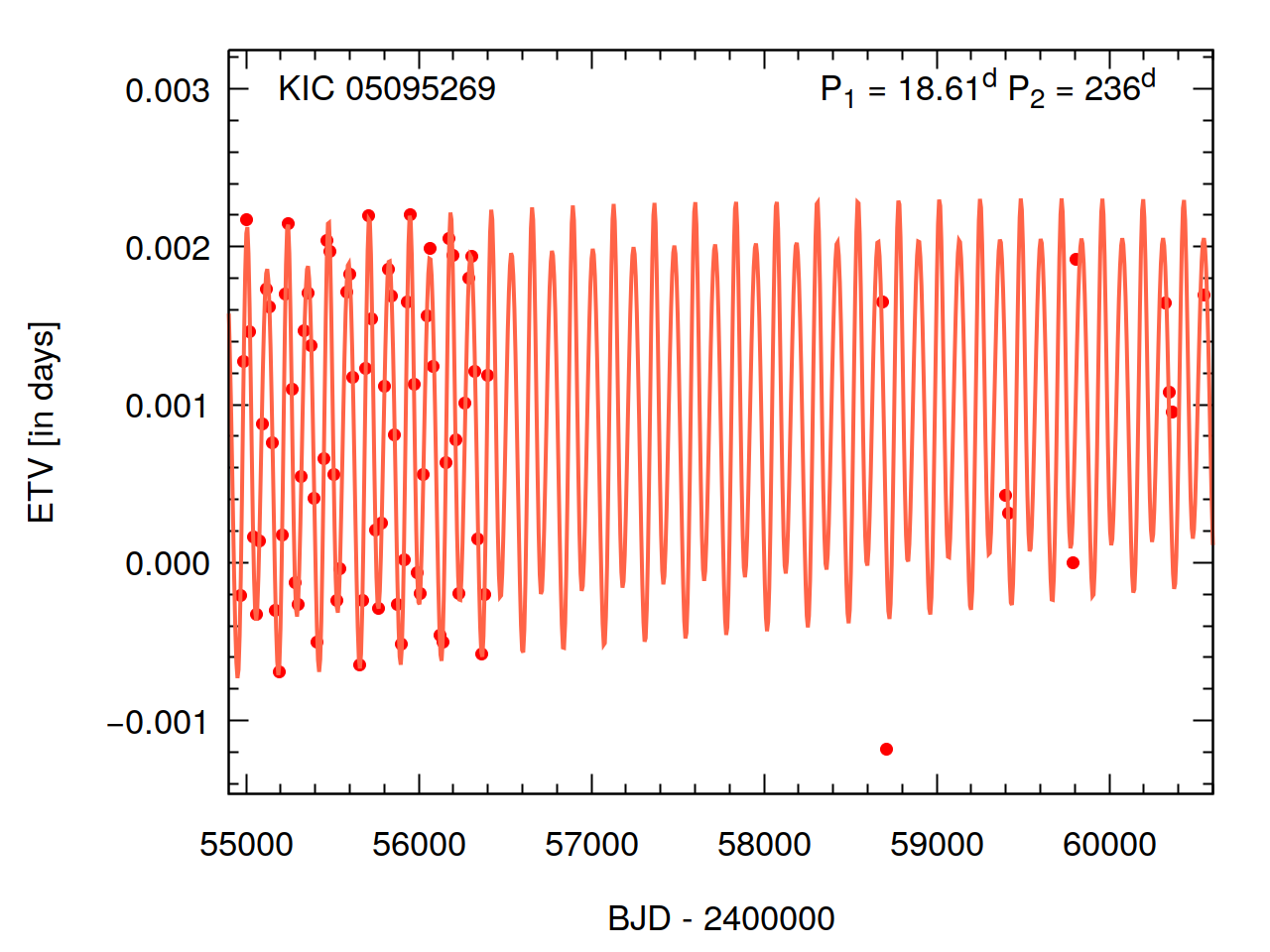}\includegraphics[width=60mm]{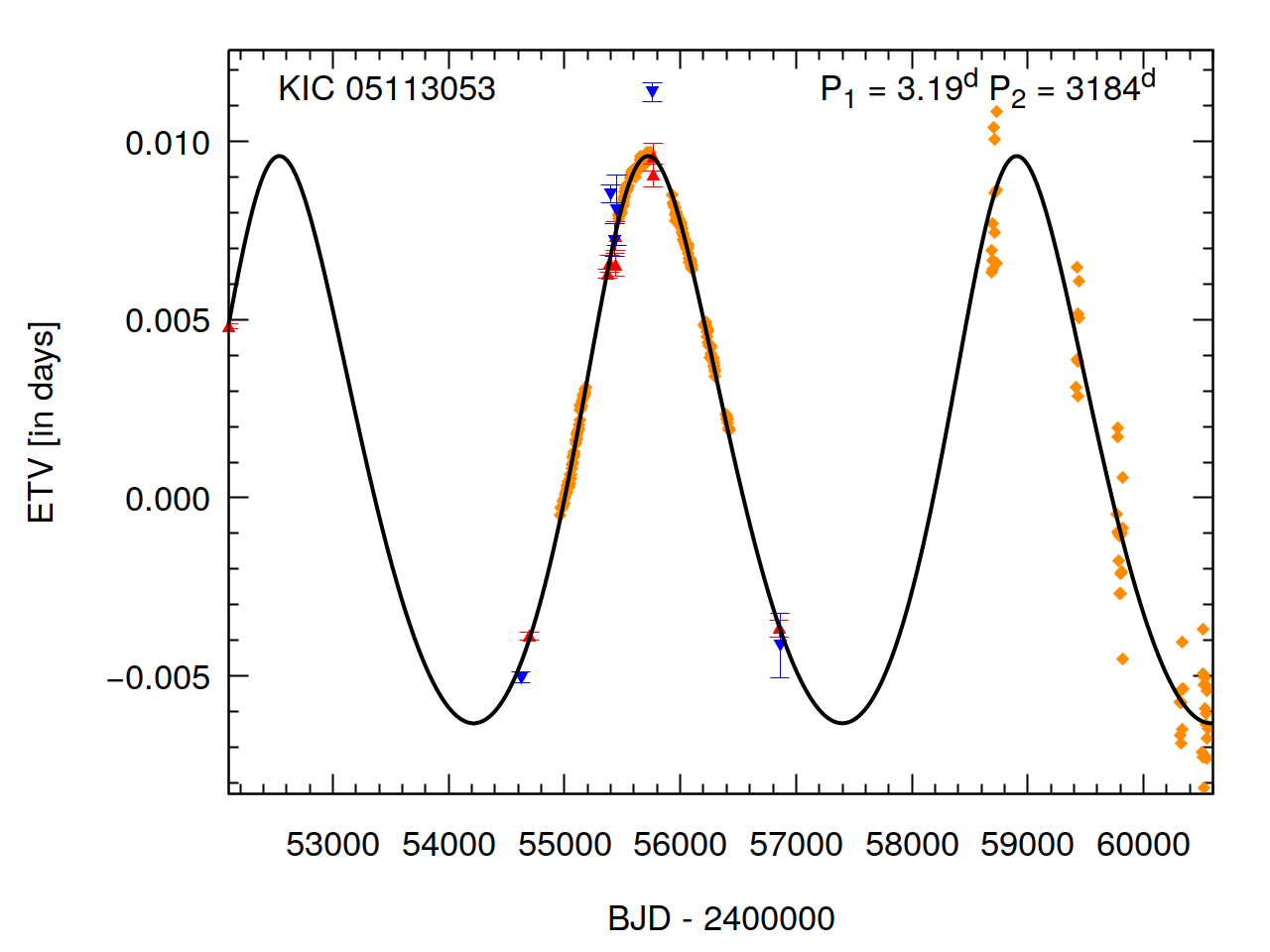}\includegraphics[width=60mm]{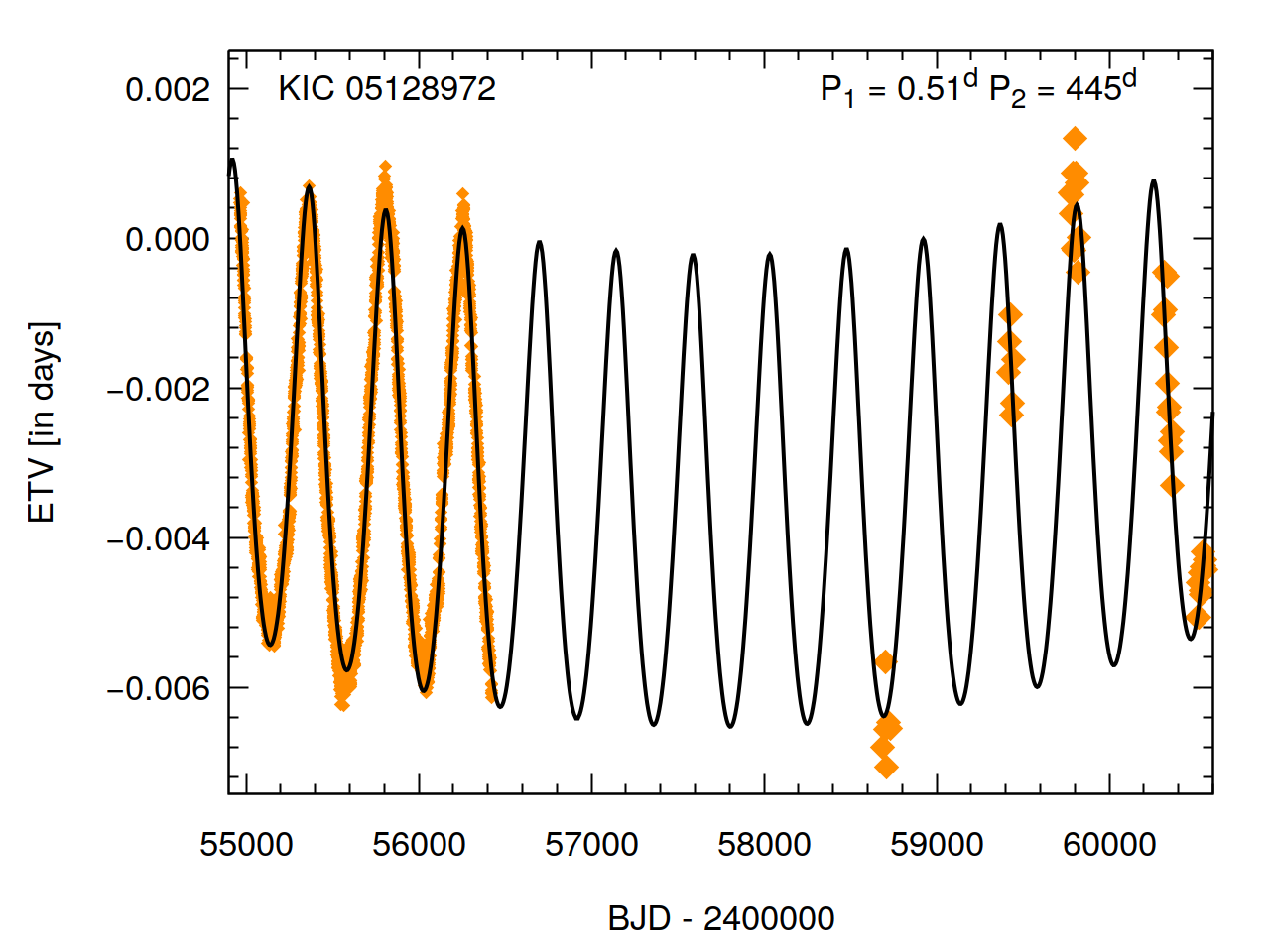}
\includegraphics[width=60mm]{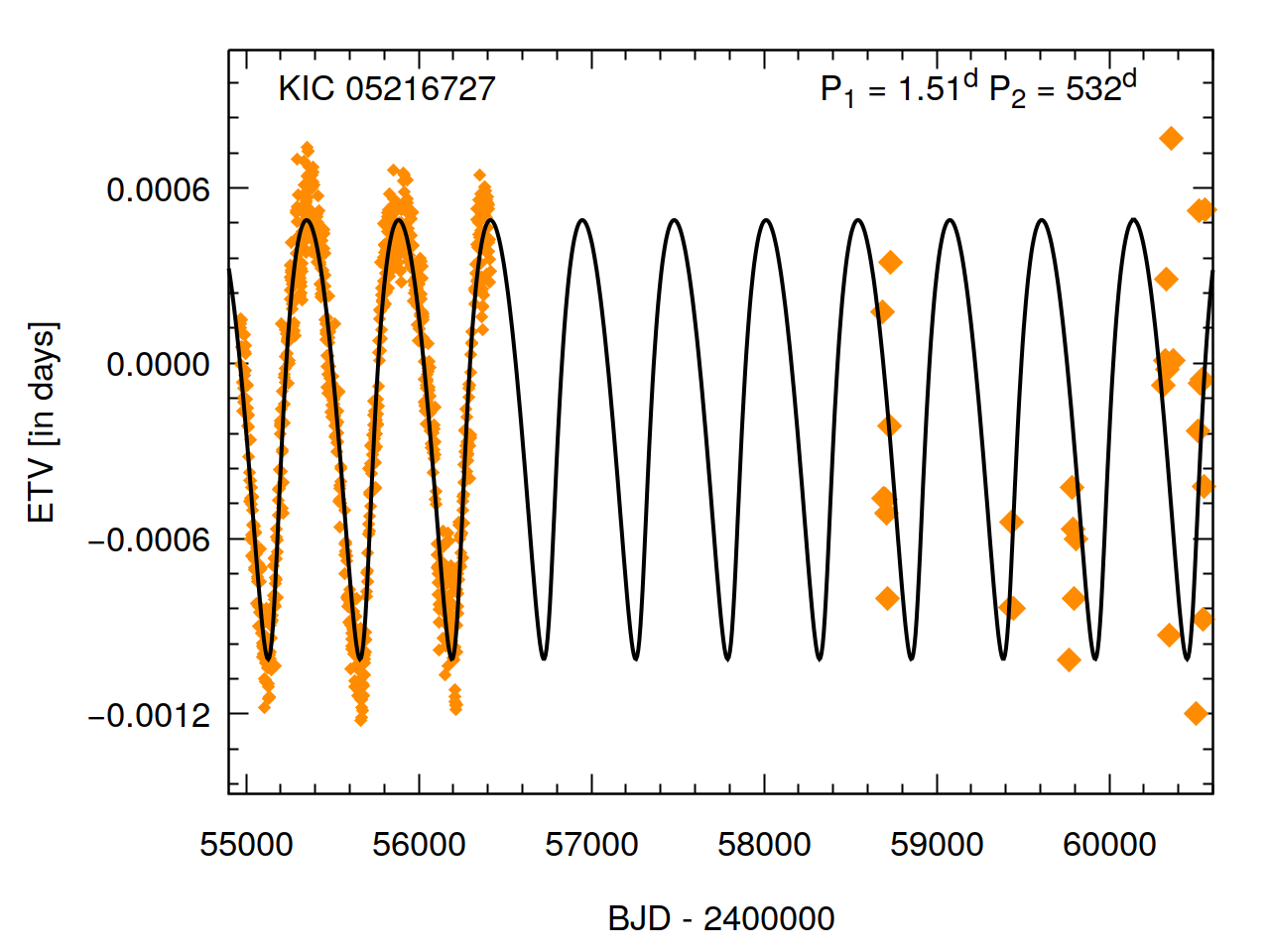}\includegraphics[width=60mm]{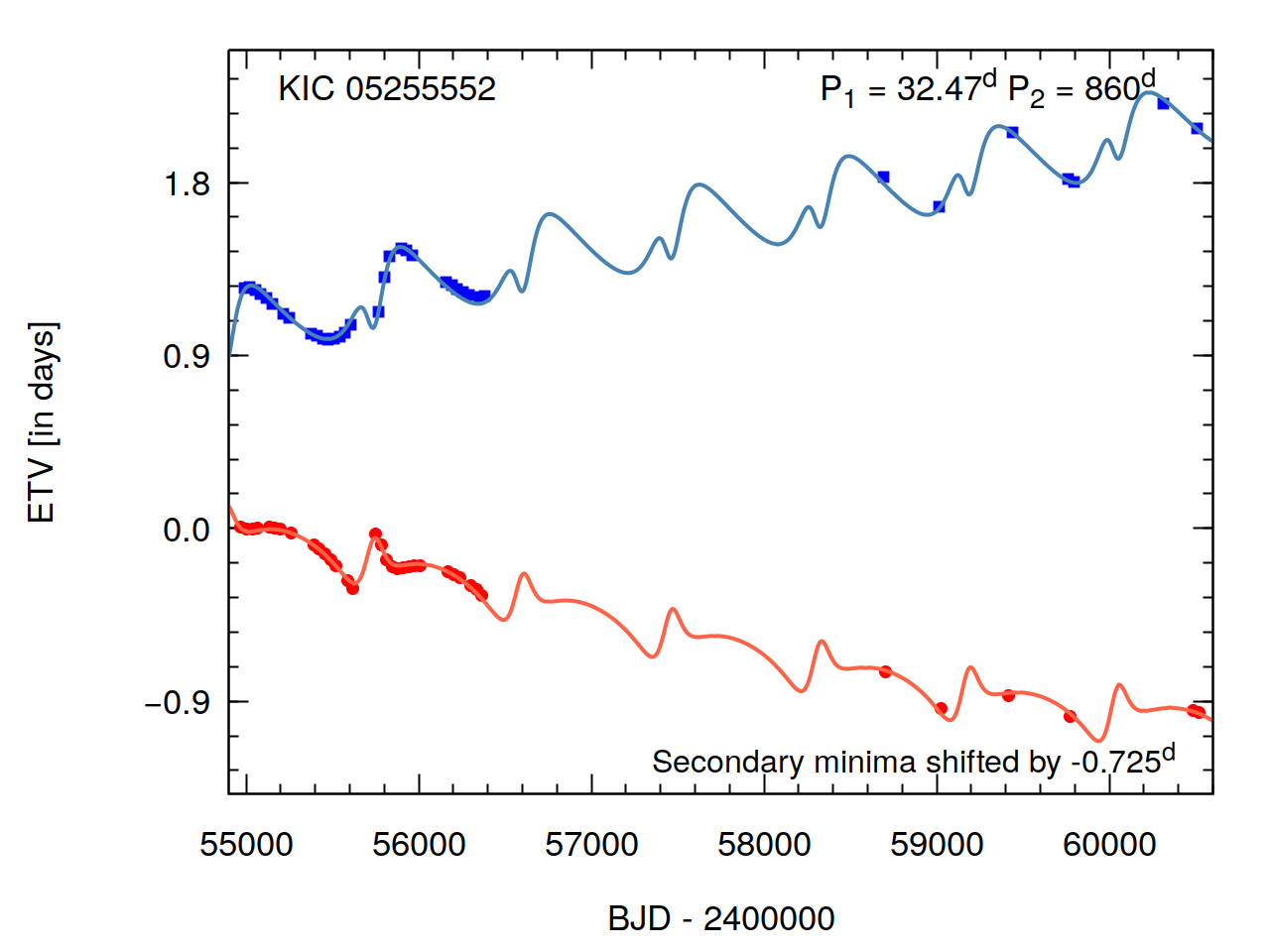}\includegraphics[width=60mm]{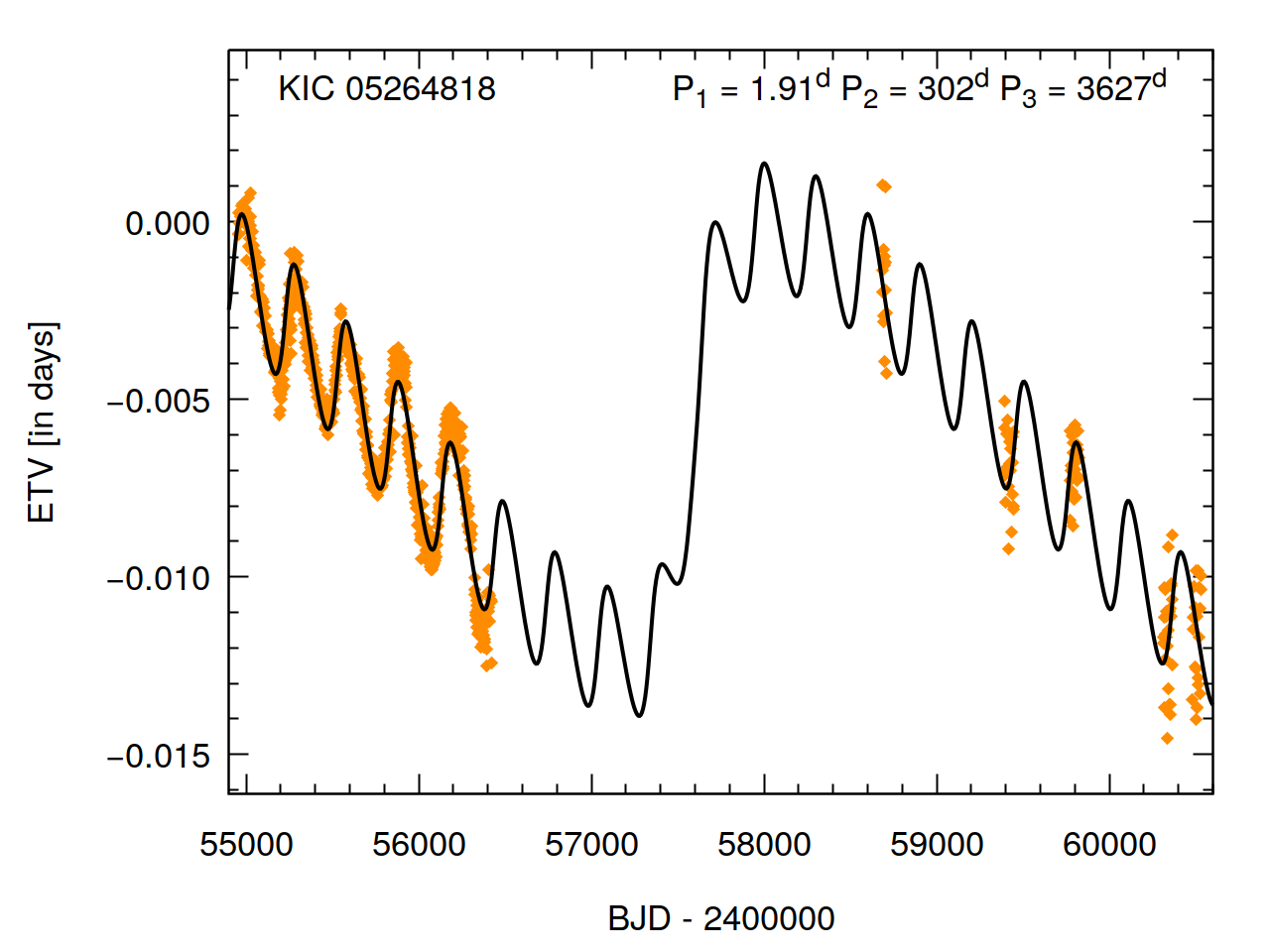}
\includegraphics[width=60mm]{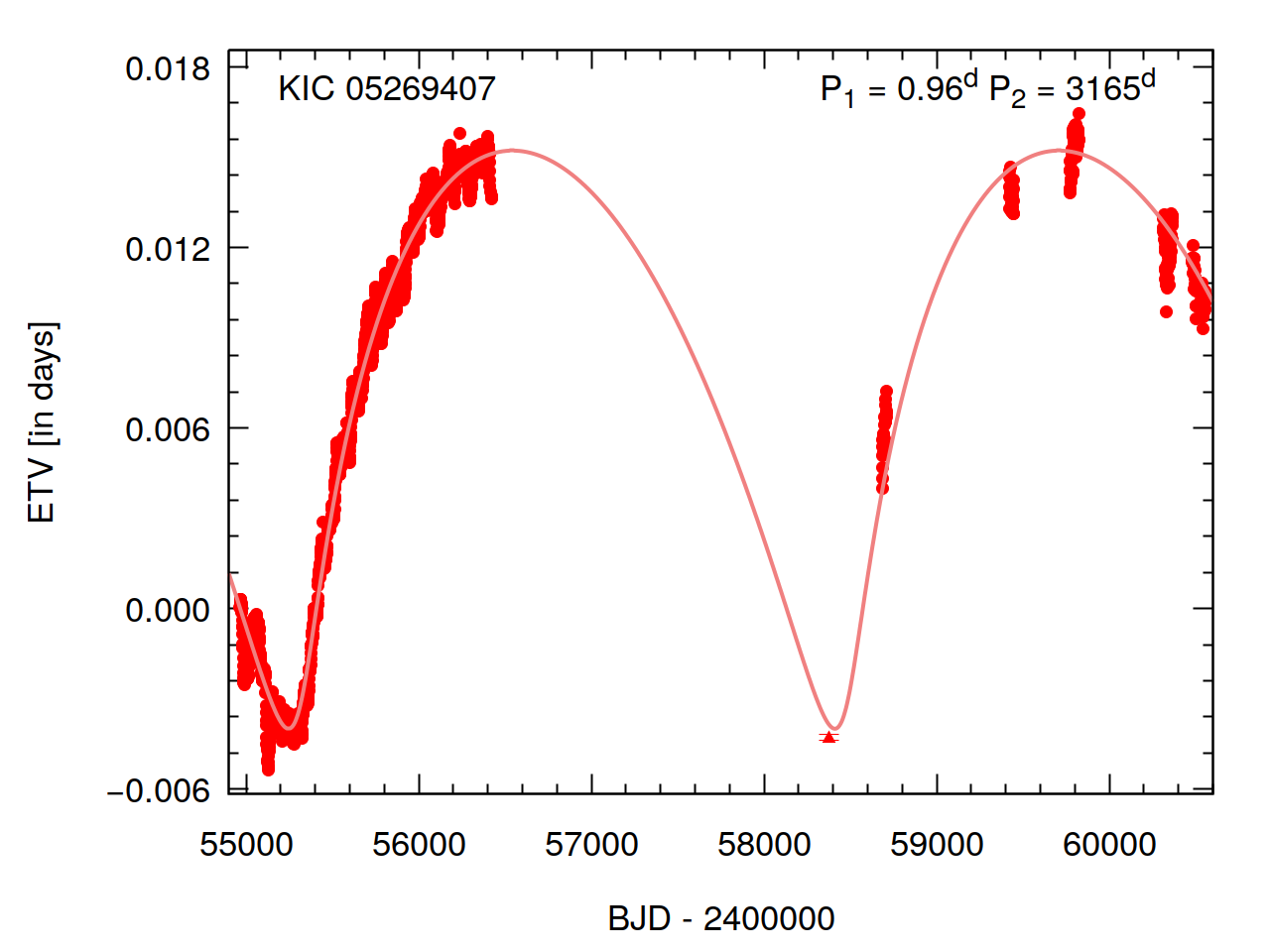}\includegraphics[width=60mm]{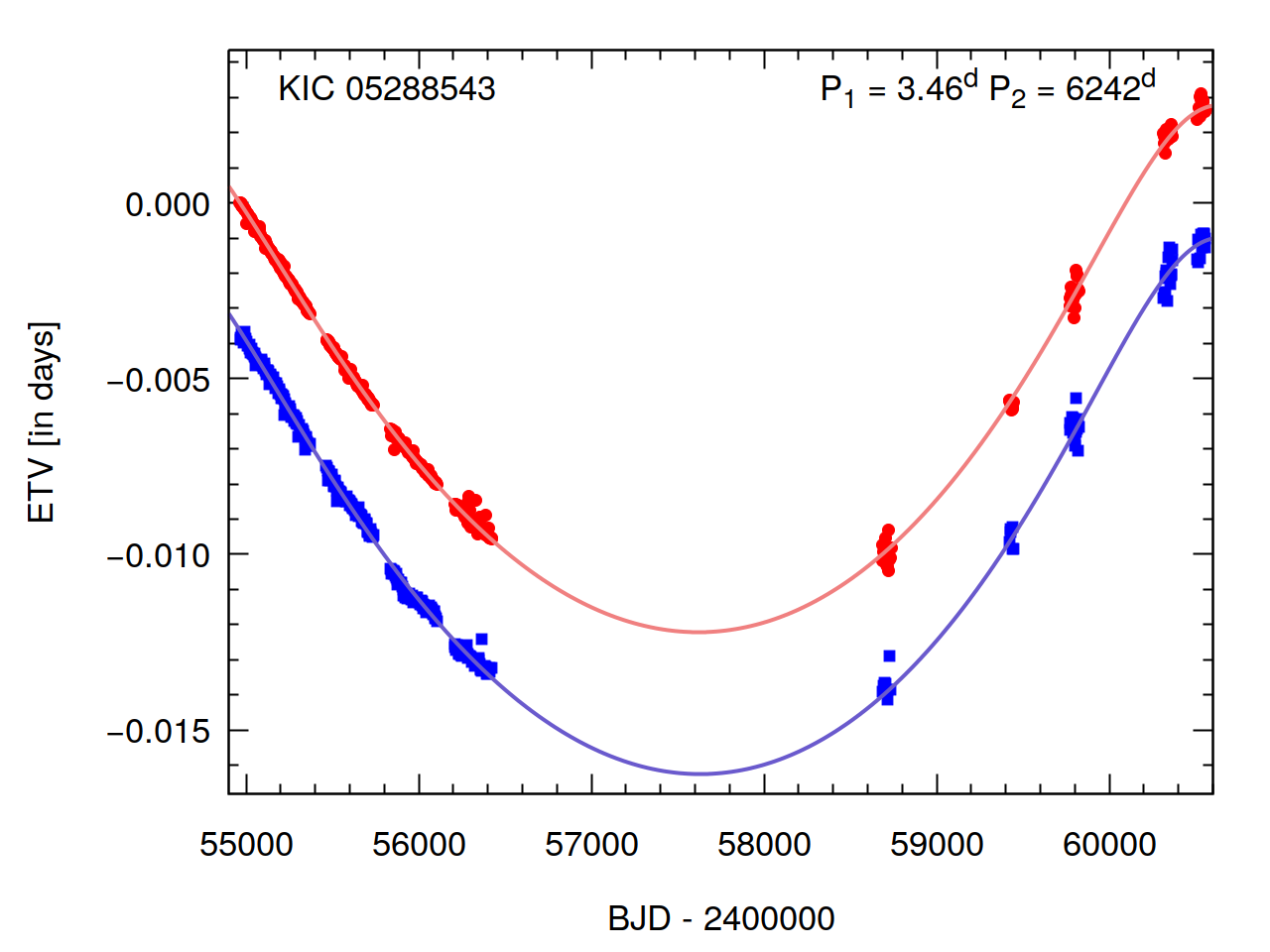}\includegraphics[width=60mm]{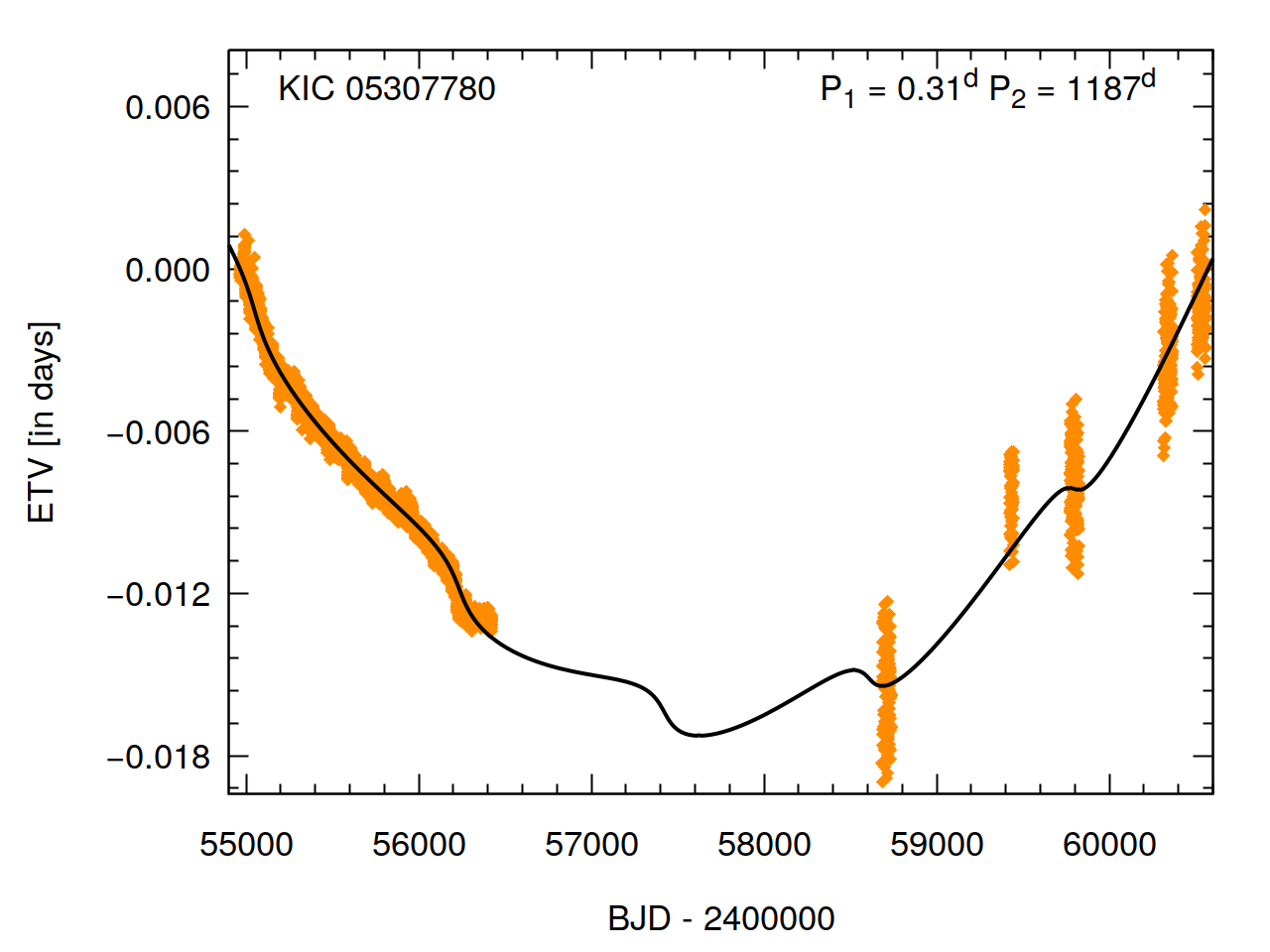}
\includegraphics[width=60mm]{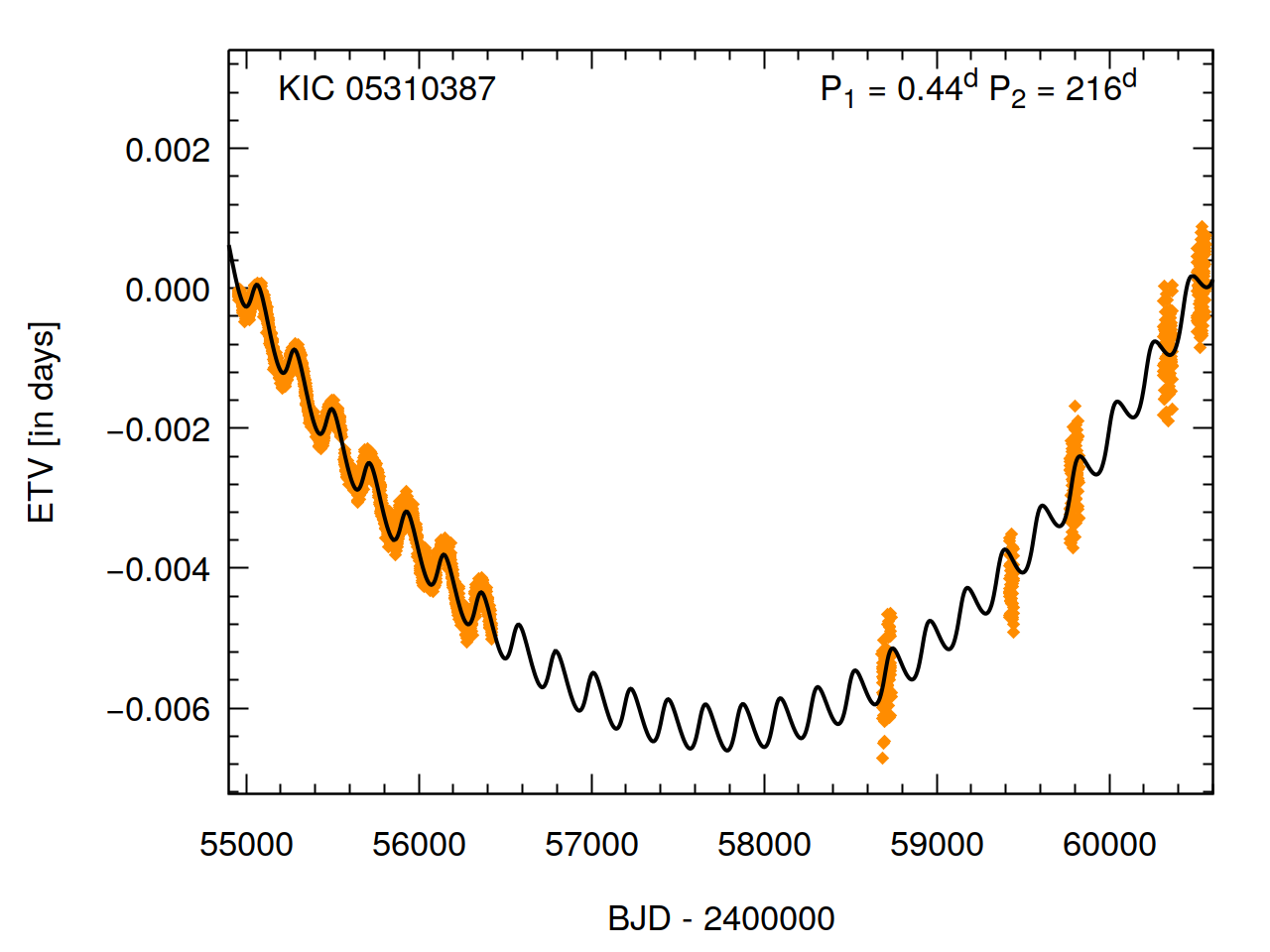}\includegraphics[width=60mm]{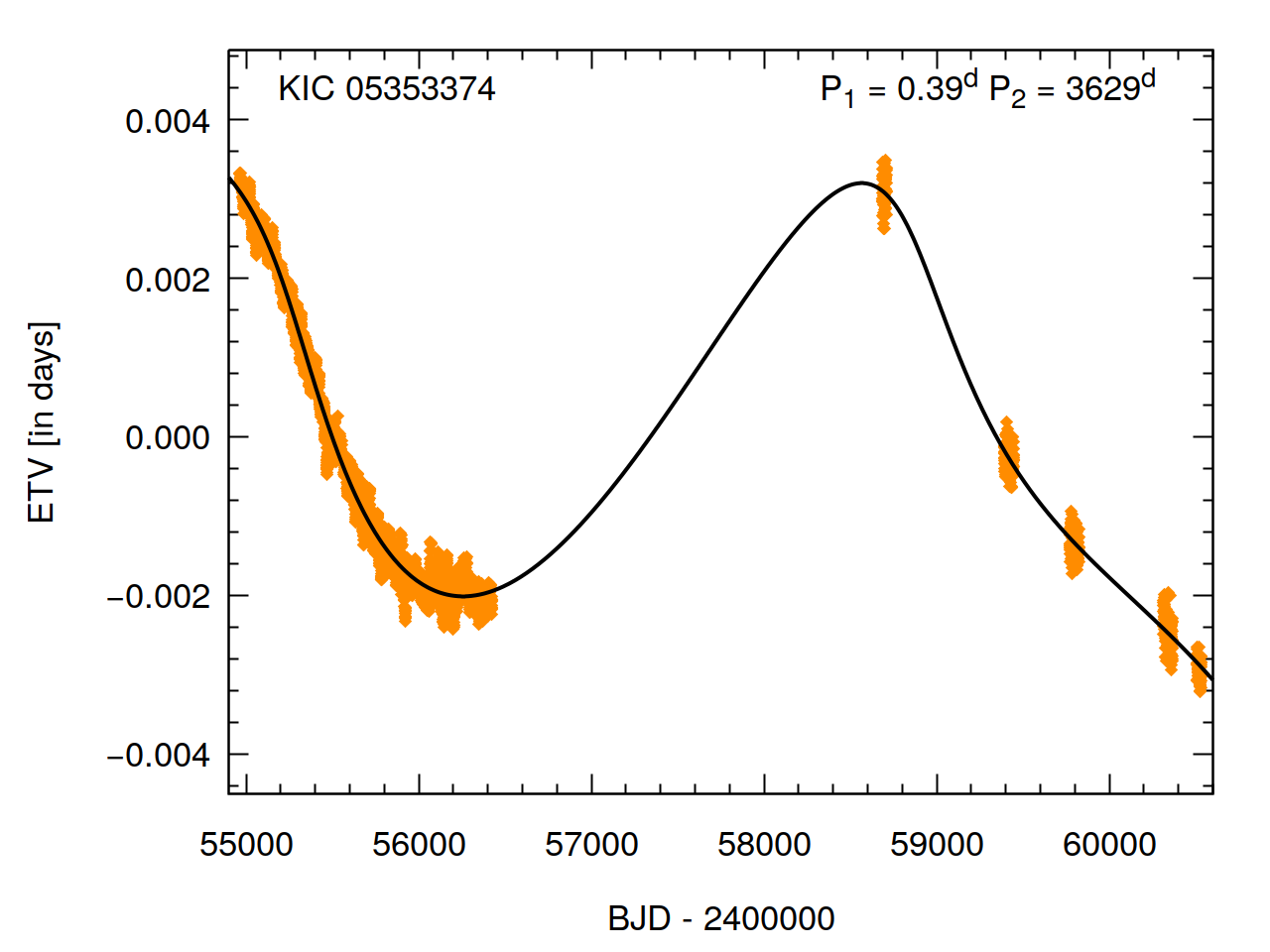}\includegraphics[width=60mm]{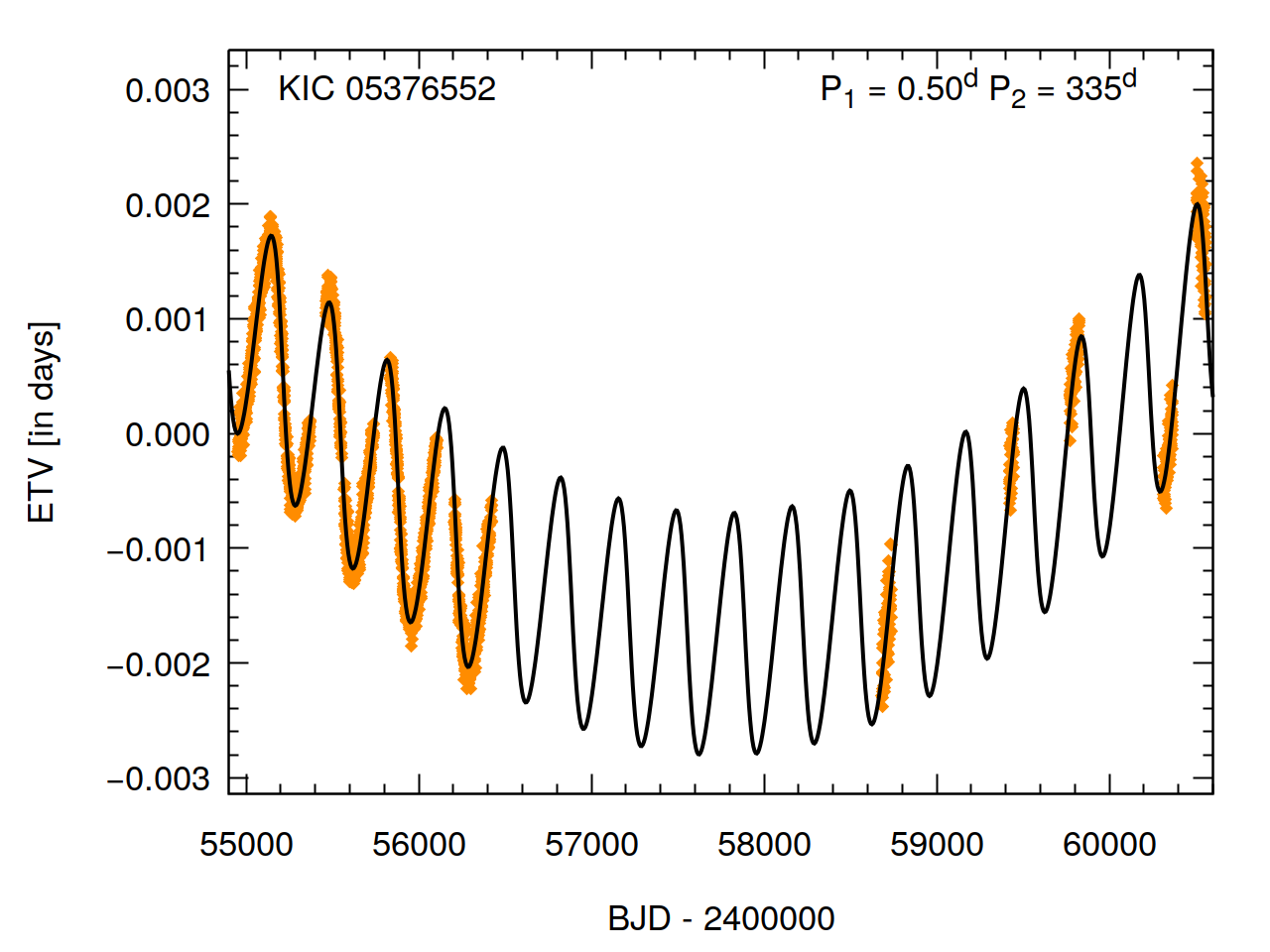}
\includegraphics[width=60mm]{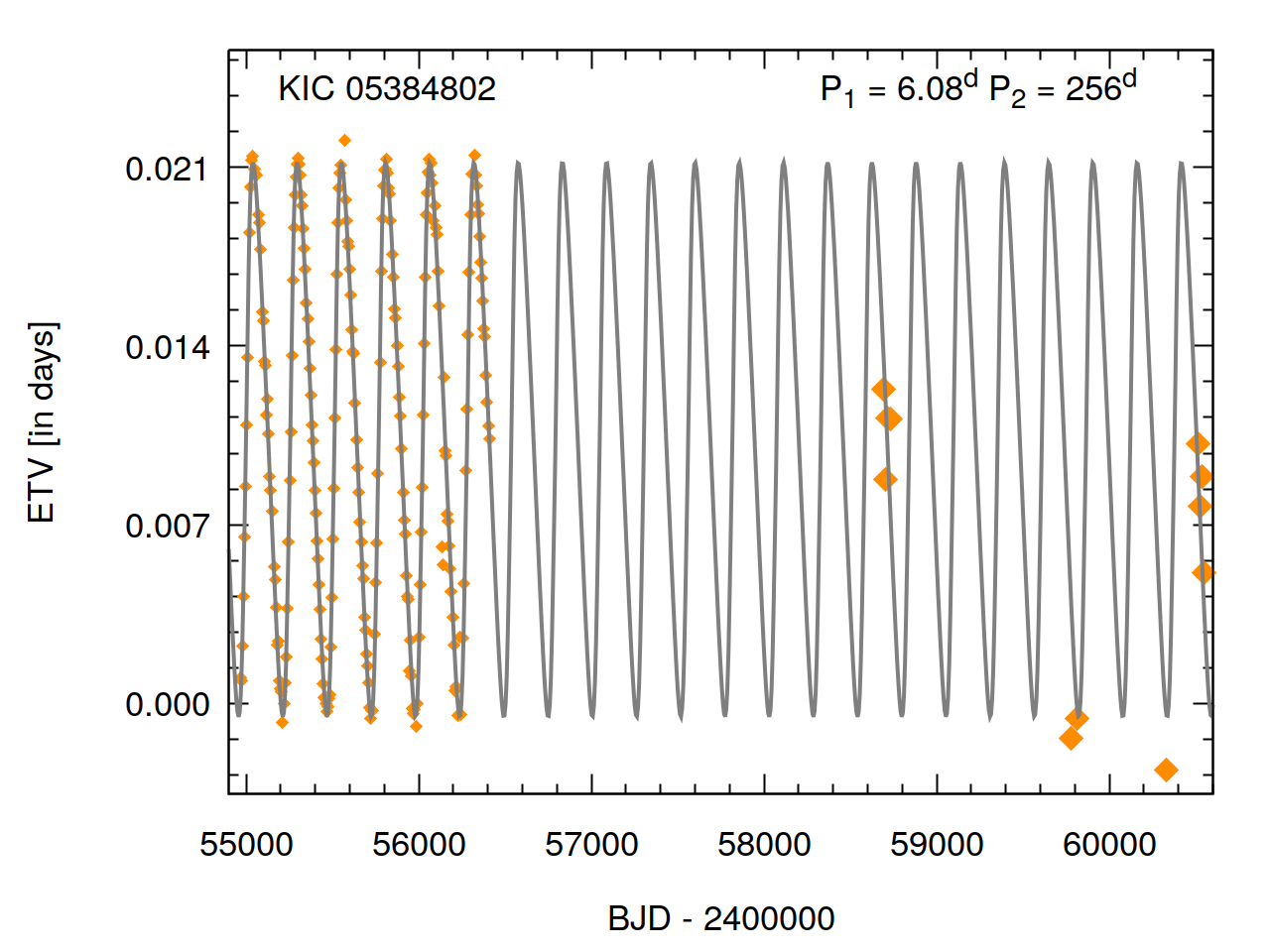}\includegraphics[width=60mm]{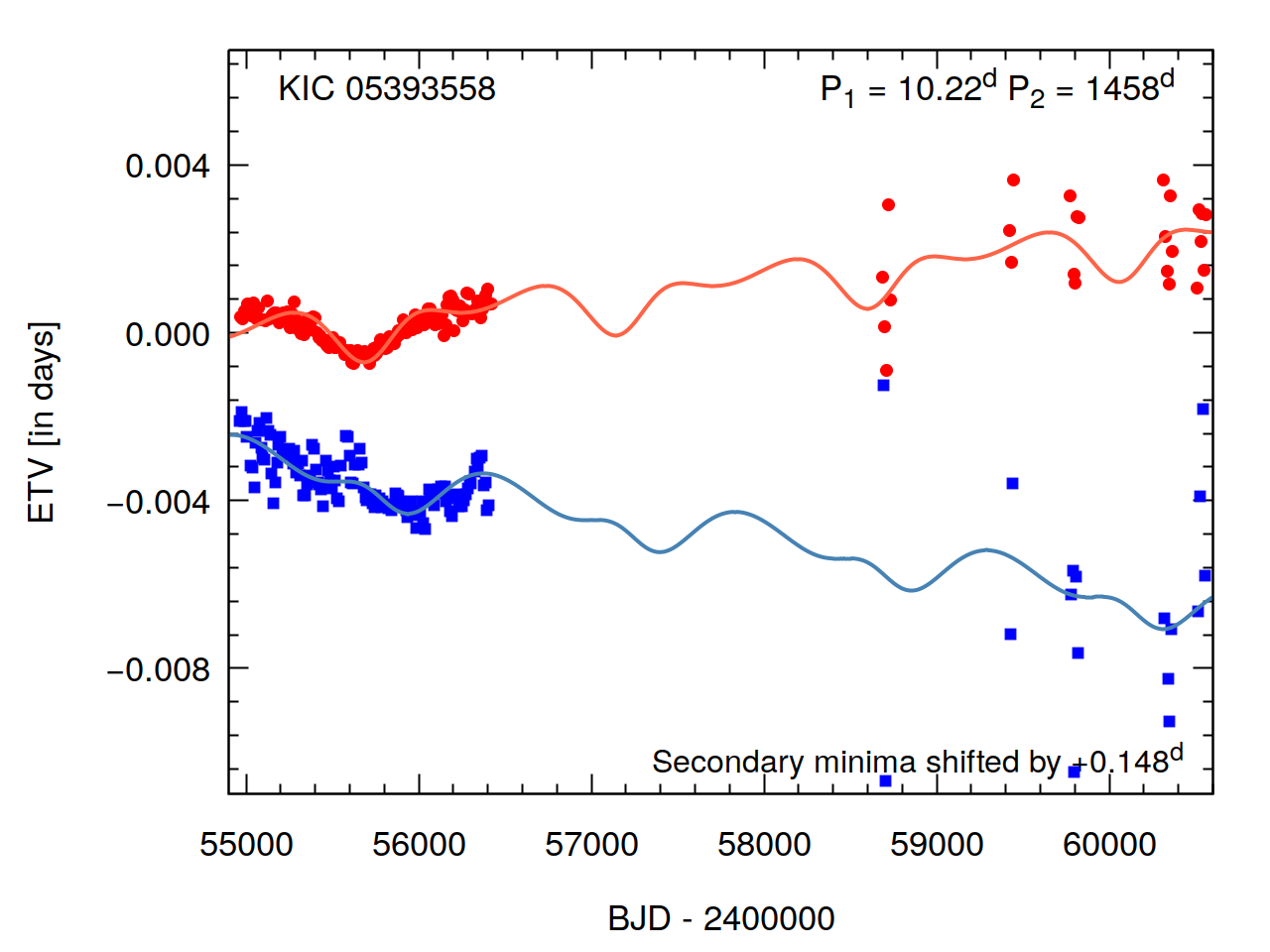}\includegraphics[width=60mm]{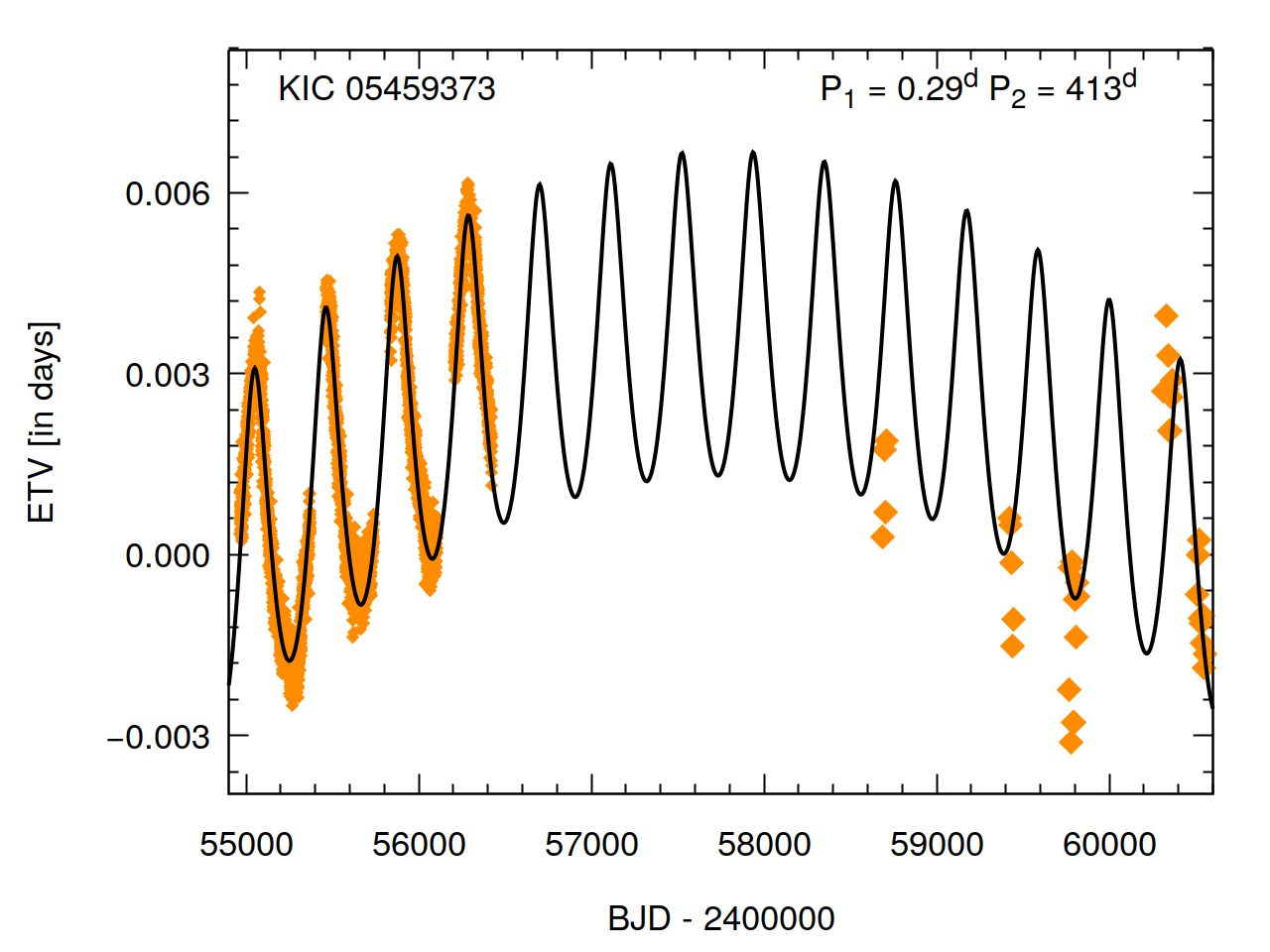}
\caption{continued.}
\end{figure*}

\addtocounter{figure}{-1}

\begin{figure*}
\includegraphics[width=60mm]{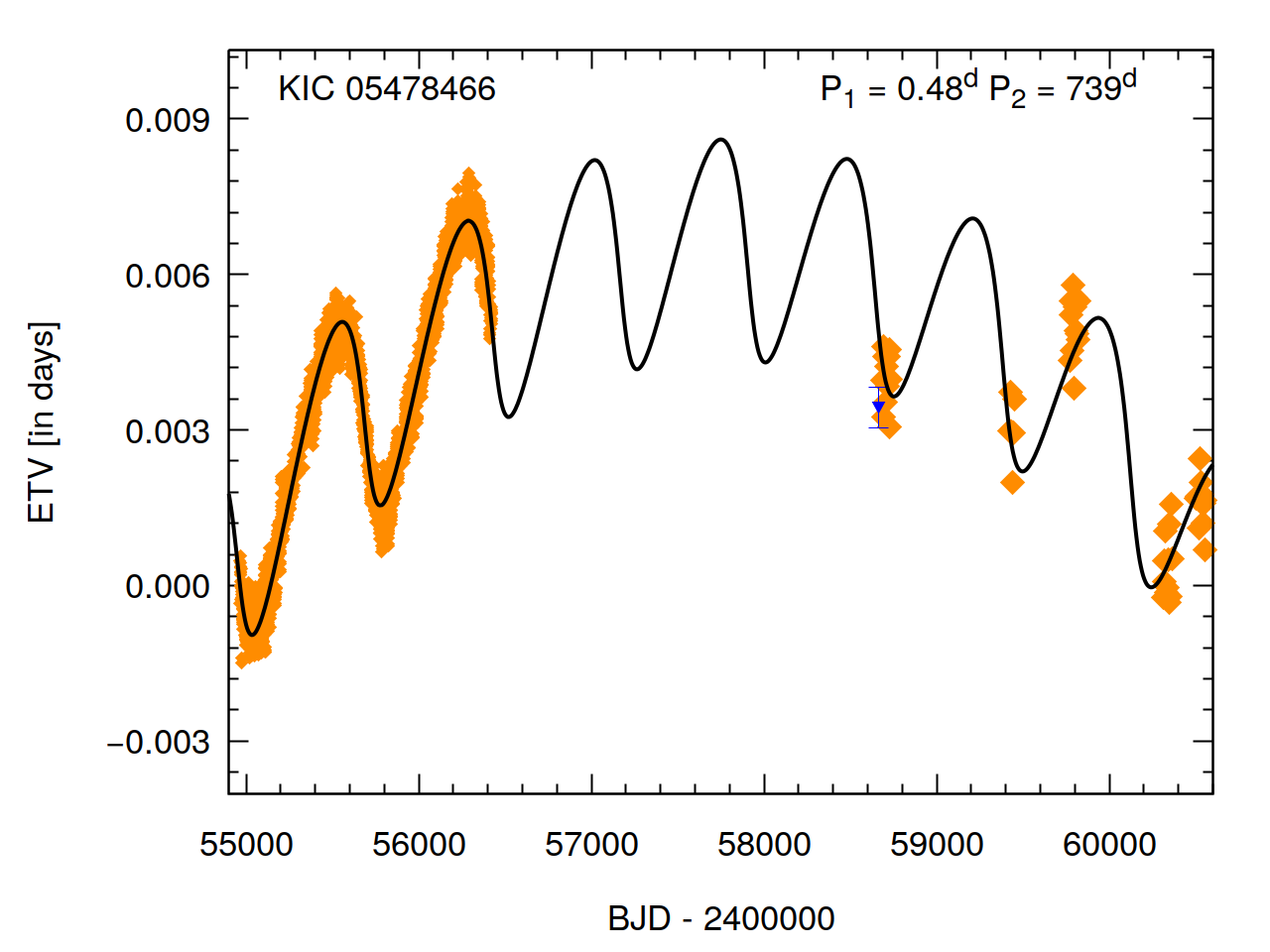}\includegraphics[width=60mm]{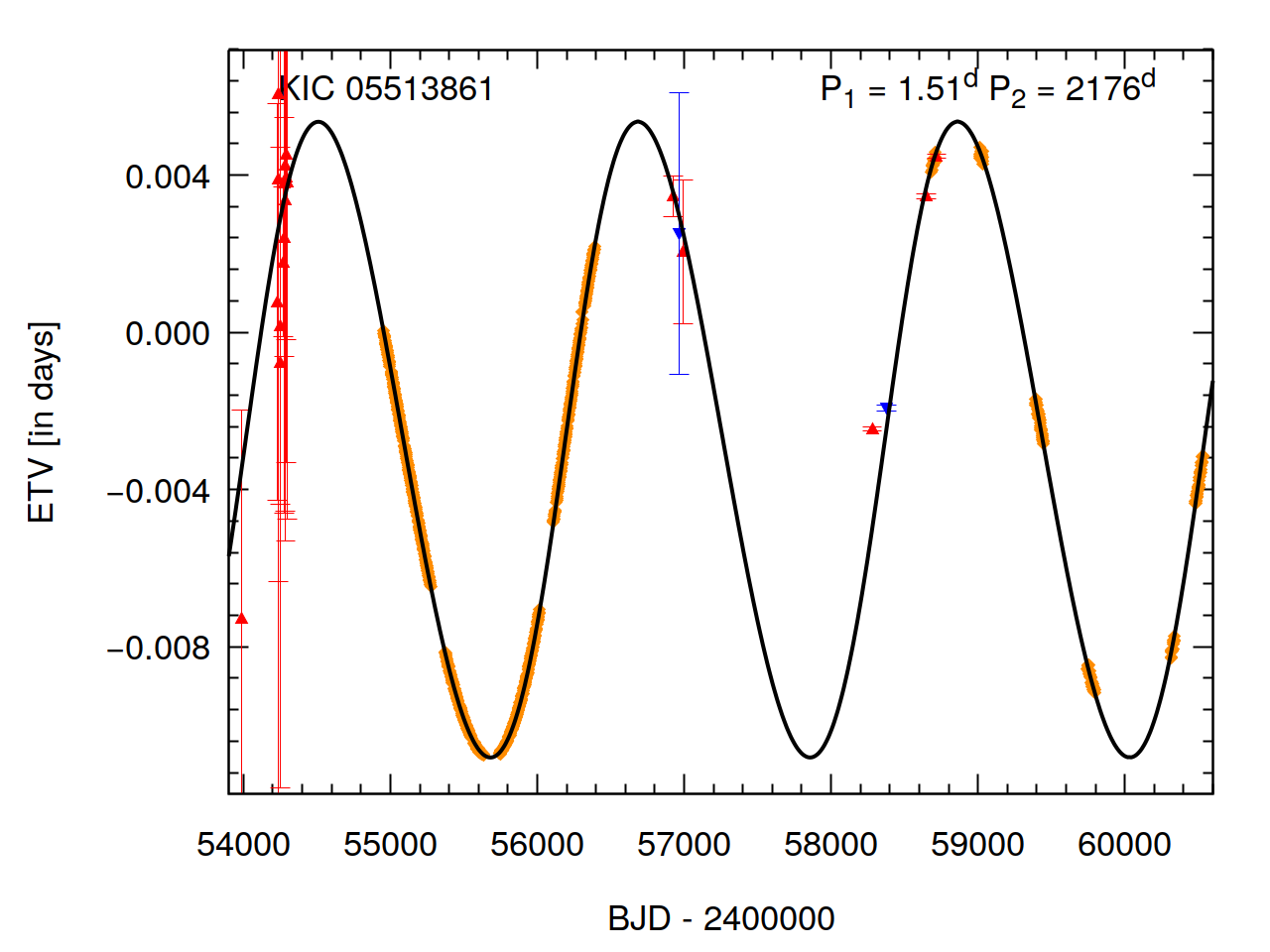}\includegraphics[width=60mm]{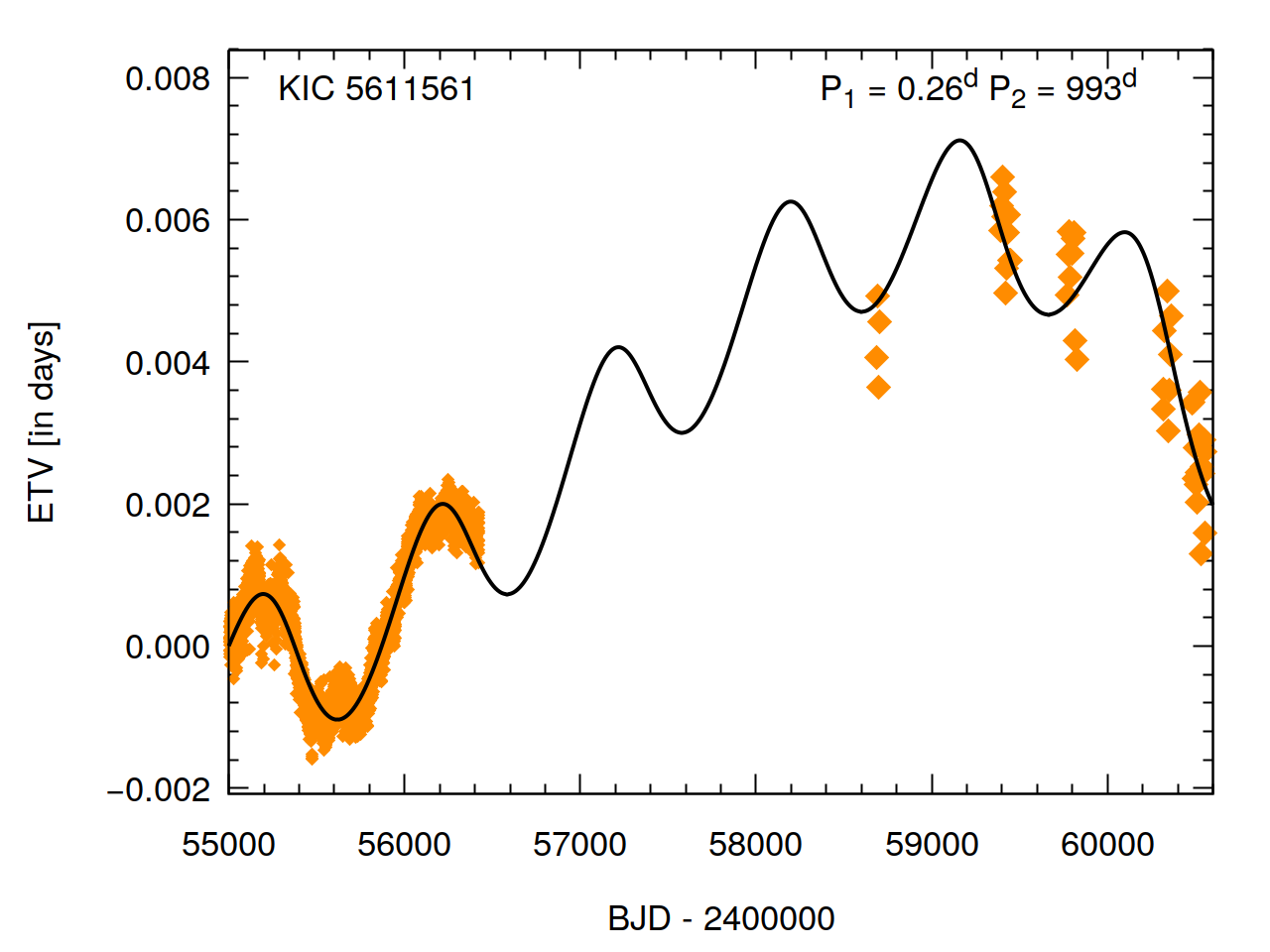}
\includegraphics[width=60mm]{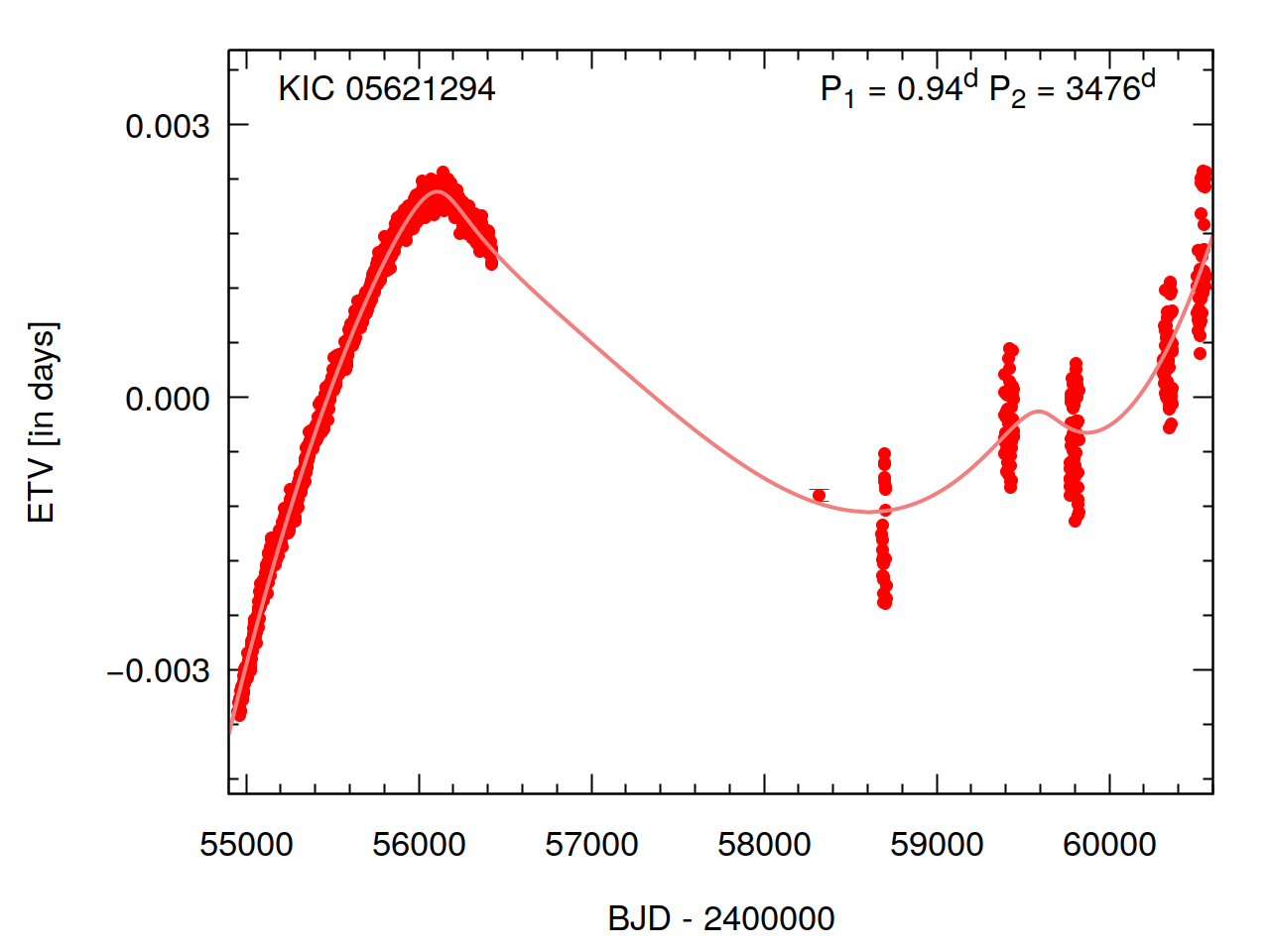}\includegraphics[width=60mm]{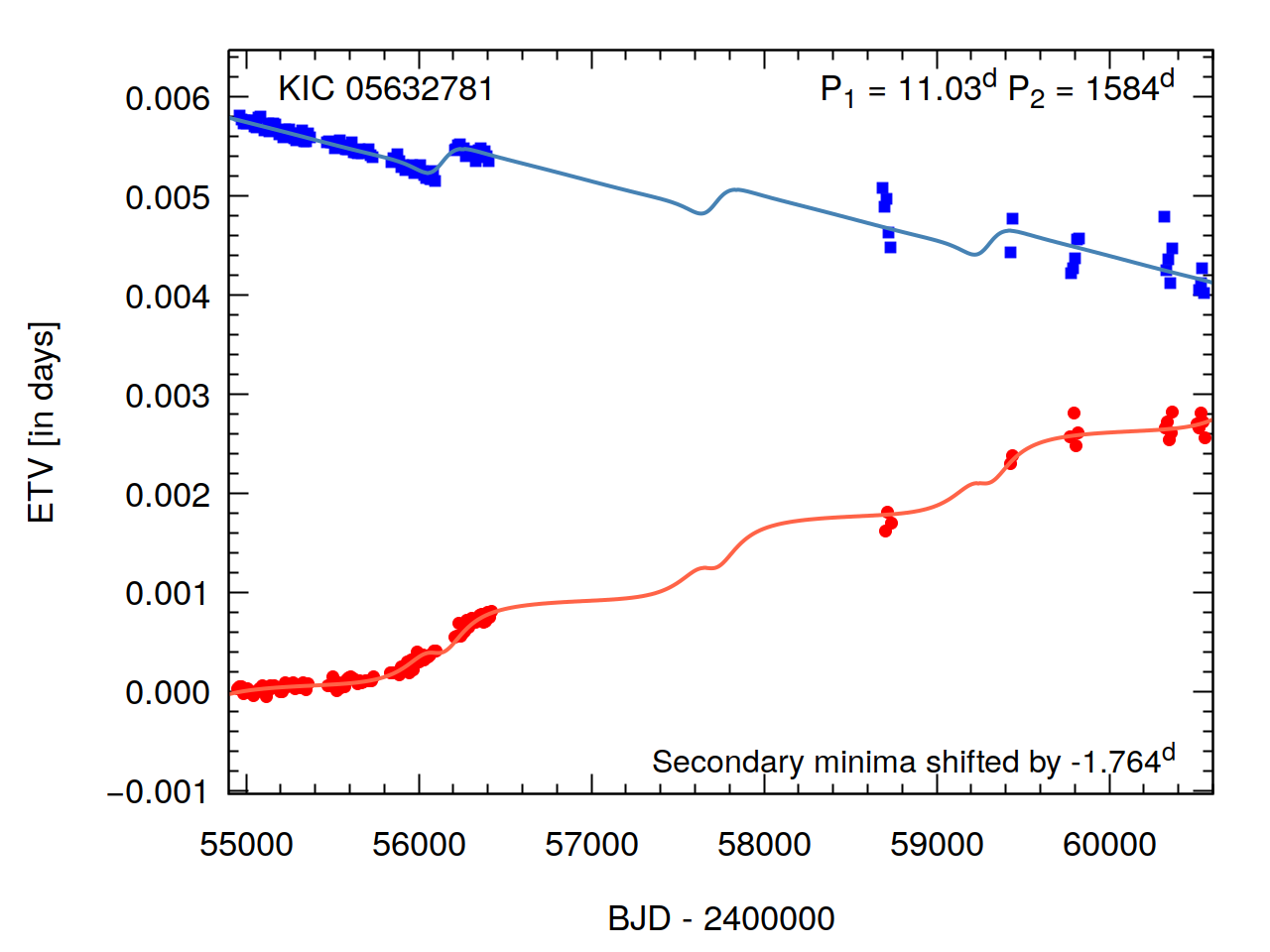}\includegraphics[width=60mm]{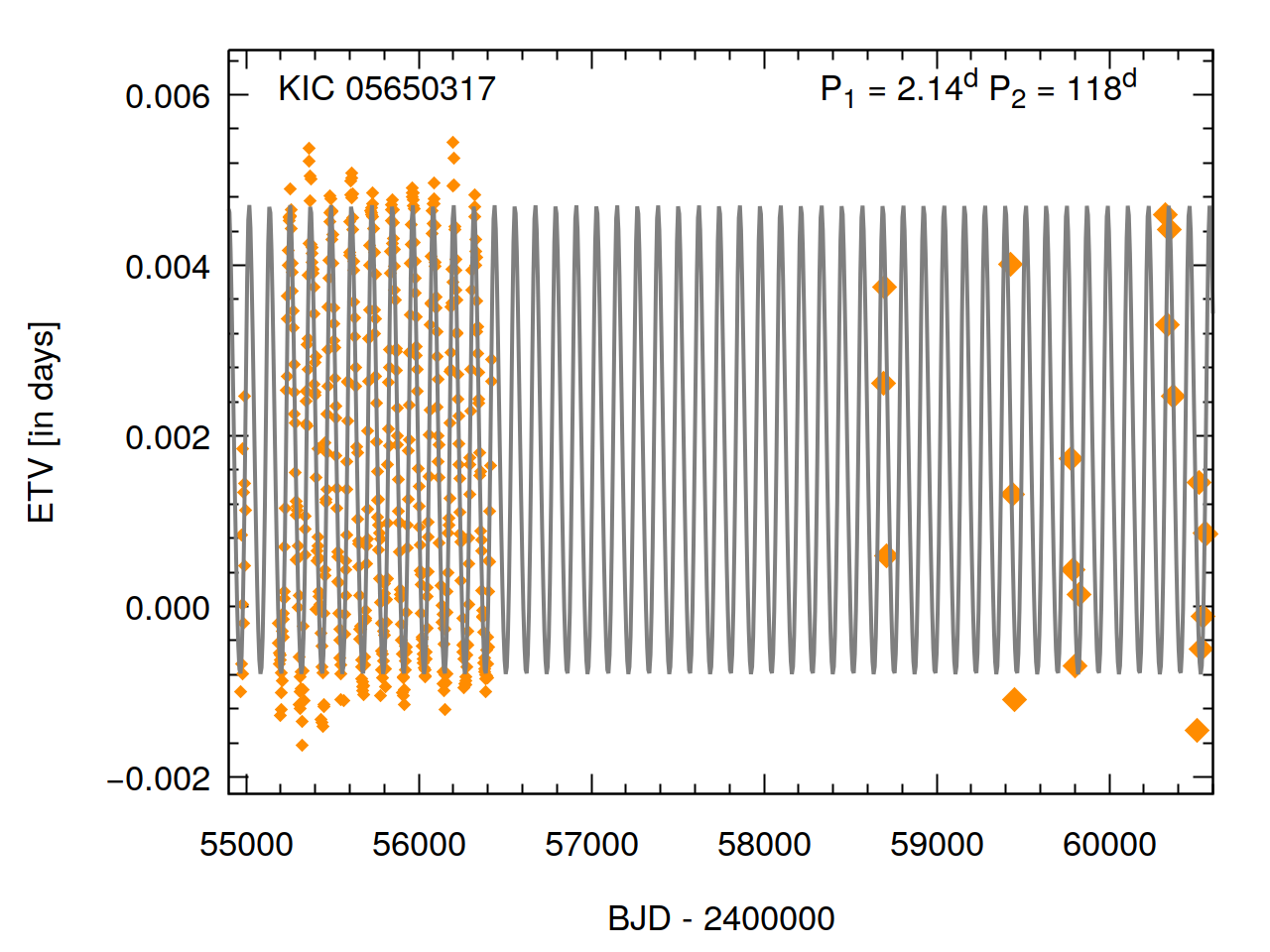}
\includegraphics[width=60mm]{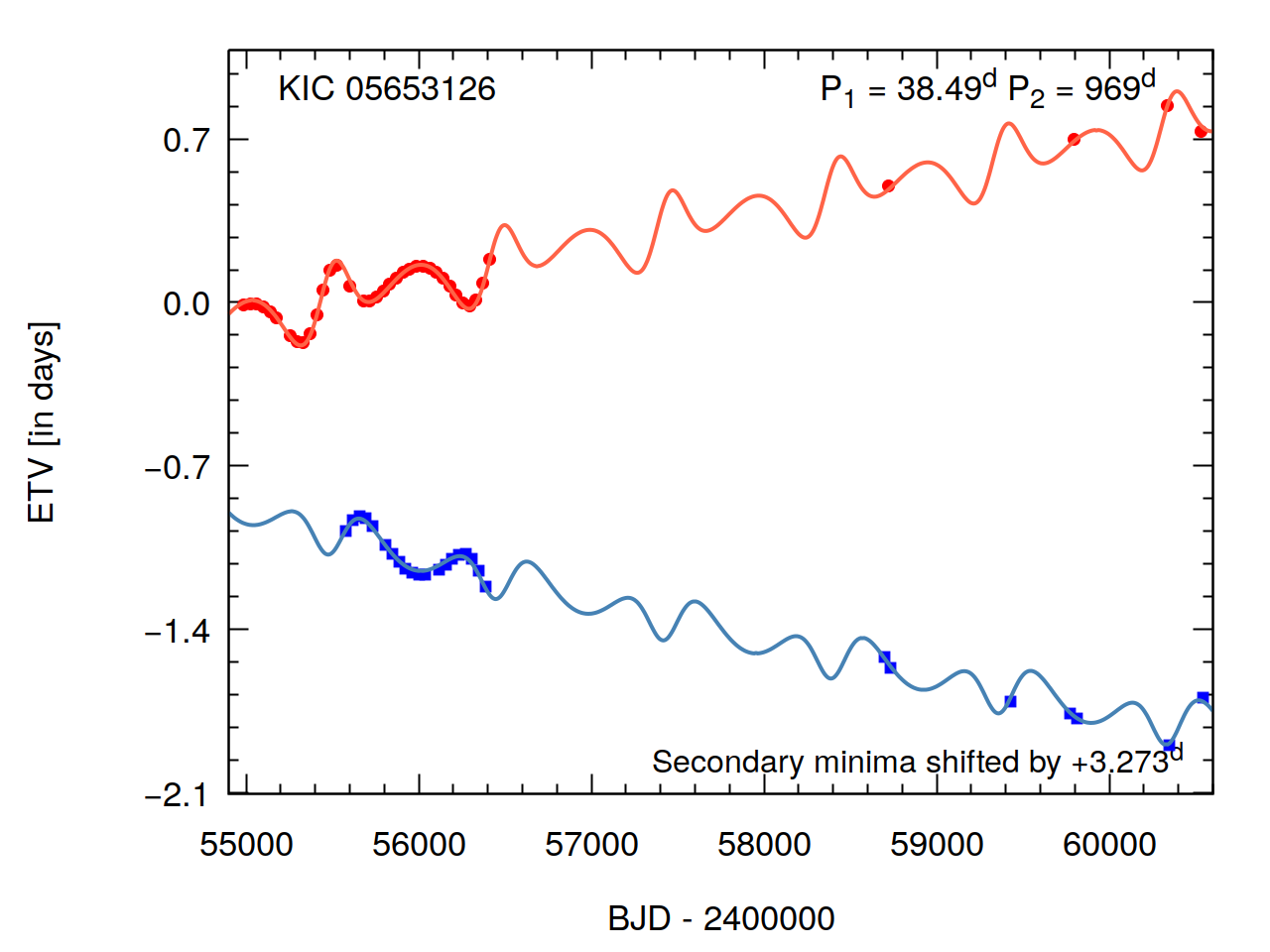}\includegraphics[width=60mm]{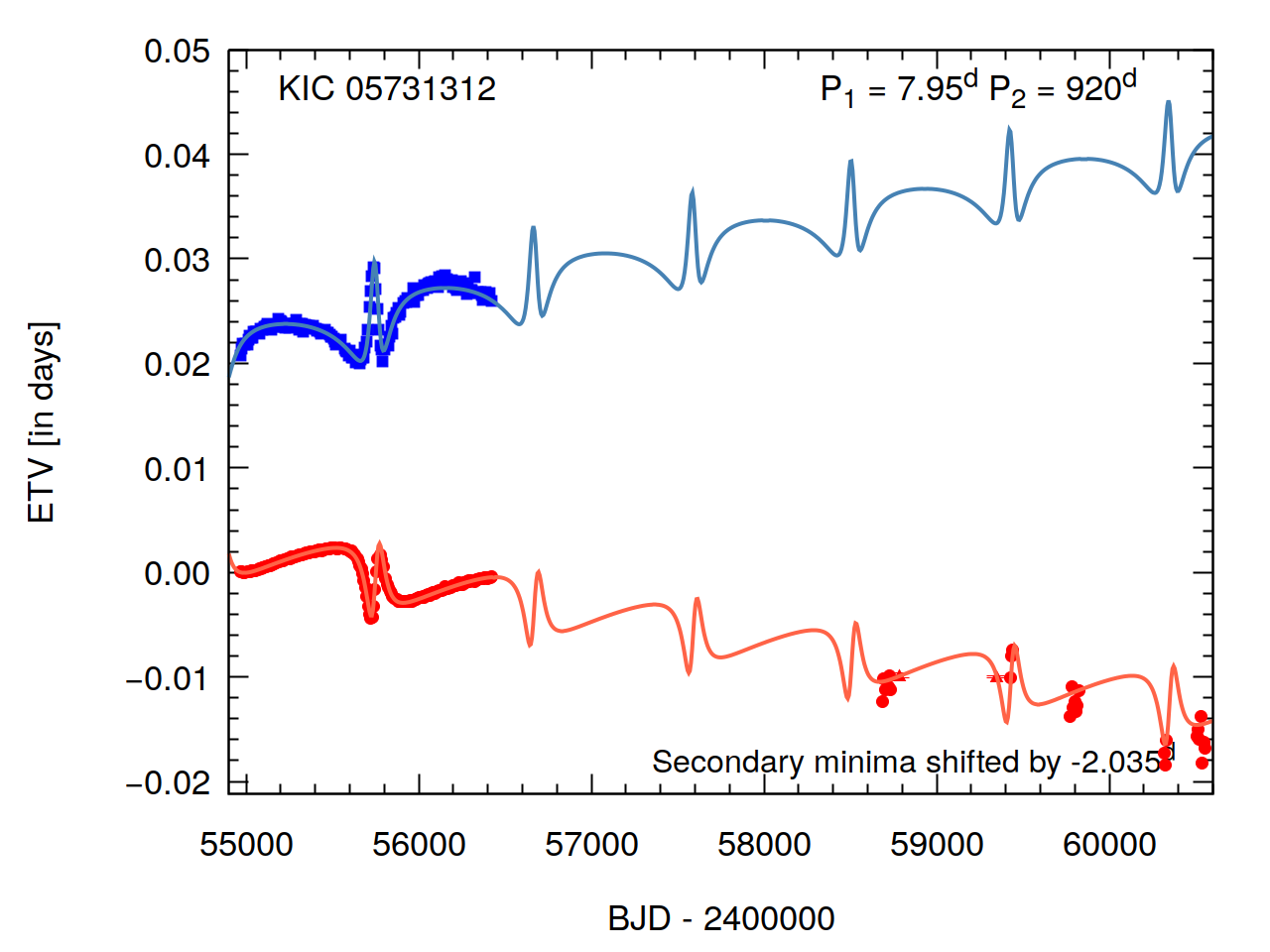}\includegraphics[width=60mm]{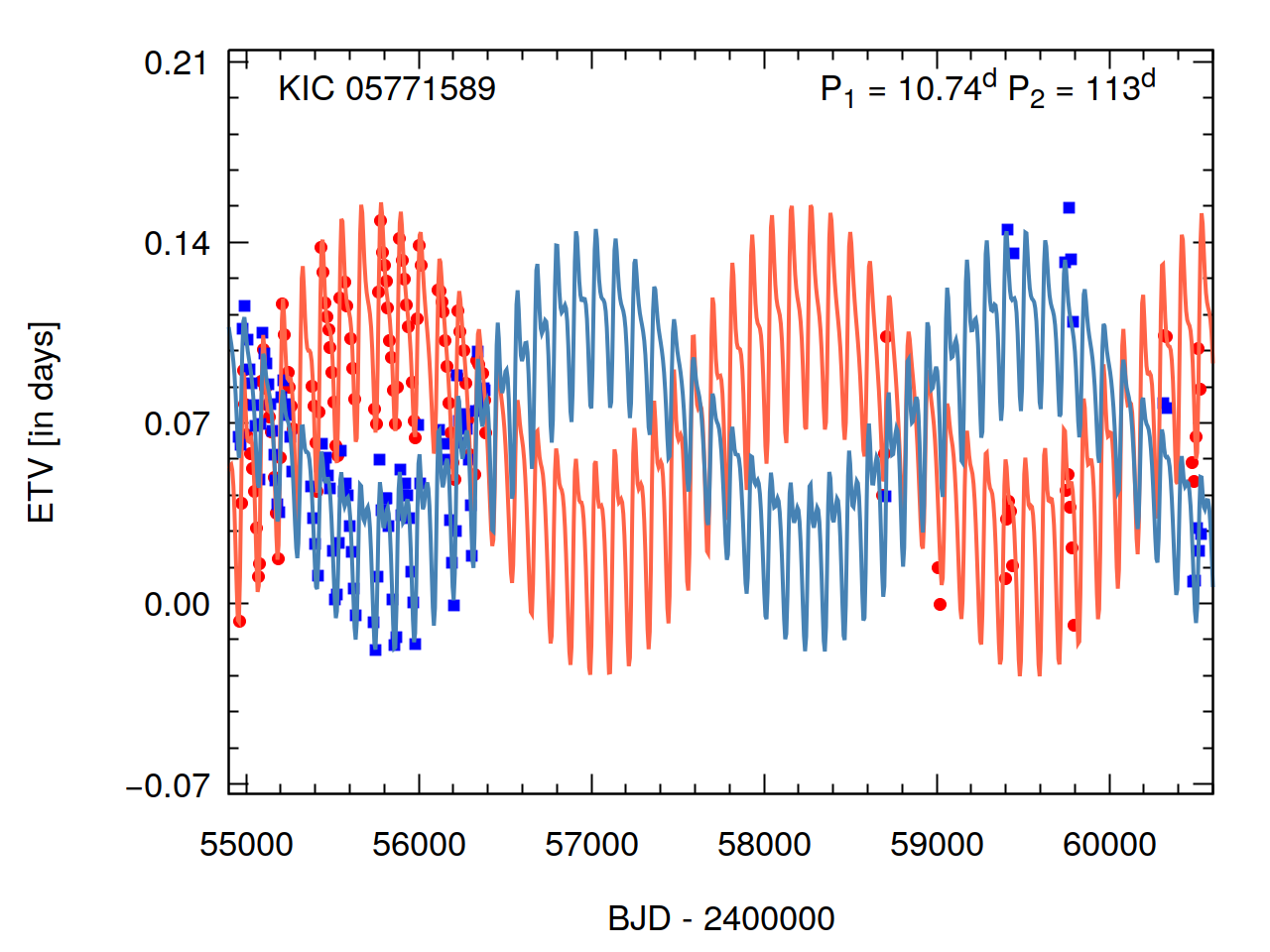}
\includegraphics[width=60mm]{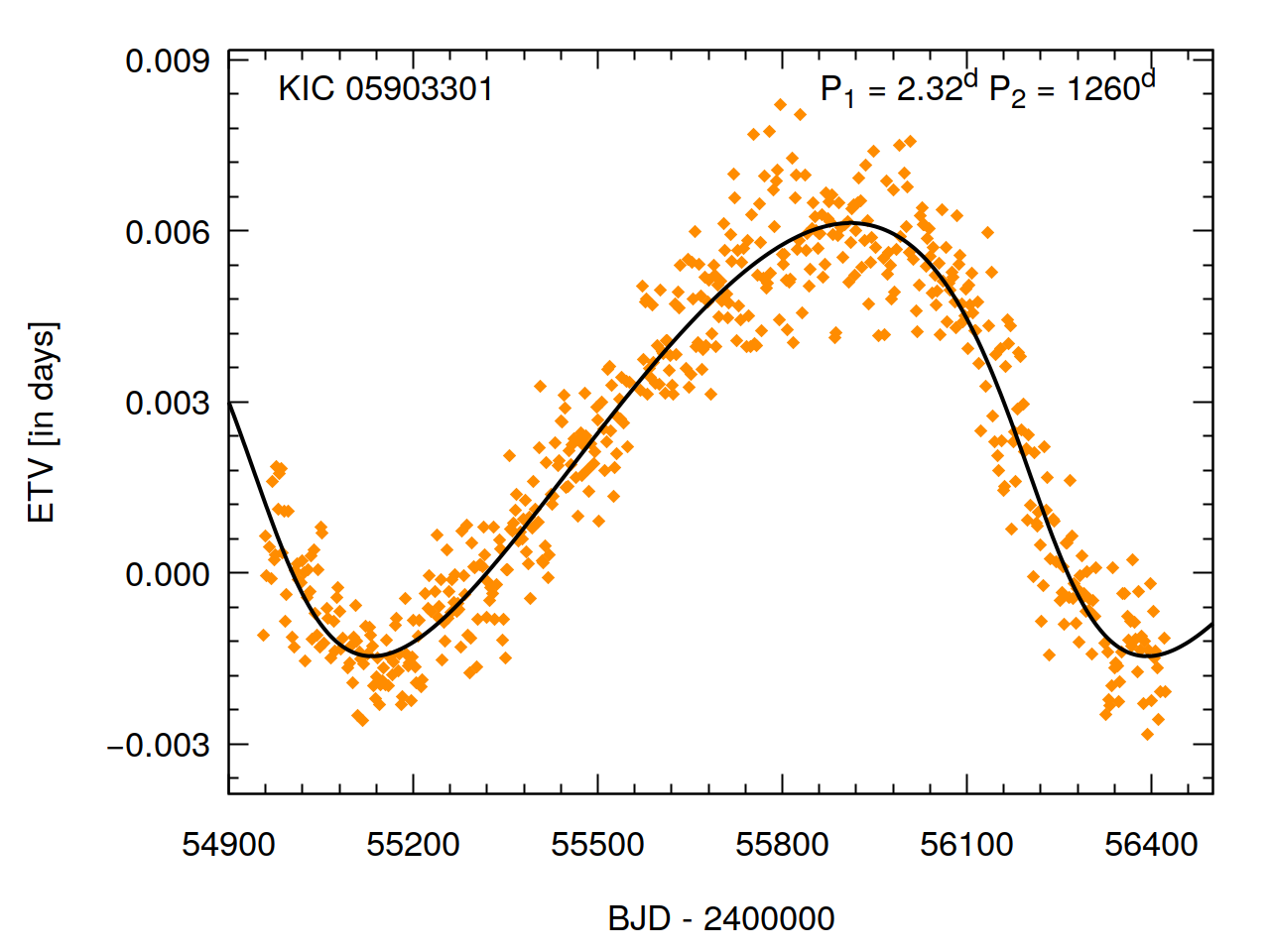}\includegraphics[width=60mm]{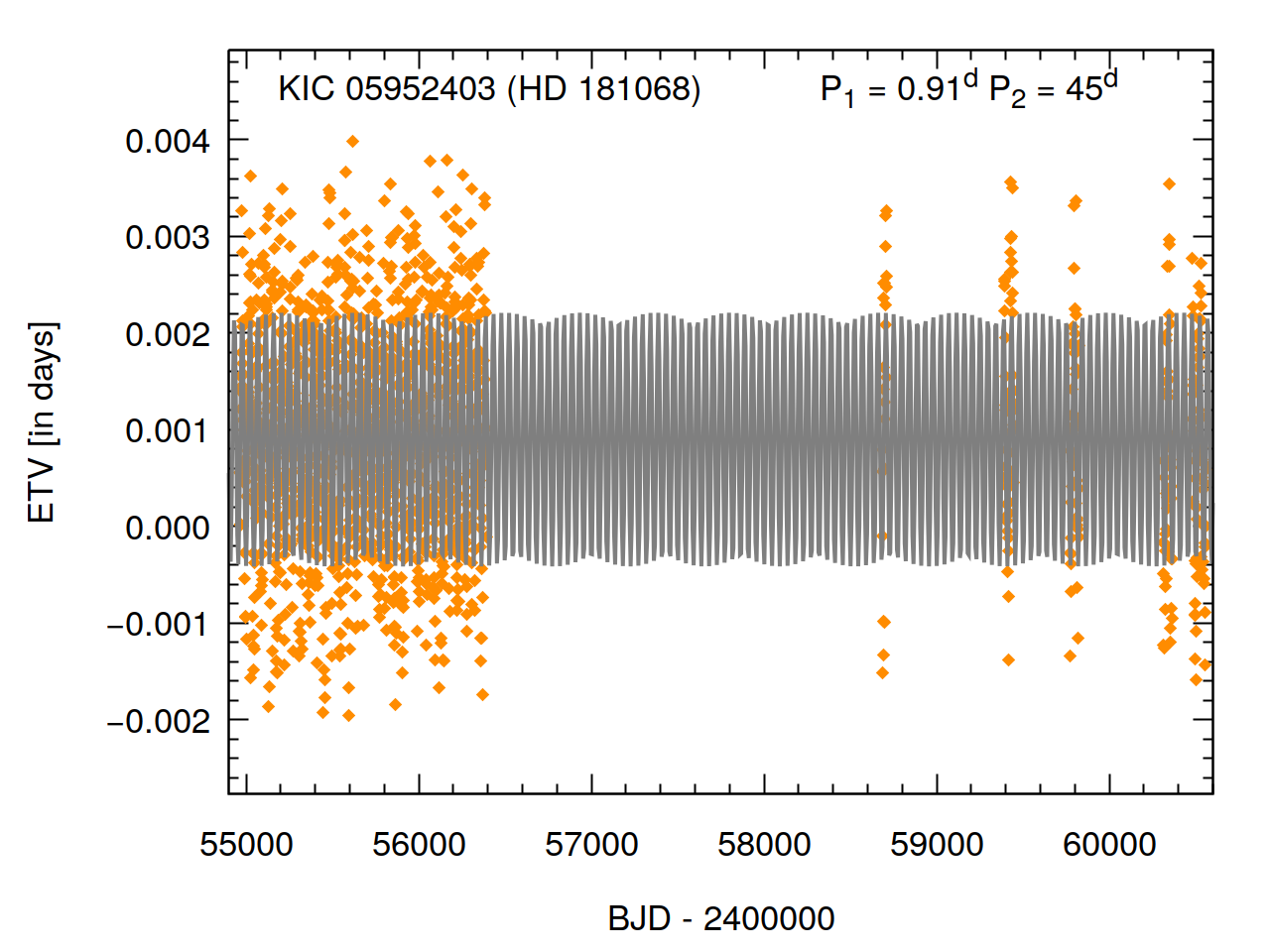}\includegraphics[width=60mm]{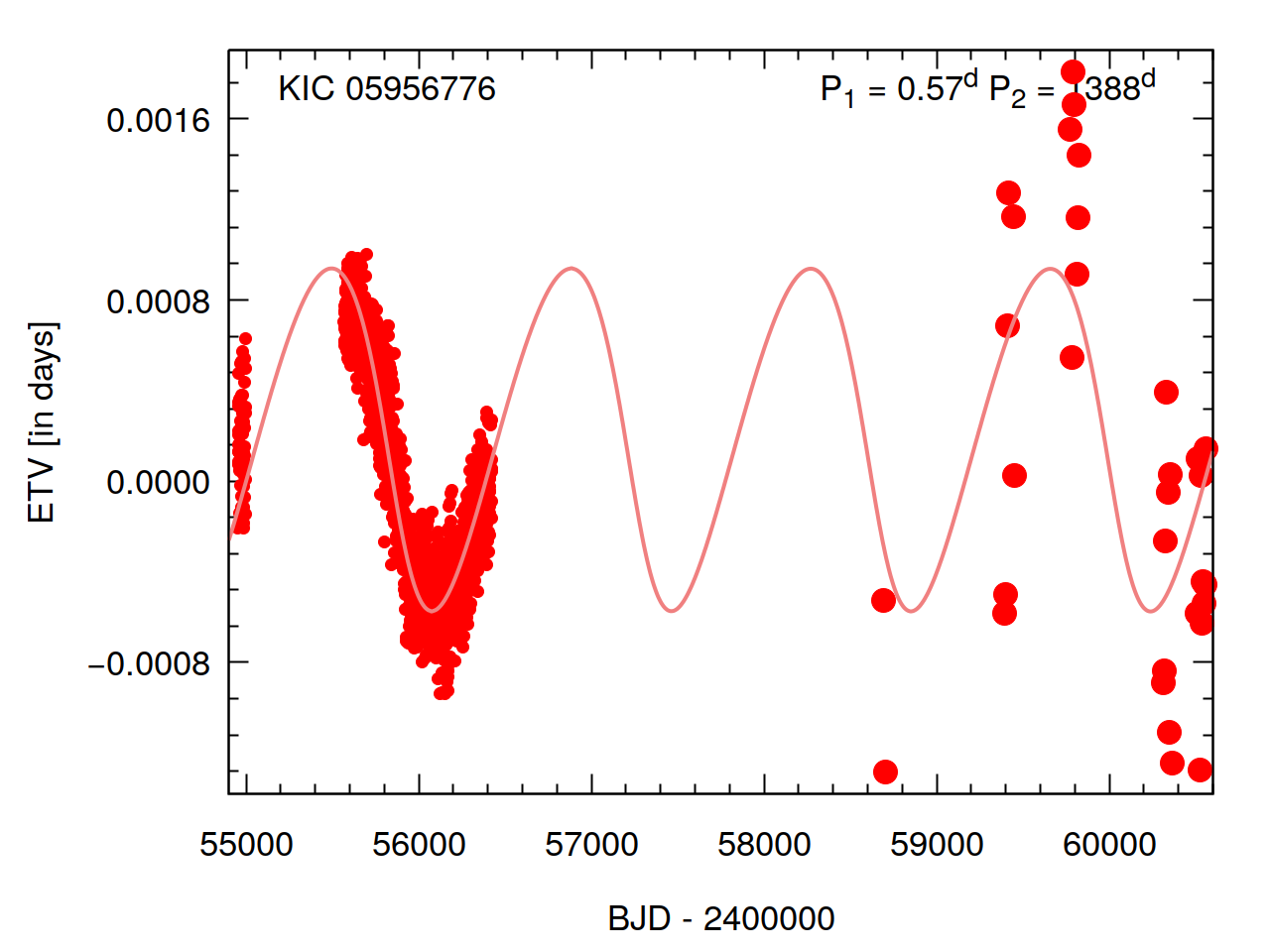}
\includegraphics[width=60mm]{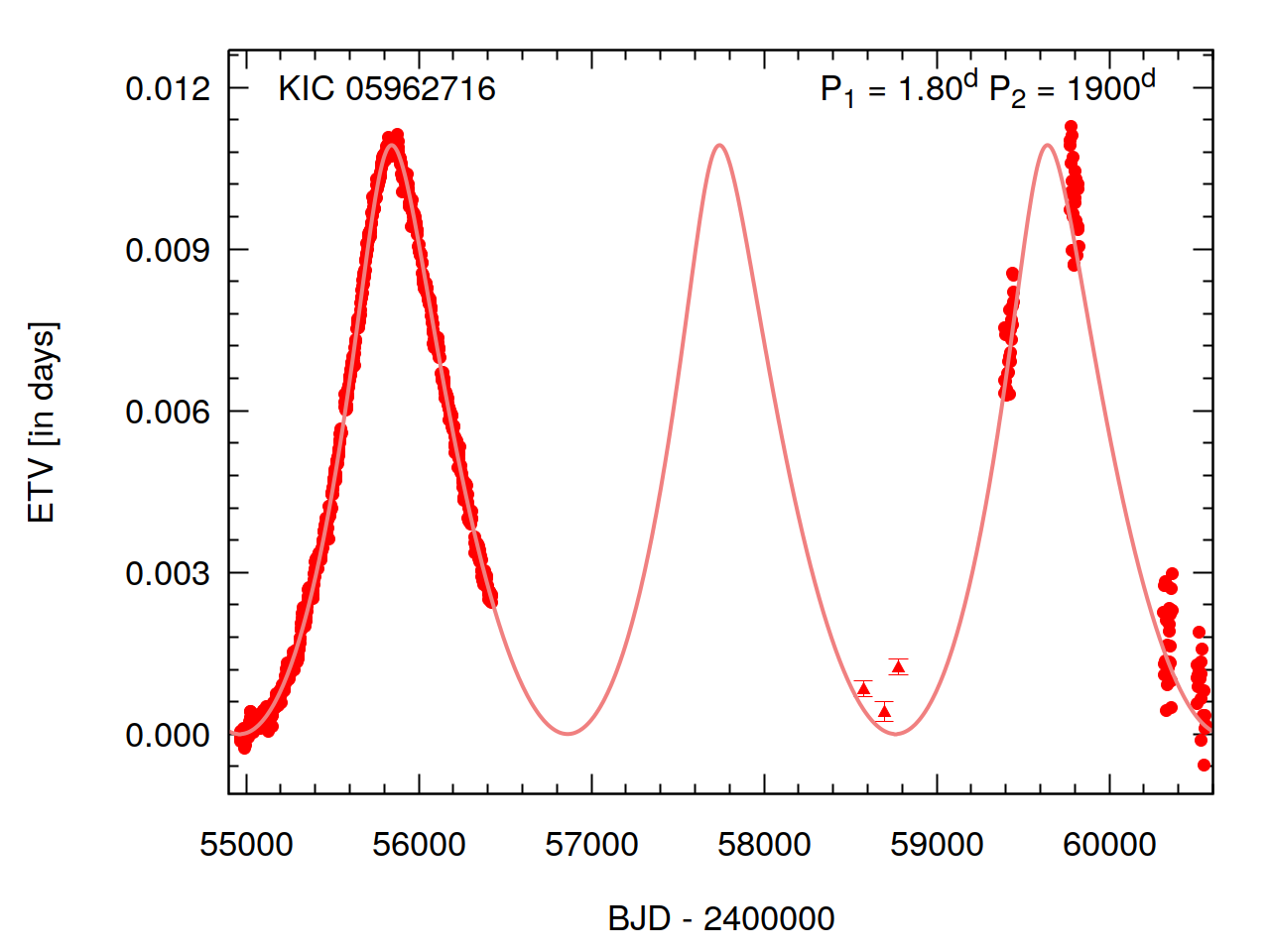}\includegraphics[width=60mm]{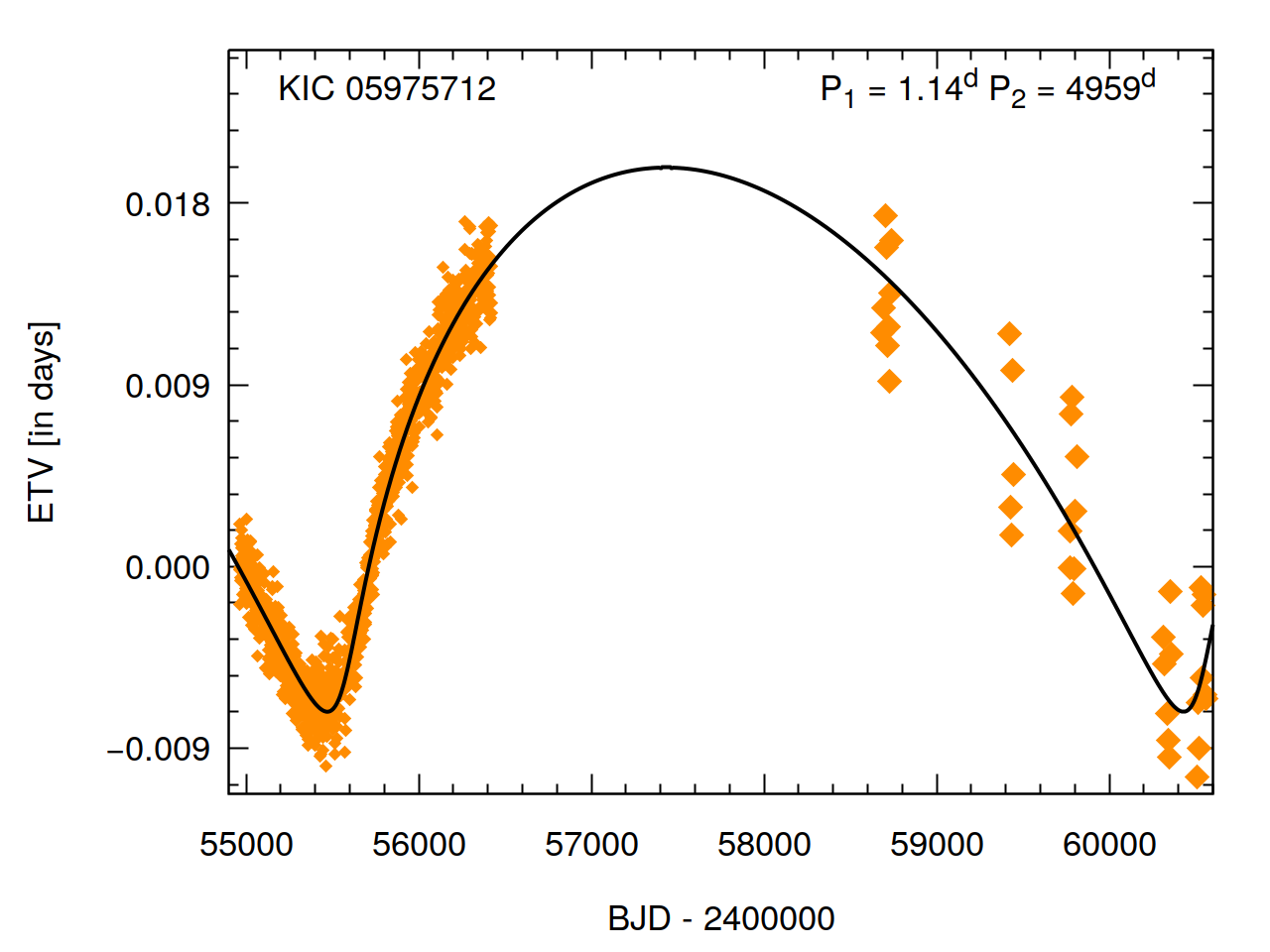}\includegraphics[width=60mm]{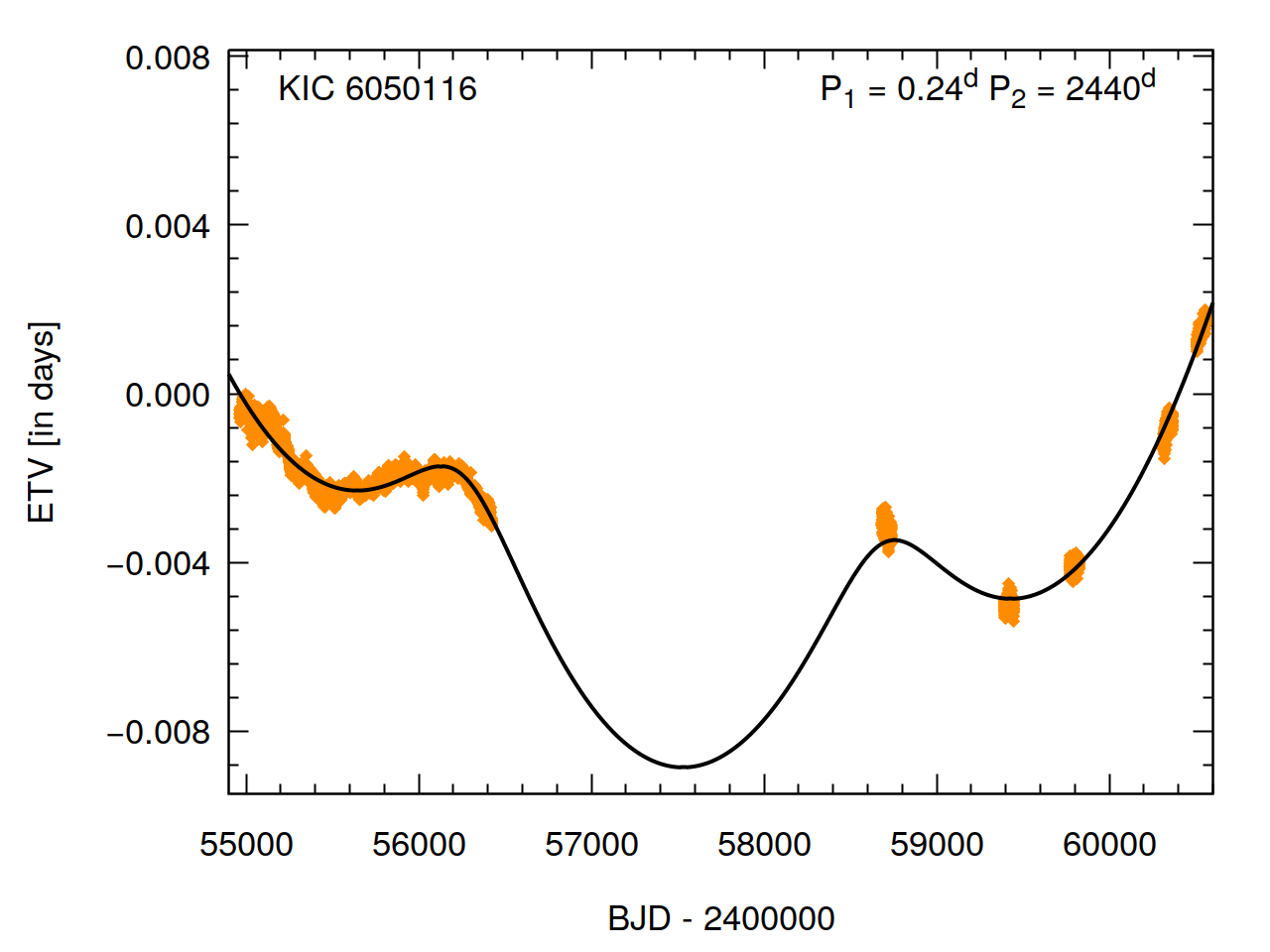}
\caption{continued.}
\end{figure*}

\addtocounter{figure}{-1}

\begin{figure*}
\includegraphics[width=60mm]{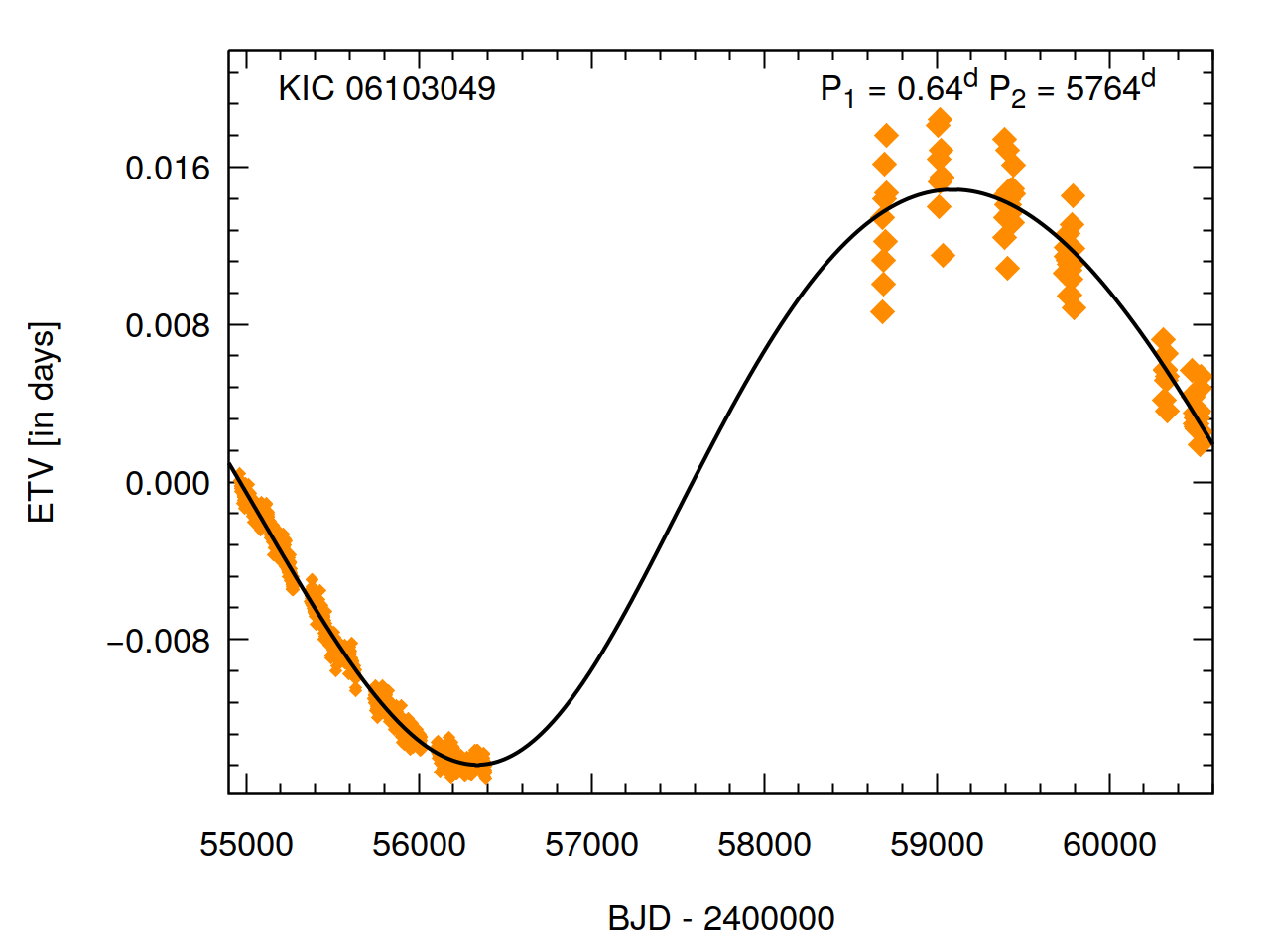}\includegraphics[width=60mm]{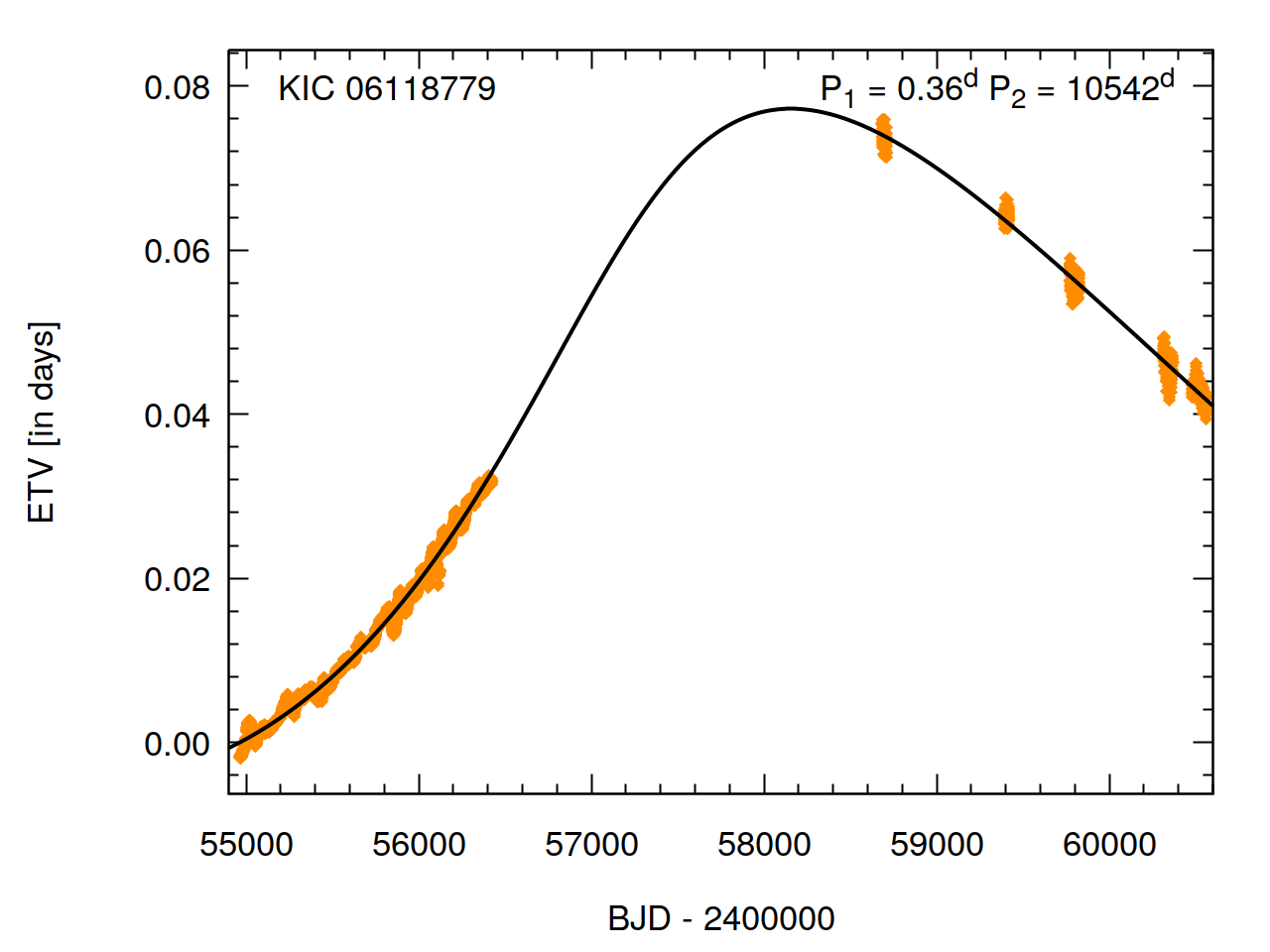}\includegraphics[width=60mm]{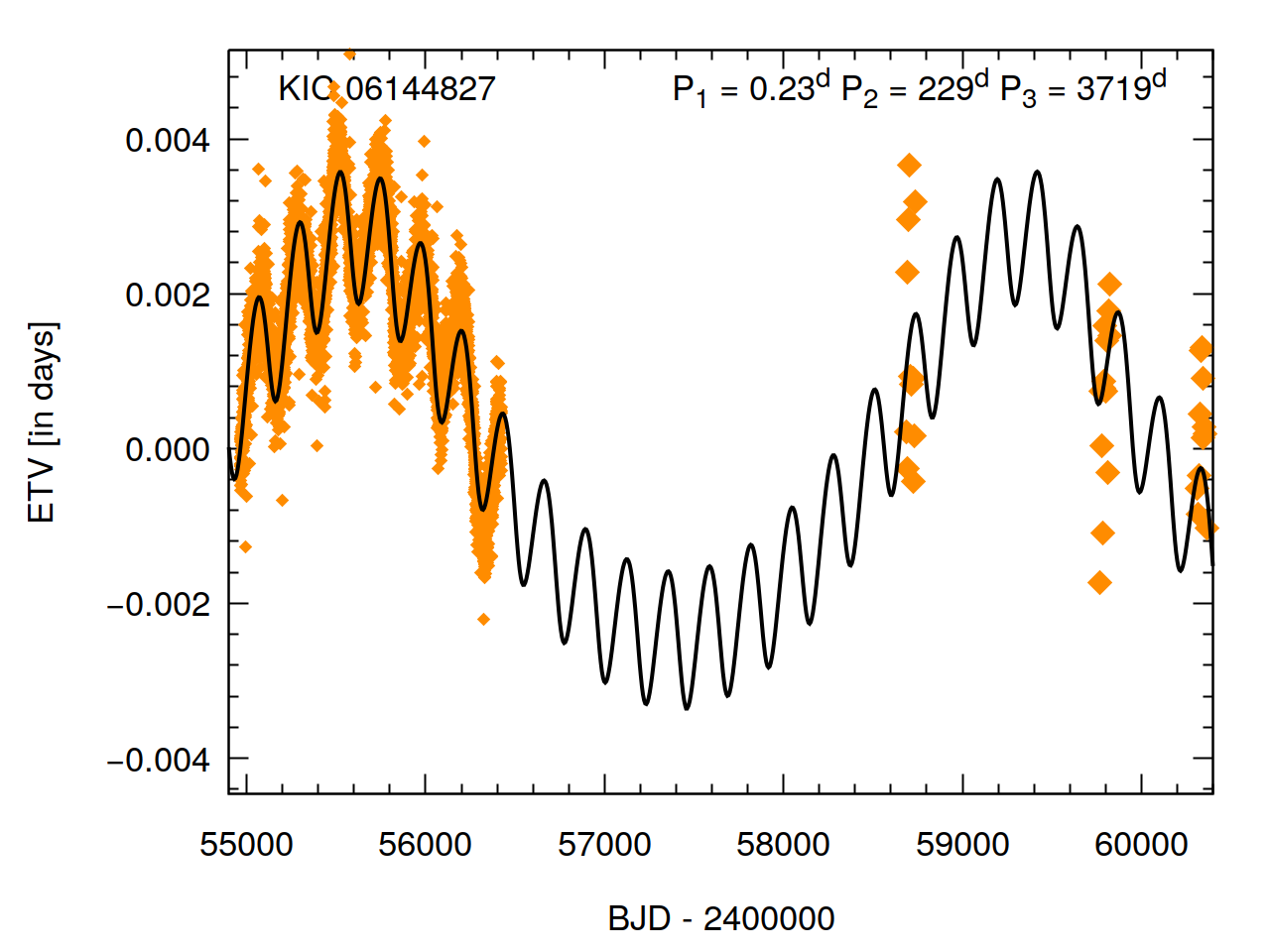}
\includegraphics[width=60mm]{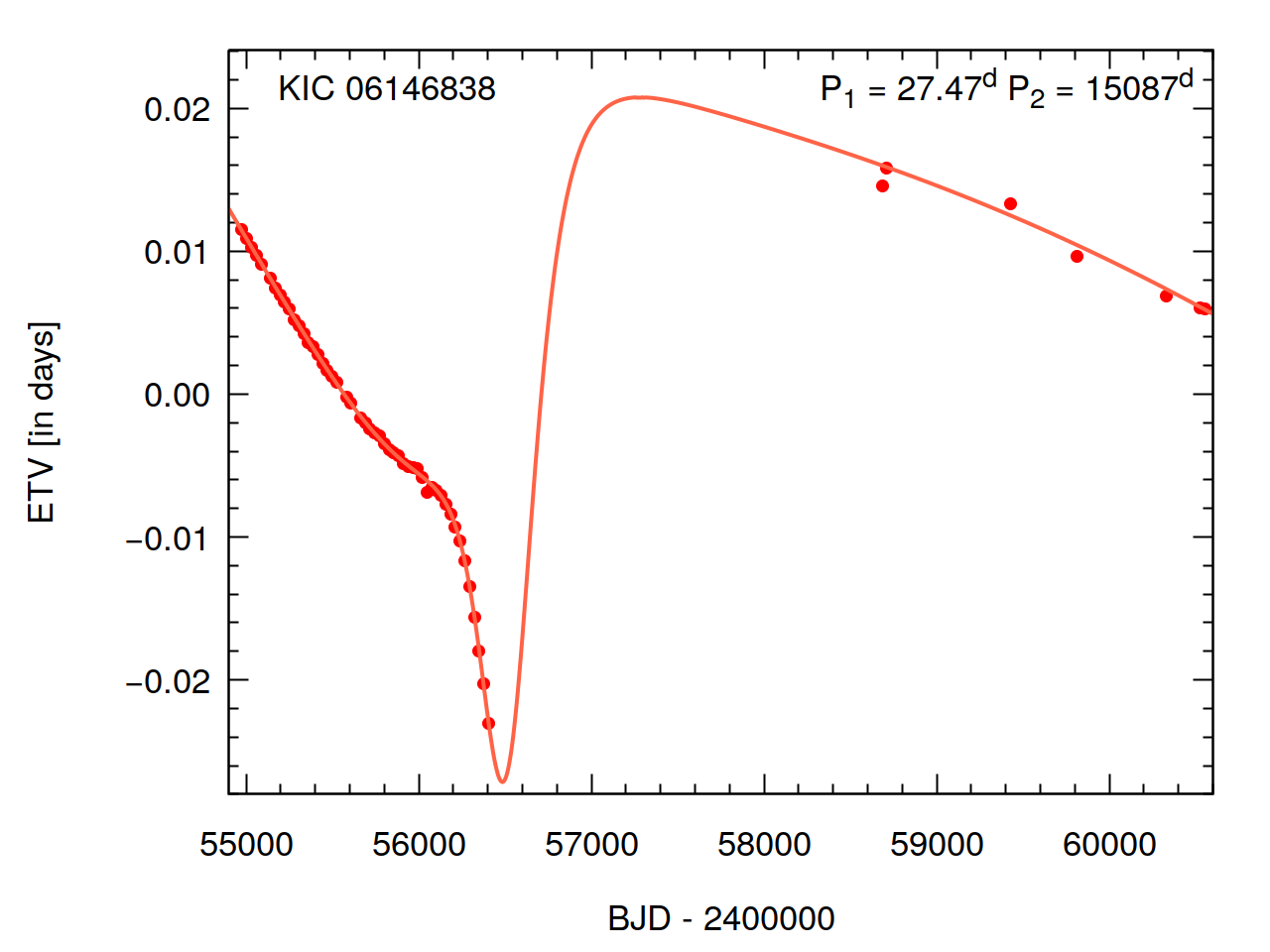}\includegraphics[width=60mm]{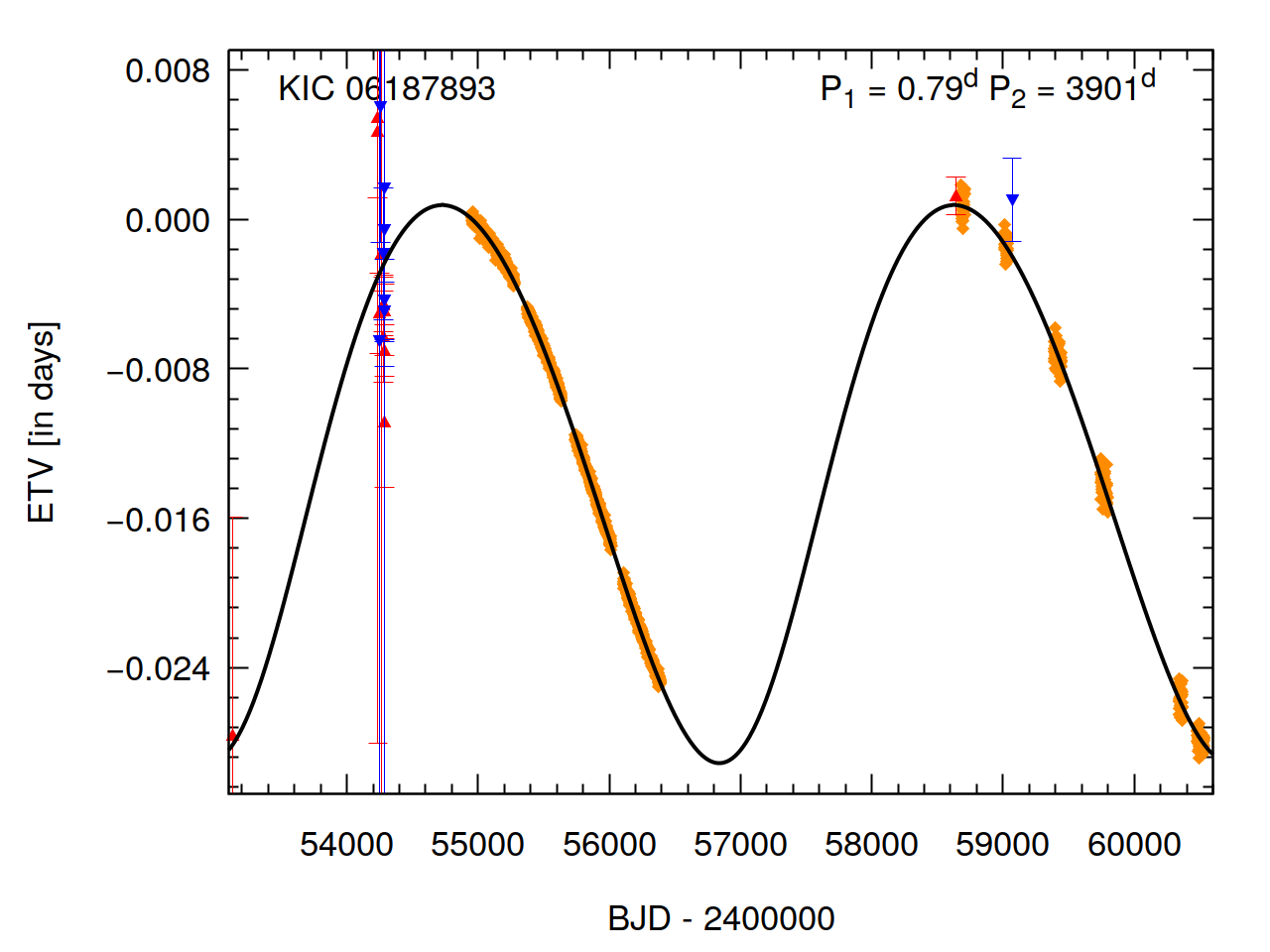}\includegraphics[width=60mm]{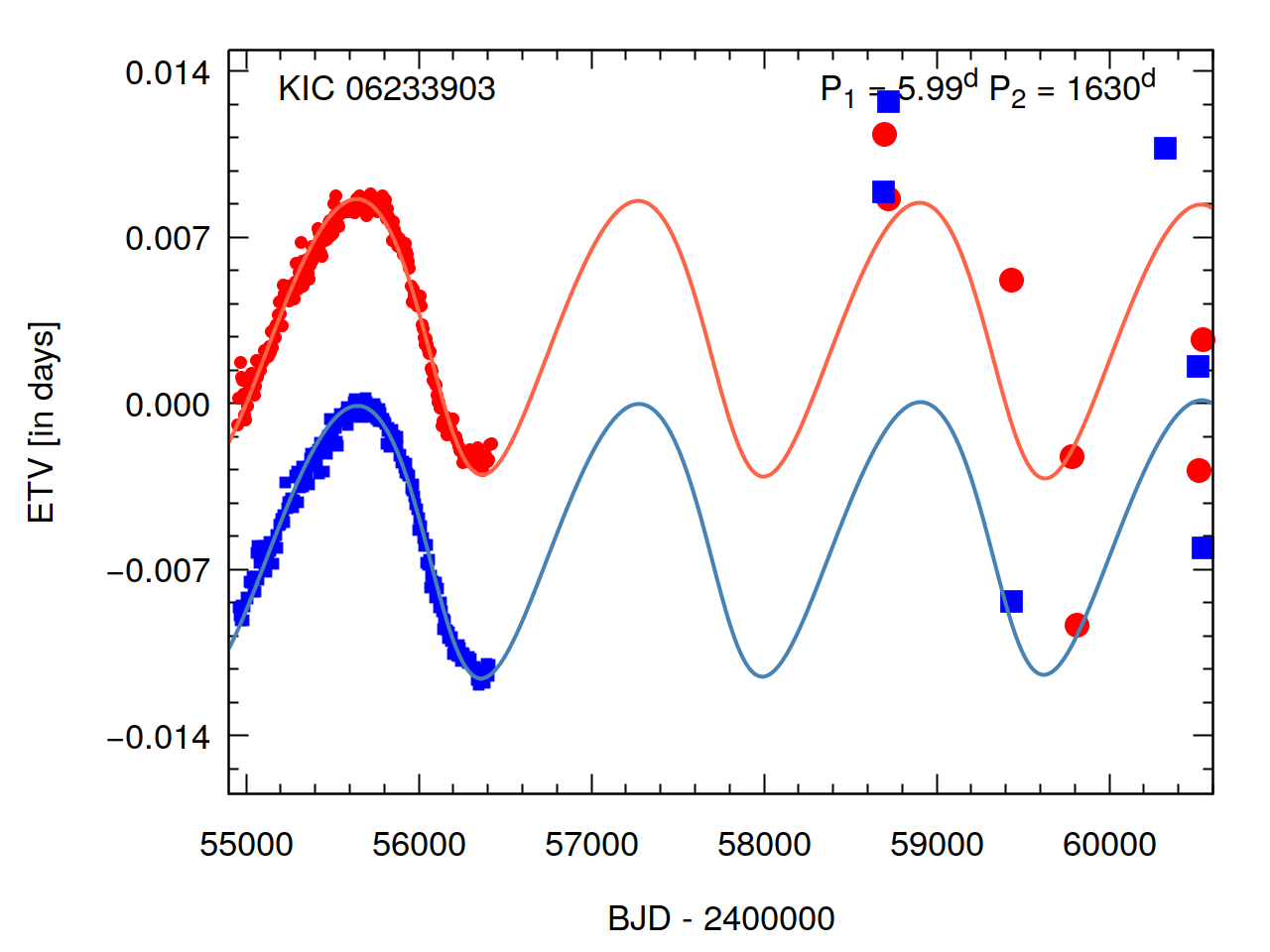}
\includegraphics[width=60mm]{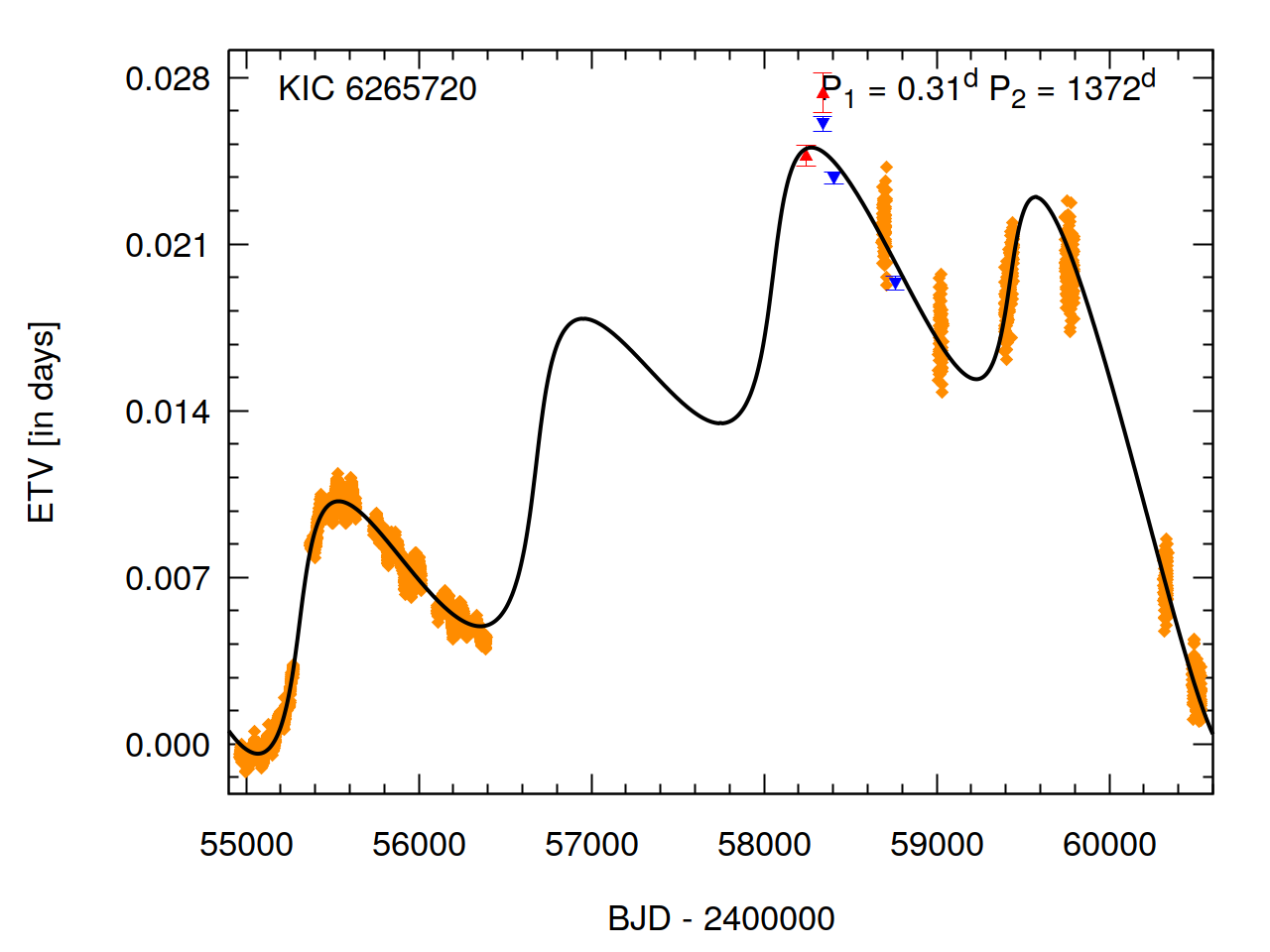}\includegraphics[width=60mm]{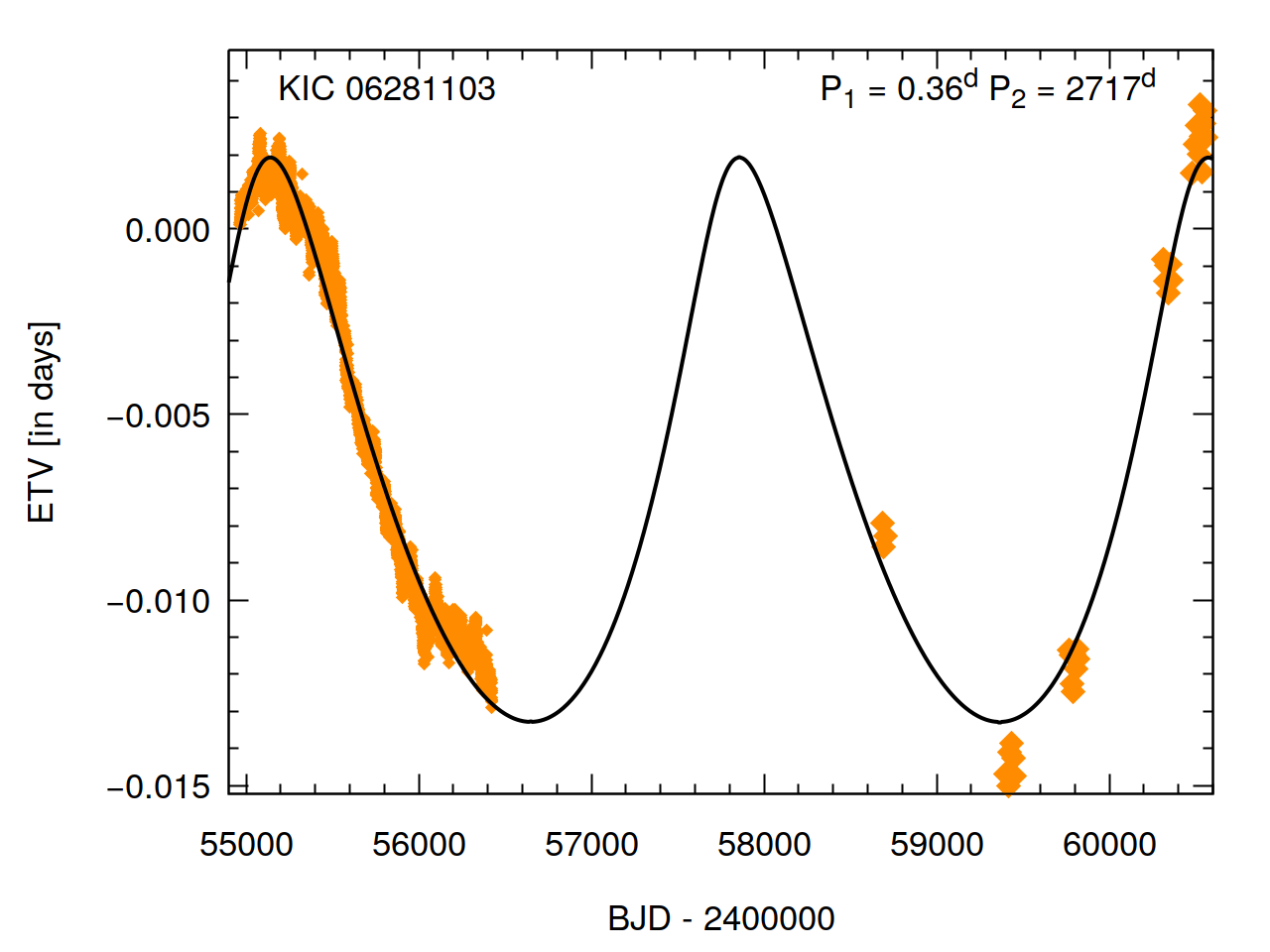}\includegraphics[width=60mm]{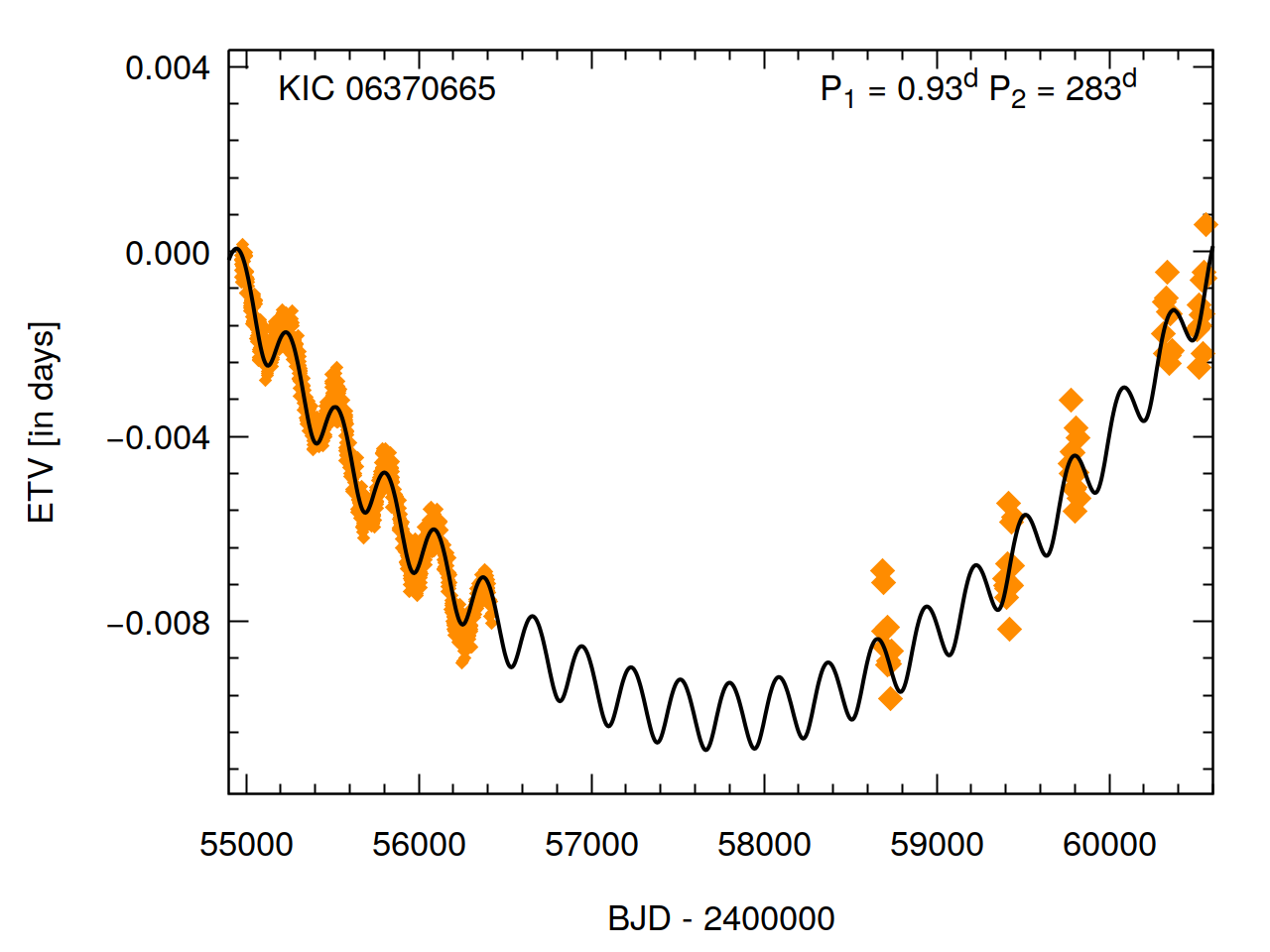}
\includegraphics[width=60mm]{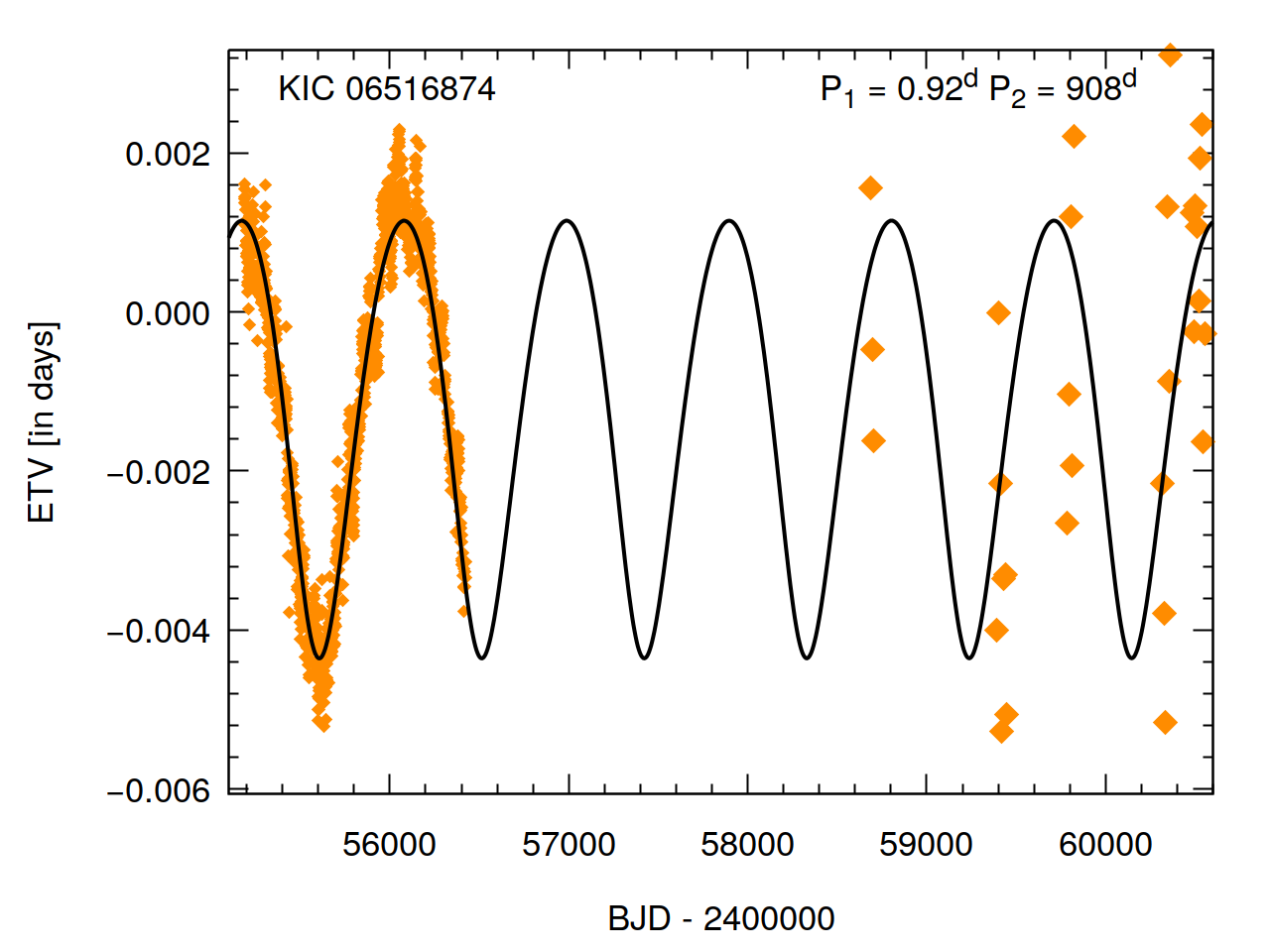}\includegraphics[width=60mm]{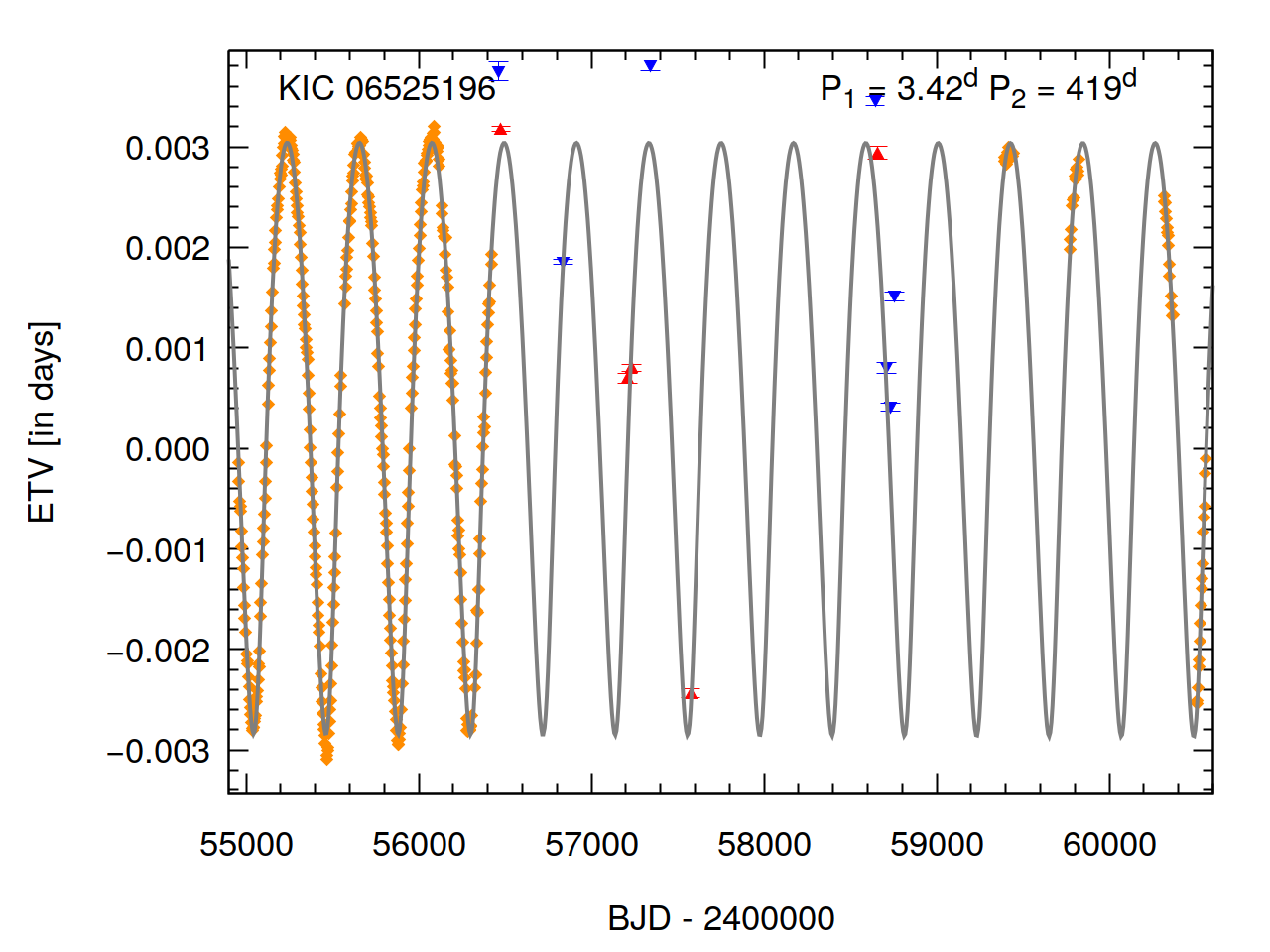}\includegraphics[width=60mm]{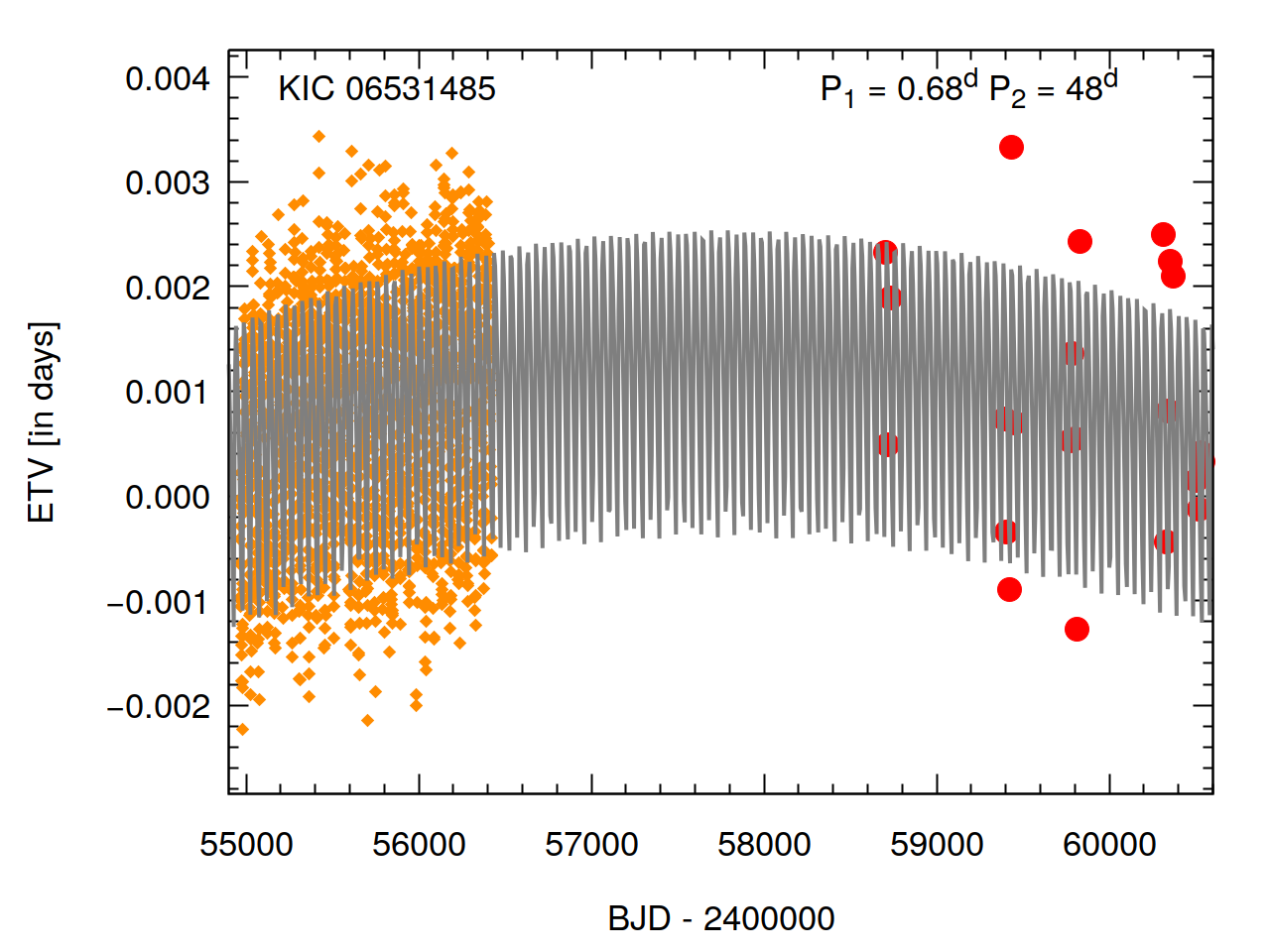}
\includegraphics[width=60mm]{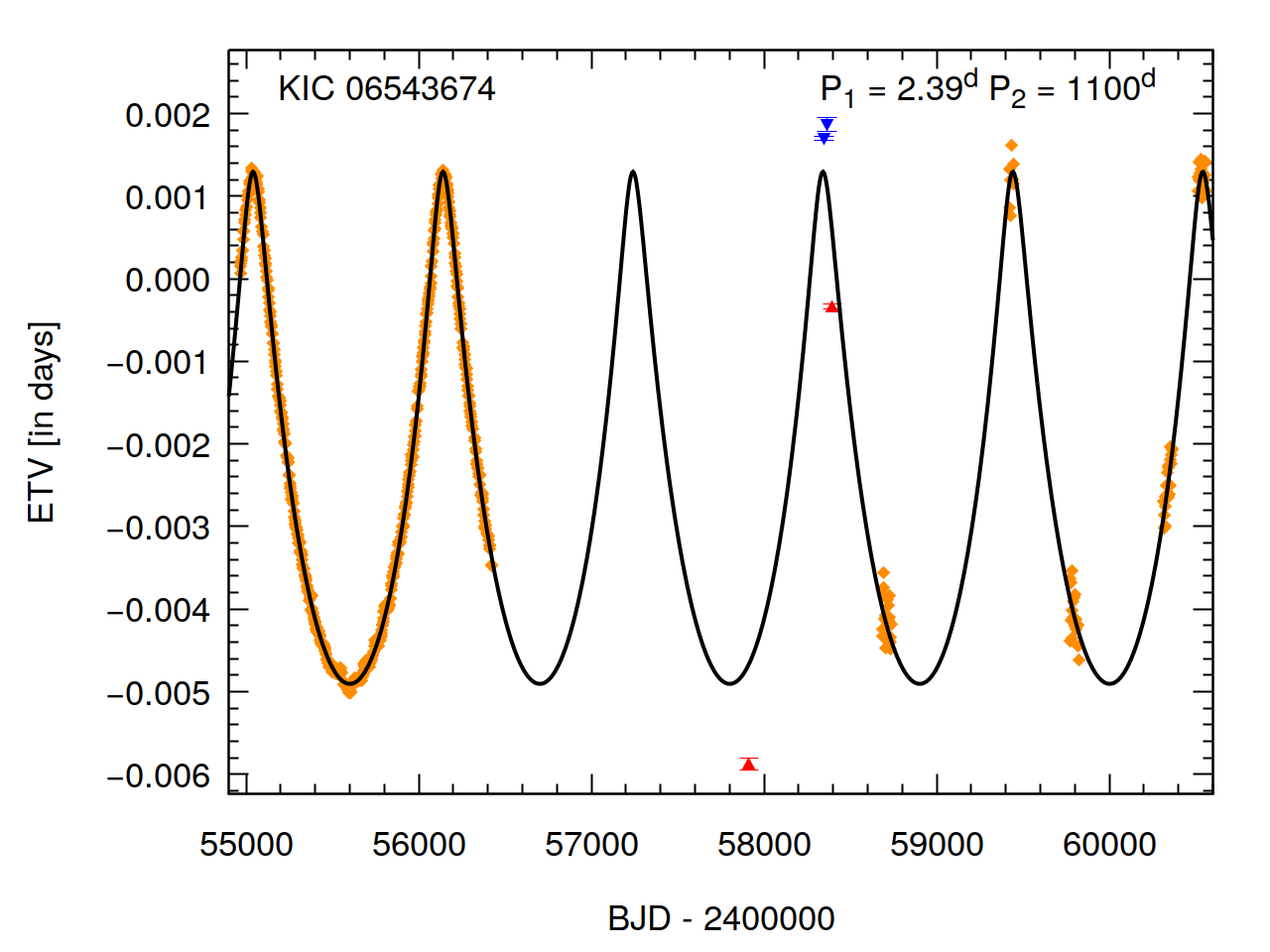}\includegraphics[width=60mm]{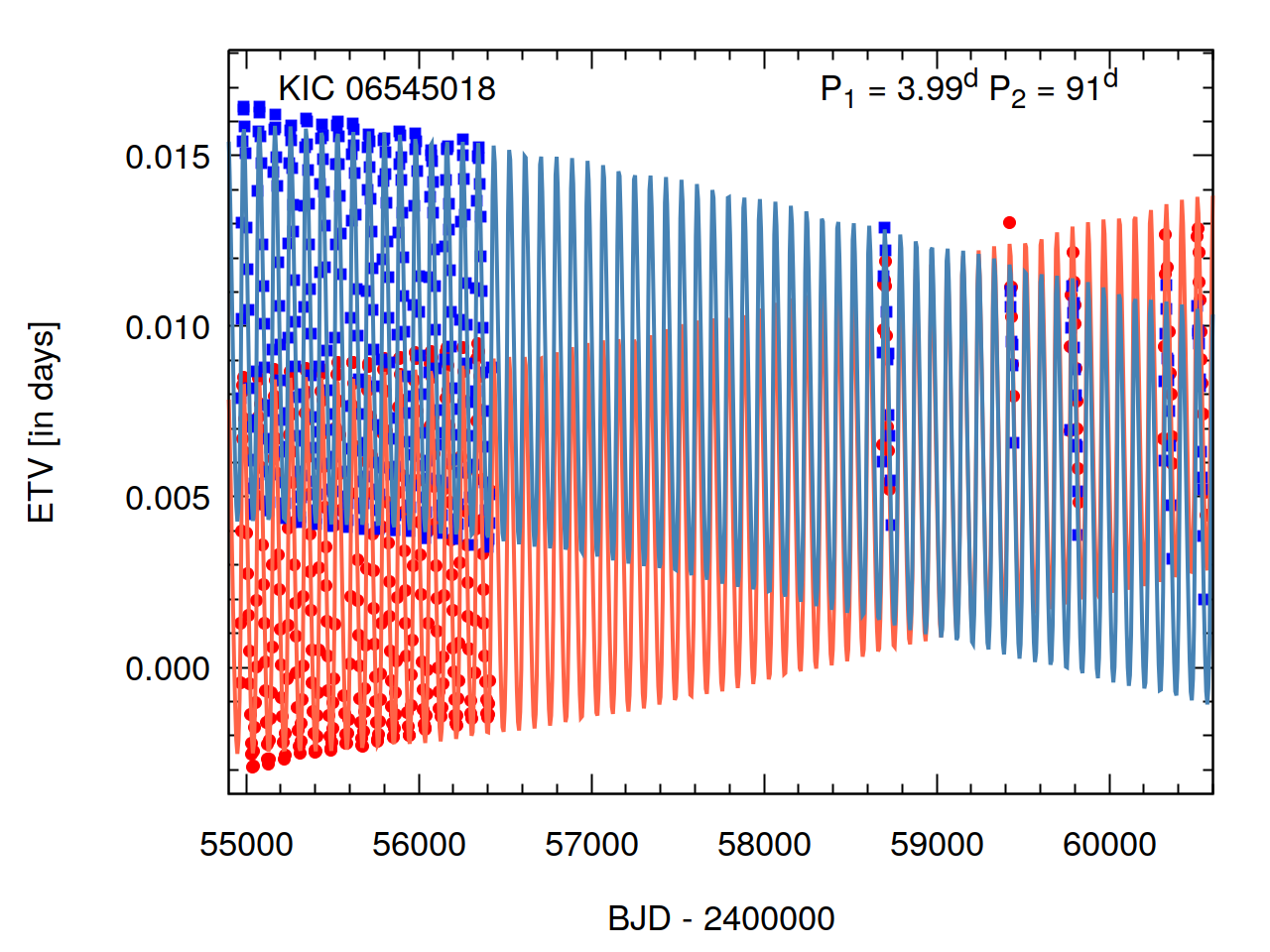}\includegraphics[width=60mm]{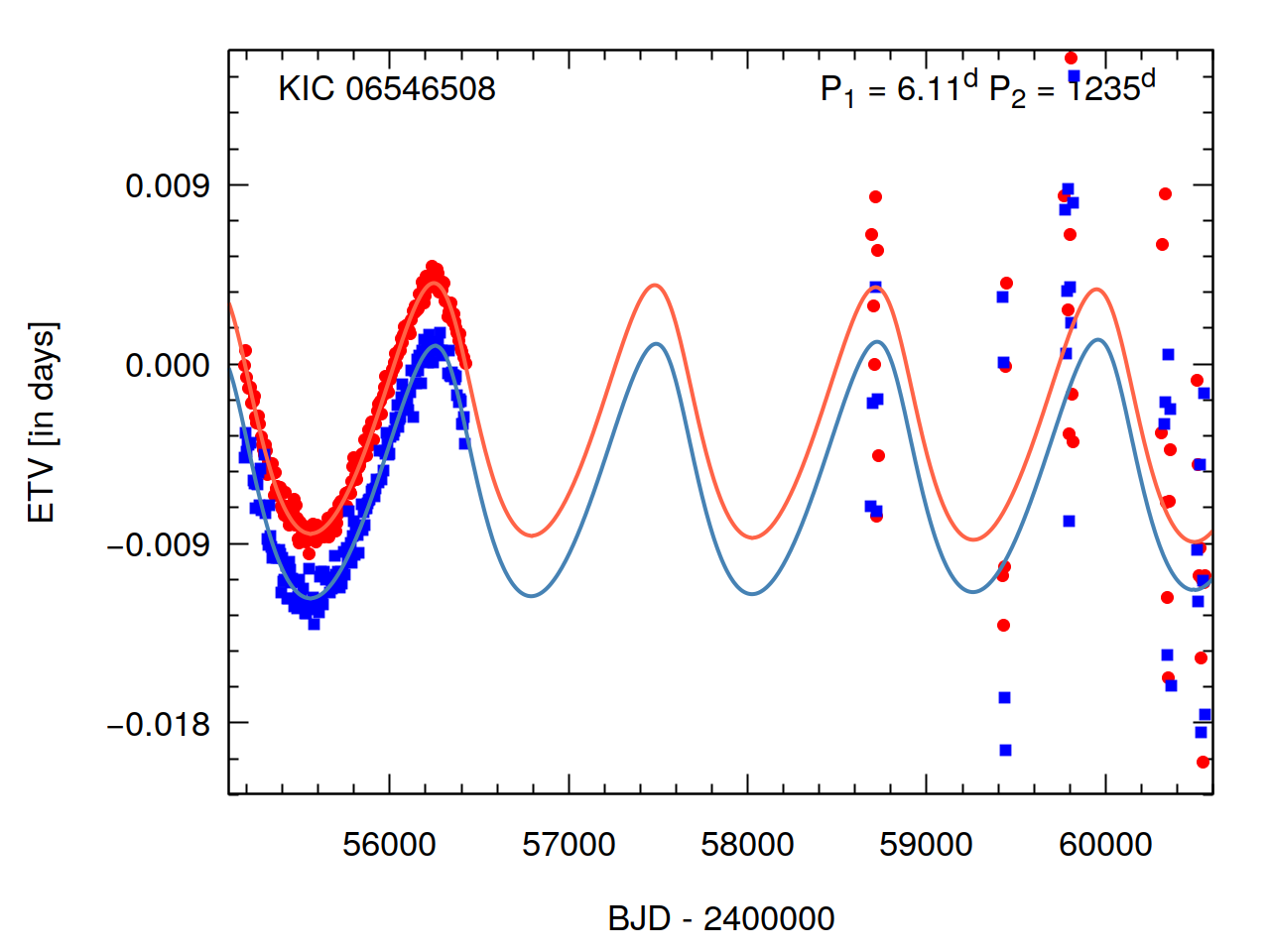}
\caption{continued.}
\end{figure*}

\addtocounter{figure}{-1}

\begin{figure*}
\includegraphics[width=60mm]{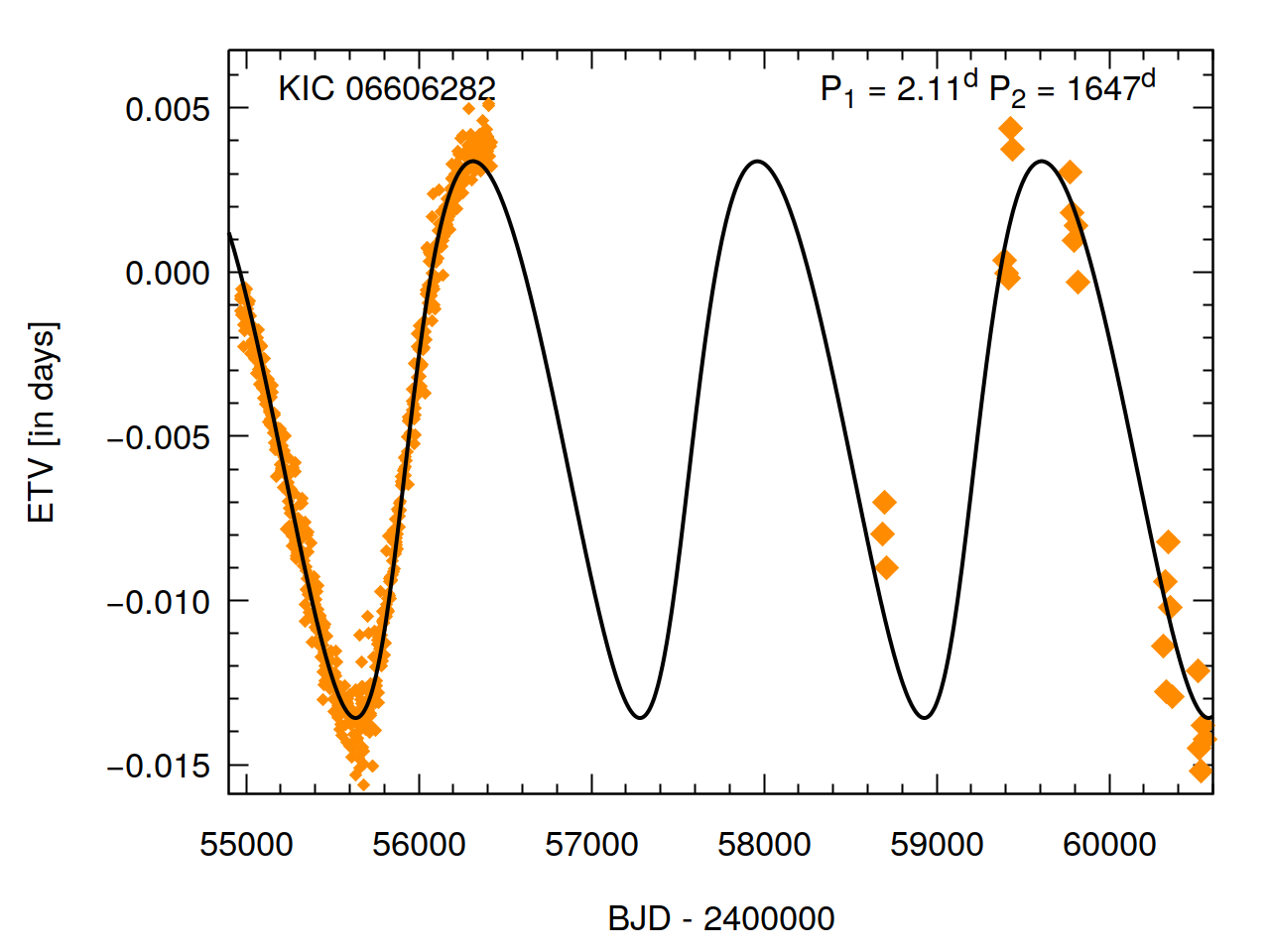}\includegraphics[width=60mm]{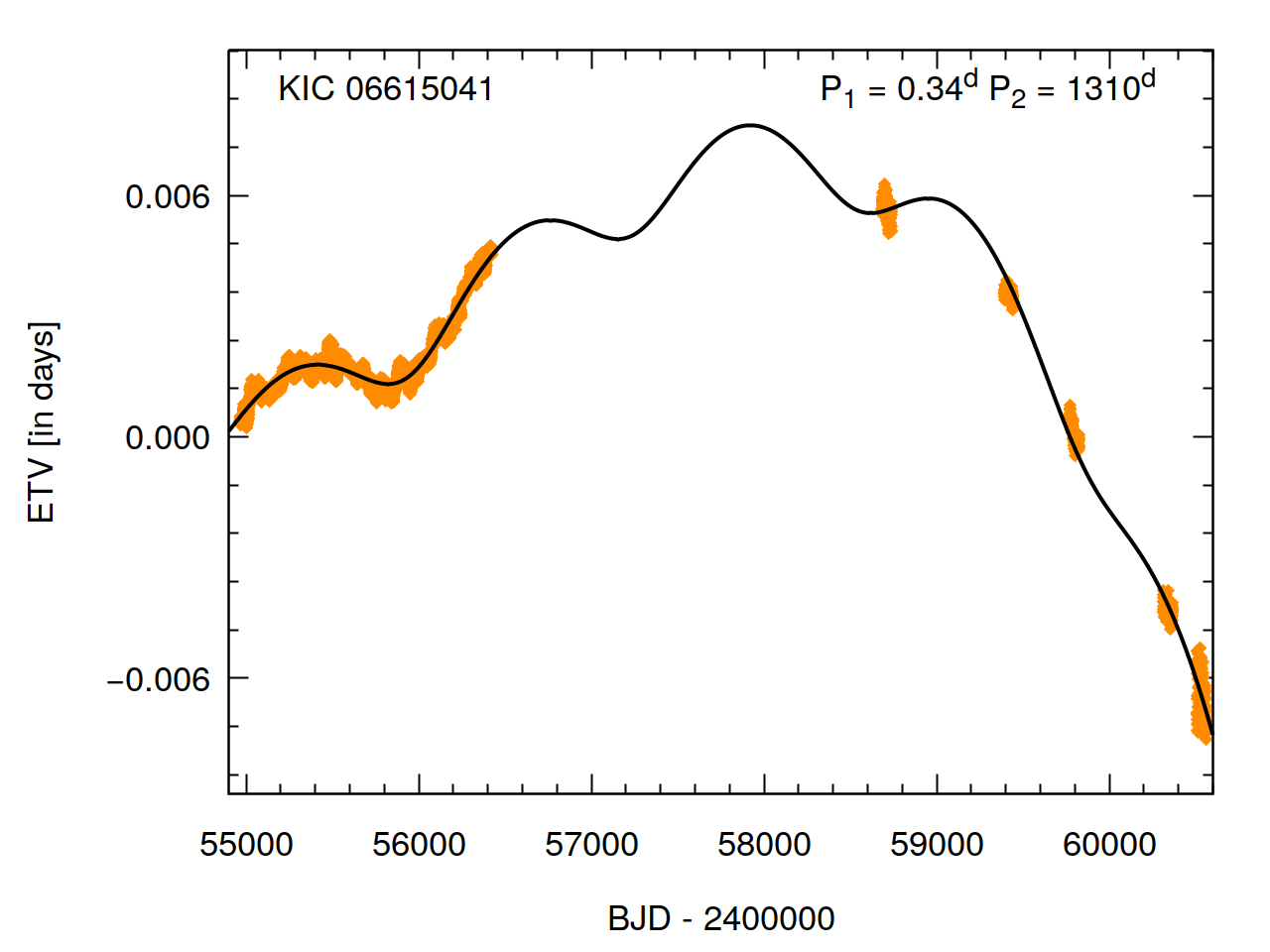}\includegraphics[width=60mm]{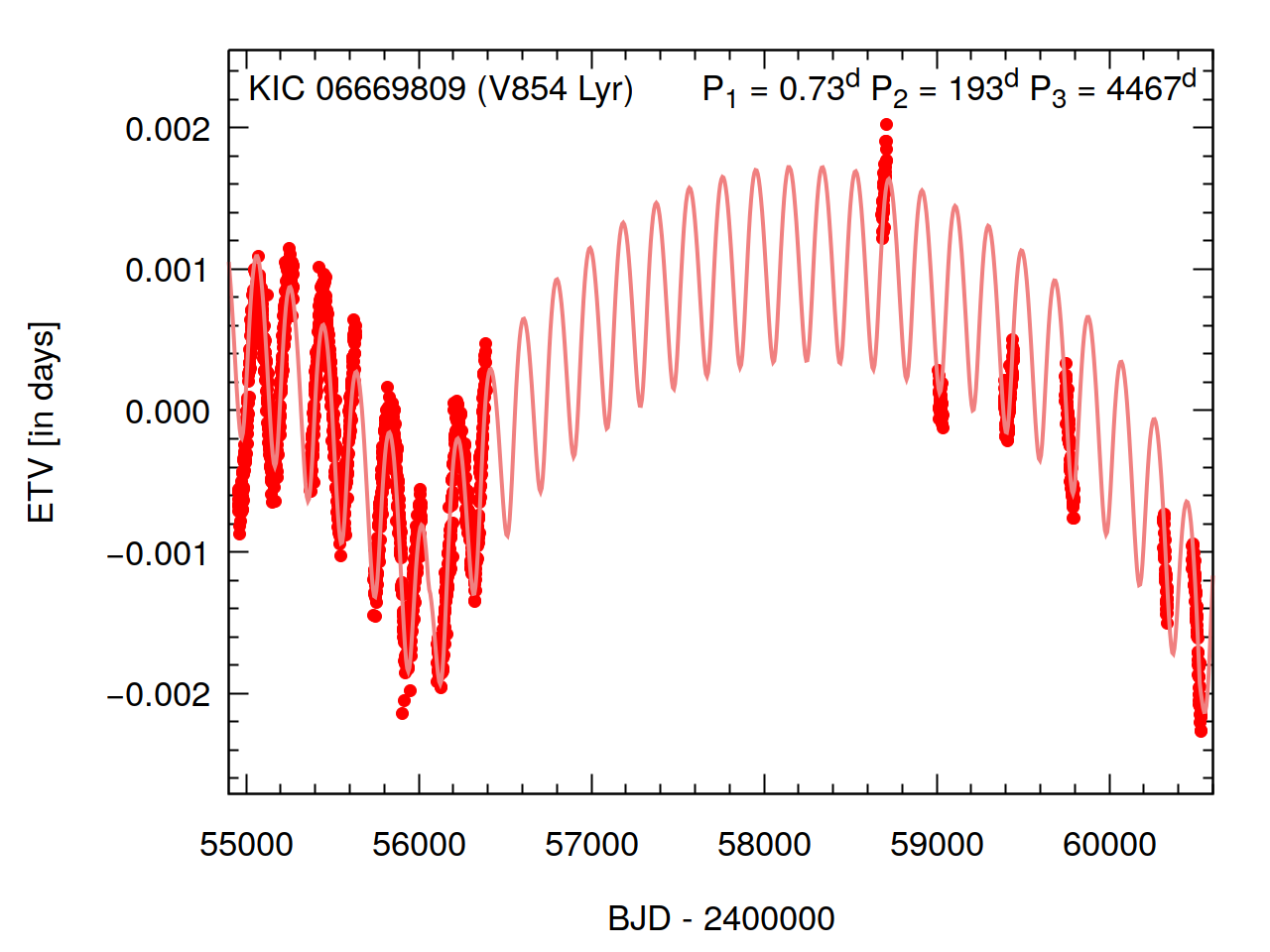}
\includegraphics[width=60mm]{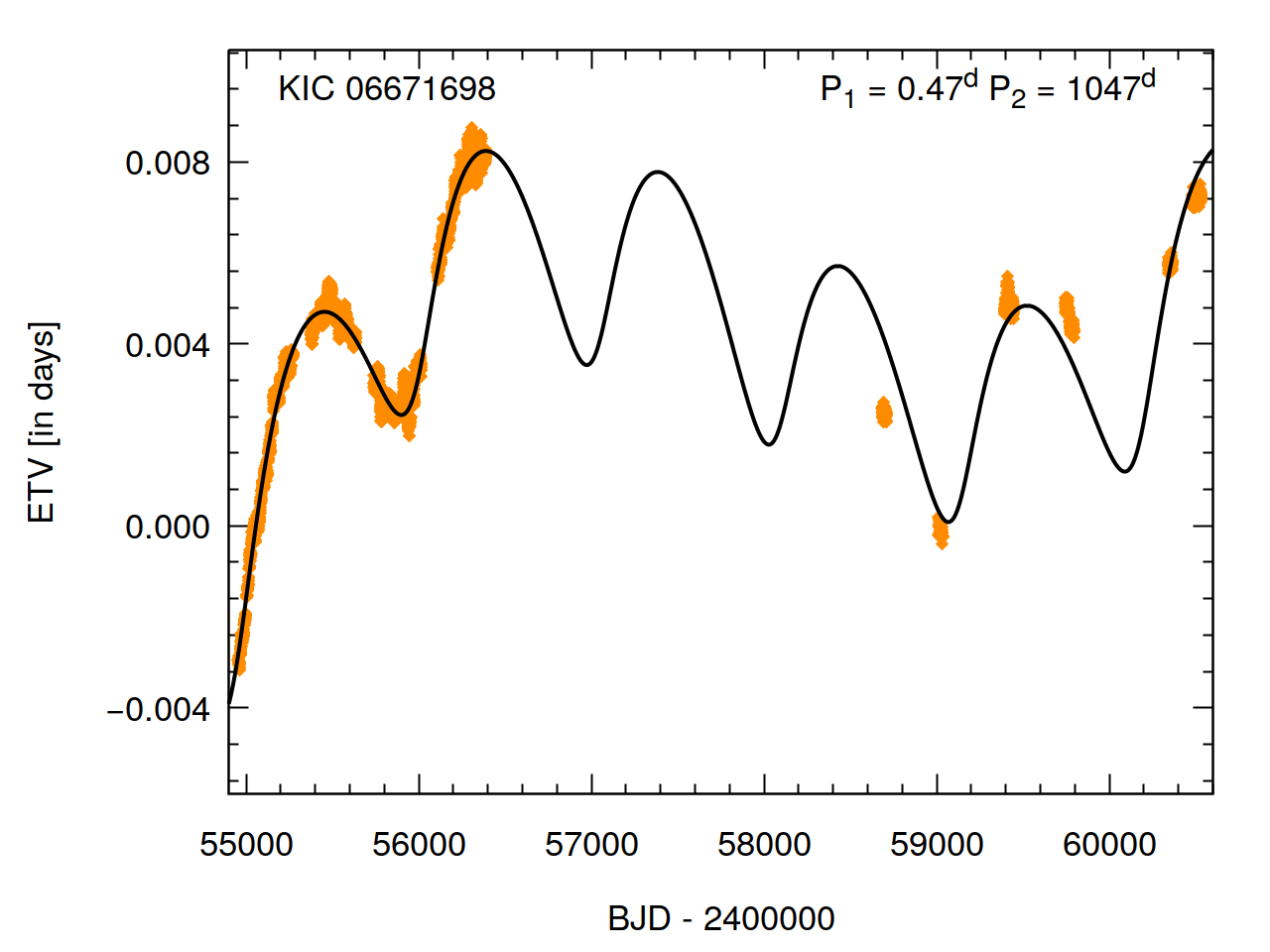}\includegraphics[width=60mm]{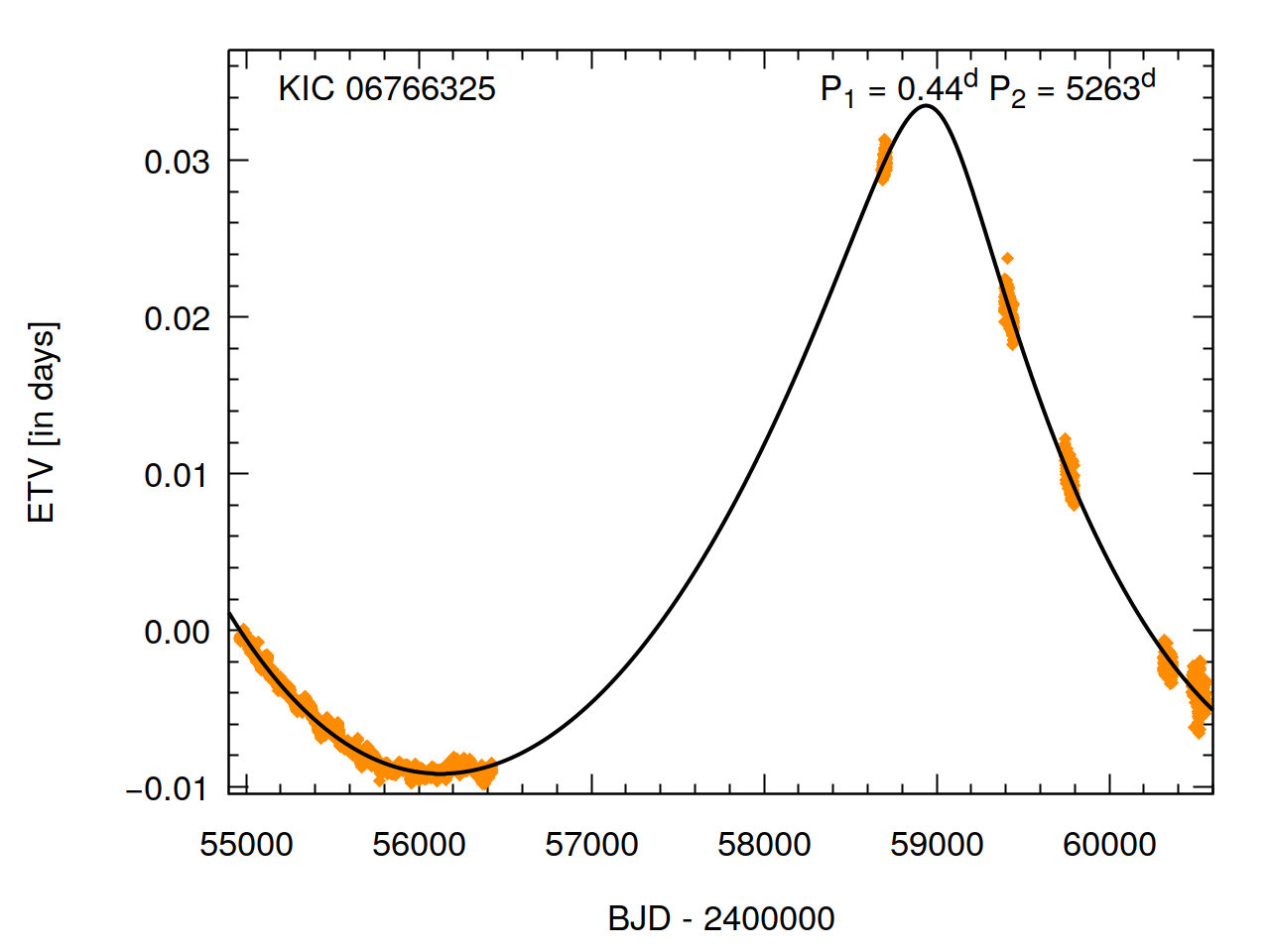}\includegraphics[width=60mm]{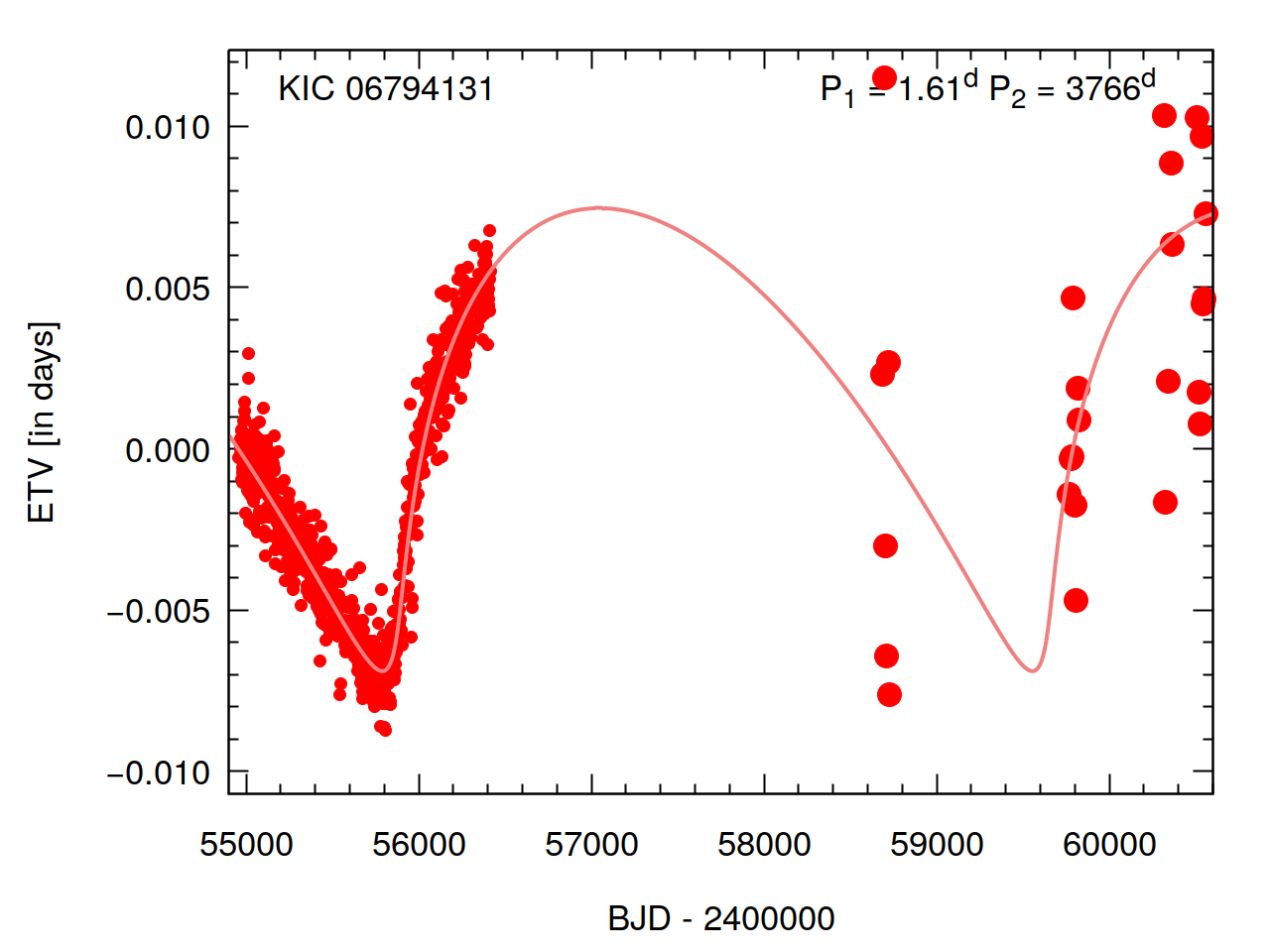}
\includegraphics[width=60mm]{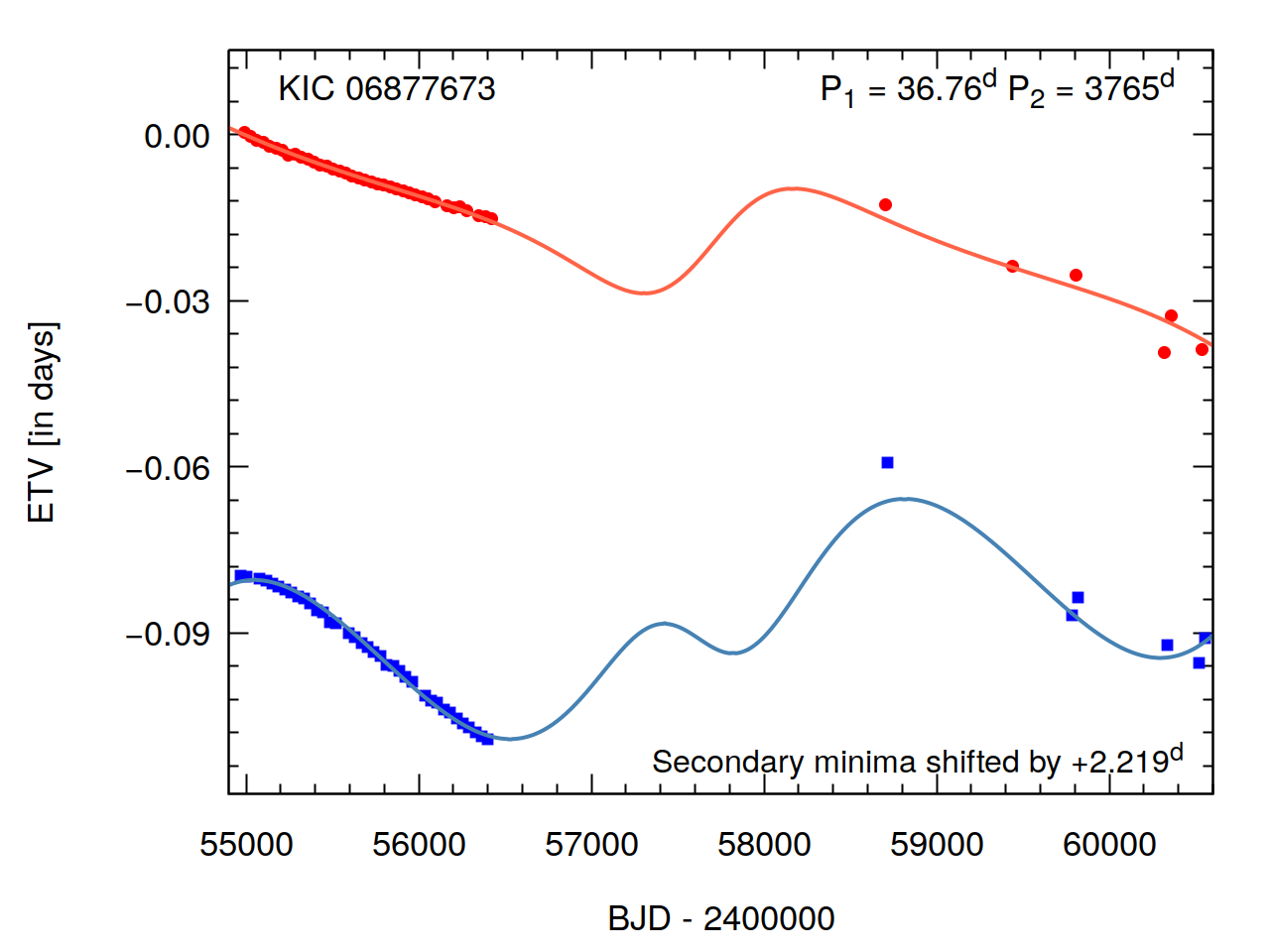}\includegraphics[width=60mm]{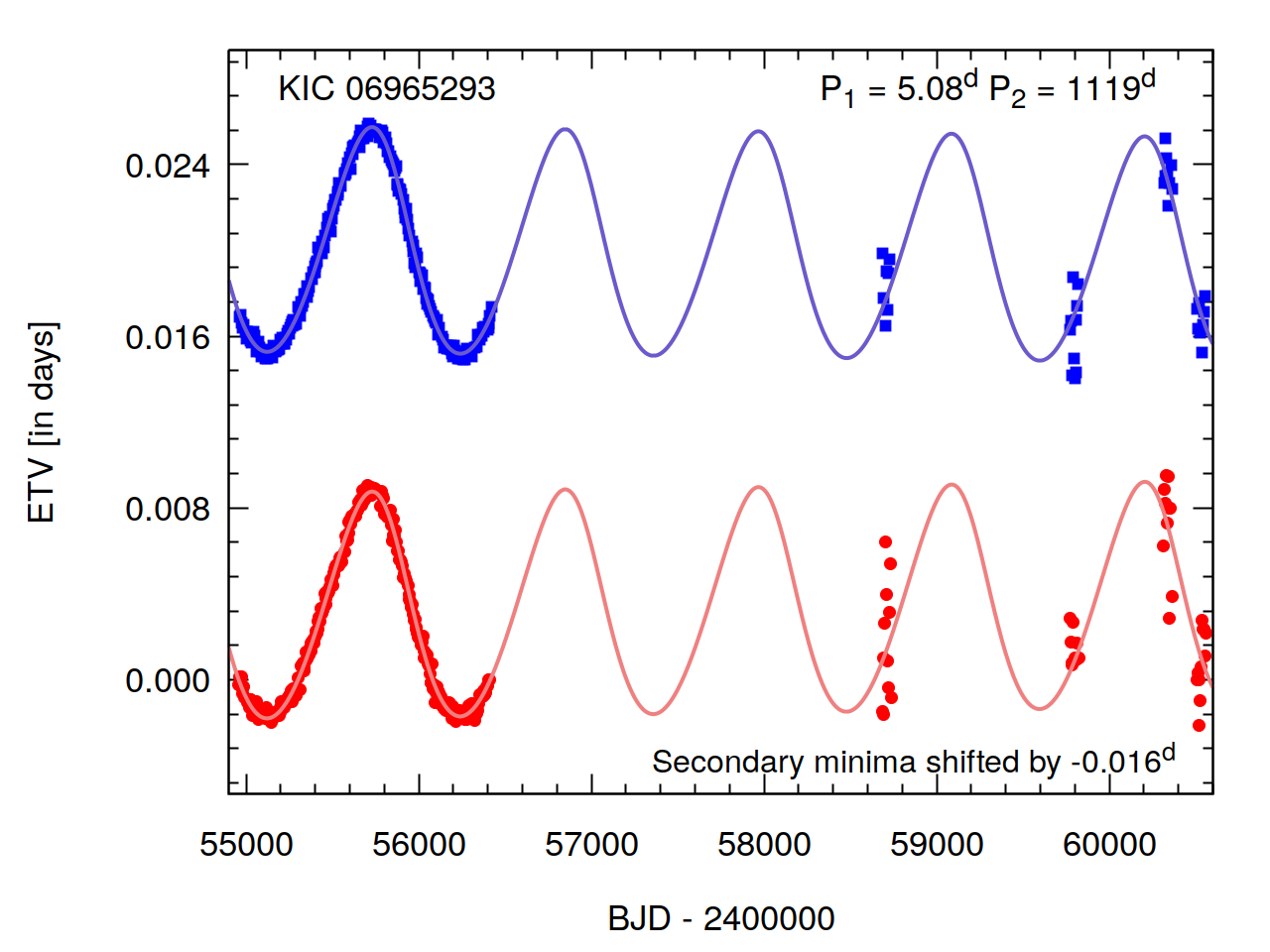}\includegraphics[width=60mm]{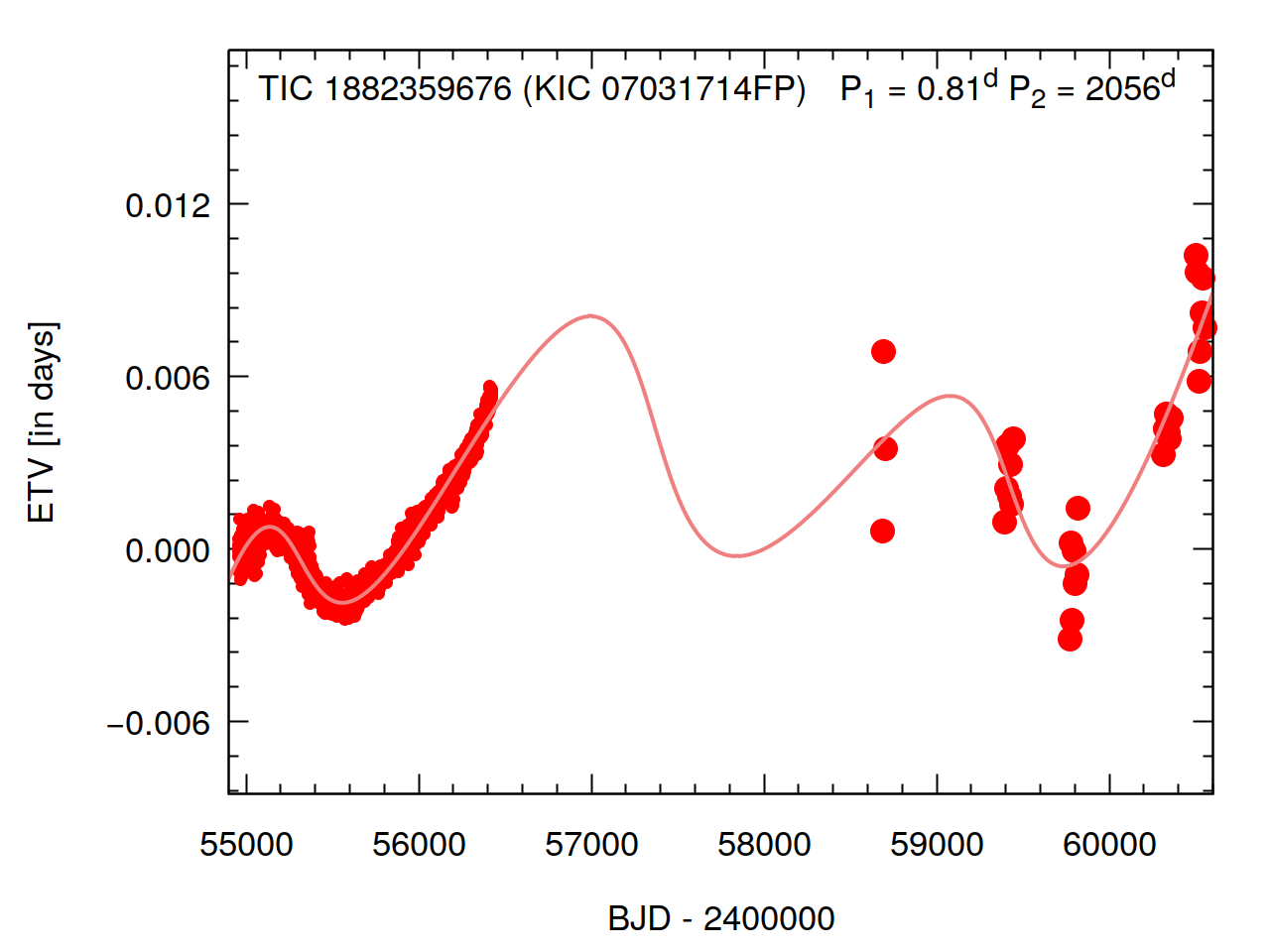}
\includegraphics[width=60mm]{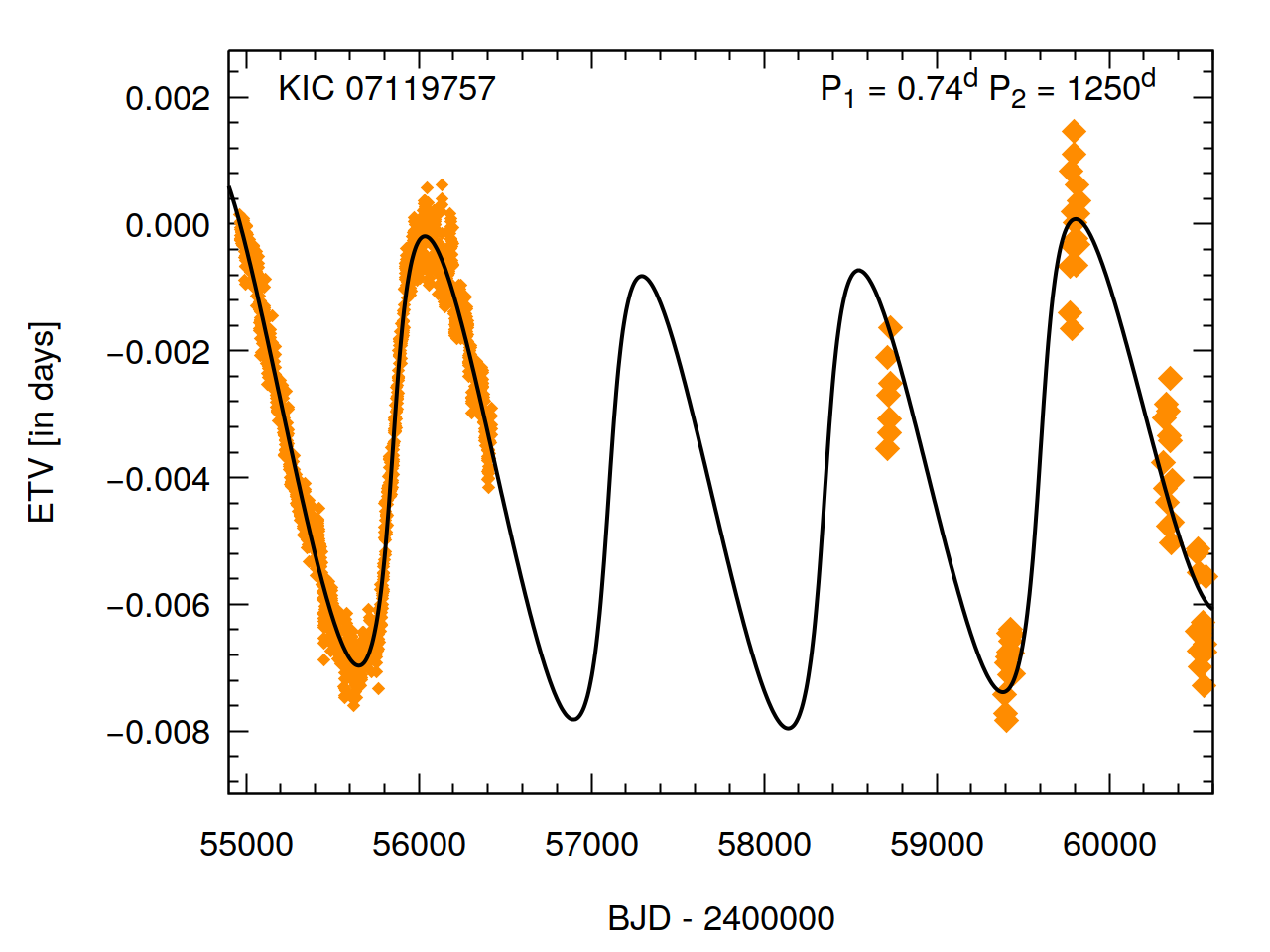}\includegraphics[width=60mm]{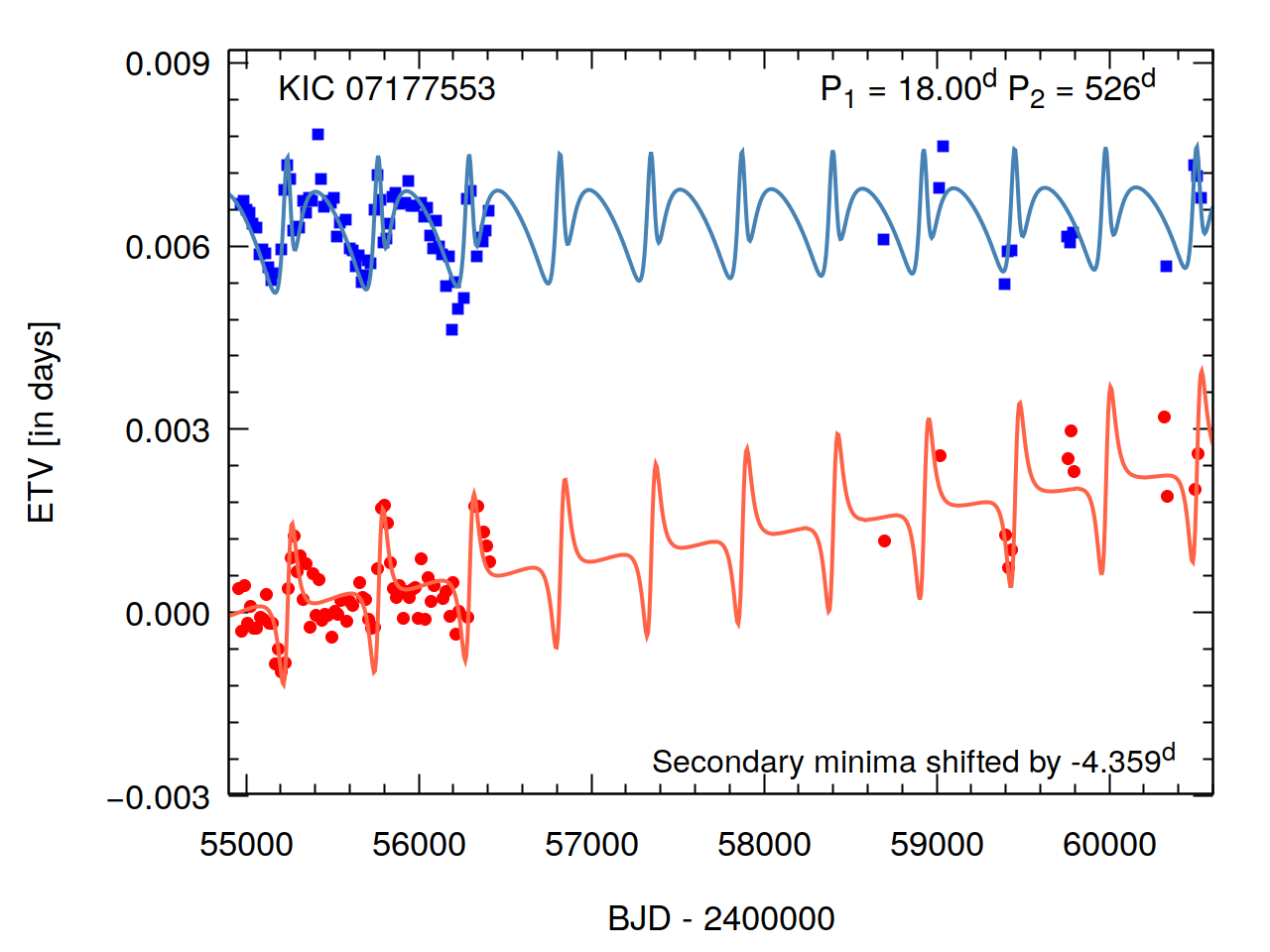}\includegraphics[width=60mm]{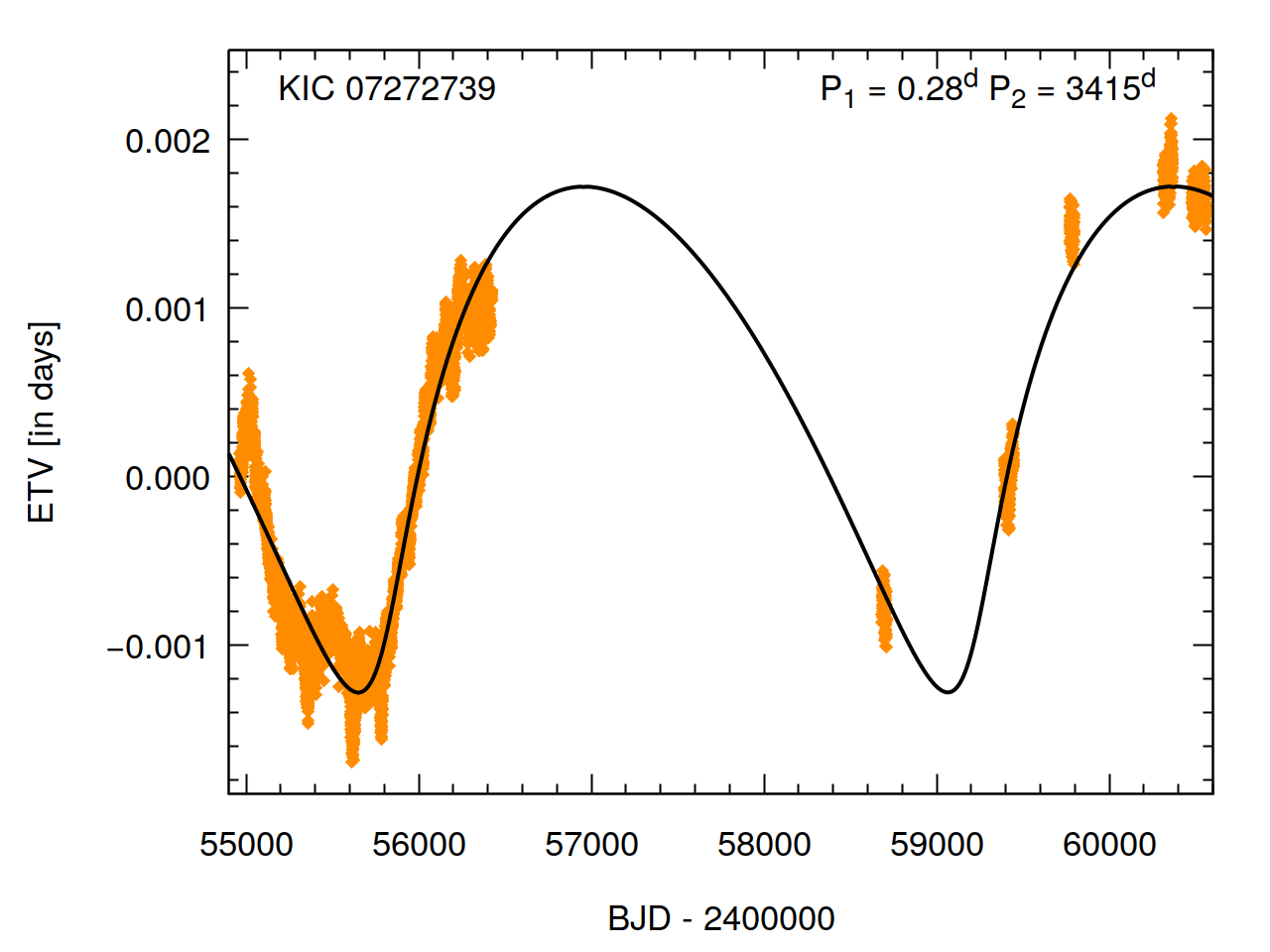}
\includegraphics[width=60mm]{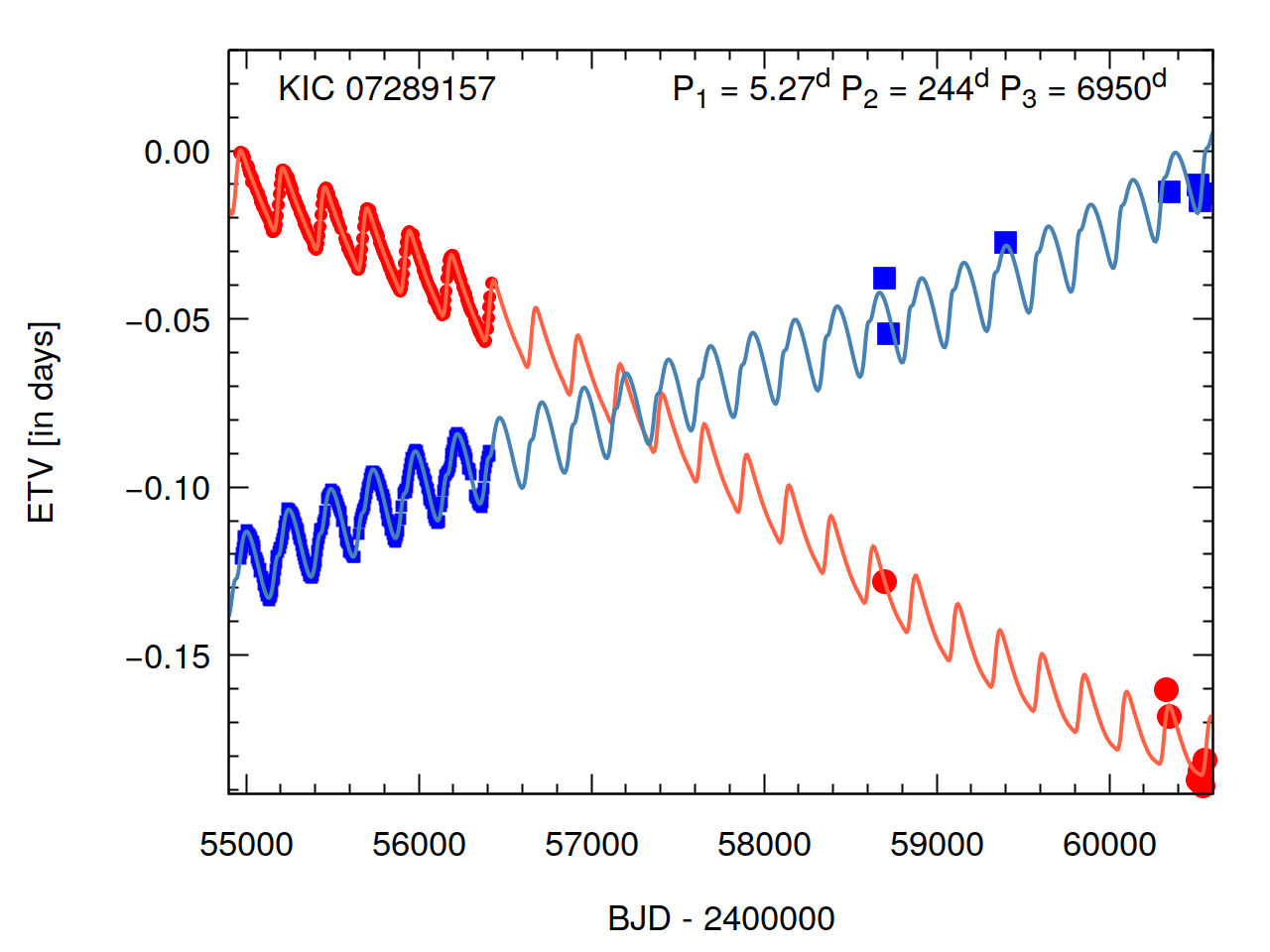}\includegraphics[width=60mm]{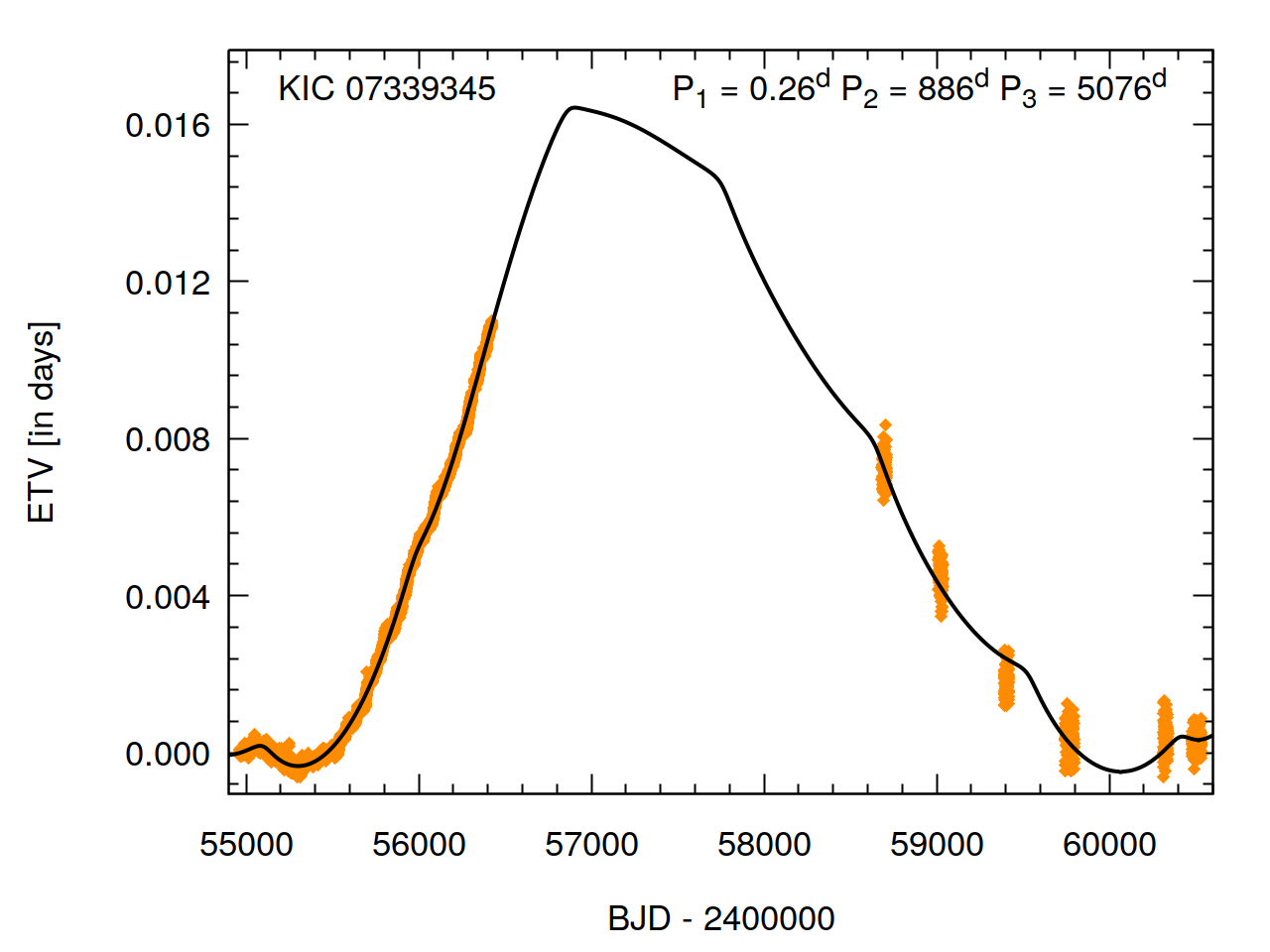}\includegraphics[width=60mm]{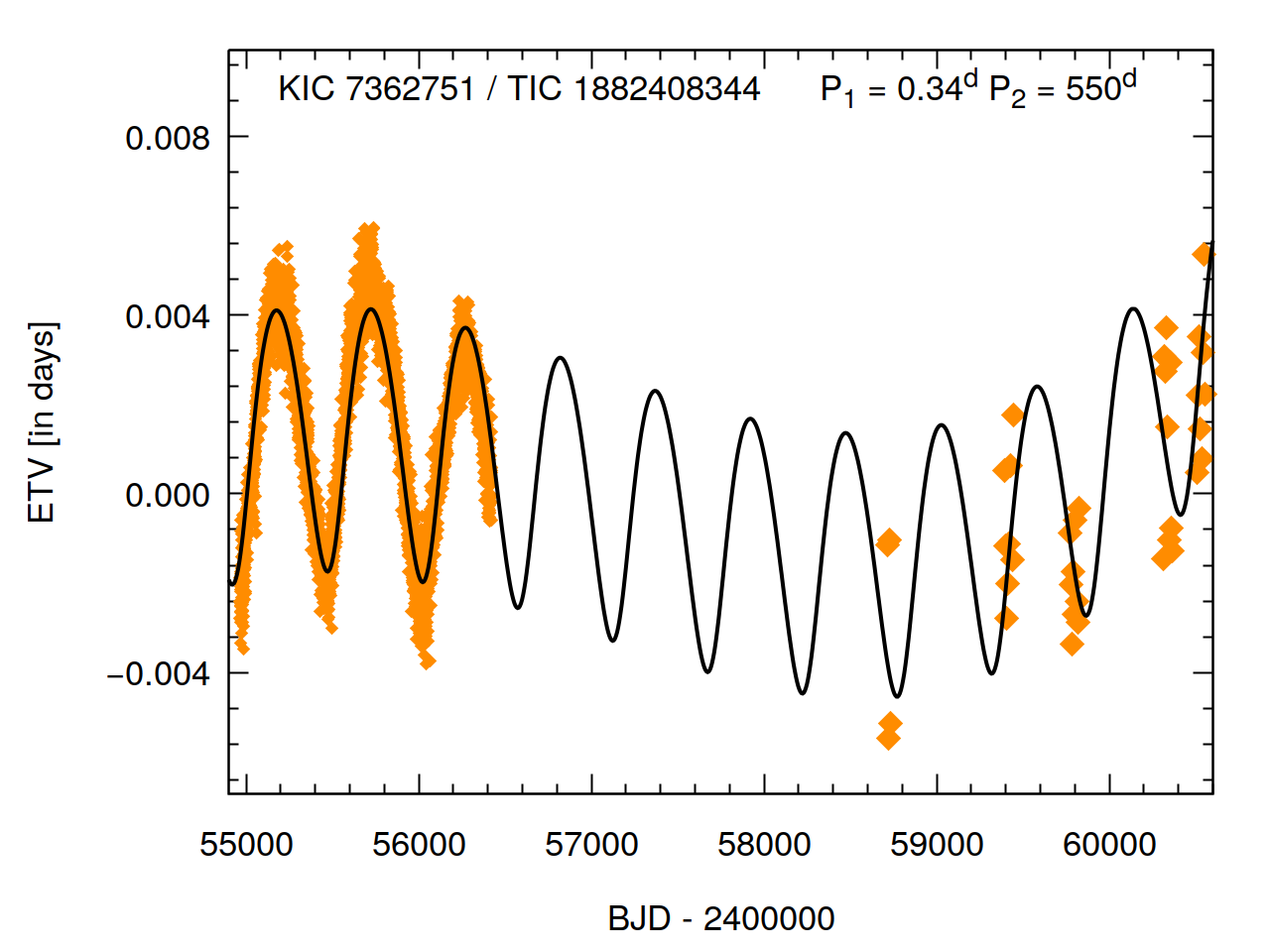}
\caption{continued.}
\end{figure*}

\addtocounter{figure}{-1}

\begin{figure*}
\includegraphics[width=60mm]{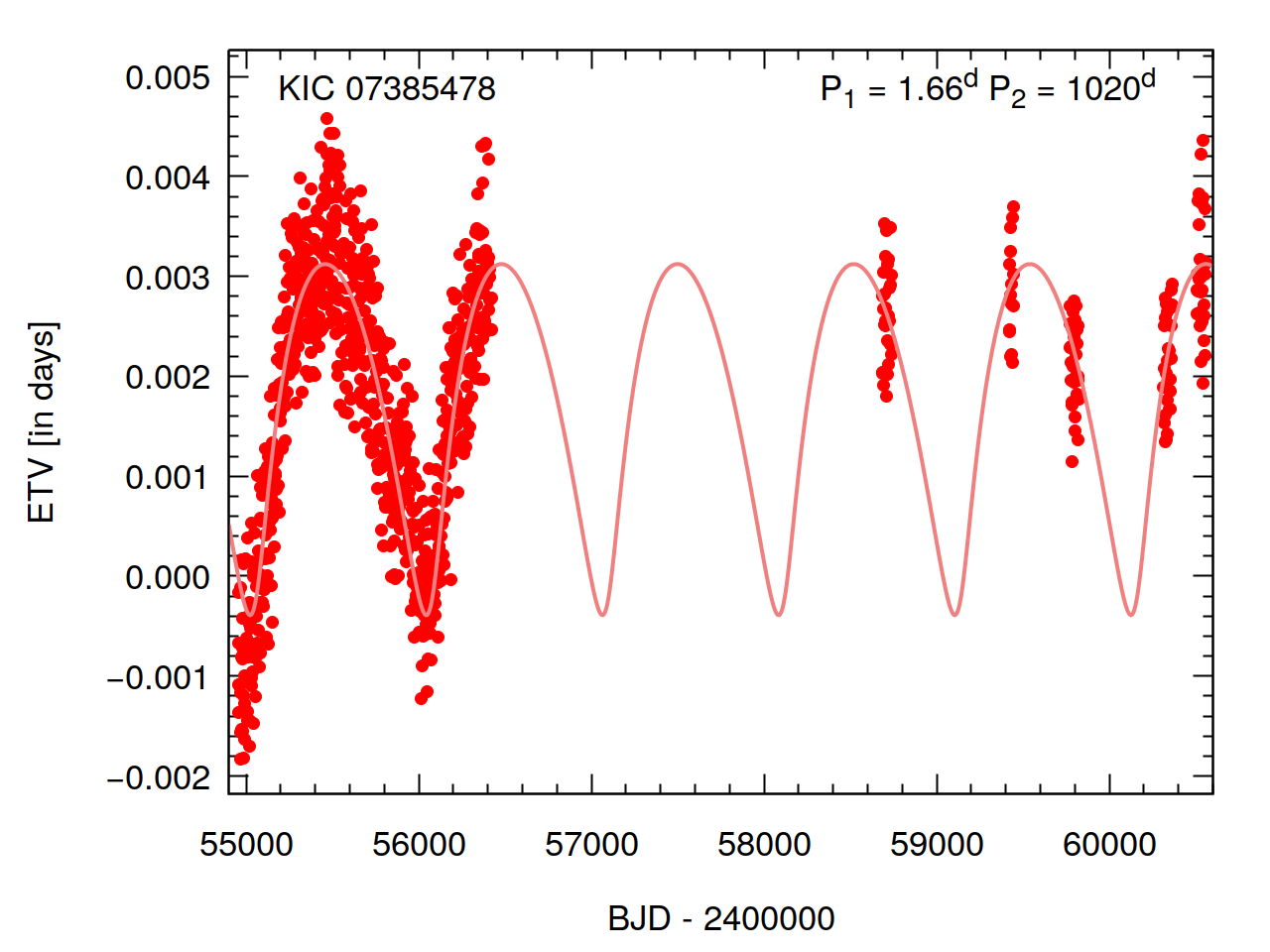}\includegraphics[width=60mm]{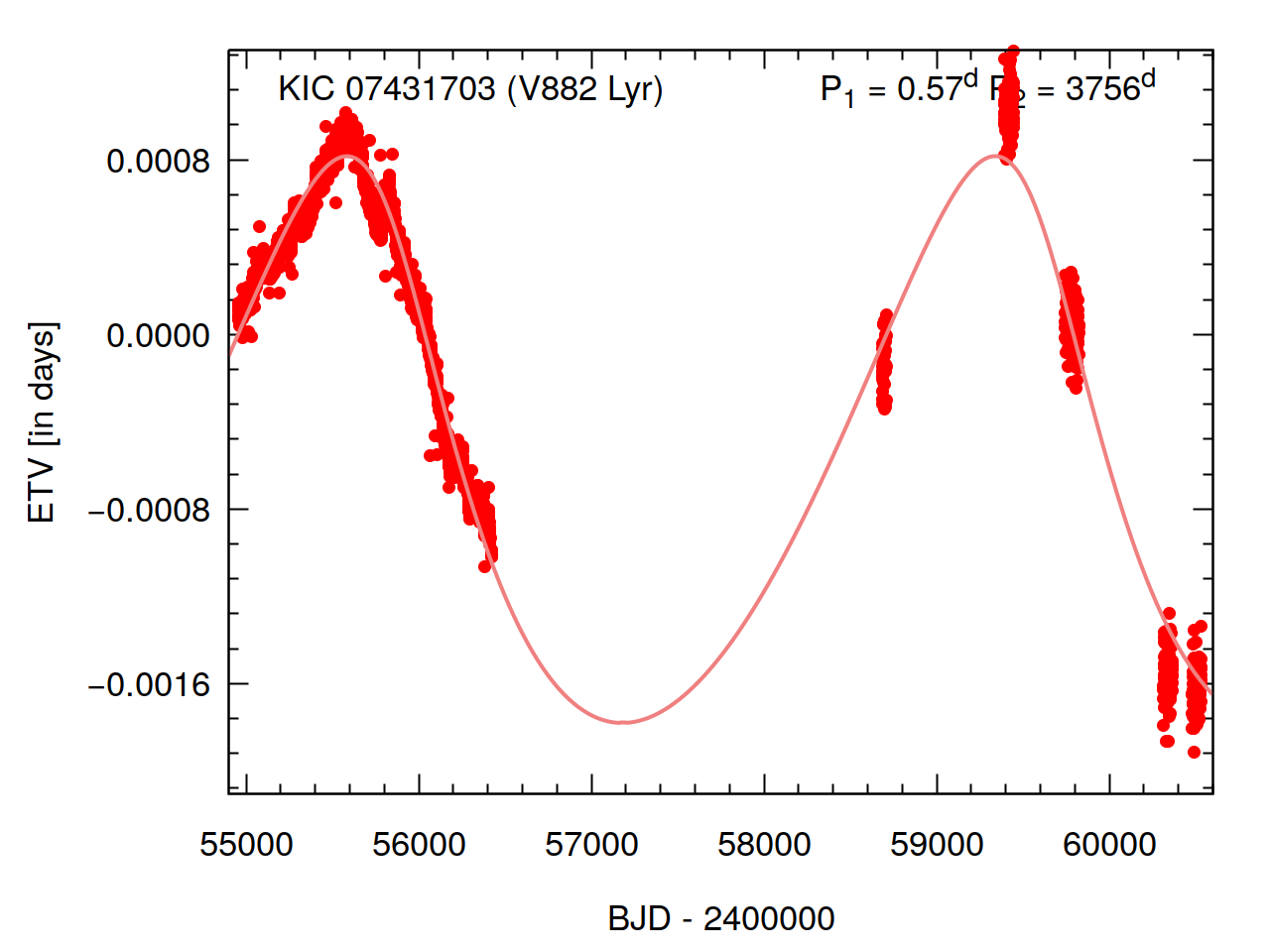}\includegraphics[width=60mm]{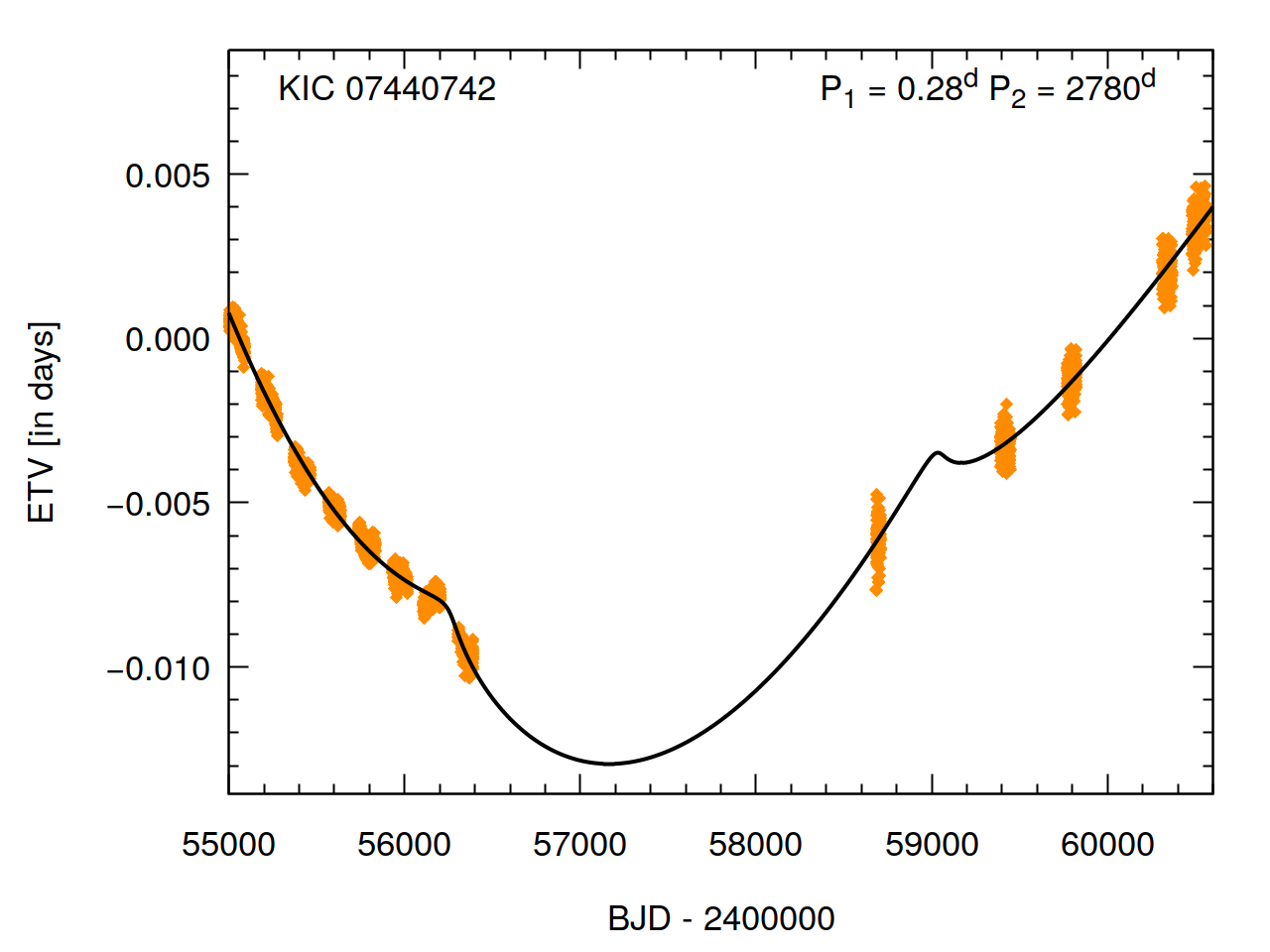}
\includegraphics[width=60mm]{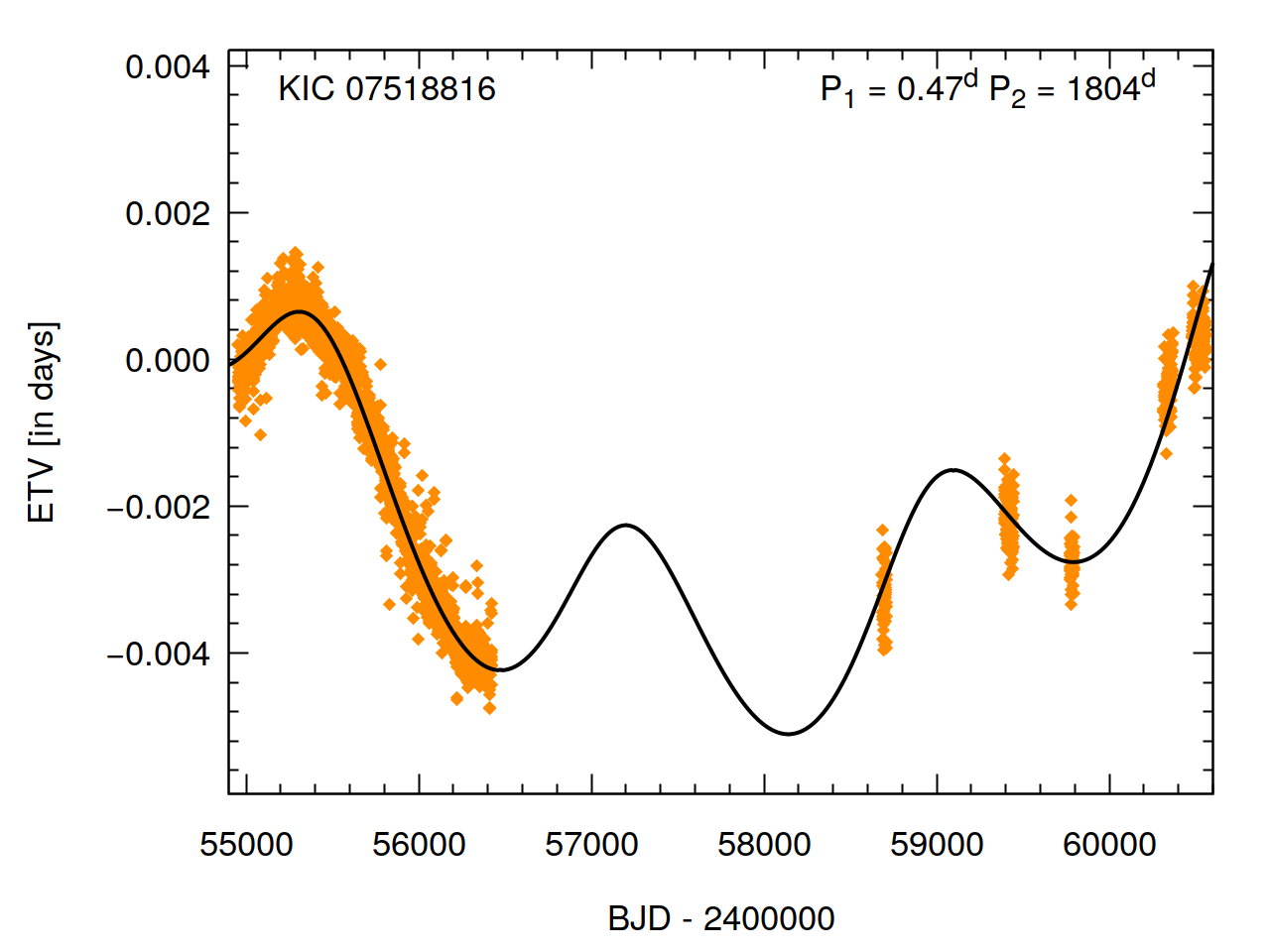}\includegraphics[width=60mm]{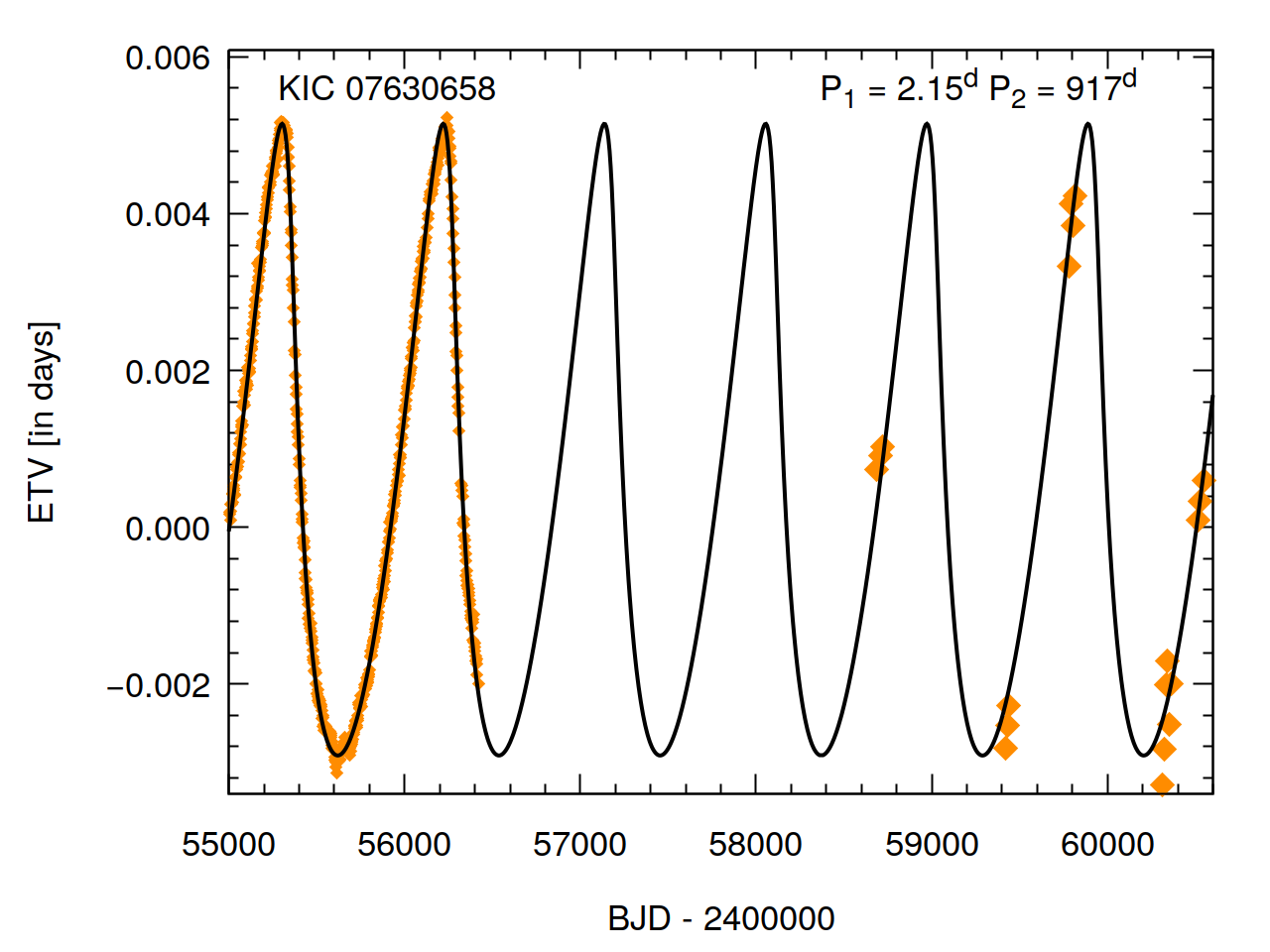}\includegraphics[width=60mm]{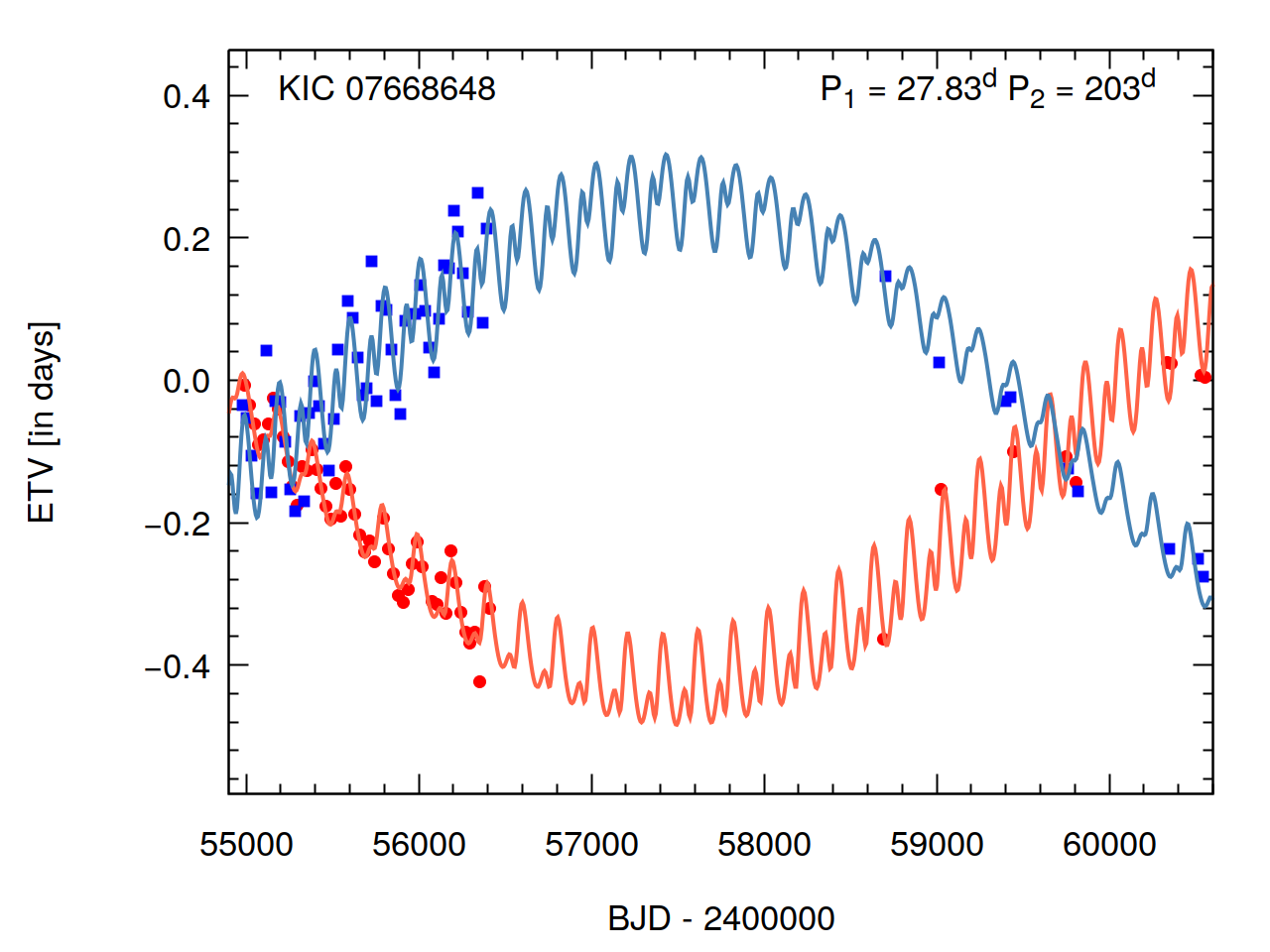}
\includegraphics[width=60mm]{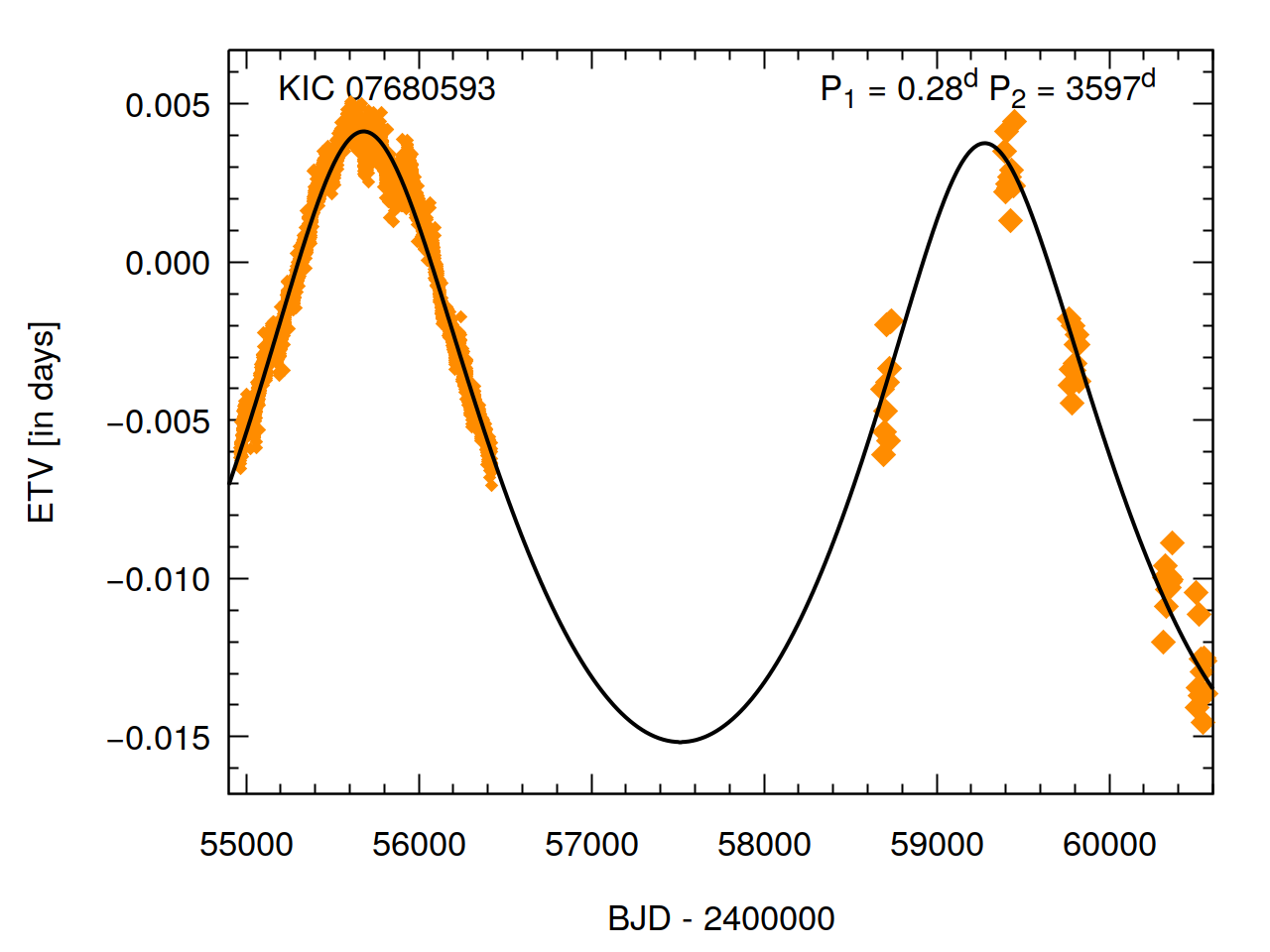}\includegraphics[width=60mm]{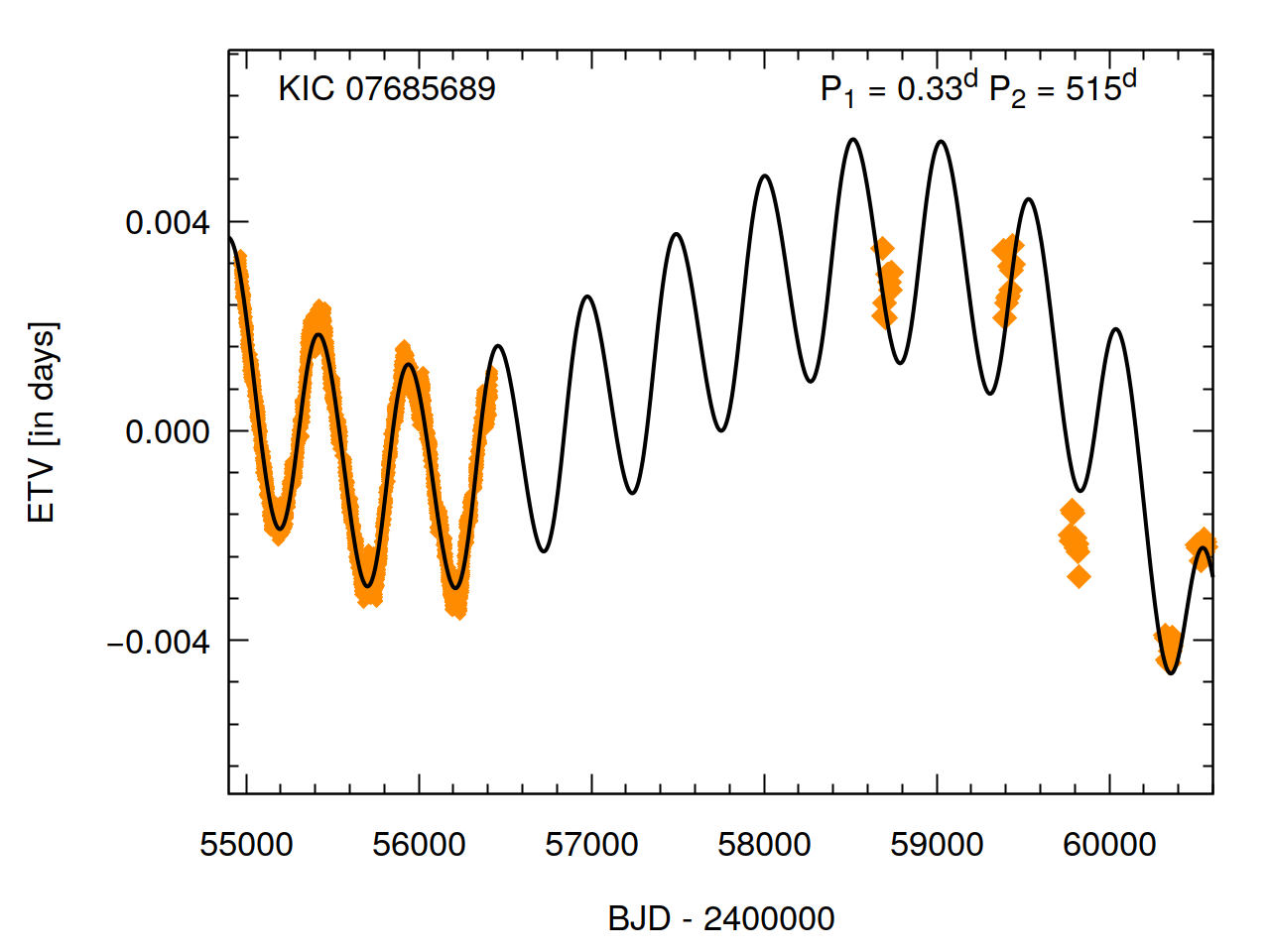}\includegraphics[width=60mm]{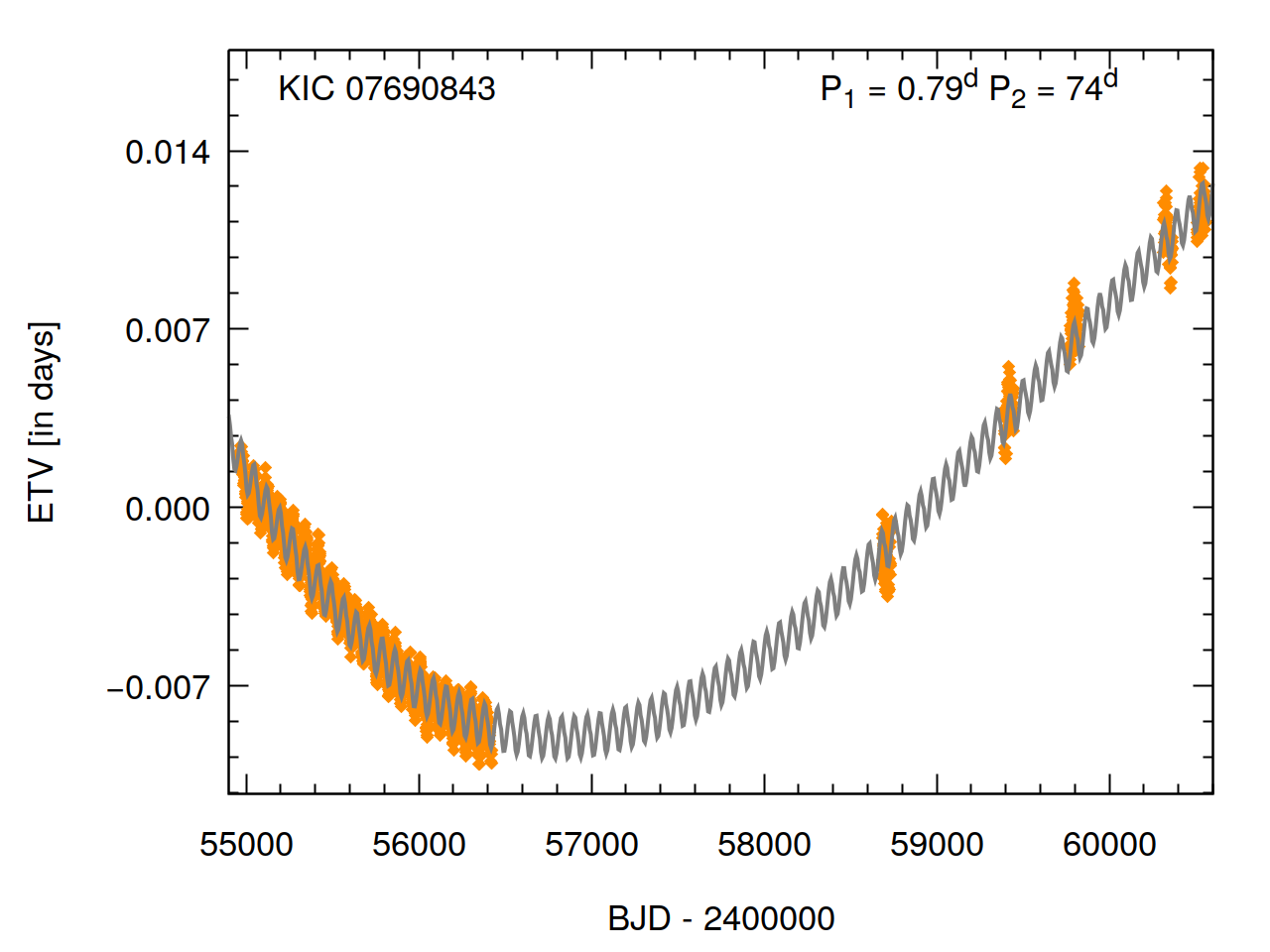}
\includegraphics[width=60mm]{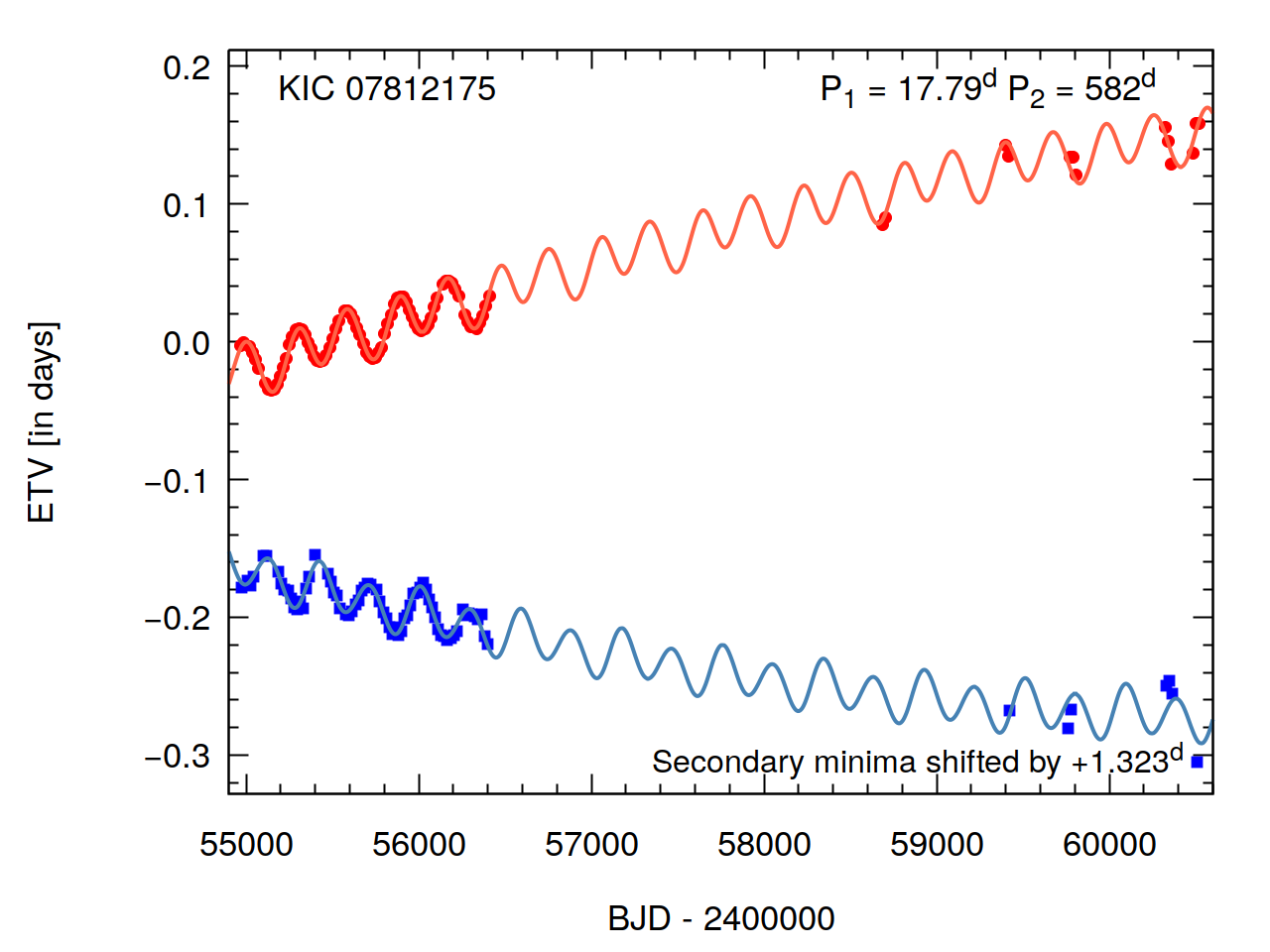}\includegraphics[width=60mm]{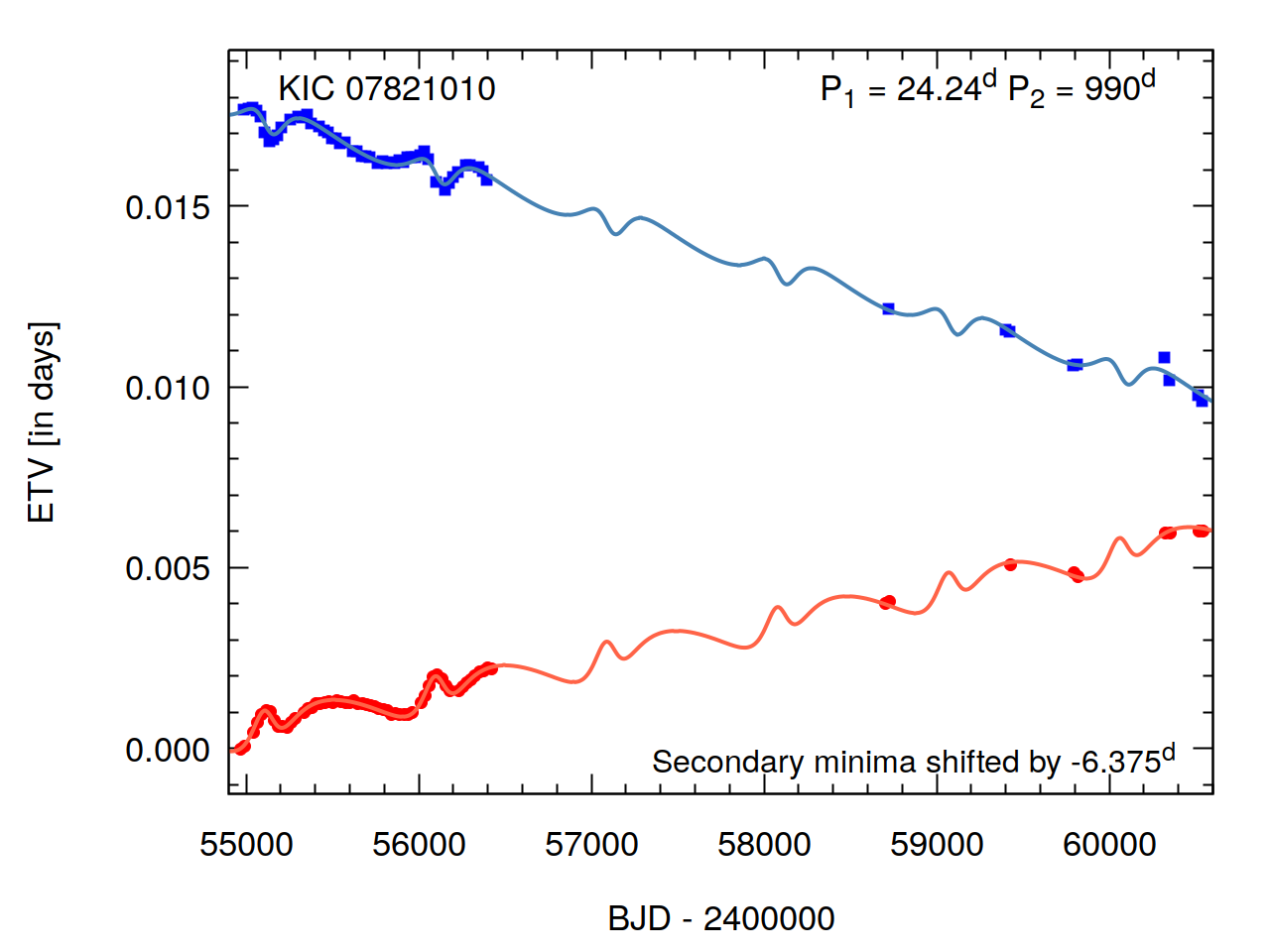}\includegraphics[width=60mm]{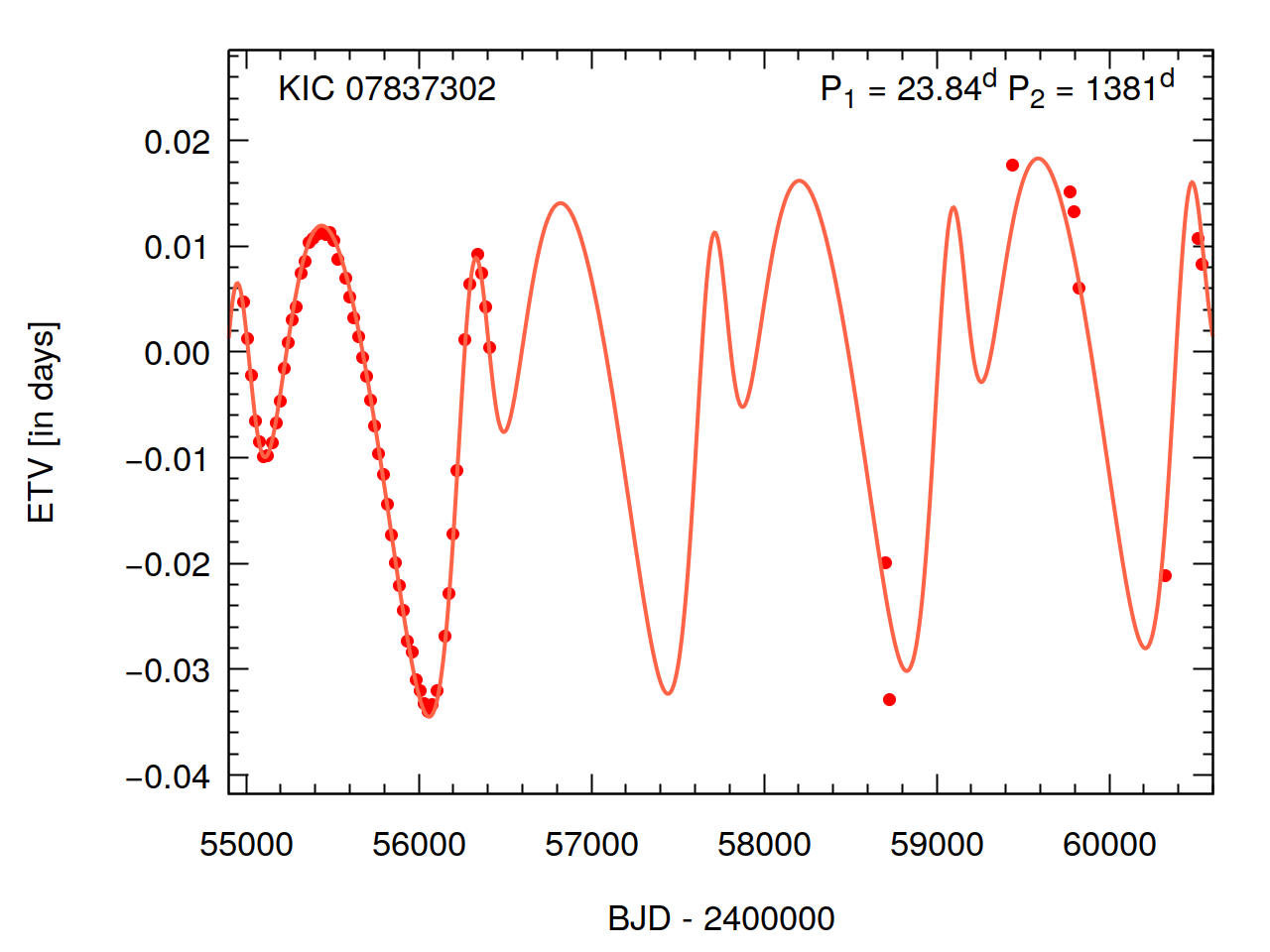}
\includegraphics[width=60mm]{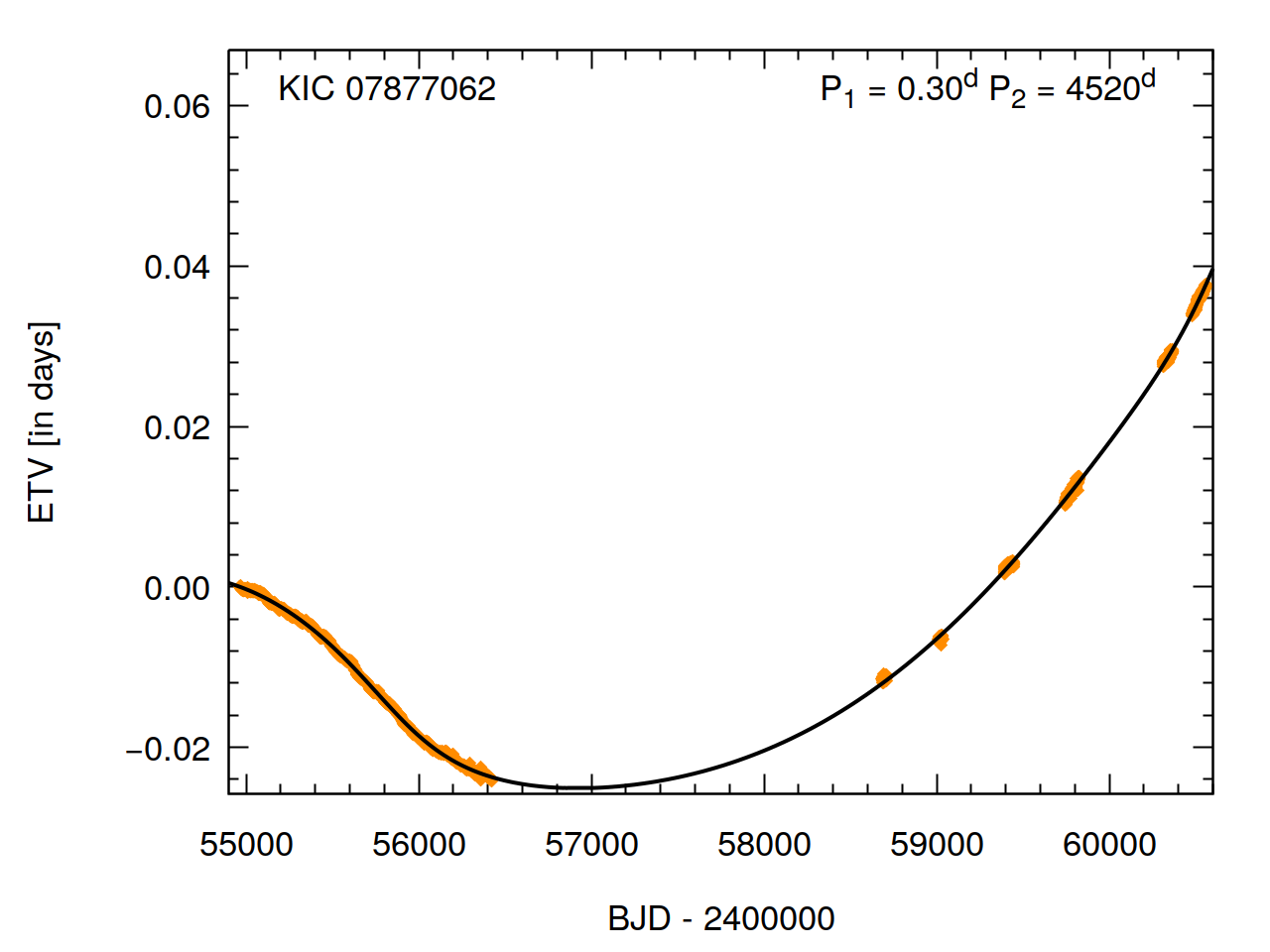}\includegraphics[width=60mm]{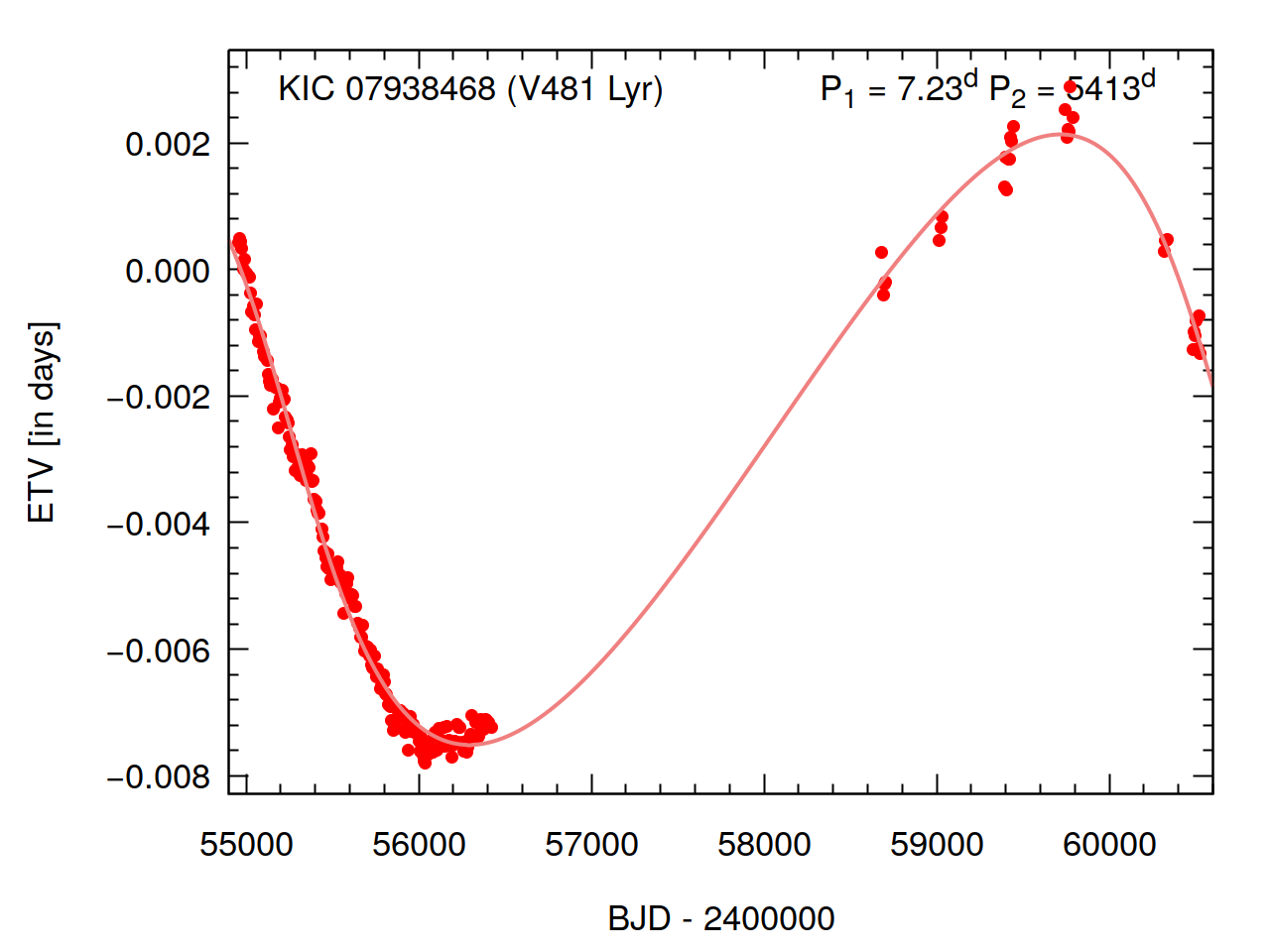}\includegraphics[width=60mm]{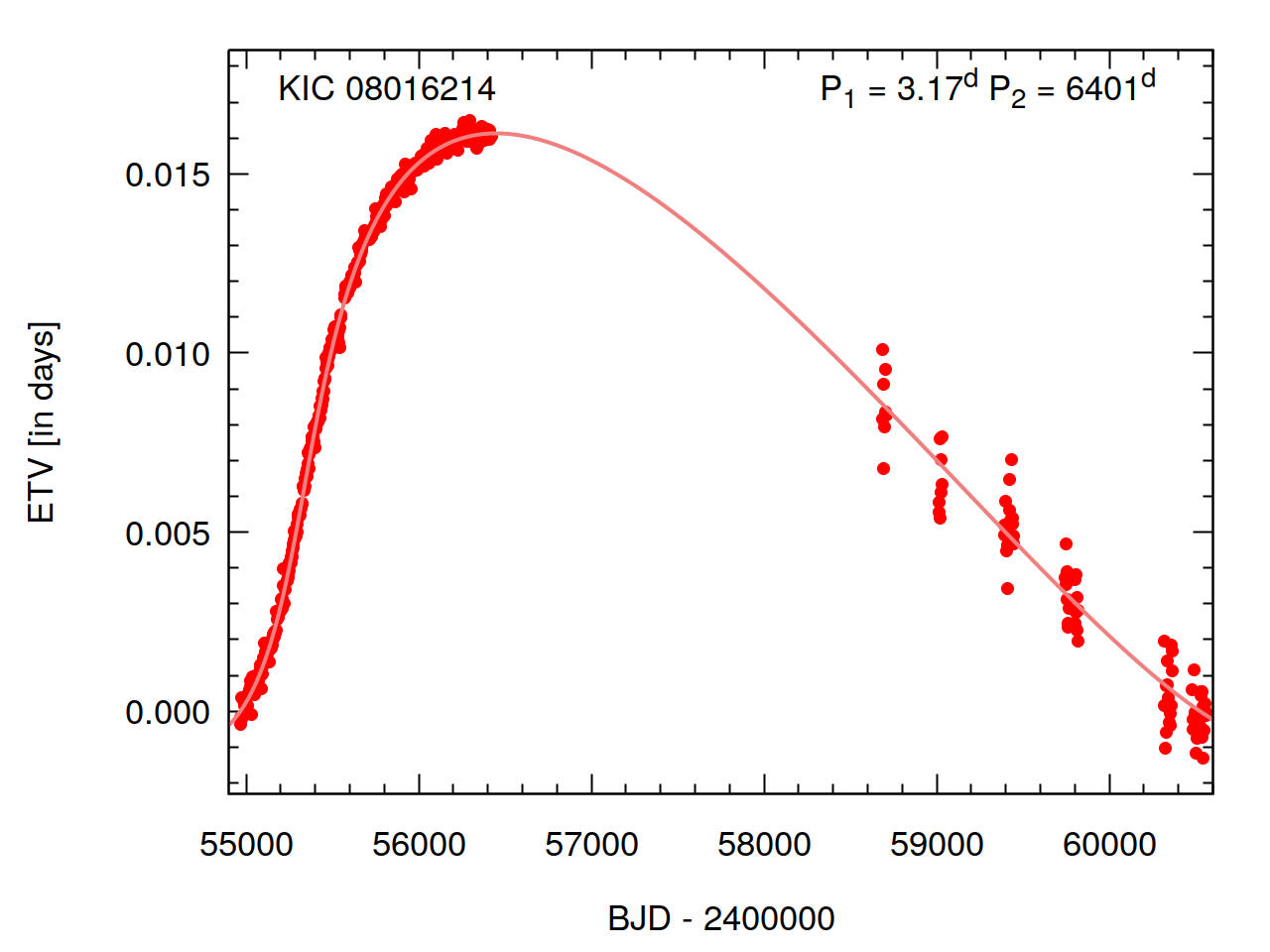}
\caption{continued.}
\end{figure*}

\addtocounter{figure}{-1}

\begin{figure*}
\includegraphics[width=60mm]{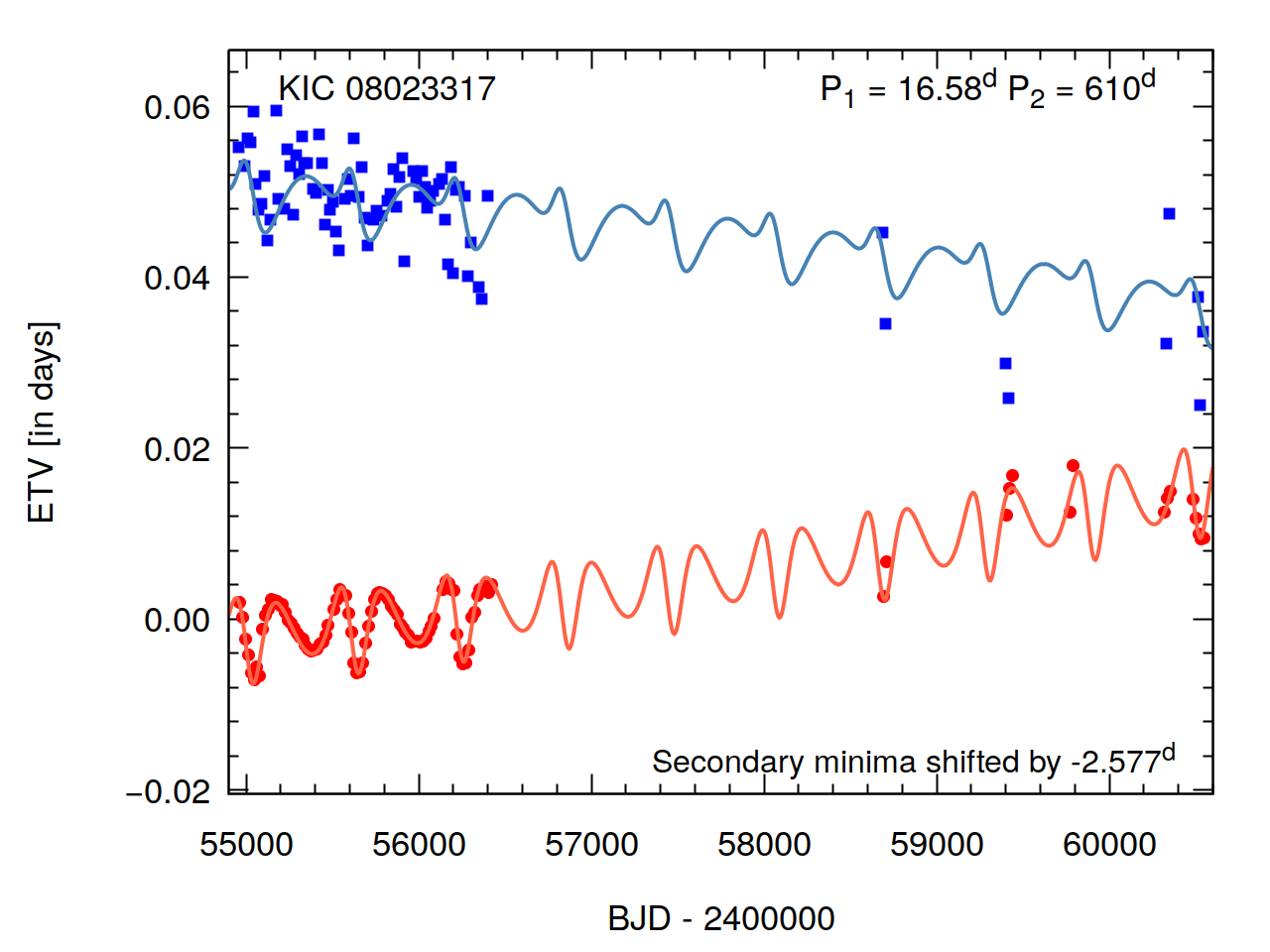}\includegraphics[width=60mm]{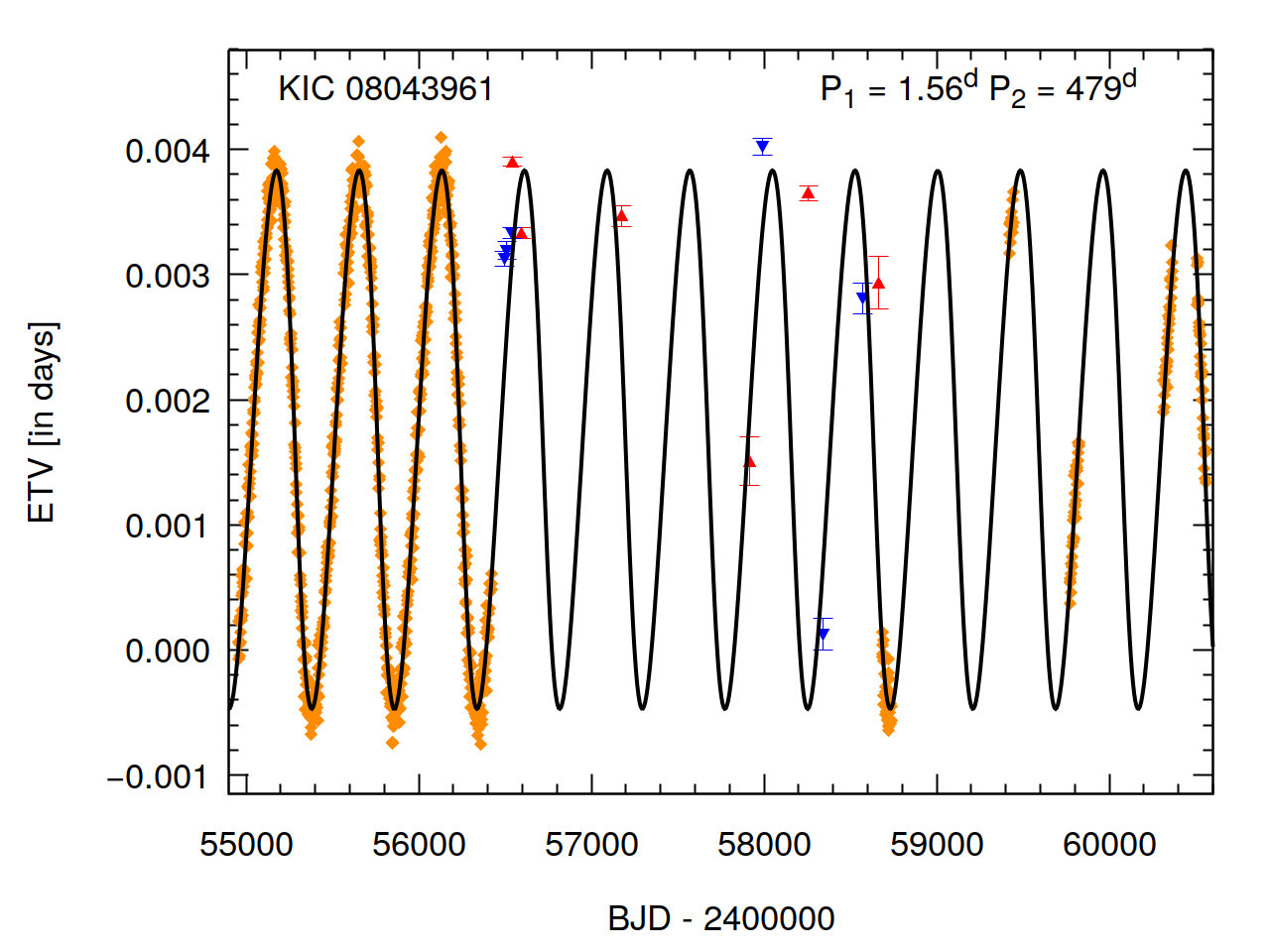}\includegraphics[width=60mm]{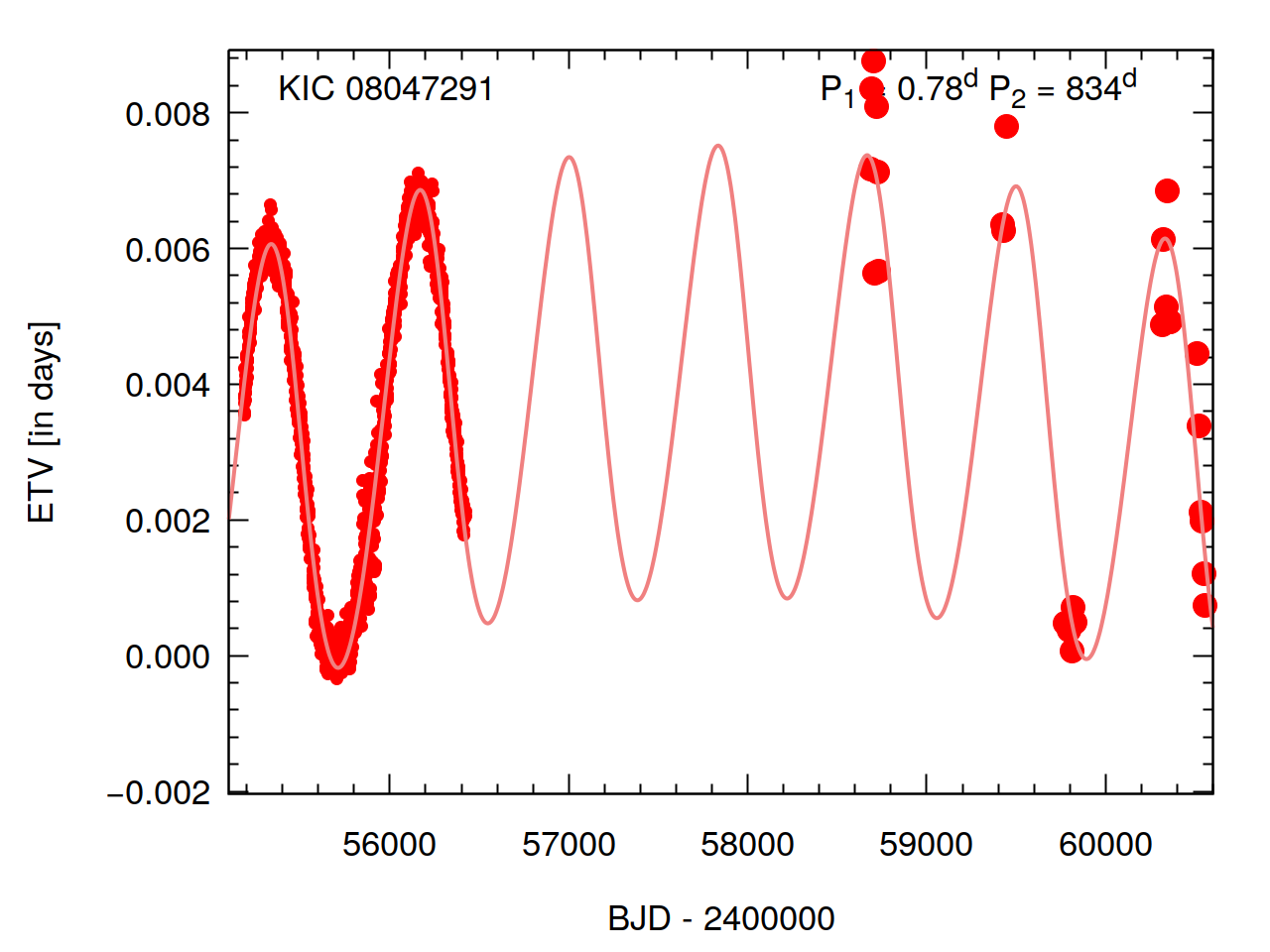}
\includegraphics[width=60mm]{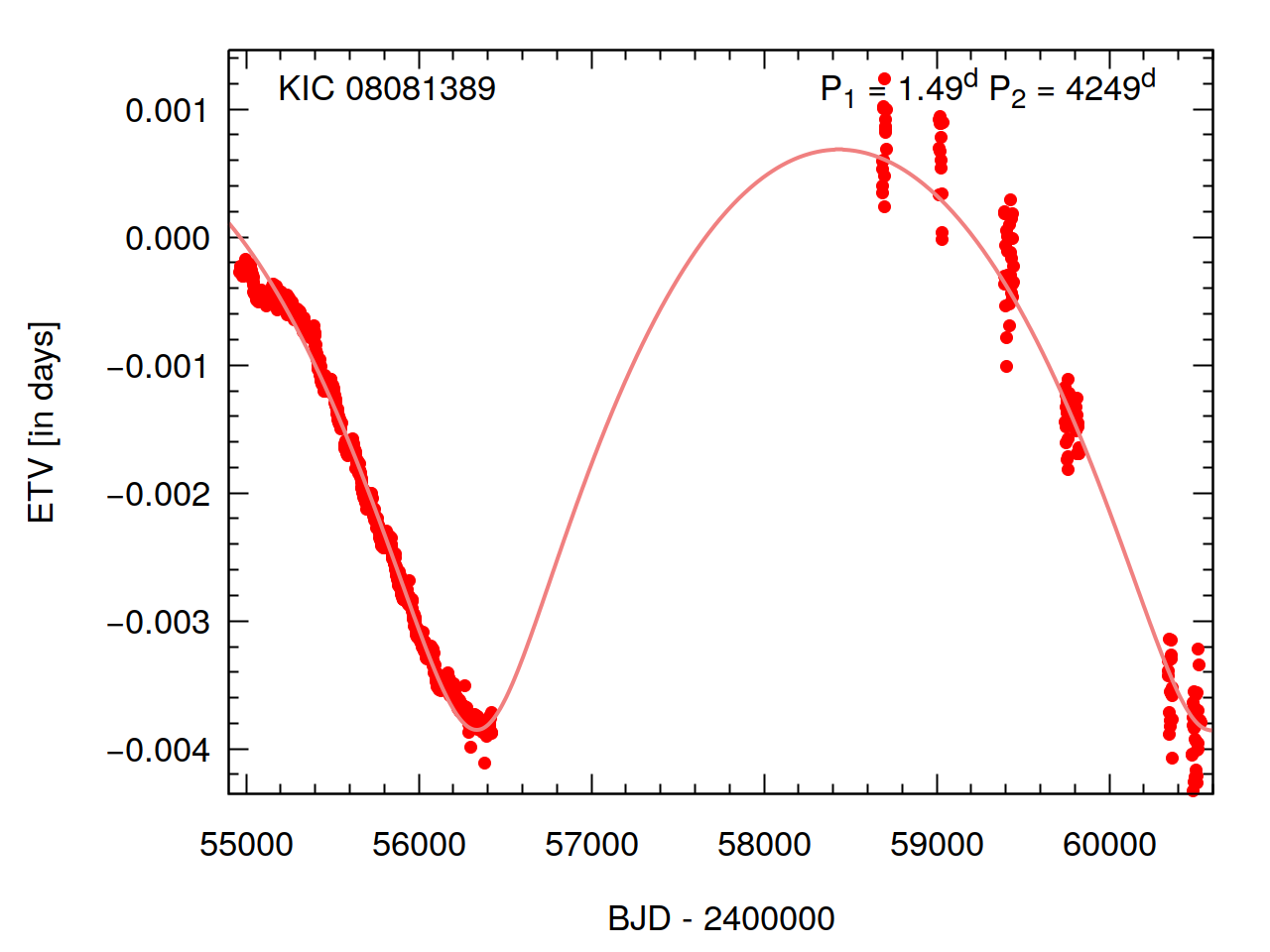}\includegraphics[width=60mm]{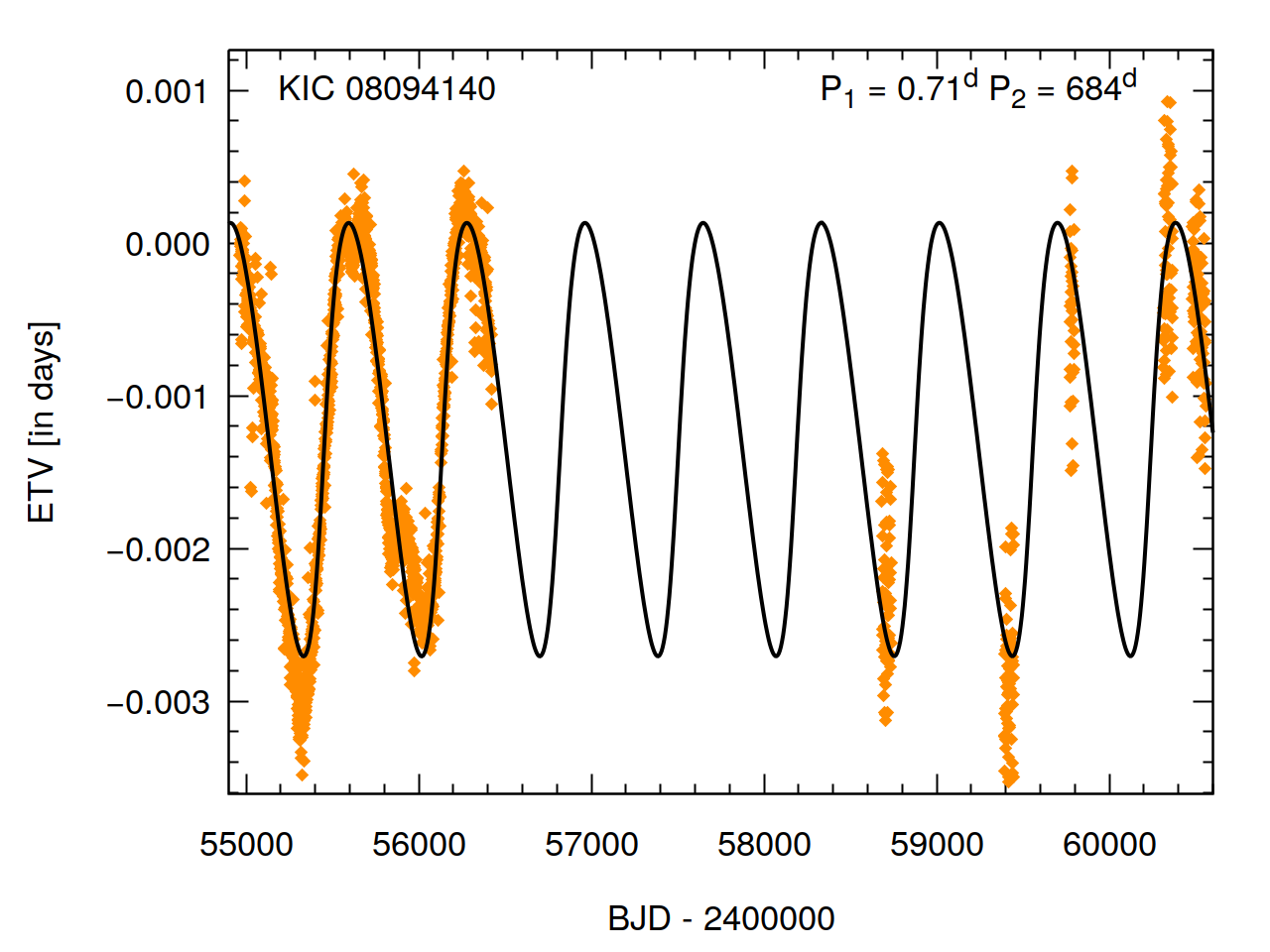}\includegraphics[width=60mm]{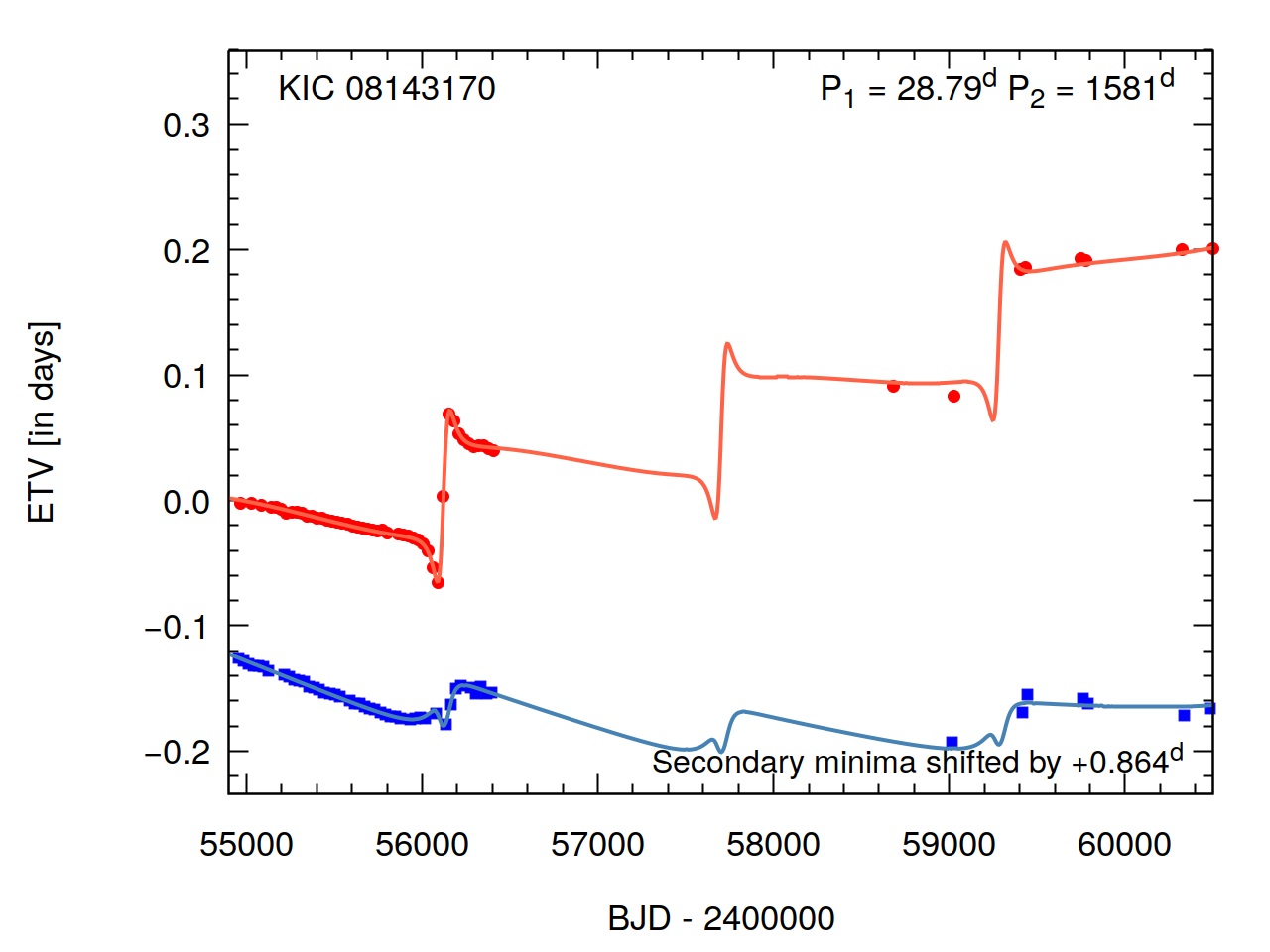}
\includegraphics[width=60mm]{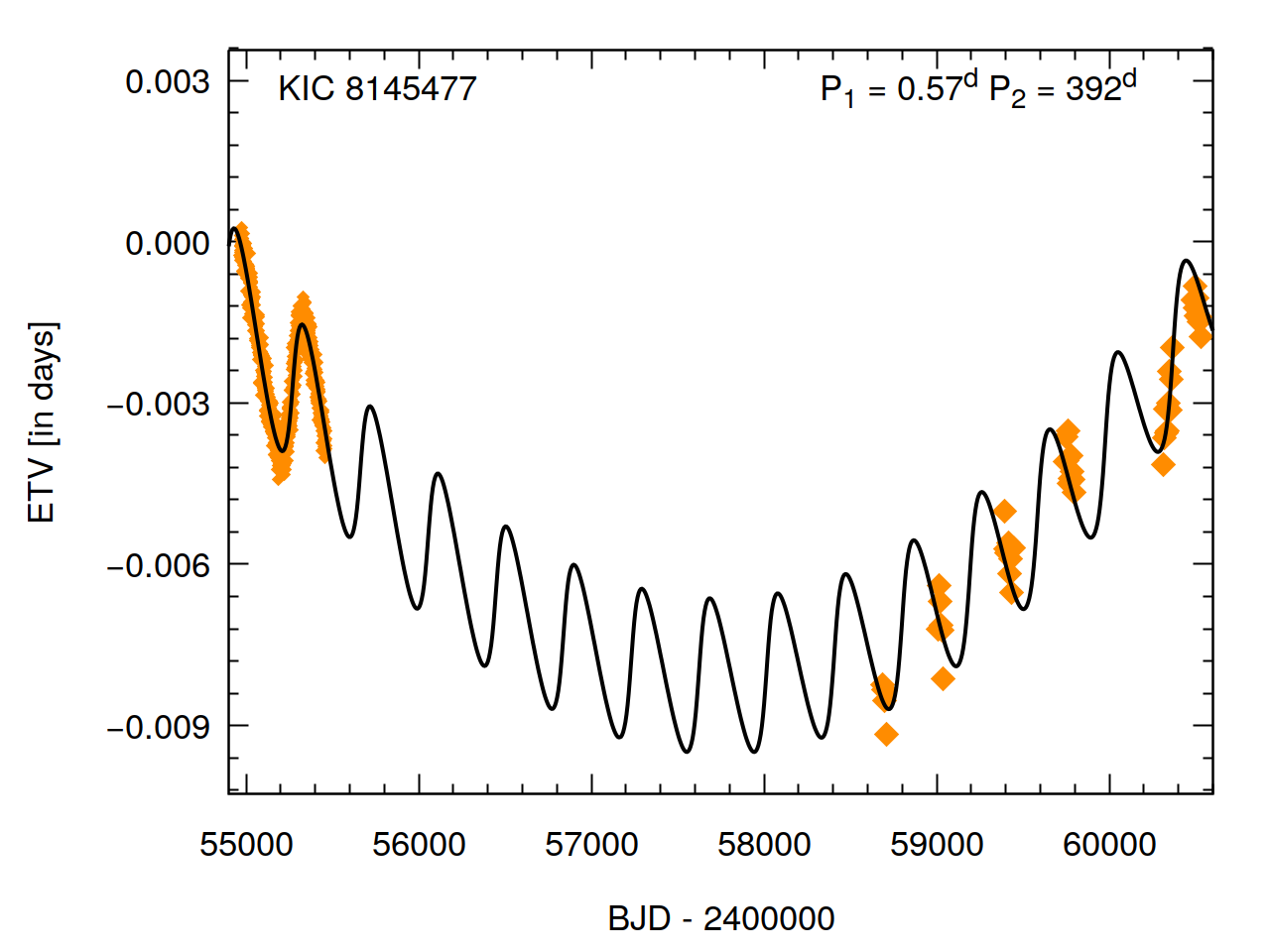}\includegraphics[width=60mm]{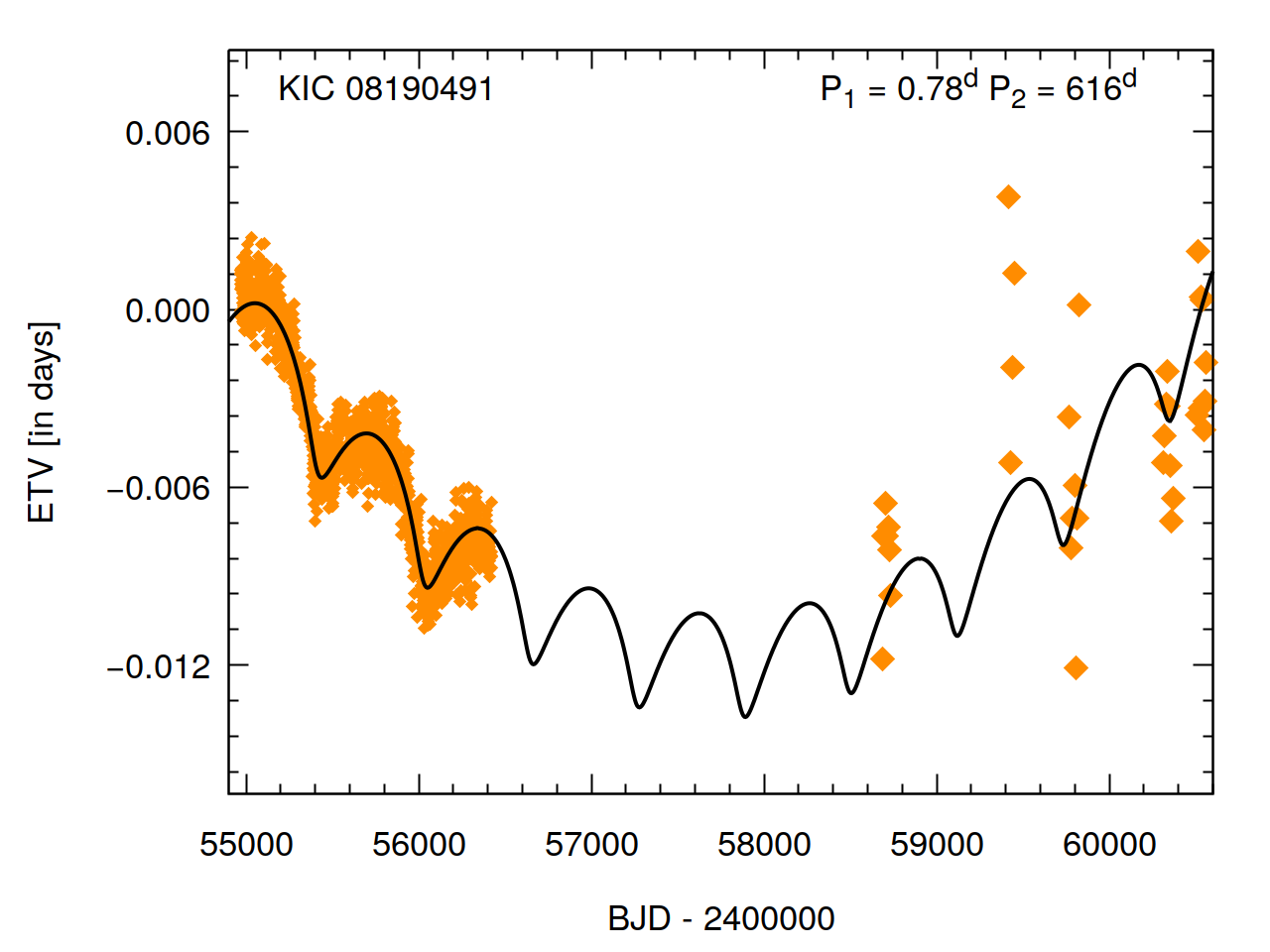}\includegraphics[width=60mm]{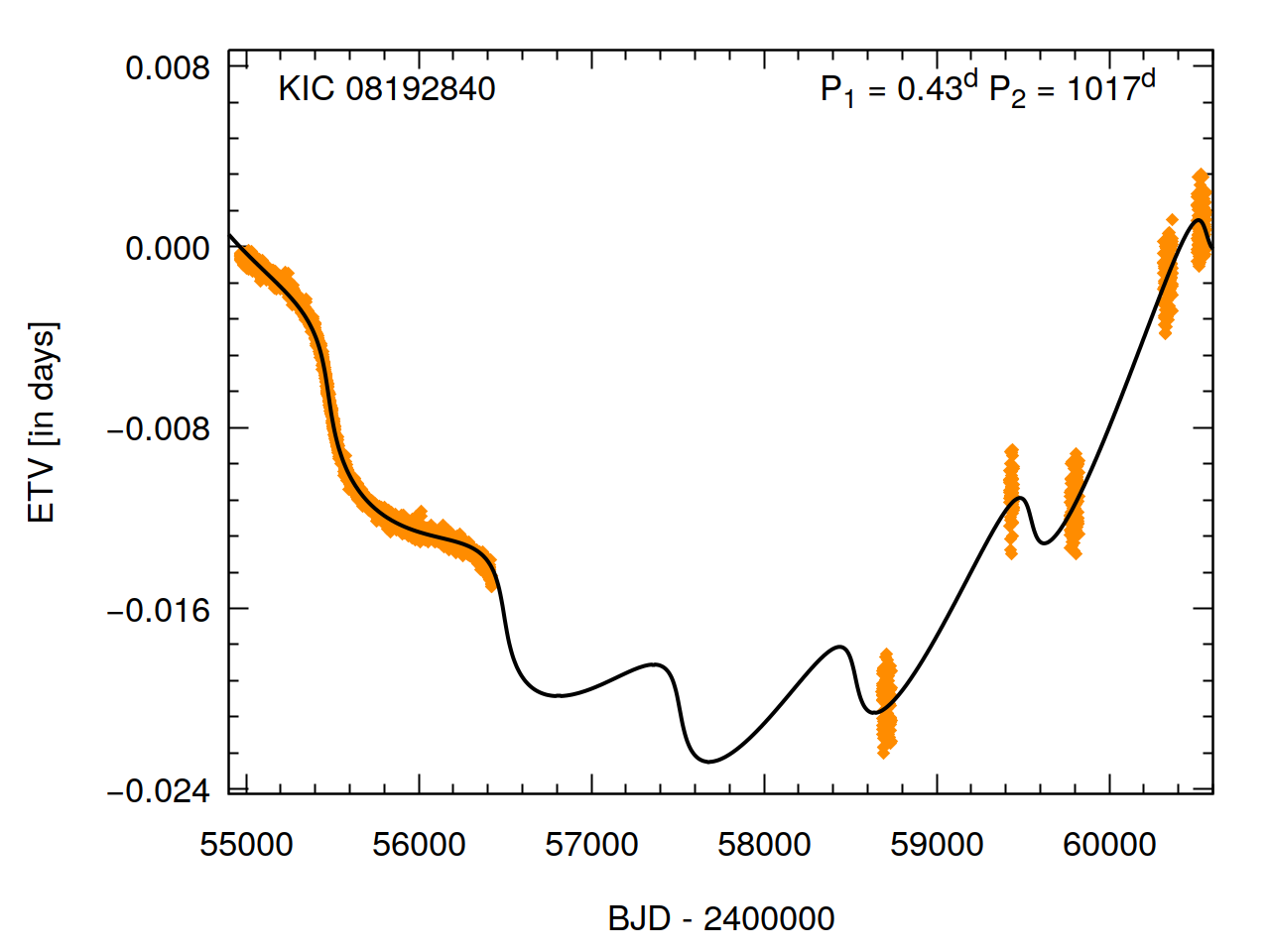}
\includegraphics[width=60mm]{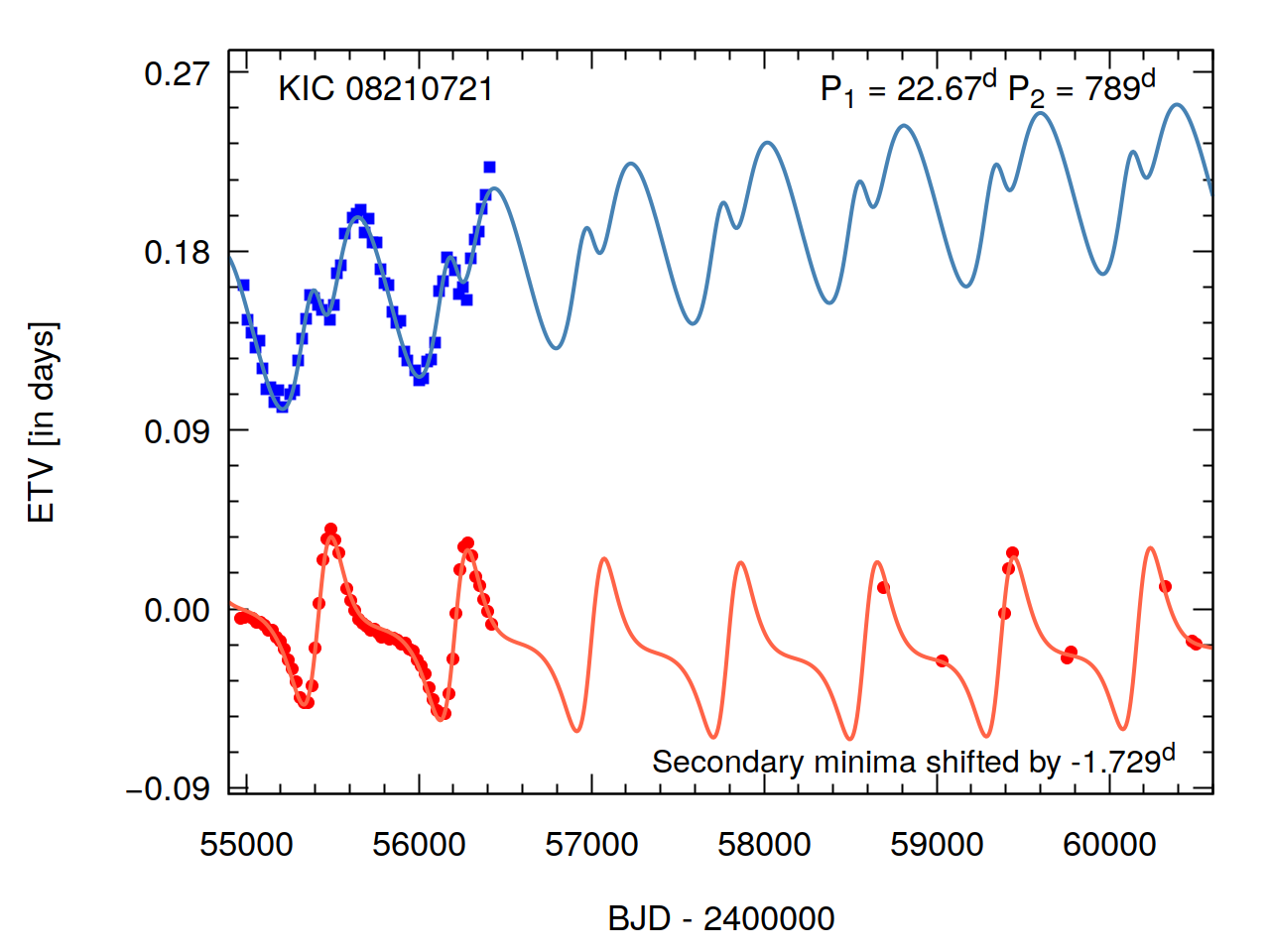}\includegraphics[width=60mm]{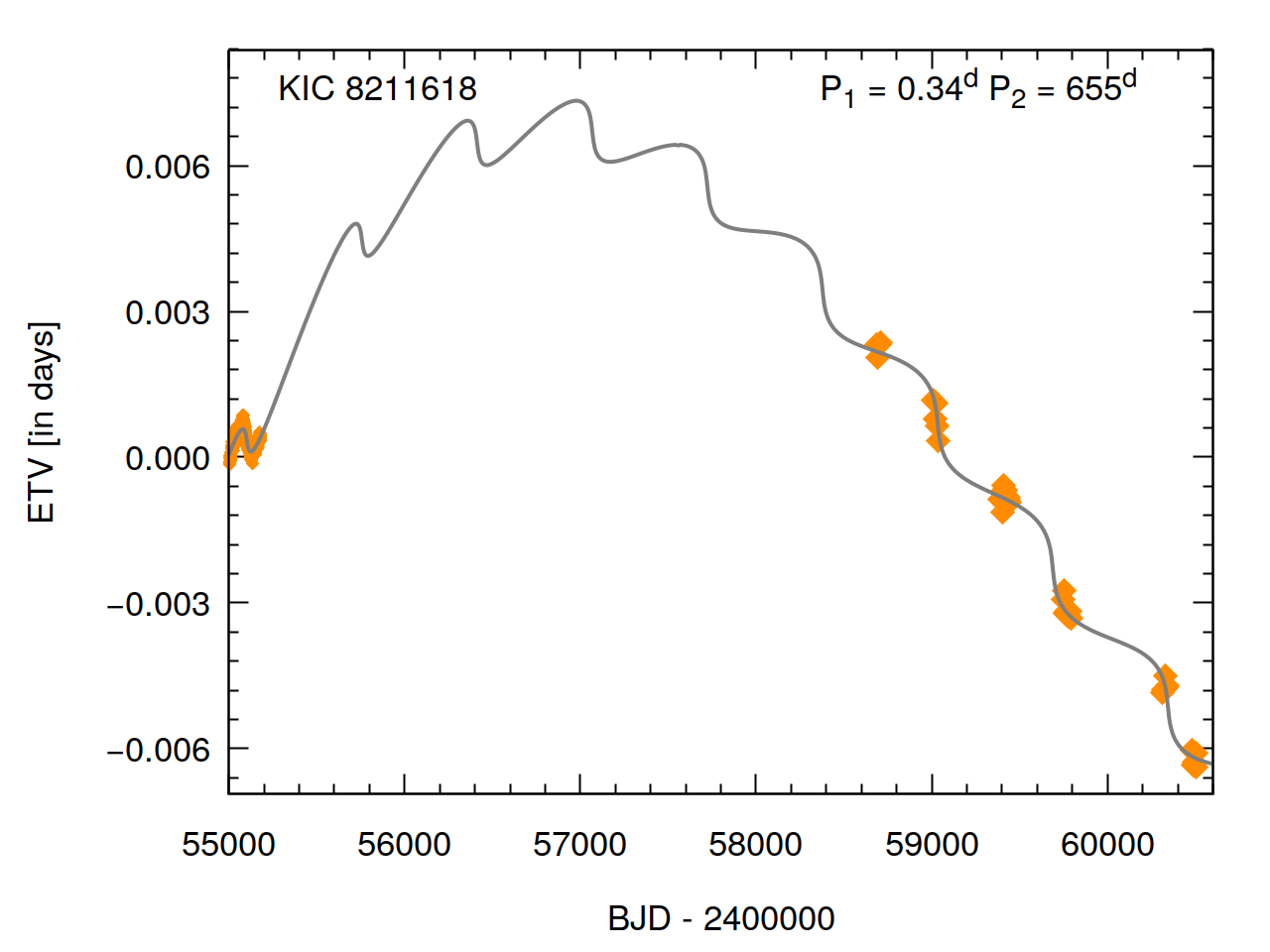}\includegraphics[width=60mm]{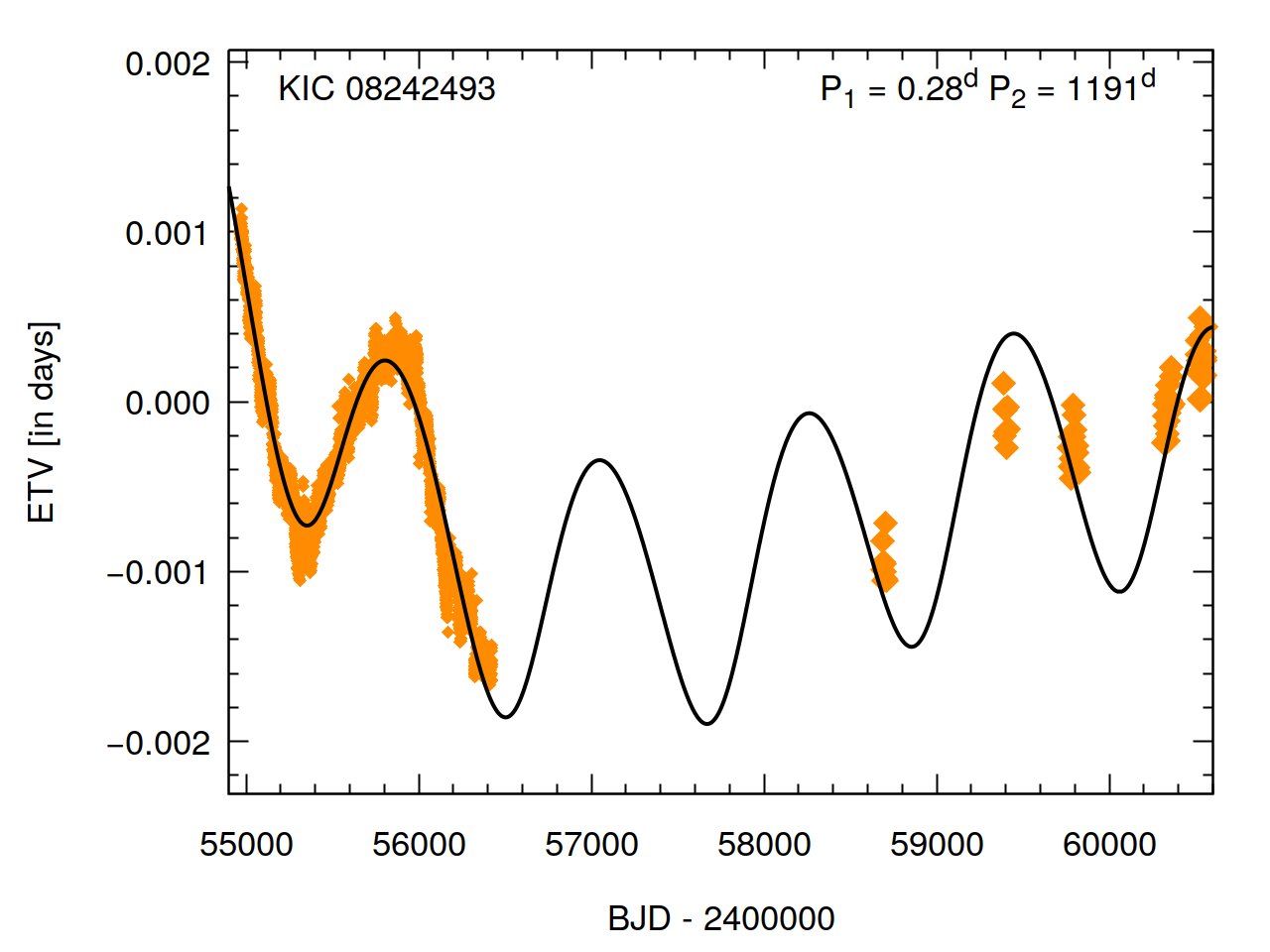}
\includegraphics[width=60mm]{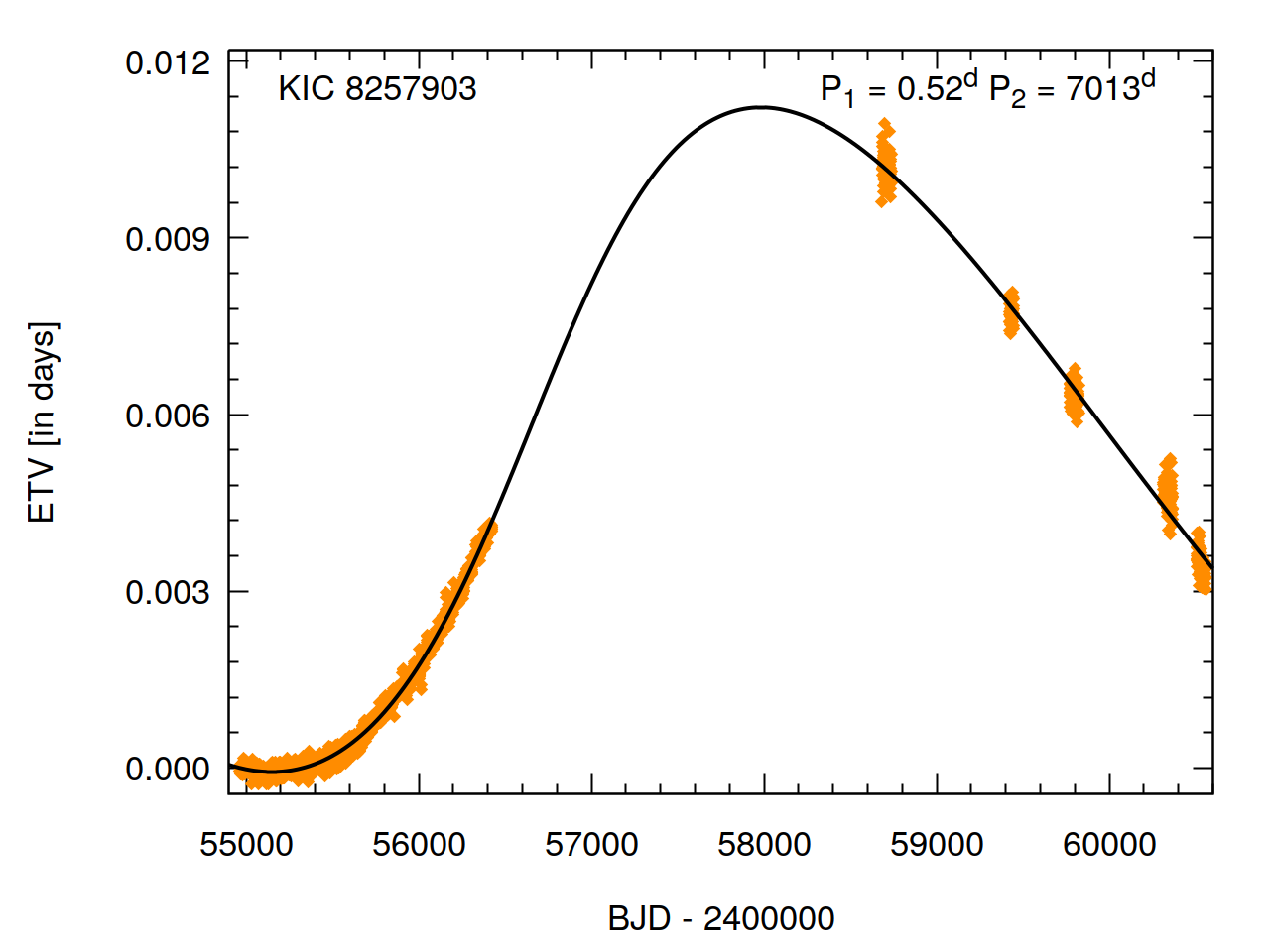}\includegraphics[width=60mm]{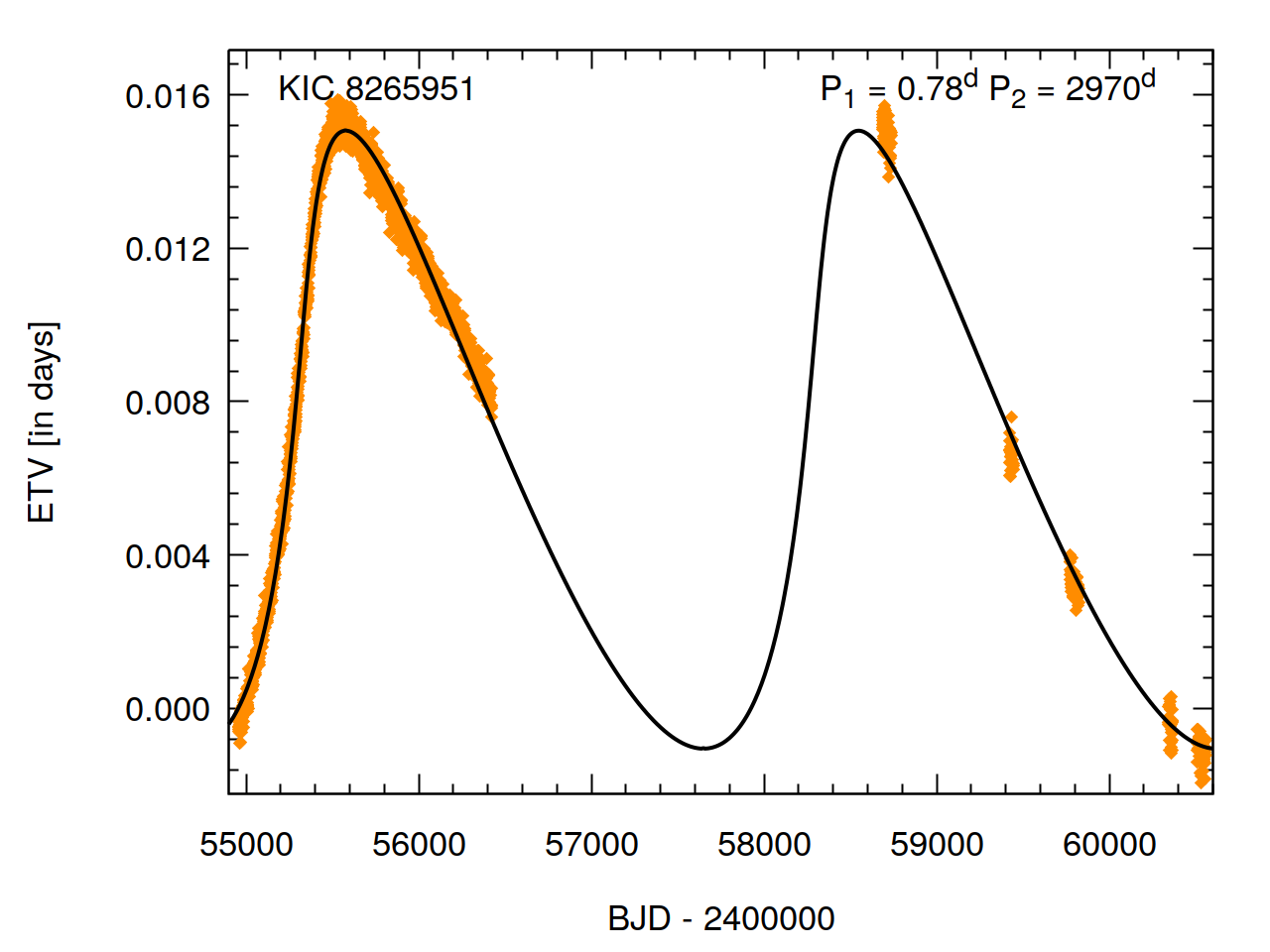}\includegraphics[width=60mm]{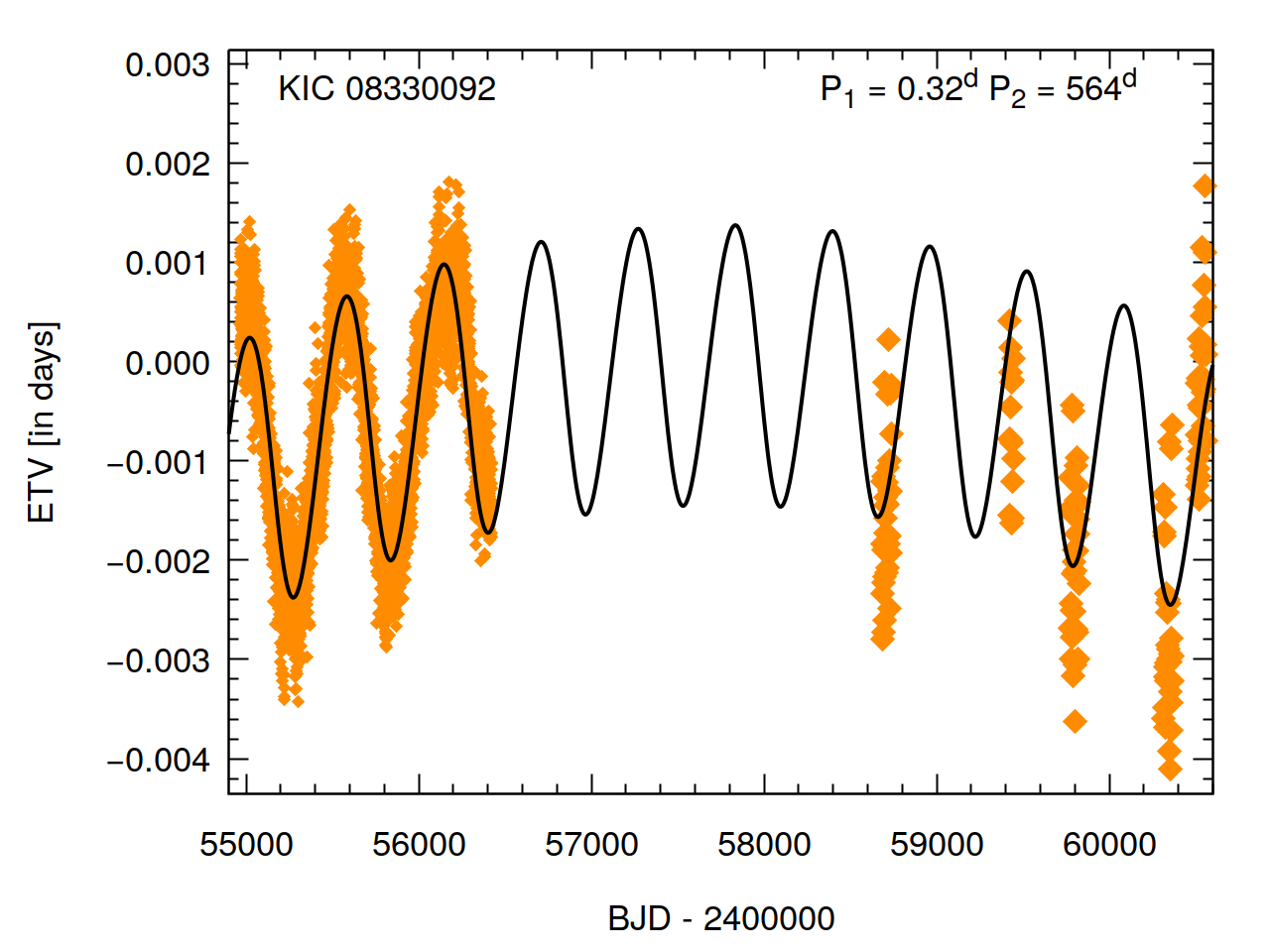}
\caption{continued.}
\end{figure*}

\addtocounter{figure}{-1}

\begin{figure*}
\includegraphics[width=60mm]{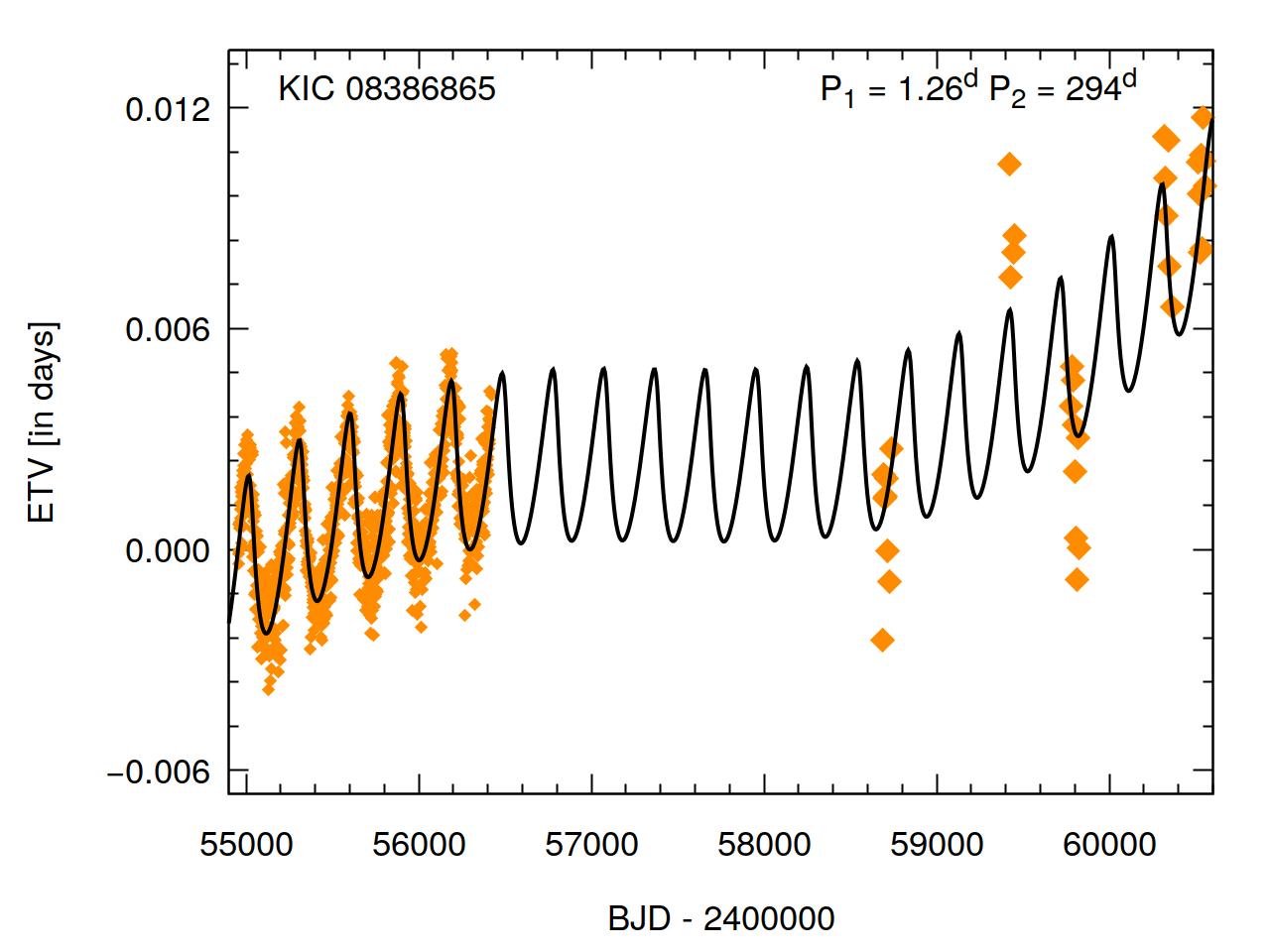}\includegraphics[width=60mm]{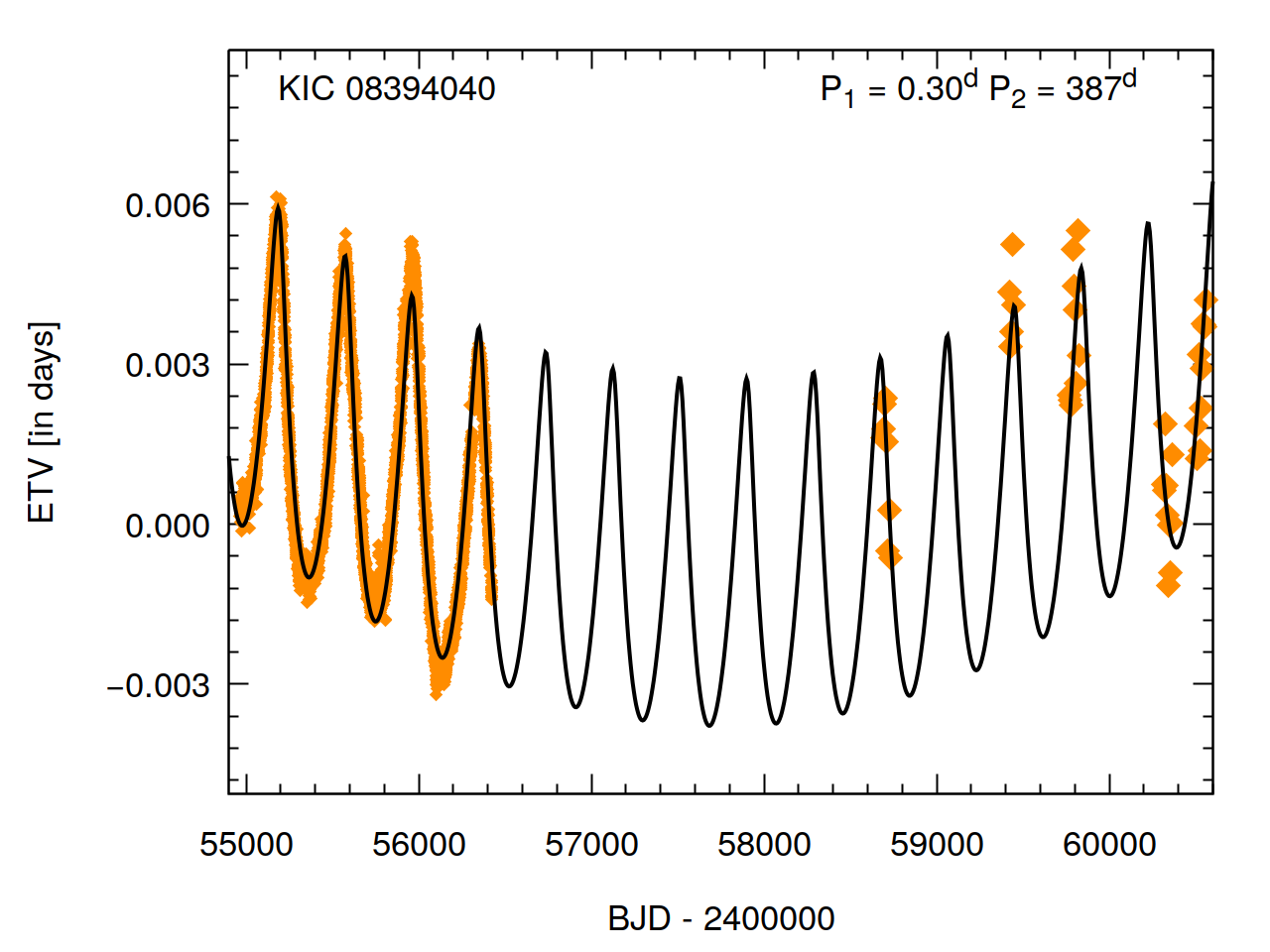}\includegraphics[width=60mm]{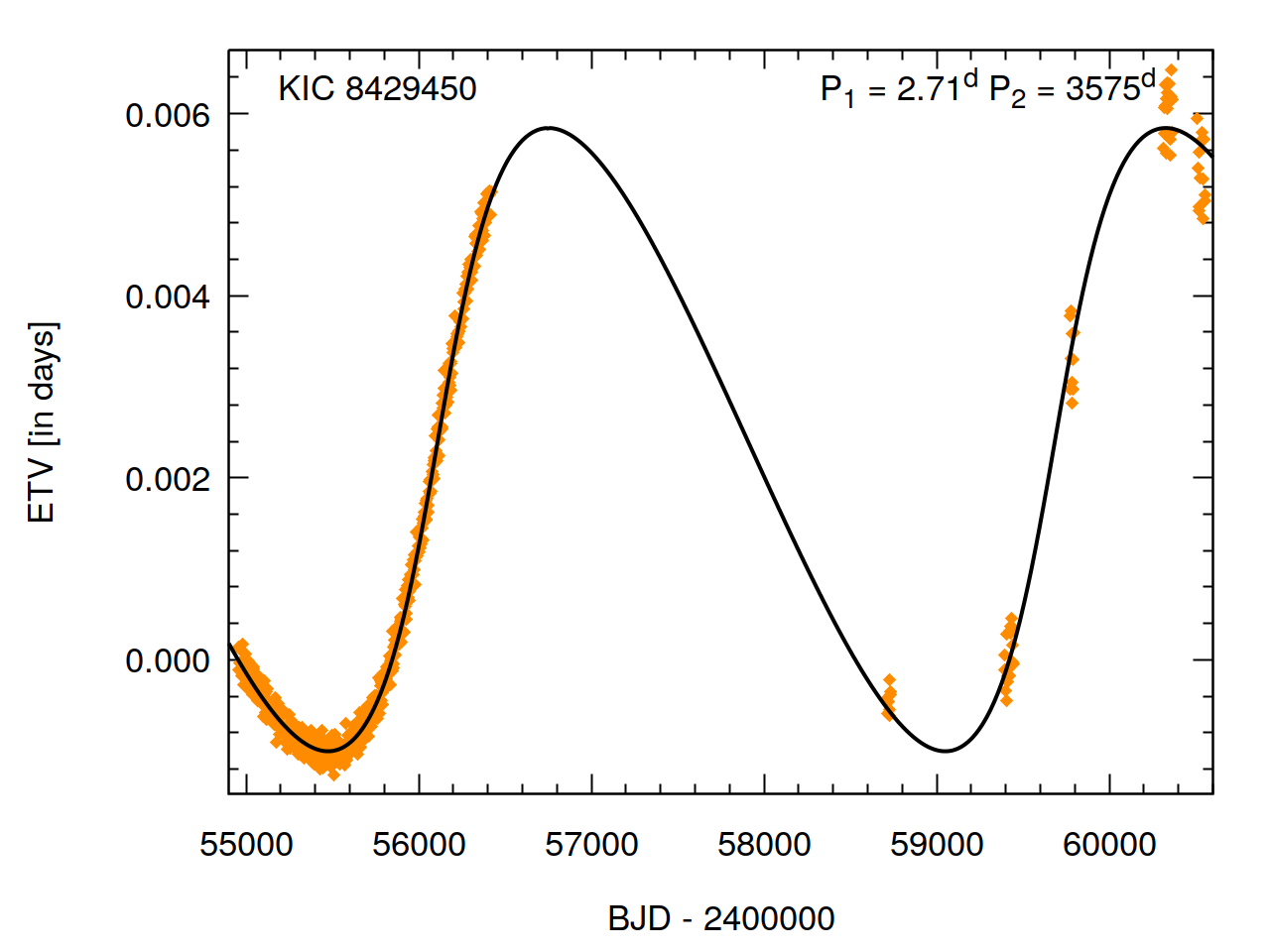}
\includegraphics[width=60mm]{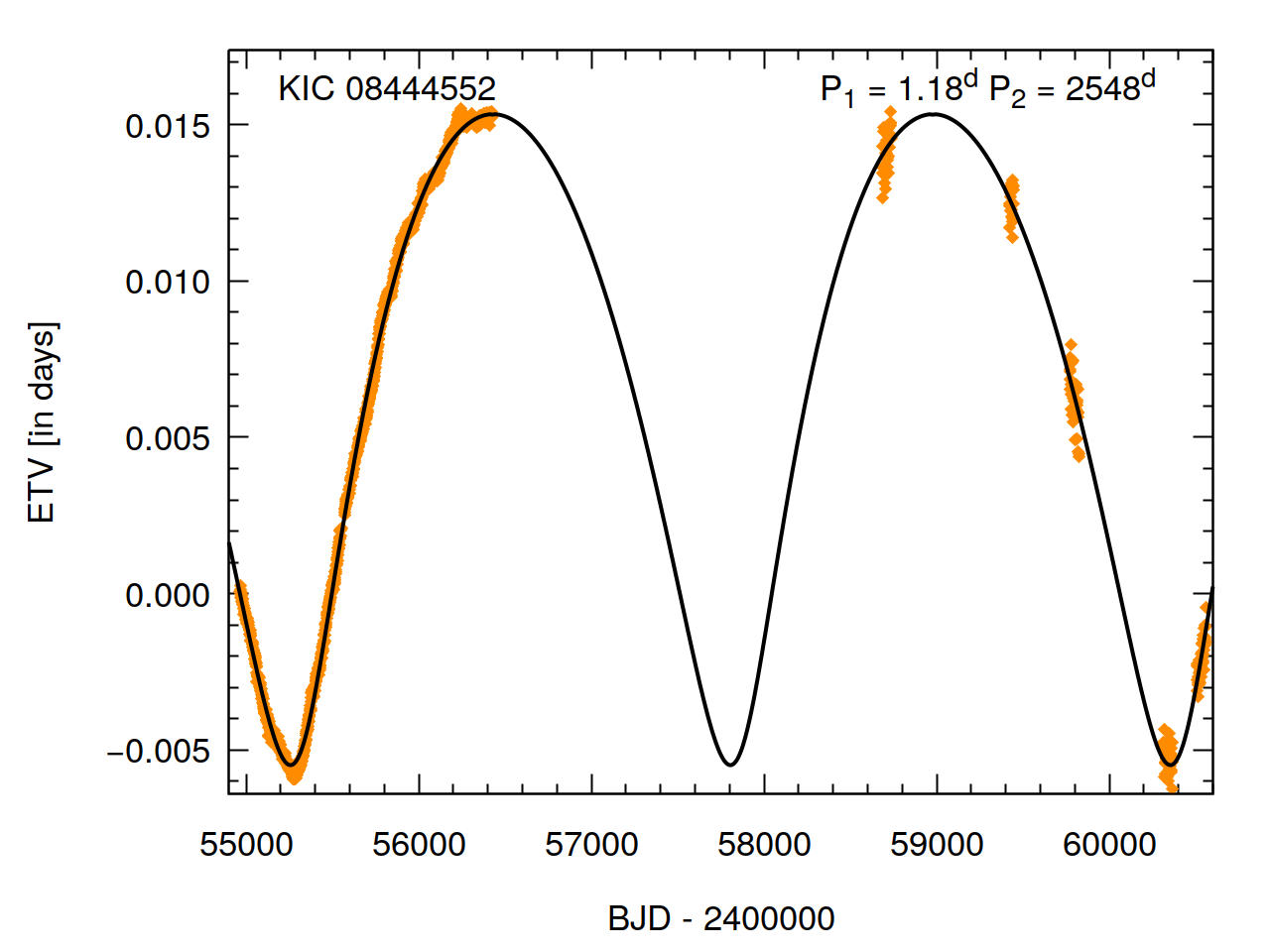}\includegraphics[width=60mm]{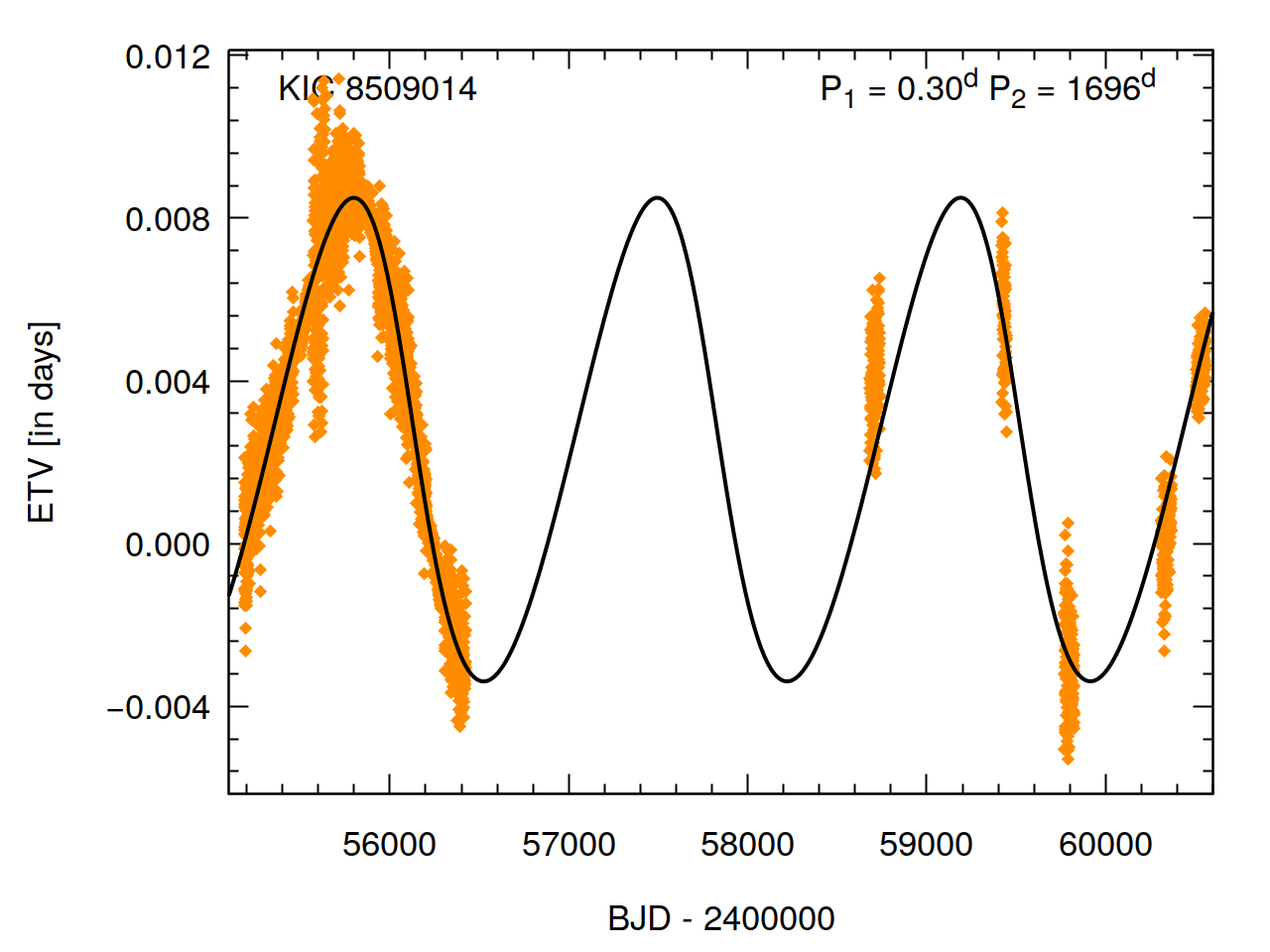}\includegraphics[width=60mm]{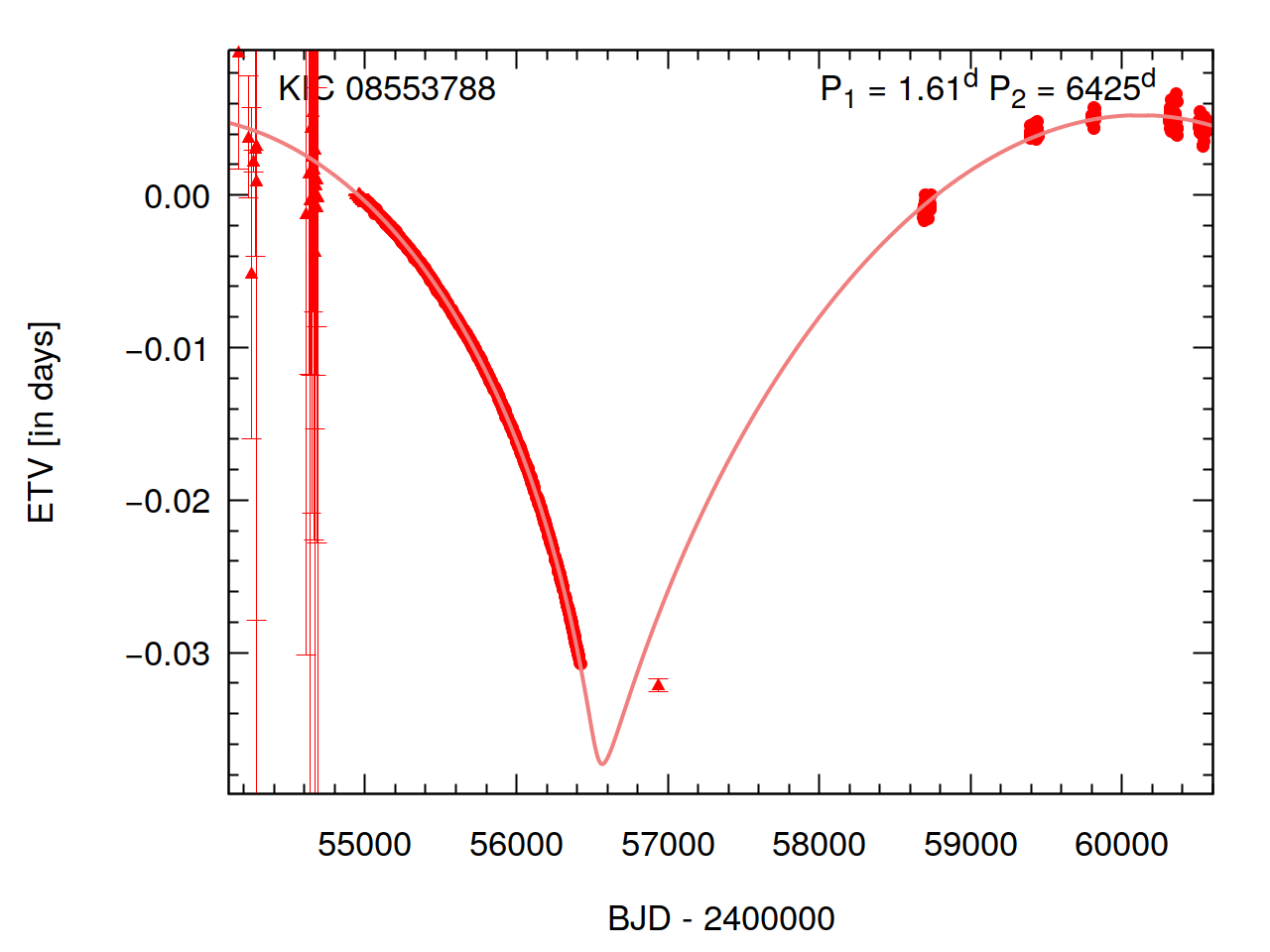}
\includegraphics[width=60mm]{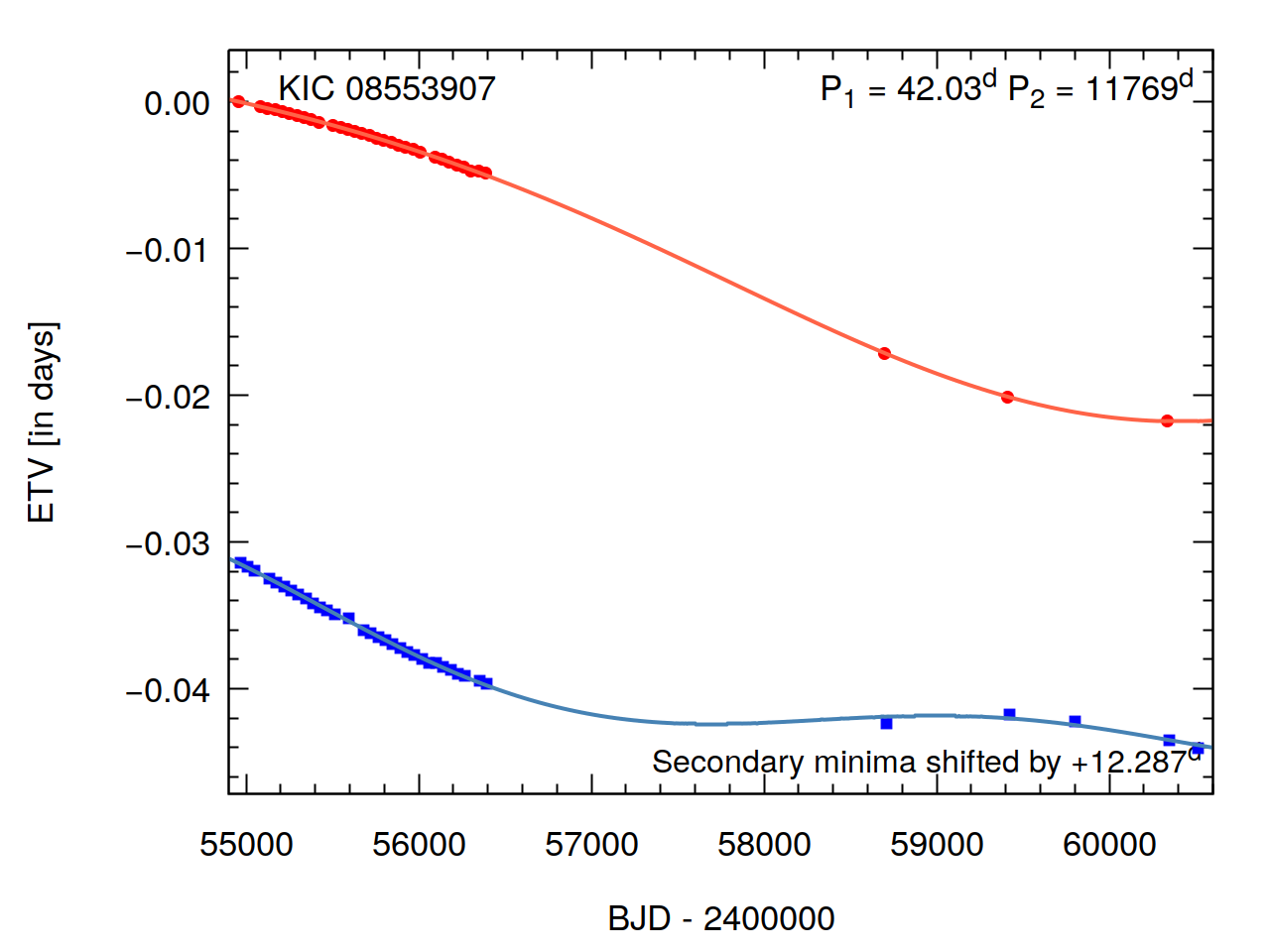}\includegraphics[width=60mm]{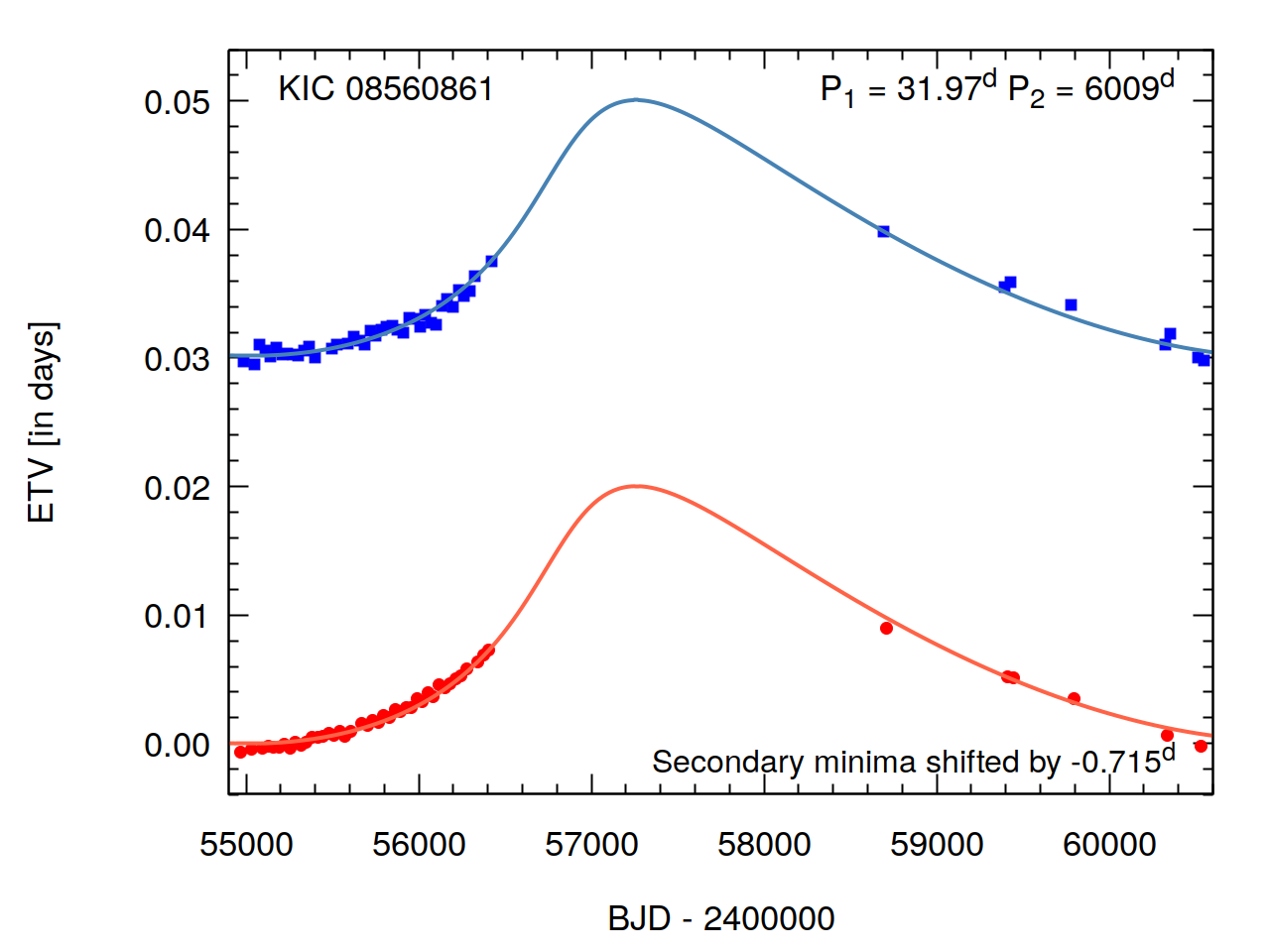}\includegraphics[width=60mm]{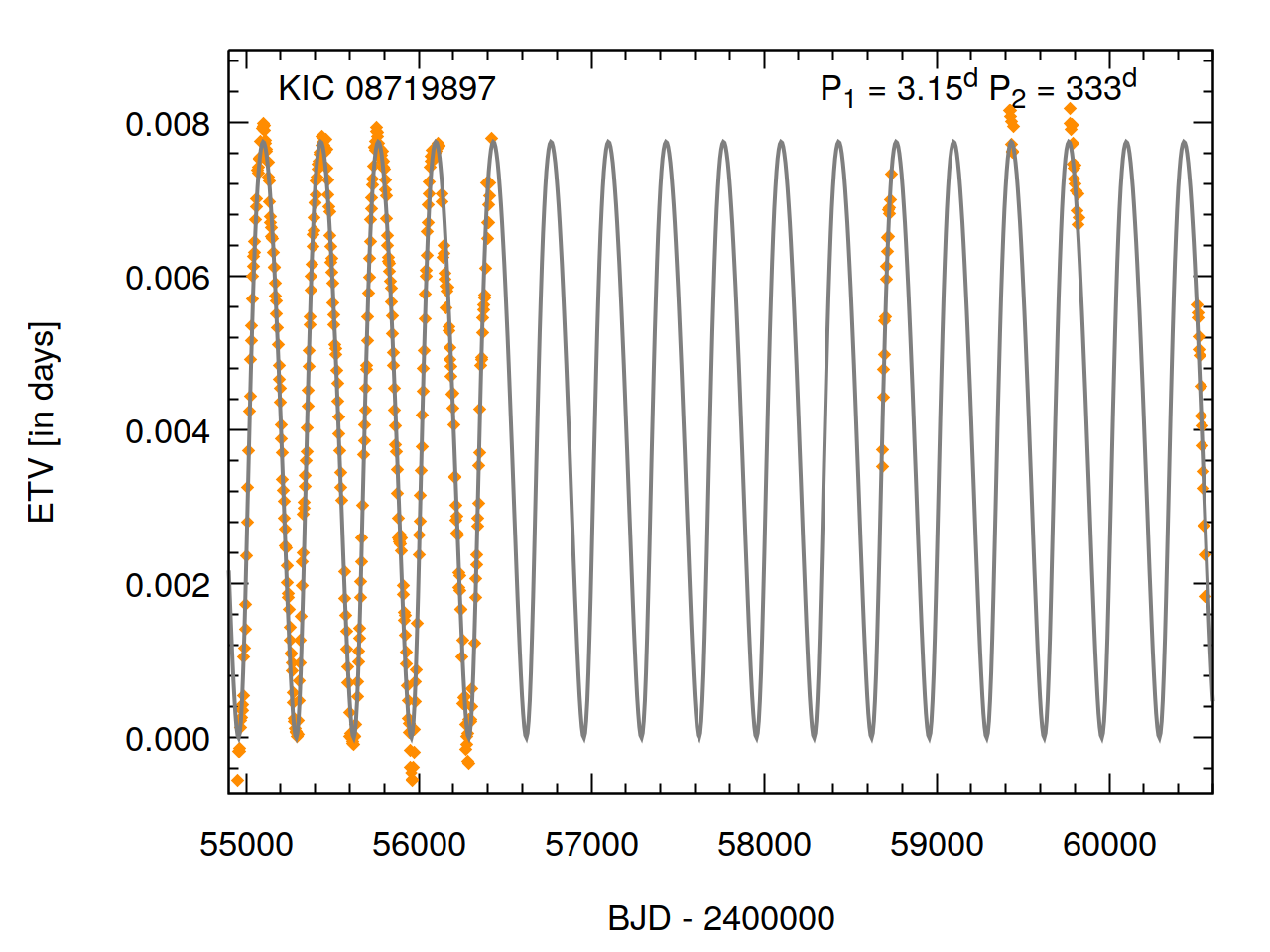}
\includegraphics[width=60mm]{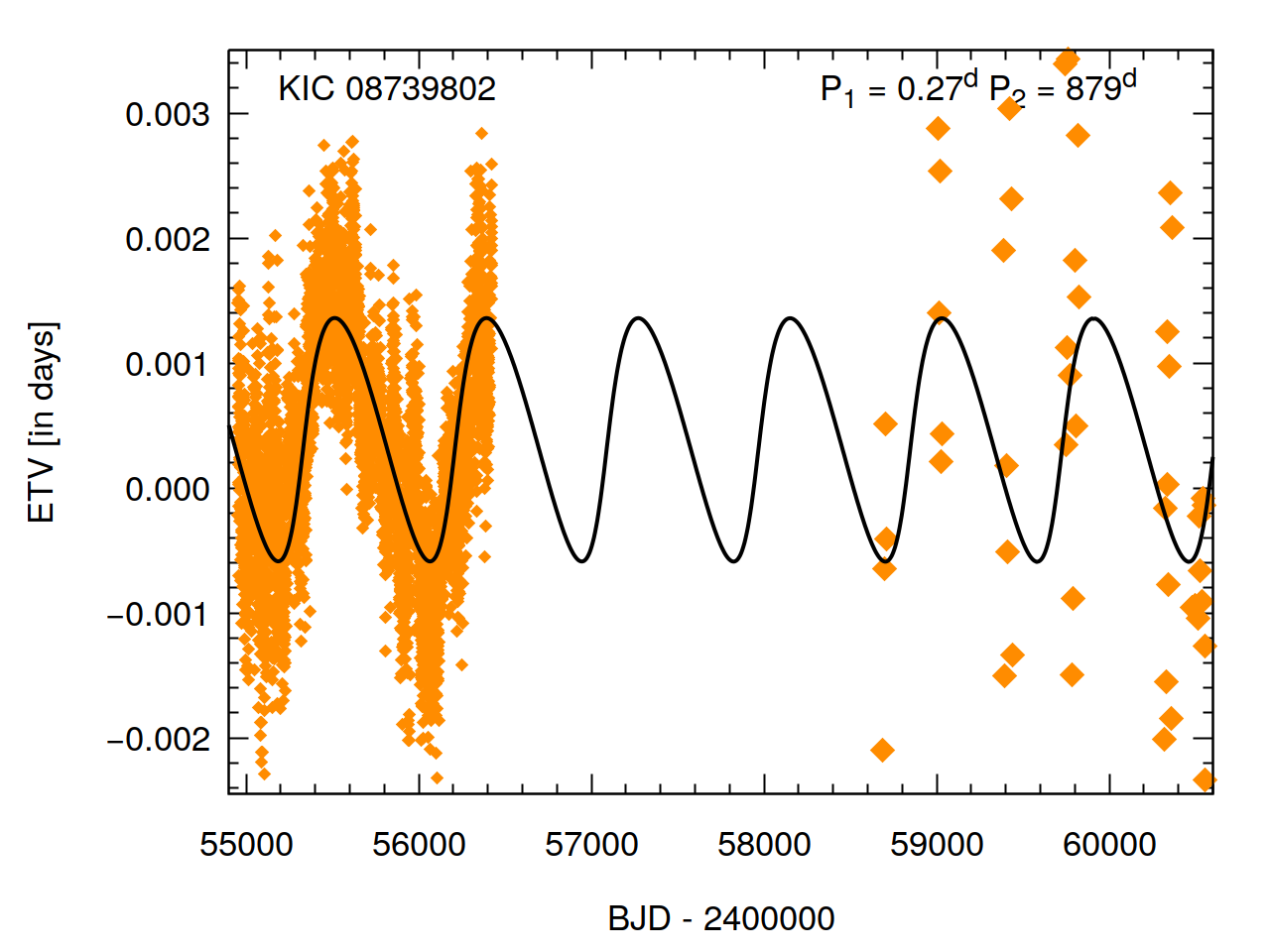}\includegraphics[width=60mm]{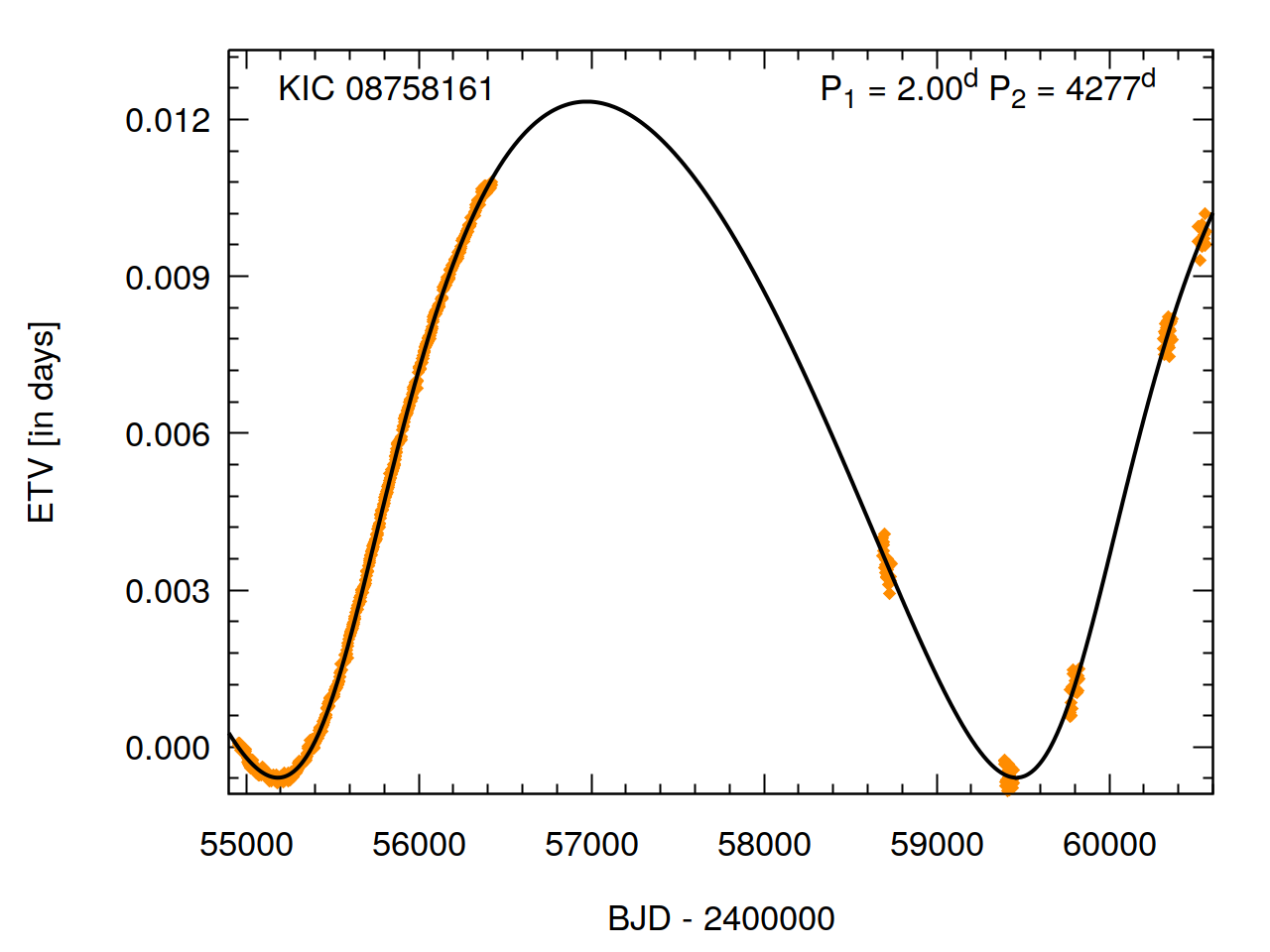}\includegraphics[width=60mm]{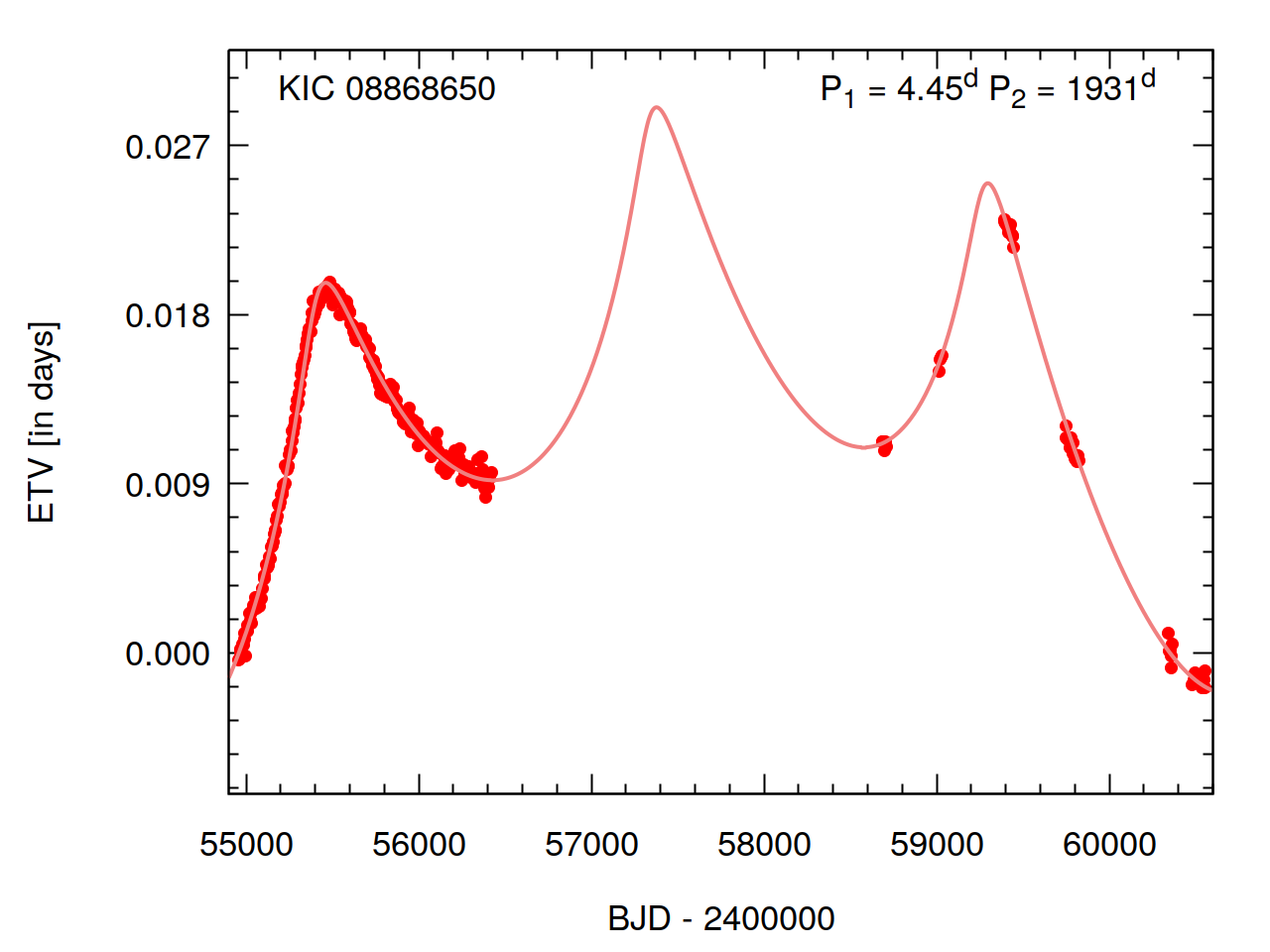}
\includegraphics[width=60mm]{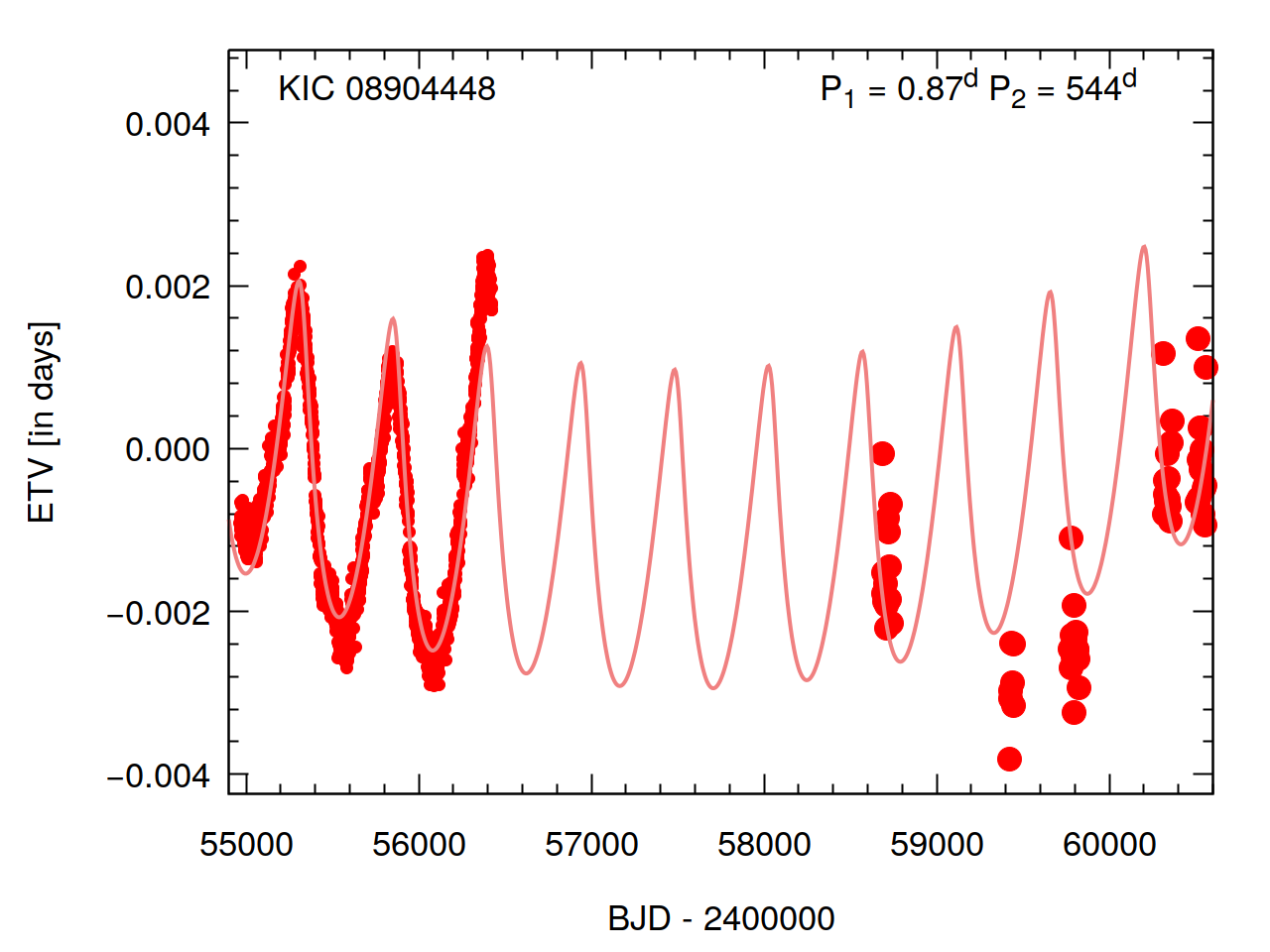}\includegraphics[width=60mm]{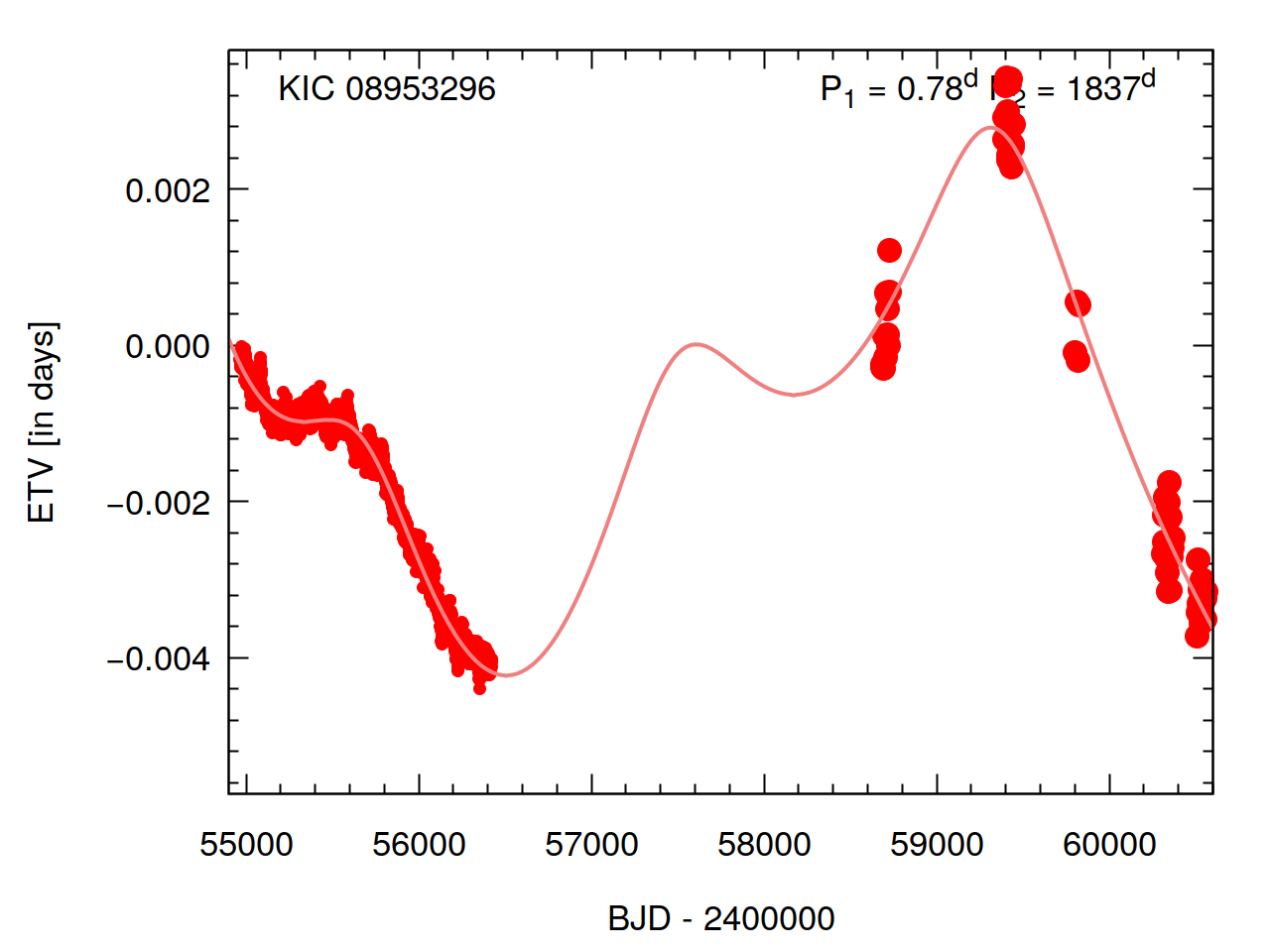}\includegraphics[width=60mm]{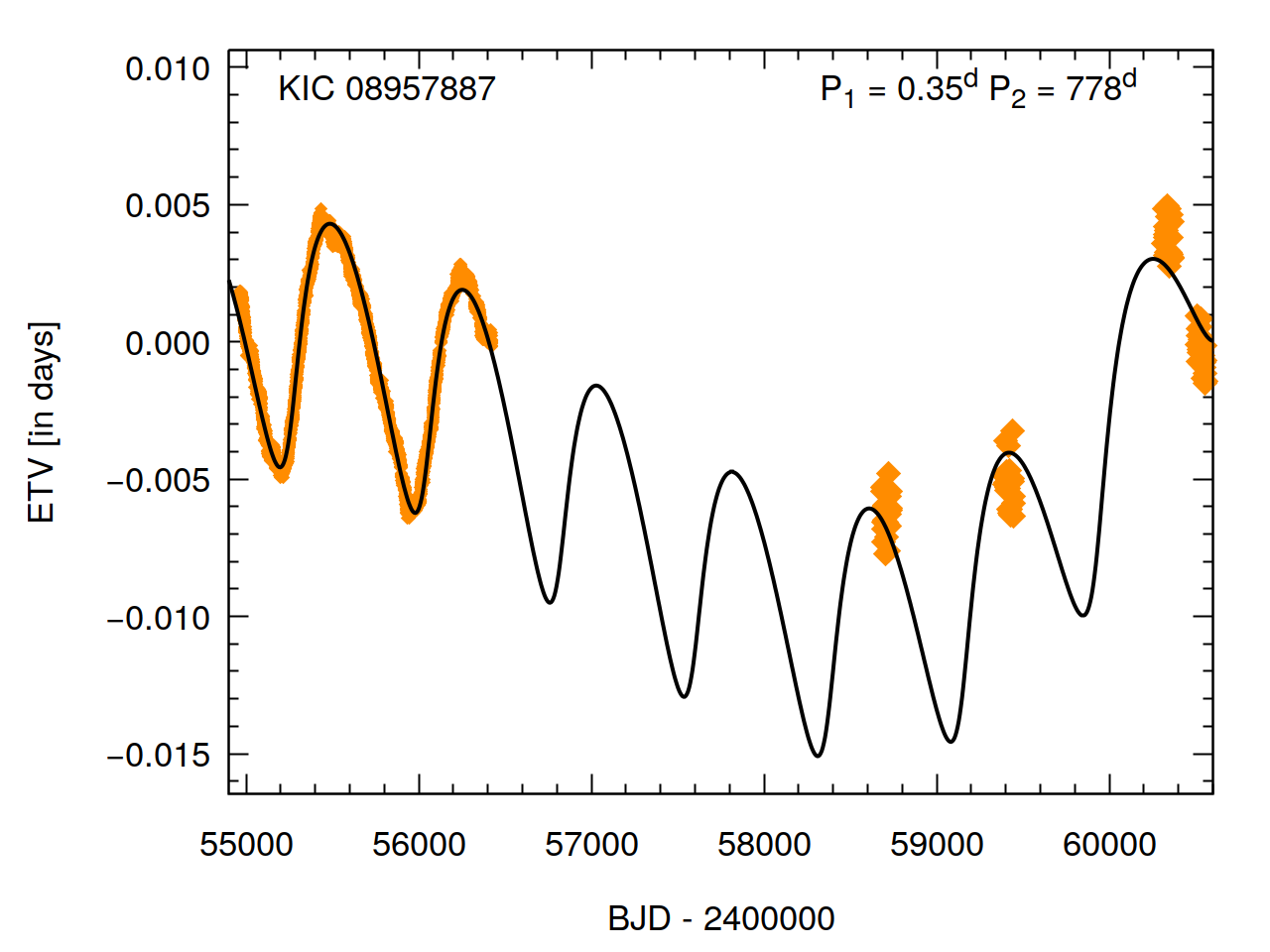}
\caption{continued.}
\end{figure*}

\addtocounter{figure}{-1}

\begin{figure*}
\includegraphics[width=60mm]{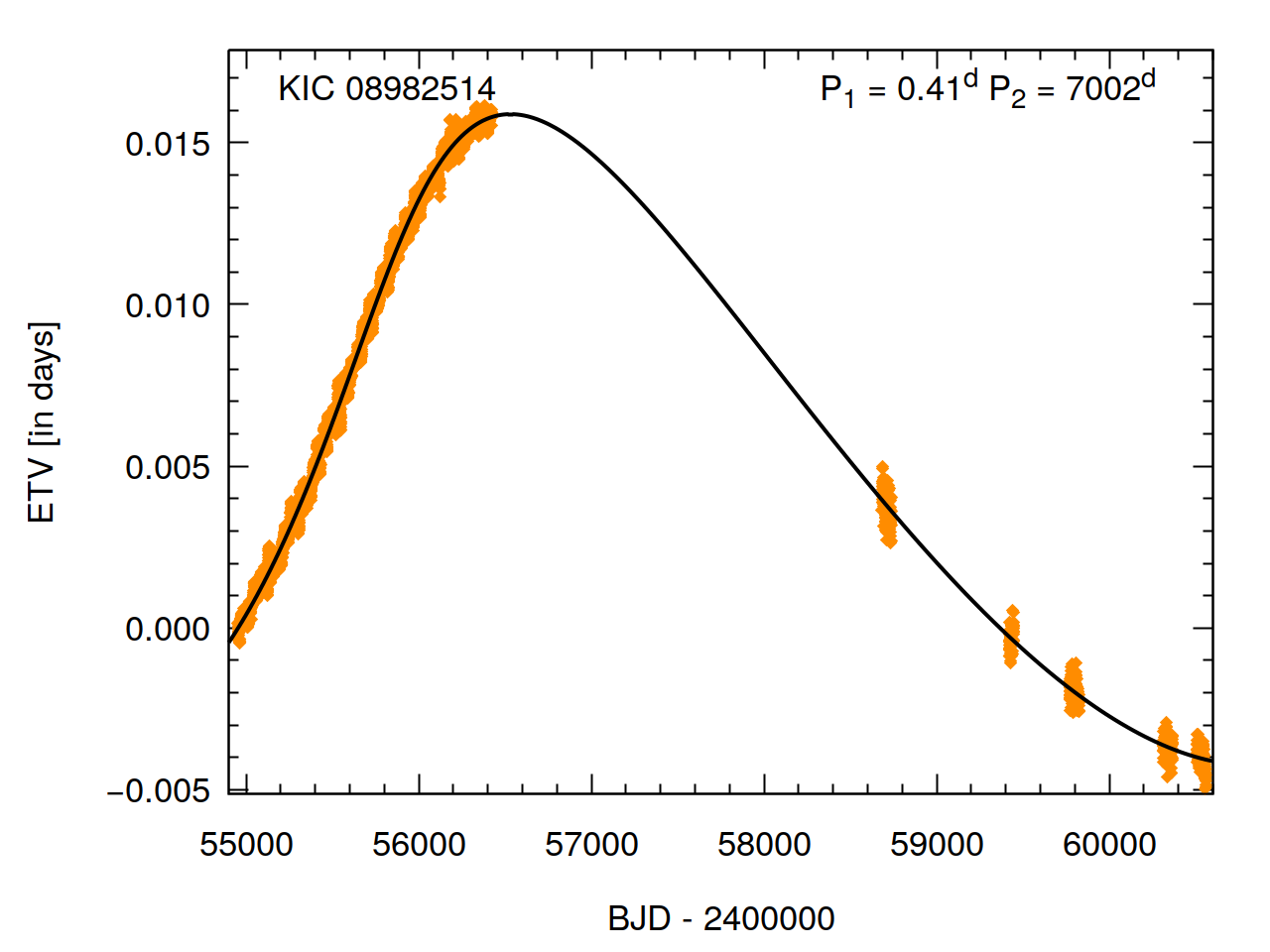}\includegraphics[width=60mm]{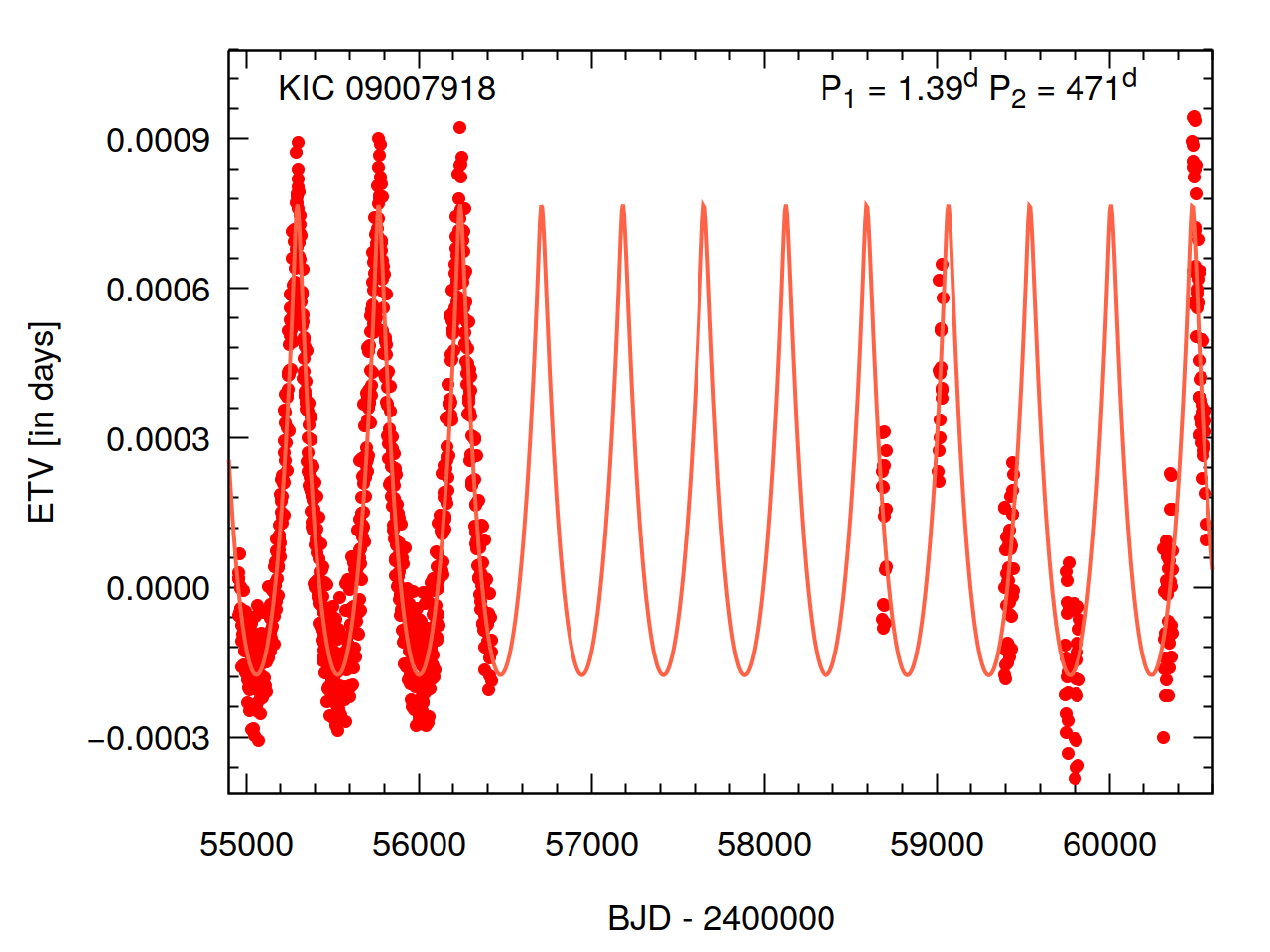}\includegraphics[width=60mm]{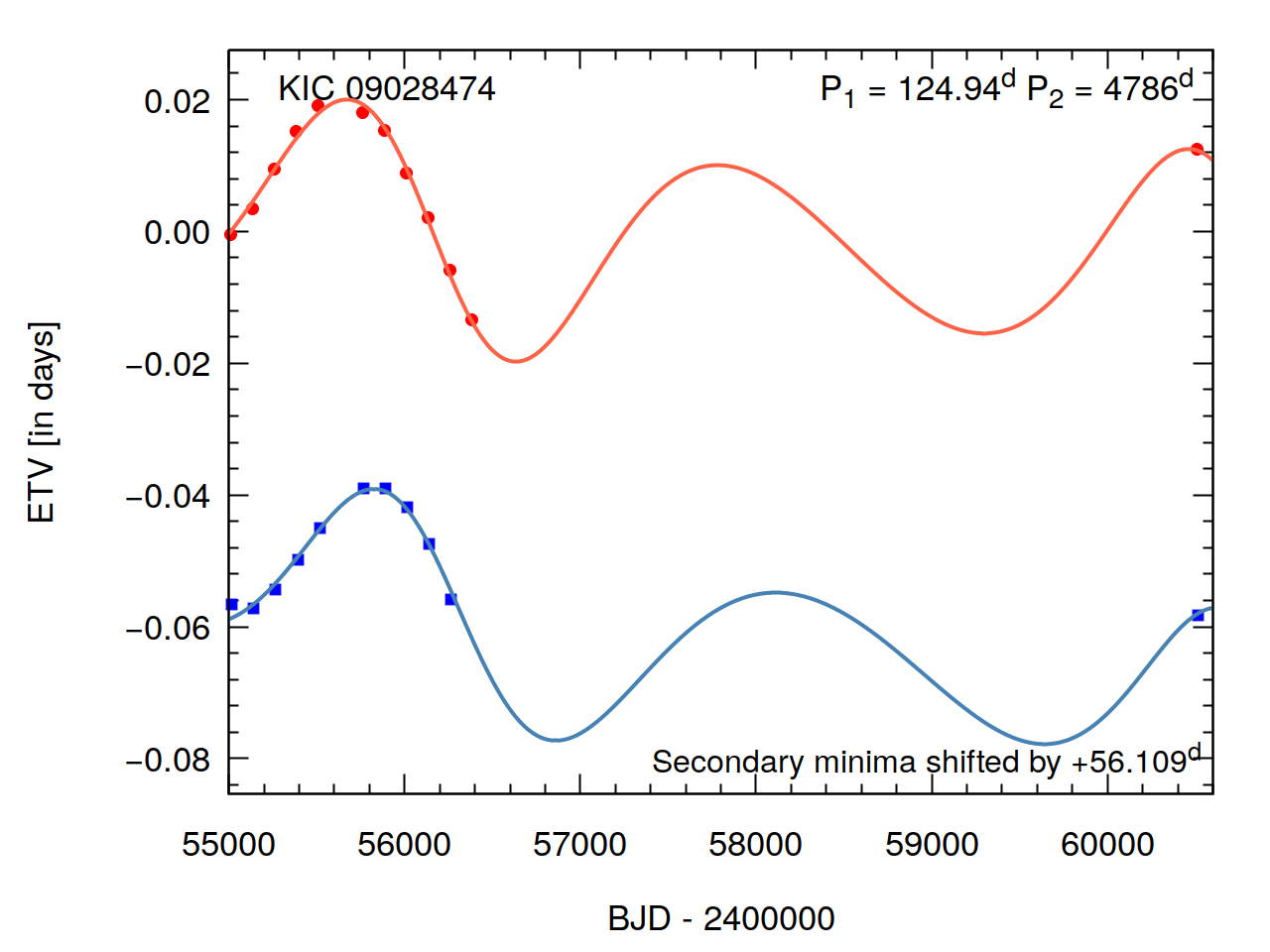}
\includegraphics[width=60mm]{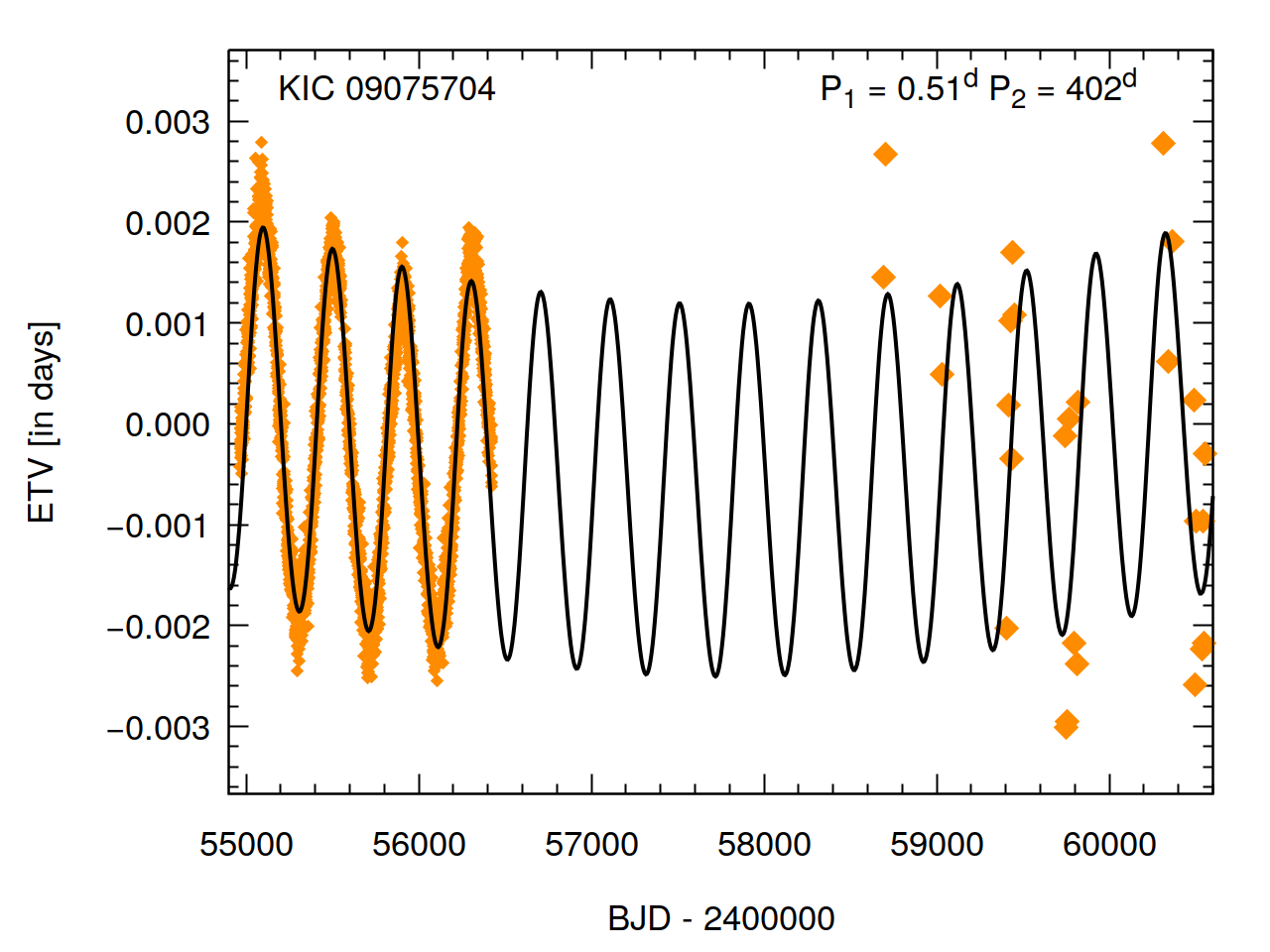}\includegraphics[width=60mm]{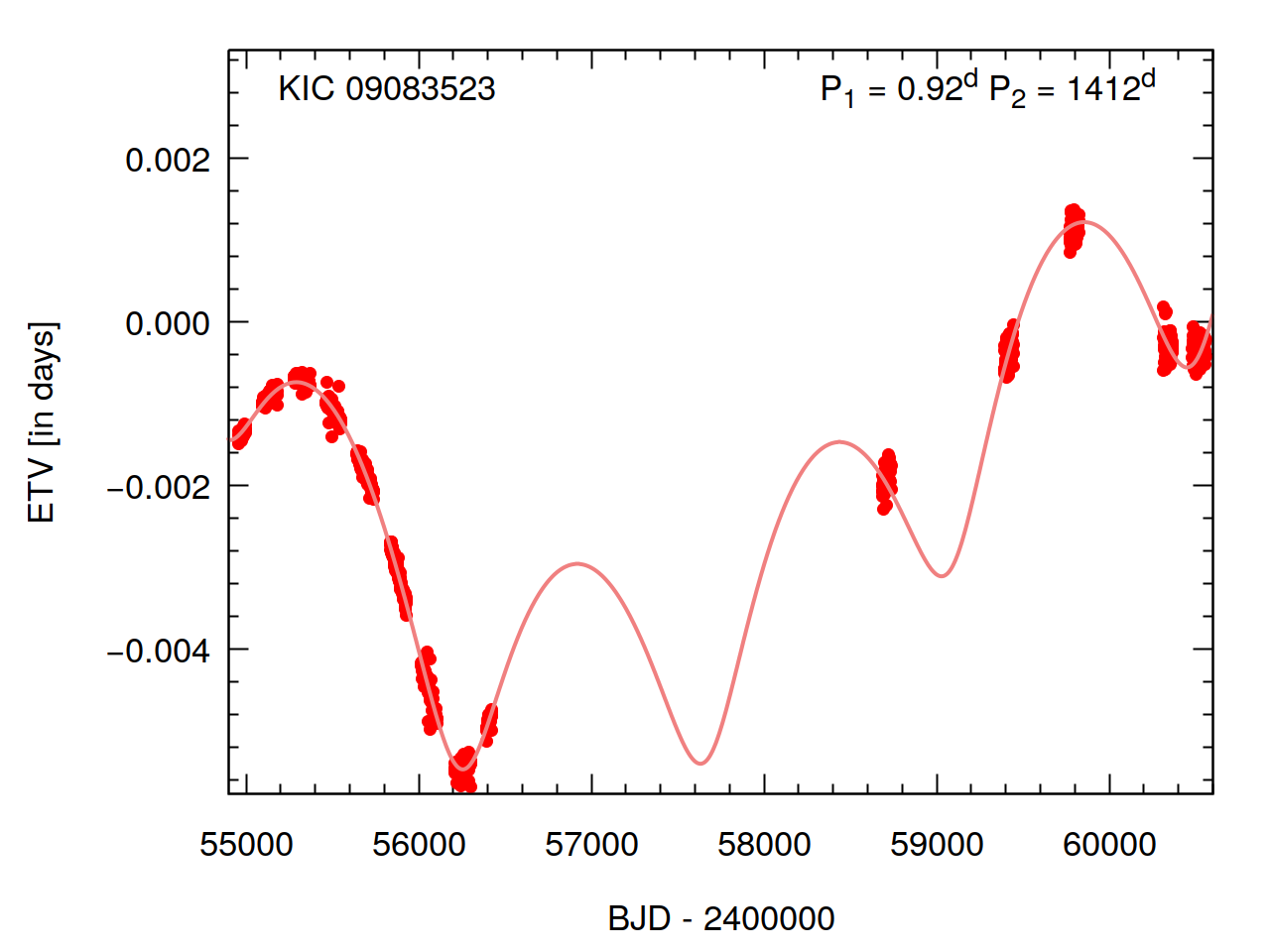}\includegraphics[width=60mm]{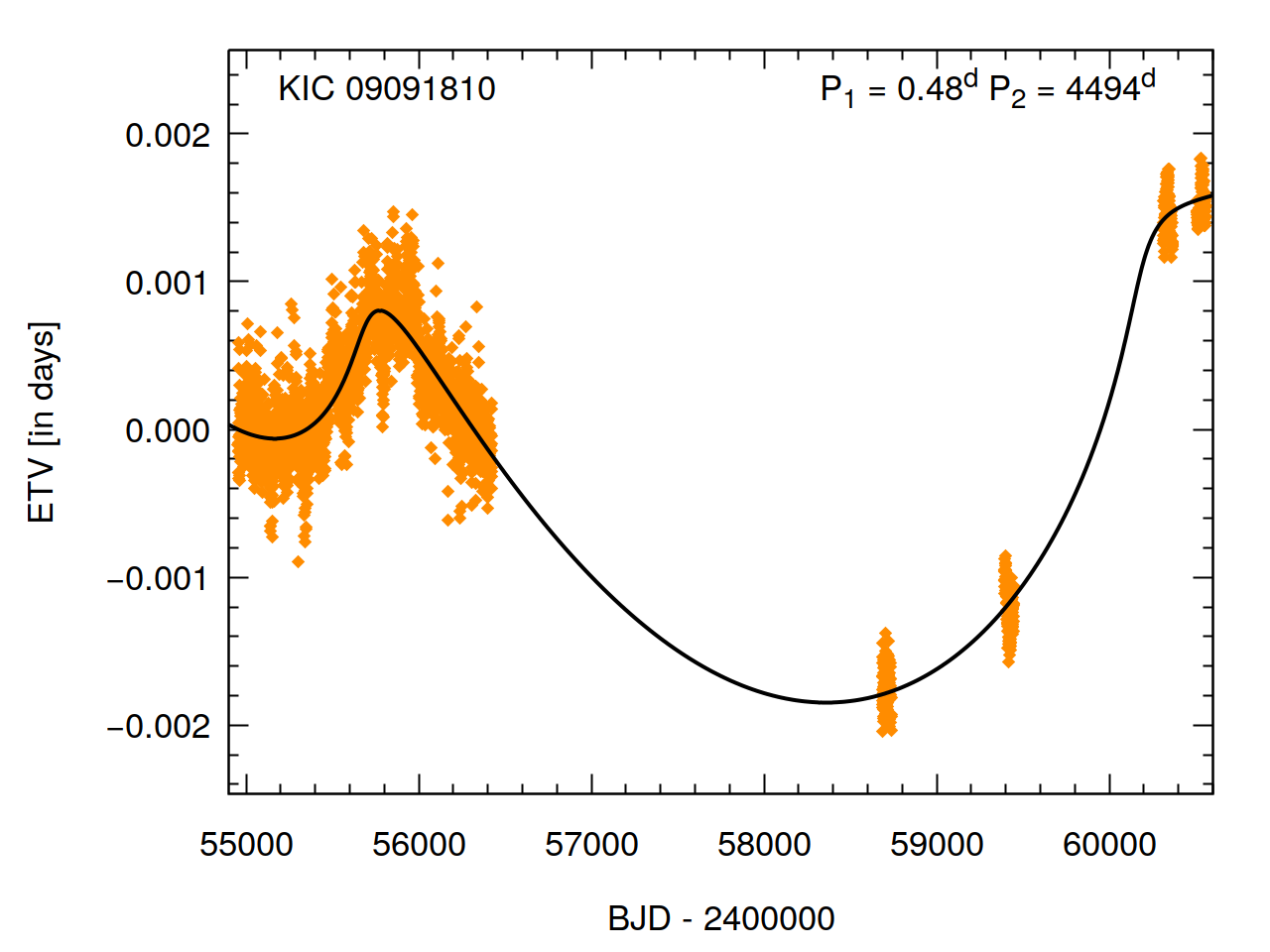}
\includegraphics[width=60mm]{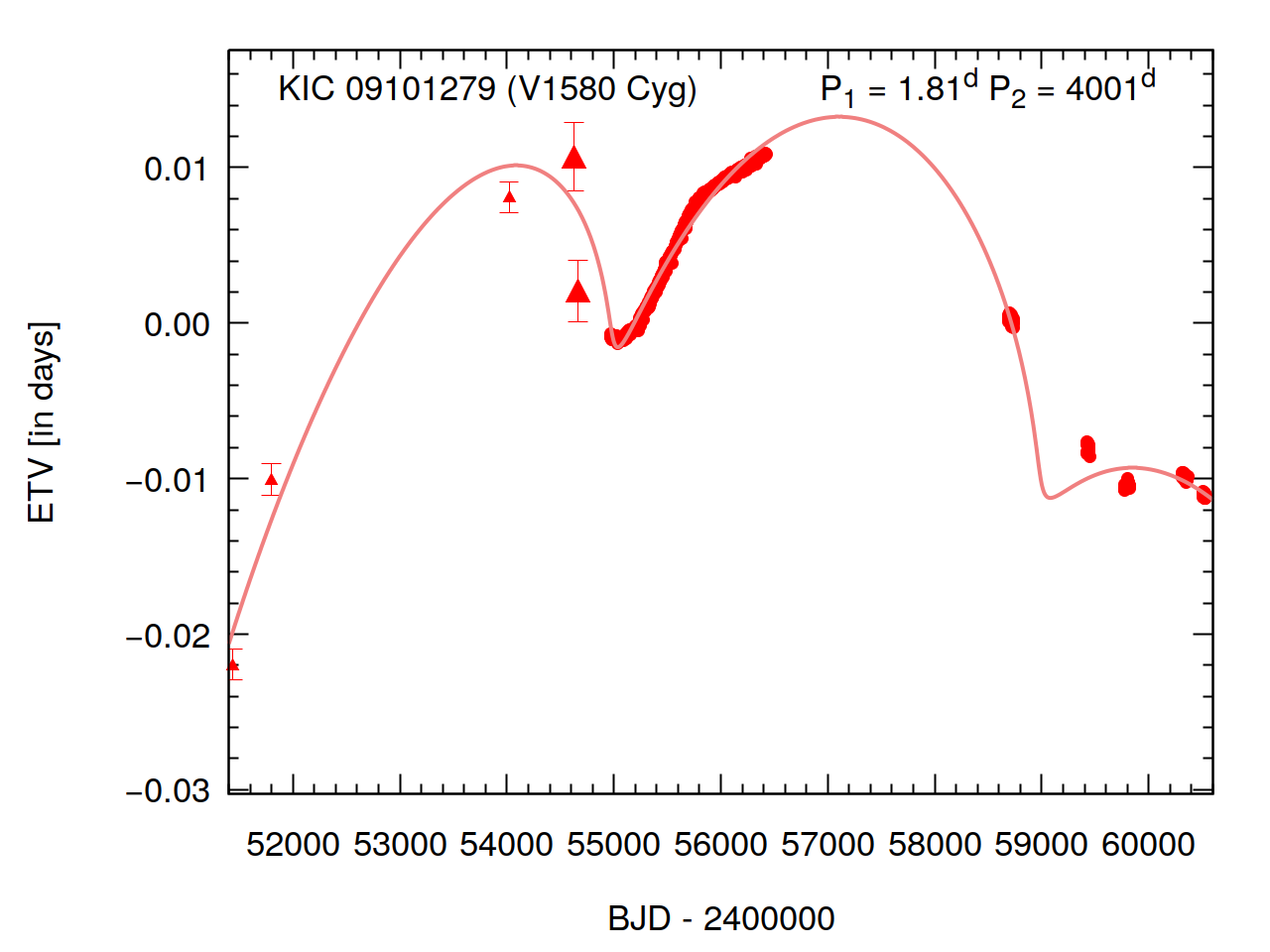}\includegraphics[width=60mm]{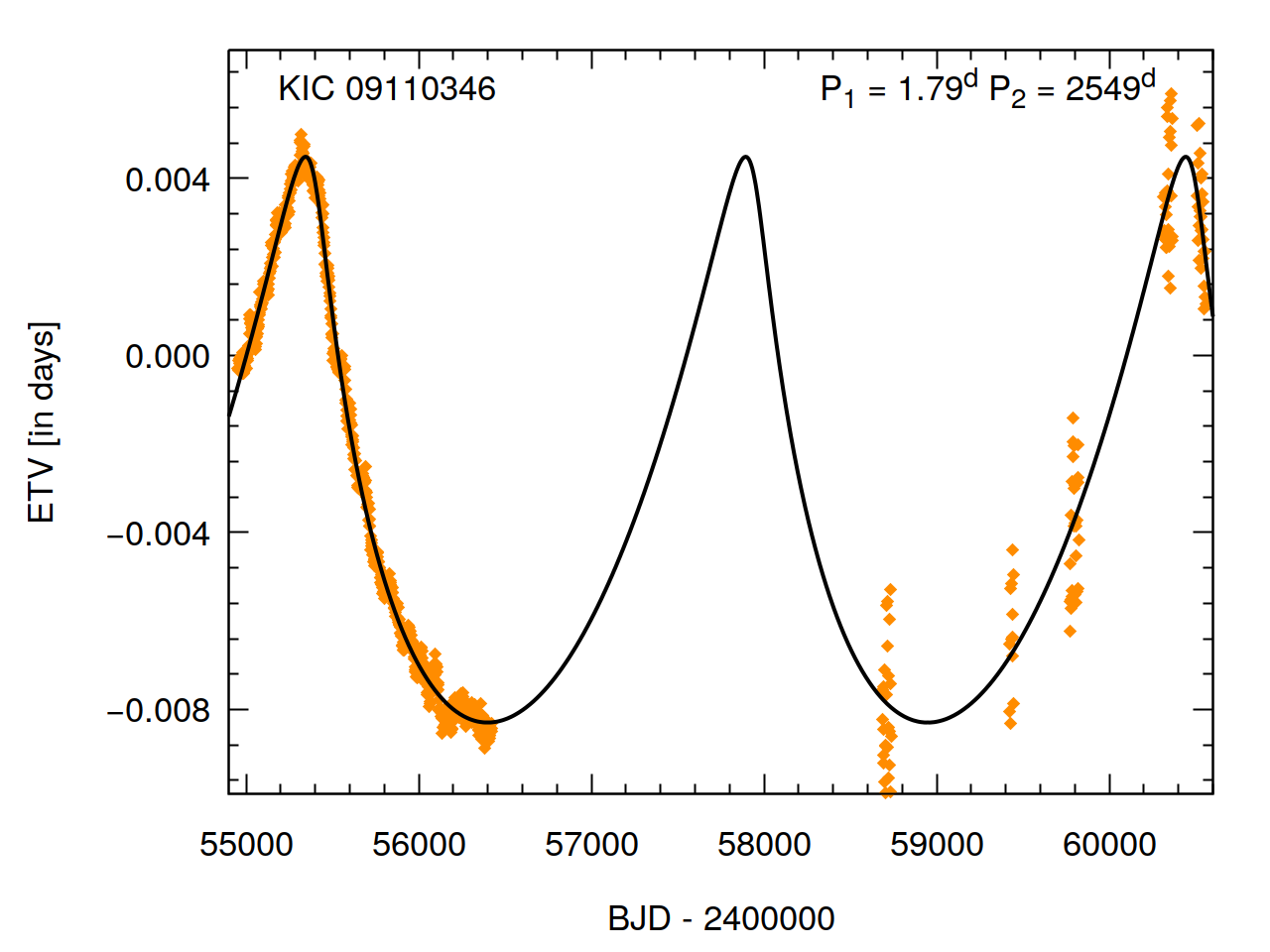}\includegraphics[width=60mm]{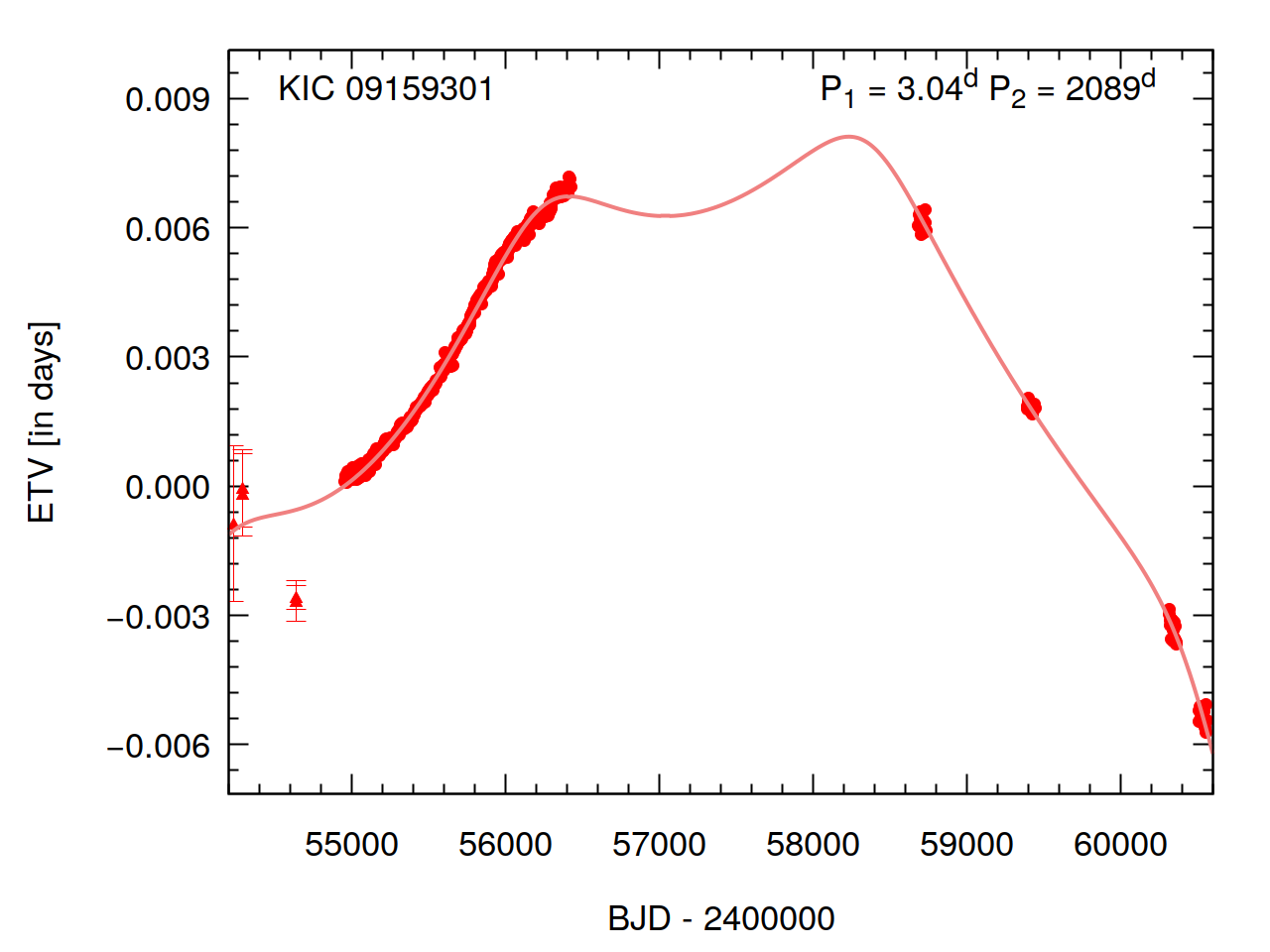}
\includegraphics[width=60mm]{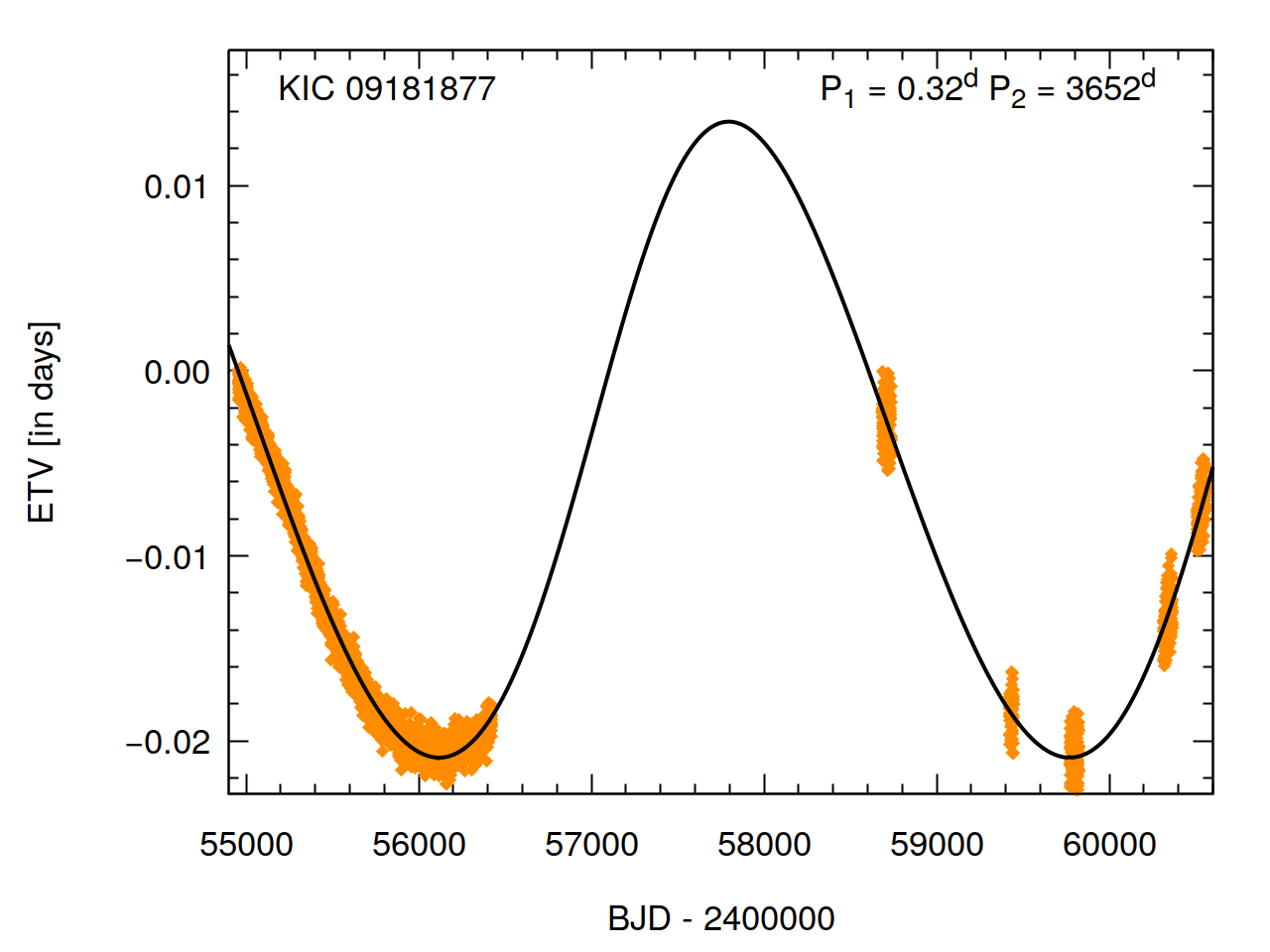}\includegraphics[width=60mm]{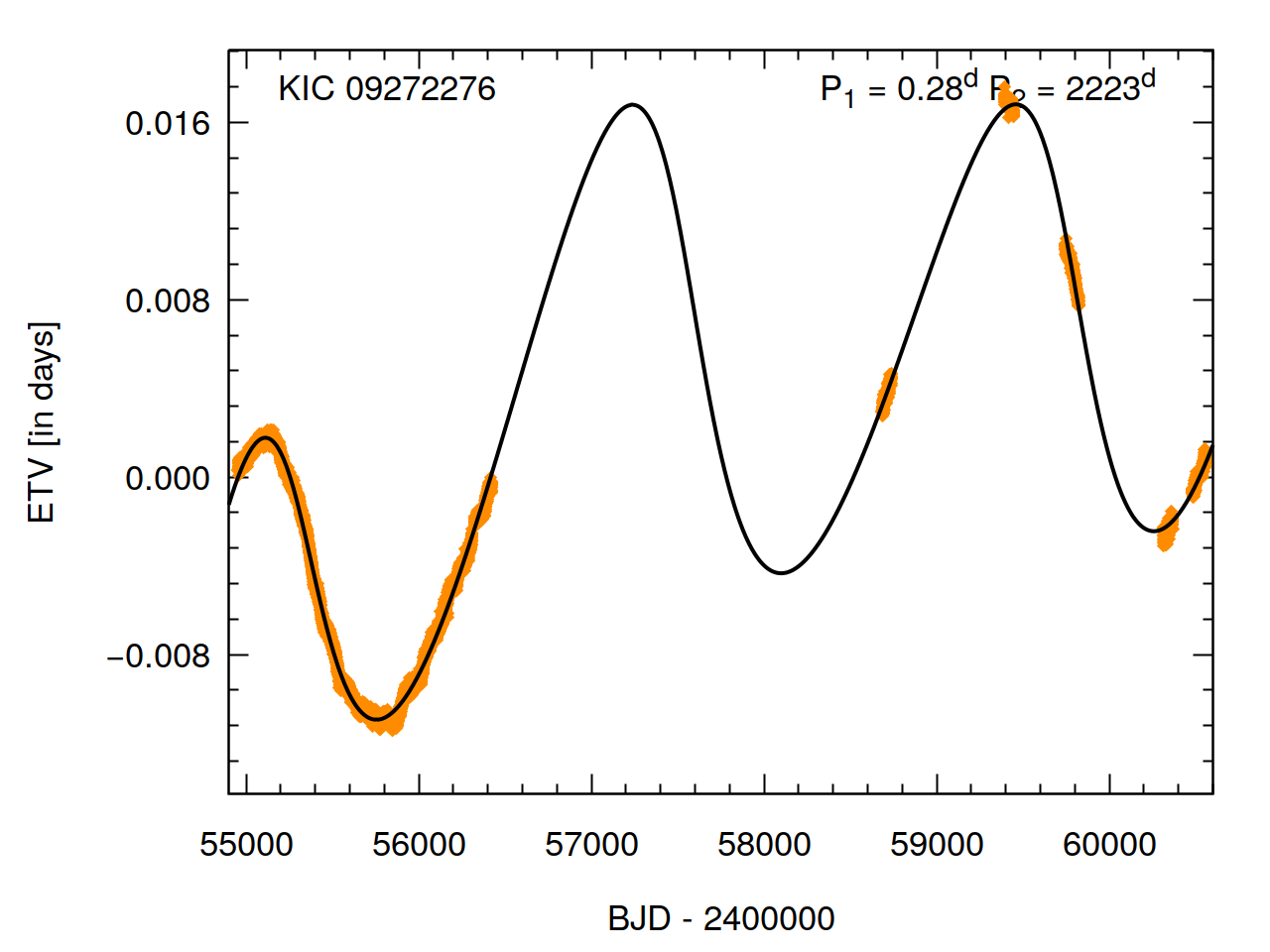}\includegraphics[width=60mm]{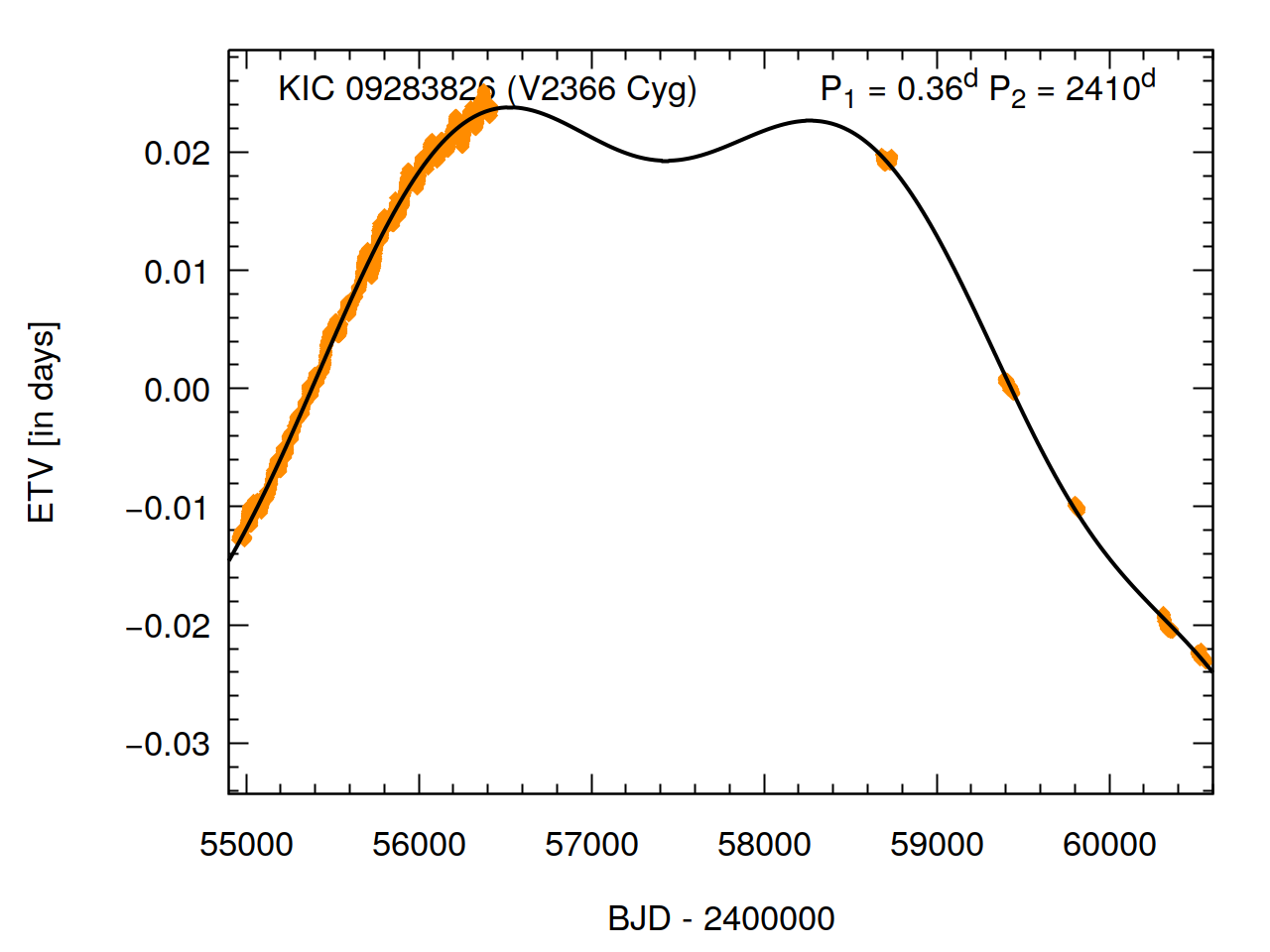}
\includegraphics[width=60mm]{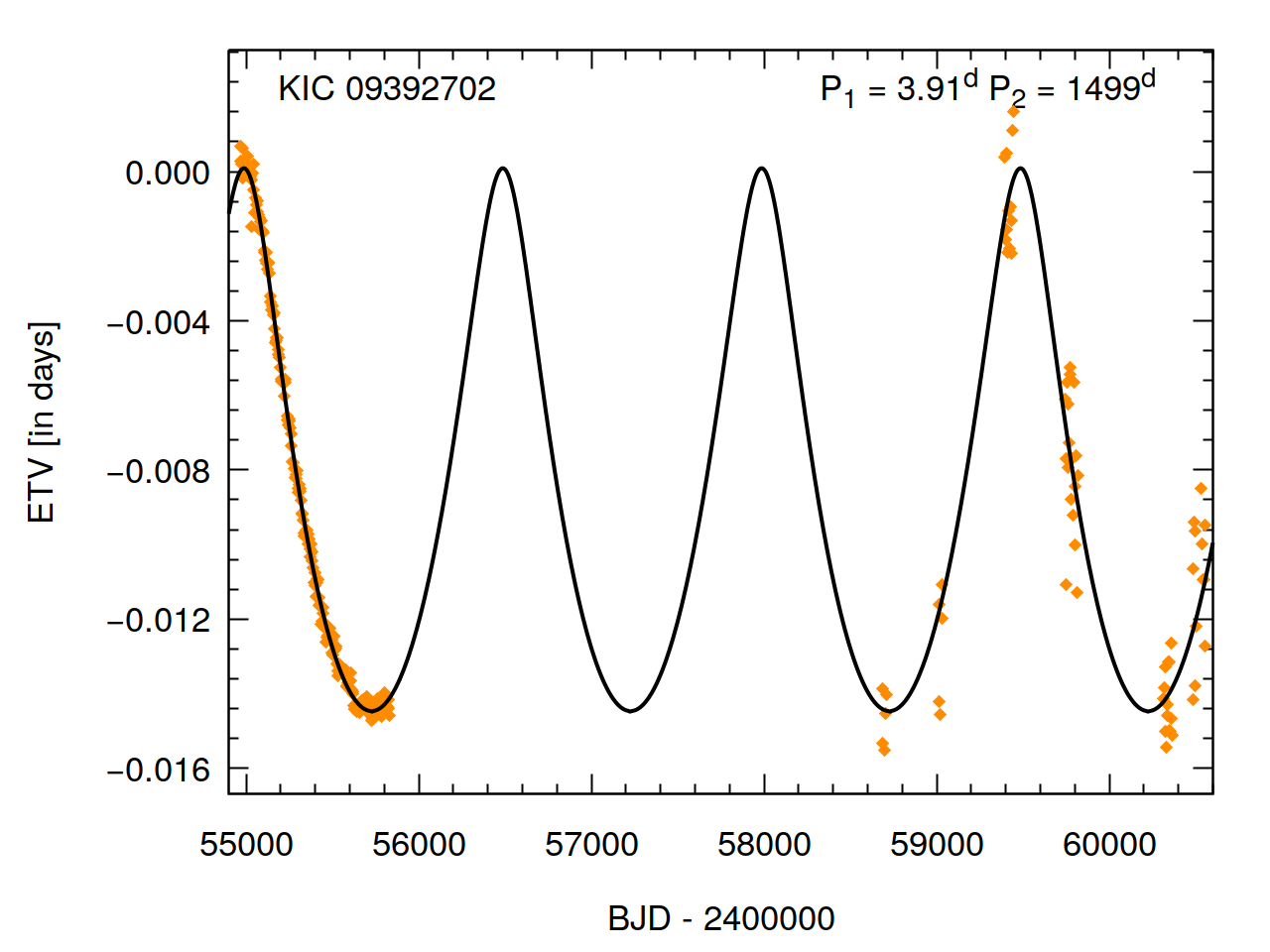}\includegraphics[width=60mm]{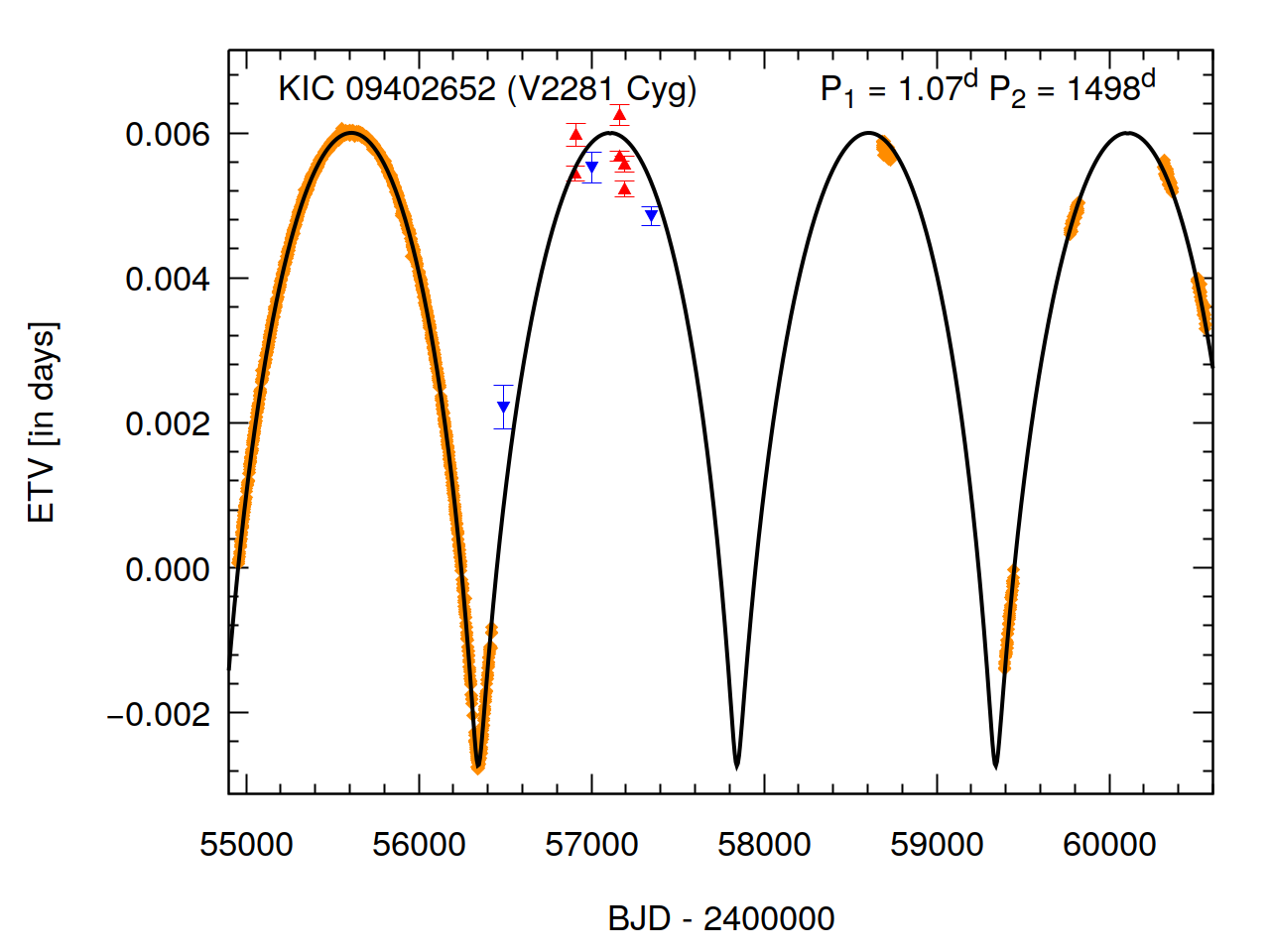}\includegraphics[width=60mm]{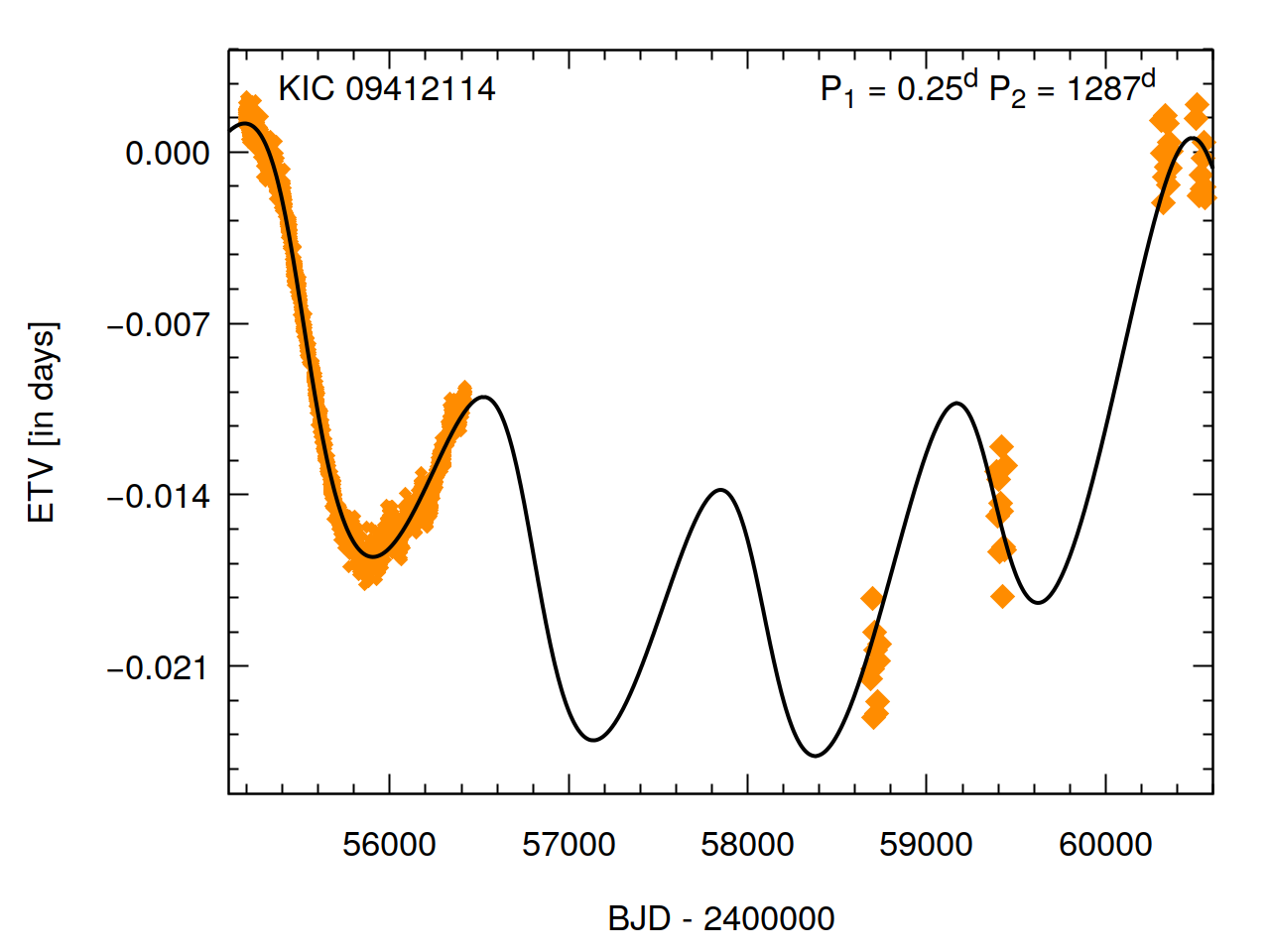}
\caption{continued.}
\end{figure*}

\addtocounter{figure}{-1}

\begin{figure*}
\includegraphics[width=60mm]{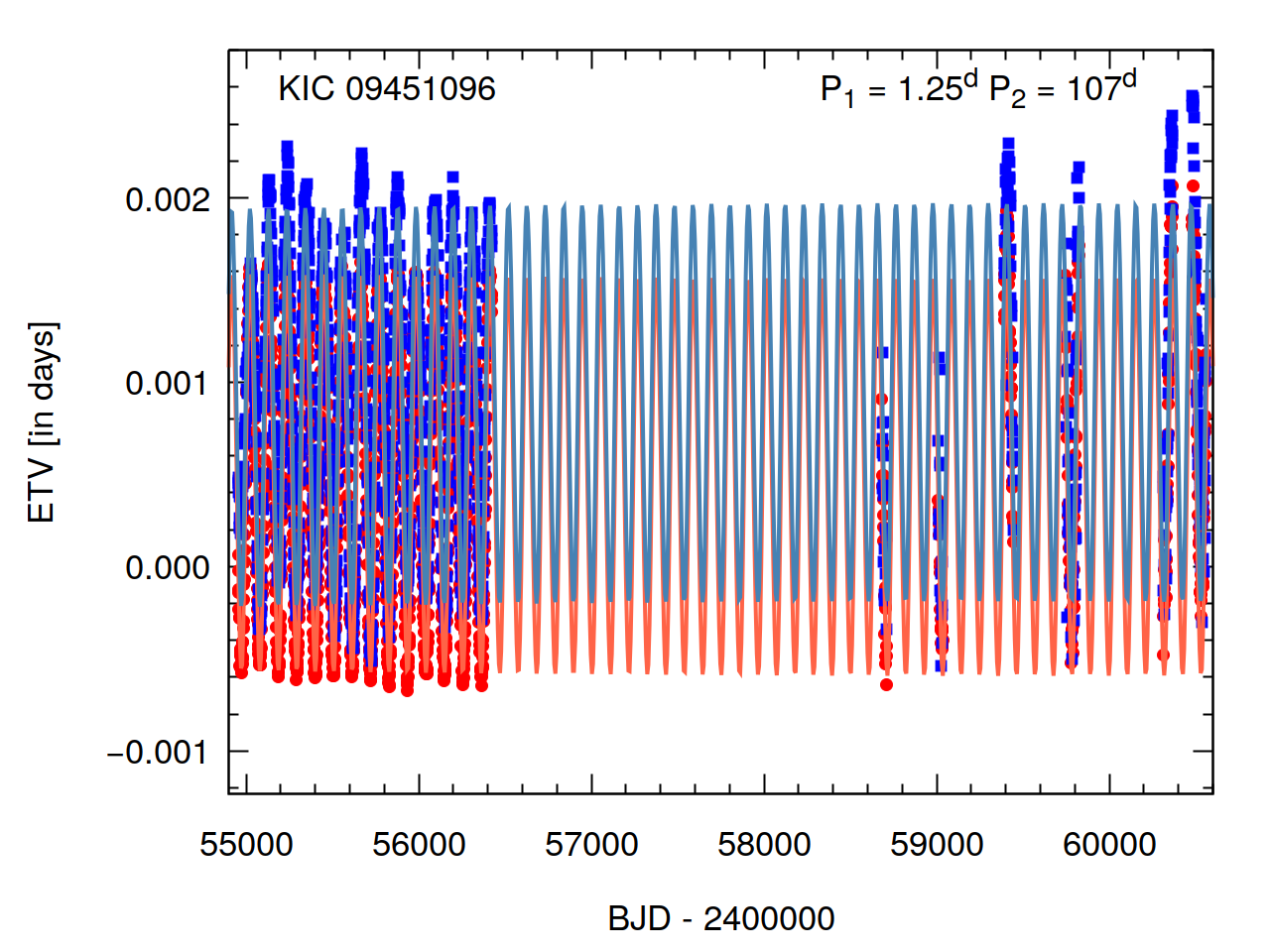}\includegraphics[width=60mm]{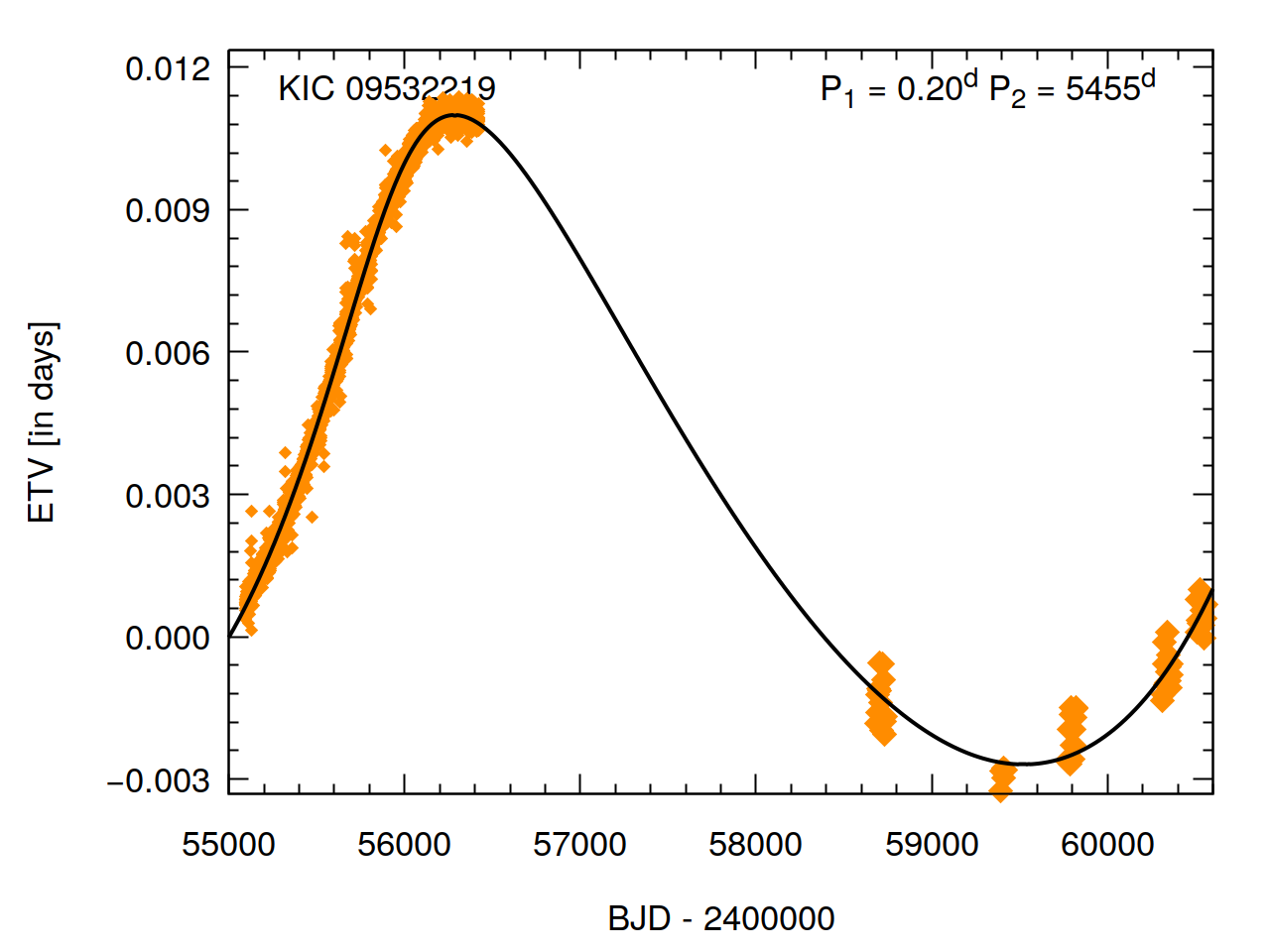}\includegraphics[width=60mm]{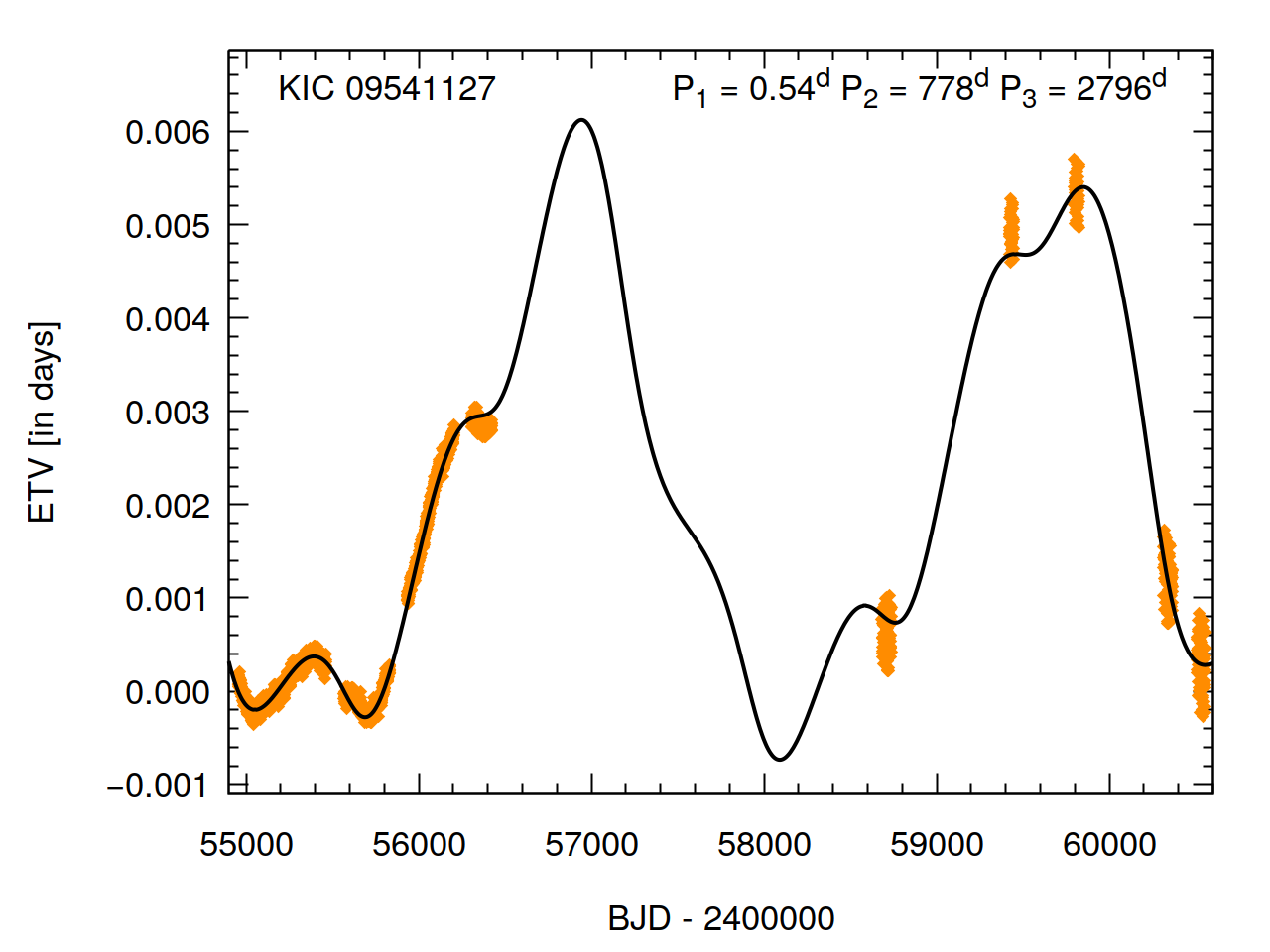}
\includegraphics[width=60mm]{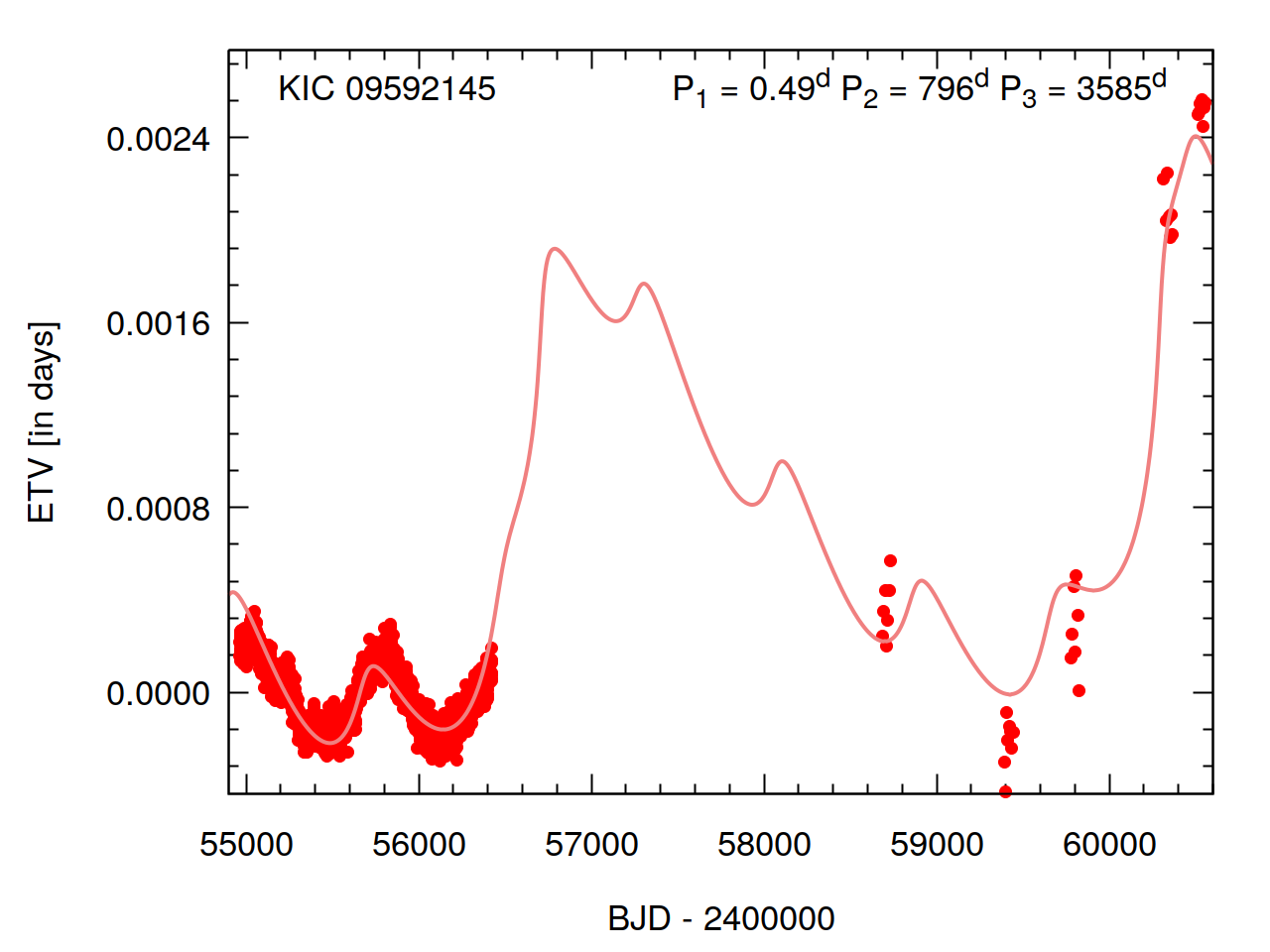}\includegraphics[width=60mm]{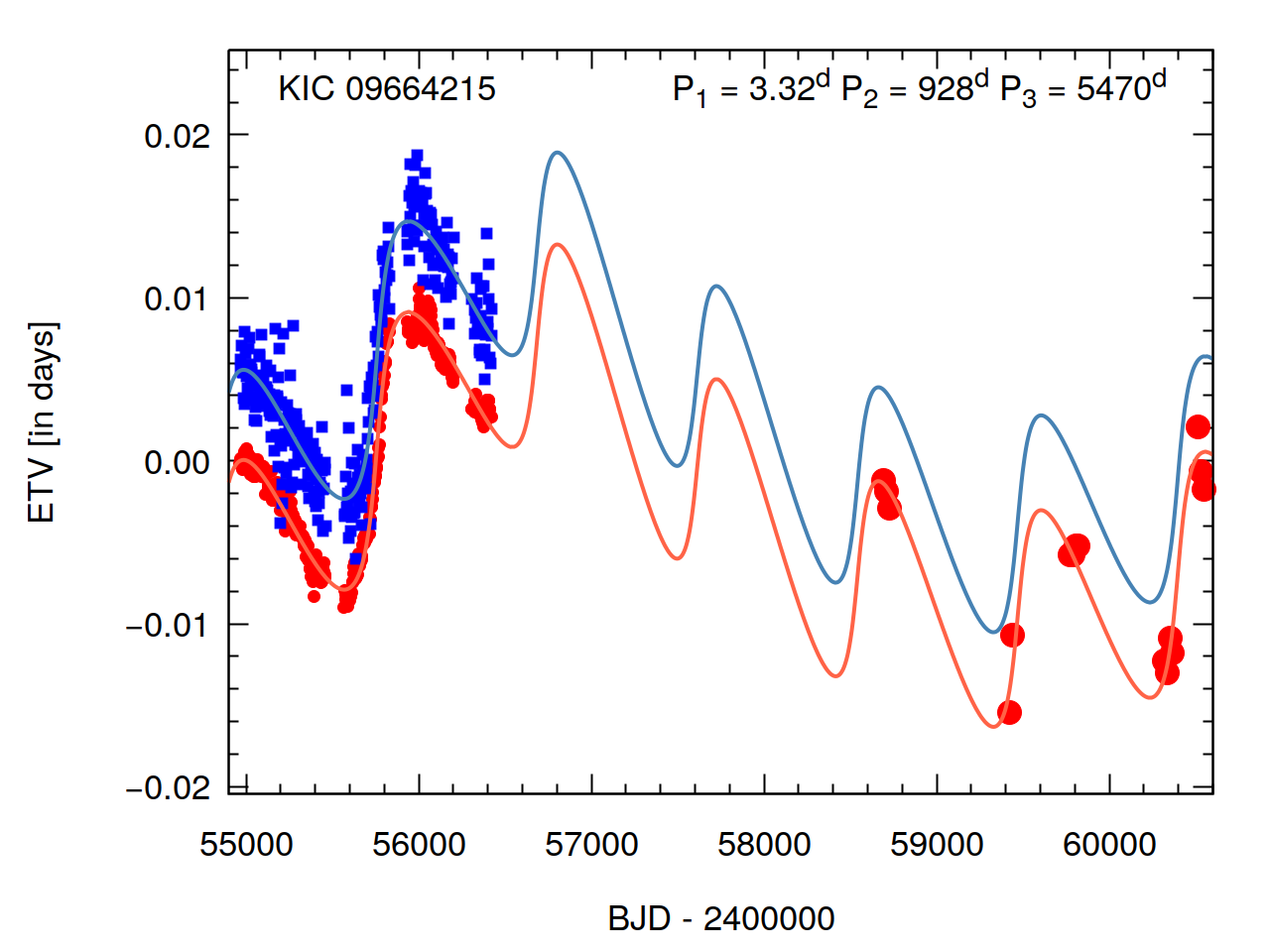}\includegraphics[width=60mm]{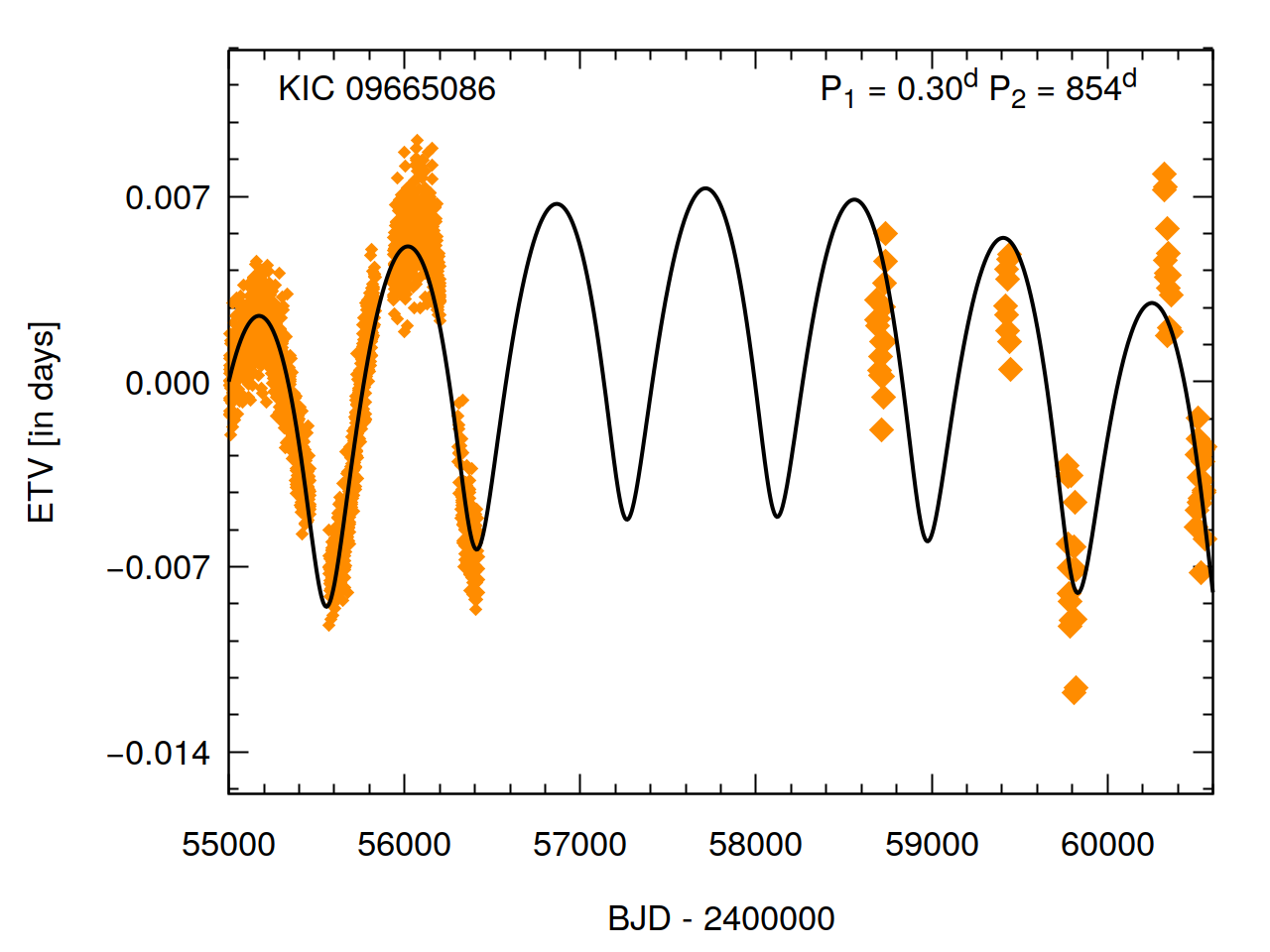}
\includegraphics[width=60mm]{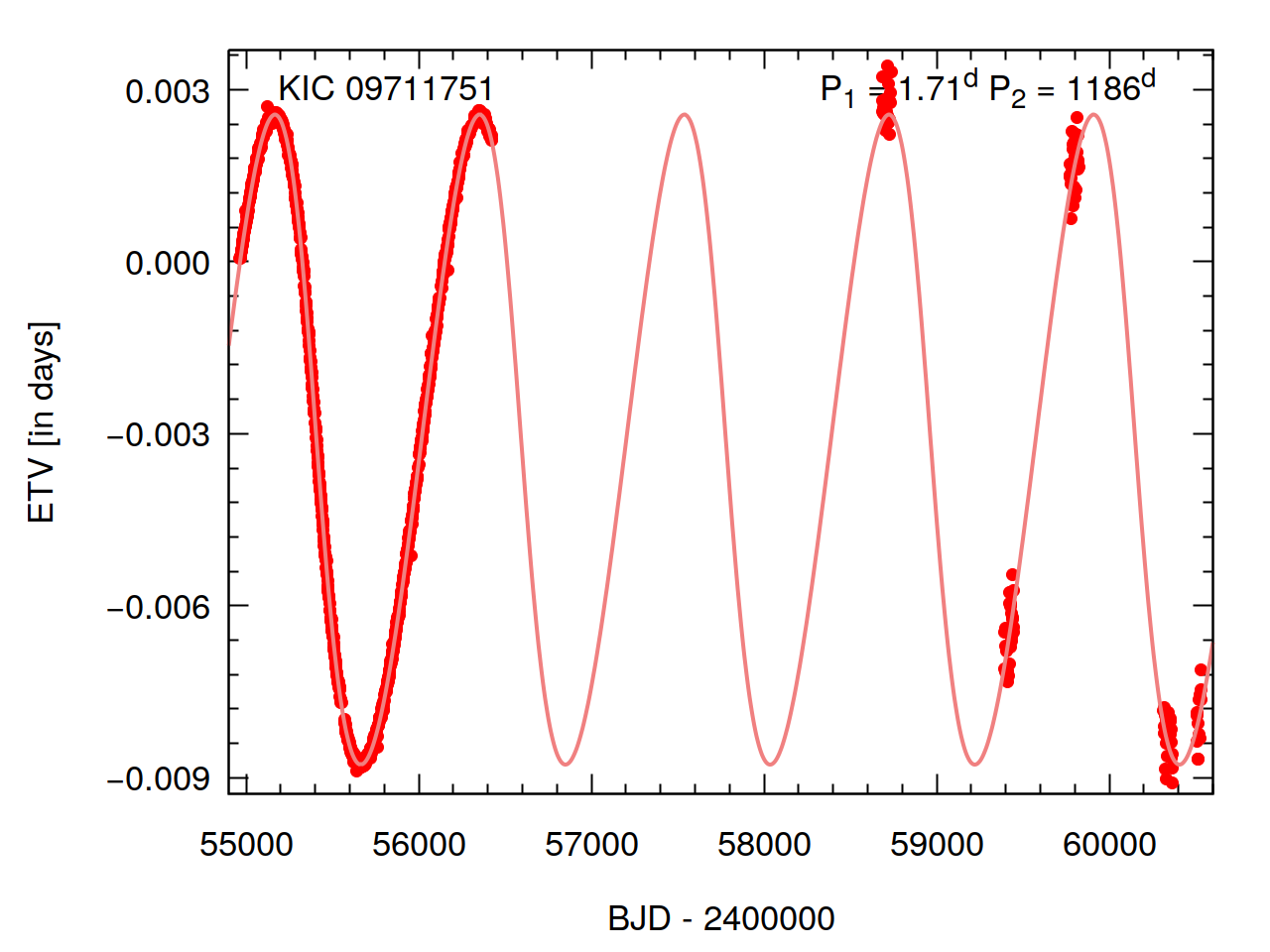}\includegraphics[width=60mm]{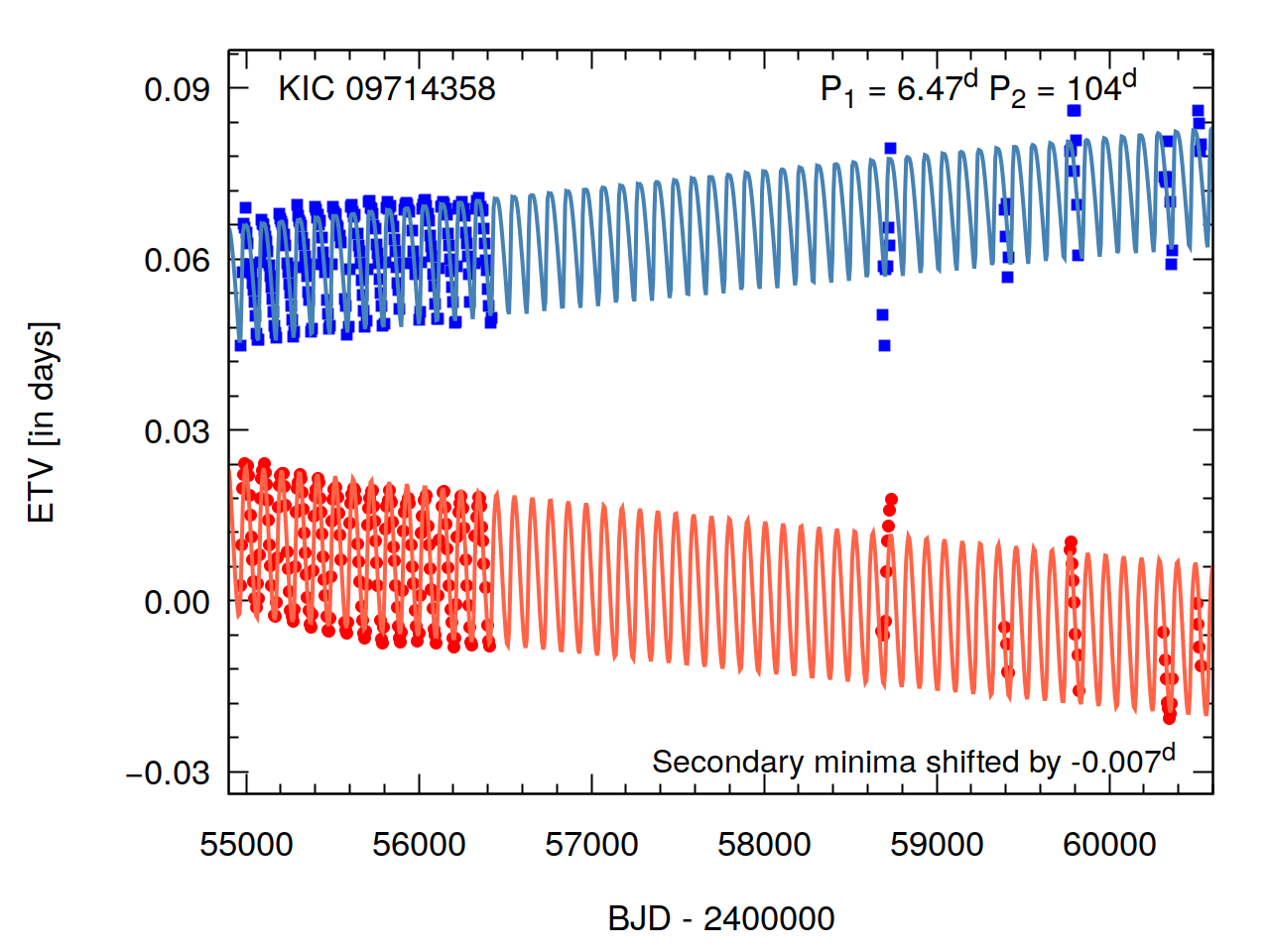}\includegraphics[width=60mm]{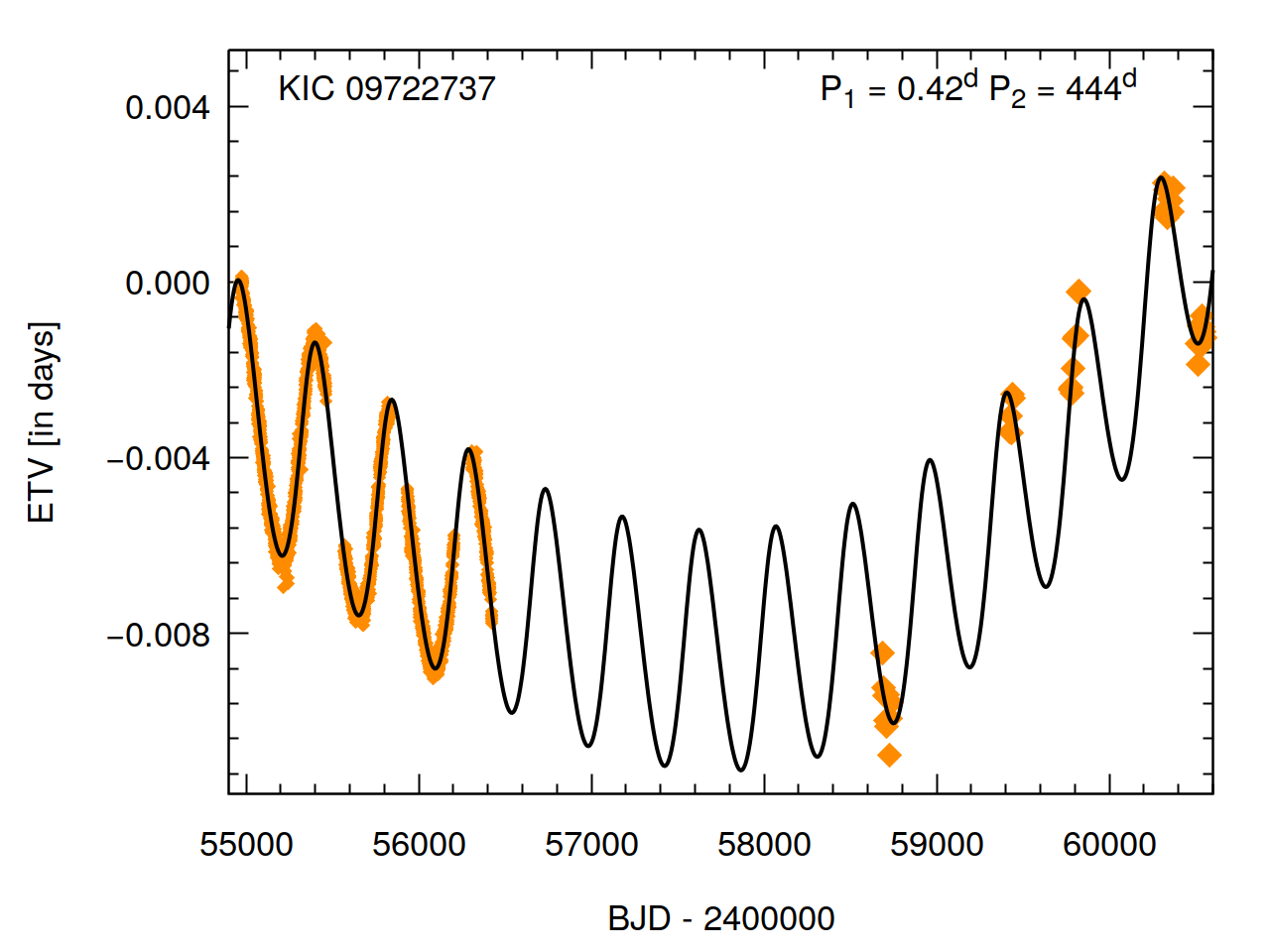}
\includegraphics[width=60mm]{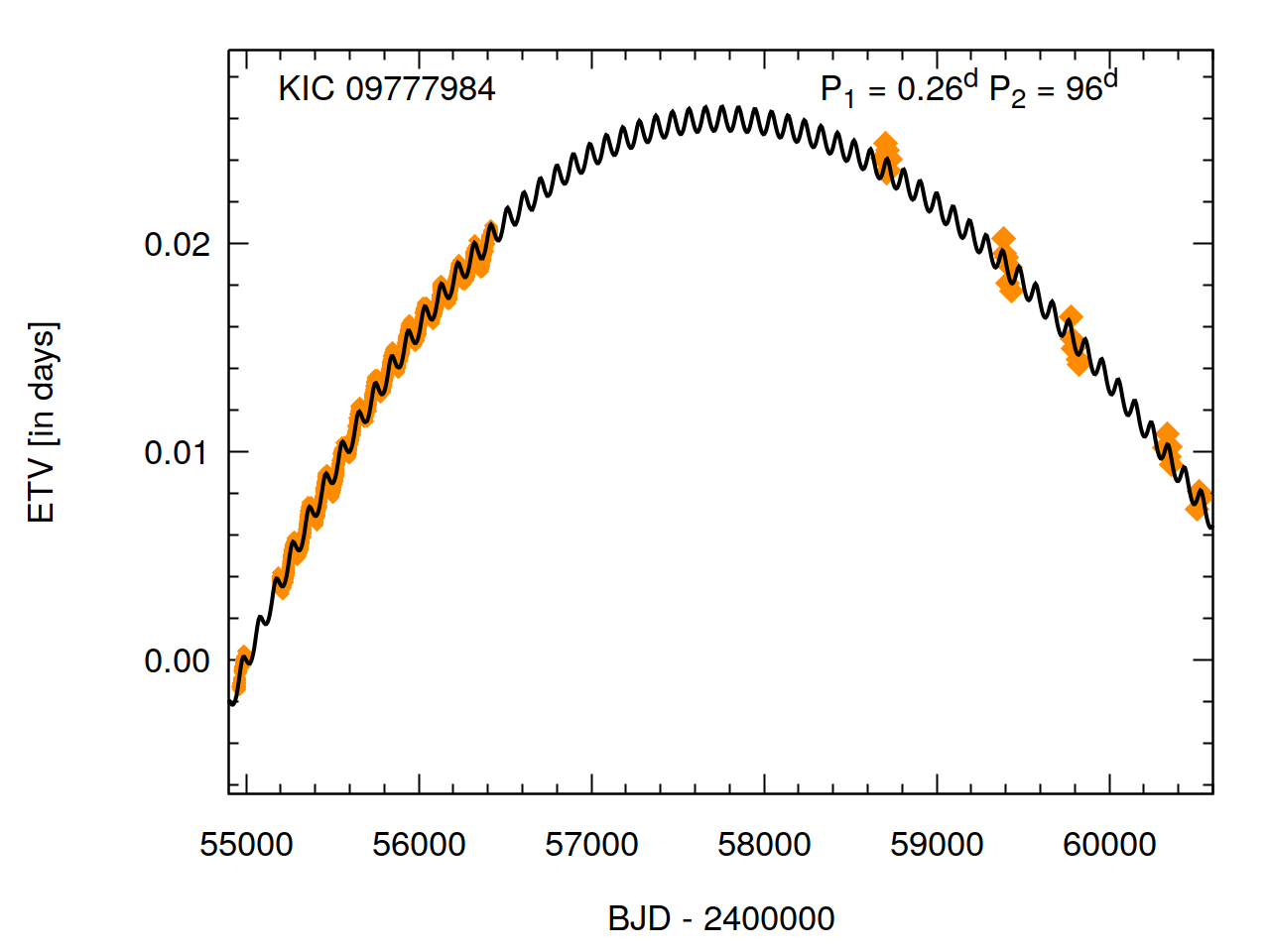}\includegraphics[width=60mm]{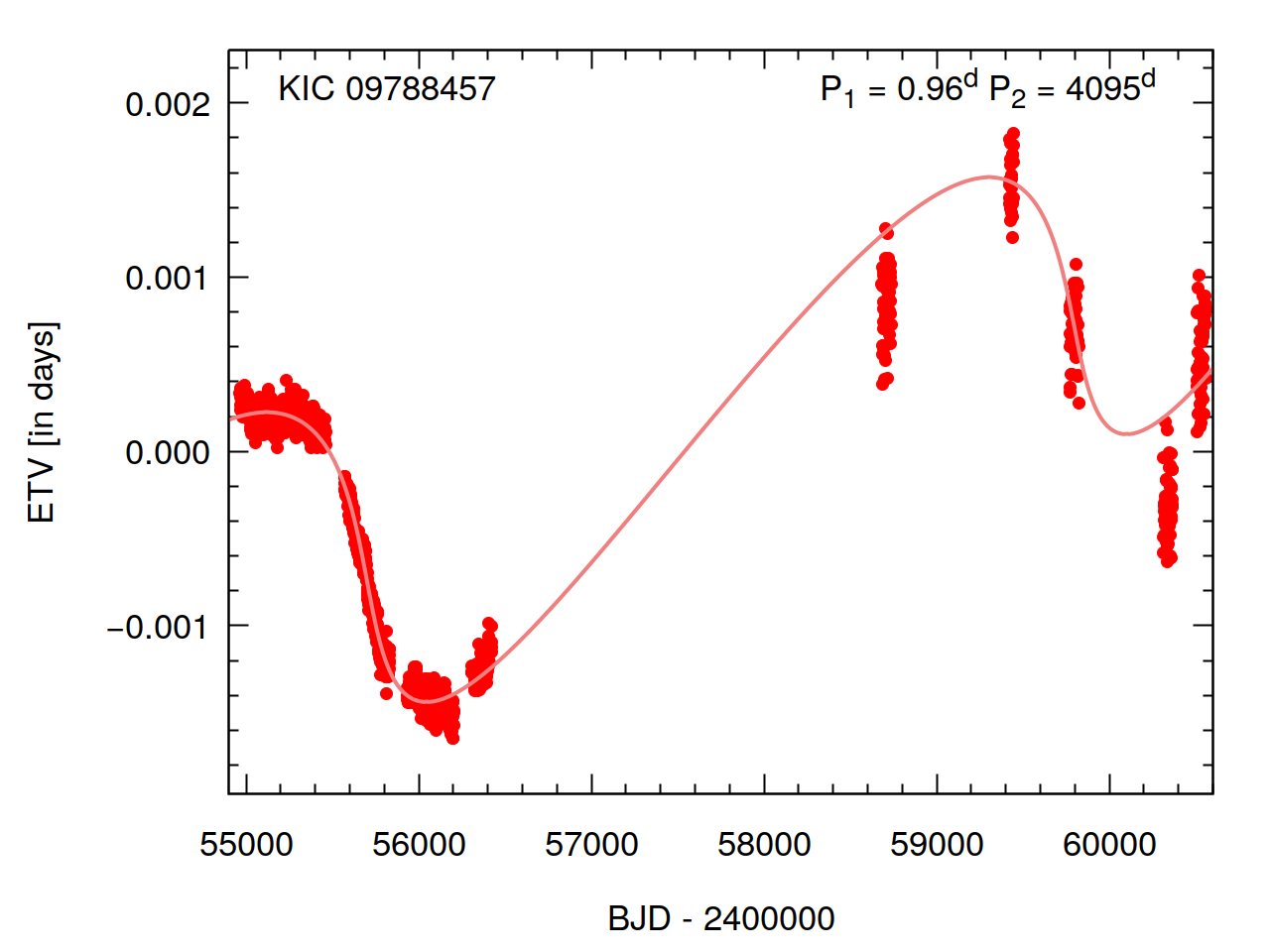}\includegraphics[width=60mm]{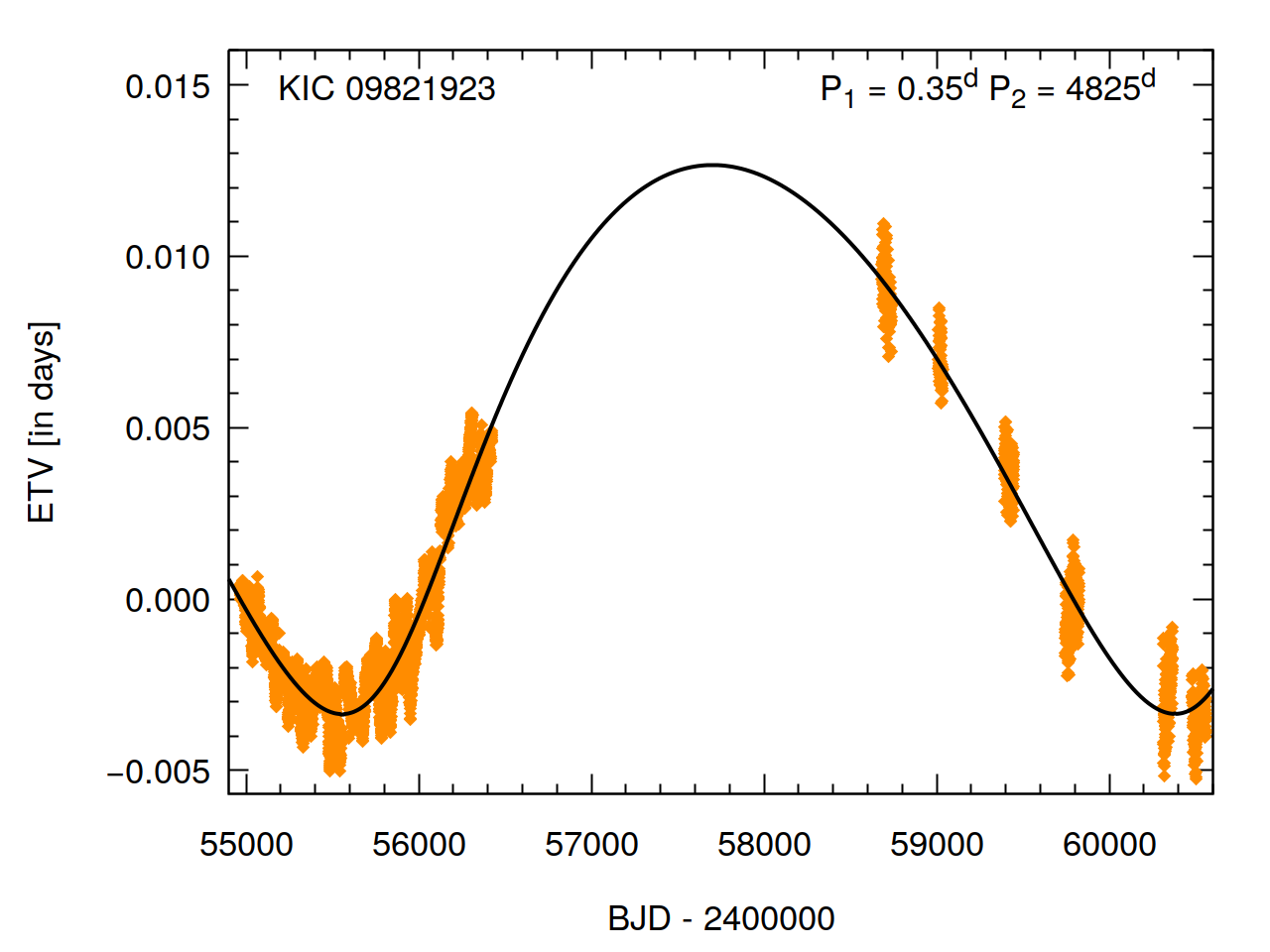}
\includegraphics[width=60mm]{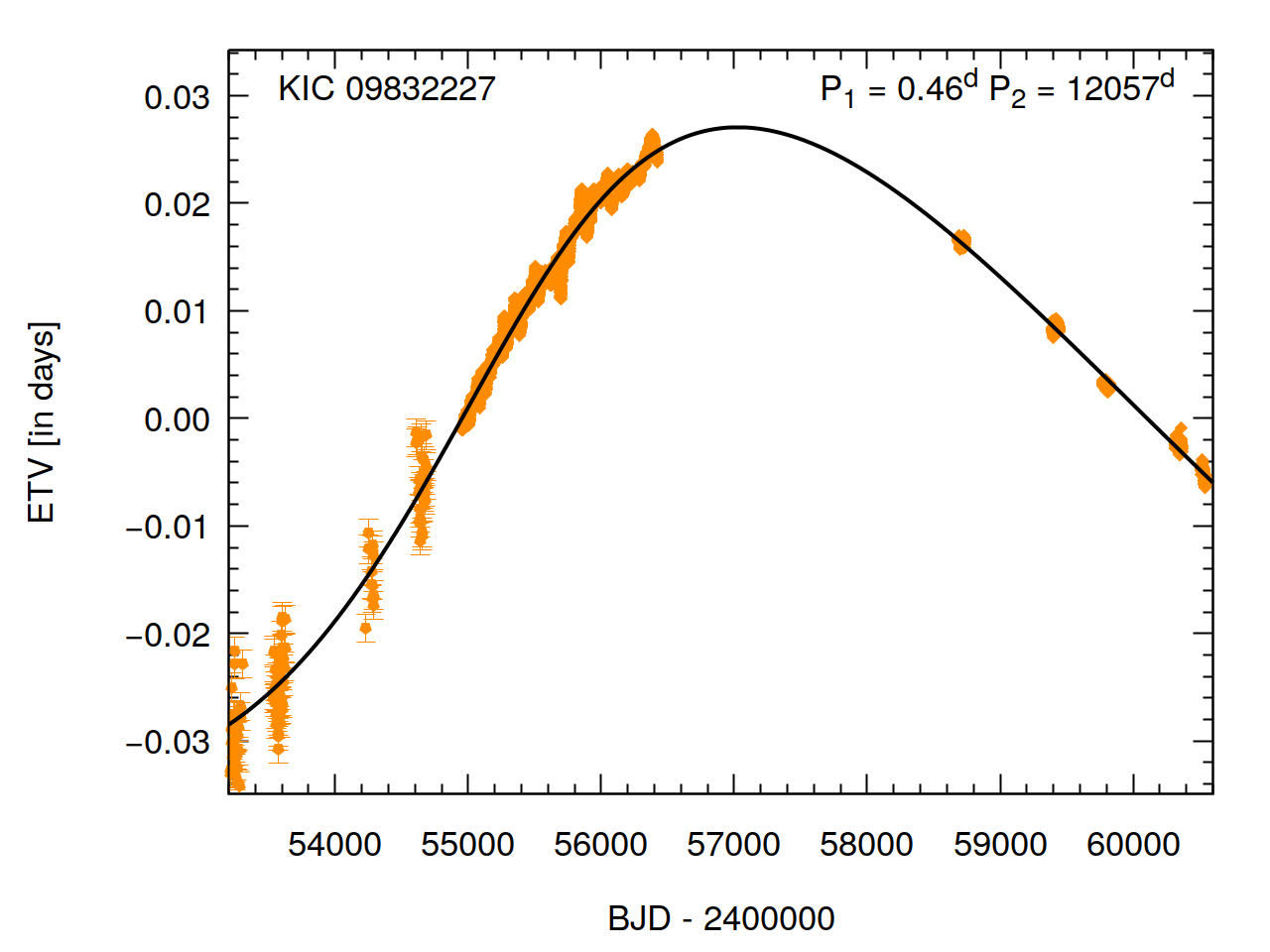}\includegraphics[width=60mm]{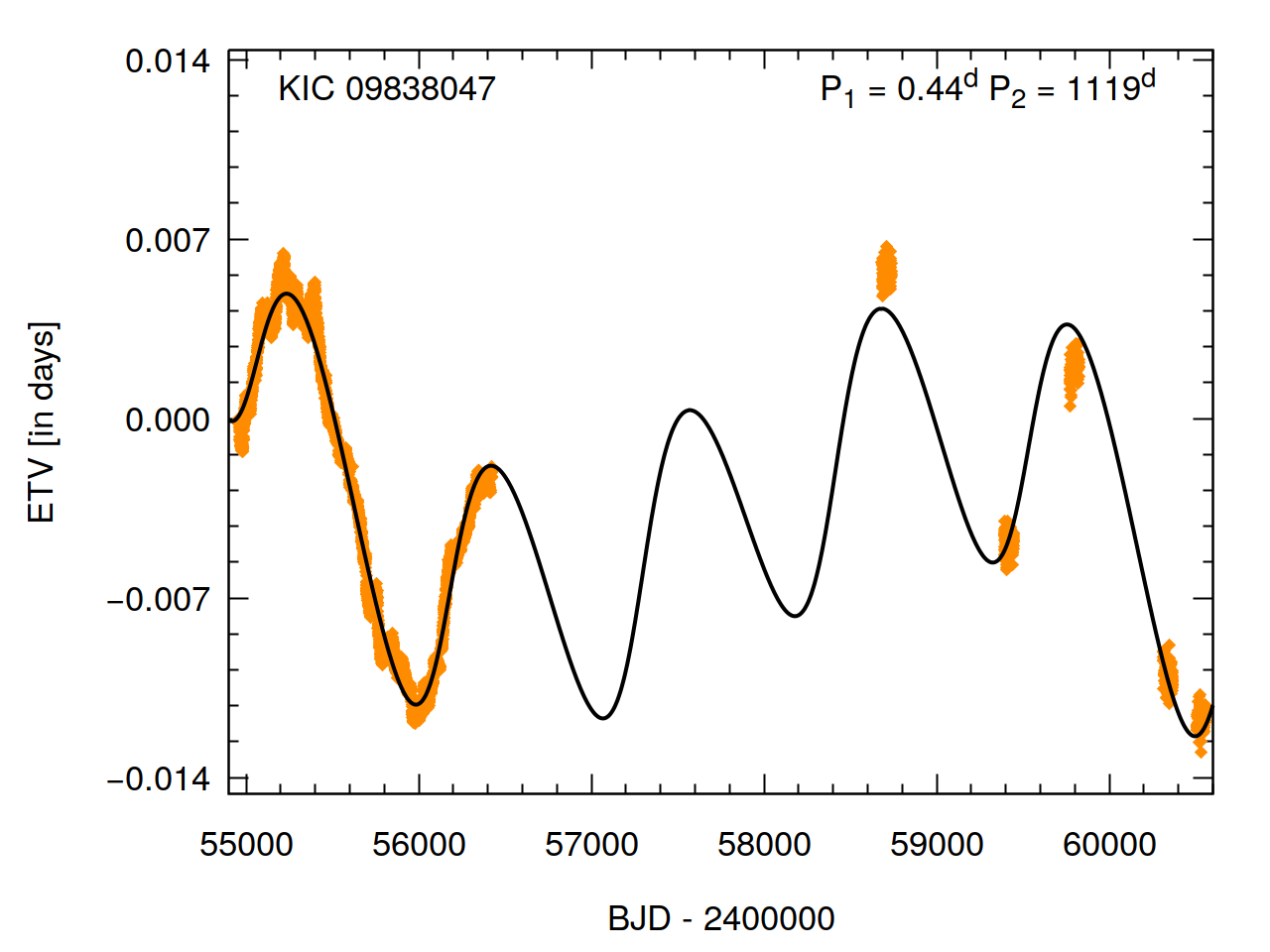}\includegraphics[width=60mm]{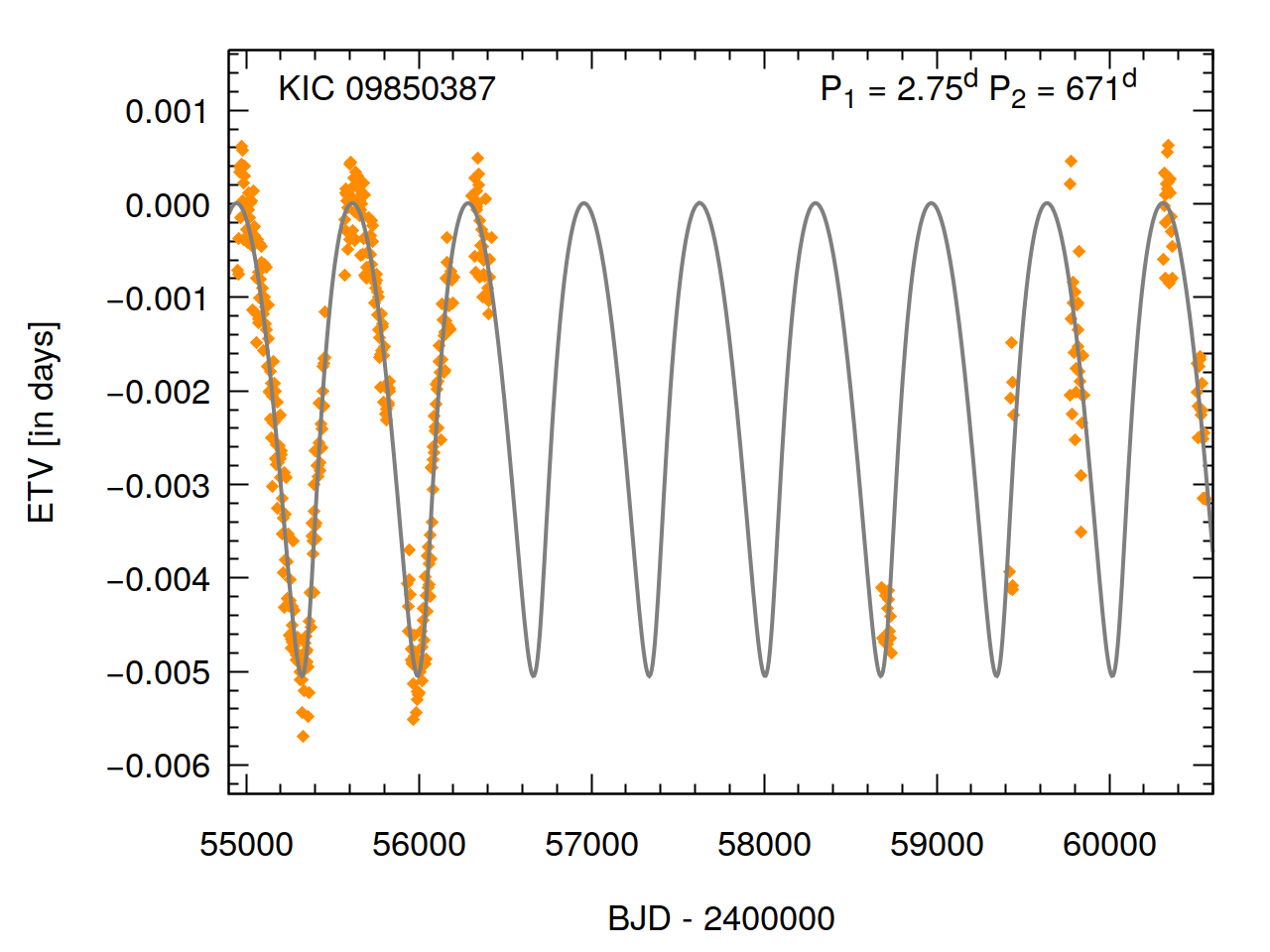}
\caption{continued.}
\end{figure*}

\addtocounter{figure}{-1}

\begin{figure*}
\includegraphics[width=60mm]{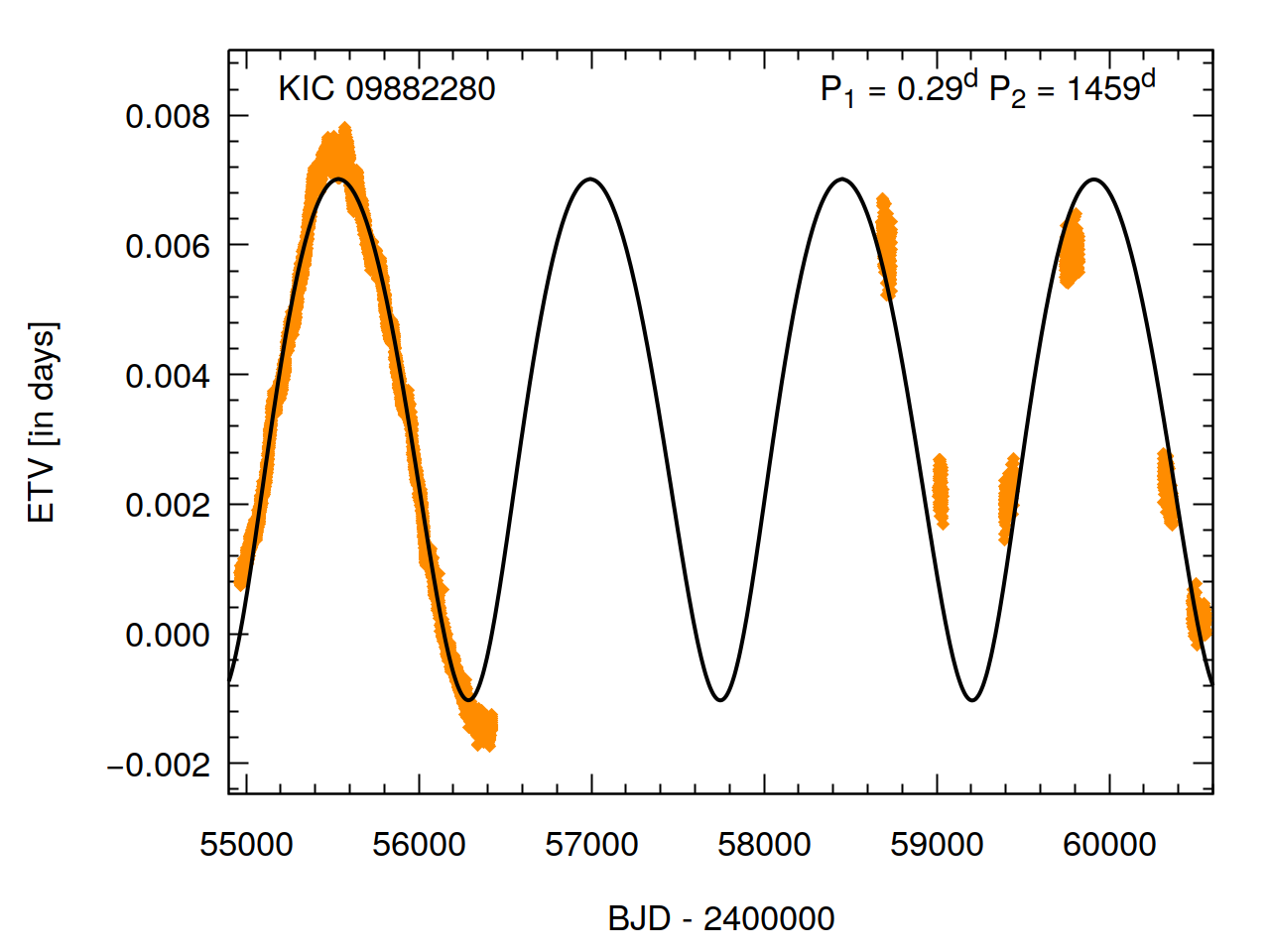}\includegraphics[width=60mm]{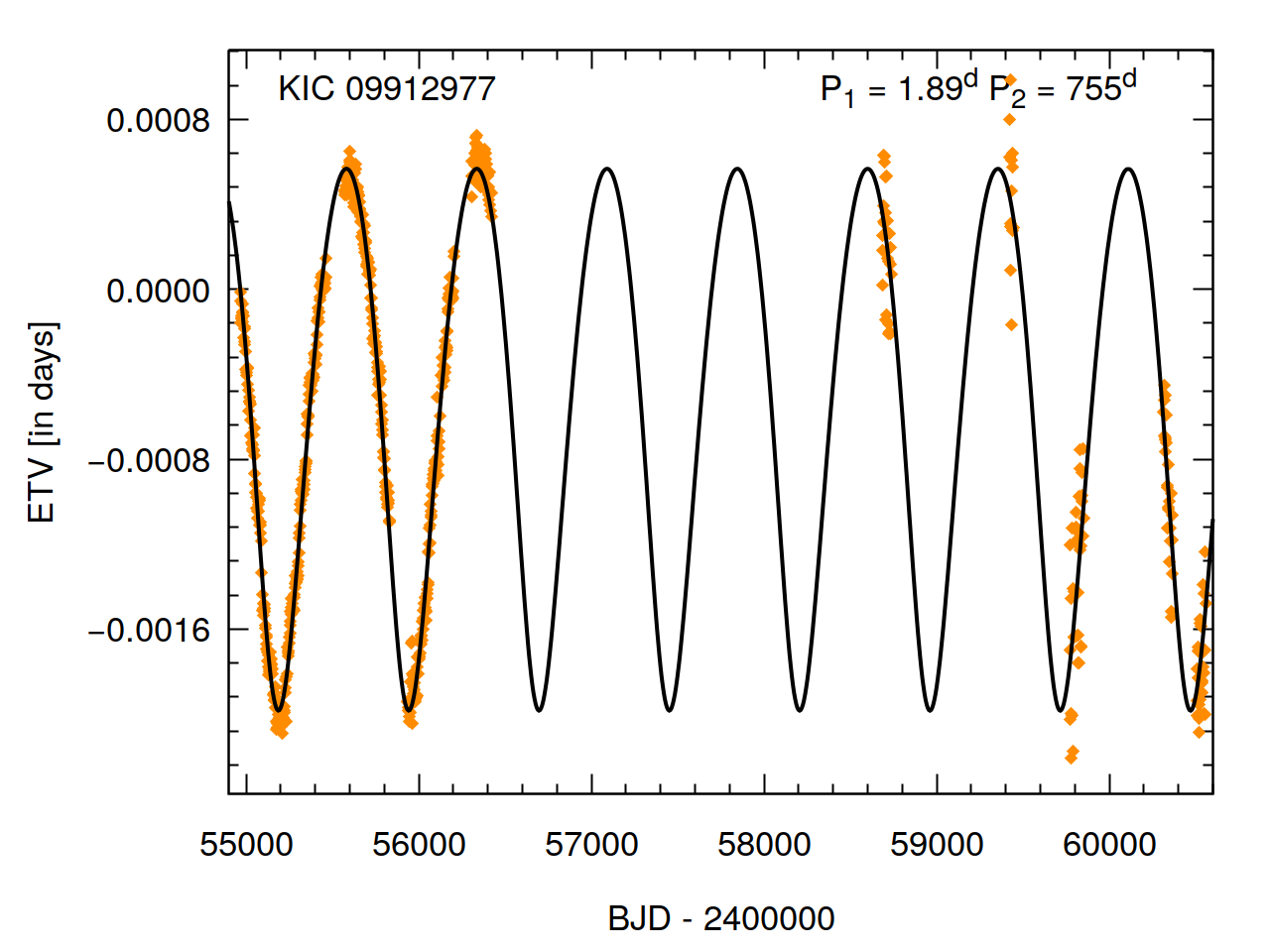}\includegraphics[width=60mm]{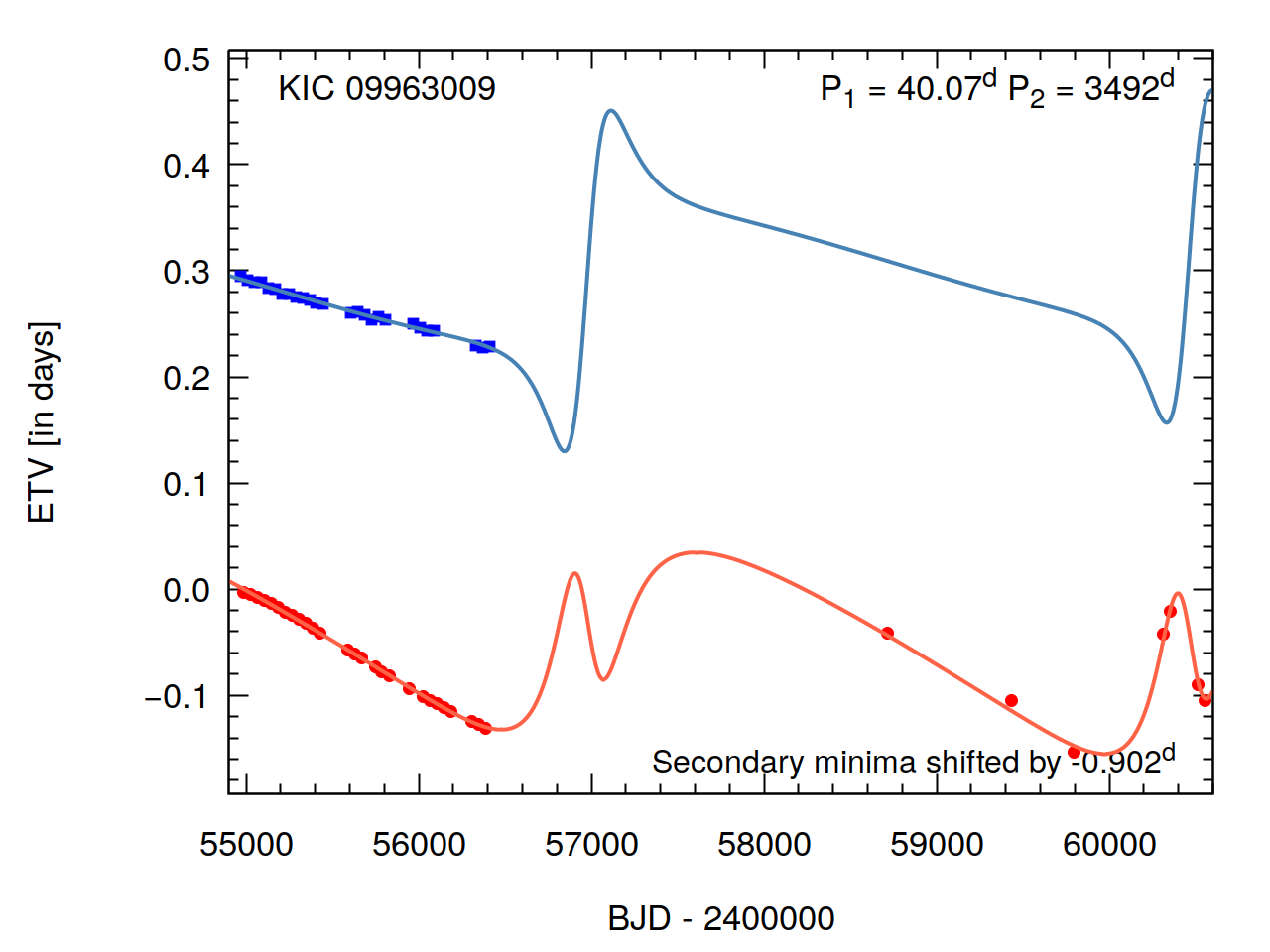}
\includegraphics[width=60mm]{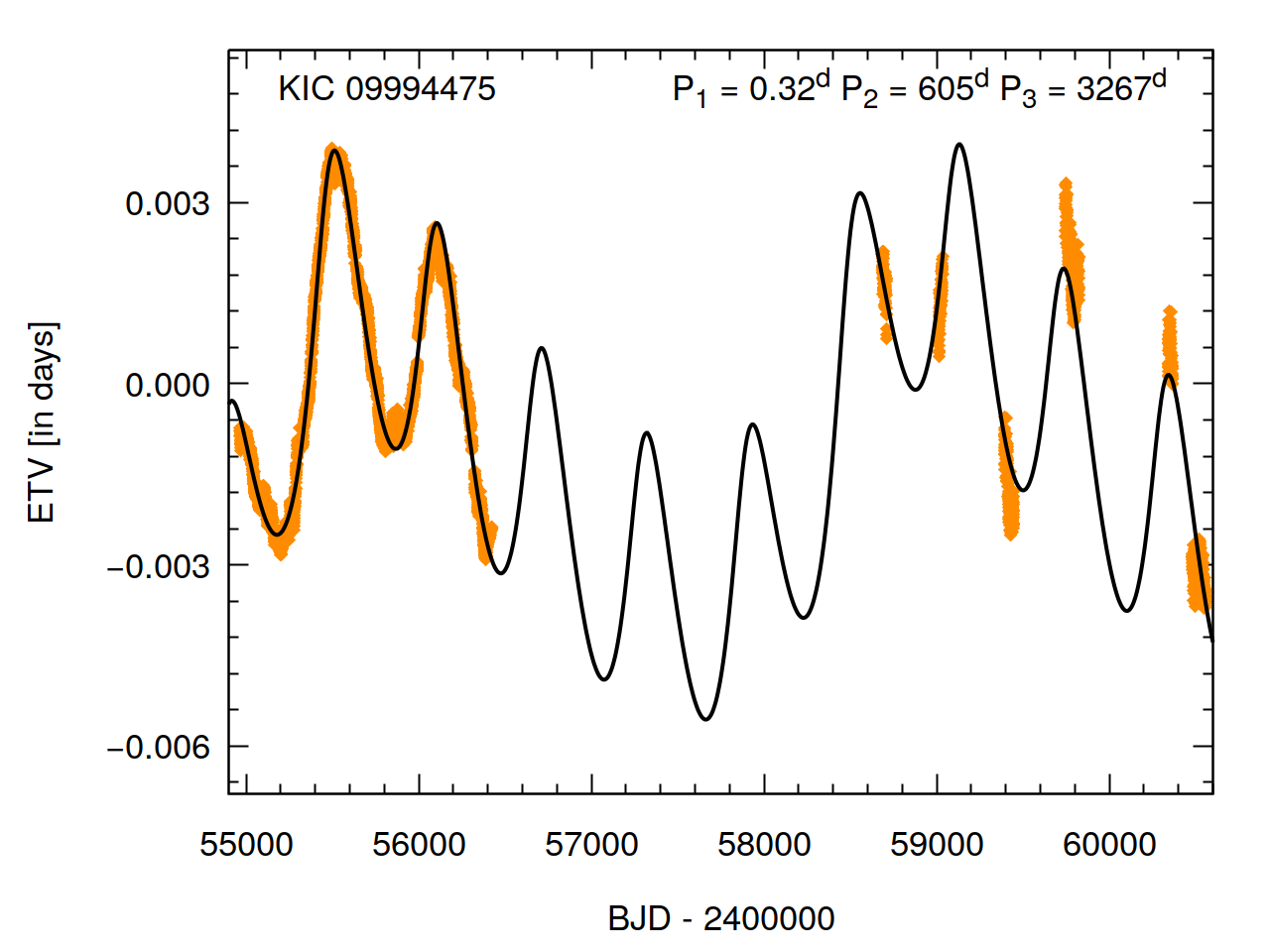}\includegraphics[width=60mm]{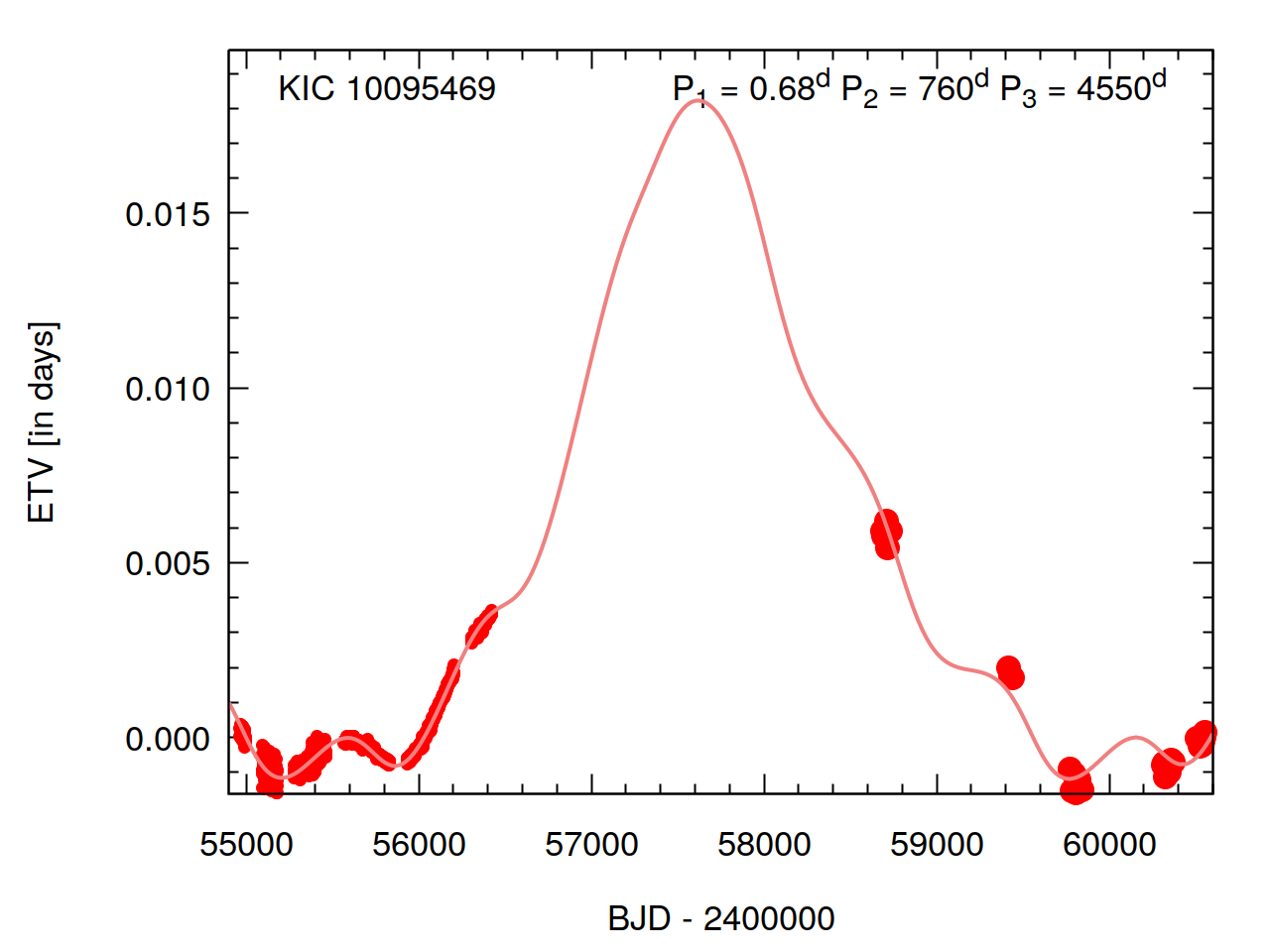}\includegraphics[width=60mm]{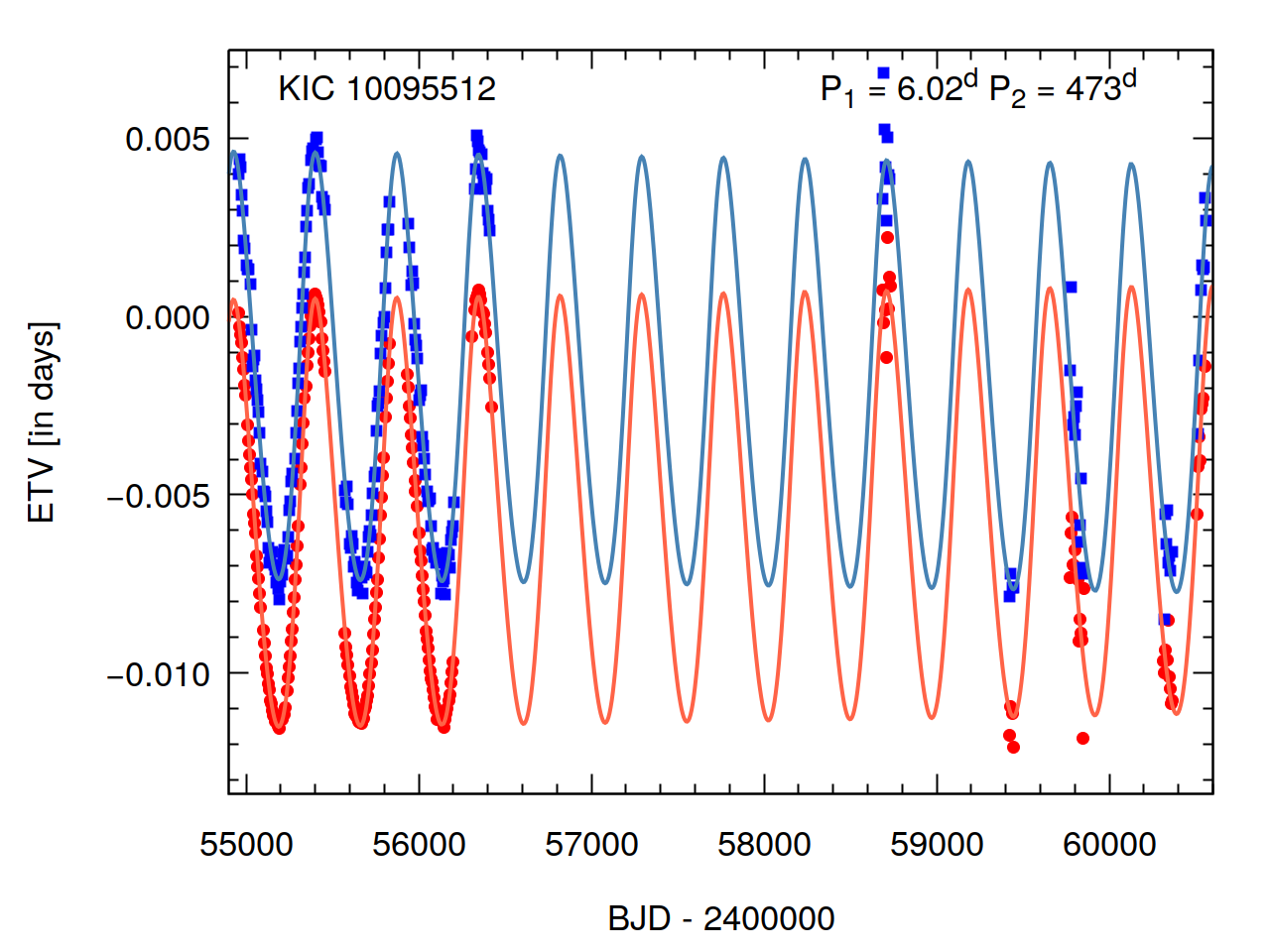}
\includegraphics[width=60mm]{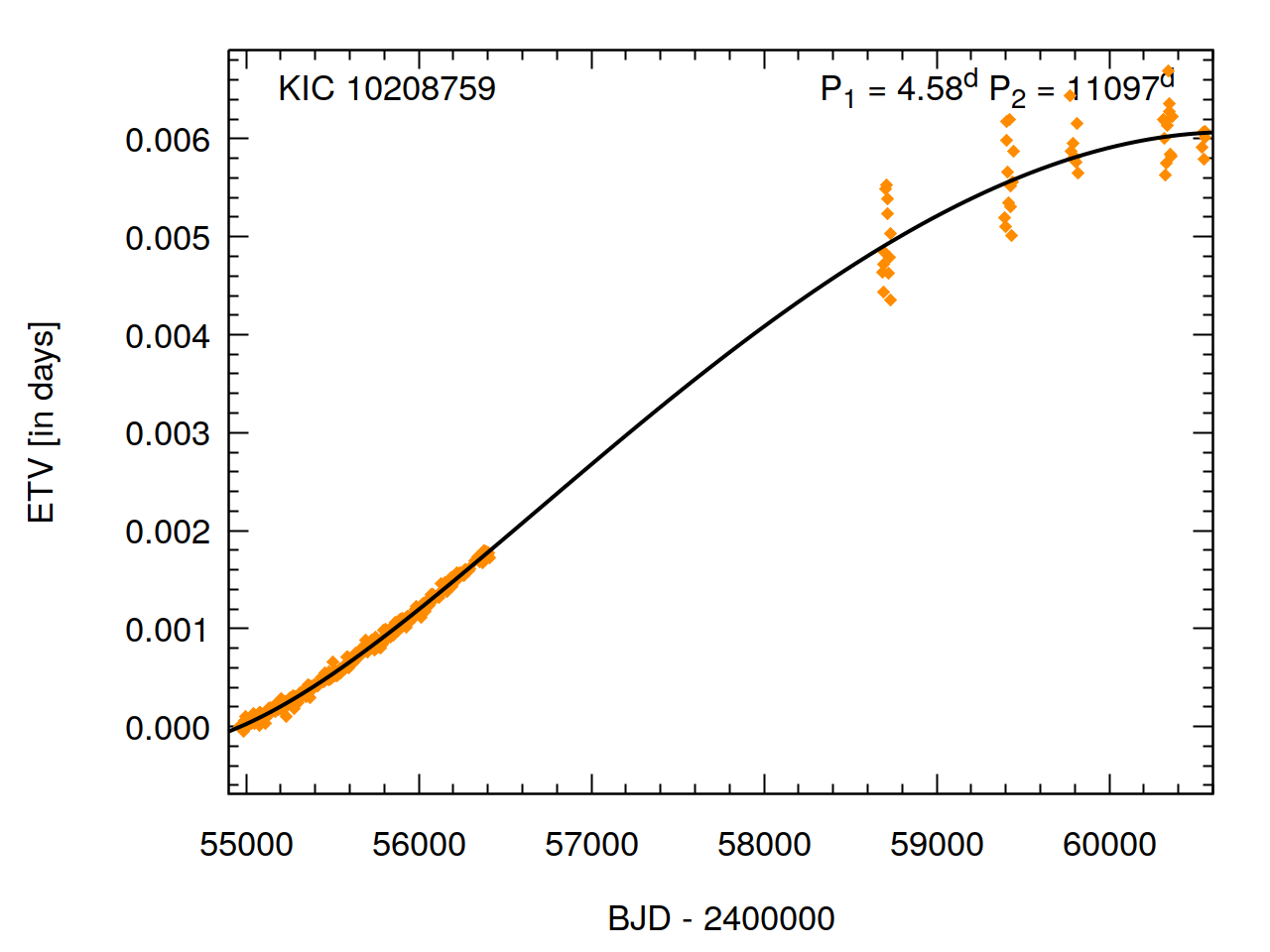}\includegraphics[width=60mm]{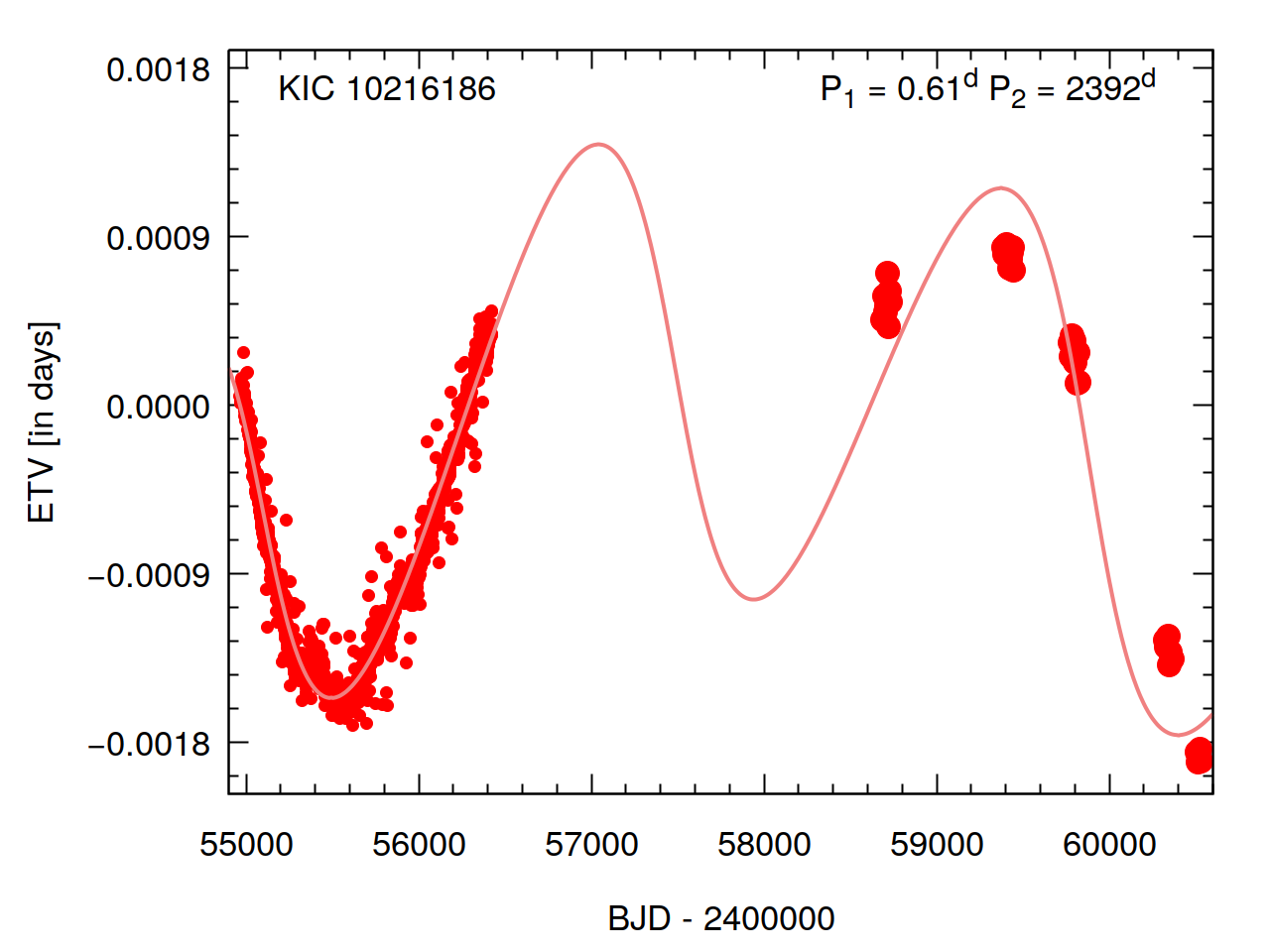}\includegraphics[width=60mm]{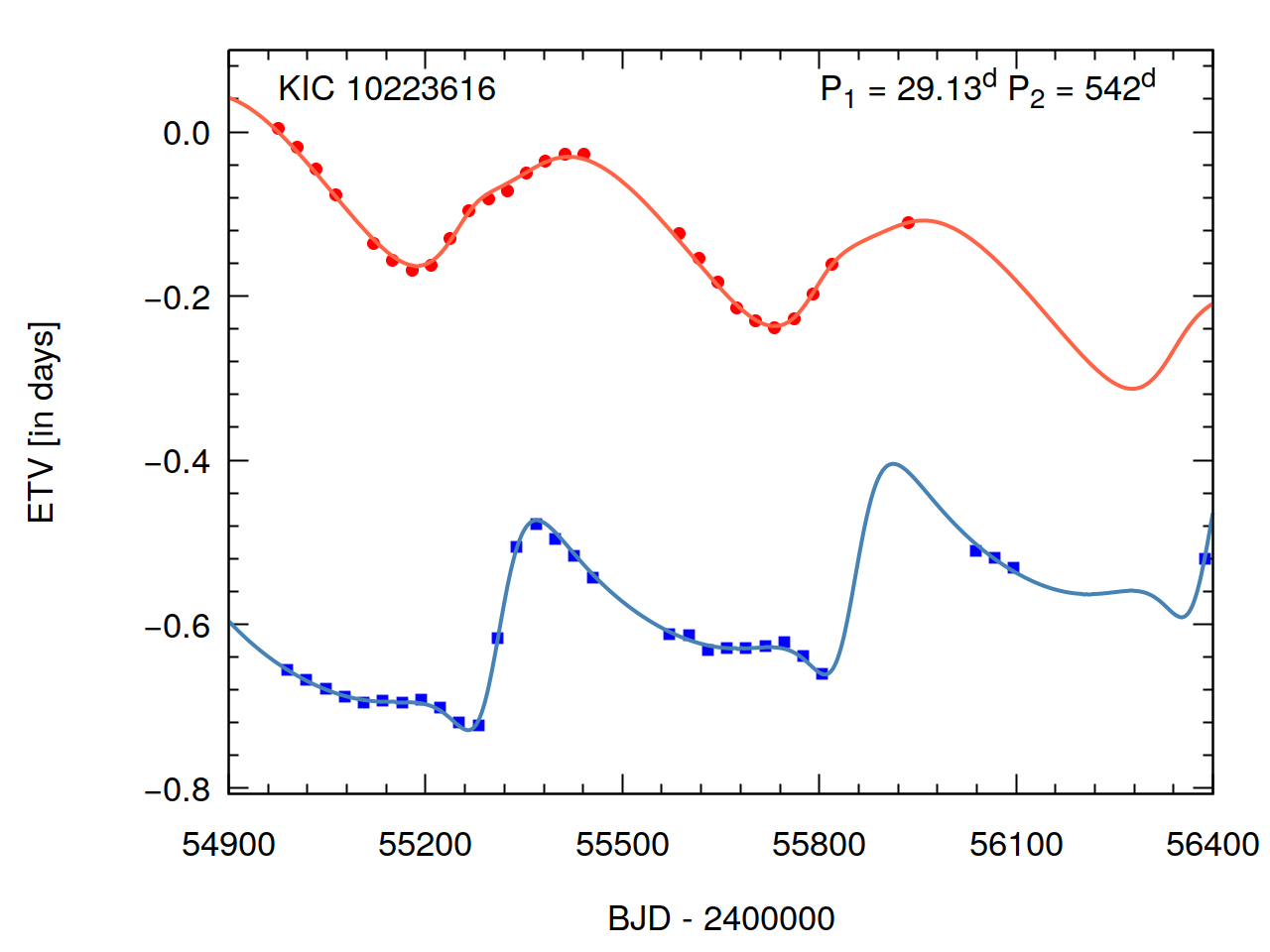}
\includegraphics[width=60mm]{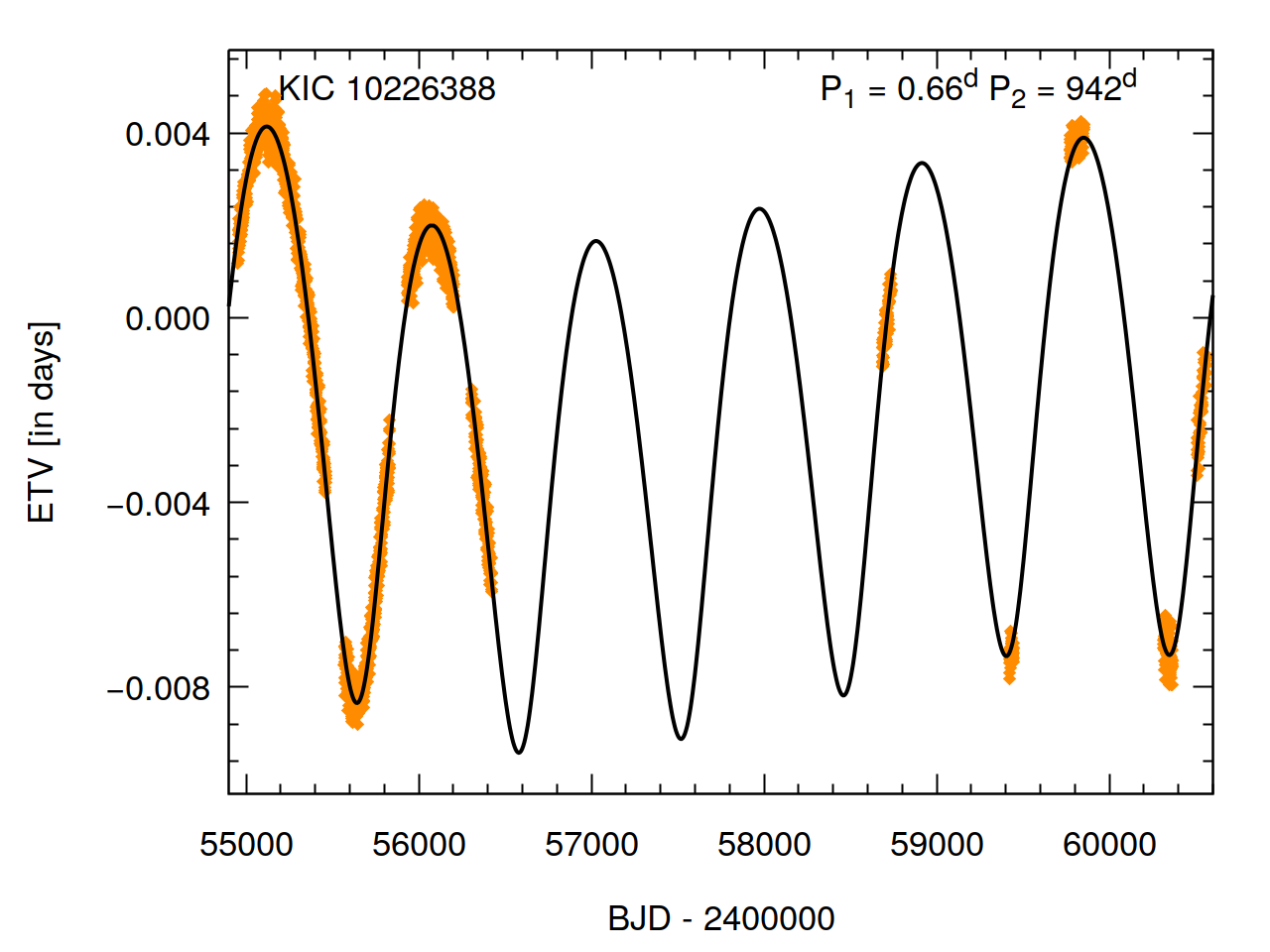}\includegraphics[width=60mm]{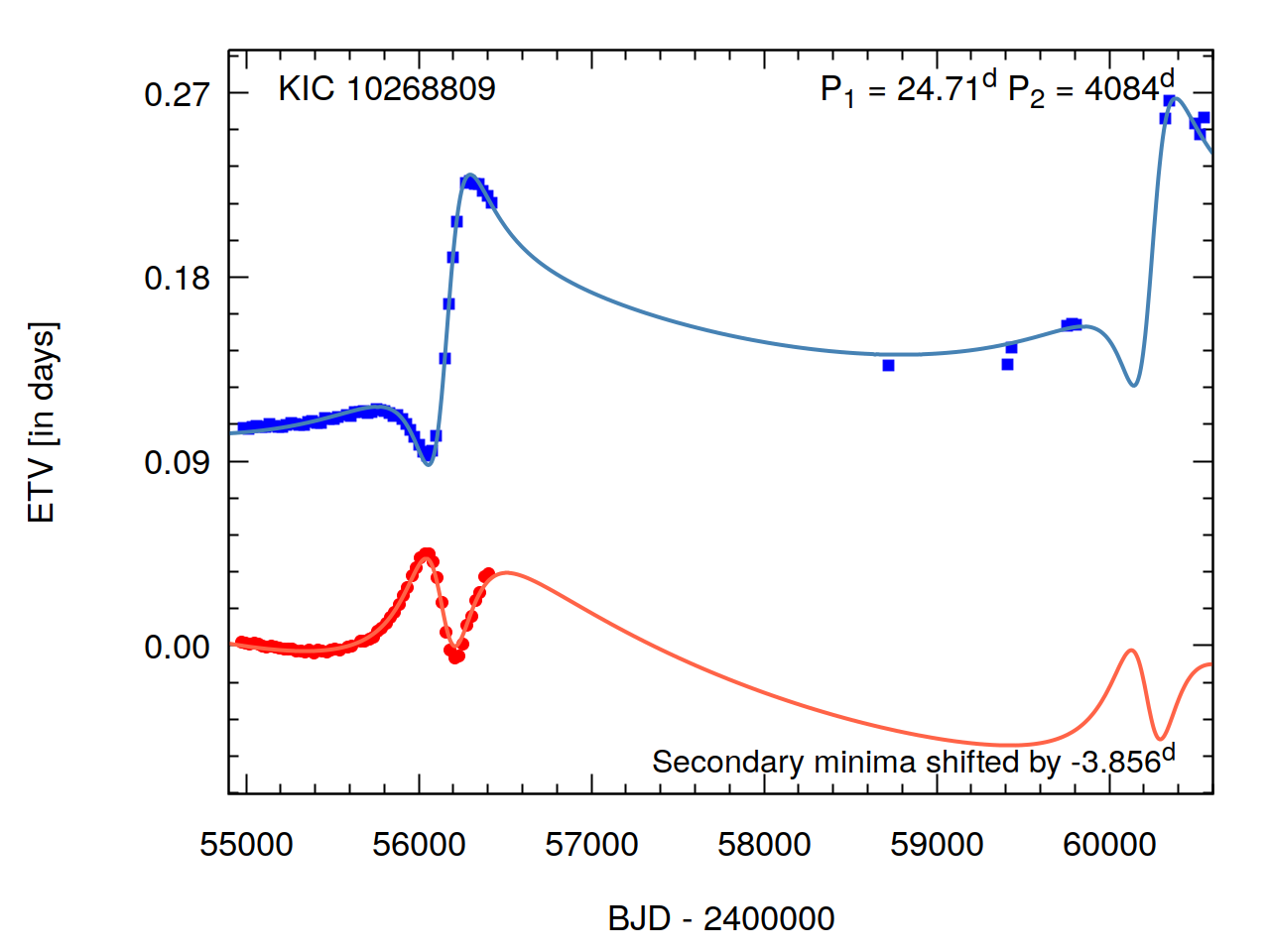}\includegraphics[width=60mm]{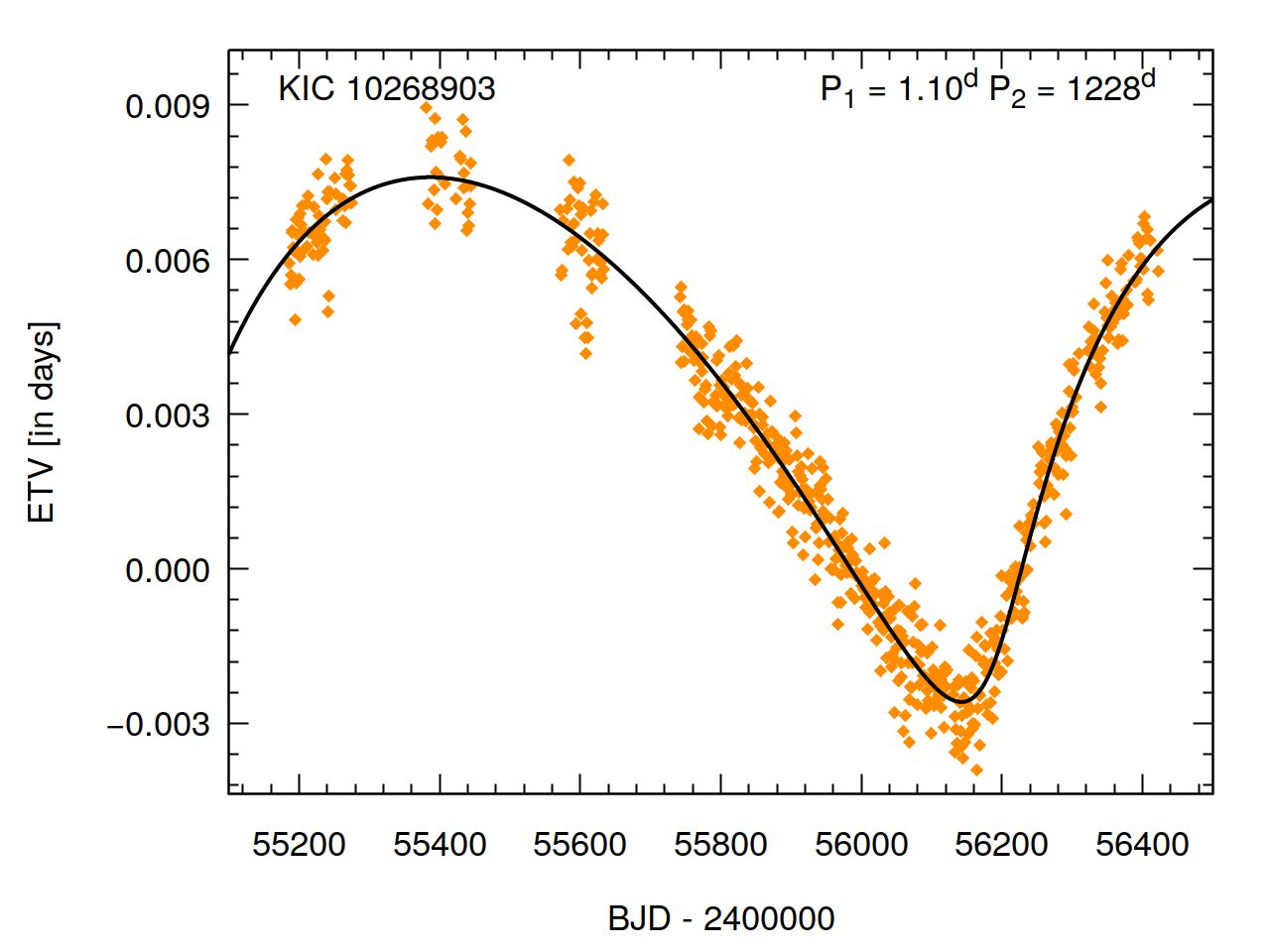}
\includegraphics[width=60mm]{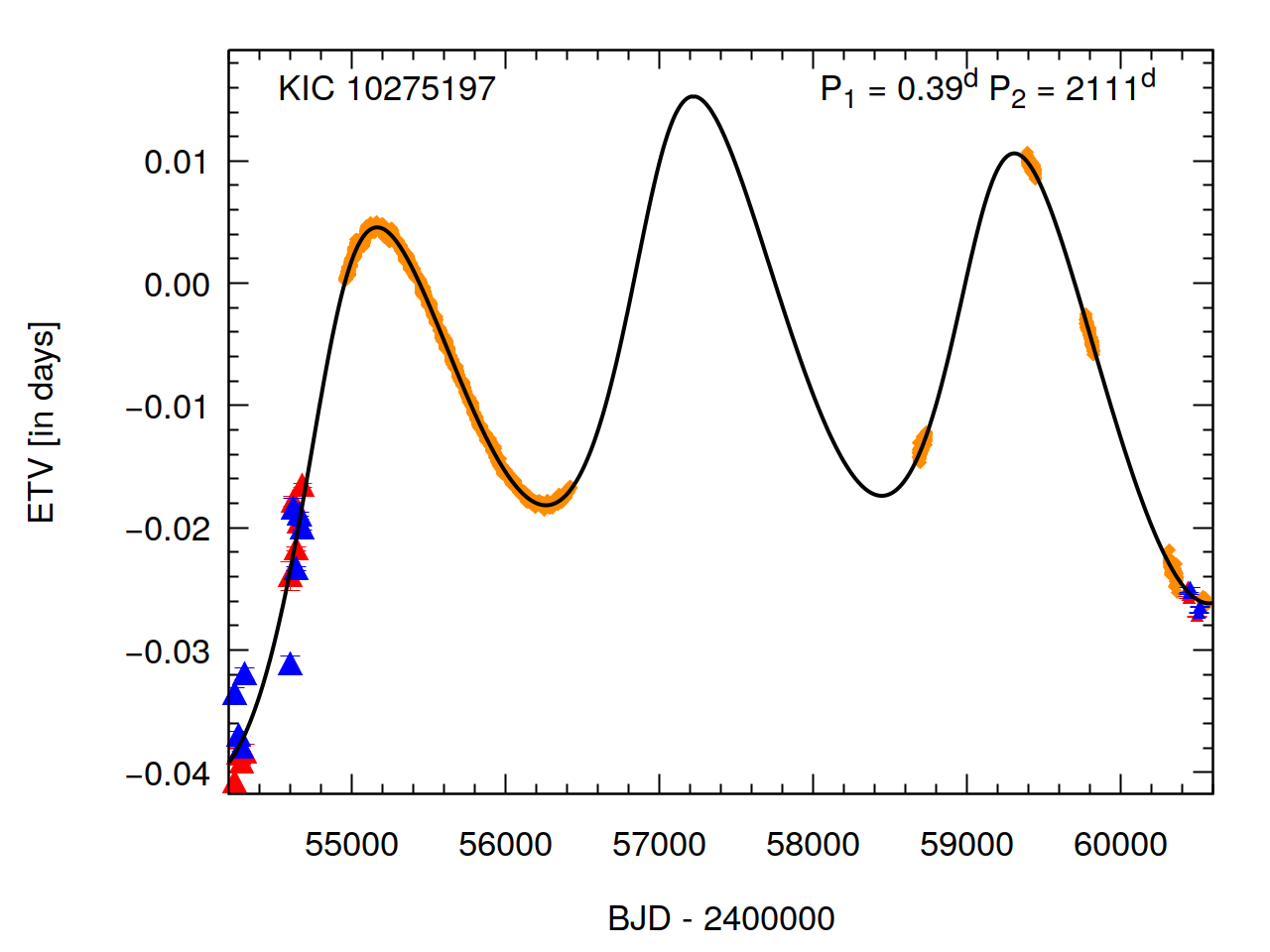}\includegraphics[width=60mm]{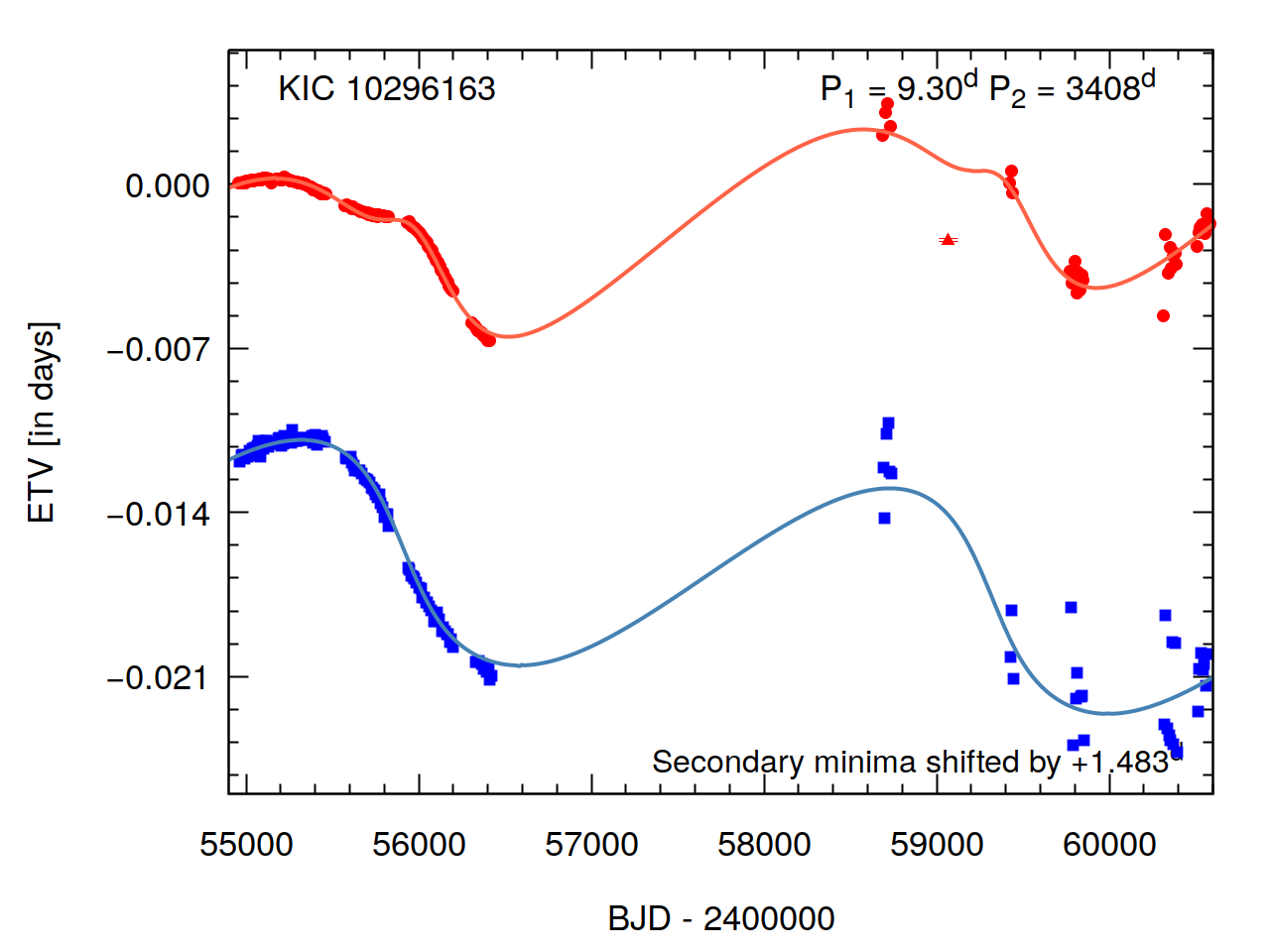}\includegraphics[width=60mm]{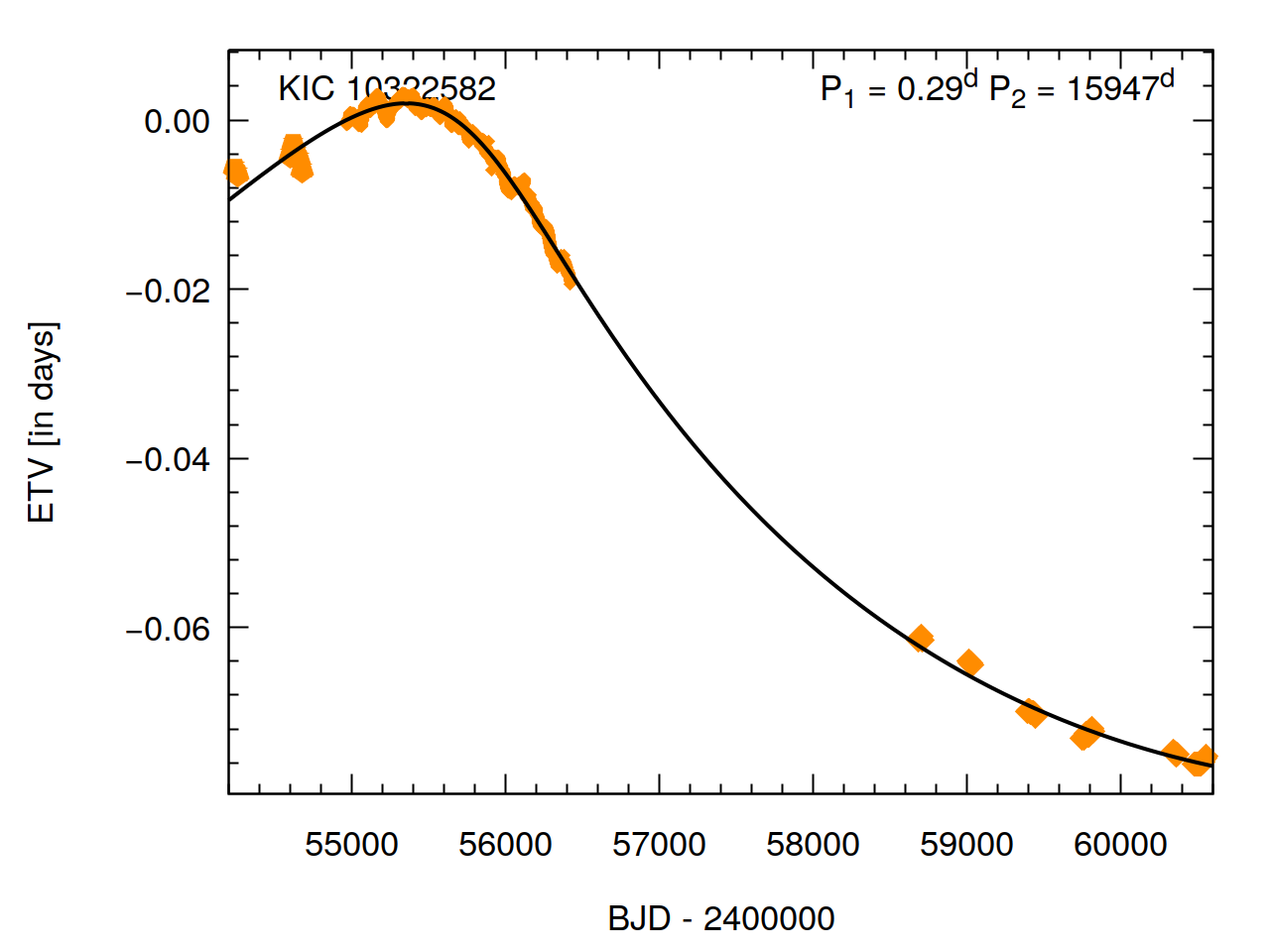}
\caption{continued.}
\end{figure*}

\addtocounter{figure}{-1}

\begin{figure*}
\includegraphics[width=60mm]{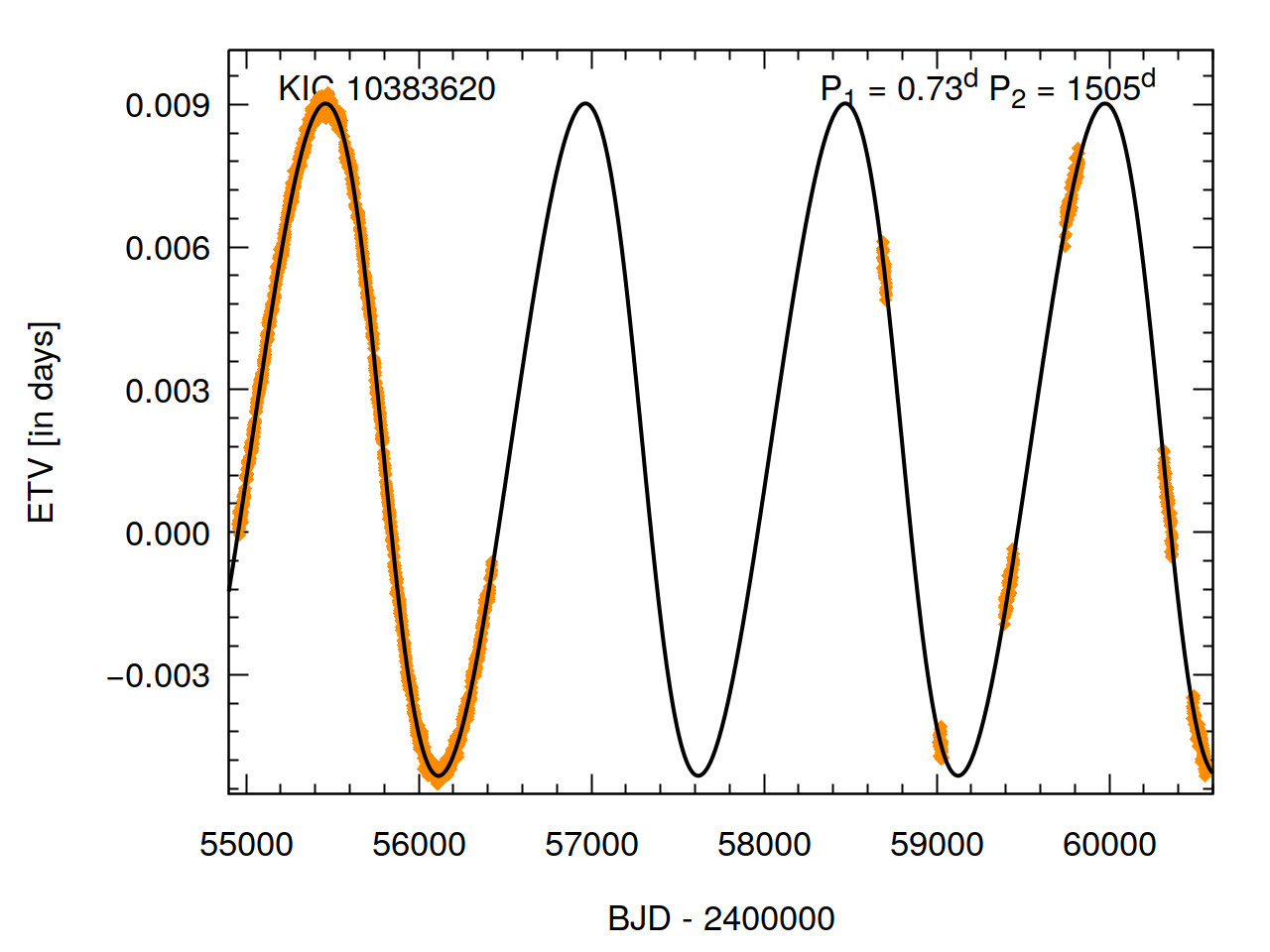}\includegraphics[width=60mm]{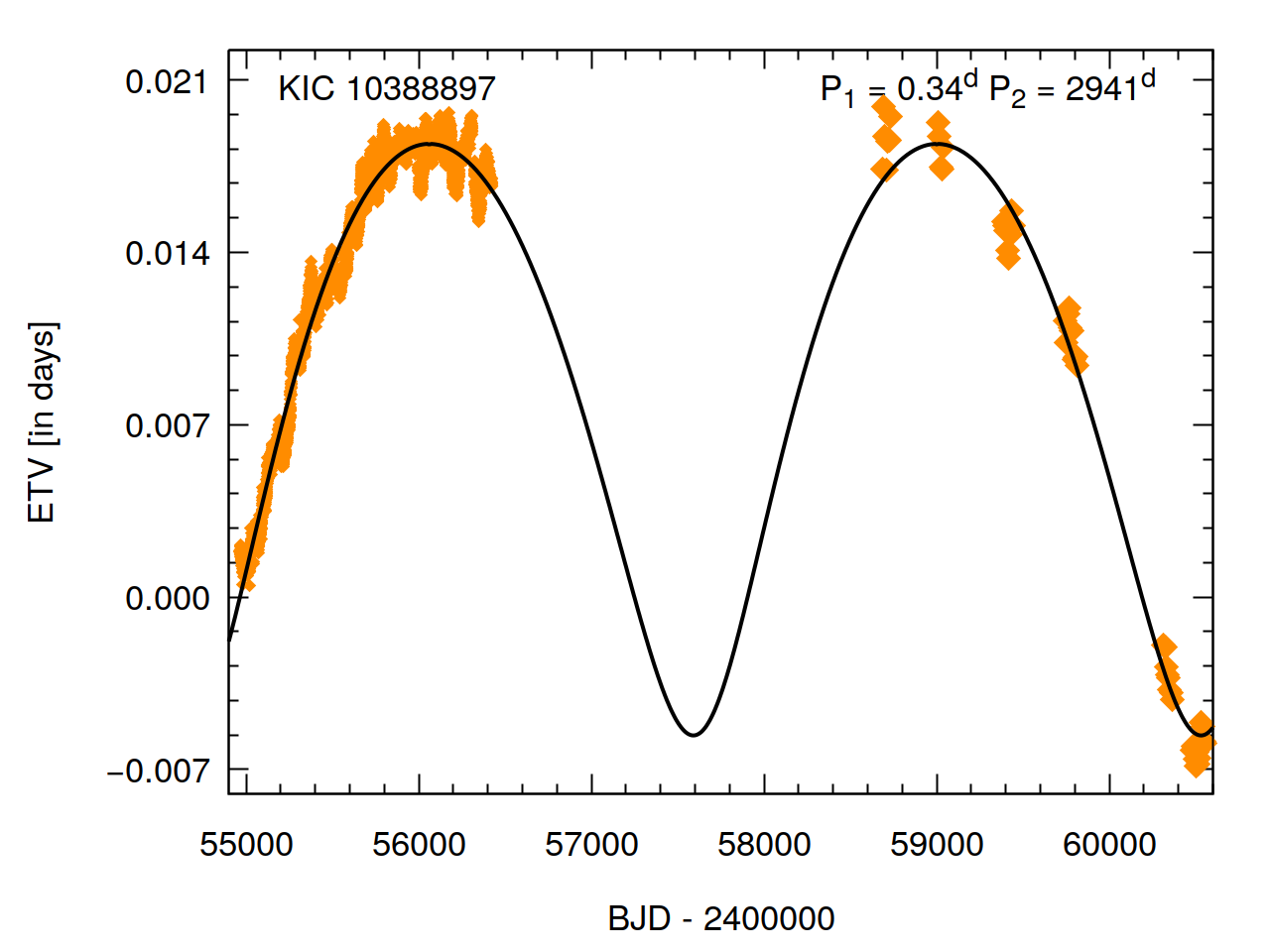}\includegraphics[width=60mm]{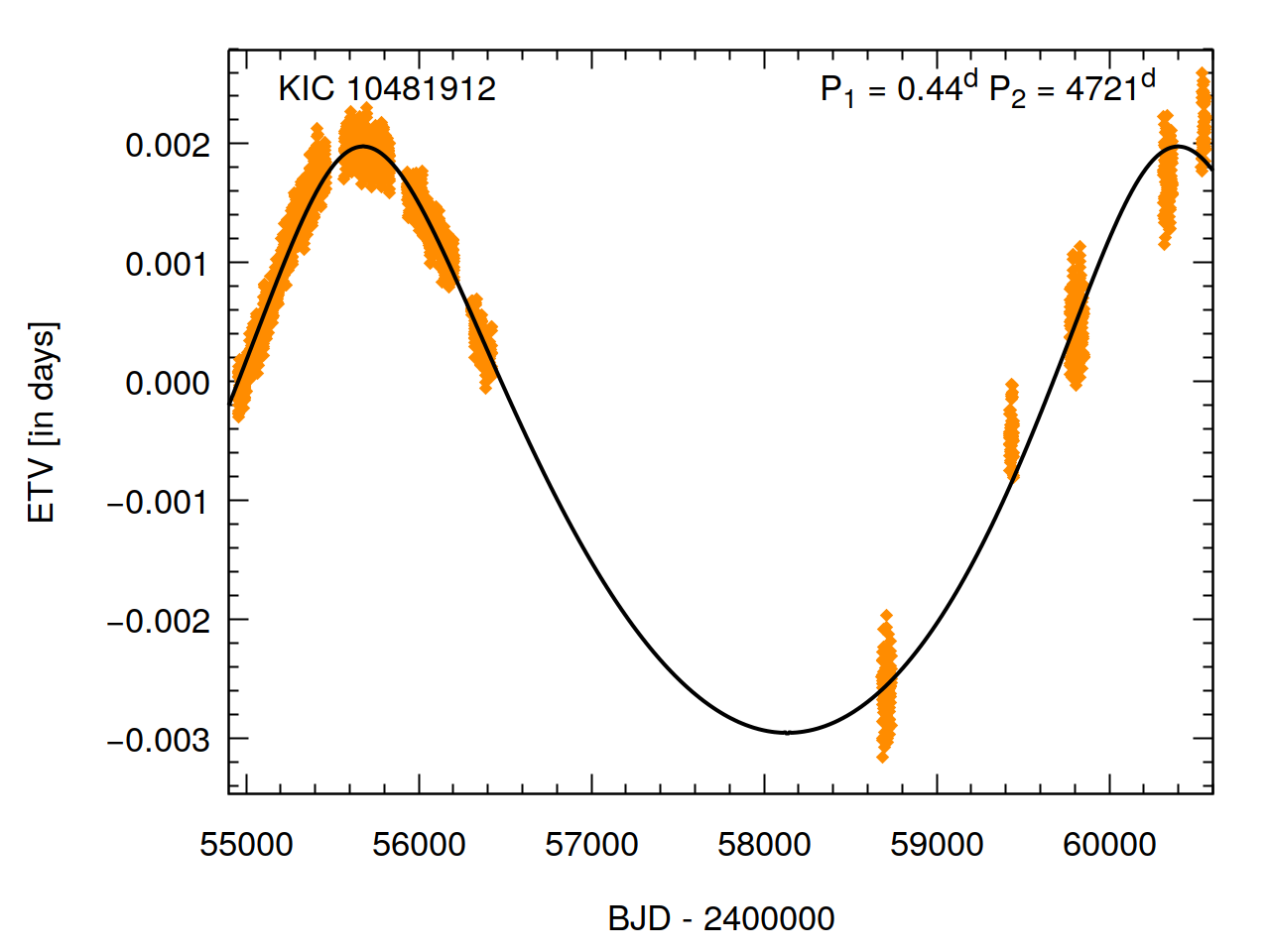}
\includegraphics[width=60mm]{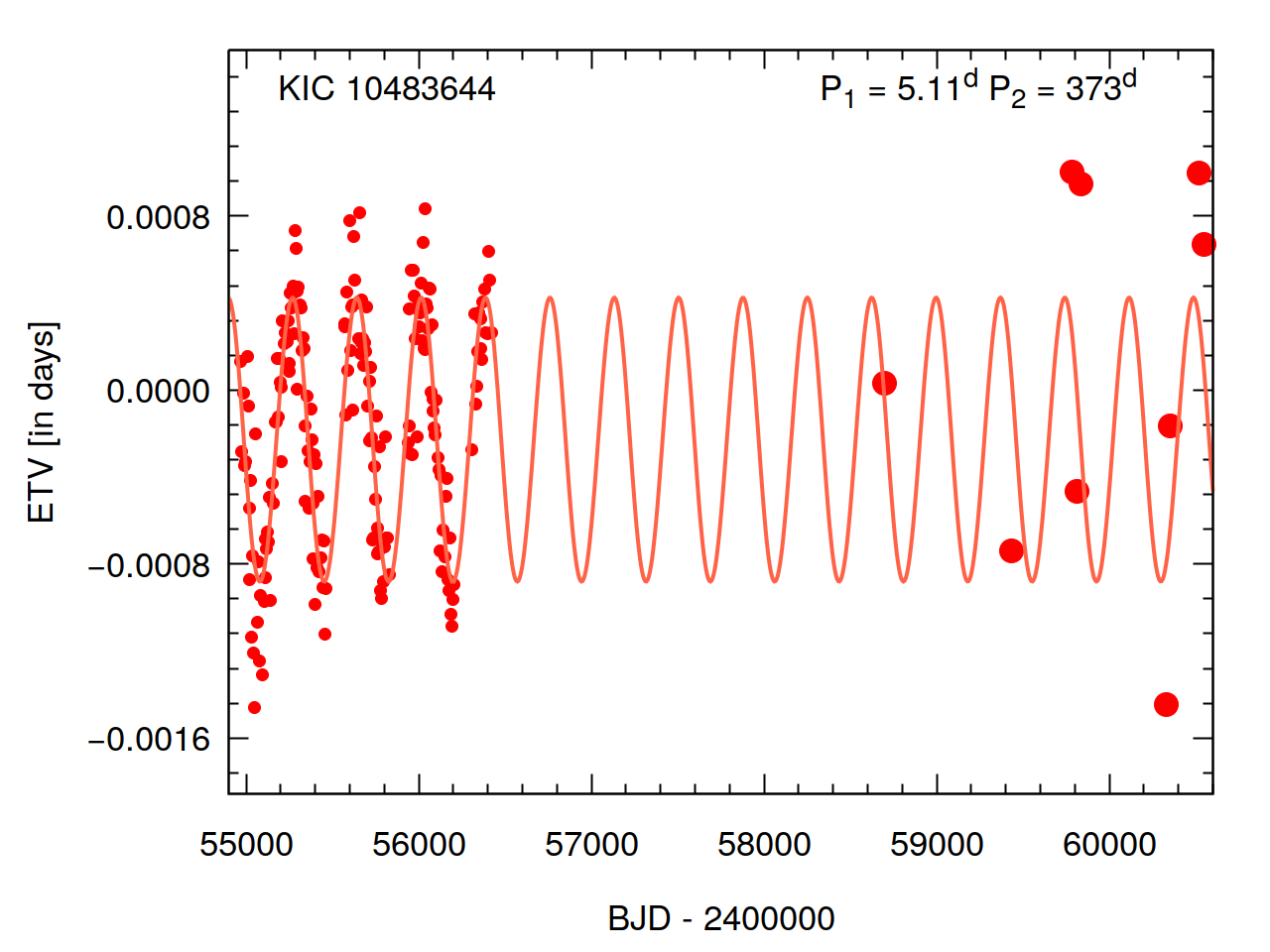}\includegraphics[width=60mm]{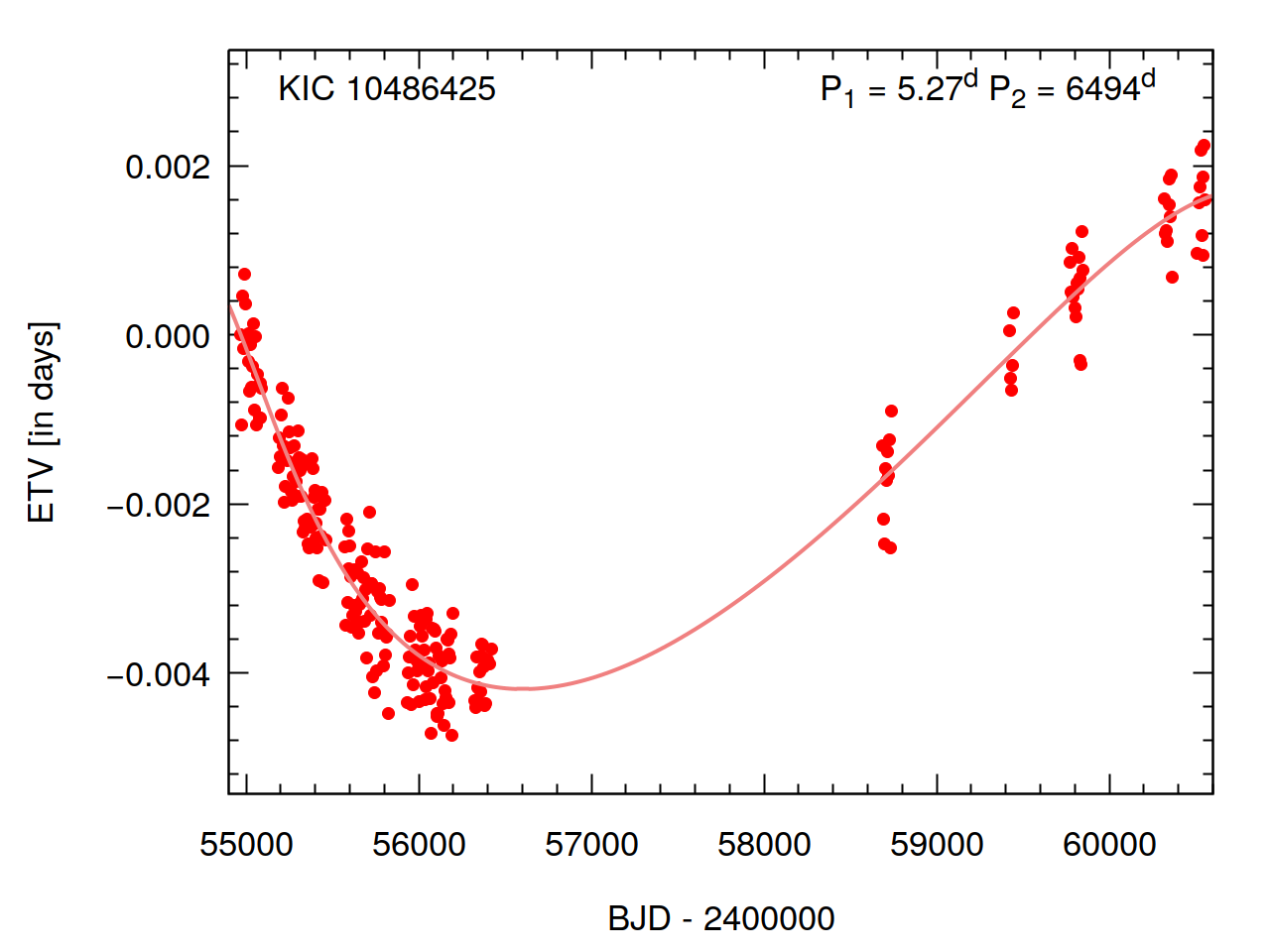}\includegraphics[width=60mm]{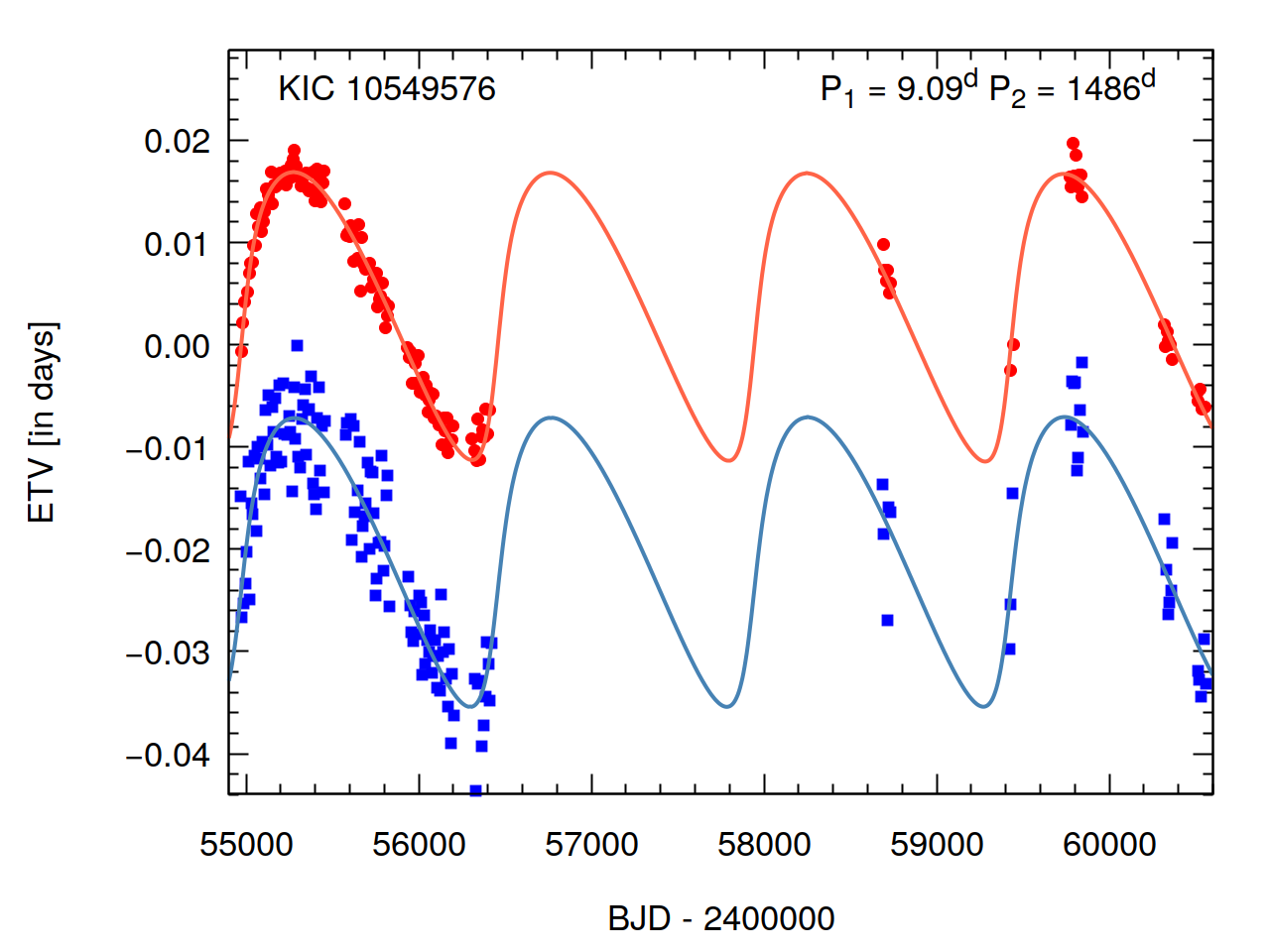}
\includegraphics[width=60mm]{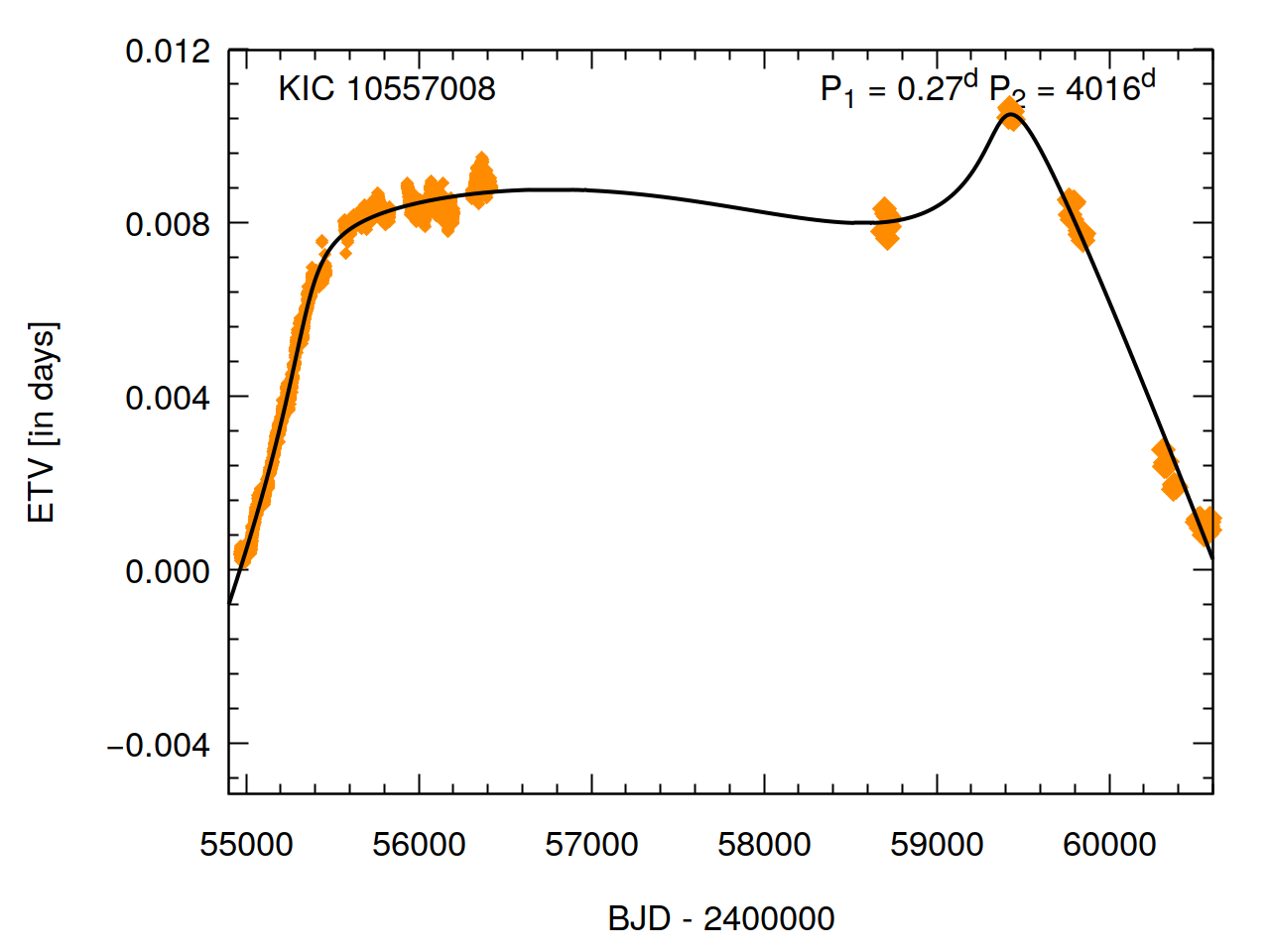}\includegraphics[width=60mm]{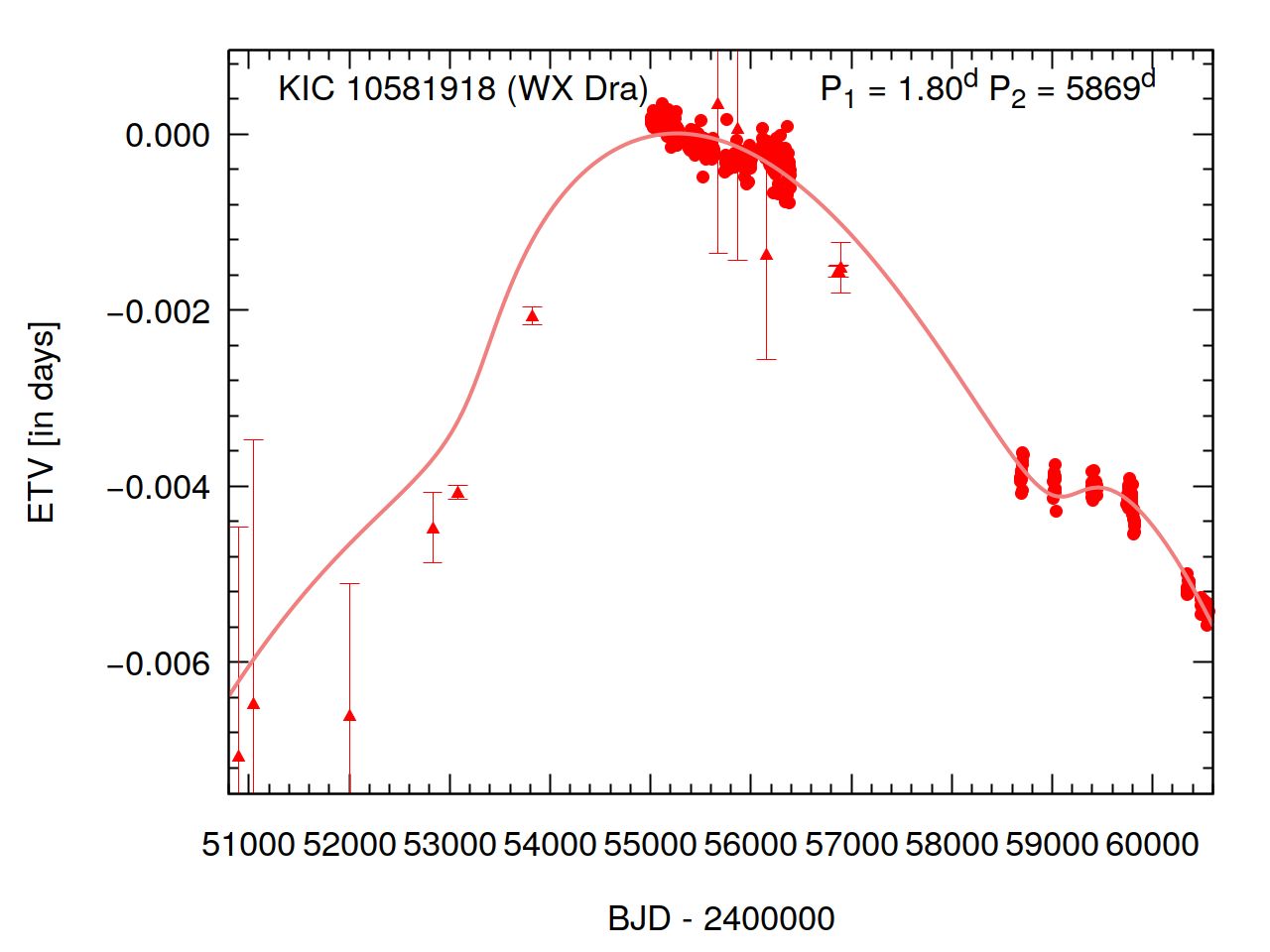}\includegraphics[width=60mm]{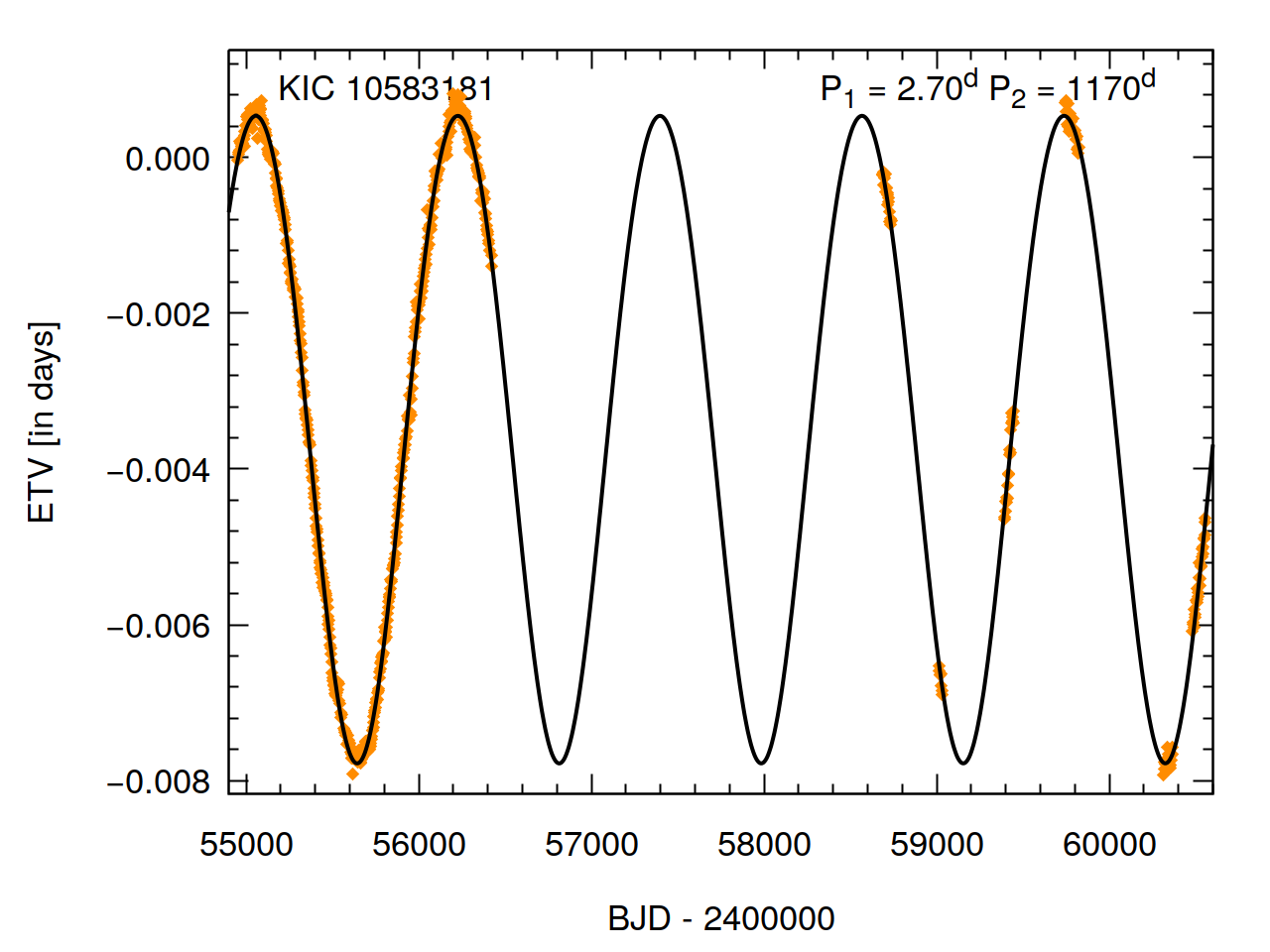}
\includegraphics[width=60mm]{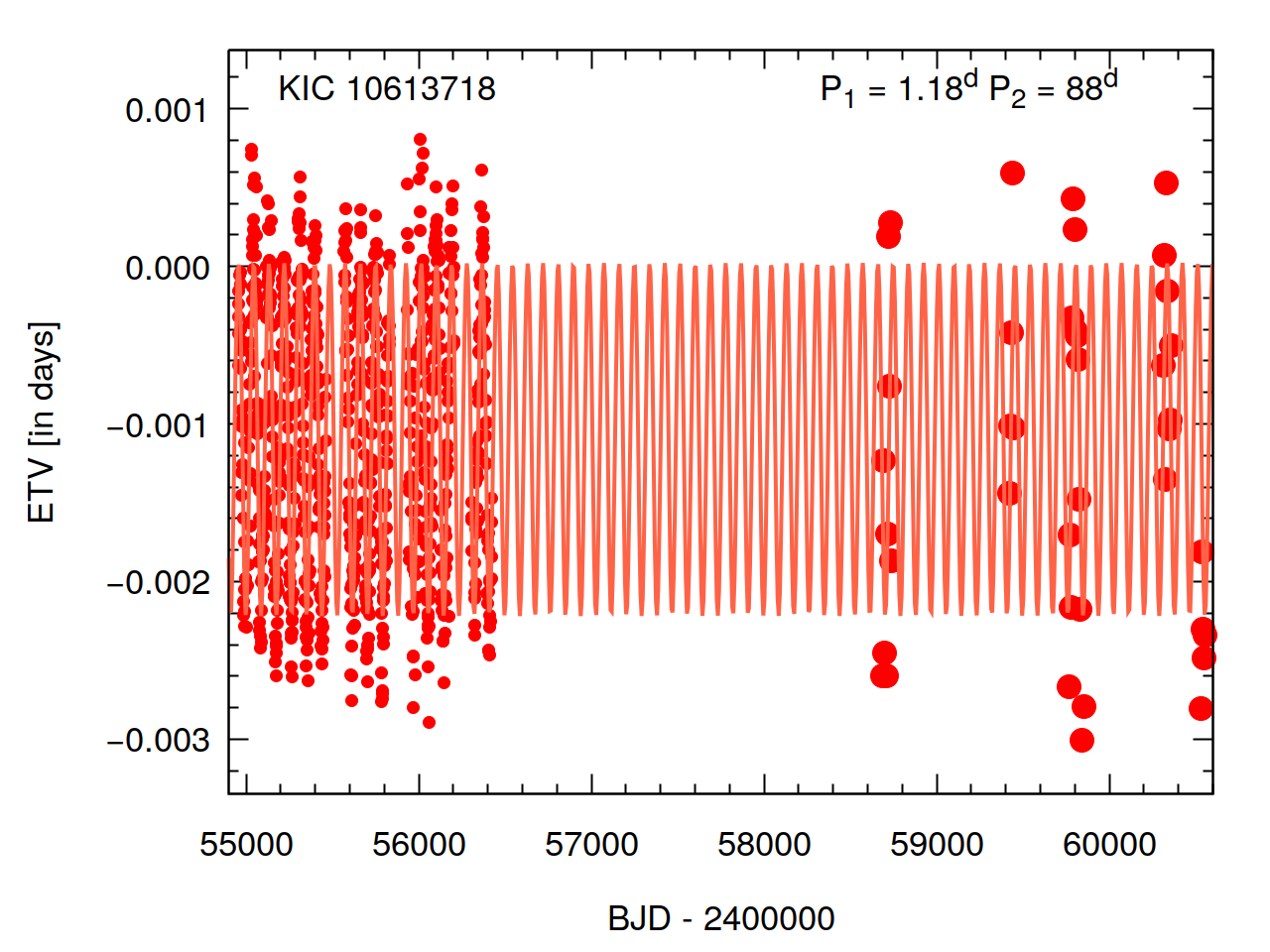}\includegraphics[width=60mm]{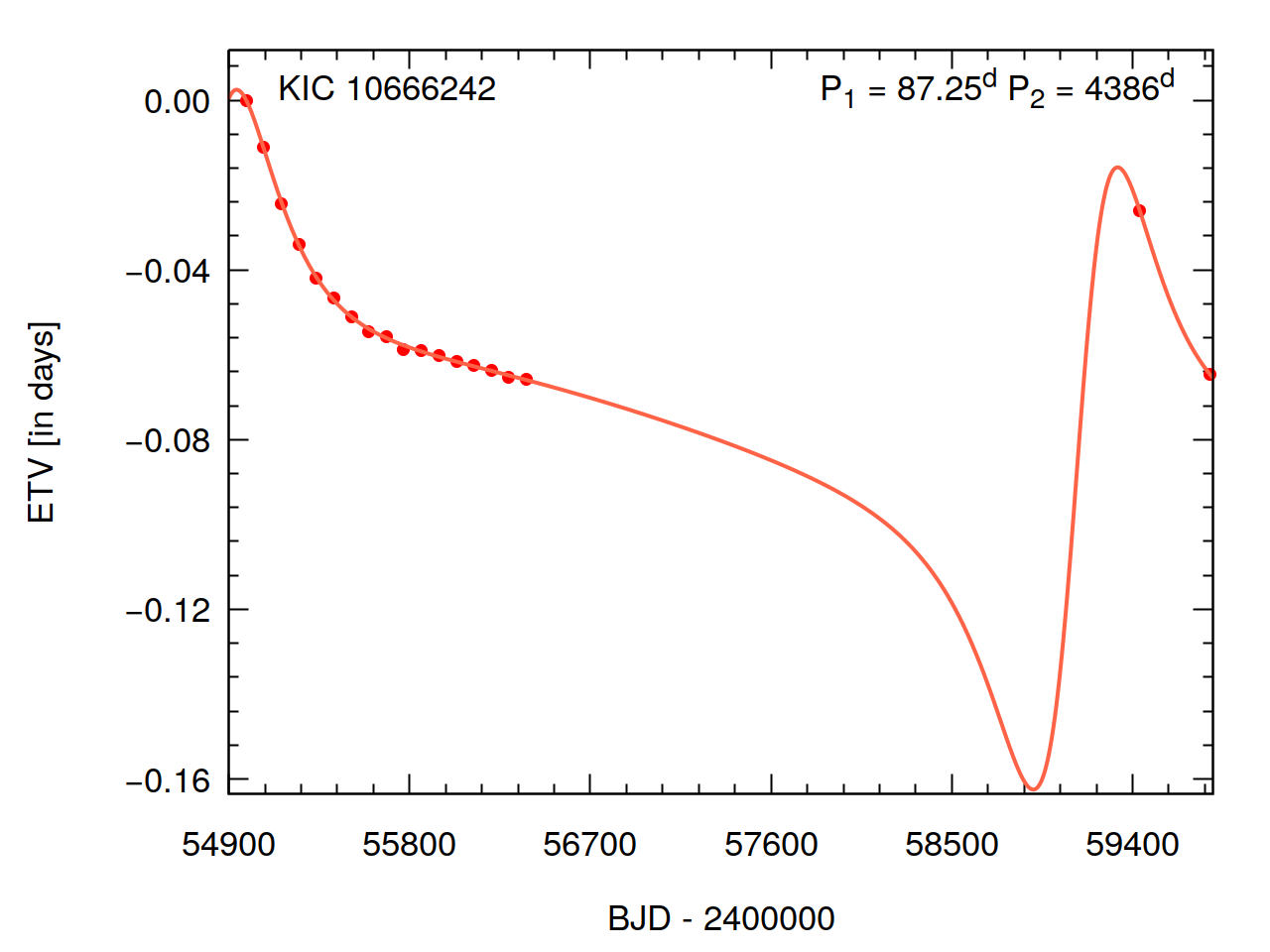}\includegraphics[width=60mm]{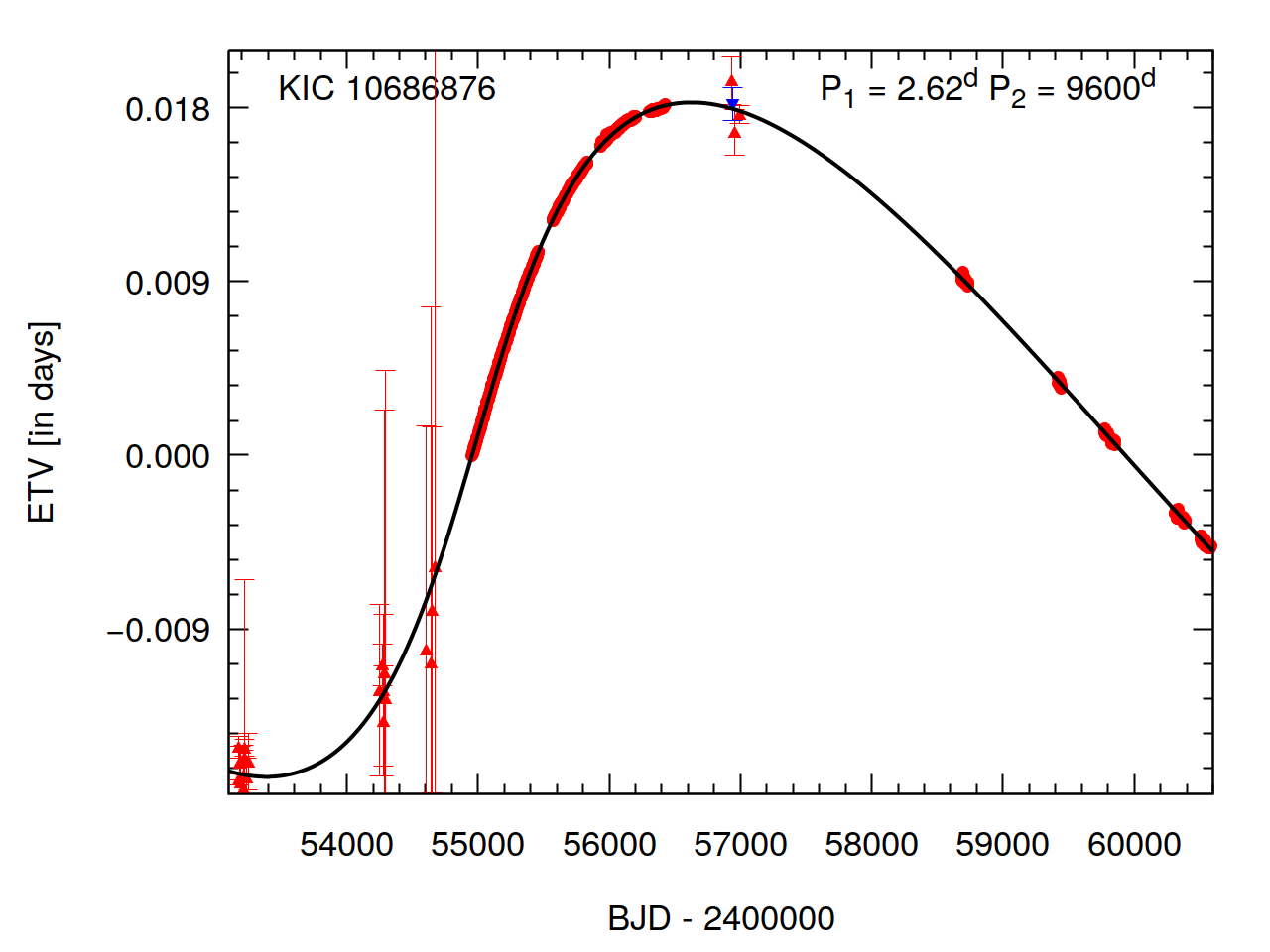}
\includegraphics[width=60mm]{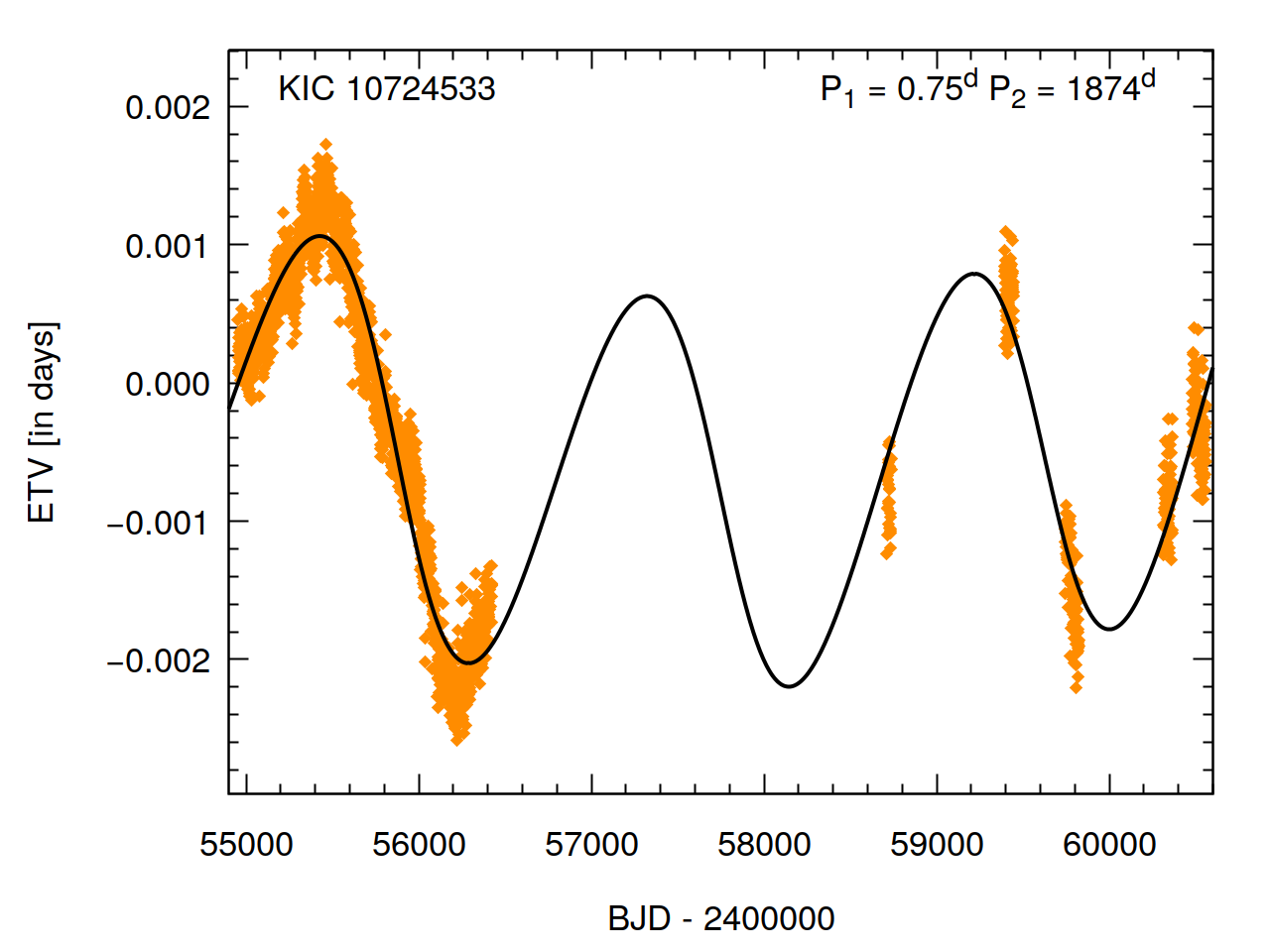}\includegraphics[width=60mm]{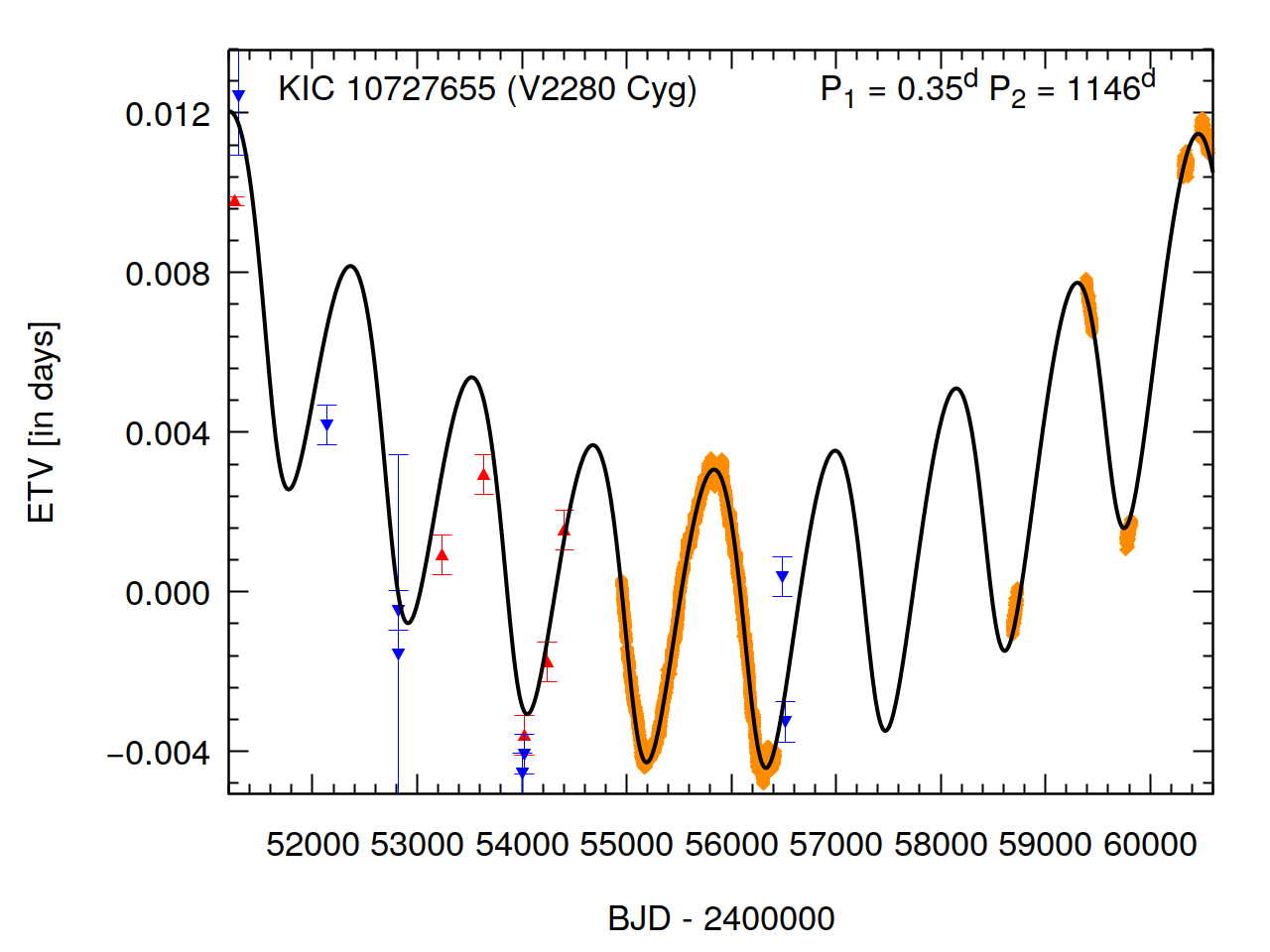}\includegraphics[width=60mm]{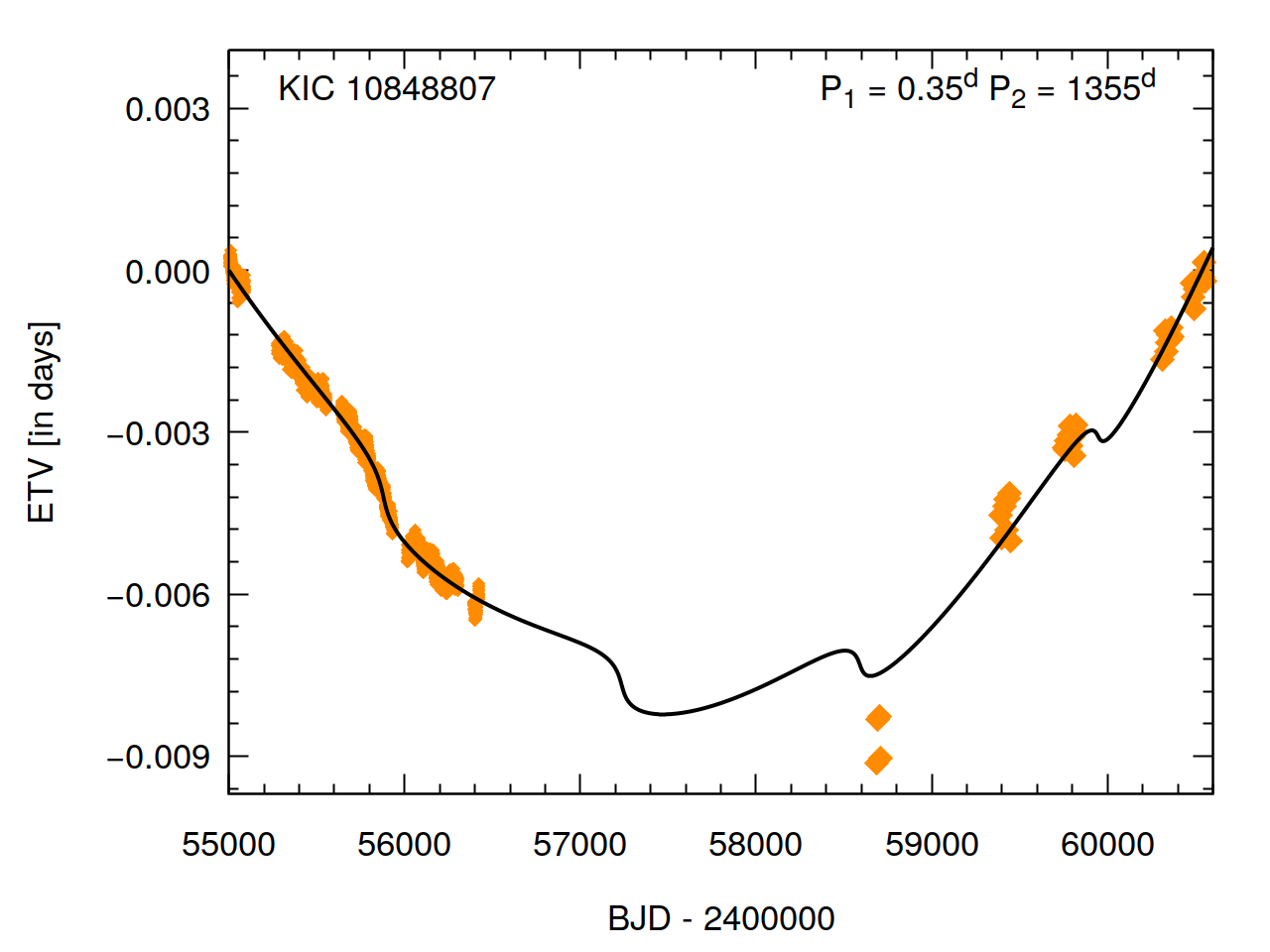}
\caption{continued.}
\end{figure*}

\addtocounter{figure}{-1}

\begin{figure*}
\includegraphics[width=60mm]{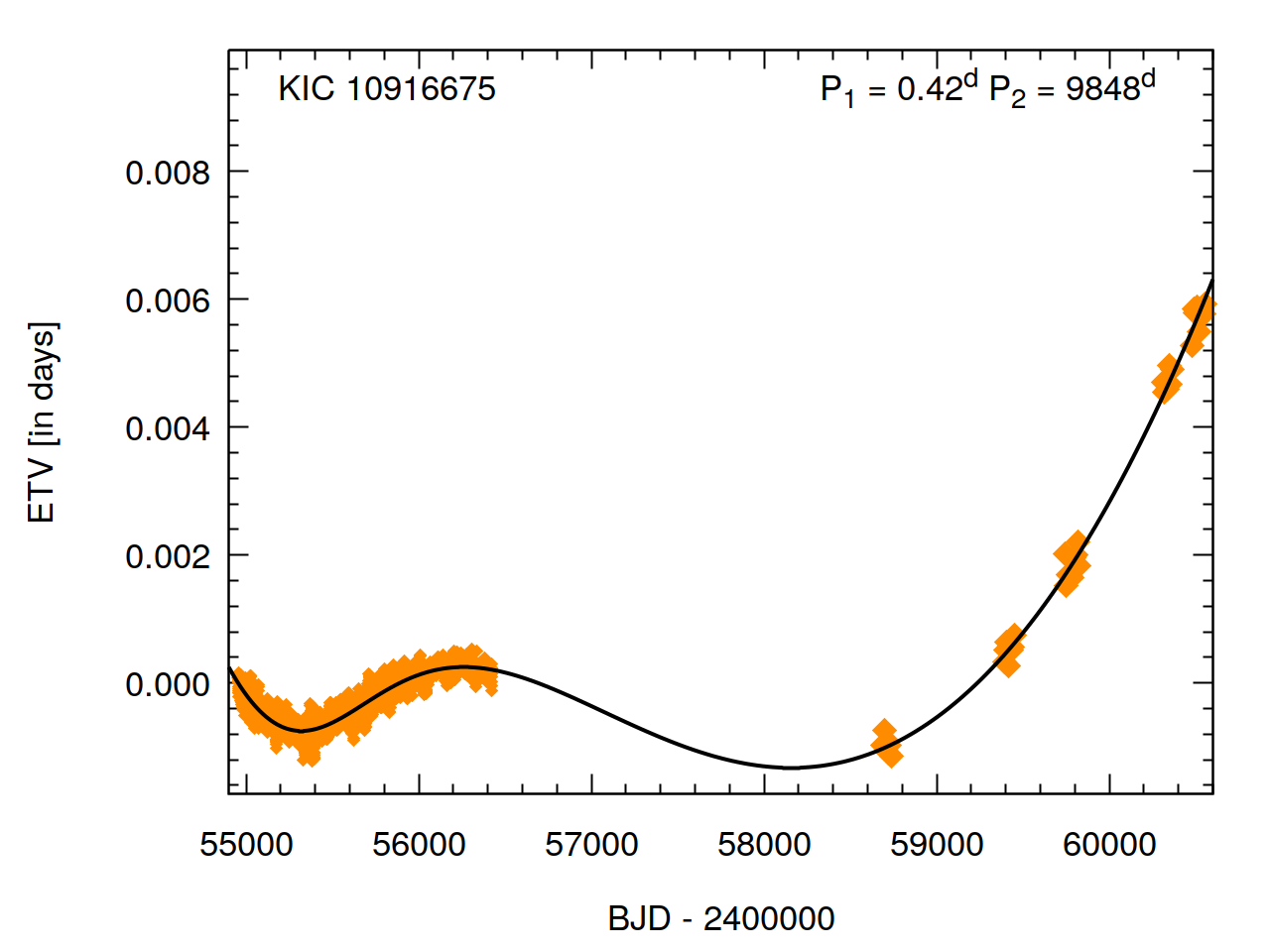}\includegraphics[width=60mm]{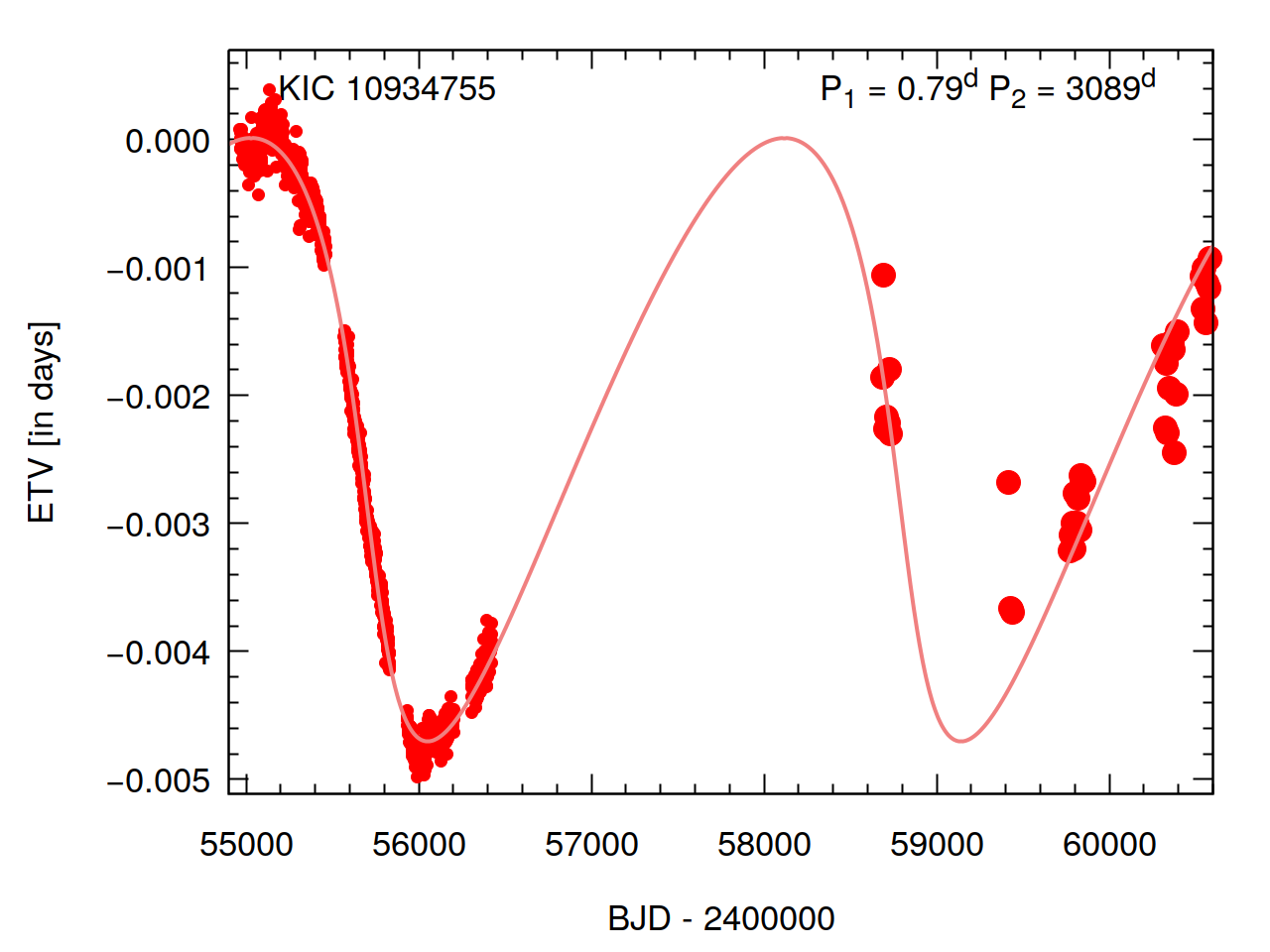}\includegraphics[width=60mm]{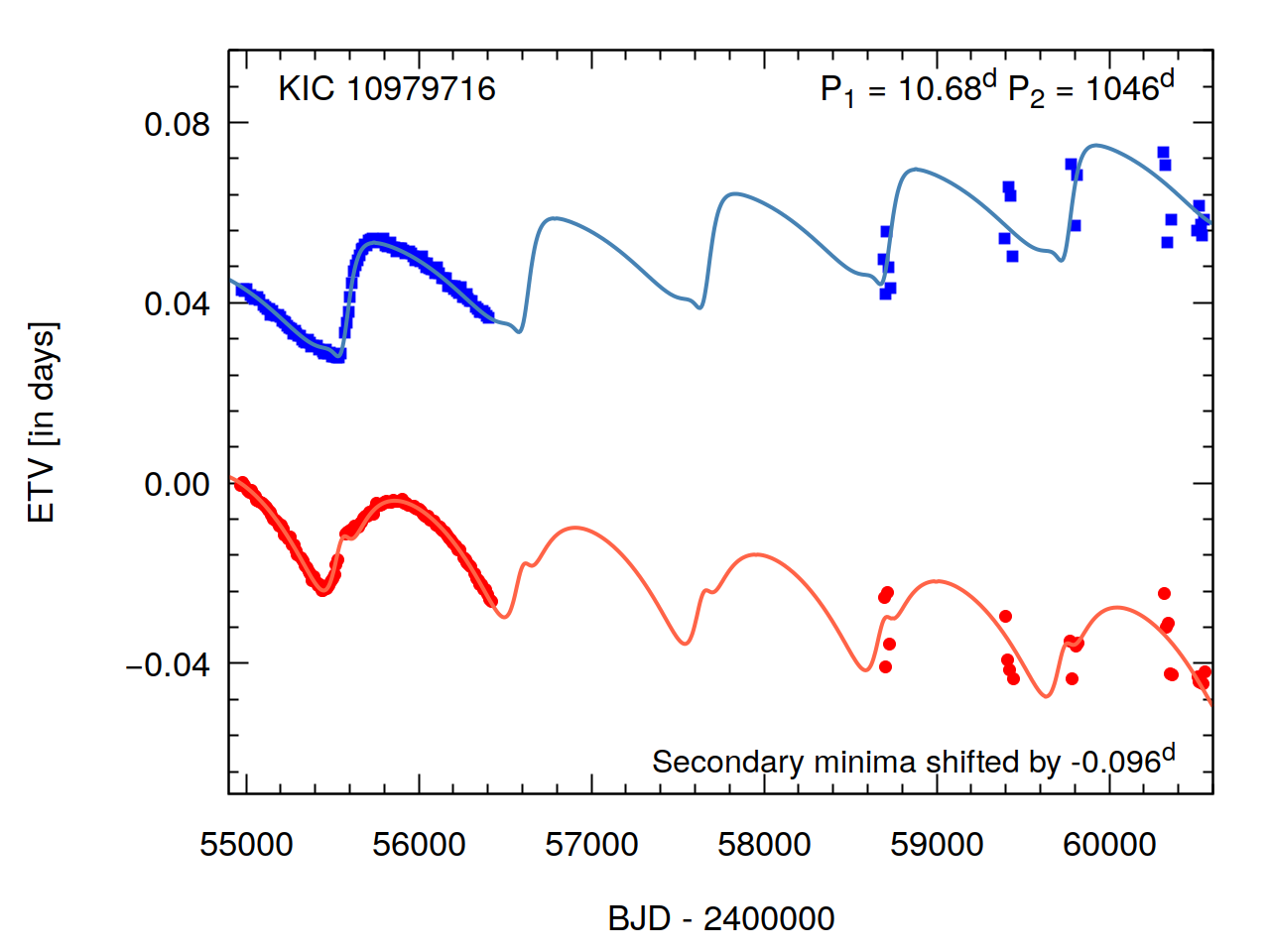}
\includegraphics[width=60mm]{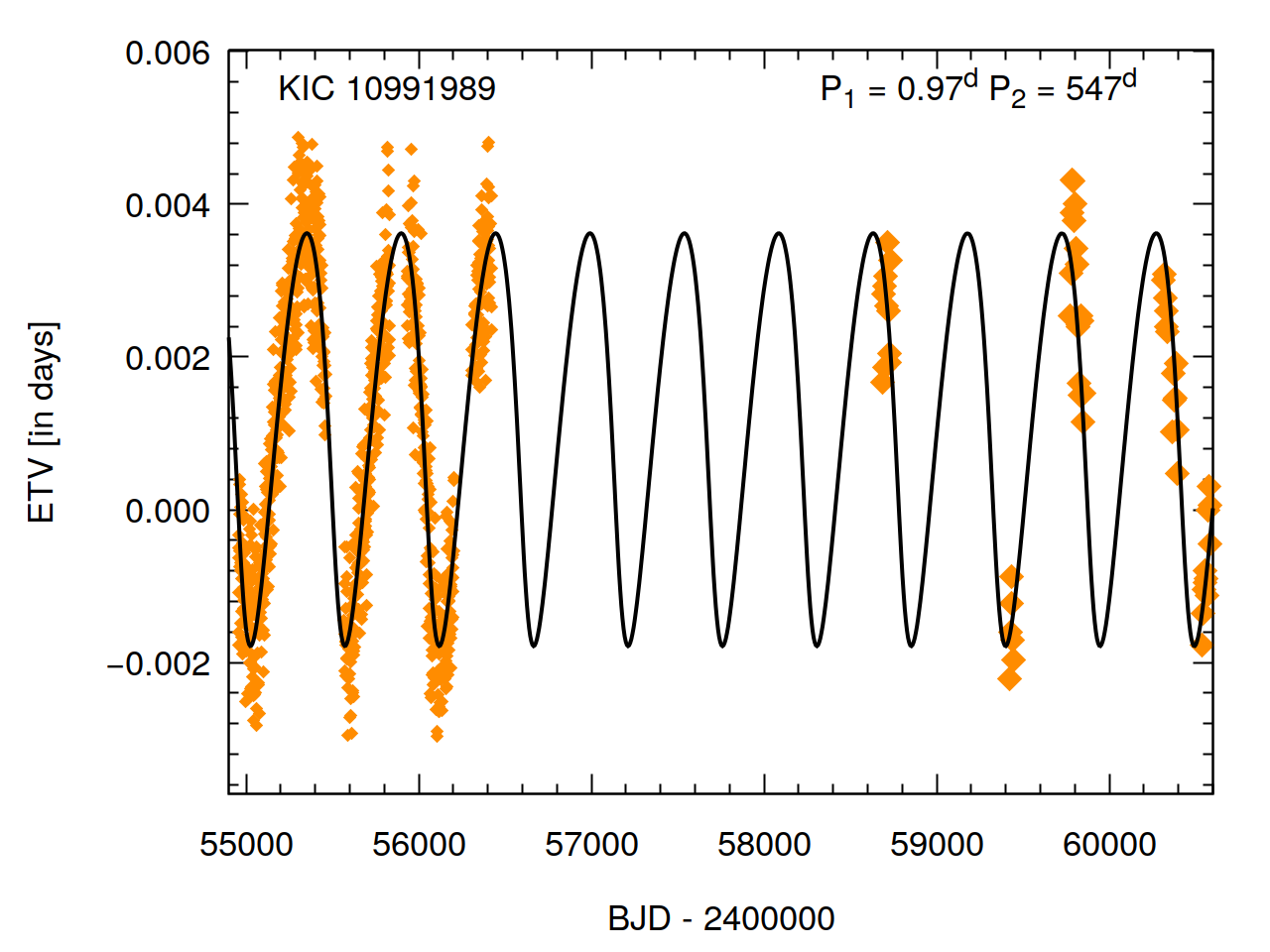}\includegraphics[width=60mm]{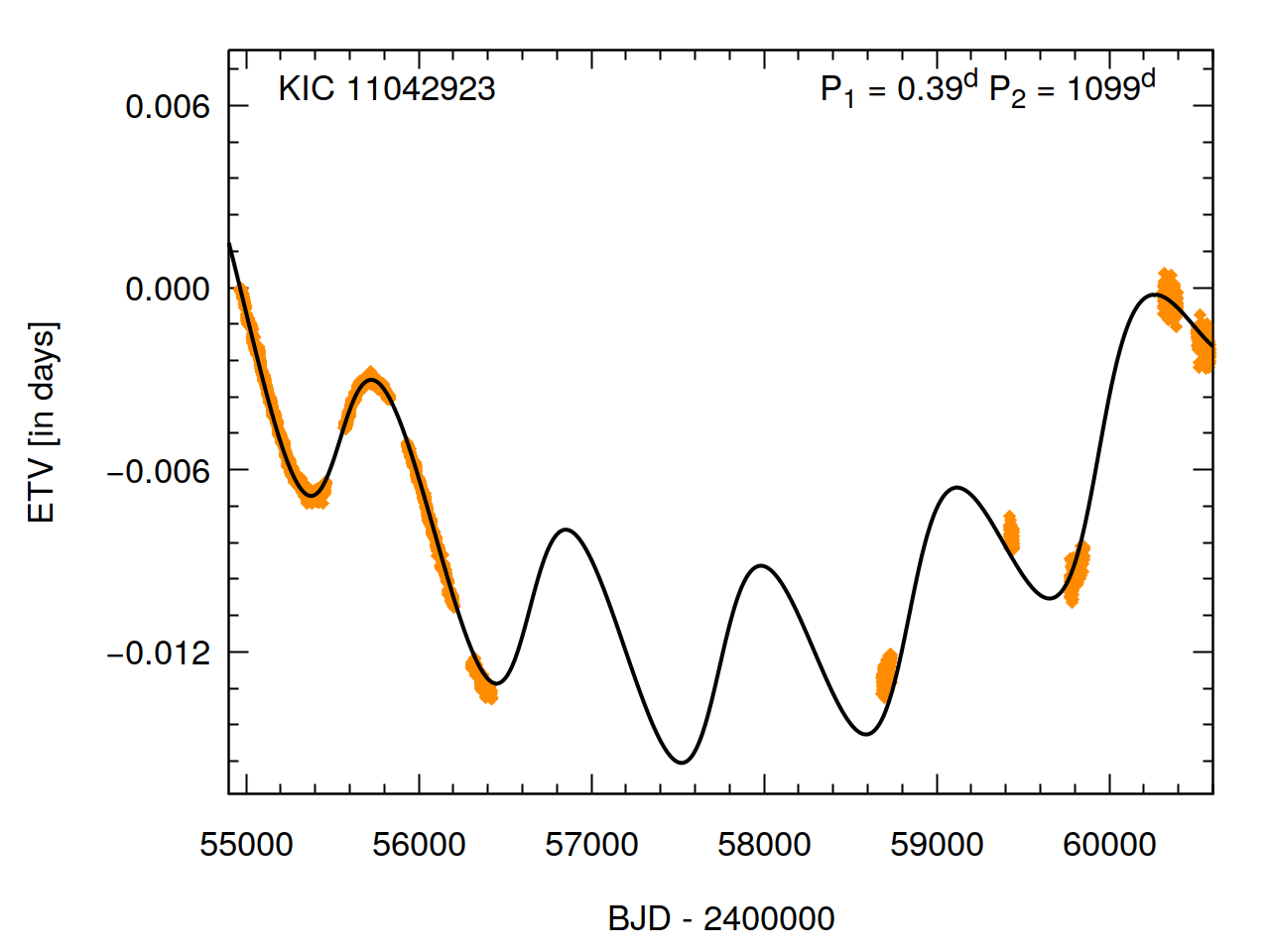}\includegraphics[width=60mm]{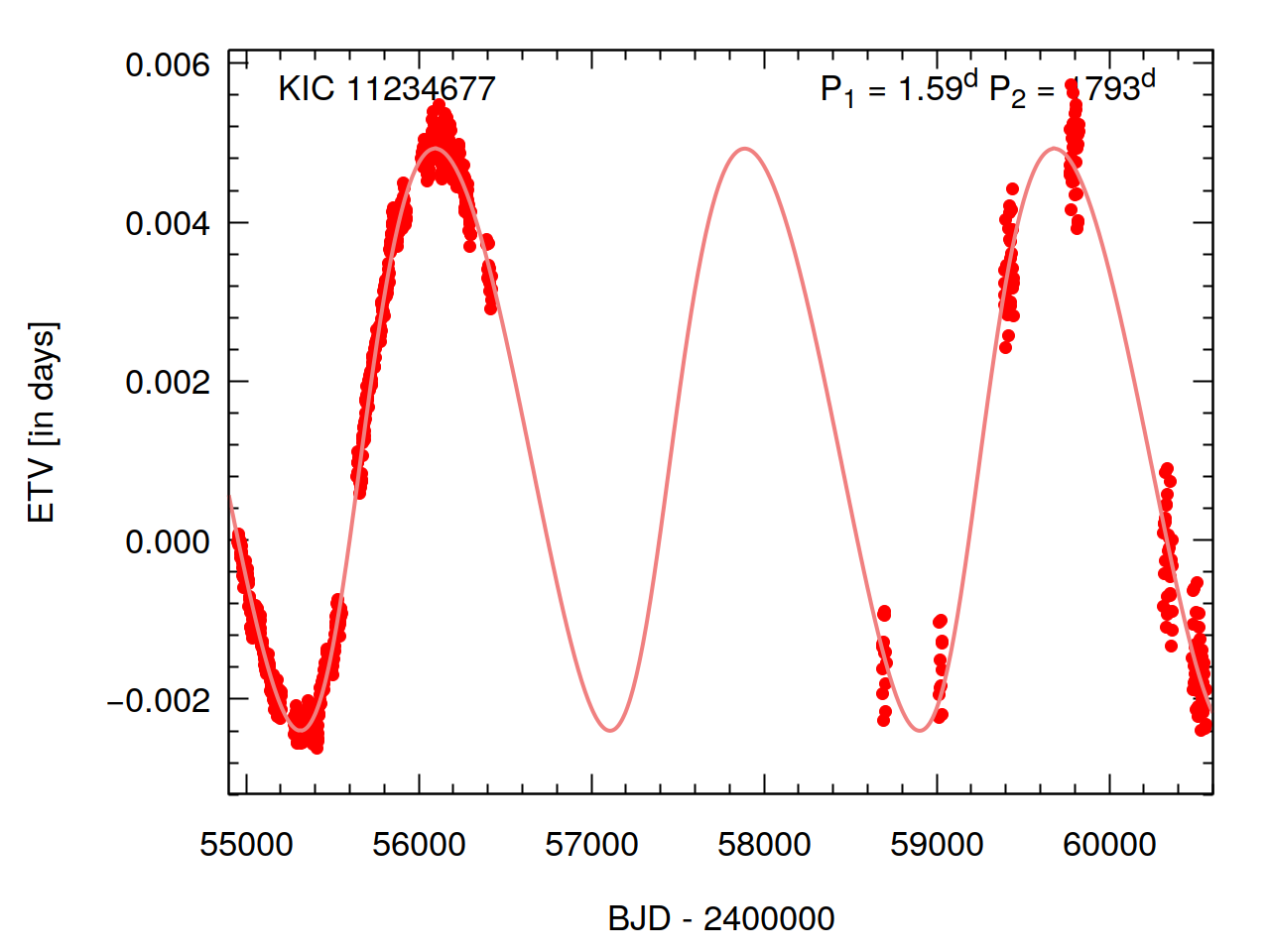}
\includegraphics[width=60mm]{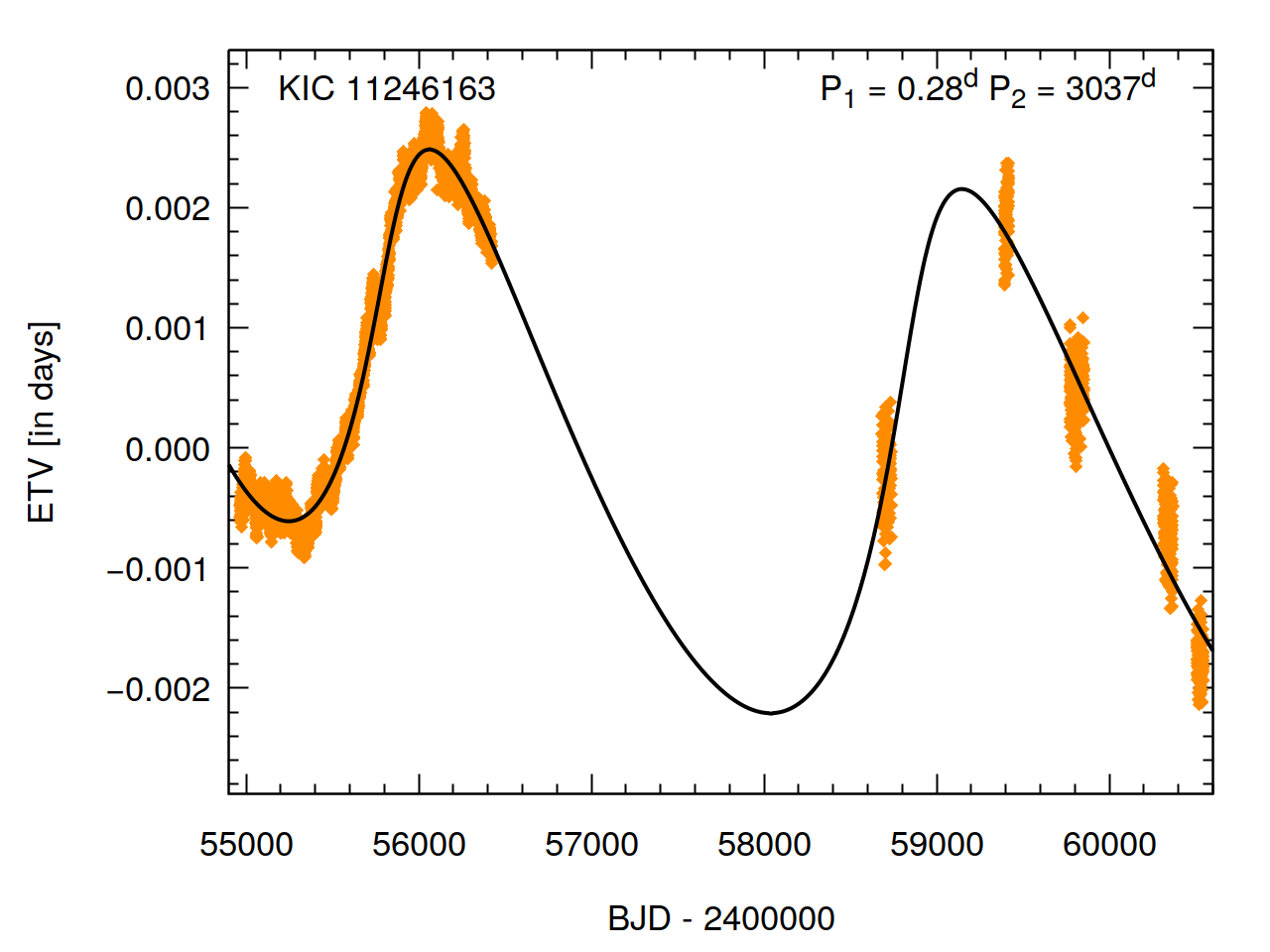}\includegraphics[width=60mm]{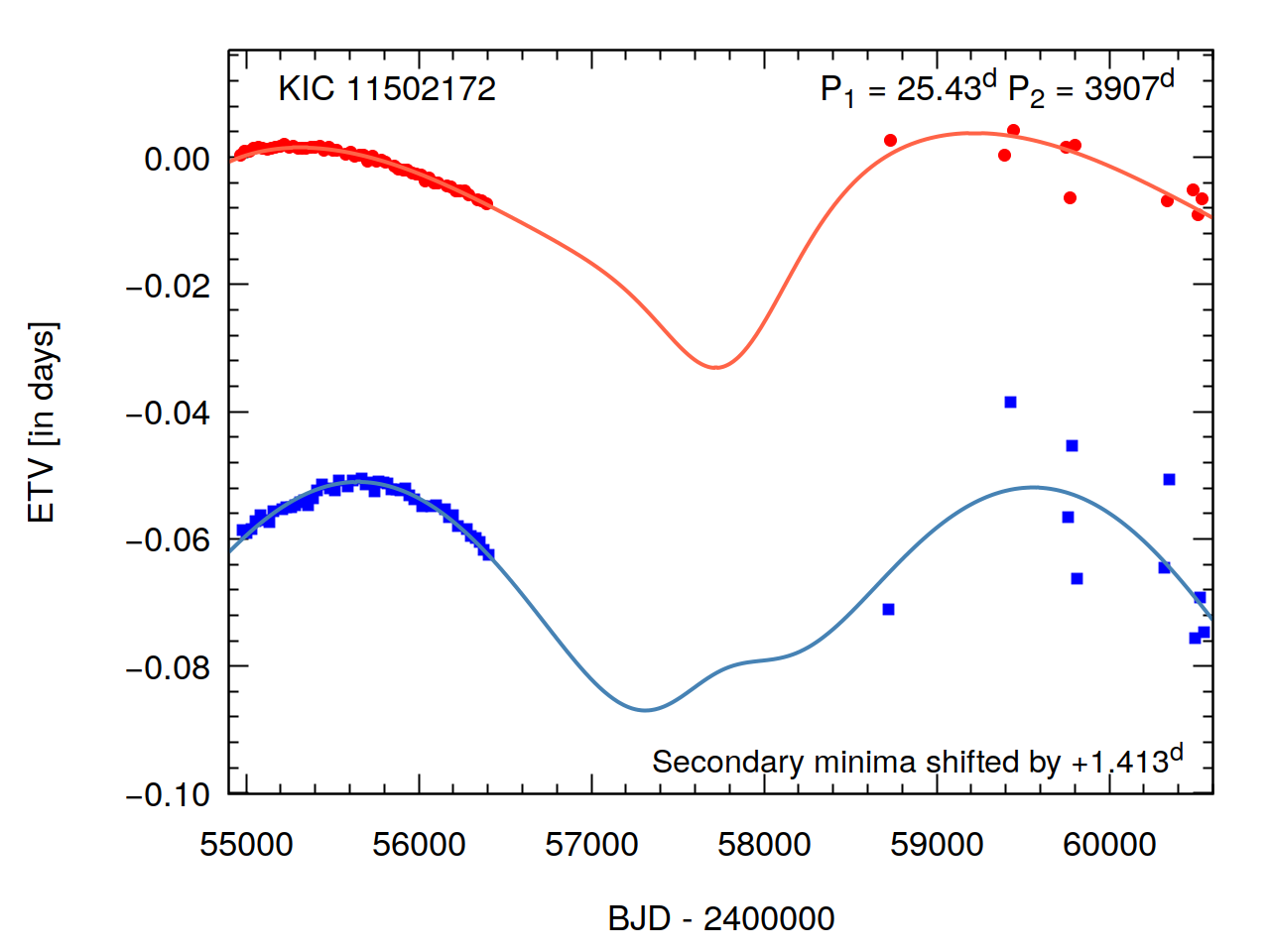}\includegraphics[width=60mm]{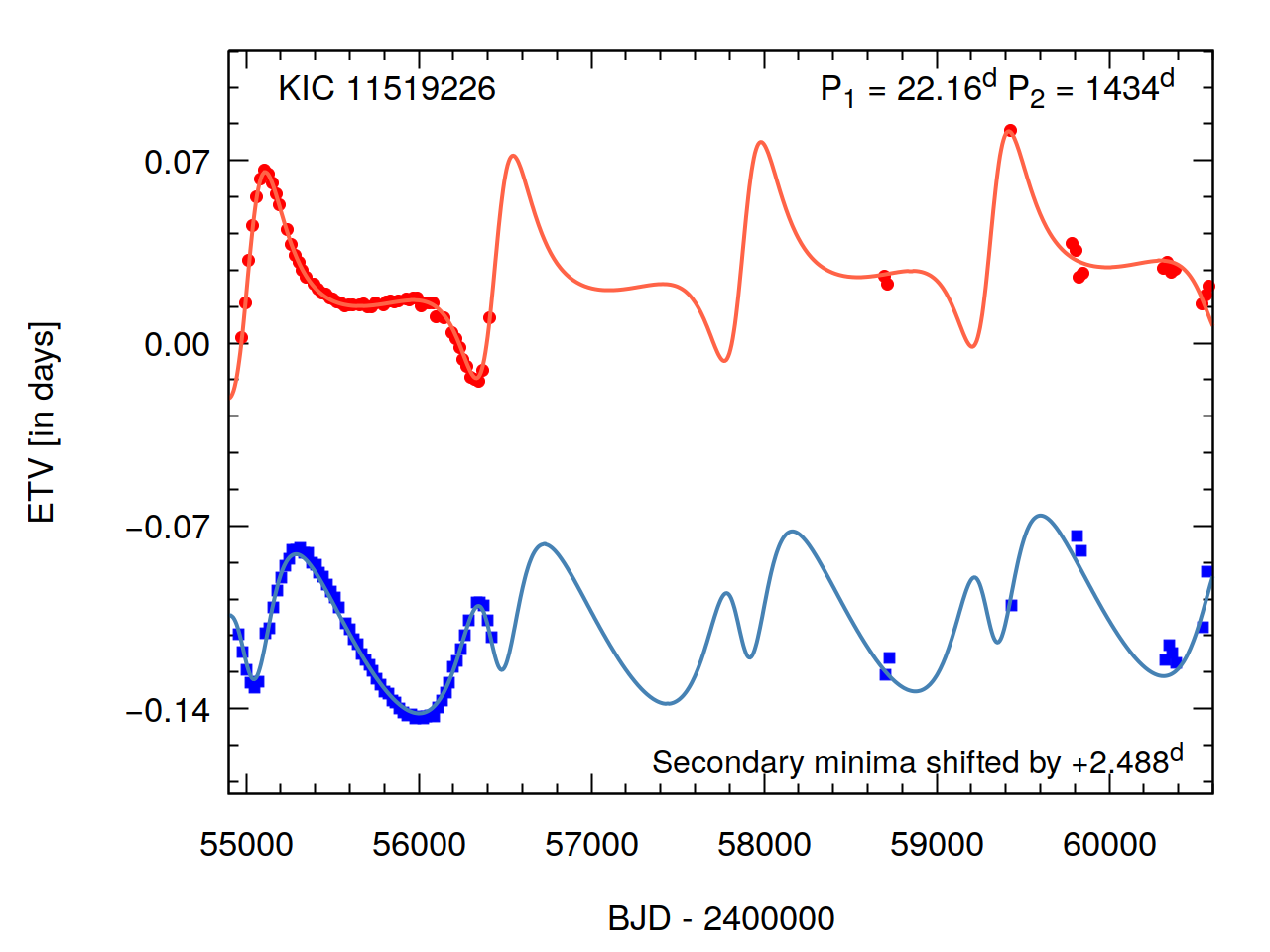}
\includegraphics[width=60mm]{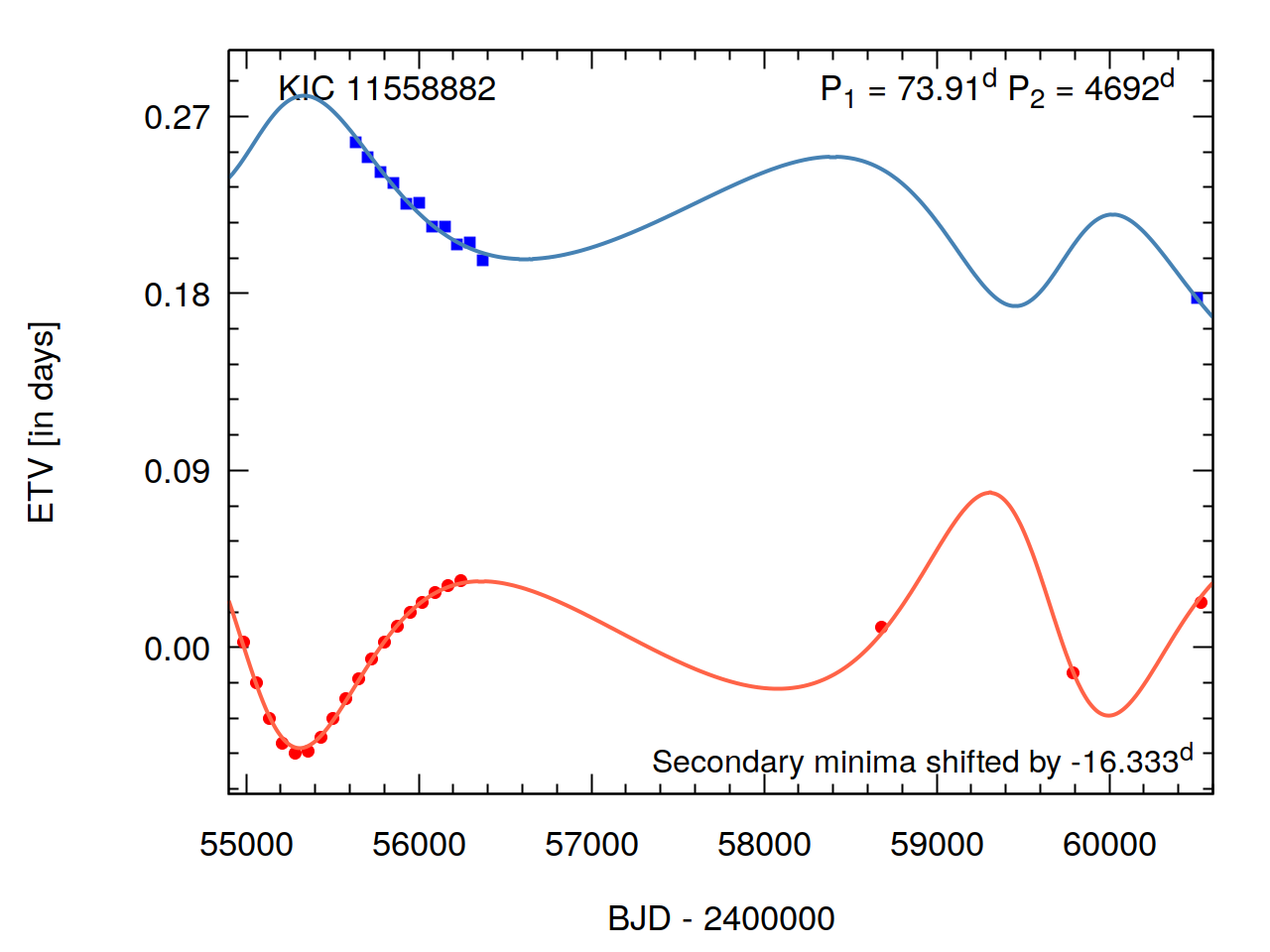}\includegraphics[width=60mm]{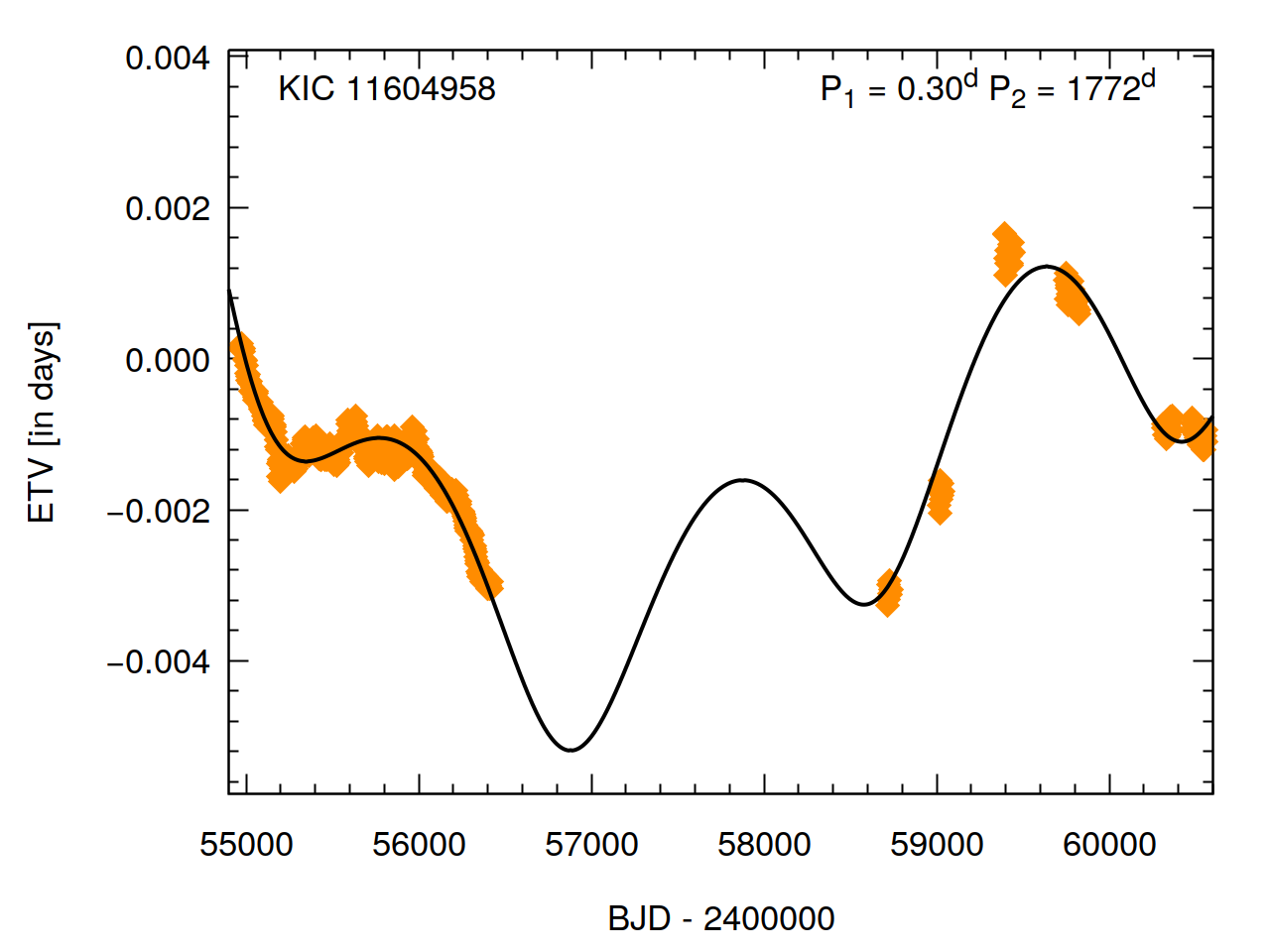}\includegraphics[width=60mm]{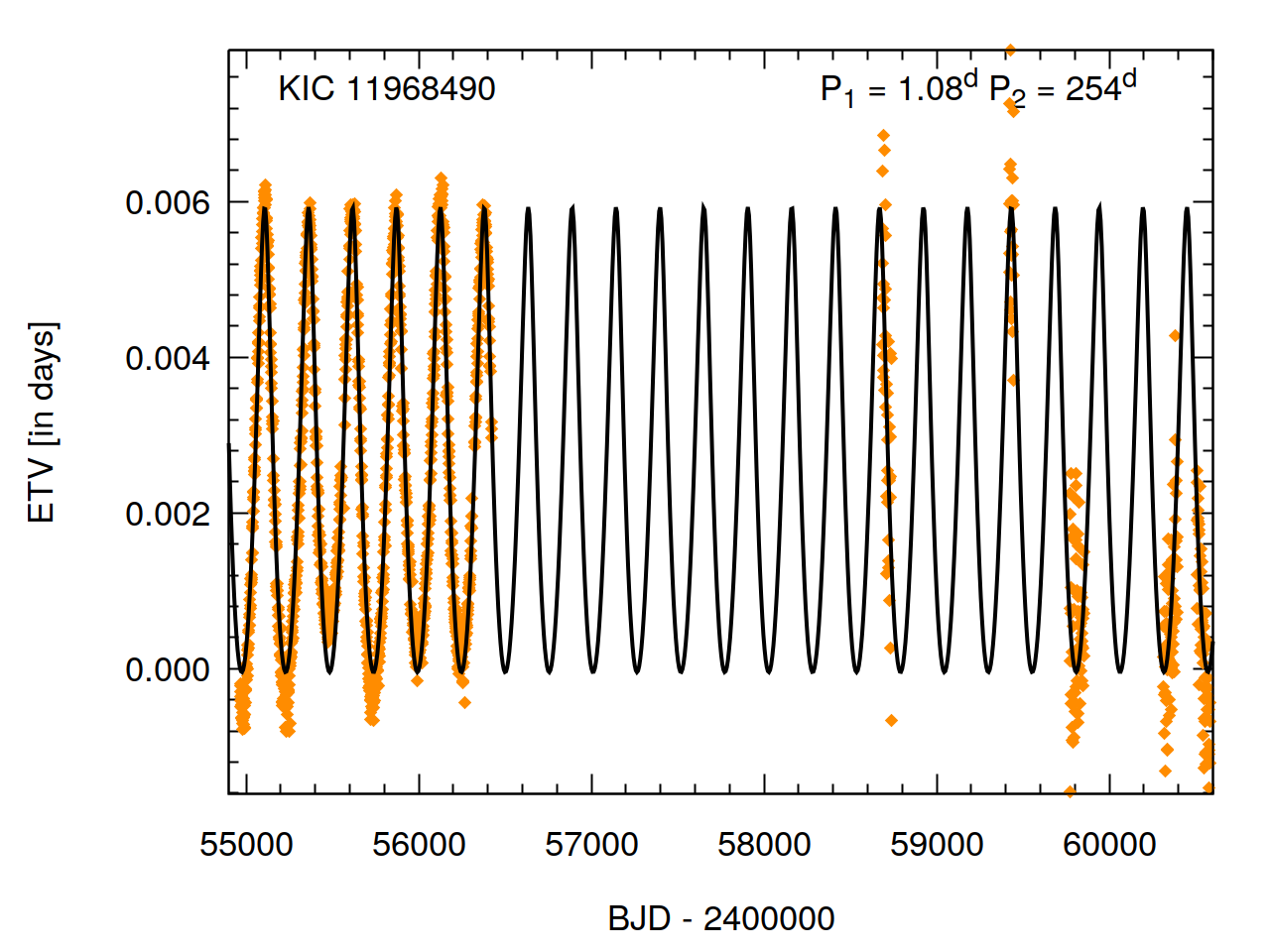}
\includegraphics[width=60mm]{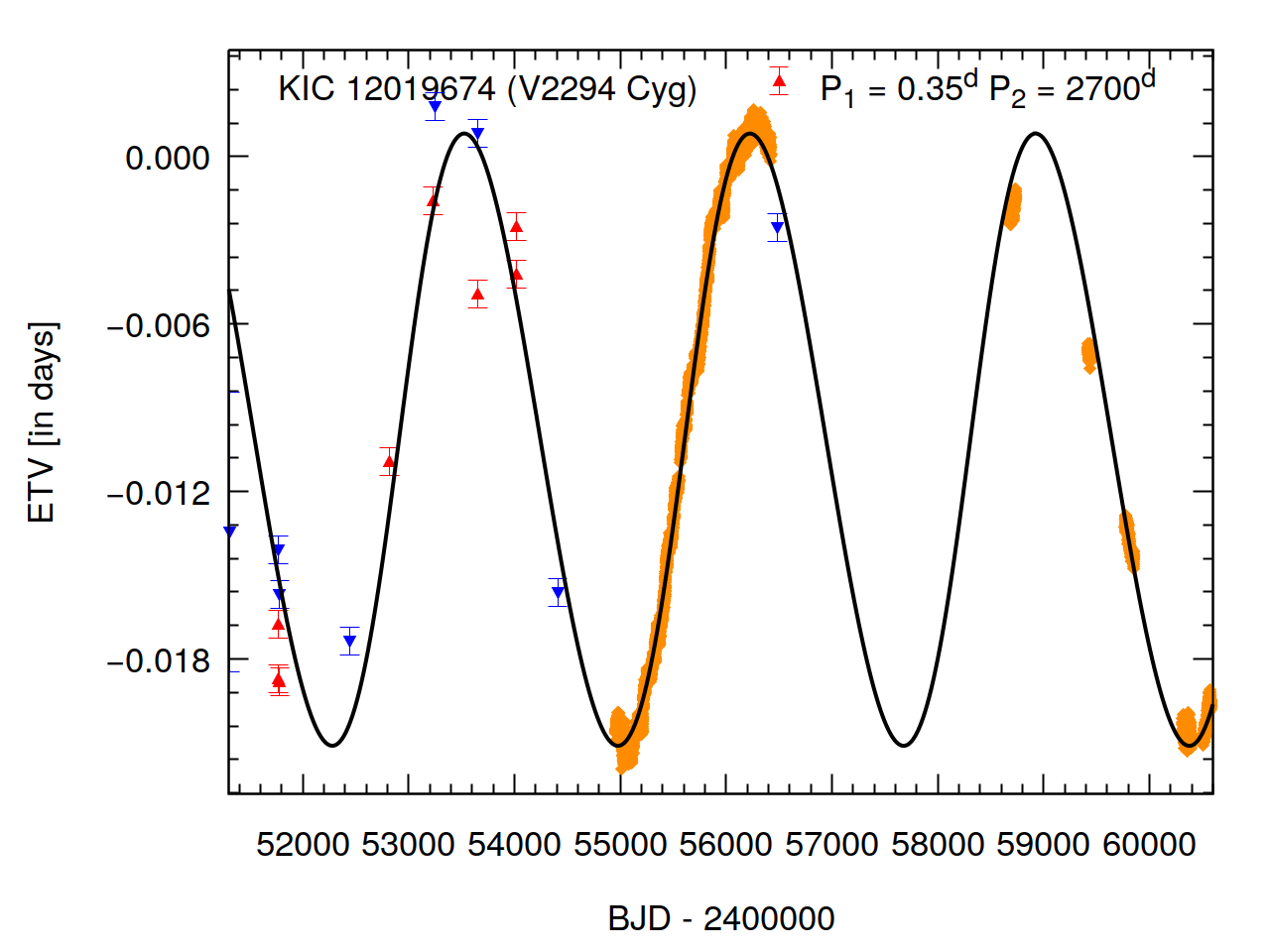}\includegraphics[width=60mm]{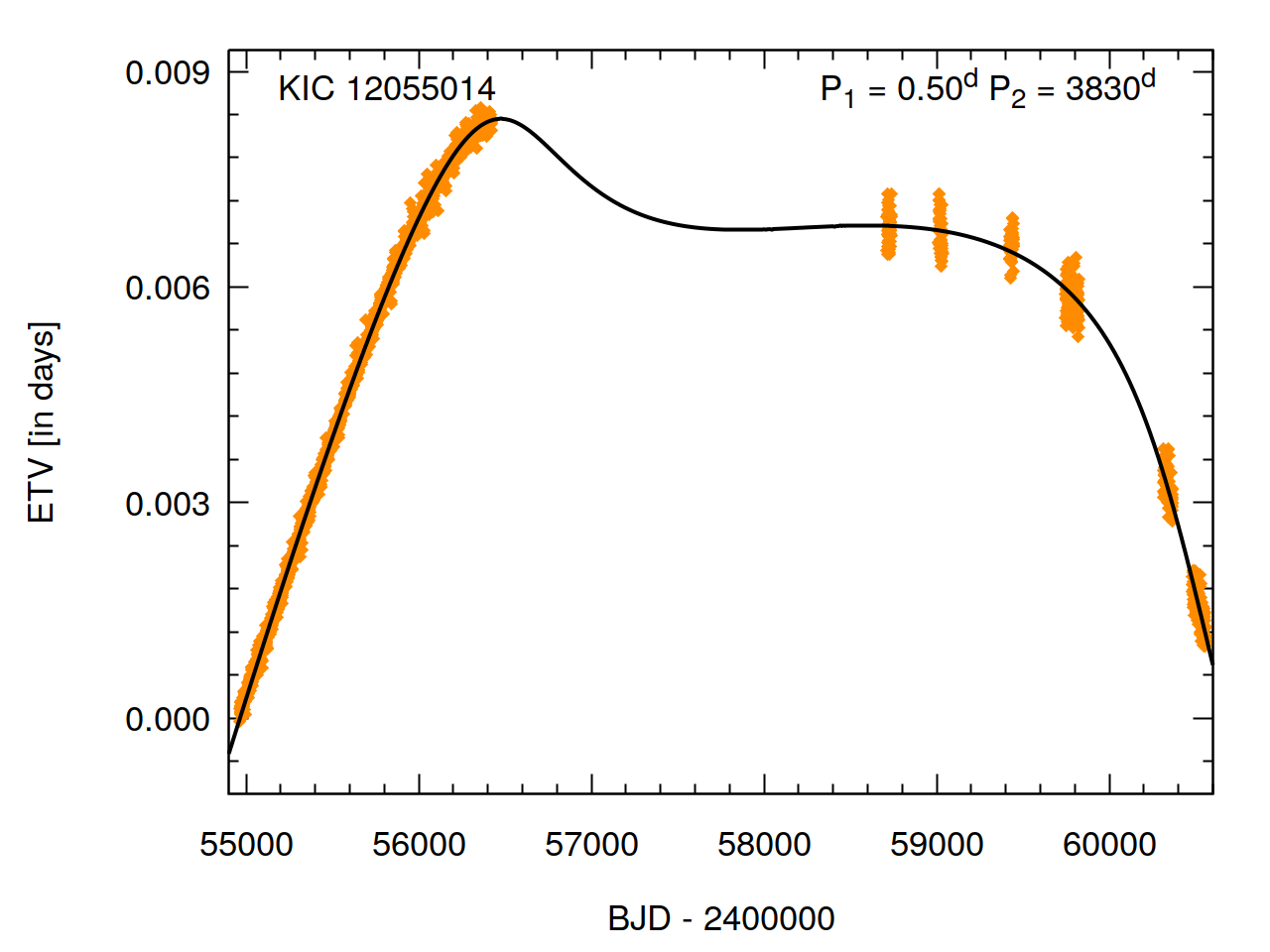}\includegraphics[width=60mm]{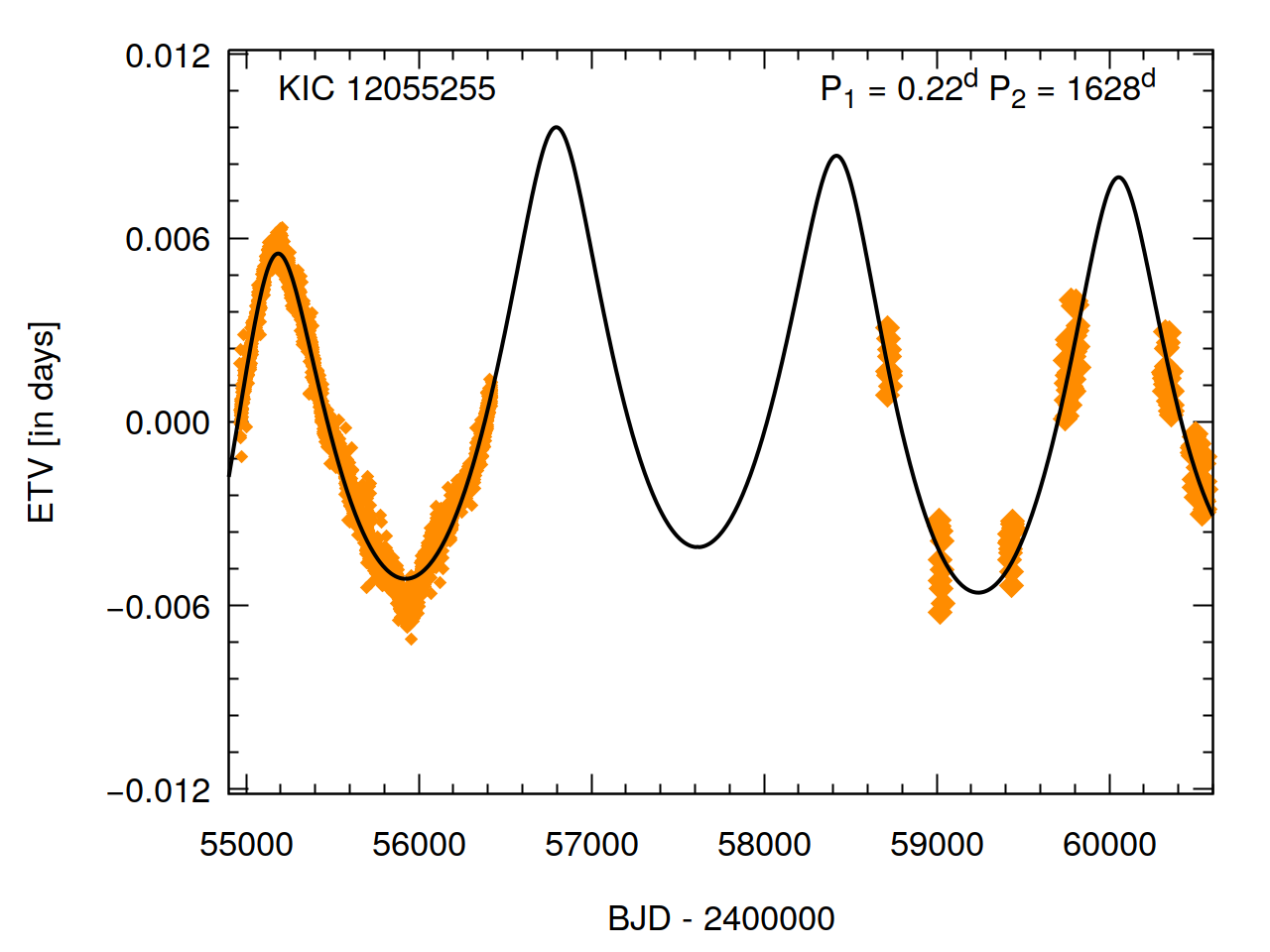}
\caption{continued.}
\end{figure*}

\addtocounter{figure}{-1}

\begin{figure*}
\includegraphics[width=60mm]{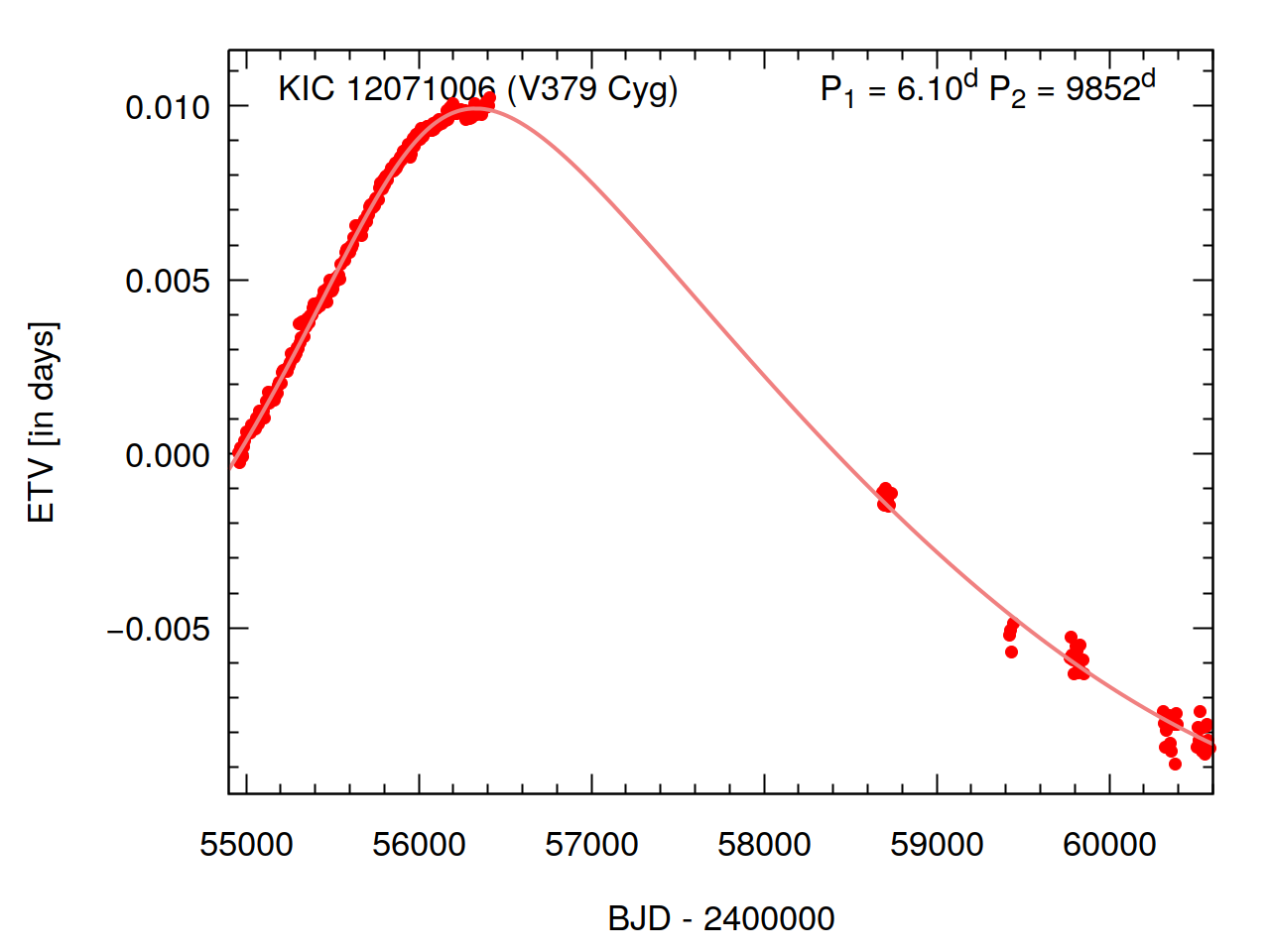}\includegraphics[width=60mm]{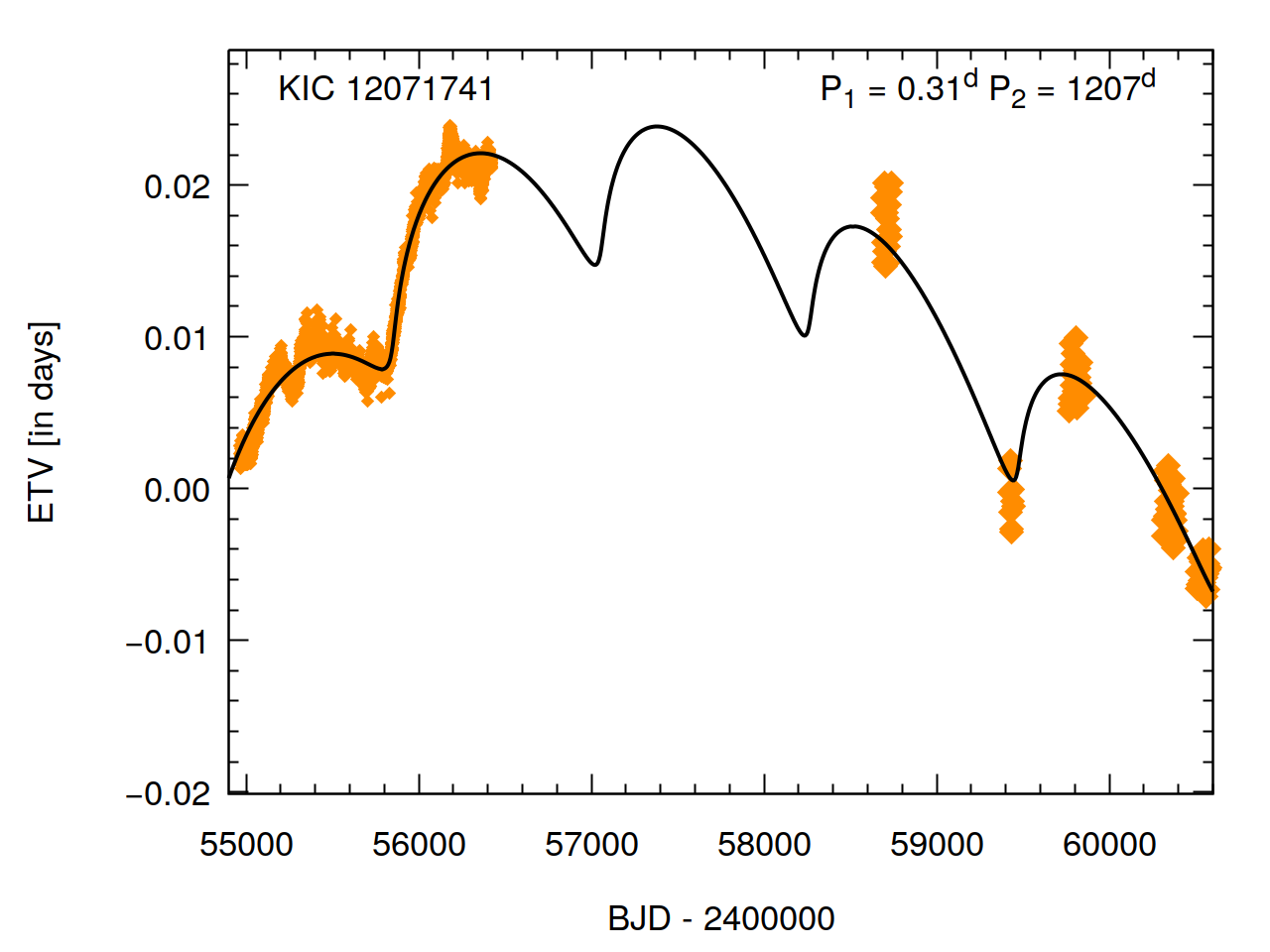}\includegraphics[width=60mm]{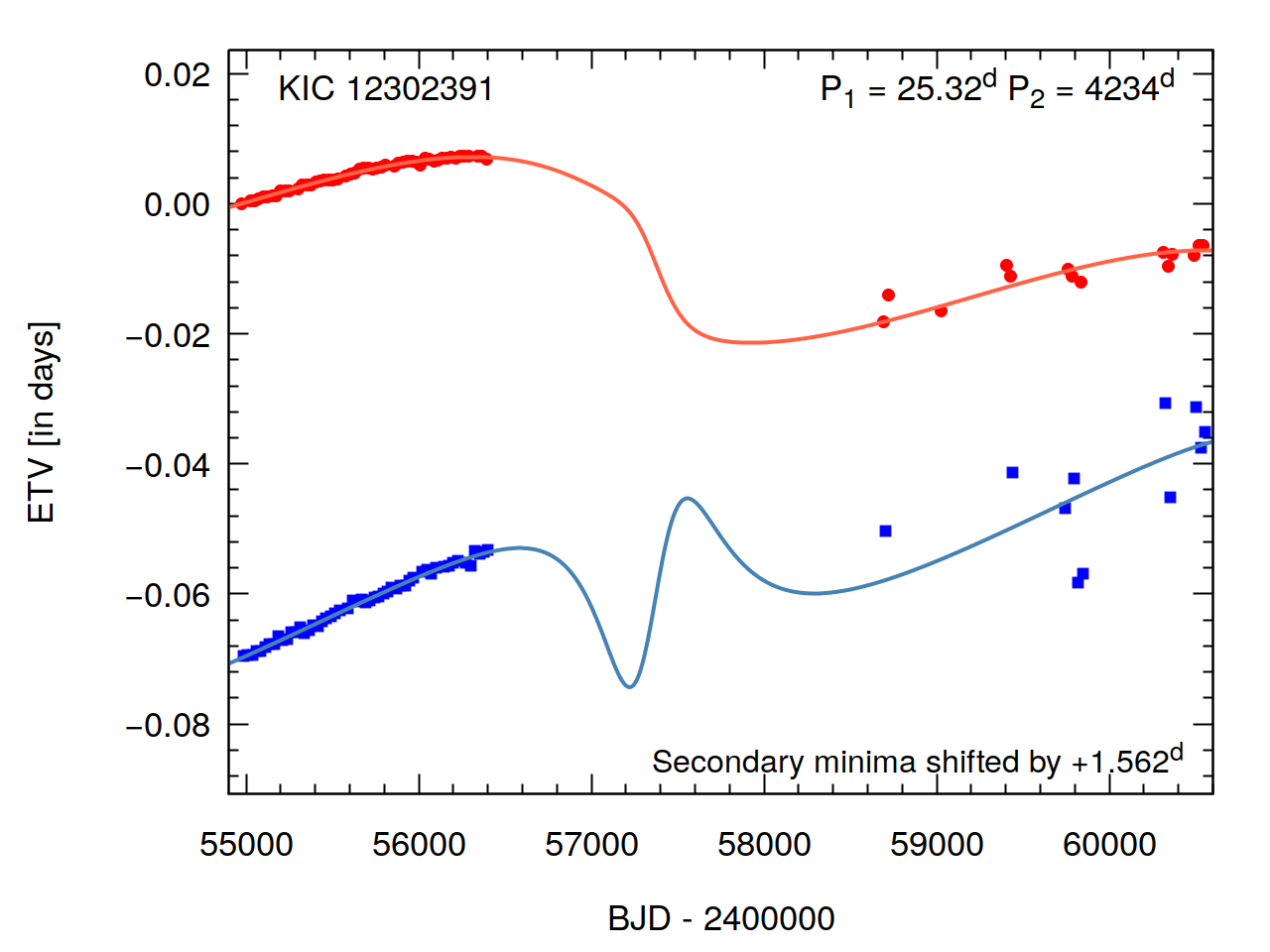}
\includegraphics[width=60mm]{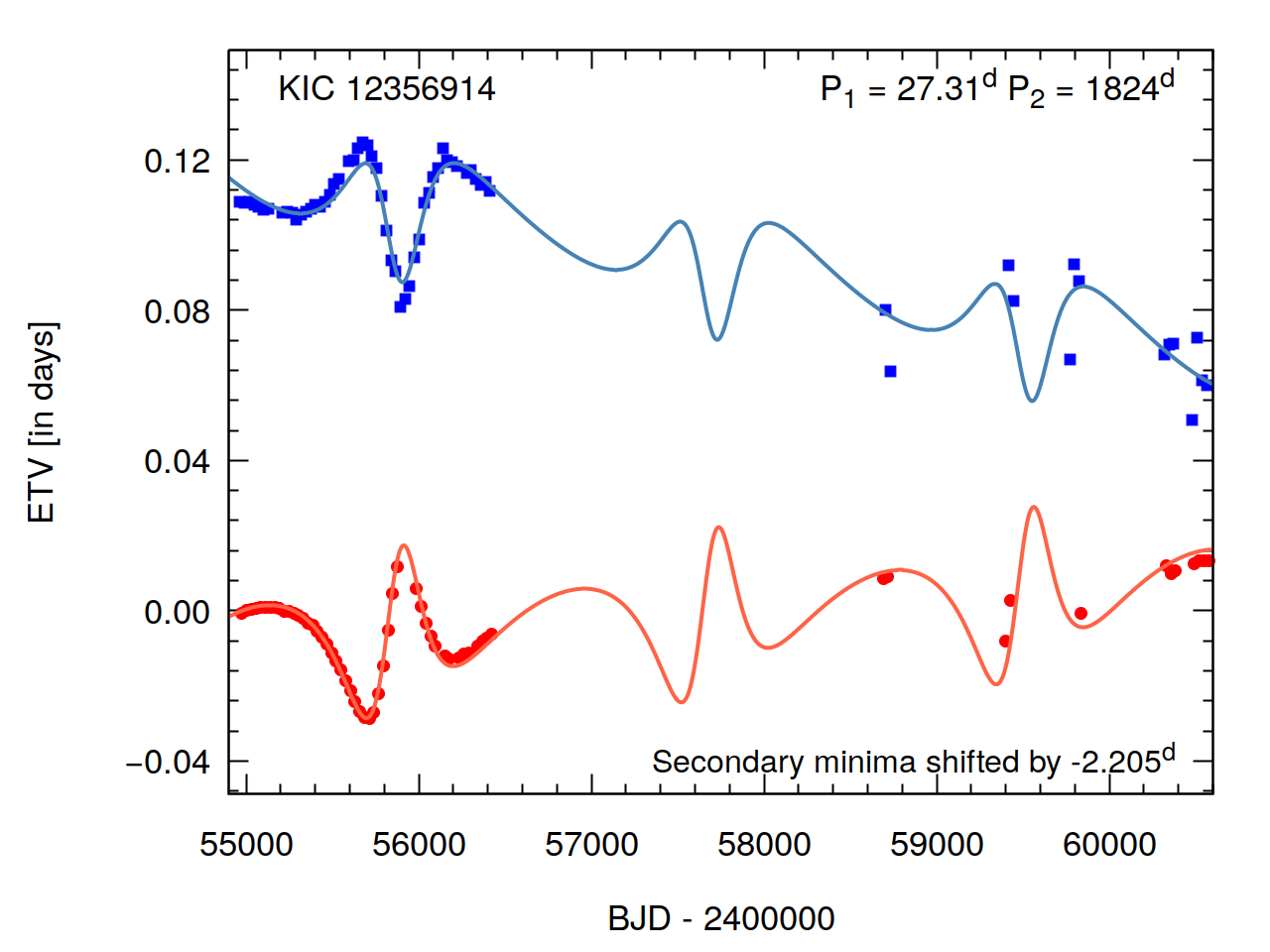}\includegraphics[width=60mm]{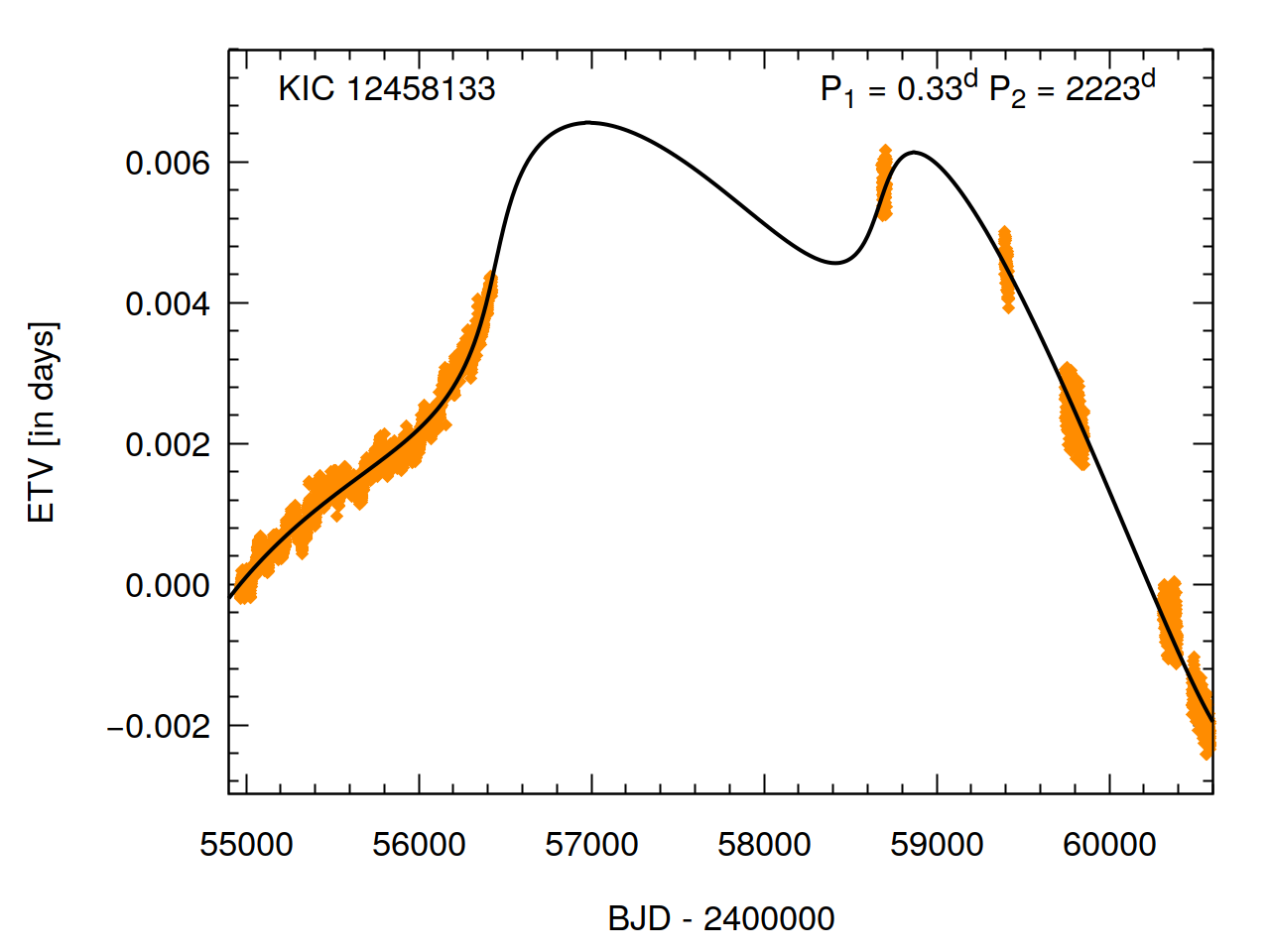}\includegraphics[width=60mm]{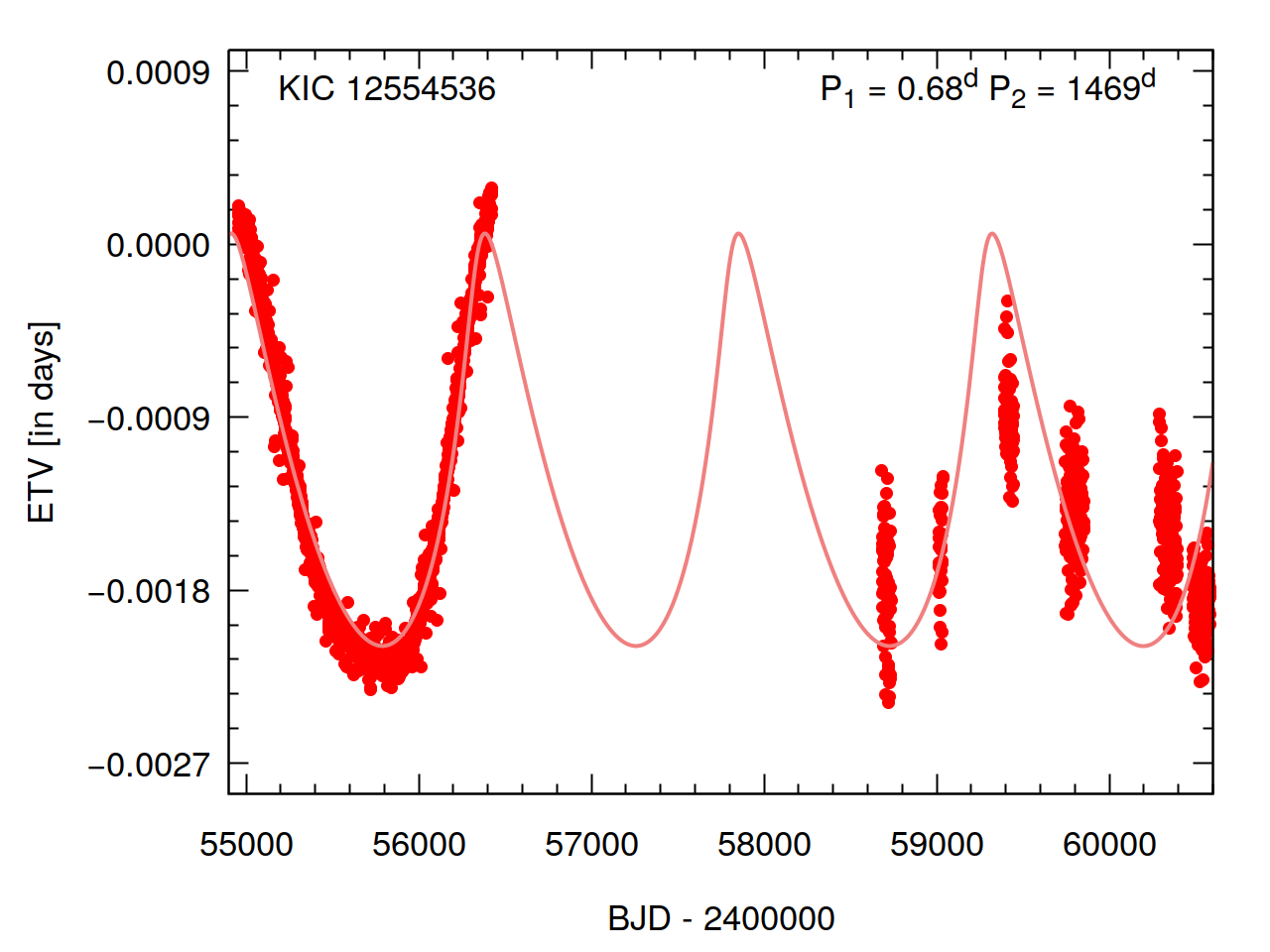}
\caption{continued.}
\end{figure*}

\end{appendix}

\end{document}